\documentclass[twoside]{book}
\usepackage{psfig,fullpage,amssymb,amsfonts,latexsym}
\usepackage{makeidx}
\newtheorem{proposition}{Proposition}[section]
\newtheorem{lemma}{Lemma}[section]
\newtheorem{theorem}{Theorem}[section]
\newtheorem{corollary}{Corollary}[section]
\newtheorem{definition}{Definition}[section]
\newtheorem{rhp}{Riemann-Hilbert Problem}[section]
\newtheorem{conjecture}{Conjecture}[section]
\newcommand{\Lip}{{\rm Lip}}
\newcommand{\op}[1]{{\cal #1}}
\newcommand{\bo}{{\cal O}}
\newenvironment{proof}{\vspace{0.1 in}
\par{\it Proof}.\,\,}{$\Box$\vspace{0.1 in}}
\newenvironment{remark}{\vspace{0.1 in}
\par$\triangleleft$\,\,{\bf Remark:}\,\,}
{$\triangleright$\vspace{0.1 in}}

\makeindex

\author{S. Kamvissis\\University of Patras \and K. T.-R. McLaughlin\\
The University of North Carolina and The University of Arizona \and
P. D. Miller\\ The University of Michigan and Monash University}
\title{Semiclassical Soliton Ensembles for the Focusing Nonlinear
Schr\"odinger Equation} \date{November 16, 2000}
\begin{document} \maketitle

\pagenumbering{roman}

\tableofcontents

\listoffigures

\listoftables

\clearpage

\thispagestyle{plain}
\section*{Abstract}
\addcontentsline{toc}{chapter}{Abstract} We present a new
generalization of the steepest descent method introduced by Deift and
Zhou \cite{DZ93} for matrix Riemann-Hilbert problems and use it to
study the semiclassical limit of the focusing nonlinear Schr\"odinger
equation with real analytic, even, bell-shaped initial data
$\psi(x,0)=A(x)$.  We provide explicit strong locally uniform
asymptotics for a sequence of exact solutions $\psi(x,t)$ corresponding
to initial data that has been modified in an asymptotically small
sense.  We call this sequence of exact solutions a semiclassical
soliton ensemble.  Our asymptotics are valid in regions of the $(x,t)$
plane where a certain scalar complex phase function can be found.  We
characterize this complex phase function directly by a finite-gap
ansatz and also via the critical point theory of a certain functional;
the latter provides the correct generalization of the variational
principle exploited by Lax and Levermore \cite{LL83} in their study of
the zero-dispersion limit of the Korteweg-de Vries equation.

For the special initial data $A(x)=A\,{\rm sech}(x)$, the scattering
data was computed explicitly for all $\hbar$ by Satsuma and Yajima
\cite{SY74}.  It turns out that for this case the modified initial
data we use in general agrees with the true initial data. Thus our
rigorous asymptotics for semiclassical soliton ensembles establish the
semiclassical limit for this initial data.

Using a genus zero ansatz for the complex phase function, we obtain
strong asymptotics of general semiclassical soliton ensembles for
small times independent of $\hbar$ in the form of a rapidly
oscillatory and slowly modulated complex exponential plane wave.  We
show how, with the help of numerical methods, the ansatz can be
verified for finite times up to a phase transition boundary curve in
the $(x,t)$-plane called the primary caustic \cite{MK98}.  Using
qualitative information obtained from the numerics concerning the mode
of failure of the genus zero ansatz at the primary caustic, we apply
perturbation theory to show that a genus two ansatz provides the
correct asymptotic description of the soliton ensemble just beyond the
caustic.  Our analysis shows that the macrostructure in the genus zero
region is governed by the exact solution of the elliptic Whitham
equations, and we obtain formulae solving this ill-posed initial-value
problem in the category of analytic initial data.  For the
Satsuma-Yajima data, our solution of the Whitham equations reproduces
that obtained many years ago by Akhmanov, Sukhorukov, and Khokhlov
\cite{ASK66}, and our rigorous semiclassical analysis places their
formal conclusions on sure footing.

\vspace{0.3 in}

\noindent{\bf
Keywords:} focusing nonlinear Schr\"odinger equation, semiclassical
limit, Riemann-Hilbert problems, minimum capacity problems.

\clearpage

\thispagestyle{plain}
\section*{Acknowledgements}
\addcontentsline{toc}{chapter}{Acknowledgements} The authors joint
work on this problem began when S.~Kamvissis and P.~D.~Miller were
Members in the School of Mathematics at the Institute for Advanced
Study during the 1997--1998 special year in Geometric Partial
Differential Equations organized by Karen Uhlenbeck, and at the same
time, K.~T.-R.~McLaughlin was visiting Princeton University.  We are
all very grateful to these institutions for their support of our
research.

S. Kamvissis acknowledges the support of the Greek General
Secretariat of Research and Technology on project number 97EL16, and
also thanks the Max Planck Institute for Mathematics in the Sciences
in Leipzig for hospitality and support in Fall 1998.
K. T.-R. McLaughlin acknowledges the support of the NSF under
Postdoctoral Fellowship grant number DMS 9508946 and also under grant
number DMS 9970328.  P.~D.~Miller was supported by the NSF under grant
number DMS 9304580 while at the IAS, and was also supported by a Logan
Fellowship while at Monash University.  

The authors also wish to thank the organizers, Pavel Bleher and
Alexander Its, of the special semester in Random Matrix Models and
their Applications held at the Mathematical Sciences Research
Institute in Spring 1999, as well as the Institute itself for
hospitality during our stays there.

Finally, we want to acknowledge several of our colleagues for useful
discussions: Percy Deift, Greg Forest, Arno Kuijlaars, Dave
McLaughlin, John Nuttall, Evguenii Rakhmanov, Herbert Stahl, and Xin
Zhou.  Also we thank Dave Levermore and Nick Ercolani for giving us a
copy of the preprint \cite{EJLM93}.

\clearpage

\pagenumbering{arabic}

\chapter{Introduction and Overview}
\label{sec:introduction}
\section[Background]{Background.}
The initial value problem for the focusing nonlinear Schr\"odinger equation 
\index{focusing nonlinear Schr\"odinger equation} is
\begin{equation}
i\hbar\partial_t\psi + \frac{\hbar^2}{2}\partial_x^2\psi + |\psi|^2\psi = 0\,,
\hspace{0.3 in} \psi(x,0)=\psi_0(x)\,.
\label{eq:IVP}
\end{equation}
We are interested in studying the behavior of solutions of this
initial value problem in the so-called semiclassical
limit\index{semiclassical limit}.  To make this precise, the initial
data is given in the form:
\begin{equation}
\psi_0(x)=A(x)\exp(iS(x)/\hbar)\,,
\label{eq:data}
\end{equation}
where $A(x)$ is a positive real amplitude function that is rapidly
decreasing for large $|x|$, and $S(x)$ is a real phase function that
decays rapidly to constant values for large $|x|$.  Studying the
semiclassical limit means: fix once and for all the functions $A(x)$
and $S(x)$, and then for each sufficiently small value of $\hbar>0$,
solve the initial value problem (\ref{eq:IVP}) subject to the initial
data (\ref{eq:data}), obtaining the solution $\psi(x,t;\hbar)$.
Describe the collection of solutions $\psi(x,t;\hbar)$ in the limit of
$\hbar\downarrow 0$.

The initial-value problem (\ref{eq:IVP}) is a key model in modern
nonlinear optical physics \index{nonlinear optics} and its
increasingly important applications in the telecommunications
\index{telecommunications} industry.  On one hand, it describes the
stationary profiles of high-intensity paraxial beams \index{paraxial
beams} propagating in materials with a nonlinear response, the
so-called Kerr effect\index{Kerr effect}.  This is the realm of {\em
spatial solitons}\index{spatial solitons}, which are envisioned as
stable beams that can form the fundamental components of an
all-optical switching system\index{optical switching}.  In this
context, the semiclassical scaling $\hbar\ll 1$ of (\ref{eq:IVP})
corresponds to the joint paraxial-ray geometrical-optics
\index{geometrical optics} limit in the presence of nonlinear effects.
On the other hand, (\ref{eq:IVP}) also describes the propagation of
(time-dependent) envelope pulses in optical fibers \index{optical
fibers} operating at carrier wavelengths in the anomalous dispersion
regime \index{anomalous dispersion} (usually infrared wavelengths near
1550 nm).  These envelope pulses are known as {\em temporal solitons}
\index{temporal solitons} and are envisioned as stable bits in a 
digital signal traveling through the fiber.  In these fiber-optic
applications, the semiclassical scaling $\hbar\ll 1$ is particularly
appropriate for modeling propagation in certain dispersion-shifted
fibers \index{dispersion-shifted fibers} that are increasingly common.
See \cite{FM98} for a careful discussion of this point leading to a
similarly scaled {\em defocusing} equation; similar arguments with
slightly adjusted parameters can lead to the focusing problem
(\ref{eq:IVP}) just as easily.  Of course in neither of these optical
applications is the small parameter actually Planck's
constant\index{Planck's constant}, but we write it as $\hbar$ in
formal analogy with the quantum-mechanical \index{quantum mechanics}
interpretation of the linear terms in (\ref{eq:IVP}) which also gives
rise to the description of the limit of interest as ``semiclassical''.

The independent variables $x$ and $t$ parametrize the semiclassical
limit, and one certainly does not expect a pointwise asymptotic
description of the solution to be uniform with respect to these
parameters.  The statement of the problem becomes more precise when
one further constrains these parameters.  For example, one might set
$x=X/\hbar$ and $t=T/\hbar$ for $X$ and $T$ fixed as $\hbar\downarrow
0$.  In this limit, several studies have suggested \cite{B96,BK99}
that for initial data with $|S(x)|$ sufficiently large the field
consists of trains of separated solitons, with the remarkable property
that there is a well-defined relationship between the soliton
amplitude and velocity (nonlinear dispersion relation)
\index{nonlinear dispersion relation} that is determined from the
initial functions $A(x)$ and $S(x)$ via the asymptotic distribution of
eigenvalues of the Zakharov-Shabat scattering
problem\index{Zakharov-Shabat scattering problem}.  In general,
solitons can have arbitrary amplitudes and velocities, so the observed
correlation is a direct consequence of the semiclassical limit.

Here, we will be concerned with a different asymptotic
parametrization.  Namely, we consider the sequence of functions
$\psi(x,t;\hbar)$ in a fixed but arbitrary compact set of the $(x,t)$
plane in the limit $\hbar\downarrow 0$.  In this scaling, the large
number of individual solitons present in the initial data are strongly
nonlinearly superposed, and interesting spatio-temporal patterns have
been observed \cite{MK98,BK99}.

This choice of scaling has several features in common with similar
limits studied in other integrable systems, {\em e.g.} the
zero-dispersion limit \index{zero-dispersion limit} of the Korteweg-de
Vries equation \index{Korteweg-de Vries equation} analyzed by Lax and
Levermore \cite{LL83}, the continuum limit of the Toda lattice
\index{Toda lattice} studied by Deift and K.~T.-R.~McLaughlin
\cite{DM98}, and the semiclassical limit of the {\em defocusing}
nonlinear Schr\"odinger equation \index{defocusing nonlinear
Schr\"odinger equation} studied by Jin, Levermore, and D.~McLaughlin
\cite{JLM99}.  In all of these cases, the challenge is to use the
machinery of the inverse-scattering transform
\index{inverse-scattering transform} to prove convergence in some
sense to a complicated asymptotic description that necessarily
consists of two disparate space and time scales.  One scale (the {\em
macrostructure}) \index{macrostructure} is encoded in the initial
data, and the other scale (the {\em microstructure})
\index{microstructure} is introduced by the small parameter (the
dispersion parameter in the Korteweg-de Vries equation, the lattice
spacing in the Toda lattice, and Planck's constant $\hbar$ in the
nonlinear Schr\"odinger equation).

In Lax and Levermore's analysis \index{Lax-Levermore analysis} of the
zero dispersion limit for the Korteweg-de Vries equation \cite{LL83},
a fundamental role was played by an explicit, albeit complicated,
formula for the exact solution of the initial-value problem for
initial data that has been modified in an asymptotically negligible
sense.  This formula directly represents the solution $u(x,t)$ of the
problem in terms of the second logarithmic derivative of a
determinant.  When the determinant is expanded as a sum of principal
minors, the minors are all positive, and the sum is shown to be
asymptotically dominated by its largest term.  This leads directly to
a discrete maximization problem in which the independent variables $x$
and $t$ appear as parameters (discrete because the number of minors is
finite but large when the dispersion parameter is small) that
characterizes the determinant up to a controllable error.  Leading
order asymptotics are obtained by letting the dispersion parameter go
to zero and observing that the discrete maximization problem goes over
into a variational problem \index{variational problem} in a space of
admissible functions.  It turns out that the weak limit of each member
of the whole hierarchy of conserved local densities for the
Korteweg-de Vries equation can be directly expressed in terms of the
solution of the variational problem and its derivatives.

In all of the problems where the method of Lax and Levermore has been
successful, the macrostructure parameters (or equivalently weak limits
of various conserved local densities) have been shown to evolve
locally in space and time as solutions of a hyperbolic system known as
the {\em Whitham equations} \index{Whitham equations} or the {\em
modulation equations}\index{modulation equations}.  The global picture
consists of several regions of the $(x,t)$ plane in each of which the
microstructure is qualitatively uniform and the macrostructure obeys a
system of modulation equations whose size (number of unknowns) is
related to the complexity of the microstructure.  The variational
method of Lax and Levermore amounts to the global analysis showing how
the solutions of the modulation equations are patched together at the
boundaries of these various regions.  By hyperbolicity and the
corresponding local well-posedness of the modulation equations, it
follows that, for example, the small-time behavior (sufficiently
small, but independent of the size of the limit parameter) of the
limit is connected with prescribed initial data in a stable fashion.

The modulation equations may be derived formally, without reference to
initial data.  For the focusing nonlinear Schr\"odinger equation,
these quasilinear equations are {\em elliptic} \cite{FL86}, which
makes the Cauchy initial-value problem for them ill-posed in common
spaces.  To illustrate this ill-posedness for the Whitham equations in
their simplest version (genus zero), one might make the assumption
that the microstructure in the solution of (\ref{eq:IVP}) resembles
the modulated rapid oscillations present in the initial data.  That
is, one could suppose that for some order one time the solution can be
represented in the form
\begin{equation}
\psi(x,t)=A(x,t)\exp(S(x,t)/\hbar)\,,
\end{equation}
where $A(x,0)=A(x)$ and $S(x,0)=S(x)$.  Setting $\rho(x,t)=A(x,t)^2$ and
$\mu(x,t)=A(x,t)^2\partial_xS(x,t)$, one finds that the initial-value problem
(\ref{eq:IVP}) implies
\begin{equation}
\partial_t\rho + \partial_x\mu = 0\,,\hspace{0.2 in}
\partial_t\mu + \partial_x\left(\frac{\mu^2}{\rho}-\frac{\rho^2}{2}
\right) = \frac{\hbar^2}{4}\partial_x(\rho\partial_x^2\log(\rho))\,,
\label{eq:EulerExact}
\end{equation}
with initial data $\rho(x,0)=A(x)^2$ and $\mu(x,0)=A(x)^2S'(x)$.  The
modulation equations corresponding to our assumption about the
microstructure are obtained by simply neglecting the terms that are
formally order $\hbar^2$ in these equations.  That is, one supposes
that for some finite time $\rho(x,t)$ and $\mu(x,t)$ are uniformly
close, respectively, to functions $\overline{\rho}(x,t)$ and
$\overline{\mu}(x,t)$ as $\hbar\downarrow 0$, where these latter two functions
solve the system
\begin{equation}
\partial_t\overline{\rho} + \partial_x\overline{\mu} = 0\,,\hspace{0.2 in}
\partial_t\overline{\mu} + \partial_x\left(\frac{\overline{\mu}^2}
{\overline{\rho}}-\frac{\overline{\rho}^2}{2}
\right) = 0\,,
\label{eq:Euler}
\end{equation}
with initial data $\overline{\rho}(x,0)=A(x)^2$ and
$\overline{\mu}(x,0)=A(x)^2S'(x)$.  This is a quasilinear nonlinear
system, and it is easy to check that it is of elliptic type; that is,
the characteristic velocities $\overline{\mu}/\overline{\rho}\pm
i\sqrt{\overline{\rho}}$ are complex at every point where
$\overline{\rho}$ is nonzero.  This implies that the Cauchy problem
posed here for the modulation equations is ill-posed\index{ill-posed}.  

This fact immediately makes the interpretation of the semiclassical
limit of the initial-value problem (\ref{eq:IVP}) complicated; even if
it turns out that one can prove convergence to the solutions of the
modulation equations for some initial data, it is not clear that one
can deduce anything at all about the asymptotics for ``nearby''
initial data.  In this sense, the formal semiclassical limit of
(\ref{eq:IVP}) is very unstable\index{unstable}.

One feature that both the hyperbolic and elliptic modulation equations
have in common is the possibility of singularities that develop in
finite time from smooth initial data.  This singularity formation
\index{singularity formation} seems physically correct in the context 
of spatial optical solitons, where the Kerr effect has been known for
some time to lead to self-focusing \index{self-focusing} of light
beams, and in two transverse dimensions (the independent variable
$x$), to the total collapse \index{collapse} of the beam in finite
propagation distance (the independent variable $t$).  As long ago as
1966, this led Akhmanov, Sukhorukov, and Khokhlov \cite{ASK66} to
propose a certain exact solution of the modulation equations
(\ref{eq:Euler}) as a model for the self-focusing phenomenon in one
transverse dimension.  They did not try to solve any initial-value
problem for these equations; indeed they were clearly aware of the
ellipticity of the system (\ref{eq:Euler}) and the coincident
ill-posedness of its Cauchy problem.  Rather, they introduced a clever
change of variables (some insight into their possible reasoning was
proposed by Whitham
\cite{W74}) and obtained a set of two real equations implicitly
defining two real unknowns as functions of $x$ and $t$.  After the
fact, they noted that their solution matched onto the initial data
$A(x)=A\,{\rm sech}(x)$ and $S(x)\equiv 0$.  The original paper of
Akhmanov, Sukhorukov, and Khokhlov contains drawings of the solution
at various times up to the formation of a finite-amplitude singularity
({\em i.e.} the singularity forms in the derivatives) at the time
$t=t_{\rm crit}=1/(2A)$.  The authors even plotted their solution
beyond the singularity, showing the onset of multivaluedness.  They
understood that the model solution cannot possibly be valid beyond the
singularity, and in the physical context of interest in their study,
ascribed this as much to the breakdown of the paraxial approximation
\index{paraxial approximation} leading to the nonlinear Schr\"odinger 
equation (\ref{eq:IVP}) as a beam propagation \index{beam propagation}
model in the first place as to the failure of the formal geometrical
optics \index{geometrical optics} (semiclassical) limit for
(\ref{eq:IVP}).

As is the case in all of the problems for which the method of Lax and
Levermore has been successful, careful analysis of the semiclassical
limit $\hbar\downarrow 0$ for (\ref{eq:IVP}) is possible in principle
because the problem can be solved for each $\hbar$ by the
inverse-scattering transform\index{inverse-scattering transform}, as
was first shown by Zakharov and Shabat \cite{ZS72}.  The small
parameter necessarily enters the problem both in the
forward-scattering \index{forward-scattering} step and in the
inverse-scattering \index{inverse-scattering} step.  Significantly,
the analysis of the semiclassical limit for (\ref{eq:IVP}) is
frustrated in both steps.  In the forward-scattering step, the
difficulties are related to the nonselfadjointness
\index{nonselfadjointness} of the scattering problem associated with
(\ref{eq:IVP}).  By contrast, in each of the cases mentioned above
where calculations of this type were successfully carried out, the
associated scattering problem is selfadjoint.  In the
inverse-scattering step, the difficulties are related to the limit
being attained by a kind of furious cancellation in which no single
term in the expansion of the solution is apparently dominant.  In
fact, in Zakharov and Shabat's paper \cite{ZS72} there appears an
explicit formula for the function $\rho(x,t)$ solving
(\ref{eq:EulerExact}) that is qualitatively very similar to that
solving the Korteweg-de Vries equation \index{Korteweg-de Vries
equation} and taken as the starting point in Lax and Levermore's
analysis.  When $t=0$, this formula has all of the properties required
by the Lax-Levermore theory.  Namely the determinant can be expanded
as a sum of positive terms, which is controlled by its largest term as
$\hbar\downarrow 0$.  This calculation is carried out in the paper of
Ercolani, Jin, Levermore, and MacEvoy \cite{EJLM93}.  But when $t$ is
fixed at any nonzero value, the principal minors lose their positive
definiteness, and it can no longer be proved that the sum is dominated
by its largest term.  If the weak limit exists, then all that can be
said from this approach is that it arises out of subtle cancellation.
In particular, from this point of view it appears that there is no
obvious variational principle characterizing the limit.

\section[Approach and Summary of Results]{Approach and summary of results.}
This paper is primarily concerned with the semiclassical analysis of
the inverse-scattering \index{inverse-scattering} step.  For
simplicity, we restrict attention from the start to the case of
initial data that satisfy $S(x)\equiv 0$.  In this case it was
observed already in Zakharov and Shabat's paper \cite{ZS72} that while
not strictly selfadjoint for any $\hbar>0$ the scattering problem
formally goes over into a semiclassically scaled selfadjoint linear
Schr\"odinger operator \index{Schr\"odinger operator} in the limit
$\hbar\downarrow 0$.  In
\cite{EJLM93}, this observation was exploited to propose WKB formulae
\index{WKB formulae} that were subsequently used to study the 
zero-dispersion limit of the modified Korteweg-de Vries
equation\index{modified Korteweg-de Vries equation}, an equation
associated with the same scattering problem as (\ref{eq:IVP}), but
whose inverse-scattering \index{inverse-scattering} step is more
straightforward because there is no cancellation of the type mentioned
above (as pointed out above, this cancellation is also absent for the
focusing nonlinear Schr\"odinger problem when $t=0$, and the
calculations in
\cite{EJLM93} hold in this case as well).  The WKB approximation
amounts to the neglect of the reflection coefficient \index{reflection
coefficient} and the replacement of the true eigenvalues with a
sequence of purely imaginary numbers that are obtained from an
explicit Bohr-Sommerfeld type quantization rule\index{Bohr-Sommerfeld
quantization rule}.  These WKB formulae have not to date been
rigorously established; their justification in
\cite{EJLM93} rests upon the fact that they reproduce the exact
initial data when $t$ is set to zero in the inverse-scattering step.
There is, however, one function $A(x)$ for which all of the exact
scattering data is known (assuming $S(x)\equiv 0$) exactly:
$A(x)=A\,{\rm sech}(x)$.  The spectrum corresponding to this potential
in the nonselfadjoint Zakharov-Shabat scattering problem
\index{Zakharov-Shabat scattering problem} was computed exactly for
all $\hbar$ by Satsuma and Yajima \cite{SY74} and published in 1974.
At face value this is a remarkable coincidence: the same initial data
for which Akhmanov, Sukhorukov, and Khokhlov found (after the fact!)
that they had an exact solution of the modulation equations turns out
to be data for which the forward-scattering problem was later shown to
be exactly solvable for all $\hbar$.  Some additional special cases of
potentials where the the spectrum can be obtained exactly for all
$\hbar$, including some cases with $S(x)\not\equiv 0$, have been
recently found by Tovbis and Venakides \cite{TV00}.

It turns out that the exact scattering data for the special initial
condition $\psi_0(x)=A\,{\rm sech}(x)$ coincides with the formal WKB
approximation to the scattering data, as long as one restricts
attention to a particular sequence of positive values of
$\hbar\in\{\hbar_N\}$ converging to zero.  For these special values of
$\hbar$, the initial data is exactly reflectionless, there are exactly
$N$ eigenvalues all purely imaginary, and also the distance between
the most excited state (the eigenvalue with the smallest magnitude)
and the continuous spectrum is exactly half of the distance between
each adjacent pair of eigenvalues.  In particular, for
$\hbar=\hbar_N$, there is no error incurred in reconstructing the
corresponding solution of (\ref{eq:IVP}) using inverse-scattering
theory {\em without reflection coefficient}; the true solution for
these values of $\hbar$ is a pure ensemble of $N$ solitons.

In this paper, we will develop a method that yields detailed strong
asymptotics for the inverse-scattering problem corresponding to the
scattering data briefly described above.  Since this scattering data
is the true scattering data corresponding to the Satsuma-Yajima
potential\index{Satsuma-Yajima potential}, our results imply rigorous
asymptotics for the corresponding initial-value problem
(\ref{eq:IVP}).  But since the scattering data for this case agrees
with its WKB approximation, we prefer to approach the problem from the
more general perspective of computing rigorous asymptotics for the
inverse problem corresponding to a general family of WKB scattering
data.  Thus, our approach to the semiclassical limit for initial-value
problem (\ref{eq:IVP}) for quite general data satisfying $S(x)\equiv
0$ is essentially the familiar step of introducing modified
reflectionless initial data \index{modified initial data} whose
scattering data is that predicted by the formal WKB approximation.
This sort of modification was the first step in the pioneering work of
Lax and Levermore
\cite{LL83}.  Of course, for the Satsuma-Yajima initial data, no
modification is necessary as long as $\hbar\in\{\hbar_N\}$.


The main idea that allows our analysis of the inverse-scattering
problem to proceed for $t\neq 0$ where the Lax-Levermore method
\index{Lax-Levermore analysis} fails is to avoid the direct connection
of the discrete scattering data with the solution of the problem via
an explicit determinant formula and instead to introduce an
intermediate object, namely an appropriately normalized eigenfunction
of the Zakharov-Shabat scattering problem\index{Zakharov-Shabat
scattering problem}.  In general, this eigenfunction satisfies a
certain matrix Riemann-Hilbert problem \index{Riemann-Hilbert problem}
with poles encoding the discrete spectrum and a jump on the real axis
of the eigenvalue corresponding to the reflection coefficient
\index{reflection coefficient} on the continuous 
spectrum\index{continuous spectrum}.  The
solution of the nonlinear Schr\"odinger equation is in turn obtained
from the solution of this Riemann-Hilbert problem.  This is the
essential content of inverse-scattering theory
\cite{FT87}.  While it of course turns out that in the reflectionless
case the Riemann-Hilbert problem may be explicitly solved in terms of
meromorphic functions and ratios of determinants, leading to the
formula that is the starting point for Lax-Levermore type analysis,
there is some advantage to ignoring this explicit solution and instead
trying to obtain uniform asymptotics for the eigenfunction that is the
solution of the Riemann-Hilbert problem.  Only in studying this
intermediate problem do we recover a variational principle that is a
generalization of the one from Lax and Levermore's method.

The method we develop in this paper to study the asymptotic behavior
of the eigenfunction generalizes the steepest descent method
\index{steepest descent method} for matrix Riemann-Hilbert problems
first proposed by Deift and Zhou in
\cite{DZ93}, and subsequently developed and further applied in several
papers \cite{DVZ94,DZ95}.  The generalization of the steepest descent
method that we will present below has its basic features in common
with the recent application of the method to the Korteweg-de Vries
equation in \cite{DVZ97}, with recent applications in the theory of
orthogonal polynomials \index{orthogonal polynomials} and random matrices
\index{random matrices}\cite{DKMVZ97,DKMVZ98A,DKMVZ98B}, and also with 
some applications to long-time asymptotics \index{long-time
asymptotics} for soliton-free initial data in the focusing nonlinear
Schr\"odinger equation \cite{K95,K96}.  These latter papers make use
of an idea that was first introduced in
\cite{DVZ94} --- using the special choice of a complex phase function
\index{complex phase function} to enable the asymptotic reduction of 
the Riemann-Hilbert problem to a simple form.  Our work generalizes
this approach because it turns out that an appropriate complex phase
function typically does not exist at all relative to a given contour
in the complex plane, unless this contour satisfies some additional
conditions.  In fact, we will show that the existence of an
appropriate complex phase function {\em selects portions of the
contour on which the Riemann-Hilbert problem should be posed to begin
with}.  In this sense, the generalization of the method proposed in
\cite{DVZ94} that we present here further develops the analogy with the
classical asymptotic method of steepest descent; the problem must be
solved on a particular contour in the complex plane.  In problems
previously treated by the steepest descents method of Deift, Zhou,
{\em et. al.}, the problem of finding this special contour has simply
not arisen because there is an obvious contour, often implied by the
selfadjointness of a related scattering problem, for which the
additional conditions that select the contour are {\em automatically}
satisfied.  The specification of this special contour can be given a
variational interpretation that is the correct generalization of the
Lax-Levermore variational principle\index{Lax-Levermore
analysis!generalization of}.

Among our primary results are:
\begin{enumerate}
\item 
Strong, leading-order semiclassical asymptotics \index{strong
asymptotics} for solutions of the focusing nonlinear Schr\"odinger
equation corresponding to sequences of initial data whose spectral
data is reflectionless \index{reflectionless} and has discrete
spectrum \index{discrete spectrum} obtained from a Bohr-Sommerfeld
quantization rule\index{Bohr-Sommerfeld quantization rule}.  These
asymptotics are valid even after wave breaking\index{wave breaking},
and come with a rigorous error bound.  The explicit model we obtain
--- to which the semiclassical solutions are asymptotically close
pointwise in $x$ and $t$ --- displays qualitatively different behavior
before and after wave breaking, and in particular exhibits violent
oscillations after breaking confirming phenomena that have been
observed in numerical experiments.
\item
Formulae explicitly involving the initial data that solve the elliptic
Whitham modulation equations.  These formulae consequently provide the
complete solution to the initial-value problem for the Whitham
equations \index{Whitham equations!solution of the initial-value
problem for} in the category of analytic initial data.
\item
The characterization of the caustic curves \index{caustics} in the
$(x,t)$-plane where the nature of the microstructure changes suddenly.
We also provide what amount to ``connection formulae''
\index{connection formulae} describing the phase transition 
\index{phase transition} that occurs at the caustic.  
In particular our analysis shows that at first
wave breaking there is a spontaneous transition from fields with
smooth amplitude (genus zero) to oscillatory fields with intermittent
concentrations in amplitude (genus two).
\item
A significant extension of the steepest descents method for asymptotic
analysis of Riemann-Hilbert problems introduced by Deift and Zhou.
For problems with analytic jump matrices, we show how the freedom of
placement of the jump contour in the complex plane can be
systematically exploited to asymptotically reduce the norms of the
singular integral operators involved in the solution of the
Riemann-Hilbert problem.  Ultimately this expresses the solution as an
explicit contribution modified by a Neumann series involving small
bounded operators.
\item
A new generalization of Riemann-Hilbert methods allowing the analysis
of inverse-scattering problems in which there is an asymptotic
accumulation of an unbounded number of solitons.
\item
An interpretation of our asymptotic solution of the Riemann-Hilbert
problem in terms of a new variational principle that generalizes the
quadratic programming problem of Lax and Levermore, and explicitly
encodes the contour selection mechanism.  This interpretation also
makes a strong connection with approximation theory where variational
problems of the same type appear when one tries to find sets of
minimal weighted Green's capacity \index{minimal weighted Green's
capacity} in the plane.
\item
A proof that the systematic selection of an appropriate contour is
guaranteed to succeed under certain generic conditions.  Finding
correct the contour amounts to solving a problem of geometric function
theory\index{geometric function theory}, namely the construction of
``trajectories of quadratic differentials''\index{quadratic
differentials!trajectories of}.  We show that the existence of such
trajectories is an open condition with respect to the independent
variables $x$ and $t$.
\end{enumerate}

\section[Outline and Method]{Outline and method.}
We begin in Chapter~\ref{sec:solitonRHP} by expressing the function
$\psi(x,t;\hbar_N)$ in terms of the solution of a holomorphic matrix
Riemann-Hilbert problem posed relative to a contour that surrounds the
locus of accumulation of eigenvalues but is otherwise arbitrary {\em a
priori}.  The scattering data is introduced in
Chapter~\ref{sec:scatteringdata}, where we present the formal WKB
formulae for initial data satisfying $S(x)\equiv 0$ and appropriate
functions $A(x)$.  We carry out some detailed asymptotic calculations
starting from the WKB approximations to the discrete eigenvalues that
we will require later, and we compare these general calculations with
the specific exact formulae of Satsuma and Yajima.  With this WKB data in
hand, we proceed in
Chapter~\ref{sec:asymptoticanalysis} to study the asymptotics of the 
inverse-scattering problem for this (generally) approximate data.  We
introduce in \S\ref{sec:phase} a certain complex scalar phase
function, and show in
\S\ref{sec:conditions} how to choose it to capture the essentially
wild asymptotic behavior of the solution of the Riemann-Hilbert
problem.  Factoring off a proper choice of the complex phase leads to
a simpler Riemann-Hilbert problem whose leading-order asymptotics can
be described explicitly.  In
\S\ref{sec:outersolve} we solve this leading-order Riemann-Hilbert
problem, the {\em outer model problem}, in terms of Riemann theta
functions (and in fact at first in terms of exponentials).  Subject
to proving the validity of this asymptotic reduction, the
solution $\psi(x,t;\hbar_N)$ is then also given at leading order in
terms of theta functions and exponentials.  

Assuming the existence of the complex phase function on an appropriate
contour, we continue with some detailed local analysis in
\S\ref{sec:inner}, building local approximations near certain
exceptional points in the complex plane.  
Patching these local approximations together with the outer
approximation yields a uniform approximation of the solution of the
Riemann-Hilbert problem that we prove is valid in
\S\ref{sec:error}.  

This detailed error analysis is completely vacuous {\em unless} we can
establish the existence of the complex phase function and its support
contour.  We carry out this construction in Chapter~\ref{sec:ansatz} using a
modification of the finite-gap ansatz familiar from the Lax-Levermore
method.  Temporarily tossing out the inequalities that the phase
function must ultimately satisfy, we show how to write down equations
for the endpoints of the bands and gaps along the contour and how the
bands of the contour can be viewed as heteroclinic orbits
\index{heteroclinic orbits} of a particular explicit differential
equation for contours in the complex plane (or trajectories of a
quadratic differential).  Some of the conditions we impose on the
endpoints of the bands and gaps are precisely those that are necessary
for the existence of the correct number of heteroclinic orbits.  There
is a finite-gap ansatz corresponding to any number of bands and gaps,
and the idea is to choose this number so that the phase function
satisfies certain inequalities as well.  This choice then determines
the local complexity (genus of the Riemann theta
function\index{Riemann theta function}) of the approximate solution of
the initial-value problem (\ref{eq:IVP}).  In
\S\ref{sec:modulation} we show that in fixed neighborhoods of fixed
$x$ and $t$, the macrostructure parameters of the solutions (moduli of
an associated hyperelliptic Riemann surface\index{hyperelliptic
Riemann surface}) satisfy a quasilinear system of partial differential
equations that we believe to be the elliptic modulation (Whitham)
equations for multiphase wavetrains\index{multiphase wavetrains}
\cite{FL86}.

In Chapter~\ref{sec:genuszero}, we investigate the simplest possible
ansatz ({\em i.e.} genus zero), showing that for small time
independent of $\hbar$ it does indeed satisfy all necessary
inequalities.  For the Satsuma-Yajima initial data this completes the
proof of convergence to the semiclassical limit for small time,
ultimately justifying the geometrical optics approximation made by
Akhmanov, Sukhorukov, and Khokhlov in 1966.  For semiclassical soliton
ensembles corresponding to more general data, we still obtain rigorous
strong asymptotics, but the connection to initial data is more
tenuous.  The asymptotics formally recover the initial data and the
successful ansatz persists for small time, but the error in our scheme
of essentially uniformly approximating the eigenfunction in the
complex plane of the eigenvalue breaks down near $t=0$, when the
regions of the complex plane where the description of the
eigenfunction requires detailed local analysis come into contact with
the locus of accumulation of poles.  On the other hand, we know that
asymptotics for $t=o(1)$ can be obtained by bypassing the
Riemann-Hilbert problem and applying the Lax-Levermore method to the
determinant solution formula \cite{EJLM93}.  Of course, even if the
error is controlled uniformly near $t=0$, the error present at $t=0$
in cases where the WKB approximation is not exact can in principle be
amplified by this unstable problem in ways that are not possible in
``selfadjoint'' integrable problems where the semiclassical limit is
``hyperbolic''.
Using a computer program to construct the ansatz for finite times (as
opposed to a perturbative calculation based at $t=0$) we verify the
ansatz in the special case of the Satsuma-Yajima data right up to the
phase transition to more complicated local behavior termed the
``primary caustic'' in
\cite{MK98}.  These computer simulations clearly demonstrate both the
selection of the special contour and the breakdown of the ansatz when
inequalities fail and/or integral curves of the differential equation
determining the contour bands become disconnected.  We use
perturbation theory in Chapter~\ref{sec:genus2} to show that when the genus
zero ansatz fails at the primary caustic, the genus two ansatz takes
over.  At such a transition, the smooth wave field ``breaks'' and
gives way to a hexagonal spatiotemporal lattice of maxima.

The conditions that we use to specify the complex phase function will
be naturally obtained in Chapter~\ref{sec:variational} as the Euler-Lagrange
variational conditions \index{Euler-Lagrange variational conditions}
describing a particular type of critical point for a certain
functional related to potential theory \index{potential theory} in the
upper half-plane.  This makes the problem of computing the
semiclassical limit equivalent to solving a certain problem of extreme
Green's capacity, and establishing regularity properties of the
solution.  Solving the variational problem can be given the physical
interpretation of finding unstable electrostatic equilibria
\index{unstable electrostatic equilibria} of a certain system of
electric charges under the influence of an externally applied field
which has an attractive component that is exactly the potential of the
WKB eigenvalue distribution.  The significance of variational problems
in the characterization of singular limits of solutions of completely
integrable partial differential equations was first established by Lax
and Levermore
\cite{LL83} for the Korteweg-de Vries equation, and the method was
subsequently extended to the Toda lattice \cite{DM98} and the entire
hierarchy of the {\em defocusing} nonlinear Schr\"odinger equation
\cite{JLM99}.  

The calculations presented in \S\ref{sec:inner} and \S\ref{sec:error}
rely on certain technical details of the Fredholm theory of
Riemann-Hilbert problems posed in H\"older spaces \index{H\"older
spaces} and small-norm theory for Riemann-Hilbert problems in $L^2$
that we present in the appendices.  In particular, the H\"older theory
that we summarize unites some very classical results of the Georgian
school of Muskelishvili {\em et. al.} with the treatment of matrix
Riemann-Hilbert problems posed on self-intersecting contours given by
Zhou \cite{Z89}.  The H\"older theory appears to have fallen by the
wayside in inverse-scattering applications, possibly because these
problems are often posed from the start in $L^p$ or Sobolev spaces.
However, in local analysis one is always dealing with explicit
piecewise-analytic jump relations on piecewise-smooth contours, and at
the same time one requires uniform control on the solutions right up
to the contour.  In such cases, the compactness required for Fredholm
theory comes almost for free (and significantly in a
contour-independent way) in H\"older spaces at the cost of an
arbitrarily small loss of smoothness.  At the same time, once
existence is established in a H\"older space, the required control up
to the contour is built-in as a property of the solution.  On the
other hand, in the bigger $L^p$ or Sobolev spaces compactness depends
on a rational approximation argument that can be a lot of work to
establish (and in particular it seems that the argument must be
tailored for each particular contour configuration).  And then having
established existence in these spaces one must put in extra effort to
obtain the required control up to the contour, with special care
needing to be taken near self-intersection points.

In summary, our primary mathematical techniques include:
\begin{enumerate}
\item
Techniques for the asymptotic analysis of matrix Riemann-Hilbert
problems, including the steepest descent method of Deift and Zhou.
\item
The Fredholm theory of Riemann-Hilbert problems in the class of
functions with H\"older continuous boundary values on
self-intersecting contours.
\item
The use of Cauchy integrals \index{Cauchy integrals} (or Hilbert
transforms\index{Hilbert transforms}) to solve certain scalar
boundary-value problems for sectionally analytic functions in the
plane.
\item
Careful perturbation theory to establish the semiclassical limit for
small times, and then to study the phase transition that occurs at a
caustic curve in the $(x,t)$-plane.
\item
Some theory of logarithmic potentials with external fields.
\end{enumerate}

\section[Special Notation]{Special notation.}
We will use several different branches of the
logarithm\index{logarithm!branches of}, distinguished one from another
by notation.  We will only use the lowercase $\log(z)$ to refer to a
generic branch (cut anywhere) when it makes no difference in an
expression, that is, when it appears in an exponent or when its real
part is considered.  The uppercase $\mbox{Log}(z)$ always refers to
the standard cut of the principal branch, defined for $z\in {\mathbb
C}\setminus {\mathbb R}_-$ by the integral
\begin{equation}
\mbox{Log}(z):=\int_1^z\frac{dw}{w}\,.
\label{eq:principalbranch}
\end{equation}
All other branches of the function $\log(\lambda-\eta)$ considered as
a function of $\lambda$ for $\eta$ fixed will be written with notation
like $L_\eta^s(\lambda)$.  Each of these is also defined for
$z=\lambda-\eta$ by (\ref{eq:principalbranch}), but with a particular
well-defined branch cut in the $\lambda$ plane that is associated with
the logarithmic pole $\eta$ and the superscript $s$.  Each of these
branches will be clearly defined when it first appears in the text.
Exponential functions will {\em always} refer to the principal branch:
$a^u=e^{u\,{\rm Log}(a)}$.

We will use the Pauli matrices \index{Pauli matrices} throughout the
paper.  They are defined as follows:
\begin{equation}
\sigma_1:=\left[\begin{array}{cc}0&1\\1&0\end{array}\right]\,,
\hspace{0.3 in}
\sigma_2:=\left[\begin{array}{cc}0&-i\\i&0\end{array}\right]\,,
\hspace{0.3 in}
\sigma_3:=\left[\begin{array}{cc}1 & 0\\0&-1\end{array}\right]\,.
\end{equation}

\chapter{Holomorphic 
Riemann-Hilbert Problems for Solitons}
\label{sec:solitonRHP}
The initial value problem (\ref{eq:IVP}) is solvable for arbitrary
$\hbar$ because the focusing nonlinear Schr\"odinger equation can be
represented as the compatibility condition \index{compatibility
condition} for two systems of linear ordinary differential equations:
\begin{equation}
\hbar\partial_x\left[\begin{array}{c}u_1\\u_2\end{array}\right]=
\left[\begin{array}{cc}-i\lambda & \psi\\-\psi^* & i\lambda\end{array}
\right]\left[\begin{array}{c}u_1\\u_2\end{array}\right]\,,
\label{eq:linearx}
\end{equation}
\begin{equation}
i\hbar\partial_t\left[\begin{array}{c}u_1\\u_2\end{array}\right]=
\left[\begin{array}{cc} \lambda^2 - |\psi|^2/2 & i\lambda\psi -
\hbar\partial_x\psi/2\\-i\lambda\psi^*-\hbar\partial_x\psi^*/2 &
-\lambda^2 +|\psi|^2/2\end{array}\right]
\left[\begin{array}{c}u_1\\u_2\end{array}\right]\,,
\label{eq:lineart}
\end{equation}
where $\lambda$ is an arbitrary complex parameter.  The compatibility
condition for (\ref{eq:linearx}) and (\ref{eq:lineart}) does not
depend on the value of $\lambda$, and is equivalent to the nonlinear
Schr\"odinger equation.  

The $N$-soliton solutions \index{$N$-soliton solutions} of the
nonlinear Schr\"odinger equation are those complex functions
$\psi(x,t)$ for which there exist simultaneous column vector solutions
of (\ref{eq:linearx}) and (\ref{eq:lineart}) of the particularly
simple form:
\begin{equation}
\begin{array}{rcl}
{\bf u}^+(x,t,\lambda)&=&\displaystyle\left[\begin{array}{c}
\displaystyle\sum_{p=0}^{N-1}A_p(x,t)\lambda^p\\\\
\displaystyle \lambda^N + \sum_{p=0}^{N-1}B_p(x,t)\lambda^p
\end{array}\right]\exp(i(\lambda x +\lambda^2 t)/\hbar)\,,\\\\
{\bf u}^-(x,t,\lambda)&=&\displaystyle\left[\begin{array}{c}
\displaystyle\lambda^N +\sum_{p=0}^{N-1}C_p(x,t)\lambda^p\\\\
\displaystyle\sum_{p=0}^{N-1}D_p(x,t)\lambda^p\end{array}
\right]\exp(-i(\lambda x +\lambda^2 t)/\hbar)\,,
\end{array}
\label{eq:form}
\end{equation}
satisfying the relations
\begin{equation}
\begin{array}{rcl}
{\bf u}^+(x,t,\lambda_k) &=&\gamma_k{\bf u}^-(x,t,\lambda_k)\,,\\
-\gamma_k^*{\bf u}^+(x,t,\lambda_k^*)&=&{\bf u}^-(x,t,\lambda_k^*)\,,\hspace{0.2 in}k=1,\dots,N\,,
\end{array}
\label{eq:relations}
\end{equation}
for some distinct complex numbers $\lambda_0,\dots,\lambda_{N-1}$ in the
upper half-plane and nonzero complex numbers (not necessarily distinct)
$\gamma_0,\dots,\gamma_{N-1}$.  It is easy to check that given the numbers
$\{\lambda_k\}$ and $\{\gamma_k\}$, the relations (\ref{eq:relations})
determine the coefficient functions $A_p(x,t)$, $B_p(x,t)$, $C_p(x,t)$
and $D_p(x,t)$ in terms of exponentials via the solution of a square
inhomogeneous linear algebraic system.  In Faddeev and Takhtajan
\cite{FT87} it is shown that this linear system is always nonsingular assuming
the $\{\lambda_k\}$ are distinct and nonreal and the $\{\gamma_k\}$
are nonzero.  The solution of the nonlinear Schr\"odinger equation for
which the column vectors ${\bf u}^\pm(x,t,\lambda)$ are simultaneous
solutions of (\ref{eq:linearx}) and (\ref{eq:lineart}) turns out to be
\begin{equation}
\psi(x,t)=2iA_{N-1}(x,t)\,.
\label{eq:psi}
\end{equation}

\begin{remark}
This construction is equivalent to a classical problem of rational
approximation, the construction of {\em multipoint Pad\'e
interpolants} \index{multipoint Pad\'e approximation} for entire
functions \cite{B75}.  Let $G(\lambda)$ be any polynomial satisfying
$G(\lambda_k)={\rm Log}(\gamma_k)$ and $G(\lambda_k^*)= {\rm
Log}(-1/\gamma_k^*)$.  Then, looking at the first row of
(\ref{eq:relations}), we see that we are seeking polynomials
$P_{N-1}(\lambda)$ and $Q_{N-1}(\lambda)$ both of degree $N-1$ such
that
\begin{equation}
\frac{P_{N-1}(\lambda)}{\lambda^N +Q_{N-1}(\lambda)}=\exp(G(\lambda))
\exp(-2i(\lambda x+ \lambda^2 t)/\hbar)\,,\hspace{0.2 in}
\mbox{for}\,\,\lambda=\lambda_0,\dots,\lambda_{N-1},\lambda_0^*,\dots,
\lambda_{N-1}^*
\,.
\label{eq:Pade}
\end{equation}
The coefficients of $P_{N-1}(\lambda)$ are the $\{A_p(x,t)\}$, and the
coefficients of $Q_{N-1}(\lambda)$ are the $\{C_p(x,t)\}$.  

Whereas the usual Pad\'e approximants are constructed by demanding
sufficiently high-order agreement in the asymptotic expansion of
(\ref{eq:Pade}) for large or small $\lambda$, the multipoint
approximants are constructed by demanding simple agreement of the
function values on the left and right-hand sides of (\ref{eq:Pade}) at
a sufficiently large number of distinct points.  This latter version
of the rational interpolation problem was first considered by Cauchy
and Jacobi.  The Cauchy-Jacobi problem \index{Cauchy-Jacobi problem}
in its most general form can fail to have a solution, {\em i.e.} given
a set of nodes of interpolation, there exist isolated ``unreachable''
function values.  In the context of the $N$-soliton solution problem,
however, this undesirable situation does not occur due to the complex
conjugation symmetry of the interpolation points and corresponding
symmetry properties of the assigned values at those points.
\end{remark}

A typical initial condition $\psi_0(x)$ for (\ref{eq:IVP}) will not
correspond exactly to a multisoliton solution.  As is well-known
\cite{ZS72,FT87}, the procedure for solving (\ref{eq:IVP}) generally
begins with the study the solutions of (\ref{eq:linearx}) for real
$\lambda$ and for $\psi=\psi_0(x)$.  One obtains from this analysis a
complex-valued {\em transmission coefficient} \index{transmission
coefficient} $T(\lambda)=1/a(\lambda)$, $\lambda\in{\mathbb R}$.  Now,
after the fact it turns out that the function $a(\lambda)$ has an
analytic continuation into the whole upper half-plane, and its zeros
occur at values of $\lambda$ for which (\ref{eq:linearx}) has an
$L^2({\mathbb R})$ eigenfunction.  In this sense, the study of the
scattering problem for real $\lambda$ yields results for complex
$\lambda$ by unique analytic continuation.  The function $a(\lambda)$
can be viewed as the interpretation of a Wronskian \index{Wronskian}
between two particular solutions of (\ref{eq:linearx}) that have
analytic continuations into the upper half-plane.  Thus at each $L^2$
eigenvalue $\lambda=\lambda_k$, there is a complex number $\gamma_k$
that is the ratio of these two analytic solutions.  In addition to the
transmission coefficient, one also finds a complex-valued function
$b(\lambda)$ that gives rise to a {\em reflection coefficient}
\index{reflection coefficient} $r(\lambda):=b(\lambda)/a(\lambda)$, 
$\lambda\in{\mathbb R}$.  The
main results of Zakharov and Shabat \cite{ZS72} are:
\begin{enumerate}
\item
When $\psi(x,t)$ is the solution of (\ref{eq:IVP}) with initial data
$\psi_0(x)$, then for each $t>0$ one has different coefficients in the
linear problem (\ref{eq:linearx}), and therefore the eigenvalues
\index{eigenvalues} $\{\lambda_k\}$, proportionality constants 
\index{proportionality constants} $\{\gamma_k\}$ and the
function $b(\lambda)$, can be computed independently for each $t>0$.
However, it follows from (\ref{eq:lineart}) that the eigenvalues
$\{\lambda_k\}$ and also $|b(\lambda)|$, $\lambda\in{\mathbb R}$, are
independent of $t$, and the proportionality constants $\{\gamma_k\}$
and $\arg(b(\lambda))$, $\lambda\in{\mathbb R}$ evolve simply in time.
Thus, $b(\lambda,t)=b(\lambda,0)\exp(-2i\lambda^2 t/\hbar)$ and
$\gamma_k(t)=\gamma_k(0)\exp(-2i\lambda_k^2 t/\hbar)$.
\item
The function $\psi(x,t)$ can be reconstructed at later times $t>0$ in
terms of the discrete spectrum $\{\lambda_k\}$, $\{\gamma_k\}$, and
the function $b(\lambda)$.
\end{enumerate}
If for the initial condition $\psi_0(x)$ we have
$b(\lambda)\equiv 0$, then the step of reconstructing the solution of
the initial value problem (\ref{eq:IVP}) is essentially what we have
already described.  Namely, one solves the linear equations
(\ref{eq:relations}) for the coefficient $A_{N-1}(x,t)$ and then the
solution of (\ref{eq:IVP}) is given by (\ref{eq:psi}).  Note that $N$
is the number of $L^2$ eigenvalues for $\psi_0(x)$ in the upper
half-plane.

In general, the reconstruction of $\psi$ from the scattering data can
be recast in terms of the solution of a matrix-valued meromorphic {\em
Riemann-Hilbert problem}\index{Riemann-Hilbert problem!meromorphic}.
One seeks (for each $x$ and $t$, which play the role of parameters) a
matrix-valued function ${\bf m}(\lambda)$ of $\lambda$ that is jointly
meromorphic in the upper and lower half-planes and for which
\begin{enumerate}
\item
${\bf m}(\lambda)\rightarrow {\mathbb I}$ in each half-plane as
$\lambda\rightarrow \infty$.
\item
The singularities of ${\bf m}(\lambda)$ are completely specified.
There are simple poles at the eigenvalues $\{\lambda_k\}$ and the
complex conjugates with residues of a certain specified type (see
below).
\item
On the real axis $\lambda\in{\mathbb R}$, there is the {\em jump
relation}
\begin{equation}
{\bf m}_+(\lambda)=
{\bf m}_-(\lambda){\bf v}(\lambda)\,,
\hspace{0.2 in}
{\bf m}_\pm(\lambda):=
\lim_{\epsilon\downarrow 0}{\bf m}(\lambda\pm i\epsilon)
\label{eq:updowndef}
\end{equation}
where ${\bf v}(\lambda)$ is a certain {\em jump matrix} \index{jump
matrix} built out of $r(\lambda)$ and depending {\em explicitly} on
$x$ and $t$.  The jump matrix becomes the identity matrix for
$b(\lambda)\equiv 0$.
\end{enumerate}

However, if the boundary values ${\bf m}_\pm(\lambda)$ are continuous, and
if $b(\lambda)\equiv 0$, then it is easy to see that the solution ${\bf
m}(\lambda)$ must be a rational function of $\lambda$.  This is
the case we will now develop in more detail.  
Let $J=\pm 1$ be a free parameter.
From the column vectors
${\bf u}^\pm(x,t,\lambda)$, we build a matrix solution of
(\ref{eq:linearx}):
\begin{equation}
\Psi(\lambda):=
({\bf u}^-(x,t,\lambda),{\bf u}^+(x,t,\lambda))
\sigma_1^{\frac{1-J}{2}}\mbox{diag}\left(\prod_{j=1}^N(\lambda-\lambda_j)^{-1},
\prod_{j=1}^N(\lambda-\lambda_j^*)^{-1}\right)\sigma_1^{\frac{1-J}{2}}
\exp(i\sigma_3\lambda^2 t/\hbar)\,.
\end{equation}
This special matrix solution of (\ref{eq:linearx}) is called a {\em
Jost solution}\index{Jost solution}.  Note that $\Psi(\lambda)$ would
also satisfy (\ref{eq:lineart}) if it were not for the exponential
factor in this formula.  The reason for this exponential factor is
that the Jost solution matrix has simple large $x$ asymptotics that
are, to leading order, independent of $t$.  Indeed, if we define a
matrix ${\bf m}(\lambda)$ by
\begin{equation}
{\bf m}(\lambda):= \Psi(\lambda)\exp(i\sigma_3\lambda x/\hbar)\,,
\label{eq:Psi-to-m}
\end{equation}
then we find using (\ref{eq:relations}) that for all fixed complex $\lambda$
different from the eigenvalues $\{\lambda_k\}$ and their complex
conjugates, ${\bf m}(\lambda)$ is a uniformly bounded function of $x$
that satisfies ${\bf m}(\lambda)\rightarrow{\mathbb I}$ as
$x\rightarrow J\infty$.  {\em Thus, the parameter $J$ merely indicates
whether the Jost solution matrix is normalized at $x=+\infty$ or
$x=-\infty$.}  

\begin{remark}
In the general case when $b(\lambda)$ does not necessarily vanish
identically, the Jost solution matrix is defined for all $\lambda\in
{\mathbb C}$ as the unique matrix solution $\Psi(\lambda)$ of
(\ref{eq:linearx}) that satisfies the two conditions:
\begin{equation}
\begin{array}{rl}
\mbox{\bf Normalization:} &
\Psi(\lambda)\exp(i\sigma_3\lambda x/\hbar)\rightarrow {\mathbb I}\,,
\hspace{0.2 in}\mbox{as}\hspace{0.2 in} x\rightarrow J\infty\\\\
\mbox{\bf Boundedness:} & 
\sup_{x\in{\mathbb R}}\|\Psi(\lambda)\|<\infty\,.
\end{array}
\end{equation}
The boundedness condition is superfluous when $\lambda\in{\mathbb R}$,
but is absolutely necessary for uniqueness when $\Im(\lambda)\neq 0$.
The definition (\ref{eq:Psi-to-m}) yields a matrix-valued function of
$\lambda$ that is meromorphic in the upper and lower half-planes.  For
$\lambda\in{\mathbb R}$ however, the three matrices ${\bf
m}(\lambda)$, ${\bf m}_+(\lambda)$, and ${\bf m}_-(\lambda)$ ({\em
cf.} (\ref{eq:updowndef})) are generally all different unless
$b(\lambda)\equiv 0$.
\end{remark}

Continuing with the pure soliton case of $b(\lambda)\equiv 0$, we can
deduce from the explicit form (\ref{eq:form}) of the vectors ${\bf
u}^\pm(x,t,\lambda)$ and from the relations (\ref{eq:relations})
that ${\bf m}(\lambda)$ solves the following problem.
\begin{rhp}[Meromorphic problem]
Given the discrete data $\{\lambda_k\}$ and $\{\gamma_k\}$, find a
matrix ${\bf m}(\lambda)$ with the following two properties:
\begin{enumerate}
\item
{\bf Rationality:} 
${\bf m}(\lambda)$ is a rational function of $\lambda$, with simple
poles confined to the eigenvalues $\{\lambda_k\}$ and the complex conjugates.
At the singularities:
\begin{equation}
\begin{array}{rcl}
\displaystyle
\begin{array}{c}\\\rm Res\\\scriptstyle \lambda=\lambda_k\end{array}
{\bf m}(\lambda)&=&
\displaystyle\lim_{\lambda\rightarrow\lambda_k}{\bf m}(\lambda)
\sigma_1^{\frac{1-J}{2}}\left[\begin{array}{cc}0 & 0\\c_k(x,t) & 0\end{array}\right]
\sigma_1^{\frac{1-J}{2}}\,,\\\\
\displaystyle
\begin{array}{c}\\\rm Res\\\scriptstyle \lambda=\lambda_k^*\end{array}
{\bf m}(\lambda)&=&
\displaystyle\lim_{\lambda\rightarrow\lambda_k^*}{\bf m}(\lambda)
\sigma_1^{\frac{1-J}{2}}
\left[\begin{array}{cc}0 & -c_k(x,t)^*\\0 & 0\end{array}\right]
\sigma_1^{\frac{1-J}{2}}\,,
\end{array}
\label{eq:residueconds}
\end{equation}
for $k=0,\dots,N-1$, with
\begin{equation}
c_k(x,t):=\left(\frac{1}{\gamma_k}\right)^J
\frac{\displaystyle\prod_{n=0}^{N-1}(\lambda_k-\lambda_n^*)}
{\displaystyle\prod_{\scriptstyle n=0\atop
\scriptstyle n\neq k}^{N-1}(\lambda_k-\lambda_n)} 
\exp(2iJ(\lambda_k x +\lambda_k^2
t)/\hbar)\,.
\label{eq:residuecoeffs}
\end{equation}
\item
{\bf Normalization:}
\begin{equation}
{\bf m}(\lambda)
\rightarrow{\mathbb I}\,,\hspace{0.2 in}\mbox{as}\hspace{0.2 in}
\lambda\rightarrow\infty\,.
\end{equation}
\end{enumerate}
\label{rhp:m}
\end{rhp}
Whereas we deduced the two properties characterizing Riemann-Hilbert
Problem~\ref{rhp:m} from the explicit construction of the vector
solutions ${\bf u}^\pm(x,t,\lambda)$, it is not difficult to see that
these two properties actually characterize the matrix function ${\bf
m}(\lambda)$ uniquely.

\begin{proposition}
The meromorphic Riemann-Hilbert Problem~\ref{rhp:m} corresponding to
the discrete data $\{\lambda_k\}$ and $\{\gamma_k\}$ has a unique
solution whenever the $\lambda_k$ are distinct in the upper half-plane
and the $\gamma_k$ are nonzero.  The function defined from the
solution by
\begin{equation}
\psi:=2i\lim_{\lambda\rightarrow\infty}\lambda m_{12}(\lambda)
\label{eq:psiagain}
\end{equation}
is a nontrivial ($N$-soliton) solution of the focusing nonlinear
Schr\"odinger equation.
\end{proposition}

\begin{proof}
One obtains a solution of the meromorphic Riemann-Hilbert
Problem~\ref{rhp:m} by making an ansatz of the form (\ref{eq:form})
and observing that the residue conditions (\ref{eq:residueconds}) with
(\ref{eq:residuecoeffs}) are equivalent to (\ref{eq:relations}), that
is, by reversing our steps.  The solvability of the linear system for
the coefficients implied by (\ref{eq:relations}) is guaranteed by the
distinctness of the $\lambda_k$ and the conditions $\gamma_k\neq 0$
\cite{FT87}.  Under an ansatz of the form (\ref{eq:form}), the
relation (\ref{eq:psiagain}) is equivalent to (\ref{eq:psi}).  Note
that the same solution of the nonlinear Schr\"odinger equation is
obtained from the matrix ${\bf m}(\lambda)$ for {\em both cases}
$J=\pm 1$ by the formula (\ref{eq:psiagain}).  Uniqueness follows from
Liouville's theorem\index{Liouville's theorem}.
\end{proof}

Thus, we may drop the explicit algebraic relations
(\ref{eq:relations}) and instead view the meromorphic Riemann-Hilbert
Problem~\ref{rhp:m} as the fundamental
characterization of the $N$-soliton solutions of the focusing
nonlinear Schr\"odinger equation.  

\begin{remark}
In the general case when $b(\lambda)\not\equiv 0$, the meromorphic
Riemann-Hilbert problem is altered.  One only insists that ${\bf m}(\lambda)$
be piecewise meromorphic in the upper and lower half-planes, and that
the boundary values taken from above and below on the real $\lambda$-axis
satisfy a jump relation ({\em cf.} (\ref{eq:updowndef})) with a matrix
${\bf v}(\lambda)$ built out of $r(\lambda)$ and going over into the
identity matrix when $b(\lambda)\equiv 0$ (and thus $r(\lambda)\equiv 0$).
When $r(\lambda)\not\equiv 0$, the corresponding meromorphic Riemann-Hilbert
problem cannot be solved by algebraic operations alone, and 
in general the solution can be obtained by 
solving a system of integral equations.  But even in this more general
case, the function $\psi(x,t)$ defined by (\ref{eq:psiagain}) satisfies
the nonlinear Schr\"odinger equation.
\end{remark}

Another very important property of the matrix ${\bf m}(\lambda)$ is
the following ``reflection symmetry'' \index{reflection symmetry} in
the real axis:
\begin{equation}
{\bf m}(\lambda^*)=\sigma_2{\bf m}(\lambda)^*\sigma_2\,.
\end{equation}
On the right-hand side and in other similar formulae, the star denotes
componentwise complex conjugation; the matrix is not transposed.

\begin{figure}[h]
\begin{center}
\mbox{\psfig{file=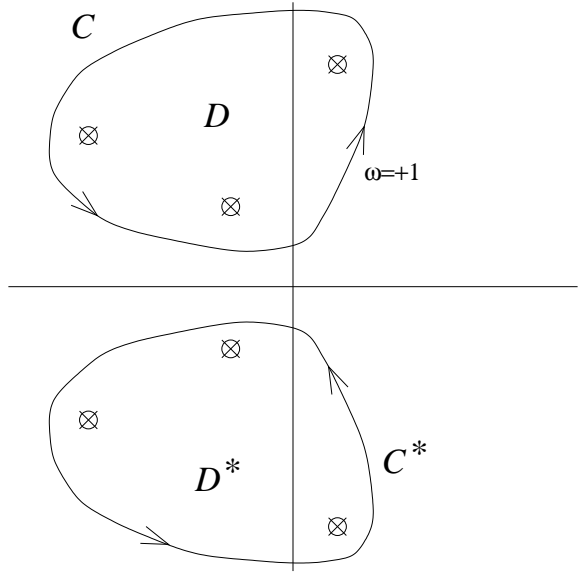,width=3 in}}
\end{center}
\caption{\em The complex $\lambda$-plane with three eigenvalues
$\lambda_k$ in the upper half-plane, their complex conjugates, and the
contours $C$, $C^*$, and domains $D$, $D^*$.  The orientation in the
figure is $\omega=+1$.}
\label{fig:generalcontours_sym}
\end{figure}

We now show how to convert the meromorphic Riemann-Hilbert problem
into a sectionally holomorphic Riemann-Hilbert
problem\index{Riemann-Hilbert problem!holomorphic}, {\em i.e.} how to
remove the poles from the problem.  The reader can find a similar
construction in \cite{DKKZ96}.  Let $C$ be a simple closed contour
that is the boundary of a simply-connected domain $D$ in the upper
half-plane that contains all of the eigenvalues $\{\lambda_k\}$.  We
assign to $C$ an orientation $\omega$ ($\omega=+1$ means
counterclockwise, and $\omega=-1$ means clockwise), and when this
orientation is important ({\em i.e.} in contour integration and
specifying Riemann-Hilbert jump relations), we will write the contour
as $C_\omega$.  By $C^*$ and $D^*$ we mean the corresponding complex
conjugate sets in the lower half-plane, and when we write $[C\cup
C^*]_\omega$, we mean that both loops share the same orientation
$\omega$.  See Figure~\ref{fig:generalcontours_sym}.

Next, we will need to interpolate the proportionality constants at the
eigenvalues.  Choose a constant $Q$ and a function $X(\lambda)$
analytic in $D$ so that
\begin{equation}
\gamma_k=Q\exp(X(\lambda_k)/\hbar)\,,\hspace{0.3 in}k=0,\dots,N-1\,.
\end{equation}
In general, $X(\lambda)$ could be systematically constructed as an
interpolating polynomial of degree $\sim N$.  In other circumstances
(see below) the phases $\gamma_k$ can be highly correlated so that for
very large $N$ one can choose for $X(\lambda)$ a polynomial of low
degree (or another simple expression).  Note that the
interpolant of the $\gamma_k$ is not
necessarily unique; for each $K$ in some indexing set there is a
distinct pair $(Q_K,X_K(\lambda))$ such that for all $j$,
$\gamma_j=Q_K\exp(X_K(\lambda_j)/\hbar)$.  We will make use of this
freedom later; for now we just carry the subscript $K$.

With the help of the interpolant of the proportionality
constants\index{proportionality constants!interpolant of}, we define a
new matrix ${\bf M}(\lambda)$ for $\lambda\in {\mathbb C}\setminus
(C\cup C^*)$ in the following way.  First, for all $\lambda\in D$, set
\begin{equation}
{\bf M}(\lambda):={\bf m}(\lambda)\sigma_1^{\frac{1-J}{2}}
\left[\begin{array}{cc}
1 & 0\\\\
\displaystyle -\left(\frac{1}{Q_K}\right)^J\left(\prod_{n=1}^N
\frac{\lambda-\lambda_n^*}{\lambda-\lambda_n}\right)\exp\left(\frac{J}{\hbar}
(2i\lambda x+
2i\lambda^2 t -X_K(\lambda))\right) & 1\end{array}\right]
\sigma_1^{\frac{1-J}{2}}\,.
\label{eq:Blaschke}
\end{equation}
Next, for all $\lambda\in D^*$, set
\begin{equation}
{\bf M}(\lambda):=\sigma_2{\bf M}(\lambda^*)^*\sigma_2\,.
\label{eq:under}
\end{equation}
Finally, for all $\lambda\in {\mathbb C}\setminus (\overline{D}\cup
\overline{D}^*)$ simply set 
\begin{equation}
{\bf M}(\lambda):={\bf m}(\lambda)\,.
\label{eq:outside}
\end{equation}

It is straightforward to verify that by our choice of interpolants,
and the Blaschke factor \index{Blaschke factor} appearing in
(\ref{eq:Blaschke}), that ${\bf M}(\lambda)$ {\em has no poles in $D$
or $D^*$ and hence is sectionally holomorphic in the complex $\lambda$
plane}.  By definition, we have preserved the reflection symmetry of
${\bf m}(\lambda)$ so that for all $\lambda\in {\mathbb C}
\setminus (C\cup C^*)$ we have:
\begin{equation}
{\bf M}(\lambda^*)=\sigma_2{\bf M}(\lambda)^*\sigma_2\,.
\end{equation}
The matrix ${\bf M}(\lambda)$ has continuous boundary values from
either side on $C$ and $C^*$.  To describe these, let the left
(respectively right) side of the oriented contour $[C\cup C^*]_\omega$ be
denoted by ``$+$'' (respectively ``$-$'').  For $\lambda\in [C\cup
C^*]_\omega$ define
\begin{equation}
{\bf M}_\pm(\lambda):=\lim_{\begin{array}{c}\scriptstyle\mu\rightarrow\lambda
\\\scriptstyle\mu\in\pm\mbox{ side of }[C\cup C^*]_\omega\end{array}}
{\bf M}(\mu)\,,
\end{equation}
that is, the nontangential limits from the left and right sides.
Then, using the fact that ${\bf m}(\lambda)$ is analytic on $C\cup
C^*$ and the piecewise definition of ${\bf M}(\lambda)$ given by
(\ref{eq:Blaschke}), (\ref{eq:under}), and (\ref{eq:outside}), 
we find
\begin{equation}
\begin{array}{rcll}
{\bf M}_+(\lambda)&=&{\bf M}_-(\lambda){\bf
v}_{\bf M}(\lambda)\,,&\lambda\in C_\omega\,,\\\\
{\bf M}_+(\lambda)&=&{\bf M}_-(\lambda)
\sigma_2{\bf v}_{\bf M}(\lambda^*)^*\sigma_2\,.
&\lambda\in [C^*]_\omega\,,
\end{array}
\label{eq:jumps}
\end{equation}
where for $\lambda\in C$,
\begin{equation}
{\bf
v}_{\bf M}(\lambda):=\sigma_1^{\frac{1-J}{2}}\left[\begin{array}{cc}
1 & 0\\\\ \displaystyle
-\omega\left(\frac{1}{Q_K}\right)^{J}\left(\prod_{n=0}^{N-1}
\frac{\lambda-\lambda_n^*}{\lambda-\lambda_n}\right)
\exp\left(\frac{J}{\hbar}(2i\lambda x+2i\lambda^2 t
-X_K(\lambda))\right) & 1
\end{array}\right]\sigma_1^{\frac{1-J}{2}}\,.
\label{eq:vM}
\end{equation}
Note that the orientation \index{orientation} choice $\omega=\pm 1$ is
arbitrary, leading to the {\em same} matrix ${\bf M}(\lambda)$.  We
allow for both possibilities of orientation for later convenience.
Also, observe that if one introduces a discrete measure in the complex
plane by
\begin{equation}
d\mu = \sum_{k=0}^{N-1} \left[\hbar\delta_{\lambda_k^*} -
\hbar\delta_{\lambda_k}
\right]\,,
\label{eq:discretetrue}
\end{equation}
then for any branch of the logarithm,
\begin{equation}
\prod_{k=0}^{N-1}\frac{\lambda-\lambda_k^*}{\lambda-\lambda_k} = \exp\left(
\frac{1}{\hbar}\int \log(\lambda-\eta)\,d\mu(\eta)\right)\,.
\end{equation}
The jump matrices are therefore conveniently written in terms of phases
\begin{equation}
\alpha := \int \log(\lambda-\eta)\,d\mu(\eta) +J\cdot 
(2i\lambda x + 2i\lambda^2 t -X_K(\lambda))\,\,\,\pmod{2\pi i\hbar}\,.
\end{equation}

Suppose the eigenvalues $\{\lambda_k\}$ and proportionality constants
$\{\gamma_k\}$ are given along with an appropriate interpolation
$Q_K\exp(X_K(\lambda)/\hbar)$ of the $\gamma_k$ and a smooth closed
contour $C$ enclosing the eigenvalues in the upper half-plane.  We
define a Riemann-Hilbert problem as follows:
\begin{rhp}[Holomorphic problem]
Given the eigenvalues $\{\lambda_k\}$, the interpolant
$Q_K\exp(X_K(\lambda)/\hbar)$, and the oriented contour $C_\omega$,
find a matrix function ${\bf M}(\lambda)$ that satisfies:
\begin{enumerate}
\item
{\bf Analyticity:} ${\bf M}(\lambda)$ is analytic in each component of
${\mathbb C}\setminus (C\cup C^*)$.
\item
{\bf Boundary behavior:} ${\bf M}(\lambda)$ assumes continuous boundary
values on $C\cup C^*$.
\item
{\bf Jump conditions:} The boundary values taken on $[C\cup C^*]_\omega$
satisfy the relations (\ref{eq:jumps}) with ${\bf v}_{\bf M}(\lambda)$
given explicitly by (\ref{eq:vM}).
\item
{\bf Normalization:} ${\bf M}(\lambda)$ is normalized at infinity:
\begin{equation}
{\bf M}(\lambda)\rightarrow {\mathbb I}\mbox{  as  }
\lambda\rightarrow\infty\,.
\end{equation}
\end{enumerate}
\label{rhp:M}
\end{rhp}

\begin{proposition}
The holomorphic Riemann-Hilbert Problem~\ref{rhp:M} has a unique
solution ${\bf M}(\lambda)$ whenever the $\lambda_k$ are distinct and
nonreal, and the $\gamma_k$ are nonzero.  The function defined by
\begin{equation}
\psi :=2i\lim_{\lambda\rightarrow\infty}\lambda M_{12}(\lambda)\,,
\label{eq:fieldatinfinity}
\end{equation}
is independent of the value of the index $J$, as well as the
particular choice of loop contour $C$ and interpolant index $K$,
and is the $N$-soliton solution of the focusing nonlinear Schr\"odinger
equation corresponding to the discrete data $\{\lambda_k\}$ and
$\{\gamma_k\}$.
\end{proposition} 

\begin{proof}
The existence part of the proof of this proposition follows from the
corresponding existence result for the meromorphic Riemann-Hilbert
Problem~\ref{rhp:m} whose solution ${\bf m}(\lambda)$ yields a
solution ${\bf M}(\lambda)$ of the holomorphic problem by the
definitions (\ref{eq:Blaschke}), (\ref{eq:under}), and
(\ref{eq:outside}).  The uniqueness part of the proof follows from the
continuity of the boundary values and Liouville's
theorem\index{Liouville's theorem}: the ratio ${\bf
M}^{(1)}(\lambda){\bf M}^{(2)}(\lambda)^{-1}$ of any two solutions is
analytic in ${\mathbb C}\setminus (C\cup C^*)$ and continuous in
${\mathbb C}$.  Therefore this ratio is entire and from the
normalization condition we learn that ${\bf M}^{(1)}(\lambda)\equiv
{\bf M}^{(2)}(\lambda)$.
\end{proof}

Note that it is possible to allow $C$ to meet the
real axis at one or more isolated points $u_k\in{\mathbb R}$, as long
as at each $u_k$ the incoming and outgoing parts of $C$ make nonzero
angles with the real axis and with each other.  The contour $C$ should
thus meet the axis in ``corners'' (if at all).

\begin{remark}
The holomorphic Riemann-Hilbert Problem~\ref{rhp:M}
with $J=+1$ is equivalent to that for $J=-1$ in the sense that they
yield the same solution $\psi$ of the nonlinear Schr\"odinger equation
via (\ref{eq:fieldatinfinity}).  Note however, that the solution
matrix ${\bf M}(\lambda)$ for $J=+1$ is {\em not} the same as that for
$J=-1$, even though they have similar leading-order asymptotics at
$\lambda=\infty$.  On the other hand, if $J$ is given a fixed value,
then the full matrix solution ${\bf M}(\lambda)$ of the
Riemann-Hilbert problem corresponding to a contour $C$ and interpolant
$K$ agrees identically with that corresponding to a contour $C'$ and
interpolant $K'$ outside any circle containing both $C\cup C^*$ and
$C'\cup {C'}^*$.

From the point of view of ease of analysis, the different formulations
of the inverse problem corresponding to different choices of $C$, $J$,
and $K$, although all equivalent, are not necessarily all equally
valuable.  For given values of $x$ and $t$, it may turn out that one
formulation of the Riemann-Hilbert problem results in a jump matrix
that is very close to the identity in some parts of the complex plane,
while in another equivalent formulation the jump matrix is large and
oscillatory, with the solution being obtained by a kind of
cancellation.  The picture to have in mind here is that of evaluating
a particularly complicated algebraic expression that one wants to
evaluate and show is small.  It may turn out that the expression can
be written as a sum of residues of a contour integral in several
different ways.  In one integral it may turn out that the path of
integration can be deformed in such a way that the integrand is
uniformly very small, which is clearly useful in analysis.  At the
same time, it may be the case in another (equivalent!) integral the
integrand cannot be made uniformly small by any deformation, and the
small result is always achieved by cancellation and consequently is
more difficult to deduce.  This is not an exact analogy, since matrix
Riemann-Hilbert problems are not solved by direct contour integration,
but it may help to illustrate the utility of having several possible
formulations of the problem at hand.
\end{remark}

\chapter{Semiclassical Soliton Ensembles}
\label{sec:scatteringdata}
In this paper, we will present a technique for studying the behavior,
near fixed $x$ and $t$, of multisoliton solutions for which the number
of solitons $N$ is large but for which the solitons are highly {\em
phase-correlated}\index{solitons!phase-correlated}.  This means that for large
$N$ the discrete measure $d\mu$ converges and at the same time the
interpolants $X_K(\lambda)$ converge and take on a simple limiting
form.

\section[WKB Formulae]{Formal WKB formulae for even, 
bell-shaped\index{bell-shaped},
real-valued initial conditions.}  This highly-correlated situation
arises naturally when one considers the semiclassical limit of the
sequence of initial value problems (\ref{eq:IVP}) for initial data of
the form (\ref{eq:data}).  Because $\hbar$ is present explicitly in
the scattering problem (\ref{eq:linearx}) and generally in the initial
data as well (if $S(x)\not\equiv 0$), the scattering data will depend
on $\hbar$, and in particular the number of $L^2$ eigenvalues is a
function of $\hbar$.  Unfortunately, the question of how to rigorously
extract all relevant asymptotic properties of the spectrum for
(\ref{eq:linearx}) in the limit of $\hbar\downarrow 0$ given fixed
functions $A(x)$ and $S(x)$ remains wide open (but see \cite{M00} for
some recent ideas in this direction).

On the other hand, some progress has been made in the {\em formal}
analysis of the nonselfadjoint Zakharov-Shabat scattering problem
(\ref{eq:linearx}) in the semiclassical limit.  For example, the
calculations in \cite{EJLM93} suppose that $S(x)\equiv 0$ and then
exploit the fact that as $\hbar$ converges to zero, the eigenvalue
problem (\ref{eq:linearx}) appears to go over into that of a
semiclassically scaled selfadjoint Schr\"odinger operator
\index{Schr\"odinger operator} with a nonselfadjoint and
energy-dependent but bounded and small (formally $\bo(
\hbar^2)$) correction (in fact this was observed already in the paper
of Zakharov and Shabat \cite{ZS72}).  This observation suggests that
WKB formulae for the Schr\"odinger operator might be valid, and in
particular that the reflection coefficient $r(\lambda)$ is
exponentially small for $\lambda\neq 0$ and that the discrete
eigenvalues accumulate on the imaginary axis in the $\lambda$-plane
with a certain asymptotic density.  The true discrete eigenvalue
measure $d\mu$ defined by (\ref{eq:discretetrue}) is presumed to
converge in the sense of weak-$*$ convergence \index{weak-$*$
convergence} of measures to a measure $d\mu_0^{\rm WKB}$, which in the
case of functions $A(x)$ having a single local maximum (without loss
of generality we take it to occur for $x=0$) for which there are
always exactly two turning points \index{turning points} is given by
the formula
\begin{equation}
d\mu_0^{\rm WKB}(\eta):=\rho^0(\eta)\chi_{[0,iA]}(\eta)\,d\eta +
\rho^0(\eta^*)^*\chi_{[-iA,0]}(\eta)\,d\eta\,,
\label{eq:WKBdensity}
\end{equation}
with
\begin{equation}
\rho^0(\eta):= \frac{\eta}{\pi}\int_{x_-(\eta)}^{x_+(\eta)}\frac{dx}{
\sqrt{A(x)^2 +\eta^2}}
=
\frac{1}{\pi}\frac{d}{d\eta}\int_{x_-(\eta)}^{x_+(\eta)}\sqrt{A(x)^2+\eta^2}\,
dx\,,
\label{eq:WKBformula}
\end{equation}
for $\eta\in (0,iA)$, where $x_\pm(\eta)$ are the two real turning
points.  In (\ref{eq:WKBdensity}), $A=A(0)$ is the maximum amplitude of
$A(x)$ and the imaginary segments $(-iA,0)$ and $(0,iA)$ are both
considered to be oriented from bottom to top to define the
differential $d\eta$.  

Note that the function $\rho^0(\eta)$ defined by (\ref{eq:WKBformula})
is well-behaved at $\eta=0$ (in the sense of limit from positive
imaginary values) if the function $A(x)$ decays sufficiently rapidly
for large $x$ (exponential is sufficient), and at $\eta=iA$ if $A(x)$
has nonvanishing curvature at its peak at $x=0$.  We assume both of
these conditions on $A(x)$ in all that follows.  Now, the formal WKB
method does not only provide a guess for the weak-$*$ limit of the
discrete eigenvalue measures, but in fact it defines approximations to
the discrete eigenvalues themselves for each value of $\hbar$.  These
are numbers $\lambda_{\hbar,n}^{\rm WKB}$ lying on the positive
imaginary axis in $(0,iA)$ that satisfy the Bohr-Sommerfeld
quantization condition\index{Bohr-Sommerfeld quantization rule}:
\begin{equation}
-\int_{\lambda_{\hbar,n}^{\rm WKB}}^{iA}\rho^0(\eta)\,d\eta = \hbar (n+1/2)\,,
\end{equation}
for $n=0,1,2,\dots,N-1$, where $N$ is the greatest integer such that
\begin{equation}
N\le -\frac{1}{\hbar}\int_0^{iA}\rho^0(\eta)\,d\eta + \frac{1}{2}\,.
\end{equation}
Corresponding to these approximations there is a discrete measure $d\mu_\hbar^{\rm WKB}$ defined by the formula ({\em cf.} (\ref{eq:discretetrue}))
\begin{equation}
d\mu_\hbar^{\rm WKB}(\eta):=\sum_{k=0}^{N-1}\left[\hbar\delta_{\lambda_{\hbar,k}^{{\rm WKB}*}}-\hbar\delta_{\lambda_{\hbar,k}^{\rm WKB}}\right]\,.
\label{eq:discreteWKB}
\end{equation}
The weak-$*$ convergence of these discrete measures to $d\mu_0^{\rm
WKB}$ is a direct matter to establish and analyze, in contrast with
the convergence of the discrete measures of the true eigenvalues.  We
carry out a detailed convergence analysis of these approximate discrete
measures in \S\ref{sec:spectralasymptotics}.

The ``ground state'' \index{state!ground} eigenvalue
$\lambda_{\hbar,0}^{\rm WKB}$ is characterized by
\begin{equation}
\int_{\lambda_{\hbar,0}^{\rm WKB}}^{iA}\rho^0(\eta)\,d\eta = 
-\frac{\hbar}{2}\,,
\end{equation}
and for symmetry it will be useful to choose a sequence of values of
$\hbar$ converging to zero so that the ``most excited state''
\index{state!most excited} eigenvalue $\lambda_{\hbar,N-1}^{\rm WKB}$ similarly satisfies
\begin{equation}
\int_0^{\lambda_{\hbar,N-1}^{\rm WKB}}\rho^0(\eta)\,d\eta = -\frac{\hbar}{2}\,.
\end{equation}
Thus, we can find a sequence of values $\hbar=\hbar_N$ so that for
each $N=1,2,3,\dots$ there are exactly $N$ WKB eigenvalues and the
ground state and most excited state are equidistant from the endpoints
of the imaginary interval $[0,iA]$ with respect to the measure
$-\rho^0(\eta)\,d\eta$.  This distance from the endpoints is exactly
half of the distance between each of the eigenvalues (with respect to
the same measure).

If in addition to having a single local maximum, and satisfying the
decay and curvature conditions mentioned above, the function $A(x)$ is
also even in $x$, then the proportionality constant $\gamma_k$
associated with each eigenvalue $\lambda_k$, purely imaginary or not,
is always equal to either plus or minus one.  This follows from two
facts.  First, note that since the matrix in (\ref{eq:linearx}) is
trace-free, the Wronskian \index{Wronskian} determinant of any two
solutions for the same value of $\lambda$ is independent of $x$.
Because the Wronskian of two bound states at the same value of
$\lambda$ necessarily vanishes as $x\rightarrow \pm\infty$, this
implies that the $L^2({\mathbb R})$ eigenspace for a given $\lambda$
is at most 1-dimensional.  This fact holds for completely arbitrary
potentials $A(x)$ and $S(x)$.  Second, when $S(x)\equiv 0$ and
$A(x)=A(-x)$, then whenever $(u_1(x),u_2(x))^T$ satisfies
(\ref{eq:linearx}) for some $\lambda$, then so does
$(v_1(x),v_2(x))^T:= (u_2(-x),u_1(-x))^T$.  Since bound states are
nondegenerate, for $S(x)\equiv 0$ and $A(x)=A(-x)$ each bound state
must be an eigenvector of this involution, whose eigenvalues are $\pm
1$.  Now, if the difference between the Zakharov-Shabat problem and
the Schr\"odinger equation can be neglected for small $\hbar$, then
from the Sturm-Liouville oscillation theorem\index{Sturm-Liouville
oscillation theorem}, one expects that the proportionality constant
simply alternates between the two values $\pm 1$ from one eigenvalue
to the next along the imaginary axis.  Thus, one is led to propose an
approximate interpolation formula $\gamma_j\approx
\gamma_{\hbar,j}^{\rm WKB}:=Q_K\exp(X_K(\lambda_{\hbar,j}^{\rm
WKB})/\hbar)$ where
\begin{equation}
Q_K:= i(-1)^K\,,\hspace{0.3 in}
X_K(\lambda):=i\pi (2K+1)\int_\lambda^{iA}\rho^0(\eta)\,d\eta\,,
\label{eq:propconstsapprox}
\end{equation}
and $K$ is an arbitrary fixed integer.  This formula gives values of
the proportionality constant that vary from $1$ to $-1$ from each WKB
eigenvalue to the next, starting with $\gamma_{\hbar,0}^{\rm WKB}=1$ for the
WKB ground state $\lambda_{\hbar,0}^{\rm WKB}$.

We now collect these formal calculations into a definition for future
reference:
\begin{definition}
\label{def:WKBspectrum}
Let $A(x)$ be a positive real-valued, even, bell-shaped function of
$x$, and let $A=A(0)$ denote the maximum value.  Let the function
$\rho^0(\eta)$ be defined for $\eta\in (0,iA)$ by
(\ref{eq:WKBdensity}).  Suppose further that $A(x)$ has nonvanishing
curvature at its peak and decays sufficiently rapidly for large $x$ so
that $\rho^0(\eta)$ has a continuous extension to the closed imaginary
interval $[0,iA]$.  For each positive integer $N$, define $\hbar_N$
by
\begin{equation}
\hbar_N:=-\frac{1}{N}\int_0^{iA}\rho^0(\eta)\,d\eta\,.
\label{eq:quantumsequence}
\end{equation}
Then, the {\bf WKB scattering data} \index{WKB scattering data} of the
potential $\psi(x)=A(x)$ is defined as follows for $\hbar=\hbar_N$.
The $L^2({\mathbb R})$ eigenvalues are the set of $N$ numbers
$\lambda_{\hbar_N,n}^{\rm WKB}$ in the interval $(0,iA)$ that satisfy
\begin{equation}
-\int_{\lambda_{\hbar_N,n}^{\rm WKB}}^{iA}
\rho^0(\eta)\,d\eta = \hbar_N(n+1/2)\,,
\mbox{  for  }n=0,1,2,\dots,N-1\,.
\end{equation}
The proportionality constant $\gamma_{\hbar_N,n}^{\rm WKB}$ 
corresponding to the
eigenvalue $\lambda_{\hbar_N,n}^{\rm WKB}$ is given by
\begin{equation}
\gamma_{\hbar_N,n}^{\rm WKB}=
Q_K\exp(X_K(\lambda_{\hbar_N,n}^{\rm WKB})/\hbar_N)\,,
\end{equation}
where $K$ is any integer and 
\begin{equation}
Q_K=i(-1)^K\,,\hspace{0.3 in}X_K(\lambda)=i\pi(2K+1)\int_\lambda^{iA}\rho^0(\eta)\,d\eta\,.
\end{equation}
Finally, the reflection coefficient $r_{\hbar_N}^{\rm WKB}(\lambda)\equiv 0$.
We call the exact solution of the focusing nonlinear Schr\"odinger
equation corresponding to this set of scattering data for arbitrary
$N$ and $\hbar=\hbar_N$ the {\bf semiclassical soliton ensemble}
\index{semiclassical soliton ensemble} associated with the function $A(x)$.
\label{def:WKB}
\end{definition}

We make no attempt here to discuss the validity of these formulae from
the point of view of direct scattering theory.  Instead, we will adopt
the approach of beginning with the reflectionless approximate WKB
spectrum, and working out a completely rigorous inverse-scattering
theory for this spectral data valid for sufficiently large $N$, which
corresponds to sufficiently small $\hbar$.  In the context of the
semiclassical analysis of the initial-value problem for the focusing
nonlinear Schr\"odinger equation, our procedure amounts to an {\em a
priori} modification of the initial data, in which we replace the
$\hbar$-independent initial data $\psi(x,0)\equiv A(x)$ by a sequence
of $\hbar$-dependent initial data $\psi(x,0)\equiv A_N(x)$ which for
each $N$ is the unique potential whose {\em exact} scattering data is
the reflectionless formal WKB approximation to the scattering data of
$A(x)$ described in detail above.  There are cases of particular
functions $A(x)$ however, in which each element $A_N(x)$ of the
sequence of functions turns out to be equal to $A(x)$, which means
that the WKB approximation is exact.  For these cases, the inverse
theory that we will shortly develop does indeed provide rigorous
asymptotics for the semiclassical limit, without any further
arguments.

\begin{remark}
Although rigorous statements about the errors of the WKB approximation
are lacking for the case of $S(x)\equiv 0$, there are reasons for
confidence that the above formulae are indeed valid, and moreover the
properties of these formulae are well-understood since they are the
same as in the classical Schr\"odinger case.  By constrast, the
asymptotic behavior of the spectrum of (\ref{eq:linearx}) when
$S(x)\not\equiv 0$ is only beginning to be explored even at the
qualitative level.  For analytic potentials at least, the eigenvalues
appear to accumulate on unions of curves in the complex plane that can
be quite complicated.  See \cite{B96}, \cite{B00}, and \cite{M00} for
more details on these spectra.
\end{remark}

\section[Asymptotics of WKB Spectra]{Asymptotic properties of the discrete WKB spectrum.}
\label{sec:spectralasymptotics}
We begin by defining a particular branch of the logarithm.
\begin{definition}
Let $L_\eta^0(\lambda)$ denote the particular branch of
$\log(\lambda-\eta)$, considered as a function of $\lambda$, that
agrees with the principal branch $\mbox{Log}(\lambda-\eta)$ for
$\lambda-\eta\in{\mathbb R}_+$, and that is cut from $\lambda=\eta\in
i{\mathbb R}$ down along the imaginary axis to $-i\infty$.  In terms of
the standard cut of the principal branch, 
\begin{equation}
L^0_\eta(\lambda):={\rm
Log}(-i(\lambda-\eta))+\frac{i\pi}{2}\,.
\end{equation}
That is, we are taking $\arg(\lambda-\eta)\in (-\pi/2,3\pi/2)$.
\label{def:Lzerodef}
\end{definition}
In our study of the inverse-scattering problem, we will shortly be
interested in the integral
\begin{equation}
I^0(\lambda):=\frac{1}{\hbar}\int L_\eta^0(\lambda)d\mu_0^{\rm
WKB}(\eta)=\frac{1}{\hbar}\int_0^{iA}L_\eta^0(\lambda)\rho^0(\eta)\,d\eta
+
\frac{1}{\hbar}\int_{-iA}^0L_\eta^0(\lambda)\rho^0(\eta^*)^*\,d\eta\,,
\end{equation}
and its difference from 
\begin{equation}
I^\hbar(\lambda):=\frac{1}{\hbar}\int L_\eta^0(\lambda)
d\mu_\hbar^{\rm WKB}(\eta)\,,
\end{equation}
where $\lambda$ lies in the complex plane away from the imaginary axis
between $-iA$ and $iA$.  Note that
\begin{equation}
\exp(I^\hbar(\lambda))=\prod_{n=0}^{N-1}\frac{\lambda-\lambda_{\hbar,n}^{{\rm WKB}*}}{\lambda-\lambda_{\hbar,n}^{\rm WKB}}\,.
\end{equation}
Since $S(x)\equiv 0$, the support of the WKB eigenvalue measure is
confined to the imaginary axis, and we therefore have
$\rho^0(\eta^*)^* \equiv -\rho^0(-\eta)$.  Now, for $\eta\in[0,iA]$,
define the mass integral \index{mass integral} by
\begin{equation}
m(\eta):=-\int_0^\eta\rho^0(\xi)\,d\xi\,.
\end{equation}
Since $-i\rho^0(\eta)>0$ for all $\eta\in [0,iA]$, this function is
invertible, with an inverse that we denote by $\eta=e(m)$.  Also, let
$M:=m(iA)$, so that $m([0,iA])=[0,M]$.  Then, with a change of
variables, we find
\begin{equation}
\int_0^{iA}L_\eta^0(\lambda)\rho^0(\eta)\,d\eta = -\int_0^M L^0_{e(m)}
(\lambda)\,dm\,.
\end{equation}
Similarly, we find
\begin{equation}
\int_{-iA}^0L_\eta^0(\lambda)\rho^0(\eta^*)^*\,d\eta =
\int_0^ML^0_{-e(m)}(\lambda)\,dm\,.
\end{equation}
Therefore,
\begin{equation}
I^0(\lambda)=\frac{1}{\hbar}\int_0^M
\left[L_{-e(m)}^0(\lambda)-L_{e(m)}^0(\lambda)\right]\,dm\,.
\end{equation}

Note that by Definition~\ref{def:WKB} of the WKB spectrum,
$e(m_n)=\lambda_{\hbar_N,n}^{\rm WKB}$ where $m_n:=M-\hbar(n+1/2)$.
Because the sequence of values of $\hbar$ is such that the points
$m_n$ are symmetrically placed in the interval $[0,M]$, the integral
can be easily represented in the form
\begin{equation}
I^0(\lambda)=\sum_{n=0}^{N-1}I^0_n(\lambda)\,,
\end{equation}
with
\begin{equation}
I^0_n(\lambda):=\frac{1}{\hbar}\int_{m_n-\hbar/2}^{m_n+\hbar/2}
\left[L_{-e(m)}^0(\lambda)-L_{e(m)}^0(\lambda)
\right]\,dm\,.
\end{equation}
The midpoint rule \index{midpoint rule} approximation for
$I^0(\lambda)$ comes from approximating $I^0_n(\lambda)$ simply by the
value of the integrand at $m=m_n$.  That is, we write
\begin{equation}
I^0_n(\lambda)=L_{-e(m_n)}^0(\lambda) - L_{e(m_n)}^0(\lambda) +
\mbox{error terms}\,,
\end{equation}
and we are reminded that $e(m_n)=\lambda_{\hbar_N,n}^{\rm WKB}$ and 
$-e(m_n)=\lambda_{\hbar_N,n}^{{\rm WKB}*}$.
Let us introduce the notation 
\begin{equation}
\tilde{I}_n(\lambda):= I^0_n(\lambda) -
\left[L_{-e(m_n)}^0(\lambda)-L_{e(m_n)}^0(\lambda)\right]\,,
\end{equation}
and 
\begin{equation}
\tilde{I}(\lambda):=\sum_{n=0}^{N-1}\tilde{I}_n(\lambda)=I^0(\lambda)-
I^\hbar(\lambda)\,.
\end{equation}
The following paragraphs will describe the asymptotic behavior of
$\tilde{I}(\lambda)$ as $\hbar_N\downarrow 0$, for various regimes of
$\lambda$.

\subsection{Asymptotic behavior for $\lambda$ fixed.}
For $\lambda$ fixed, the midpoint rule is accurate to second order in
$\hbar=\hbar_N$.  To see this, expand the integrand as follows:
\begin{equation}
\begin{array}{rcl}
\displaystyle L_{-e(m)}^0(\lambda)-L_{e(m)}^0(\lambda) &=& 
\displaystyle L_{-e(m_n)}^0(\lambda)-L_{e(m_n)}^0(\lambda) \\\\
&&\displaystyle \,\,+\,\,
\frac{2e'(m_n)\lambda}{\lambda^2-e(m_n)^2}(m-m_n)
\\\\
&&\displaystyle
\,\,+\,\, 
\int_{m_n}^m d\zeta \,\int_{m_n}^\zeta d\xi\,
\left[\frac{2e''(\xi)\lambda^3-2e''(\xi)e(\xi)^2\lambda+4e'(\xi)^2e(\xi)\lambda}{(\lambda^2-e(\xi)^2)^2}\right]\,.
\end{array}
\label{eq:integrand}
\end{equation}
Because the interval of integration for $I^0_n(\lambda)$ is symmetric
about $m_n$, the linear term in the expansion integrates to zero.
Thus,
\begin{equation}
\tilde{I}_n(\lambda) = 
\frac{1}{\hbar}\int_{m_n-\hbar/2}^{m_n+\hbar/2}dm\,
\int_{m_n}^m d\zeta \,\int_{m_n}^\zeta d\xi\,
\left[\frac{2e''(\xi)\lambda^3-
2e''(\xi)e(\xi)^2\lambda+4e'(\xi)^2e(\xi)\lambda}
{(\lambda^2-e(\xi)^2)^2}\right]\,.
\end{equation}
For $\lambda$ in the upper half-plane uniformly bounded away from the
imaginary segment $[0,iA]$, the denominator of the quadratic term is
bounded away from zero.  From our assumptions on $A(x)$, $e'(\cdot)$
and $e''(\cdot)$ are uniformly bounded functions.  Of course we
automatically have $|e(\cdot)|\le A$.  Under these conditions, we
easily get a bound on the quantity in square brackets that is uniform
with respect to $n$:
\begin{equation}
\left|\frac{2e''(\xi)\lambda^3-2e''(\xi)e(\xi)^2\lambda+4e'(\xi)^2e(\xi)\lambda}{(\lambda^2-e(\xi)^2)^2}\right|\le K_1\,.
\label{eq:K}
\end{equation}
It follows that
\begin{equation}
\left|\tilde{I}_n(\lambda)\right|
\le \frac{2K_1}{\hbar}\int_{m_n}^{m_n+\hbar/2}dm\, \int_{m_n}^m
d\zeta\,\int_{m_n}^\zeta d\xi =
\frac{2K_1}{\hbar}\frac{1}{6}\left(\frac{\hbar}{2}\right)^3=\frac{K_1\hbar^2}{24}\,.
\label{eq:crudeerror}
\end{equation}
Since for sufficiently small $\hbar$, there exists a constant $K_2$
such that $N\le K_2/\hbar$, summing over $n$ gives
\begin{equation}
\left|\tilde{I}(\lambda)
\right|\le \frac{K_1K_2\hbar}{24}\,.
\end{equation}
It is the relative error that is of second order in $\hbar$ here,
since the absolute error is order $\hbar$ and $I^0(\lambda)$ itself is
of order $\hbar^{-1}$.

\subsection{Letting $\lambda$ approach the origin.}
This estimate fails if $\lambda$ approaches the origin.  We can
improve upon the estimate by assuming first of all that we are dealing
with values of $\lambda=|\lambda|e^{i\theta}$ in the upper half-plane
such that $|\cos(\theta)|\ge \delta>0$.  That is, we prevent
$\lambda$ from coming within a symmetrical sector containing the
imaginary axis.

The idea is to refine the estimate (\ref{eq:K}).  We will use the
following estimates of the denominators.  First, because $\lambda$
lies in the upper half-plane and $e(\xi)$ is positive imaginary,
\begin{equation}
|\lambda+e(\xi)|\ge |\lambda-e(\xi)|\,.
\end{equation}
Furthermore, we get both
\begin{equation}
|\lambda-e(\xi)|\ge \delta |\lambda| \mbox{  and  }
|\lambda-e(\xi)|\ge \delta |e(\xi)|\,.
\end{equation}
Also, we will use the partial fraction expansion
\begin{equation}
\frac{2e''(\xi)\lambda^3-2e''(\xi)e(\xi)^2\lambda + 4e'(\xi)^2e(\xi)\lambda}
{(\lambda^2-e(\xi)^2)^2}=
\frac{e''(\xi)}{\lambda+e(\xi)}+\frac{e''(\xi)}{\lambda-e(\xi)}-
\frac{e'(\xi)^2}{(\lambda+e(\xi))^2}+\frac{e'(\xi)^2}{(\lambda-e(\xi))^2}\,.
\label{eq:rewrite}
\end{equation}

From this we see that one estimate of the integrand in
(\ref{eq:integrand}) is
\begin{equation}
\left|\frac{2e''(\xi)\lambda^3-2e''(\xi)e(\xi)^2\lambda + 
4e'(\xi)^2e(\xi)\lambda}{(\lambda^2-e(\xi)^2)^2}\right|
\le \frac{2E_2}{\delta|\lambda|} + \frac{2E_1^2}{\delta^2|\lambda|^2}\,,
\end{equation}
where $E_1$ is the supremum of $|e'(\xi)|$ and $E_2$ is that of
$|e''(\xi)|$, taken over the whole interval $\xi\in (0,M)$.  As was the case
with the estimate (\ref{eq:K}), this estimate does not depend on $\xi$ nor on
$n$, so by the same reasoning as that leading to (\ref{eq:crudeerror}) we
get
\begin{equation}
\left|\tilde{I}_n(\lambda)
\right|\le \frac{E_2\hbar^2}{12\delta |\lambda|} + \frac{E_1^2\hbar^2}
{12\delta^2|\lambda|^2}\,.
\end{equation}
If we further assume that $|\lambda|$ is uniformly bounded, then we
can find a constant $C>0$ depending on $E_1$, $E_2$, $\delta$, and the
bound on $|\lambda|$ so that the right-hand side of this estimate does
not exceed $C\hbar^2/|\lambda|^2$, which gives
\begin{equation}
\left|\tilde{I}_n(\lambda)
\right|\le \frac{C\hbar^2}{|\lambda|^2}\,.
\label{eq:nearfield}
\end{equation}

On the other hand, another estimate of the integrand in (\ref{eq:integrand})
is 
\begin{equation}
\left|\frac{2e''(\xi)\lambda^3-2e''(\xi)e(\xi)^2\lambda + 
4e'(\xi)^2e(\xi)\lambda}{(\lambda^2-e(\xi)^2)^2}\right|
\le \frac{2E_2}{\delta |e(\xi)|} + \frac{2E_1^2}{\delta^2 |e(\xi)|^2}\,.
\end{equation}
Since $|e(\xi)|\le A<\infty$, we get $|e(\xi)|\ge |e(\xi)|^2/A$ and 
therefore with $D=2AE_2/\delta + 2E_1^2/\delta^2$, we get
\begin{equation}
\left|\frac{2e''(\xi)\lambda^3-2e''(\xi)e(\xi)^2\lambda + 
4e'(\xi)^2e(\xi)\lambda}{(\lambda^2-e(\xi)^2)^2}\right|
\le \frac{D}{|e(\xi)|^2}\,.
\end{equation}
Now, we use the fact that for any continuous function $g(\xi)$ we have
\begin{equation}
\begin{array}{rcl}
\displaystyle
\int_{m_n-\hbar/2}^{m_n+\hbar/2}dm\,\int_{m_n}^md\zeta\,\int_{m_n}^\zeta d\xi\,
g(\xi) &=&\displaystyle
 \int_{m_n}^{m_n+\hbar/2}dm\,\int_{m_n}^md\zeta\,\int_{m_n}^\zeta 
d\xi\, g(\xi) \\\\
&&\,\,+\,\,\displaystyle
\int_{m_n-\hbar/2}^{m_n}dm\,\int_{m}^{m_n}d\zeta\,\int_{\zeta}^{m_n} 
d\xi\, g(\xi)\,,
\end{array}
\end{equation}
where all volume elements are positively oriented in the integrals on
the right-hand side, to get an estimate like (\ref{eq:crudeerror}).  That is,
we find
\begin{equation}
\left|\tilde{I}_n(\lambda)
\right|\le \frac{D}{\hbar}\left(
\int_{m_n}^{m_n+\hbar/2}dm\,\int_{m_n}^m d\zeta\,
\int_{m_n}^\zeta \frac{d\xi}{|e(\xi)|^2} +
\int_{m_n-\hbar/2}^{m_n}dm\,\int_m^{m_n}d\zeta\,\int_\zeta^{m_n}
\frac{d\xi}{|e(\xi)|^2}\right)\,.
\end{equation}
Now since $|e(\xi)|$ is by construction an increasing function of $\xi$,
we have $|e(\xi)|\ge |e(m_n-\hbar/2)|$ in both integrals.  Thus,
\begin{equation}
\left|\tilde{I}_n(\lambda)
\right|\le \frac{D\hbar^2}{24|e(m_n-\hbar/2)|^2}\,.
\label{eq:farfield}
\end{equation}

To estimate the total error, we combine the two estimates
(\ref{eq:nearfield}) and (\ref{eq:farfield}).  In particular, pick an
integer $L$ with $0<L<N$, and use (\ref{eq:farfield}) to estimate the terms
with $0\le n\le L-1$ and (\ref{eq:nearfield}) to estimate the terms with
$n\ge L$.  Thus,
\begin{equation}
\left| \tilde{I}(\lambda)
\right|\le
\frac{D\hbar^2}{24}\sum_{n=0}^{L-1}\frac{1}{|e(m_n-\hbar/2)|^2} +
(N-L)\frac{C\hbar^2}{|\lambda|^2}\,.
\end{equation}
Now, again since $|e(\xi)|$ is an increasing function, we easily obtain
\begin{equation}
\sum_{n=0}^{L-1}\frac{1}{|e(m_n-\hbar/2)|^2}\le \frac{1}{\hbar}\int_{m_{L-1}-3\hbar/2}^{m_0-3\hbar/2}\frac{dm}{|e(m)|^2}=
\frac{1}{\hbar}\int_{M-\hbar L-\hbar}^{M-2\hbar}\frac{dm}{|e(m)|^2}\,.
\end{equation}
Also, since $|e(m)|$ is increasing and $e(0)=0$ while $e'(0)\neq 0$,
there is some constant $F>0$ such that $|e(m)|\ge F m$.  Therefore
\begin{equation}
\int_{M-\hbar L-\hbar}^{M-2\hbar}\frac{dm}{|e(m)|^2}\le
\frac{1}{F^2}\int_{M-\hbar L-\hbar}^{M-2\hbar}\frac{dm}{m^2} =
\frac{1}{F^2}\left(\frac{1}{M-\hbar L-\hbar}-\frac{1}{M-2\hbar}\right)\,.
\end{equation}
Our estimate of the total error thus becomes
\begin{equation}
\left| \tilde{I}(\lambda)\right|\le \frac{D\hbar}{24F^2}
\left(\frac{1}{M-\hbar L -\hbar}-\frac{1}{M-2\hbar}\right)
+ (N-L)\frac{C\hbar^2}{|\lambda|^2}\,.
\end{equation}

We now consider how to optimally choose the cutoff $L$.  Let
$\beta>0$, and take $L$ to be the largest integer so that $M-\hbar
L-\hbar \ge \hbar^\beta$.  The quantity $N-L$ is then of order
$\hbar^{\beta-1}$, and we thus have
\begin{equation}
\left| \tilde{I}(\lambda)\right| = 
{\cal O}(\hbar^{1-\beta}) + 
{\cal O}(\hbar^{1+\beta} |\lambda|^{-2})\,.
\end{equation}
If we assume that $|\lambda|\ge C\hbar^\gamma$ for some constant $C>0$ and
for some $\gamma\ge 0$, then the estimate becomes
\begin{equation}
\left|\tilde{I}(\lambda)\right| = 
{\cal O}(\hbar^{1-\beta}) + {\cal O}(\hbar^{1+\beta-2\gamma})\,.
\end{equation}
As long as $\gamma<1$, the best error for $0<\beta<1$ is established
via a dominant balance which occurs for $\beta=\gamma$.  This gives an
optimal estimate of
\begin{equation}
\left| \tilde{I}(\lambda)\right| = 
{\cal O}(\hbar^{1-\gamma})\,.
\end{equation}
Thus, the error that we found to be order $\hbar$ for $\lambda$
fixed, is in fact small as long as $|\lambda|\gg \hbar$.  This error
estimate is uniform for $\lambda$ big enough compared to $\hbar$ and
lying in the upper half-plane outside of any given sector containing
the imaginary axis.

\subsection{Approximations uniformly valid for $\lambda$ near the origin.}
The error $\tilde{I}(\lambda)$ is not uniformly small if $\lambda={\cal
O}(\hbar)$.  However, we may extract an additional contribution and
then show that what remains is uniformly small for $\lambda$ in some
neighborhood of the origin.

Although $\tilde{I}(\lambda)$ is not small, the total contribution of
most of the terms $\tilde{I}_n(\lambda)$ for which $|e(m_n)|$ is large
enough compared to $\hbar$ will in fact be negligible.  So again, we
introduce a cutoff integer $L$, and then according to our previous
results, we get
\begin{equation}
\tilde{I}(\lambda)=\sum_{n=0}^{L-1}\tilde{I}_n(\lambda) + 
\sum_{n=L}^{N-1}\tilde{I}_n(\lambda)\,,
\end{equation}
and we have the estimate
\begin{equation}
\left|\sum_{n=0}^{L-1}\tilde{I}_n(\lambda)
\right|\le \frac{D\hbar}{24F^2}\left(\frac{1}{M-\hbar L-\hbar}-\frac{1}{M-2\hbar}\right)\,.
\end{equation}
As before, it will be convenient to pick a number $\beta>0$ and
then choose $L$ to be the largest integer so that $M-\hbar L-\hbar\ge
\hbar^\beta$.  The above estimate then becomes a $\lambda$-independent
${\cal O}(\hbar^{1-\beta})$.

In each remaining term $\tilde{I}_n(\lambda)$, both $m_n$ and the
integration variable $m$ will be very close to zero, and it appears
that it may be prudent to expand $\tilde{I}_n(\lambda)$ to reflect
this fact.  Thus, we write
\begin{equation}
\tilde{I}_n(\lambda) = J_n(\lambda) + \tilde{J}_n(\lambda)\,,
\end{equation}
where
\begin{equation}
J_n(\lambda):=
\frac{1}{\hbar}
\int_{m_n-\hbar/2}^{m_n+\hbar/2}
\left[L^0_{-e'(0)m}(\lambda)-L^0_{e'(0)m}(\lambda)\right]\,dm 
- \left[L^0_{-e'(0)m_n}(\lambda)-L^0_{e'(0)m_n}(\lambda)\right]\,.
\label{eq:Jdef}
\end{equation}

We begin by estimating $\tilde{J}_n(\lambda)$ for 
$L\le n \le N-1$.  Explicitly,
\begin{equation}
\begin{array}{rcl}
\tilde{J}_n(\lambda)&=&\displaystyle
\frac{1}{\hbar}\int_{m_n-\hbar/2}^{m_n+\hbar/2}
\left[L^0_{-e(m)}(\lambda)-L^0_{-e'(0)m}(\lambda)-L^0_{e(m)}(\lambda)+
L^0_{e'(0)m}(\lambda)\right]\,dm\\\\&&\displaystyle
\,\, -\,\, \left[
L^0_{-e(m_n)}(\lambda)-L^0_{-e'(0)m_n}(\lambda)-L^0_{e(m_n)}(\lambda)+
L^0_{e'(0)m_n}(\lambda)\right]\,.
\end{array}
\end{equation}
Recall Definition \ref{def:Lzerodef} of $L^0_\eta(\lambda)$.  One
estimate of $\tilde{J}_n(\lambda)$ comes from grouping the logarithms
in pairs as follows.  Imagining either $z=m$ or $z=m_n$, and using
Taylor's theorem with remainder\index{Taylor's theorem with
remainder}, we have
\begin{equation}
L^0_{\pm e(z)}(\lambda)-L^0_{\pm e'(0)z}(\lambda)=\mbox{Log}\left(1\mp\frac{e''(\xi_\pm)z^2}{2(\lambda\mp e'(0)z)}\right)\,,
\end{equation}
where the $\xi_\pm$ lie between zero and $z$.  Now since $\lambda$ is
excluded from the symmetrical sector about the positive imaginary axis
subtended by an angle of $2\sin^{-1}(\delta)$, and since $e'(0)$ is
positive imaginary while $z>0$ is real,
\begin{equation}
|\lambda\mp e'(0)z|\ge \delta |e'(0)| z\,.
\end{equation}
Therefore, we find that for $z$ sufficiently small there is a constant
$G>0$ such that
\begin{equation}
\left|L^0_{\pm e(z)}(\lambda)-L^0_{\pm e'(0)z}(\lambda)\right|
\le \frac{G\sup |e''|}{2\delta |e'(0)|} z = Hz\,.
\end{equation}
Consequently, we find
\begin{equation}
\left|\tilde{J}_n(\lambda)\right|\le 4H (m_n+\hbar/2) + 4H m_n\,,
\end{equation}
because in the integral we have $m<m_n+\hbar/2$.  Additionally since
$m_n\ge \hbar/2$, we have $m_n+\hbar/2\le 2m_n$.  Therefore, we also have
\begin{equation}
\left|\tilde{J}_n(\lambda)\right|\le 12H m_n\,.
\end{equation}
So, summing over $L\le n\le N-1$, we find that
\begin{equation}
\left|\sum_{n=L}^{N-1}\tilde{J}_n(\lambda)\right|\le 12H\sum_{n=L}^{N-1}m_n
= 12H(N-L)\left[(M-\hbar L+\hbar/2)-\hbar\frac{N-L+1}{2}\right]\,.
\end{equation}
Now, $N-L$ and $N-L+1$ are both order $\hbar^{\beta-1}$, while 
$M-\hbar L+\hbar/2$ is order $\hbar^\beta$.  Therefore, we find that
both terms are of the same order and we have
\begin{equation}
\sum_{n=L}^{N-1}\tilde{J}_n(\lambda) = {\cal O}(\hbar^{2\beta-1})\,.
\end{equation}
This estimate is uniform for bounded $\lambda$ and will be small as
long as $\beta> 1/2$.

Now we return to the quantities $J_n(\lambda)$ defined by
(\ref{eq:Jdef}).  The integrals in $J_n(\lambda)$ can be evaluated exactly, and
considerable simplification follows.  In fact, if we set
$\lambda=e'(0)\hbar w$, then we have, exactly,
\begin{equation}
\exp\left(\sum_{n=L}^{N-1} J_n(\lambda)\right) = 
w^{-w}(-w)^{-w}\frac{\Gamma(1/2+w)}{\Gamma(1/2-w)}\frac{(\overline{N}+w)^{\overline{N}+w}}
{(\overline{N}-w)^{\overline{N}-w}}\frac{\Gamma(\overline{N}+1/2-w)}{\Gamma(\overline{N}+1/2+w)}\,,
\label{eq:exactly}
\end{equation}
where $\overline{N}:=N-L$.  Stirling's formula \index{Stirling's
formula} says:
\begin{equation}
\Gamma(z)=e^{-z}z^{z-1/2}\sqrt{2\pi}(1+{\cal O}(|z|^{-1}))\,,
\end{equation}
where the error is uniform with respect to direction for $z$ in any
sector $-\pi<-\phi\le \mbox{Arg}(z)\le \phi < \pi$.  If $\overline{N}$
is large, which is the same thing as saying that $\beta<1$, we can
apply this formula to (\ref{eq:exactly}), obtaining errors of the form
${\cal O}((\overline{N}+1/2\pm w)^{-1})$.  But because $w$ is
prevented from entering the symmetrical sector about the positive real
axis with opening angle $2\sin^{-1}(\delta)$, it is not difficult to
see that these terms are of order ${\cal O}(\delta/\overline{N})$, or
for $\delta>0$ fixed, simply ${\cal O}(\overline{N}^{-1})$ uniformly
for $w$ in the right half-plane without the sector of angular width
$2\sin^{-1}(\delta)$ about the positive real axis.  The upshot of
these considerations is that one finds
\begin{equation}
\exp\left(\sum_{n=L}^{N-1} J_n(\lambda)\right) = W(w)(1+{\cal
O}(\overline{N}^{-1}))\,,
\end{equation}
uniformly as $\overline{N}$ tends to infinity with $w$ outside
any fixed sector of the positive real axis of fixed angle, where
\begin{equation}
W(w):=e^{2w}w^{-w}(-w)^{-w}\frac{\Gamma(1/2+w)}{\Gamma(1/2-w)}\,.
\label{eq:Wdef}
\end{equation}
The reader is reminded that all exponential functions are defined
using the traditional cut of the principal branch of the logarithm;
thus $W(w)$ is defined for $\Re(w)>0$ and $\Im(w)\neq 0$.  There is a
cut on the positive real $w$ axis corresponding to the positive
imaginary axis in the $\lambda$-plane.  Note that
$\overline{N}={\cal O}(\hbar^{\beta-1})$.

We thus see that while $\tilde{I}(\lambda)$ itself is not small
uniformly in $\lambda$, we can write
\begin{equation}
\exp(\tilde{I}(\lambda)) = W(w)(1+{\cal O}(\hbar^{1-\beta}))\exp({\cal O}(\hbar^{2\beta-1}))\exp({\cal O}(\hbar^{1-\beta})) = W(w)(1+{\cal O}(\hbar^{1-\beta})+
{\cal O}(\hbar^{2\beta-1}))\,.
\end{equation}
The dominant balance determining the optimal exponent occurs for $\beta=2/3$,
which gives the statement
\begin{equation}
\exp(\tilde{I}(\lambda))=W(w)(1+{\cal O}(\hbar^{1/3}))\,,
\end{equation}
where the error is {\em uniformly} small for $\lambda$ in any bounded
region of the upper half-plane minus the sector about the imaginary
axis of angular width $2\sin^{-1}(\delta)$.

\subsection{Convergence theorems for discrete WKB spectra.}
The work in the above paragraphs establishes the following two theorems.
\begin{theorem}[Near-field spectral asymptotics]
\index{spectral asymptotics!near-field}
Let $\delta>0$ be given and consider $\lambda=|\lambda|e^{i\theta}$ to
lie in a bounded set $\Lambda$ in the upper half-plane such that
$|\cos(\theta)|\ge\delta$.  Then, there is a constant $B_{\rm in}>0$
such that for $N$ sufficiently large, the WKB eigenvalues satisfy
\begin{equation}
\left|\left[\prod_{n=0}^{N-1}\frac{\lambda-\lambda_{\hbar_N,n}^{{\rm WKB}*}}
{\lambda-\lambda_{\hbar_N,n}^{\rm WKB}}
\right]\exp(-I^0(\lambda))W(w)-1\right|\le B_{\rm in}\hbar_N^{1/3}\,,
\end{equation}
uniformly for $\lambda\in\Lambda$, where
$w=\lambda/(e'(0)\hbar_N)=-\rho^0(0)\lambda/\hbar_N$, and where the
canonical function $W(w)$ is defined by (\ref{eq:Wdef}).
\label{theorem:inner}
\end{theorem}

\begin{theorem}[Far-field spectral asymptotics]
\index{spectral asymptotics!far-field}
Suppose the conditions of Theorem~\ref{theorem:inner}, but also
suppose that the set $\Lambda$ contains only values $\lambda$ satisfying
$|\lambda|^{-1} ={\cal O}(\hbar_N^{-\gamma})$ for some $\gamma>0$,
({\em i.e.} $\lambda$ is bounded away from the
origin by an asymptotically small amount), then there is a constant
$B_{\rm out}>0$ such that for all $N$ sufficiently large,
\begin{equation}
\left|\left[\prod_{n=0}^{N-1}\frac{\lambda-\lambda_{\hbar_N,n}^{{\rm WKB}*}}
{\lambda-\lambda_{\hbar_N,n}^{\rm WKB}}
\right]\exp(-I^0(\lambda))-1\right|\le B_{\rm out}\hbar_N^{1-\gamma}\,.
\end{equation}
Furthermore, if $\lambda$ is held fixed as $N$ increases, and does not
lie on the imaginary segment $[-iA,iA]$, then the sectorial condition
$|\cos(\theta)|\ge\delta$ can be dropped and for large enough $N$,
\begin{equation}
\left|\left[\prod_{n=0}^{N-1}\frac{\lambda-\lambda_{\hbar_N,n}^{{\rm WKB}*}}
{\lambda-\lambda_{\hbar_N,n}^{\rm WKB}}
\right]\exp(-I^0(\lambda))-1\right|\le B_{\rm out}\hbar_N\,.
\end{equation}
\label{theorem:outer}
\end{theorem}

\section[The Satsuma-Yajima Ensemble]{The Satsuma-Yajima semiclassical soliton ensemble.} 
\label{sec:ParticularEnsemble}
There is at least one case of a real, even, single-maximum potential
where the scattering data are known {\em exactly} for all $\hbar$.  In
1974, Satsuma and Yajima
\cite{SY74} considered (essentially) the scattering problem
(\ref{eq:linearx}) for arbitrary $\hbar>0$ and for the special initial
data
\begin{equation}
A(x)=A\,\mbox{sech}(x)\,,\hspace{0.2 in} S(x)\equiv 0\,,
\end{equation}
where $A>0$ is an arbitrary constant.  For this choice, they solved
(\ref{eq:linearx}) explicitly in terms of hypergeometric functions
\index{hypergeometric functions} and obtained
\begin{equation}
b(\lambda)=i\sin(\pi A/\hbar)\,\mbox{sech}(\pi\lambda/\hbar)\,,\hspace{0.2 in}
\lambda\in{\mathbb R}\,.
\label{eq:reflection}
\end{equation}
This implies that the reflection coefficient $r(\lambda)$ vanishes identically
as long as
\begin{equation}
\hbar = \hbar_N = \frac{A}{N}\,,\hspace{0.2 in} N=1,2,3,\dots\,.
\end{equation}
Moreover, if $\hbar=\hbar_N$, then there are exactly $N$ eigenvalues, and
the relevant discrete data is given by
\begin{equation}
\lambda_k = i\hbar_N(N-k-1/2)\,,\hspace{0.2 in}
\gamma_k = (-1)^{k}\,,\hspace{0.2 in}
k=0,\dots,N-1\,.
\label{eq:SYdiscrete}
\end{equation}
When we consider the semiclassical limit in detail for this special
initial data, we will always assume in this paper that
$\hbar=\hbar_N$, that is, the parameter $\hbar$ always takes one of
the ``quantum'' values where there is no contribution to the solution
from the reflection coefficient.  Satsuma and Yajima's calculations
have recently been generalized to some other potentials by Tovbis and
Venakides \cite{TV00}.

It is interesting to compare these exact results with their WKB
approximations.  Using $A(x)=A\,{\rm sech}(x)$ in (\ref{eq:WKBdensity}), 
one finds
\begin{equation}
\rho^0(\eta)=\rho_{\rm SY}^0(\eta)\equiv i\,,\hspace{0.3 in} \eta\in (0,iA)\,.
\end{equation}
Therefore, it is easy to check directly that in this case the WKB
approximations to the true eigenvalues for $\hbar=\hbar_N$ turn out to
be exact:
\begin{equation}
\lambda_{\hbar_N,n}^{\rm WKB} = \lambda_n  \mbox{  for  } A(x)=A\,{\rm sech}(x)\,.
\end{equation}
Also, since $\hbar=\hbar_N$ implies that $b(\lambda)\equiv 0$, the
true scattering data is reflectionless, as is the approximate WKB
scattering data, {\em i.e.} $r(\lambda)\equiv r^{\rm
WKB}(\lambda)\equiv 0$.

For the Satsuma-Yajima initial data \index{Satsuma-Yajima potential}
with $\hbar=\hbar_N$, we may interpolate the true proportionality
constants $\{\gamma_j\}$ at the true eigenvalues $\{\lambda_j\}$ by
very simple expressions.  Thus, using (\ref{eq:SYdiscrete}) we have
the exact expressions
\begin{equation}
\gamma_j = Q_{{\rm SY},K}\exp(X_{{\rm SY},K,K'}(\lambda_j)/\hbar)\,, 
\end{equation}
where
\begin{equation}
Q_{{\rm SY},K}=i(-1)^K\,,\hspace{0.2 in}
X_{{\rm SY},K,K'}(\lambda)=(2K+1)\pi\lambda-(2K'+1)i\pi A\,,
\end{equation}
with $K$ and $K'$ being arbitrary integers that index the
interpolants.  The key feature of these exact formulae in relation to
the semiclassical limit is that the number $Q_{{\rm SY},K}$ and the
analytic function $X_{{\rm SY},K,K'}(\lambda)$ are {\em independent of
$\hbar$}.  Now, when $\hbar=\hbar_N=A/N$, we see that the function
$Q_{{\rm SY},K}\exp(X_{{\rm SY},K,K'}(\lambda)/\hbar)$ is independent
of the parametrizing integer $K'$.  However, the parametrizing integer
$K$ enters in a more useful way (this will become clear in
\S\ref{sec:postponing}).  For later convenience,
we will from now on set $K'=K$ and write $X_{{\rm SY},K}(\lambda)$ for
$X_{{\rm SY},K,K}(\lambda)$.  For $\hbar=\hbar_N$, and for the special
initial data we are considering, the indexing set of interpolants is
just $\mathbb Z$.  Note that for the Satsuma-Yajima data,
we have the exact relation ({\em cf.} equation
(\ref{eq:propconstsapprox}))
\begin{equation}
X_{{\rm SY},K}(\lambda)=i\pi (2K+1)\int_\lambda^{iA}\rho_{\rm SY}^0(\eta)\,d\eta\,.
\end{equation}
Thus, the WKB formula for the proportionality constants, although
unjustified in general, is not only asymptotic but {\em exact} for all
$\hbar=\hbar_N$ in this special case.

So, for the special case of the Satsuma-Yajima initial data
$\psi(x,0)=A\,{\rm sech}(x)$, and for the sequence of ``quantum''
values \index{quantum sequence of values of $\hbar$} of
$\hbar=\hbar_N=A/N$, the true scattering data agrees exactly with its
WKB approximation.  This means that the solution $\psi(x,t)$ to the
focusing nonlinear Schr\"odinger equation with initial data
$\psi(x,0)=A\,{\rm sech}(x)$ is in this case exactly what we have
called the semiclassical soliton ensemble corresponding to the
function $A(x)=A\,{\rm sech}(x)$.  Therefore the rigorous asymptotics
that we will develop below for semiclassical soliton ensembles
corresponding to quite general analytic functions $A(x)$ will provide
without any further argument the rigorous semiclassical asymptotic
description of the solution to the initial value problem
(\ref{eq:IVP}) with the special initial data $\psi_0(x)=A\,{\rm
sech}(x)$.

\chapter{Asymptotic Analysis of the Inverse Problem}
\label{sec:asymptoticanalysis}
In this chapter, we will study the asymptotic behavior, in the limit
of $N$ tending to infinity, of semiclassical soliton ensembles
\index{semiclassical soliton ensemble} corresponding to analytic, even, 
bell-shaped functions $A(x)$ with nonzero curvature at the peak and
sufficient decay for large $x$, so that the density $\rho^0(\eta)$
defined by (\ref{eq:WKBdensity}) has an analytic extension off of the
imaginary interval $(0,iA)$.  This means that we are going to study
Riemann-Hilbert Problem
\ref{rhp:M} for the matrix ${\bf M}(\lambda)$, posed 
with discrete data $\{\lambda_{\hbar_N,0}^{\rm
WKB},\dots,\lambda_{\hbar_N,N-1}^{\rm WKB}\}$ and
$\{\gamma_{\hbar_N,0}^{\rm WKB},\dots,\gamma_{\hbar_N,N-1}^{\rm
WKB}\}$ defined in terms of $A(x)$ by
Definition~\ref{def:WKBspectrum}, in the limit as $N$ tends to
infinity.

\begin{figure}
\begin{center}
\mbox{\psfig{file=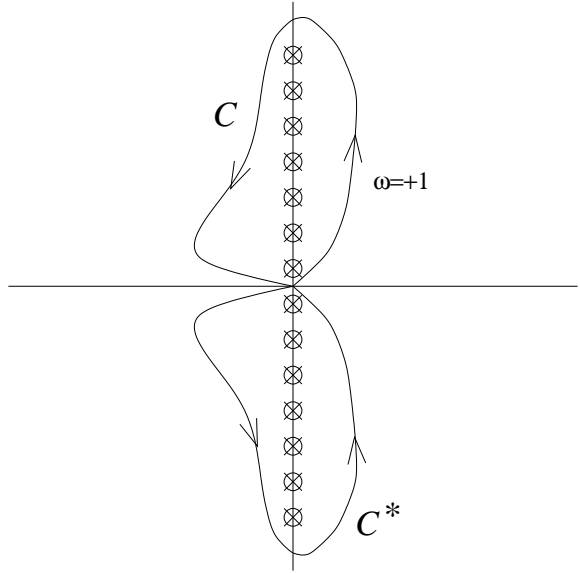,width=3 in}}
\end{center}
\caption{\em Examples of contours $C$ and $C^*$ appropriate for studying
a semiclassical soliton ensemble in the limit $N\uparrow\infty$.}
\label{fig:specificcontours_2}
\end{figure}

In our analysis, we would like to consider using a contour $C$ that is
held fixed as $N$ grows and $\hbar_N$ becomes small.  Because the set
of complex numbers $\{\lambda_{\hbar_N,n}^{\rm WKB}\}$ has zero as an
accumulation point, we need to ensure that $C\subset{\mathbb C}_+$
passes through zero, and in fact we will need this to occur with some
nonzero angle with respect to the imaginary axis (cusps will not be
allowed).  See Figure~\ref{fig:specificcontours_2}.

\section[Introducing the Complex Phase]{Introducing the complex phase.}
\label{sec:phase}
As pointed out in the first remark in Chapter~\ref{sec:solitonRHP}, the
conditions (\ref{eq:relations}) essentially describe a problem of
rational interpolation \index{rational interpolation} of entire
(exponential) functions; the numerator and denominator of the rational
interpolant are the polynomials appearing in (\ref{eq:form}).  These
polynomials are then related to the matrix elements of the solution
${\bf M}(\lambda)$ of the Riemann-Hilbert problem.  As
$\hbar_N\downarrow 0$, and correspondingly as $N\uparrow\infty$, the
degree of interpolation increases, and one expects the behavior of the
interpolants to become increasingly wild (oscillatory and/or large in
magnitude) away from the points of collocation.

In studying the Riemann-Hilbert problem for ${\bf M}(\lambda)$ in
this limit, it is essential to capture this wild behavior.  We therefore
suppose that the matrix ${\bf M}(\lambda)$ can be written in the form
\begin{equation}
{\bf M}(\lambda)={\bf
N}(\lambda)\exp(g(\lambda)\sigma_3/\hbar)\,,
\label{eq:changeofvariables}
\end{equation}
where $g(\lambda)$ is some scalar function, yet to be determined,
called the {\em complex phase}\index{complex phase function}.  At this
point, the matrix ${\bf N}(\lambda)$ is just the quotient in
(\ref{eq:changeofvariables}).  The complex phase function is intended
to capture the wild behavior of ${\bf M}(\lambda)$ as $\hbar\downarrow
0$.  {\em The guiding principle is that $g(\lambda)$ should be chosen
so that the jump matrices for ${\bf N}(\lambda)$ implied by the change
of variables (\ref{eq:changeofvariables}) can be approximated by
``simple'' jump matrices, and so that the error terms involved in the
approximation can be controlled.}

There are several elementary constraints we place on the complex phase
function $g(\lambda)$:
\begin{definition}
The elementary properties of a complex phase function $g(\lambda)$ are:  
\begin{enumerate}
\item
{\bf No $\hbar$ Dependence:}  $g(\lambda)$ is independent of $\hbar$.
\item
{\bf Analyticity:}  $g(\lambda)$ is analytic for
$\lambda\in {\mathbb C}\setminus (C\cup C^*)$.
\item
{\bf Decay:} $g(\lambda)\rightarrow 0$ as
$\lambda\rightarrow\infty$.
\item
{\bf Boundary Behavior:} $g(\lambda)$ assumes
continuous boundary
values from both sides of $C\cup C^*$.
\item
{\bf Symmetry:}
$g(\lambda^*)+g(\lambda)^* = 0$ for
all $\lambda\in {\mathbb C}\setminus (C\cup C^*)$.
\end{enumerate}
\label{def:gbasic}
\end{definition}

Note that the analyticity in a deleted neighborhood of infinity and
the decay at infinity together imply analyticity at infinity ({\em
i.e.} $g$ has a series representation in positive powers of
$\lambda^{-1}$ convergent for sufficiently large $|\lambda |$) and
therefore uniformity of the decay with respect to direction.  The
symmetry condition on $g(\lambda)$ ensures that the
matrix ${\bf N}(\lambda)$ inherits the reflection
symmetry \index{reflection symmetry}
\begin{equation}
{\bf N}(\lambda^*)=
\sigma_2{\bf N}(\lambda)^*\sigma_2\,.
\end{equation}

Let a function $g(\lambda)$ satisfying the five conditions enumerated
above be given.  
For $\lambda\in C_\omega$, define
the functions
\begin{equation}
\theta(\lambda):=iJ(g_+(\lambda)-
g_-(\lambda))\,,
\label{eq:thetadef}
\end{equation}
and
\begin{equation}
\phi(\lambda)
:=
 \int L^0_\eta(\lambda)\,d\mu_{\hbar_N}^{\rm WKB}(\eta) 
+
J\cdot\left( 2i\lambda x + 2i\lambda^2 t - X_K(\lambda) -
g_+(\lambda) -
g_-(\lambda)\right)\,,
\end{equation}
where $L^0_\eta(\lambda)$ is explained in
Definition~\ref{def:Lzerodef}, $d\mu_{\hbar_N}^{\rm WKB}(\eta)$ is the
discrete eigenvalue measure defined by (\ref{eq:discreteWKB})
corresponding to the WKB eigenvalues for $\hbar=\hbar_N$, and
$X_K(\lambda)$ is the interpolant of the WKB proportionality constants
\index{proportionality constants!interpolant of}defined in
Definition~\ref{def:WKBspectrum}.  From these functions, build a
matrix for $\lambda\in C$:
\begin{equation}
{\bf v}_{\bf N}(\lambda):=\sigma_1^{\frac{1-J}{2}}
\left[\begin{array}{cc}
\exp(i\theta(\lambda)/\hbar) & 0 \\\\
-((-1)^Ki)^{-J}\omega\exp(\phi(\lambda)/\hbar) & 
\exp(-i\theta(\lambda)/\hbar)
\end{array}\right]\sigma_1^{\frac{1-J}{2}}
\,.
\label{eq:vN}
\end{equation}
Then, in place of the holomorphic Riemann-Hilbert Problem~\ref{rhp:M}
for ${\bf M}(\lambda)$, we may consider the following Riemann-Hilbert
problem\index{Riemann-Hilbert problem!phase-conjugated}.
\begin{rhp}[Phase-conjugated problem]
Given a complex phase function $g(\lambda)$ satisfying the conditions
of Definition~\ref{def:gbasic}, find a matrix function 
${\bf N}(\lambda)$ that satisfies the following:
\begin{enumerate}
\item{\bf Analyticity:}
${\bf N}(\lambda)$ is analytic for
$\lambda\in{\mathbb C}\setminus(C\cup C^*)$.
\item
{\bf Boundary behavior:} ${\bf N}(\lambda)$ assumes continuous boundary
values on $C\cup C^*$.
\item
{\bf Jump conditions:}  The boundary values taken on $C\cup C^*$ satisfy
\begin{equation}
\begin{array}{rcll}
{\bf N}_+(\lambda)&=&{\bf N}_-(\lambda){\bf v}_{\bf N}(\lambda)
\,,&\hspace{0.2 in}\lambda\in C_\omega\,,\\
{\bf N}_+(\lambda)&=&{\bf N}_-(\lambda)\sigma_2{\bf v}_{\bf N}(\lambda^*)^*
\sigma_2\,,&\hspace{0.2 in}
\lambda\in [C^*]_\omega\,,
\end{array}
\end{equation}
where ${\bf v}_{\bf N}(\lambda)$ is defined by (\ref{eq:vN}).
\item
{\bf Normalization:}
${\bf N}(\lambda)$ is normalized at infinity:
\begin{equation}
{\bf N}(\lambda)\rightarrow{\mathbb I}\mbox{ as }
\lambda\rightarrow\infty\,.
\end{equation}
\end{enumerate}
\label{rhp:N}
\end{rhp}

\begin{proposition}
For each given complex phase function $g(\lambda)$ satisfying
Definition~\ref{def:gbasic}, the phase-conjugated Riemann-Hilbert
Problem~\ref{rhp:N} has a unique solution, and is equivalent to the
holomorphic Riemann-Hilbert Problem~\ref{rhp:M}.
\end{proposition}

\begin{proof}  One finds a solution of the phase-conjugated problem by solving
the holomorphic Riemann-Hilbert Problem~\ref{rhp:M} for ${\bf
M}(\lambda)$ and then obtains ${\bf N}(\lambda)$ from
(\ref{eq:changeofvariables}).  The analyticity and boundary behavior
follow from the analogous properties of ${\bf M}(\lambda)$ and
$g(\lambda)$.  The jump conditions for ${\bf N}(\lambda)$ are verified
using the formula (\ref{eq:changeofvariables}) and the boundary
conditions satisfied by ${\bf M}(\lambda)$, taking into account the
discrepancy in boundary values of $g(\lambda)$ along the contour and
the symmetry of $g(\lambda)$.  Finally the normalization condition
follows from the corresponding property of ${\bf M}(\lambda)$ and the
decay of $g(\lambda)$.  Therefore, ${\bf N}(\lambda)$ so defined
solves the phase-conjugated Riemann-Hilbert problem.  Uniqueness of
solutions for the Riemann-Hilbert Problem~\ref{rhp:N} follows as
before from Liouville's theorem \index{Liouville's theorem} using the
continuity of the boundary values and the normalization at infinity.
Clearly the whole procedure can be reversed, and the unique solution
of the holomorphic Riemann-Hilbert Problem~\ref{rhp:M} can be obtained
from the solution ${\bf N}(\lambda)$ of the phase-conjugated
Riemann-Hilbert Problem~\ref{rhp:N} by the same formula,
(\ref{eq:changeofvariables}).
\end{proof}

The next goal will be to find a ``good'' contour $C$ and then use the
additional freedom afforded by the choice of the
complex phase function $g(\lambda)$ so that as $\hbar$ tends to zero,
the phase-conjugated Riemann-Hilbert Problem~\ref{rhp:N} for ${\bf
N}(\lambda)$ takes on a particularly simple form, namely one that can
be solved exactly.

\section[Conditions on the Complex Phase]{Representation as a complex 
single-layer potential.  Passing to the continuum limit.  Conditions
on the complex phase leading to the outer model problem.}
\label{sec:conditions}
We begin this section with an {\em ad hoc} assumption about
$g(\lambda)$ that will be justified only in
Chapter~\ref{sec:genuszero}.  Recall that the contour $C$ lives in the
upper half-plane and meets the origin in a corner point.  Therefore,
in a sufficiently small neighborhood of the origin, $C$ is the union
of two smooth arcs that join at $\lambda=0$ at some angle.  Near the
origin there are two kinds of behavior we will consider.
\begin{definition}
A complex phase function with parity $\sigma$ \index{complex phase
function!parity of} is a function $g^\sigma(\lambda)$ satisfying the
basic conditions of Definition~\ref{def:gbasic} such that for
$\sigma=+1$ there is a sufficiently small neighborhood $U$ of the
origin in which $g^\sigma(\lambda)$ is analytic on the part of $C\cap
U$ in the left half-plane and has a nonconstant difference in its
boundary values on the part of $C\cap U$ in the right half-plane.  For
$\sigma=-1$, the roles of the left and right half-planes are reversed.
\end{definition}
In particular, this means that the domain of analyticity of
$g^\sigma(\lambda)$ is simply connected due to a (possibly small) gap
in its contour of discontinuity on one side of the origin or the
other.  The two cases, $\sigma=+1$ and $\sigma=-1$ yield, for each
value of $x$ and $t$, (as well as for each value of $J$ and $K$)
different functions $g^\sigma(\lambda)$, and hence most important
quantities such as $\phi(\lambda)$, $\theta(\lambda)$, ${\bf
N}(\lambda)$, and ${\bf v}_{\bf N}(\lambda)$ will inherit this
dependence on $\sigma$.  Thus, from now on we write
$\phi(\lambda)=\phi^\sigma(\lambda)$,
$\theta(\lambda)=\theta^\sigma(\lambda)$, ${\bf N}(\lambda)={\bf
N}^\sigma(\lambda)$, and ${\bf v}_{\bf N}(\lambda)={\bf v}^\sigma_{\bf
N}(\lambda)$.

We now introduce a representation of $g^\sigma(\lambda)$ as the
complex single-layer potential \index{complex single-layer potential}
of a measure supported on $C\cup C^*$.  First, we define a new branch
of the logarithm.
\begin{definition}
Suppose $\eta\in C\cup C^*$.  Let $L^{C,\sigma}_\eta(\lambda)$
be the branch of $\log(\lambda-\eta)$ that is given by the principal
branch integral (\ref{eq:principalbranch}) but when considered as a
function of $\lambda$ is cut from the point $\lambda=\eta$
backwards, using the orientation $\sigma$,
along $C$ (if $\eta\in C$) and $C^*$ to $\lambda=0$ and then along the
negative real axis to $-\infty$ for $\sigma=+1$ or along the positive
real axis to $+\infty$ for $\sigma=-1$.  See Figure~\ref{fig:LogC}.
\begin{figure}[h]
\begin{center}
\mbox{\psfig{file=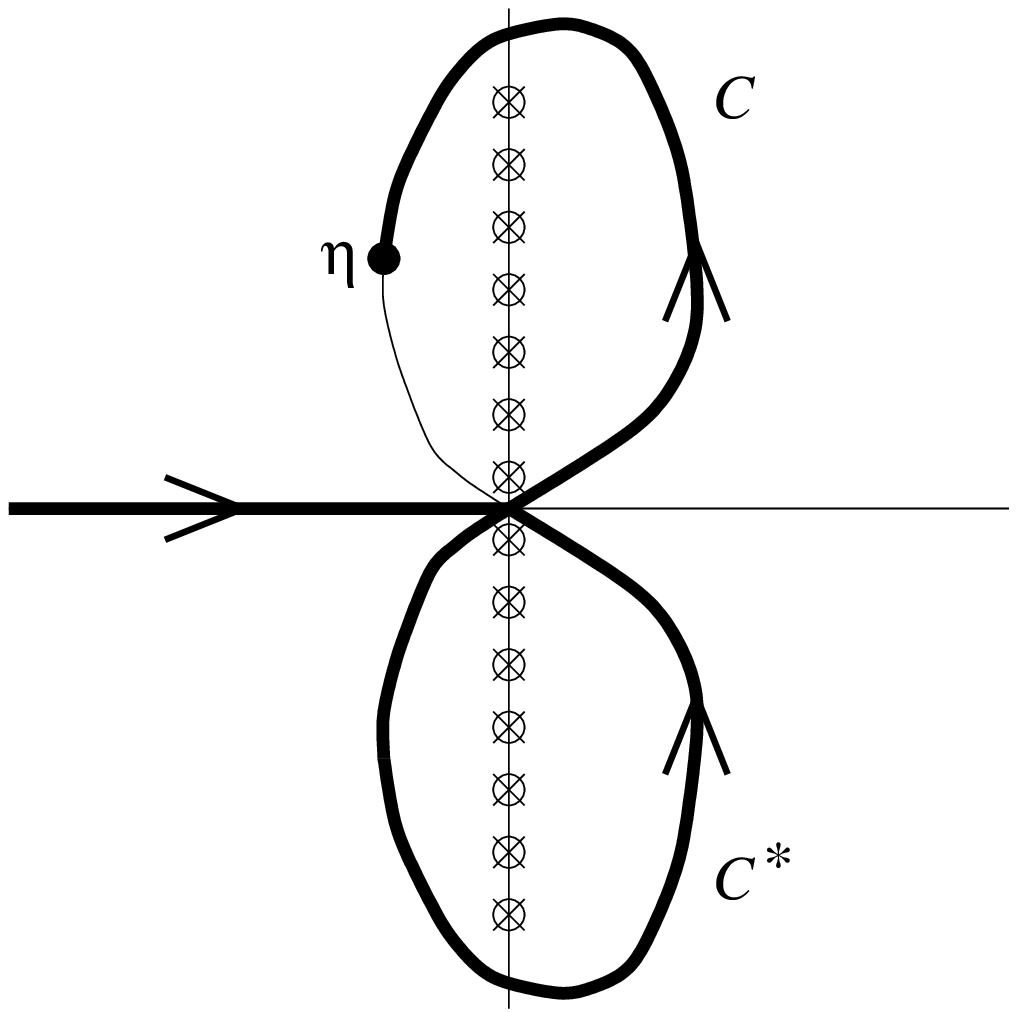,width=2.5 in}
\hspace{0.3 in}\psfig{file=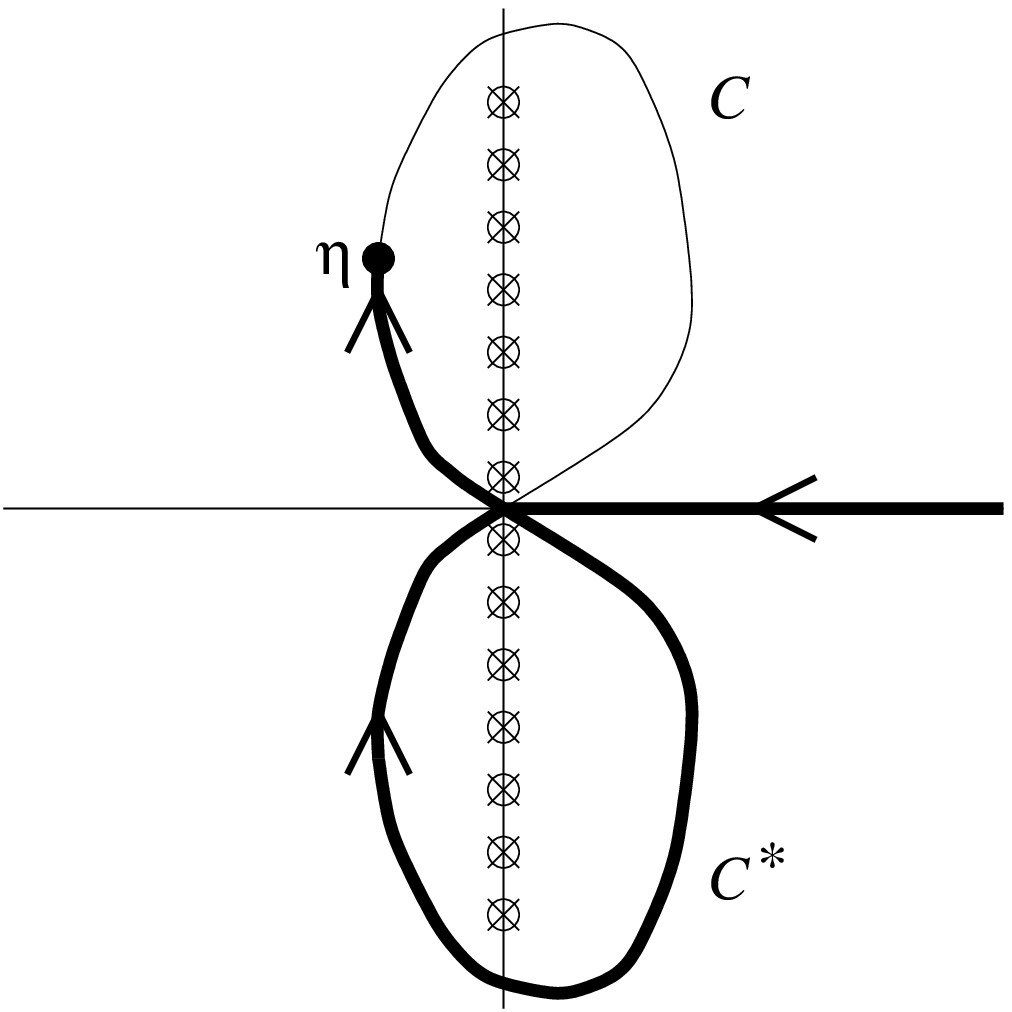,width=2.5 in}}
\end{center}
\caption{\em The branch cut of the functions
$L^{C,\sigma}_\eta(\lambda)$ in the $\lambda$-plane is determined by
the value of $\eta$, the shape of the contour $C$, and the orientation
$\sigma$.  The cut is illustrated in bold in these diagrams, and for
$\sigma=+1$ (left diagram) it connects to infinity along the negative
real axis, while for $\sigma=-1$ (right diagram) it connects to
infinity along the positive real axis.  In each case, the arrows
indicate the orientation $\sigma$ of $C$ and $C^*$.}
\label{fig:LogC}
\end{figure}
\label{def:LCdef}
\end{definition}
Without loss of generality, we represent $g^\sigma(\lambda)$ in the
form of a contour integral along $[C\cup C^*]_\sigma$ (both components
have orientation $\sigma$)
\begin{equation}
g^\sigma(\lambda)=\frac{J}{2}\int_{[C\cup
C^*]_\sigma}L^{C,\sigma}_\eta(\lambda)\rho^\sigma(\eta)\,d\eta\,,
\label{eq:repnofg}
\end{equation}
for some complex-valued {\em density function} $\rho^\sigma(\eta)$.
The basic symmetry and decay conditions from
Definition~\ref{def:gbasic} then require that we assume
\begin{equation}
\int_{[C\cup C^*]_{\sigma}}\rho^\sigma(\eta)\,d\eta = 0\,,\hspace{0.2
in} \mbox{and}\hspace{0.2 in}
\rho^\sigma(\eta^*)=\rho^\sigma(\eta)^*\,,\hspace{0.1
in} \eta\in C\cup C^*\,.
\label{eq:conditionsonrho}
\end{equation}
Thus, in particular, it is sufficient to determine
$\rho^\sigma(\eta)$ for $\eta\in C$.  If we define for
$\lambda\in [C\cup C^*]_\sigma$
\begin{equation}
g^\sigma_\pm(\lambda)=\lim_{\begin{array}{c}\scriptstyle
\mu\rightarrow\lambda\\\scriptstyle
\mu\in\pm\mbox{ side of }[C\cup C^*]_\sigma\end{array}}
g^\sigma(\mu)\,,
\end{equation}
then we have for $\lambda\in C\cup C^*$
\begin{equation}
\theta^\sigma(\lambda)=iJ(g^\sigma_+(\lambda)-g^\sigma_-(\lambda))=
-\pi \int_\lambda^0
\rho^\sigma(\eta)\,d\eta\,,
\label{eq:jumpofg}
\end{equation}
with the path of integration lying on $[C\cup C^*]_\sigma$.  More
precisely, if $\lambda\in C$ then the path of integration in
(\ref{eq:jumpofg}) continues from $\eta=\lambda$ along $C_\sigma$ to
$\eta=0$, whereas if $\lambda\in C^*$ then the path of integration in
(\ref{eq:jumpofg}) begins at $\eta=\lambda$, continues along $[C^*]_\sigma$ to
$\eta=0$, and then includes {\em all} of $C_\sigma$.

The fact that this representation (\ref{eq:repnofg}) of the complex
phase function is general can be seen by solving (\ref{eq:jumpofg})
for the density $\rho^\sigma(\eta)$ in terms of a given
absolutely continuous jump
$g^\sigma_+(\lambda)-g^\sigma_-(\lambda)$ on
$[C\cup C^*]_\sigma$ that satisfies the symmetry condition.  This condition
guarantees that the density $\rho^\sigma(\eta)$ we compute
will satisfy (\ref{eq:conditionsonrho}).  The difference between
$g^\sigma(\lambda)$ and the integral on the right-hand side
of (\ref{eq:repnofg}) is then an entire function of $\lambda$ that
vanishes at infinity, and is therefore zero.

Using the representation of $g^\sigma(\lambda)$ as a complex
potential, we now describe how to choose the density function
$\rho^\sigma(\eta)$ to obtain a simple Riemann-Hilbert problem in the
limit $\hbar\downarrow 0$.  Having broken symmetry in the two distinct
alternatives $\sigma=\pm 1$ in the phase function, we now do the same
in the Riemann-Hilbert problem by fixing the arbitrary orientation of
the contours supporting the jumps as a matter of convenience: from now
on we choose to set 
\begin{equation}
\omega:=\sigma\,.  
\end{equation}
This choice will allow us to ultimately describe the asymptotics in
\S\ref{sec:outersolve} using formulae that do not depend on $\sigma$.

We first propose a ``nearby'' Riemann-Hilbert problem.  Given a
complex phase function $g^\sigma(\lambda)$, define for $\lambda\in
C_\sigma$ the scalar function
\begin{equation}
\begin{array}{rcl}
\displaystyle\tilde{\phi}^\sigma(\lambda)
&:=&\displaystyle
\int_{0}^{iA} L^0_\eta(\lambda)\rho^0(\eta)\,d\eta  +
\int_{-iA}^0 L^0_\eta(\lambda)\rho^0(\eta^*)^*\,d\eta \\\\
&&\displaystyle\hspace{0.2 in}
+\,\,
J\cdot\left(
2i\lambda x + 2i\lambda^2 t - (2K+1)i\pi \int_\lambda^{iA}\rho^0(\eta)\,d\eta - 
g^\sigma_+(\lambda) -
g^\sigma_-(\lambda)\right)\,,
\end{array}
\label{eq:phitildedef}
\end{equation}
and for $\lambda\in C$ then define a matrix function by
\begin{equation}
{\bf v}_{\tilde{\bf N}}^\sigma(\lambda):=
\sigma_1^{\frac{1-J}{2}}\left[\begin{array}{cc}
\exp(i\theta^\sigma(\lambda)/\hbar) & 0 \\\\
-((-1)^K i)^{-J}\sigma\exp(\tilde{\phi}^\sigma(\lambda)/\hbar) &
\exp(-i\theta^\sigma(\lambda)/\hbar)\end{array}\right]
\sigma_1^{\frac{1-J}{2}}\,.
\end{equation}
Note that $\tilde{\phi}^\sigma(\lambda)$ is what one would get by
replacing the discrete measure $d\mu_{\hbar_N}^{\rm WKB}$ in the
formula for $\phi^\sigma(\lambda)$ by its weak-$*$ limit $d\mu_0^{\rm
WKB}$.

\begin{rhp}[Formal continuum limit]
\index{Riemann-Hilbert problem!formal continuum limit}
Given a complex phase function $g^\sigma(\lambda)$, find a matrix
function $\tilde{\bf N}^\sigma(\lambda)$ satisfying:
\begin{enumerate}
\item
{\bf Analyticity:} $\tilde{\bf N}^\sigma(\lambda)$ is analytic for
$\lambda\in{\mathbb C}\setminus(C\cup C^*)$.
\item
{\bf Boundary behavior:} $\tilde{\bf N}^\sigma(\lambda)$ assumes continuous
boundary values on $C\cup C^*$.
\item
{\bf Jump conditions:} The boundary values taken on $C\cup C^*$ satisfy
\begin{equation}
\begin{array}{rcll}
\tilde{\bf N}^\sigma_+(\lambda)&=&\tilde{\bf N}^\sigma_-(\lambda)
{\bf v}^\sigma_{\tilde{\bf N}}(\lambda)\,,&\hspace{0.2 in}
\lambda\in C_\sigma\,,\\\\
\tilde{\bf N}^\sigma_+(\lambda)&=&\tilde{\bf N}^\sigma_-(\lambda)
\sigma_2{\bf v}^\sigma_{\tilde{\bf N}}(\lambda^*)^*\sigma_2\,,&\hspace{0.2 in}
\lambda\in [C^*]_\sigma\,.
\end{array}
\end{equation}
\item
{\bf Normalization:} $\tilde{\bf N}^\sigma(\lambda)$ is normalized at
infinity:
\begin{equation}
\tilde{\bf N}^\sigma(\lambda)\rightarrow{\mathbb I}\mbox{ as }
\lambda\rightarrow\infty\,.
\end{equation}
\end{enumerate}
\label{rhp:tildeN}
\end{rhp}
Thus, we have introduced a {\em new} Riemann-Hilbert problem where we
have taken the ``continuum limit'' \index{continuum limit} of the
discrete eigenvalue measure.  According to
Theorem~\ref{theorem:outer}, the error in replacing the discrete sums
by integrals amounts to an $\bo (\hbar_N)$ error for fixed $\lambda$.
Thus, the matrix ${\bf v}_{\tilde{\bf N}}^\sigma(\lambda)$ is an $\bo
(\hbar_N)$ accurate approximation to the original jump matrix ${\bf
v}_{\bf N}^\sigma(\lambda)$ for all fixed $\lambda$ on the contours.
But as we have seen in
\S\ref{sec:spectralasymptotics} the approximation necessarily breaks
down in neighborhoods of radius $\bo (\hbar_N)$ near $\lambda=0$, and in
this region other corrections will need to be taken into account (see
\S\ref{sec:origin}).

\begin{proposition}
If the continuum limit Riemann-Hilbert Problem~\ref{rhp:tildeN} has
a solution $\tilde{\bf N}^\sigma(\lambda)$, then the solution is unique
and satisfies the symmetry relation 
\begin{equation}
\tilde{\bf N}^\sigma(\lambda^*)=\sigma_2\tilde{\bf N}^\sigma(\lambda)^*
\sigma_2\,.
\end{equation}
\end{proposition}

\begin{proof}
Uniqueness follows from the continuity of boundary values and the
normalization condition via Liouville's theorem\index{Liouville's
theorem}.  The symmetry of the solution then follows from the
corresponding symmetry of the jump relations and uniqueness.
\end{proof}

\begin{remark}
Note that it is by no means clear that the Riemann-Hilbert
Problem~\ref{rhp:tildeN} for $\tilde{\bf N}^\sigma(\lambda)$ has any
solution at all.  Unlike the problem for ${\bf N}^\sigma(\lambda)$,
this Riemann-Hilbert problem has not been obtained from an explicit
rational matrix by a well-defined sequence of transformations.
Rather, it is simply introduced as a reasonable asymptotic model for
the Riemann-Hilbert problem of interest.  We will bypass questions of
existence of solutions to the the continuum limit Riemann-Hilbert
Problem~\ref{rhp:tildeN} at this time because its role is only to lead
us, at a formal level, to an approximation of ${\bf
N}^\sigma(\lambda)$ that we will {\em prove} is valid in
\S\ref{sec:error}.
\end{remark}


The following two conditions are fundamental for the viability of our
asymptotic analysis, as well as for the actual determination of the
complex phase function $g^\sigma(\lambda)$ and the contour $C$:
\begin{equation}
\index{measure reality condition}
\mbox{\bf Measure Reality Condition:  }
\rho^\sigma(\eta)\,d\eta\in{\mathbb R}\,,\hspace{0.2 in}
\eta\in [C\cup C^*]_\sigma\,.
\label{eq:measurerealitycondition}
\end{equation}
\begin{equation}
\index{variational inequality condition}
\mbox{\bf Variational Inequality Condition:  }
\Re(\tilde{\phi}^\sigma(\lambda))
\le 0\,,\hspace{0.1 in}\lambda\in C\,.
\label{eq:ineq}
\end{equation}
The measure reality condition (\ref{eq:measurerealitycondition}) can
be understood from the following heuristic argument.  Note that
$\tilde{\bf N}^\sigma(\lambda)$ and the corresponding jump matrix
${\bf v}_{\tilde{\bf N}}^\sigma(\lambda)$ have determinant one.  This
implies that, if the complex phase $g^\sigma(\lambda)$ is to be chosen
so that the matrix $\tilde{\bf N}^\sigma(\lambda)$ has uniformly
bounded elements in the limit of $\hbar\downarrow 0$, then it is
necessary for the jump matrix to also be bounded as $\hbar\downarrow
0$.  This is only possible if the function $\theta^\sigma(\lambda)$ is
{\em real-valued} for $\lambda\in C$, which further constrains the
density function $\rho^\sigma(\eta)$ by requiring that
(\ref{eq:measurerealitycondition}) hold true.  The variational
inequality condition (\ref{eq:ineq}) is so named for reasons that will
be explained in detail in Chapter~\ref{sec:variational}.  Heuristically,
this condition can be understood in a similar fashion: if
(\ref{eq:ineq}) were to fail somewhere, then the jump matrix ${\bf
v}_{\tilde{\bf N}}^\sigma(\lambda)$ would have terms that are exponentially
large for $\hbar$ small --- a situation to be avoided.

\begin{remark}
Note that the measure reality condition
(\ref{eq:measurerealitycondition}) depends as much on the choice of
the oriented contour $C_\sigma$ via the complex-valued differential
$d\eta$ as on the complex-valued density function $\rho^\sigma(\eta)$.
Also, notice that reality of $\rho^\sigma(\eta)\,d\eta$, when
considered along with the second of the conditions
(\ref{eq:conditionsonrho}), actually implies a stronger version of the
first of the conditions (\ref{eq:conditionsonrho}): the integral over
$[\gamma\cup\gamma^*]_\sigma$, where $\gamma$ is any subcontour of
$C$, vanishes.
\end{remark}

For reasons that will become clear later, we will want to admit the
possibility that the variational inequality (\ref{eq:ineq}) is not
strict everywhere in $C$.  We do, however, assume for simplicity that
the subset of $C$ where the inequality fails to be strict forms a
system of closed subintervals of the regular curve $C$, whose topology
is defined, say, by the arc length parametrization.  If $\lambda\in C$
is in this system of closed subintervals, then we will say that
$\lambda$ lies in a {\em band}.  Any other value of $\lambda\in C\cup
C^*$ (that is, where the variational inequality (\ref{eq:ineq}) is
strict if $\lambda\in C$), is said to lie in a {\em gap}.  
\begin{definition}[Bands and gaps]
A band \index{band} is a maximal connected component of the system
of closed subintervals of $C$ where $\Re(\tilde{\phi}^\sigma(\lambda))\equiv
0$.  By symmetry, we say that $\lambda^*$ lies in a band if $\lambda$
does (although not in the same band).  A gap \index{gap} is
a maximal connected component of $C$ minus the union of the
bands, an open interval of $C$ in which the strict inequality
$\Re(\tilde{\phi}^\sigma(\lambda))<0$ holds.  By symmetry, $\lambda^*$ lies
in a gap if $\lambda$ does.  We will always assume the number of bands
and gaps on $C$ to be finite.
\end{definition}

If $\lambda\in C$ lies in a gap, then the off-diagonal entry
$-((-1)^Ki)^{-J}\sigma\exp(\tilde{\phi}^\sigma(\lambda)/\hbar)$ in the
jump matrix ${\bf v}_{\tilde{\bf N}}^\sigma(\lambda)$ is exponentially
small as $\hbar\downarrow 0$ with $\lambda$ held fixed.  The diagonal
entries are bounded but become increasingly oscillatory with respect
to any fixed parametrization of $C$ as $\hbar$ tends to zero.  These
wild oscillations in the jump matrix will lead to growth of
$\tilde{\bf N}^\sigma(\lambda)$ as one moves away from $C$, and
consequently it will be impossible to obtain uniform control of the
solution in the complex $\lambda$-plane unless it is further assumed
that $\theta^\sigma(\lambda)$ is {\em constant} (independent of
$\lambda\in C$) throughout each gap.  The constant value may be
different in each gap.  By the definition (\ref{eq:jumpofg}) of the
function $\theta^\sigma(\lambda)$, we are therefore assuming that
\begin{equation}
\Re(\tilde{\phi}^\sigma(\eta))<0\,,\hspace{0.1
in}\mbox{and}\hspace{0.1 in} \rho^\sigma(\eta)\equiv 0\,,\hspace{0.2
in} \eta\,\,\mbox{in a gap of}\,\,C\,.
\end{equation}

On the other hand, if $\lambda\in C$ lies in a band, then
$\Re(\tilde{\phi}^\sigma(\lambda))\equiv 0$, but the matrix
does not appear to simplify more as $\hbar$ tends to zero.  At this
point, we want to resist the temptation to take
$\theta^\sigma(\lambda)$ to be constant in the bands as well
as in the gaps, since we would then have
$\rho^\sigma(\eta)\equiv 0$ for all $\eta\in C$ which would
imply $g^\sigma(\lambda)\equiv 0$, thus defeating the
purpose of introducing the complex phase.  Instead, we proceed by
factoring the jump matrices as follows:
\begin{equation}
{\bf v}_{\tilde{\bf N}}^\sigma(\lambda)={\bf
a}^{\sigma,+}(\lambda) {\bf t}^\sigma(\lambda){\bf
a}^{\sigma,-}(\lambda)\,,
\end{equation}
where
\begin{equation}
{\bf a}^{\sigma,\pm}(\lambda):=
\sigma_1^{\frac{1-J}{2}}\left[\begin{array}{cc}
1 & -i^J(-1)^K\sigma\exp(-\tilde{\phi}^\sigma(\lambda)/\hbar)
\exp(\pm i\theta^\sigma(\lambda)/\hbar)\\\\
0 & 1\end{array}\right]\sigma_1^{\frac{1-J}{2}}\,,
\end{equation}
and
\begin{equation}
{\bf t}^\sigma(\lambda)=
\sigma_1^{\frac{1-J}{2}}\left[\begin{array}{cc}
0 & i^J(-1)^K\sigma\exp(-\tilde{\phi}^\sigma(\lambda)/\hbar)\\\\
i^J(-1)^K\sigma\exp(\tilde{\phi}^\sigma(\lambda)/\hbar)& 0
\end{array}
\right]\sigma_1^{\frac{1-J}{2}}\,.
\end{equation}
Let $I_k^+$ denote one of the bands on $C$.  Suppose that in the band
$I_k^+$ the functions
$\tilde{\phi}^\sigma(\lambda)$ and
$\theta^\sigma(\lambda)$ are the restrictions to the band of
two functions $q_k(\lambda)$ and $r_k(\lambda)$
respectively, each of which is analytic in $\lambda$ in a neighborhood
of the interior of the band.  

\begin{definition}
We denote by $q_k(\lambda)$ the analytic extension of
$\tilde{\phi}^\sigma(\lambda)$ from the interior of the band $I_k^+$,
when such an extension exists.  Likewise we denote by $r_k(\lambda)$
the analytic extension of $\theta^\sigma(\lambda)$ from the band
$I_k^+$.
\label{def:qr}
\end{definition}

\begin{figure}[h]
\begin{center}
\mbox{\psfig{file=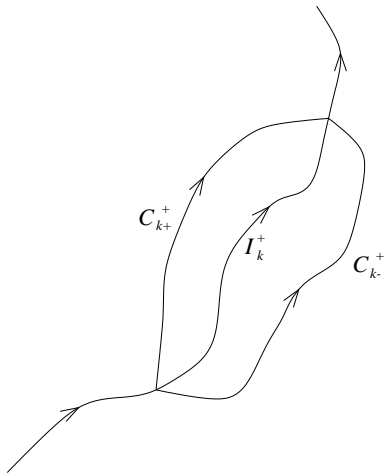,width=2 in}}
\end{center}
\caption{\em The lens-shaped \index{lens} region enclosed by the oriented 
contours $C^+_{k+}$ and $C^+_{k-}$ about the band $I^+_k\subset C$.}
\label{fig:lens}
\end{figure}

Let $C^+_{k+}$ (respectively
$C^+_{k-}$) denote a contour connecting the two endpoints of the
band, sharing the same orientation as $C_\sigma$, and lying within
the domain of analyticity of $q_k$ and $r_k$ to the left
(respectively right) of the band (see Figure~\ref{fig:lens}).
We think of $C^+_{k+}$ and $C^+_{k-}$ as being independent of $\hbar$,
but lying sufficiently close to $I_k^+$ to allow us to draw
conclusions concerning the behavior of the analytic functions $q_k$
and $r_k$ on these contours from the Cauchy-Riemann equations
\index{Cauchy-Riemann equations} on $I^+_k$ as we will describe below.

With these definitions, we may now make a change of variables from the
matrix $\tilde{\bf N}^\sigma(\lambda)$ to a new matrix ${\bf
O}^\sigma(\lambda)$.  Let $C^-_{k\pm}$ denote the complex-conjugate
contours $(C^+_{k\pm})^*$ which get their orientation not from the
conjugation operation, but from $[C^*]_\sigma $.  Define the matrix
${\bf O}^\sigma(\lambda)$ by:
\begin{equation}
\begin{array}{rcl}
{\bf O}^\sigma(\lambda)&:=&\displaystyle
\tilde{\bf N}^\sigma(\lambda)
\sigma_1^{\frac{1-J}{2}}\left[\begin{array}{cc}
1 & i^J(-1)^K\sigma
\exp(-q_k(\lambda)/\hbar)\exp(-ir_k(\lambda)/\hbar)\\\\
0 & 1\end{array}\right]\sigma_1^{\frac{1-J}{2}}\,,\\\\
&&\hspace{0.3 in}
\mbox{ for }\lambda\mbox{ in between } C^+_{k+}\mbox{ and }I^+_k\,,
\\\\
{\bf O}^\sigma(\lambda)&:=&\displaystyle\tilde{\bf N}^\sigma(\lambda)
\sigma_1^{\frac{1-J}{2}}\left[\begin{array}{cc}
1 & -i^J(-1)^K\sigma
\exp(-q_k(\lambda)/\hbar)\exp(ir_k(\lambda)/\hbar)\\\\
0 & 1\end{array}\right]\sigma_1^{\frac{1-J}{2}}\,,\\\\
&&\hspace{0.3 in}
\mbox{ for }\lambda\mbox{ in between } C^+_{k-}\mbox{ and }I^+_k\,,\\\\
{\bf O}^\sigma(\lambda)&:=&\sigma_2{\bf
O}^\sigma(\lambda^*)^*\sigma_2\\\\
&&\hspace{0.3 in}
\mbox{ for }\lambda^*\mbox{ in between } C^+_{k-}\mbox{ and }
C^+_{k+}\,,\\\\  
{\bf O}^\sigma(\lambda)&:=&\tilde{\bf N}^\sigma(\lambda)\,,\\\\
&&\hspace{0.3 in}\mbox{ otherwise. }
\end{array}
\label{eq:Odefine}
\end{equation}
Here, $k$ ranges over all bands of $C$.  This change of variables, in
conjunction with Riemann-Hilbert Problem~\ref{rhp:tildeN} suggests
a new problem posed on the contour $C\cup C^*$ with ``lenses''.
\begin{rhp}[Open lenses]
\index{Riemann-Hilbert problem!open lenses}
Given a complex phase function $g^\sigma(\lambda)$, find a matrix function 
${\bf O}^\sigma(\lambda)$ satisfying:
\begin{enumerate}
\item
{\bf Analyticity:} 
${\bf O}^\sigma(\lambda)$ is analytic for $\lambda\in{\mathbb C}$
minus $C\cup C^*\cup \{\mbox{lens boundaries}\}$.
\item
{\bf Boundary behavior:} ${\bf O}^\sigma(\lambda)$ assumes boundary
values from each connected component of the domain of analyticity that
are uniformly continuous, including at corner points corresponding to
self-intersections of the contour $C\cup C^*\cup \{\mbox{lens
boundaries}\}$.
\item
{\bf Jump conditions:}  For $\lambda$ in any gap of $C_\sigma$,
\begin{equation}
{\bf O}^\sigma_+(\lambda)={\bf O}^\sigma_-(\lambda){\bf
v}^\sigma_{\tilde{\bf N}}(\lambda)\,.
\end{equation}
For $\lambda$ in the band $I_k^+$ of $C_\sigma$,
\begin{equation}
{\bf O}^\sigma_+(\lambda)={\bf O}^\sigma_-(\lambda){\bf t}^\sigma(\lambda)\,.
\end{equation}
For $\lambda$ in either of the lens boundaries $C^+_{k\pm}$ of any
band of $C_\sigma$,
\begin{equation}
{\bf O}^\sigma_+(\lambda)={\bf O}^\sigma_-(\lambda)
\sigma_1^{\frac{1-J}{2}}
\left[\begin{array}{cc}
1 & -i^J(-1)^K\sigma\exp(-q_k(\lambda)/\hbar)
\exp(\mp ir_k(\lambda)/\hbar)\\\\
0 & 1\end{array}\right]\sigma_1^{\frac{1-J}{2}}\,.
\end{equation}
Finally, for $\lambda^*$ on any of the above contours we have
\begin{equation}
{\bf O}^\sigma_+(\lambda)={\bf O}^\sigma_-(\lambda)\sigma_2{\bf v}_{\bf O}^\sigma(\lambda^*)^*\sigma_2\,,
\end{equation}
where the orientation used to define the
boundary values is induced by $[C^*]_\sigma$, and where
${\bf v}_{\bf O}^\sigma(\lambda)$ is the jump matrix defined on
the contours in the upper half-plane by ${\bf v}_{\bf
O}^\sigma(\lambda):={\bf O}_-^\sigma(\lambda)^{-1}{\bf
O}_+^\sigma(\lambda)$.
\item
{\bf Normalization:}
${\bf O}^\sigma(\lambda)$ is normalized at infinity:
\begin{equation}
{\bf O}^\sigma(\lambda)\rightarrow{\mathbb I}\mbox{ as }\lambda\rightarrow
\infty\,.
\end{equation}
\end{enumerate}
\label{rhp:O}
\end{rhp} 

\begin{proposition}
The Riemann-Hilbert Problem~\ref{rhp:O} has at most one solution ${\bf
O}^\sigma(\lambda)$, which necessarily satisfies the symmetry
\begin{equation}
{\bf O}^\sigma(\lambda^*)=\sigma_2{\bf O}^\sigma(\lambda)^*\sigma_2\,.
\end{equation}
The Riemann-Hilbert Problem~\ref{rhp:O} is equivalent to the Riemann-Hilbert
Problem~\ref{rhp:tildeN} for the matrix $\tilde{\bf N}^\sigma(\lambda)$.
\end{proposition}

\begin{proof}
Uniqueness and symmetry are proved as for $\tilde{\bf
N}^\sigma(\lambda)$.  The equivalence is established by the explicit
triangular change of variables via the definition (\ref{eq:Odefine});
the transformation is clearly invertible, and it is a direct
calculation to show that any solution of the Riemann-Hilbert
Problem~\ref{rhp:tildeN} leads via (\ref{eq:Odefine}) to a solution of
the Riemann-Hilbert Problem~\ref{rhp:O}.
\end{proof}

\begin{remark}
On notation.  Throughout this section we are introducing a sequence of
Riemann-Hilbert problems.  Whenever the unknown in a new problem is
related to the unknown in a previous problem by an explicit
transformation, we denote the new unknown with a new letter.  Whenever
the unknown in a new problem is not directly related to that of the
previous problem, but the jump matrices are an {\em ad hoc}
approximation of the previous jump matrices, the new unknown is
written with the same letter as the old, but with a tilde.  Thus, each
appearance of a new tilde denotes a new formal approximation that will
need to be justified later.
\end{remark}

Now we will argue that with two additional constraints on
$\rho^\sigma(\lambda)$ in the bands, the jump relations in the
Riemann-Hilbert Problem~\ref{rhp:O} simplify dramatically as $\hbar$
tends to zero.  We know that the analytic functions $q_k(\lambda)$ are
purely imaginary while $r_k(\lambda)$ are purely real for $\lambda\in
I^+_k$.  In particular, it follows from the Cauchy-Riemann equations
that for $C^+_{k\pm}$ sufficiently close to $I^+_k$, the real part of
$-ir_k(\lambda)$ will be strictly negative on $C^+_{k+}$ except at the
endpoints and at the same time the real part of $ir_k(\lambda)$ will
be strictly negative on $C^+_{k-}$ except at the endpoints, {\em if
the real-valued function $\theta^\sigma(\lambda)$ is assumed to be
strictly decreasing along $I^+_k$ with its orientation $\sigma$}.  If
we apply similar arguments to the analytic functions $q_k(\lambda)$,
we see that the only sure way to prevent the real part of
$-q_k(\lambda)$ from being positive either on $C^+_{k+}$ or on
$C^+_{k-}$ is to insist that {\em the imaginary part of the function
$\tilde{\phi}^\sigma(\lambda)$ is constant along $I^+_k$}.  Therefore,
we are assuming
\begin{equation}
\tilde{\phi}^\sigma(\lambda)= \mbox{an imaginary
constant}\,,\hspace{0.1 in} \mbox{and}\hspace{0.1
in}\rho^\sigma(\lambda)\,d\lambda < 0\,,\hspace{0.2 in}
\lambda\,\,\mbox{in a band of}\,\,C\,.
\label{eq:bandcriterion}
\end{equation}
The latter conditions are equivalent to the required monotonicity of
$\theta^\sigma(\lambda)$, and they give us another interpretation of
the bands.  Their union is the support of the real measure
$\rho^\sigma(\eta)\,d\eta$ in the complex plane.  Under the condition
(\ref{eq:bandcriterion}), then it is clear that as $\hbar\downarrow
0$, the jump matrices on the outsides $C^+_{k\pm}$ and $C^-_{k\pm}$ of
the lenses converge pointwise to the identity matrix, uniformly in any
neighborhood that does not contain the endpoints.  At the same time,
the jump for the matrix ${\bf O}^\sigma(\lambda)$ in each band is a
constant (with respect to $\lambda$) matrix whose elements remain
bounded as $\hbar$ tends to zero.  Note that the jump matrix in each
band, the restriction of ${\bf t}^\sigma(\lambda)$, may be a different
constant in each band.

We collect these observations into a definition for future reference.
\begin{definition}
An admissible density function \index{density function!admissible}
for $g^\sigma(\lambda)$
is a complex-valued function $\rho^\sigma(\eta)$ defined on a loop
contour $C$ in the upper half-plane and its complex-conjugate $C^*$
such that the following five conditions are satisfied:
\begin{enumerate}
\item
$\rho^\sigma(\eta)\,d\eta$ is a real differential for $\eta\in
C_\sigma$, and $\rho^\sigma(\eta^*)=\rho^\sigma(\eta)^*$.
\item
The support of $\rho^\sigma(\eta)\,d\eta$ consists of a finite system of
intervals, the bands, whose complement in $C\cup C^*$ is the system of
gaps, and for $\sigma=+1$ (respectively $\sigma=-1$) the origin is
contained in a band that emerges only in the right half-plane
(respectively left half-plane).
\item
In the interior of each band of $C_\sigma$, the function
$\tilde{\phi}^\sigma(\lambda)$ evaluates to an imaginary constant,
possibly different in each band, and the differential
$\rho^\sigma(\eta)\,d\eta$ is strictly negative.
\item
In the interior of each gap of $C$, where $\rho^\sigma(\eta)\,d\eta$
vanishes, we have the strict inequality
$\Re(\tilde{\phi}^\sigma(\lambda))<0$.
\item
The restriction of the complex-valued function $\rho^\sigma(\lambda)$
to the interior of each band of $C\cup C^*$ has an analytic
continuation in some lens-shaped neighborhood on either side of
the band.
\end{enumerate}
\label{def:admissiblerho}
\end{definition}
If for some indices $J$, $K$, and $\sigma$ a contour $C$ and an
admissible density function $\rho^\sigma(\eta)$ for $\eta\in C\cup
C^*$ can be found, then the jump relations in the Riemann-Hilbert
problem for ${\bf O}^\sigma(\lambda)$ become very simple
asymptotically as $\hbar\downarrow 0$, at least away from the interval
endpoints.

\begin{remark}
The reader will observe that an elementary consequence of the
definitions (\ref{eq:thetadef}) of $\theta^\sigma(\lambda)$ and
(\ref{eq:phitildedef}) of $\tilde{\phi}^\sigma(\lambda)$ is that the
function $\theta^\sigma(\lambda)+i\tilde{\phi}^\sigma(\lambda)$, while
defined on the contour $C\cup C^*$, is the boundary value of a
function analytic on the ``minus'' side of the whole contour $[C\cup
C^*]_\sigma $, and similarly that
$\theta^\sigma(\lambda)-i\tilde{\phi}^\sigma(\lambda)$ is the boundary
value of a function analytic on the ``plus'' side of $[C\cup
C^*]_\sigma $.  On the outside of the contour loops, the region of
analyticity is in fact the entire exterior of $C\cup C^*$, while on
the inside of the loops the region of analyticity excludes only the
support of the asymptotic eigenvalue measure $\rho^0(\eta)\,d\eta$ on
the imaginary axis.

This fact has interesting implications if a complex phase function can
be found corresponding to an admissible density function
$\rho^\sigma(\eta)$ as described in
Definition~\ref{def:admissiblerho}.  In each band
$\tilde{\phi}^\sigma(\lambda)$ is an imaginary constant while in each
gap $\theta^\sigma(\lambda)$ is a real constant.  It follows that, for
example, the restrictions of $\tilde{\phi}^\sigma(\lambda)$ to two
different gaps extend to a given side of the contour as two analytic
functions of $\lambda$ {\em whose difference is a constant}.
Similarly, the restrictions of $\theta^\sigma(\lambda)$ to two
different bands have analytic extensions that differ only by a
constant.  Differentiating with respect to $\lambda$, one then sees
from the definition (\ref{eq:thetadef}) that the density function
$\rho^\sigma(\lambda)$ is the same analytic function in each band, and
that this function has an extension to the whole complex
$\lambda$-plane except the gaps of $C\cup C^*$ and the support of the
measure $\rho^0(\eta)\,d\eta$.  Thus, in particular, the third
condition in Definition~\ref{def:admissiblerho} implies the fifth.

A very important consequence of the fact that the function
$\tilde{\phi}^\sigma(\lambda)$ has an analytic continuation to either
side of each gap is that {\em the gap segments of the contour $C$ may
be deformed slightly with the endpoints held fixed without violating
the strict inequality $\Re(\tilde{\phi}^\sigma)<0$ on the interior of
the gap}.  On the other hand, the band segments of $C$ cannot be
freely deformed at all without violating the condition that the
differential $\rho^\sigma(\eta)\,d\eta$ should be real.

These continuation arguments also give relations between the analytic
extension of $\theta^\sigma(\lambda)$ from a band $I_k^+$ and
the analytic extension of $\tilde{\phi}^\sigma(\lambda)$ from
a gap $\Gamma_j^+$.  If $\theta^\sigma_j$ denotes the constant
value of $\theta^\sigma(\lambda)$ in the gap $\Gamma_j^+$ and
$\tilde{\phi}^\sigma_k$ denotes the constant value of
$\tilde{\phi}^\sigma(\lambda)$ in the band $I_k^+$, then
the continuations of these two functions to the left (``plus'' side)
of $C_\sigma$ are simply related:
\begin{equation}
\theta^\sigma(\lambda)-i\tilde{\phi}^\sigma_k \equiv
\theta^\sigma_j-i\tilde{\phi}^\sigma(\lambda)\,.
\label{eq:plusside}
\end{equation}
Likewise, continuing these two functions to the right (``minus'' side)
of $C_\sigma$ one finds:
\begin{equation}
\theta^\sigma(\lambda)+i\tilde{\phi}^\sigma_k \equiv
\theta^\sigma_j+i\tilde{\phi}^\sigma(\lambda)\,.
\label{eq:minusside}
\end{equation}
These relations will be particularly useful in the local analysis
that must be undertaken near a point $\lambda_k$ separating a band
from a gap.
\end{remark}

\begin{figure}[h]
\begin{center}
\mbox{\psfig{file=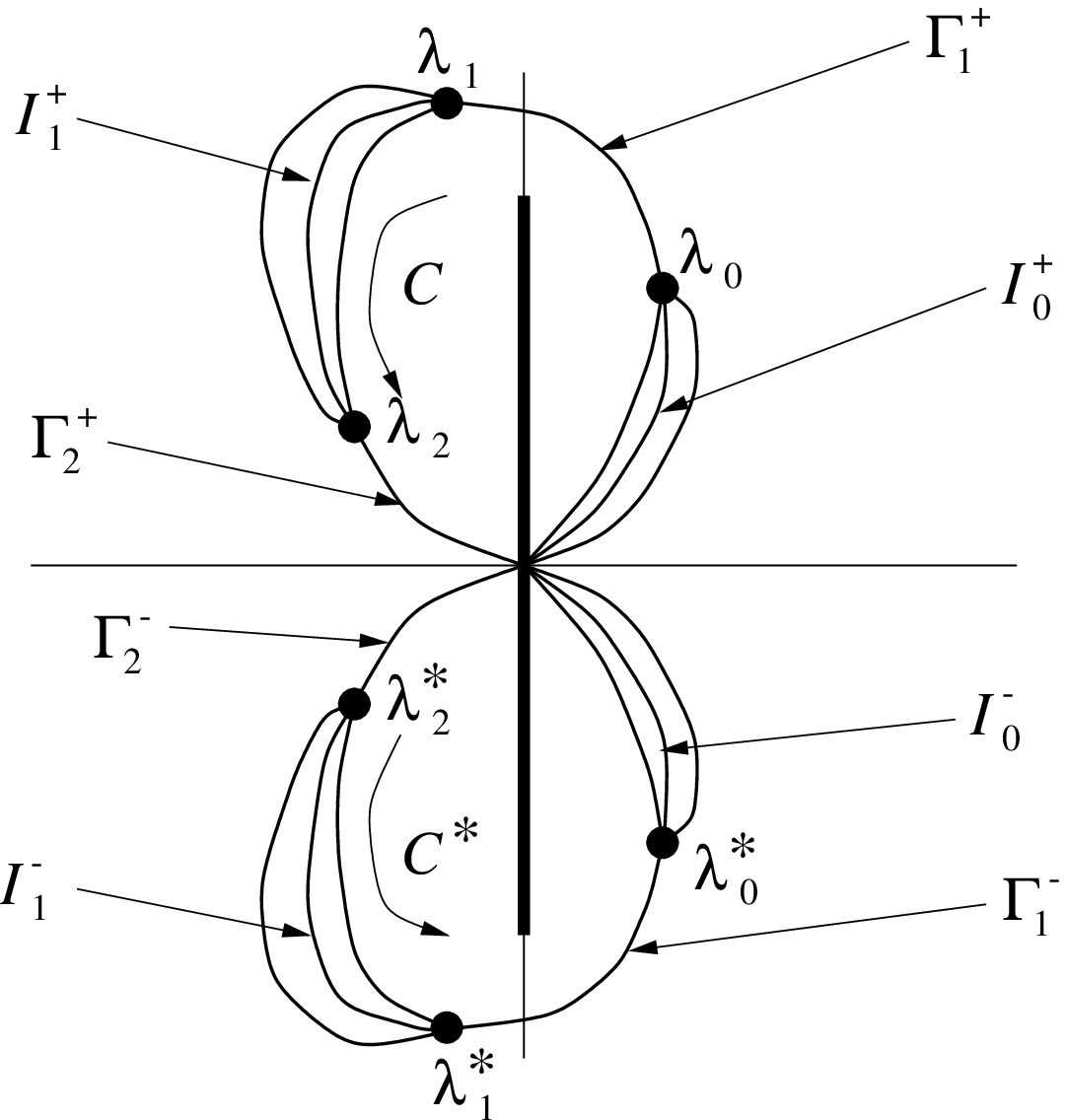,width=2.5 in}\hspace{0.3 in}
\psfig{file=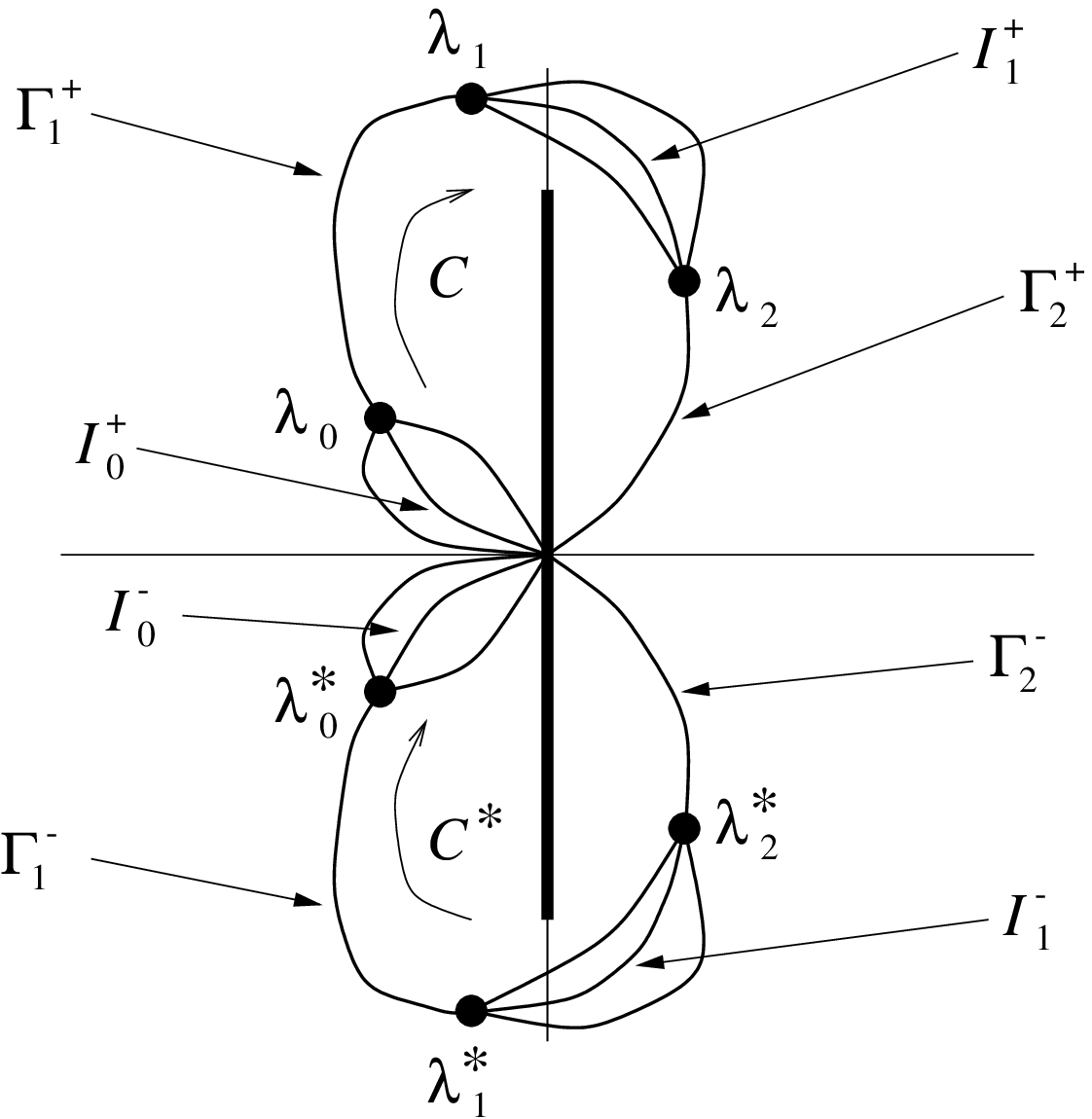,width=2.5 in}}
\end{center}
\caption{\em Notational conventions for the bands and gaps illustrated
for $\sigma=+1$ (left) and 
$\sigma=-1$ (right).}
\label{fig:jumpsforO}
\end{figure}

Note that, by the very meaning of the index $\sigma$ on the complex
phase function $g^\sigma(\lambda)$ as introduced at the beginning of
this chapter, the origin $\lambda=0$ is a boundary point between a
band and a gap of $C$.  If we consider $C_\sigma$ as an oriented loop
beginning (and ending) at the origin, then by definition the band
occupies the initial part of $C_\sigma$, while the final part of
$C_\sigma$ is a gap.  This leads us to introduce some notation for the
bands and gaps.  For some even nonnegative integer $G$, the bands on
$C$ are labeled, in order of the orientation $\sigma$ starting from
$\lambda=0$: $I_0^+, I_1^+, I_2^+,\dots,I_{G/2}^+$.  The gaps
interlacing these bands are labeled in order along $C_\sigma$ as
$\Gamma_k^+$, $k=1,\dots,G/2+1$. The endpoints of the bands,
enumerated along $C_\sigma$ are denoted $0,\lambda_0,\dots,\lambda_G$.
On $[C^*]_\sigma$, we have by symmetry bands $I_k^-=I_k^{+*}$ for
$k=0,\dots,G/2$ and gaps $\Gamma_k^-=\Gamma_k^{+*}$.  By convention,
we will set $I_0:=I_0^+\cup I_0^-$.  With this notation, the
orientation of each band and gap is expressed in terms of the
endpoints $\{\lambda_k\}$ is a way that avoids direct reference to
$\sigma$.  Namely, each band $I_k^+$ is an oriented contour segment
from $\lambda_{2k-1}$ to $\lambda_{2k}$, except for $I_0^+$ which is
an oriented contour from the origin to $\lambda_0$.  Similarly,
$I_k^-$ is an oriented contour from $\lambda_{2k}^*$ to
$\lambda_{2k-1}^*$, except for $I_0^-$ which goes from $\lambda_0^*$
to the origin.  The gap $\Gamma_k^+$ begins at $\lambda_{2k-2}$ and
ends at $\lambda_{2k-1}$, except for $\Gamma_{G/2+1}^+$, which ends at
the origin.  Similarly, $\Gamma_k^-$ begins at $\lambda_{2k-1}^*$ and
ends at $\lambda_{2k-2}^*$, except for $\Gamma_{G/2+1}^-$, which
begins at the origin.  These conventions regarding the contour where
${\bf O}^\sigma(\lambda)$ has jumps are illustrated in
Figure~\ref{fig:jumpsforO}.

\begin{remark}
The complex numbers $\lambda_0,\dots,\lambda_G$ should not be confused
with $L^2({\mathbb R})$ eigenvalues of the Zakharov-Shabat eigenvalue
problem or their WKB approximations which we are denoting by
$\lambda_{\hbar_N,n}^{\rm WKB}$.
\end{remark}

As $\hbar\downarrow 0$, the jump matrices for ${\bf
O}^\sigma(\lambda)$ simplify as described above subject to the
availability of an appropriate complex phase function
$g^\sigma(\lambda)$.  Assuming the existence of such a
$g^\sigma(\lambda)$, we now yield to the temptation of the pointwise
asymptotics of the jump matrices described above to propose on an {\em
ad hoc} basis a new model Riemann-Hilbert problem for an approximation
$\tilde{\bf O}(\lambda)$ to ${\bf O}^\sigma(\lambda)$.  In the gap
$\Gamma_k^+\subset C$, let $\theta_k$ denote the real constant value
of the function $J\cdot\theta^\sigma(\lambda)$.  Note that it follows
from (\ref{eq:jumpofg}) and the conditions imposed on
$\rho^\sigma(\lambda)$ in Definition~\ref{def:admissiblerho} that
$\theta_{G/2+1}\equiv 0$.  In the band $I_k^+\subset C$ for $k\ge 0$
let $i\alpha_k$ denote the imaginary constant value of
$J\cdot\tilde{\phi}^\sigma(\lambda)$ in that band.
\begin{rhp}[Outer model problem]
\index{Riemann-Hilbert problem!outer model problem}
Find a matrix function $\tilde{\bf O}(\lambda)$ satisfying:
\begin{enumerate}
\item
{\bf Analyticity:}  $\tilde{\bf O}(\lambda)$ is analytic for $
\lambda\in {\mathbb C}\setminus
((C\setminus \Gamma_{G/2+1}^+)\cup
(C^*\setminus\Gamma_{G/2+1}^-))$.
\item
{\bf Boundary behavior:} $\tilde{\bf O}(\lambda)$ assumes boundary
values that are continuous except at the endpoints of the bands and
gaps, where at worst inverse fourth-root singularities are admitted.
\item
{\bf Jump conditions:} For $\lambda\in I_k^+\cup I_k^-$, and $k=0,\dots,G/2$,
\begin{equation}
\tilde{\bf O}_+(\lambda)=\tilde{\bf O}_-(\lambda)
\left[\begin{array}{cc} 0 & -i\exp(-i\alpha_k/\hbar)\\\\
-i\exp(i\alpha_k/\hbar) & 0 \end{array}\right]\,.
\end{equation}
For $\lambda\in\Gamma_k^+\cup \Gamma_k^-$, and $k=1,\dots,G/2$,
\begin{equation}
\tilde{\bf O}_+(\lambda)=\tilde{\bf O}_-(\lambda)\left[\begin{array}{cc}
\exp(i\theta_k/\hbar) & 0 \\\\0 & \exp(-i\theta_k/\hbar)\end{array}\right]\,.
\end{equation}
\item
{\bf Normalization:}  $\tilde{\bf O}(\lambda)$ is normalized at infinity:
\begin{equation}
\tilde{\bf O}(\lambda)\rightarrow{\mathbb I}
\mbox{ as }\lambda\rightarrow\infty\,.
\end{equation}
\end{enumerate}
\label{rhp:model}
\end{rhp}

\begin{remark}
The necessity of dropping the condition of uniform continuity of the
boundary values will be explained in \S\ref{sec:outersolve}.
\end{remark}
\begin{remark}
Note that under the assumption that
\begin{equation}
i^J(-1)^K\sigma = -i\,,
\label{eq:JKsigma}
\end{equation}
which we will see in Chapter~\ref{sec:ansatz} is sufficient for our purposes
(one can think of this relation as defining, say, the interpolant
index $K$ in terms of $J$ and $\sigma$), the jump matrices for
$\tilde{\bf O}(\lambda)$ are obtained from those for ${\bf
O}^\sigma(\lambda)$ simply by explicitly computing pointwise
leading-order asymptotics as $\hbar\downarrow 0$.  Moreover, with this
choice of parameters, the pointwise convergence yields the
Riemann-Hilbert Problem~\ref{rhp:model} in which the parameter
$\sigma$ no longer explicitly appears.
\end{remark}

\begin{remark}
The formal continuum limit Riemann-Hilbert Problem~\ref{rhp:tildeN}
and the open lenses Riemann-Hilbert Problem~\ref{rhp:O} will not
be solved directly in this paper.  In particular, details of behavior of
boundary values {\em etc.} are included for completeness of
presentation but will not immediately constrain our analysis.  Both of
these problems are posed primarily as intermediate steps in passing from
the phase-conjugated Riemann-Hilbert Problem~\ref{rhp:N} for which we
have existence and uniqueness and the outer model Riemann-Hilbert
Problem~\ref{rhp:model} for which we will prove existence and
uniqueness.  But then, we will use the explicit solution for
$\tilde{\bf O}(\lambda)$ together with some local models we will obtain
in \S\ref{sec:inner} to build an approximation to ${\bf N}^\sigma(\lambda)$
that we will {\em prove} is uniformly valid in \S\ref{sec:error}.  
\end{remark}

Given the pointwise convergence of the jump matrices, one might expect
that a solution of the Riemann-Hilbert Problem~\ref{rhp:model} might
yield a good approximation of ${\bf O}^\sigma(\lambda)$, that by an
explicit change of variables approximates $\tilde{\bf
N}^\sigma(\lambda)$, and thus ${\bf N}^\sigma(\lambda)$.  The same
caveats hold for the relationship of the Riemann-Hilbert
Problem~\ref{rhp:model} to the Riemann-Hilbert Problem~\ref{rhp:O} for
${\bf O}^\sigma(\lambda)$ as mentioned when we introduced $\tilde{\bf
N}^\sigma(\lambda)$ in place of ${\bf N}^\sigma(\lambda)$.  Namely, it
is not clear at the moment whether there exists a solution at all, or
whether it should be a good approximation to ${\bf O}^\sigma(\lambda)$
anywhere in the complex plane.  We put aside the justification of
these formal approximations, as that will come later in
\S\ref{sec:error}.  We now turn to the issue of existence of solutions
for the model Riemann-Hilbert Problem~\ref{rhp:model}.

\section[The Outer Model Problem]{Exact solution of the outer model problem.}
\label{sec:outersolve}
The model Riemann-Hilbert Problem~\ref{rhp:model} for the matrix
$\tilde{\bf O}(\lambda)$ is the result of finding as described above
an appropriate contour $C$ and density function $\rho^\sigma(\lambda)$
for some choice of indices $J$, $K$, and $\sigma$ satisfying
(\ref{eq:JKsigma}), and then neglecting small elements of the
resulting jump matrices.  This will be a good approximation of the
jump matrices everywhere except near the endpoints $\lambda_k$ and
$\lambda_k^*$, and near $\lambda=0$.  The same sort of thing can be
said about the ``continuum limit'' approximation made when
substituting the matrix $\tilde{\bf N}^\sigma(\lambda)$ for ${\bf
N}^\sigma(\lambda)$; the approximation of the jump matrices is only
good outside of a small neighborhood near $\lambda=0$.  Reversing the
approximation steps, we might optimistically expect the matrix
$\tilde{\bf O}(\lambda)$ to provide a good approximation to ${\bf
N}^\sigma(\lambda)$ everywhere in the complex $\lambda$-plane except
near $\lambda=0$ and the endpoints.  The choice of terminology in
calling the Riemann-Hilbert Problem~\ref{rhp:model} an ``outer'' model
problem borrows from the theory of matched asymptotic
expansions\index{matched asymptotic expansions}.  In order to obtain a
uniformly valid leading-order asymptotic description of ${\bf
N}^\sigma(\lambda)$, we will need to develop ``inner'' model problems
as well to describe the behavior in the neighborhoods where the outer
approximation fails.  We will carry out this program in
\S\ref{sec:inner}.

In this section, however, we will solve the outer model problem.  The
reader will observe that similar Riemann-Hilbert problems are solved
in
\cite{DIZ97,DKMVZ98A,DKMVZ98B}, but here we will emphasize a slightly
different, more algebro-geometrical point of
view\index{algebro-geometric point of view}.  We do this for two
concrete reasons.  First, we want to clearly motivate the use of
various tools and techniques from Riemann surface theory that we will
need (and that were used in \cite{DIZ97,DKMVZ98A,DKMVZ98B}) by
explicitly introducing a Riemann surface and building functions on it,
rather than working on a complex plane with cuts.  Second, we want to
strengthen the connection between the Riemann-Hilbert approach to
semiclassical theory for integrable systems on the one hand and the
self-consistent Whitham modulation theory developed by Dubrovin,
Novikov, Krichever, {\em et al.} on the other hand.  The latter theory
has a strong algebro-geometric flavor, and a central role is played by
the {\em Baker-Akhiezer function}\index{Baker-Akhiezer function}, a
unique and canonically defined function on a given Riemann surface
that can be constructed using the Riemann theta function of the
surface and certain Abelian integrals\index{Abelian integrals}.  One
of the results of this section will be an explicit construction of the
slowly modulating Baker-Akhiezer function via the Riemann-Hilbert
problem for $\tilde{\bf O}(\lambda)$.  We will continue to assume
throughout this section that $G$ is an even integer satisfying $0\le G
<\infty$.


The main advantage here is that the jump matrix \index{jump
matrix!piecewise constant} is now piecewise constant as a function of
$\lambda$.  On the other hand, the Riemann-Hilbert
Problem~\ref{rhp:model} for $\tilde{\bf O}(\lambda)$ has one important
flaw: the jump matrix is not continuous at the endpoints $\lambda_k$.
This implies that all solutions will blow up as
$\lambda\rightarrow\lambda_k$.
However, we will be able to find a solution such that near each
endpoint $\lambda_k$, the elements of $\tilde{\bf O}(\lambda)$ grow
like $(\lambda-\lambda_k)^{-1/4}$, and for which the boundary values
are smooth on any open subset of the contour not containing any
endpoints.  All other matrix functions with the same domain of
analyticity and satisfying the same jump relations almost everywhere
will be proportional via a meromorphic matrix-valued function with all
singularities confined to the contour.  Thus, the condition that a
solution of the Riemann-Hilbert Problem~\ref{rhp:model} should have at
worst inverse fourth root singularities at all endpoints will pick out
the only solution that has finite boundary values between the
endpoints and at the same time treats all endpoints on a symmetrical
basis.

\begin{remark}
The simple fact that the jump matrices for $\tilde{\bf O}(\lambda)$
are piecewise constant \index{jump matrix!piecewise constant} also
suggests that special functions \index{special functions} will play a
role in the solution.  To see this, consider the derivative
$\partial_\lambda
\tilde{\bf O}(\lambda)$ of the solution.  This matrix shares the same
domain of analyticity as its primitive, and also satisfies the same
jump relations on each segment of the contour where the jump matrix is
constant.  This means that the quotient $\partial_\lambda\tilde{\bf
O}(\lambda)\cdot\tilde{\bf O}(\lambda)^{-1}$ is a meromorphic function on
the complex plane, with poles at the endpoints $\lambda_k$ and their
conjugates, and vanishing as $\lambda\rightarrow\infty$ (this follows
from the normalization condition).  It follows that the elements of the
matrix $\tilde{\bf O}(\lambda)$ satisfy a $2\times 2$ linear system of
ordinary differential equations in $\lambda$ with rational coefficients.
\end{remark}

\subsection{Reduction to a problem in 
function theory on hyperelliptic curves.}
Suppose that $h(\lambda)$ is an analytic function in the finite
$\lambda$-plane wherever $\tilde{\bf O}(\lambda)$ is supposed to be
analytic, taking on continuous boundary values {\em that are uniformly
bounded}, and that satisfies:
\begin{equation}
\begin{array}{rclll}
h_+(\lambda)-h_-(\lambda)&=&-
\theta_k\,,&
\lambda\in\Gamma_k^+\cup\Gamma_{k}^-\,,&k=1,\dots,G/2\,,\\
h_+(\lambda)+h_-(\lambda)&=&-\alpha_k\,,&
\lambda\in I_k^+\cup I_{k}^-\,,&k=0,\dots,G/2\,. 
\end{array}
\end{equation}
Consider the matrix defined by
\begin{equation}
{\bf P}(\lambda):=\tilde{\bf
O}(\lambda)\exp(ih(\lambda)\sigma_3/\hbar)\,.
\end{equation}
It is straightforward to verify that the matrix ${\bf
P}(\lambda)$ has the identity matrix as the jump matrix in all
gaps $\Gamma_k^+$ and $\Gamma_k^-$.  Since the boundary values of
$\tilde{\bf O}(\lambda)$ and $h(\lambda)$ are assumed
to be continuous, it follows that ${\bf P}(\lambda)$ is in fact
analytic in the gaps.  In the bands, the jump relation becomes simply
\begin{equation}
{\bf P}_+(\lambda)=-i{\bf P}_-(\lambda)
\sigma_1\,,
\end{equation}
so the jump relation is the same in all bands.  Next, suppose that
$\beta(\lambda)$ is a scalar function analytic in the
$\lambda$-plane except at the bands, where it satisfies
\begin{equation}
\beta_+(\lambda)=i\beta_-(\lambda)\,.
\end{equation}
Suppose further for the sake of concreteness that
$\beta(\lambda)\rightarrow 1$ as $\lambda\rightarrow\infty$.
Then, setting
\begin{equation}
{\bf Q}(\lambda):=\beta(\lambda){\bf P}(\lambda)\,,
\end{equation}
we see that the jump relations for ${\bf Q}(\lambda)$ take on the
elementary form:
\begin{equation}
{\bf Q}_+(\lambda)={\bf Q}_-(\lambda)\sigma_1\,,\hspace{0.3 in}
\lambda\in \cup_k (I_k^+\cup I_k^-)\,.
\label{eq:twistjumps}
\end{equation}
Our purpose in reducing the jump relations to this universal constant
form is to be able to move, as we will see shortly, from the cut plane
onto a compact Riemann surface on which the function theory is trivial
by comparison.

Before continuing to study ${\bf Q}(\lambda)$ let us describe the
scalar functions $h(\lambda)$ and $\beta(\lambda)$.  For
$\beta(\lambda)$ we propose the formula:
\begin{equation}
\beta(\lambda)^4=\frac{\lambda-\lambda_0}{\lambda-\lambda_0^*}
\prod_{k=1}^{G/2}\frac{\lambda-\lambda^*_{2k-1}}{\lambda-\lambda_{2k-1}}
\frac{\lambda-\lambda_{2k}}{\lambda-\lambda_{2k}^*}\,,
\label{eq:beta}
\end{equation}
and for $\beta(\lambda)$ we select the branch that tends to
unity for large $\lambda$ and that is cut along the bands $I_k^+$ and
$I_k^-$.  It is easily checked that $\beta(\lambda)$ as defined
here is the only function satisfying the required jump condition and
normalization at infinity that has continuous boundary values (except
at half of the endpoints).  
To find $h(\lambda)$, we introduce the function $R(\lambda)$
defined by
\begin{equation}
R(\lambda)^2 := \prod_{k=0}^{G}(\lambda-\lambda_k)(\lambda-\lambda_k^*)\,,
\end{equation}
choosing the particular branch that is cut along the bands $I_k^+$ and
$I_k^-$ and satisfies
\begin{equation}
\frac{R(\lambda)}{\lambda^{G+1}}\rightarrow -1\,,\hspace{0.2 in}
\lambda\rightarrow \infty\,, \mbox{   or equivalently,   }
R(0_+)=\prod_{k=1}^G |\lambda_k|\,.
\end{equation}
This defines a real function, {\em i.e.} one that satisfies
$R(\lambda^*)=R(\lambda)^*$.  At the bands, we have
$R_+(\lambda)=-R_-(\lambda)$, while $R(\lambda)$ is analytic in the
gaps.  Setting
\begin{equation}
h(\lambda)=k(\lambda)R(\lambda)\,,
\label{eq:hfromhtilde}
\end{equation}
we see that $k(\lambda)$ satisfies the jump relations:
\begin{equation}
\begin{array}{rcll}
k_+(\lambda)-k_-(\lambda)&=& 
\displaystyle -\frac{\theta_n}{R(\lambda)}\,,&\lambda\in
\Gamma_n^+\cup\Gamma_n^-\,,\\\\
k_+(\lambda)-k_-(\lambda)&=&
\displaystyle -\frac{\alpha_n}{R_+(\lambda)}\,,&\lambda\in
I_n^+\cup I_n^-\,,
\end{array}
\label{eq:jumpsforhtilde}
\end{equation}
and is otherwise analytic.  Such a function is given by the Cauchy integral
\index{Cauchy integral}
\begin{equation}
k(\lambda)=\frac{1}{2\pi i}
\sum_{n=1}^{G/2}\theta_n\int_{\Gamma_n^+\cup\Gamma_n^-}
\frac{d\eta}{(\lambda-\eta)R(\eta)} +
\frac{1}{2\pi i}\sum_{n=0}^{G/2}\alpha_n
\int_{I_n^+\cup I_n^-}\frac{d\eta}{(\lambda-\eta)R_+(\eta)}\,.
\label{eq:Cauchyforhtilde}
\end{equation}
This function blows up like $(\lambda-\lambda_n)^{-1/2}$ near each
endpoint, has continuous boundary values in between the endpoints, and
vanishes like $1/\lambda$ for large $\lambda$.  It is the only such
solution of the jump relations (\ref{eq:jumpsforhtilde}).  For
concreteness, we accept exactly this solution of
(\ref{eq:jumpsforhtilde}) and construct $h(\lambda)$ by using
(\ref{eq:hfromhtilde}).  The factor of $R(\lambda)$ renormalizes
the singularities at the endpoints, so that, as desired, the boundary
values of $h(\lambda)$ are bounded continuous functions.  Near
infinity, there is the asymptotic expansion:
\begin{equation}
\begin{array}{rcl}
h(\lambda)&=&h_G\lambda^G + h_{G-1}\lambda^{G-1} +
\dots + h_1\lambda + h_0 + \bo (\lambda^{-1})\\\\
&=&p(\lambda) + \bo (\lambda^{-1})\,,
\end{array}
\label{eq:hexpansion}
\end{equation}
where all coefficients $h_j$ of the polynomial $p(\lambda)$
can be found explicitly by expanding $R(\lambda)$ and the Cauchy
integral (\ref{eq:Cauchyforhtilde}) for large $\lambda$.  It is easy
to see from the reality of $\theta_j$ and $\alpha_j$ that
$p(\lambda)$ is a polynomial with real coefficients.

Now, let us return to the matrix ${\bf Q}(\lambda)$, and
determine what properties it must have in order for $\tilde{\bf
O}(\lambda)$ to have the appropriate boundary behavior and
asymptotic behavior at infinity.  Since $\tilde{\bf
O}(\lambda)$ should be $\bo ((\lambda-\lambda_n)^{-1/4})$
at each endpoint, it follows from the behavior of
$\beta(\lambda)$ that at $\lambda_{2k}^*$ for $k=0,\dots,G/2$ and
$\lambda_{2k-1}$ for $k=1,\dots,G/2$, we need to ask that ${\bf
Q}(\lambda)$ be bounded.
Similarly, near
$\lambda_{2k}$ for $k=0,\dots,G/2$ and $\lambda_{2k-1}^*$ for
$k=1,\dots,G/2$, we need to require that ${\bf Q}(\lambda)$ blow up
no worse than an inverse square root.
Near $\lambda=\infty$, the simple asymptotic
behavior required of $\tilde{\bf O}(\lambda)$ implies that
\begin{equation}
{\bf Q}(\lambda)\exp(-ip(\lambda)\sigma_3/\hbar) = 
{\mathbb I} + \bo (\lambda^{-1})\,,\hspace{0.2 in}
\lambda\rightarrow\infty\,,
\label{eq:Qnormalization}
\end{equation}
where we recall that $p(\lambda)$ is a polynomial of degree $G$ in
$\lambda$ with coefficients expressed explicitly in terms of
the $\theta_k$ and $\alpha_k$.

In fact, the jump relation (\ref{eq:twistjumps}), the asymptotic
relation (\ref{eq:Qnormalization}), and the condition that ${\bf
Q}(\lambda)$ be holomorphic outside of the bands with boundary
values for which the only allowable singularities are of inverse
square root type near $\lambda_0, \lambda_1^*, \lambda_2,
\lambda_3^*,\dots,\lambda_G$, determine the matrix
${\bf Q}(\lambda)$ uniquely.  This leads us to pose a new problem.
\begin{rhp}[Hyperelliptic problem]
\index{Riemann-Hilbert problem!hyperelliptic problem}
Let $p(\lambda)$ be a given polynomial of degree $G$.  Find a matrix
function ${\bf Q}(\lambda)$ satisfying
\begin{enumerate}
\item
{\bf Analyticity:} ${\bf Q}(\lambda)$ is analytic for $\lambda\in
{\mathbb C}\setminus \cup_k I_k^\pm$.\item {\bf Boundary behavior:}
${\bf Q}(\lambda)$ takes continuous boundary values on $\cup_k
I_k^\pm$ except at the alternating sequence of endpoints $\lambda_0,
\lambda_1^*,
\lambda_2,\lambda_3^*,\dots,\lambda_G$, where inverse square root
singularities in the matrix elements are admitted.
\item
{\bf Jump conditions:} On the interior of each oriented band
$I_k^\pm$, the boundary values of ${\bf Q}(\lambda)$ satisfy the
canonical jump conditions (\ref{eq:twistjumps}).
\item
{\bf Normalization:} ${\bf Q}(\lambda)$ has an essential singularity
at infinity, where it is normalized so that (\ref{eq:Qnormalization})
holds.
\end{enumerate}
\label{rhp:Q}
\end{rhp}

From the above explicit transformations relating $\tilde{\bf O}$ and
$\bf Q$, we have proved the following.
\begin{proposition}
When the polynomial $p(\lambda)$ is the principal part of the Laurent expansion
\index{Laurent expansion} of $h(\lambda)$ at infinity ({\em cf.} (\ref{eq:hexpansion})), and where
$h(\lambda)$ is given in terms of the constant parameters $\alpha_k$
and $\theta_k$ by the formula $h(\lambda)=R(\lambda)k(\lambda)$ with
$k(\lambda)$ given by (\ref{eq:Cauchyforhtilde}), the Riemann-Hilbert
Problem~\ref{rhp:Q} is equivalent to the outer model Riemann-Hilbert
Problem~\ref{rhp:model}.
\end{proposition}

In view of the jump relation (\ref{eq:twistjumps}) and the continuity
of the boundary values within the bands, we may solve the
Riemann-Hilbert Problem~\ref{rhp:Q} by considering the two columns of
the matrix ${\bf Q}(\lambda)$ as two projections of a {\em
single-valued} vector function defined on a hyperelliptic Riemann
surface \index{hyperelliptic Riemann surface} $X$ that is a double
covering \index{double covering} of the complex $\lambda$-plane.  To
achieve this, introduce $X$ as two copies of the complex plane,
individually cut and then mutually identified along the bands.  From
${\bf Q}(\lambda)$, define a vector-valued function ${\bf v}(P)$ for
$P\in X$ by arbitrarily labeling one copy of the cut complex plane in
$X$ as the ``first sheet'' and the other copy as the ``second sheet'',
and then setting
\begin{equation}
{\bf v}(P) := \left\{\begin{array}{ll}
\mbox{the first column of }
{\bf Q}(\lambda) \mbox{ if $P\in $ the first sheet of $X$}\,,\\
\mbox{the second column of }
{\bf Q}(\lambda) \mbox{ if $P\in $ the second sheet of $X$}\,. 
\end{array} \right.
\label{eq:vPdef}
\end{equation}
With suitable interpretations at the cuts, each point in the $\lambda$
plane has two preimages on $X$, except for the $2G+2$ branch points
$\{\lambda_k\}$ and $\{\lambda_k^*\}$.  Denote the preimage of
$\lambda=\infty$ on the first (respectively second) sheet of $X$ by
$P=\infty_1$ (respectively $P=\infty_2$).  With the inclusion of these
two points, $X$ is a compact Riemann surface of genus
$G$\index{Riemann surface!genus of}.

The function ${\bf v}(P)$ so defined on $X$ is holomorphic on $X$
away from $P=\infty_1$, $P=\infty_2$, and the points $\lambda_0,
\lambda_1^*, \lambda_2, \lambda_3^*,\dots,\lambda_G$.  At the two
infinite points of $X$, ${\bf v}(P)$ has essential
singularities\index{essential singularities}, whereas at the other
singular points the elements of ${\bf v}(P)$ grows, in terms of the
hyperelliptic projection \index{hyperelliptic projection}
$\lambda(P)$, at worst like an inverse square root.  In view of the
double ramification of $X$ at these isolated points, we see that in
terms of {\em holomorphic charts} \index{holomorphic charts} ({\em
i.e.} as a function on the {\em complex manifold}
\index{complex manifold} $X$), ${\bf v}(P)$ {\em has at worst simple
poles at exactly half of the branching points of $X$}.  Thus, ${\bf
v}(P)$ is a meromorphic function on $X\setminus\{\infty_1,\infty_2\}$.
Its poles in this finite part of $X$ are at worst simple, and are
confined to the branch points $\lambda_0,
\lambda_1^*,\lambda_2,\lambda_3^*,\dots,
\lambda_G$.

The two scalar components of ${\bf v}(P)$ have the elementary
properties that they are meromorphic functions on
$X\setminus\{\infty_1,\infty_2\}$ with the formal sum 
\begin{equation}
{\cal
D}^0=
\lambda_0+\lambda_1^*+\lambda_2+\lambda_3^*+\dots +\lambda_G\,,
\end{equation}
as the {\em divisor} \index{divisor!of poles} of the poles.  The asymptotic
behavior near the two infinite points of $X$ is given by expansions of
the form:
\begin{equation}
\begin{array}{rcll}
v_1(P)&\sim &\exp(ip(\lambda)/\hbar)(1+\bo (\lambda^{-1}))\,,&
P\rightarrow\infty_1\,,\\
v_1(P)&\sim &\exp(-ip(\lambda)/\hbar)\bo (\lambda^{-1})\,,&
P\rightarrow\infty_2\,,\\\\
v_2(P)&\sim &\exp(ip(\lambda)/\hbar)\bo (\lambda^{-1})\,, &
P\rightarrow\infty_1\,,\\
v_2(P)&\sim &\exp(-ip(\lambda)/\hbar)(1+\bo (\lambda^{-1}))\,, &
P\rightarrow\infty_2\,.
\end{array}
\end{equation}
Now, from the Riemann-Roch theorem \index{Riemann-Roch theorem} and
the easily checked nonspeciality of the divisor
\index{divisor!nonspecial} ${\cal D}^0-\infty_2$ it follows that there
exists a one-dimensional linear space of meromorphic functions on $X$
having $G+1$ simple poles at the points $P_k$ of the divisor ${\cal
D}^0$ and a simple zero at $P=\infty_2$.  This implies that there
exists a unique meromorphic function $f_1(P)$ with these properties
and normalized so that $f_1(\infty_1)=1$.  Similarly, there exists a
unique function $f_2(P)$ on $X$ with simple poles at the same points
as $f_1(P)$, a simple zero at $P=\infty_1$, and normalized so that
$f_2(\infty_2)=1$.  Each of these functions has exactly $G$ zeros on
$X$ (counted with multiplicities) in addition to the specified zero.
Let ${\cal D}_1=P^1_1+\dots + P^1_G$ and ${\cal D}_2=P^2_1+\dots
+P^2_G$ denote the divisors of these zeros\index{divisor!of zeros}.
These divisors are necessarily nonspecial.  Define $z_1(P)$ and
$z_2(P)$ by
\begin{equation}
z_1(P):=\frac{v_1(P)}{f_1(P)}\,,\hspace{0.3 in}
z_2(P):=\frac{v_2(P)}{f_2(P)}\,.
\end{equation}
These two functions are called {\em Baker-Akhiezer
functions}\index{Baker-Akhiezer function}.  They are meromorphic
functions on the finite part of $X$ with poles confined to the
divisors ${\cal D}_1$ and ${\cal D}_2$ respectively.  Near the two
infinite points of $X$,
\begin{equation}
\begin{array}{rcll}
z_1(P)&\sim &\exp(ip(\lambda)/\hbar)(1+\bo (\lambda^{-1}))\,,&
P\rightarrow\infty_1\,,\\
z_1(P)&\sim &\exp(-ip(\lambda)/\hbar)\bo (1)\,,&
P\rightarrow\infty_2\,,\\\\
z_2(P)&\sim &\exp(ip(\lambda)/\hbar)\bo (1)\,, &
P\rightarrow\infty_1\,,\\
z_2(P)&\sim &\exp(-ip(\lambda)/\hbar)(1+\bo (\lambda^{-1}))\,, &
P\rightarrow\infty_2\,.
\end{array}
\end{equation}

Given the polynomial $p(\lambda)$, each of these functions is uniquely
determined by these elementary properties.  The algebro-geometric
argument for uniqueness goes as follows.  From the existence of any
such function with minimal degree at the points of the divisor
characterizing its admissible poles (that is, a function having poles
of the largest admissible degree at these points), the uniqueness
follows again from the Riemann-Roch theorem\index{Riemann-Roch
theorem}.  For example, if one presumes the existence of {\em two}
functions satisfying the conditions of, say, $z_1(P)$, one of which
has minimal degree at the points of ${\cal D}_1$ and constructs their
ratio with the minimal degree function in the denominator, then this
ratio is a meromorphic function on {\em all} of $X$ (the essential
singularities cancel) with degree $G$, and all poles of the ratio come
from the zeros of the denominator.  These zeros of the denominator
move around on $X$ as the parameters $x$ and $t$ vary, and it is
reasonable to assume that the motion of the degree $G$ divisor of
these zeros avoids the codimension 1 locus of divisors for which the
Abel mapping \index{Abel mapping} (see below) fails to be invertible,
the {\em special divisors}\index{divisor!special}.  In this sense, the
most abstract form of the argument holds only for generic complex
values of the parameters $x$ and $t$.  However, with some additional
information about the Riemann surface $X$ and the divisor ${\cal D}_1$
(so-called {\em reality conditions}\index{algebro-geometric reality
conditions}) it is possible to prove that as long as $x$ and $t$ are
real, the divisor of the zeros of the denominator will always be
nonspecial.  In any case, once it is known or assumed that the divisor
of zeros (which is the pole divisor of the ratio) is nonspecial, then
it follows from the Riemann-Roch theorem \index{Riemann-Roch theorem}
that the ratio is a constant function on $X$.  By the normalization at
$P=\infty_1$ (for $z_1(P)$), this ratio is exactly unity.  Of course,
from another point of view uniqueness is not really an issue at all
here, because the Riemann-Hilbert problem for $\tilde{\bf O}(\lambda)$
has itself been used to prove uniqueness.

We have just specified two functions $z_1(P)$ and $z_2(P)$ on a
Riemann surface $X$, and if we can prove that such functions exist,
then we have established the existence of a solution to the
hyperelliptic Riemann-Hilbert Problem~\ref{rhp:Q}.  We now
pursue the construction of these two functions.

\subsection{Formulae for the Baker-Akhiezer functions.}
To establish the existence part of the argument, we will now provide
formulae for the two Baker-Akhiezer functions\index{Baker-Akhiezer
function}.  There are several ingredients we will need to define.  See
Dubrovin \cite{D81} for any details we do not give here.  The first is
a {\em homology basis} \index{homology basis} on $X$.  One starts with
the system of equivalence classes of closed, noncontractable, oriented
contours on $X$, with two contours being considered equivalent if
their difference (the union with the orientation of one contour
reversed) forms the oriented boundary of a surface in $X$.  The
equivalence classes will be referred to by representatives.  Two
contour representatives of the same class are called {\em homologous}
\index{homologous cycles} cycles; 
the integral of any meromorphic differential without residues
gives the same value over any two homologous cycles.  The system of
homology classes may be viewed as a linear space with integer
coefficients.  The zero element of this space is the equivalence class
of contractable oriented loops on $X$.  It is a fundamental
topological result that this space has dimension $2G$.  A homology
basis is a basis $\{a_1,
\dots,a_G,b_1,\dots,b_G\}$ of this linear space that has certain
properties with respect to contour intersection.  Let $C_1$ and $C_2$
be two oriented closed contours on $X$.  The {\em intersection number}
\index{intersection number} $C_1\circ C_2$ is defined as the number of 
times $C_2$ crosses $C_1$
from the right of $C_1$, minus the number of times $C_2$ crosses $C_1$
from the left.  The intersection number is a skew-symmetric bilinear
class function.  A homology basis is required to have the following
properties:
\begin{equation}
a_j\circ a_k=b_j\circ b_k = 0\,,\hspace{0.3 in}
a_j\circ b_k = \delta_{jk}\,.
\end{equation}
This does not make the basis unique, even up to homology equivalence
of class representatives.  Any linear transformation of the basis
elements in the matrix group $Sp(2G,{\mathbb Z})$ will preserve the
intersection number but modify the particular basis.  Later, we will
select a particular homology basis in order to simplify the {\em
appearance} of the formulae we will write down, but of course the
results themselves, by uniqueness, are independent of this choice.
Once the homology cycles $a_j$ are fixed, the dual cycles $b_j$ are
determined by the intersection relations up to transformations of the
form
\begin{equation}
b_k\rightarrow b_k+\sum_{j=1}^G s_{kj}a_j\,,
\end{equation}
where $s_{kj}$ are integers and $s_{kj}=s_{jk}$.
Of course, for $G=0$ there are no homology cycles at all.

The next ingredient we need are the {\em normalized holomorphic
differentials}\index{differential!normalized holomorphic}.  On $X$
there is a complex $G$-dimensional linear space of holomorphic
differentials, with basis elements $\nu_k(P)$ for $k=1,\dots,G$ that
can be written in the form:
\begin{equation}
\nu_k(P)=\frac{\displaystyle\sum_{j=0}^{G-1}c_{kj}\lambda(P)^j}
{R_X(P)}\,d\lambda(P)\,,
\end{equation}
where $R_X(P)$ is a ``lifting'' of the function $R(\lambda)$
from the cut plane to $X$: if $P$ is on the first sheet of $X$ then
$R_X(P)=R(\lambda(P))$ and if $P$ is on the second sheet of
$X$ then $R_X(P)=-R(\lambda(P))$.  The coefficients $c_{kj}$
are uniquely determined by the constraint that the differentials
satisfy the normalization conditions:
\begin{equation}
\oint_{a_j}\nu_k(P) = 2\pi i\delta_{jk}\,.
\label{eq:holomorphicnormalization}
\end{equation}
From the normalized differentials, one defines a $G\times G$ matrix
$\bf H$ (the {\em period matrix}\index{period matrix}) by the formula:
\begin{equation}
H_{jk}=\oint_{b_j}\nu_k(P)\,.
\end{equation}
It is a consequence of the standard theory of Riemann surfaces that
${\bf H}$ is a symmetric matrix whose real part is {\em negative
definite}.  

Associated with the matrix $\bf H$, and therefore with the choice of
homology basis on $X$, is the {\em Riemann theta function}
\index{Riemann theta function} defined for ${\bf w}\in {\mathbb C}^G$
by the Fourier series
\begin{equation}
\Theta({\bf w}):=\sum_{{\bf n}\in {\mathbb Z}^G}\exp\,\left(\frac{1}{2}
{\bf n}^T{\bf Hn}+{\bf n}^T{\bf w}\right)\,.
\end{equation}
It is an entire function on ${\mathbb C}^G$.

Let ${\bf e}_k$ denote the standard unit vectors in ${\mathbb C}^G$
and let ${\bf h}_k$ denote the $k$th column of the matrix ${\bf H}$,
{\em i.e.} ${\bf h}_k:={\bf H}{\bf e}_k$ for $k=1,\dots,G$.  Denote by
$\Lambda\subset{\mathbb C}^G$ be the lattice generated by linear
combinations with integer coefficients of the vectors $2\pi i {\bf
e}_k$ and ${\bf h}_k$ for $k=1,\dots,G$.  That is,
\begin{equation}
\Lambda := 2\pi i{\mathbb Z}{\bf e}_1+\dots+2\pi i{\mathbb Z}{\bf e}_G +
{\mathbb Z}{\bf h}_1+\dots+{\mathbb Z}{\bf h}_G\,.
\label{eq:Lambdadef}
\end{equation}
The {\em Jacobian variety} \index{Jacobian variety} of $X$,
$\mbox{Jac}\,(X)$, is simply the complex torus ${\mathbb
C}^G/\Lambda$.  We arbitrarily fix a base point $P_0$ on $X$.  The
{\em Abel-Jacobi mapping} \index{Abel mapping} ${\bf A}:X\rightarrow
\mbox{Jac}\,(X)$ is then defined componentwise as follows:
\begin{equation}
A_k(P;P_0):=\int_{P_0}^P\nu_k(P')\,,\hspace{0.2 in}
k=1,\dots,G\,,
\end{equation}
where $P'$ is an integration variable.  The range of the mapping is in
the Jacobian because the path of integration is not specified.  The
Abel-Jacobi mapping is also defined for integral divisors
\index{divisor!integral} ${\cal D}=P_1+\dots+P_M$ by summation:
\begin{equation}
{\bf A}({\cal D};P_0):={\bf A}(P_1;P_0)+\dots +{\bf A}(P_M;P_0)\,,
\end{equation}
and finally extended to non-integral divisors
\index{divisor!non-integral} ${\cal D}={\cal D}^+-{\cal D}^-$ for
integral divisors ${\cal D}^\pm$ by ${\bf A}({\cal D};P_0):={\bf
A}({\cal D}^+;P_0)- {\bf A}({\cal D}^-;P_0)$.  If the degree of the
divisor $\cal D$ is zero then ${\bf A}({\cal D};P_0)$ is independent
of the base point $P_0$.  {\em Abel's theorem} \index{Abel's theorem}
states that if ${\cal D}^+$ catalogs the zeros, and ${\cal D}^-$ the
poles, of a meromorphic function on the compact surface $X$, then with
${\cal D}={\cal D}^+-{\cal D}^-$, ${\bf A}({\cal D};P_0)=0$ in the
Jacobian, or equivalently the integral always yields a lattice vector
in $\Lambda\subset {\mathbb C}^G$.  Note that Abel's theorem applied
to the functions $f_1(P)$ and $f_2(P)$ yields the identities
\begin{equation}
\begin{array}{rcll}
{\bf A}({\cal D}_1;P_0)&=&{\bf A}({\cal D}^0;P_0)-{\bf A}(\infty_2;P_0)\,,
&\pmod\Lambda \,,\\ {\bf A}({\cal D}_2;P_0)&=&{\bf A}({\cal
D}^0;P_0)-{\bf A}(\infty_1;P_0)\,, &\pmod\Lambda \,.
\end{array}
\label{eq:AbelsTheorem}
\end{equation}

Finally, a particularly important element of the Jacobian is the {\em
Riemann constant vector} \index{Riemann constant vector}
$\bf K$ which is defined, modulo the lattice
$\Lambda$, componentwise by
\begin{equation}
K_k:=\pi i + \frac{H_{kk}}{2}-\frac{1}{2\pi i}\sum_{j=1\atop j\neq
k}^G\oint_{a_j} \left(\nu_j(P)\int_{P_0}^P\nu_k(P')\right)\,,
\end{equation}
where the index $k$ varies between $1$ and $G$.

Next, we will need to define a certain meromorphic differential 
\index{differential!meromorphic} on
$X$.  Let $\Omega(P)$ be holomorphic away from the points $\infty_1$
and $\infty_2$, where it has the behavior
\begin{equation}
\begin{array}{rcccll}
\Omega(P) &= &dp(\lambda(P))& +& \displaystyle \bo
\left(\frac{d\lambda(P)}{\lambda(P)^{-2}}\right)\,, &P\rightarrow
\infty_1\,,\\\\ \Omega(P) &=&-dp(\lambda(P))&+&\displaystyle
\bo\left( \frac{d\lambda(P)}{\lambda(P)^{-2}}\right)\,,
&P\rightarrow \infty_2\,,
\end{array}
\label{eq:Omegadef}
\end{equation}
and made unique by the normalization conditions
\begin{equation}
\oint_{a_j}\Omega(P) = 0\,,\hspace{0.2 in} j=1,\dots,G\,.
\label{eq:meromorphicnormalization}
\end{equation}
Let the vector ${\bf U}\in {\mathbb C}^G$ be defined
componentwise by
\begin{equation}
U_j:=\oint_{b_j}\Omega(P)\,.
\end{equation}
Note that $\Omega(P)$ has no residues.

With these ingredients, we may now give formulae for the functions
$z_1(P)$ and $z_2(P)$.  First, define $y_1(P)$ and
$y_2(P)$ by choosing particular vectors ${\bf V}_m\in{\mathbb
C}^G$ corresponding to the points ${\bf A}({\cal D}_m;P_0)+{\bf
K}\in\,\mbox{Jac}\,(X)$, and then setting
\begin{equation}
y_m(P):=\frac{\Theta({\bf A}(P;P_0)-{\bf V}_m + i{\bf
U}/\hbar)}{\Theta({\bf A}(P;P_0)-{\bf V}_m)}
\exp\left(\frac{i}{\hbar}\int_{P_0}^P\Omega(P')\right)\,,
\end{equation}
where $m=1$ or $m=2$.  The path of integration in the exponent is the
same path as in the Abel-Jacobi mapping, but is otherwise unspecified.
The fact that these formulae actually define functions that do not depend
on that path follows from the transformation laws for the theta function:
\begin{equation}
\Theta({\bf w}+2\pi i {\bf e}_k)=\Theta({\bf w})\,,\hspace{0.2 in}
\Theta({\bf w}+{\bf h}_k)=\exp(-H_{kk}/2-w_k)\Theta({\bf w})\,.
\end{equation}
So, if the path of integration is augmented by adding one of the
homology cycles $a_k$, then the exponent is invariant by the
normalization condition for $\Omega(P)$.  At the same time, we
have ${\bf A}(P;P_0)\rightarrow{\bf A}(P;P_0)+2\pi i{\bf e}_k$ and by
the first transformation law we see that $y_m(P)$ is invariant.
Similarly, if the homology cycle $b_k$ is added to the path of
integration, then the exponential transforms by producing a factor of
$\exp(iU_k/\hbar)$.  And at the same time, ${\bf
A}(P;P_0)\rightarrow{\bf A}(P;P_0)+{\bf h}_k$ and by the second
transformation law we again deduce that $y_m(P)$ is invariant.
This means that the functions $y_m(P)$ are well-defined given the
choices of homology, base point $P_0$, and representatives ${\bf
V}_m$.

Now, since each divisor ${\cal D}_m$ is nonspecial, the zeros of the
denominator are exactly the points $P_1^m$, \dots, $P_G^m$.  Since the
theta function is entire and the differential $\Omega(P)$ is
holomorphic away from the points $\infty_1$ and $\infty_2$, it follows
that $y_m(P)$ is meromorphic on $X\setminus
\{\infty_1,\infty_2\}$, with poles exactly at the points of the
divisor ${\cal D}_m$.  Near the points $\infty_1$ and $\infty_2$, we
have
\begin{equation}
\begin{array}{rcll}
y_m(P)&=&\exp(ip(\lambda(P))/\hbar)\bo (1)\,, &
P\rightarrow \infty_1\,,\\
y_m(P)&=&\exp(-ip(\lambda(P))/\hbar)\bo (1)\,, &
P\rightarrow\infty_2\,,
\end{array}
\end{equation}
where the leading order term in each expansion depends on $m$.  To
obtain formulae for $z_1(P)$ and $z_2(P)$ it then suffices to
appropriately normalize the functions $y_1(P)$ and
$y_2(P)$.  So, let
\begin{equation}
\begin{array}{rcl}
N_1&:=&\displaystyle
\lim_{P\rightarrow \infty_1} y_1(P)
\exp(-ip(\lambda(P))/\hbar)\,,\\\\
N_2&:=&\displaystyle\lim_{P\rightarrow \infty_2} y_2(P)
\exp(ip(\lambda(P))/\hbar)\,.
\end{array}
\label{eq:normalizingconstants}
\end{equation}
Then, we set
\begin{equation}
z_1(P):=\frac{y_1(P)}{N_1}\,,\hspace{0.2 in}
z_2(P):=\frac{y_2(P)}{N_2}\,.
\end{equation}
These functions at last satisfy all the required conditions, and by
the Riemann-Roch argument or the equivalent uniqueness argument for
the hyperelliptic Riemann-Hilbert Problem \ref{rhp:Q} we summarized
earlier, {\em they are the only such functions}.  In particular,
$z_1(P)$ and $z_2(P)$ do not depend on the choice of homology cycles,
the choice of base point $P_0$, or the choice of representatives ${\bf
V}_m$.  While this is true, certain properties of the functions can be
elucidated by making particular convenient choices of these arbitrary
``gauge'' parameters\index{gauge!parameters}.

Thus, we have established the existence of the two Baker-Akhiezer
functions $z_1(P)$ and $z_2(P)$, which amounts to the solution of the
hyperelliptic Riemann-Hilbert Problem \ref{rhp:Q}, or equivalently the
solution of the model Riemann-Hilbert Problem \ref{rhp:model}.  But
these formulae become more effective if we break the gauge symmetry by
specifying all paths of integration concretely in the cut plane.  We
carry out this program now.

\subsection{Making the formulae concrete.}
We now develop these formulae in more detail.  First observe that for the
functions $f_1(P)$ and $f_2(P)$ we have the explicit representations:
\begin{equation}
\begin{array}{rcl}
f_1(P)&=&\displaystyle\frac{R_X(P)-(\lambda(P)-\lambda_0^*)(\lambda(P)-\lambda_1)\dots(\lambda(P)-\lambda_G^*)}{2R_X(P)}\,,\\\\
f_2(P)&=&\displaystyle\frac{R_X(P)+(\lambda(P)-\lambda_0^*)(\lambda(P)-\lambda_1)\dots(\lambda(P)-\lambda_G^*)}{2R_X(P)}\,.
\end{array}
\end{equation}
Next, we make the observation that 
\begin{equation}
\begin{array}{rcll}
\Omega(P)&=&dh(\lambda(P))\,, &P\,\,\,\mbox{on the first sheet of}\,\,\,X\,,\\
\Omega(P)&=&-dh(\lambda(P))\,,&P\,\,\,\mbox{on the second sheet of}\,\,\,X\,.
\end{array}
\label{eq:differentialidentity}
\end{equation}
To see this, one {\em defines} an Abelian differential
\index{differential!Abelian} $\hat{\Omega}(P)$ on $X$ by the
right-hand side of (\ref{eq:differentialidentity}); it is not
difficult to see from the jump relations for the scalar function $h$
that this indeed defines a meromorphic differential on the whole of
the compact surface $X$.  Next, it follows from the definition of
$\Omega(P)$ near the points $\infty_1$ and $\infty_2$ that the
difference $\Omega(P)-\hat{\Omega}(P)$ is a holomorphic differential
on $X$ because the singularities cancel.  The difference will
therefore be identically zero if it can be shown that
\begin{equation}
\oint_{a_j}(\Omega(P)-\hat{\Omega}(P))=0\,,
\end{equation}
for all $j=1,\dots,G$.  But the first term vanishes by definition of
$\Omega(P)$, and it can be shown from the jump relations for
$h(\lambda)$ that the same is true of $\hat{\Omega}(P)$,
regardless of the choice of homology basis.

\begin{figure}[h]
\begin{center}
\mbox{\psfig{file=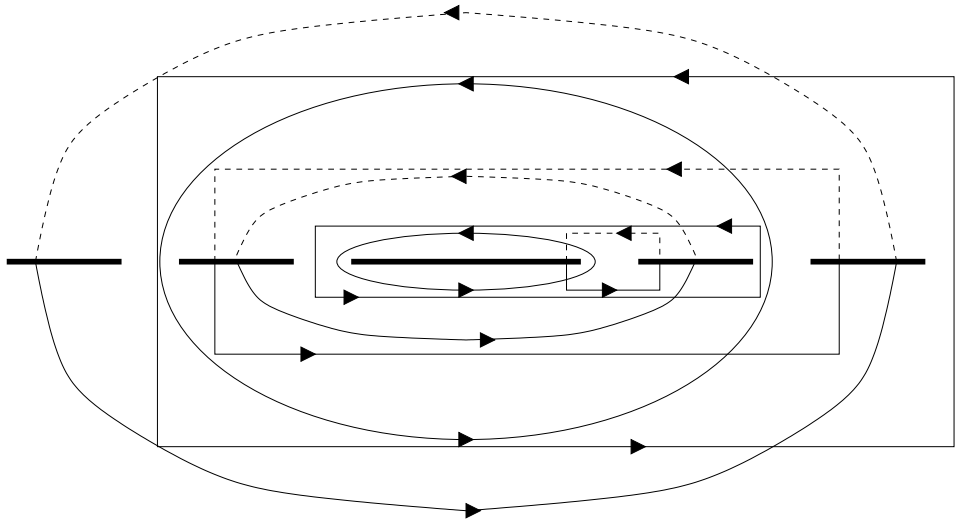,width=5 in}}
\end{center}
\caption{\em A particular choice of homology cycles, illustrated for
$G=4$, on a surface that is smoothly deformed so that the cuts lie
along a straight line.  The endpoints of the cuts are, from left to
right, $\lambda_G^*,
\dots,\lambda_0^*,\lambda_0,\dots,\lambda_G$.
Paths on the first (second) sheet are indicated with solid (dashed) lines.}
\label{fig:homologytopology}
\end{figure}

Let us specify a useful homology basis.  For topological purposes, we
can deform each sheet of $X$ so that the contour becomes a straight
line along which the endpoints occur from left to right in order:
$\lambda_G^*,\dots,\lambda_1^*,\lambda_0^*,
\lambda_0,\lambda_1,\dots,\lambda_G$.  The basis we choose is then
illustrated in Figure~\ref{fig:homologytopology}.  
The $a$ cycles appear in order $a_1,a_2,\dots$ from the inside out as
the oval paths, while the $b$ cycles appear in the same order as
rectangular paths.  Paths on the first sheet are solid, while their
continuations through the bold cuts are dashed.  Although this
illustration is for genus $G=4$, it should be obvious how the pattern
generalizes for other genera.  

\begin{figure}[h]
\begin{center}
\mbox{\psfig{file=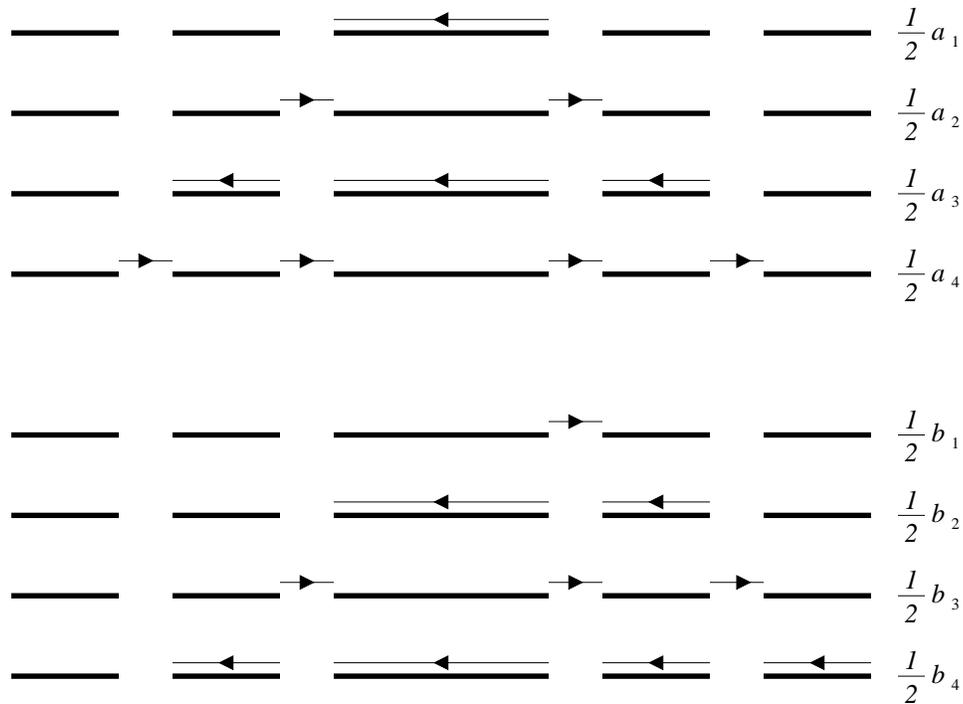,width=5 in}}
\end{center}
\caption{\em Evaluating homology cycle integrals of any antisymmetric 
differential by integrating on the first sheet along the immediate
left of the jump contour of the model Riemann-Hilbert Problem
\ref{rhp:model}.  In each line of this figure, the jump contour is
imagined as a straight line oriented from left to right and the bands
are shown in bold.  It should be clear how this picture generalizes to
other genera.}
\label{fig:cutintegrals}
\end{figure}

The point of such a choice is that it simplifies certain integrals
on the Riemann surface $X$.
\begin{definition}
By an antisymmetric differential \index{differential!antisymmetric}
$\omega(P)$ on $X$ we mean one for which whenever $\lambda$ is not a
branch point and $P^+(\lambda)$ and $P^-(\lambda)$ are the distinct
preimages on $X$ of $\lambda$ under the sheet projection mapping, then
$\omega(P^-(\lambda))=-\omega(P^+(\lambda))$.
\label{def:antisymmetric}
\end{definition}
Such antisymmetric differentials include the holomorphic differentials
$\nu_k(P)$ and the meromorphic differential $\Omega(P)$.  With the
particular choice of homology basis illustrated in
Figure~\ref{fig:homologytopology}, it is easy to express loop
integrals of any antisymmetric differential $\omega(P)$ in terms of
integrals only on the first sheet of $X$, along the jump contour of
the model Riemann-Hilbert Problem \ref{rhp:model} for $\tilde{\bf
O}(\lambda)$.  The paths of integration on the first sheet of $X$
corresponding to the cycles making up the homology basis are
illustrated in Figure~\ref{fig:cutintegrals}.
The first consequence of this choice is that the integral of any
antisymmetric differential over an $a$ cycle is equal to twice the
value of an integral over a concrete path on the first sheet of the
cut plane that has a symmetry under complex conjugation of the plane.
Given an oriented path $c$ on the cut plane, denote by $c^*$ the path
obtained by complex conjugation followed by reversal of orientation.
From Figure~\ref{fig:cutintegrals}, keeping in mind that in the figure
complex conjugation of the plane corresponds to left/right reflection,
one then sees that 
\begin{equation}
\frac{1}{2}a_j-\frac{1}{2}a_j^* = 0\,,
\end{equation}
for all $j=1,\dots,G$.  For the homology cycles $b_j$ we can find
similar relations:
\begin{equation}
\frac{1}{2}b_j+\frac{1}{2}b_j^*\equiv 0 \,,\hspace{0.2 in}
\mbox{modulo }\frac{1}{2}\{a_1,\dots,a_G\}\,.
\label{eq:symmetries}
\end{equation}
For example, $b_1+b_1^*=a_2$, $b_2+b_2^*=a_1+a_3$, and so on.  It is
easy to see that with such a choice of homology cycles, the constants
$c_{kj}$ in the holomorphic differentials are all made manifestly real
by the normalization condition (\ref{eq:holomorphicnormalization}).
Indeed, the linear equations implied by
(\ref{eq:holomorphicnormalization}) for the constants $c_{kj}$ all
have real coefficients, and the system is invertible.  Once it is
known that these constants are all real, the symmetries
(\ref{eq:symmetries}) can be used to show that 
\begin{equation}
\Im(H_{jk}) = 2\pi n_{jk}\,,
\mbox{  where  }n_{jk}\in{\mathbb Z}\,,\hspace{0.2 in}j,k=1,\dots,G\,.
\label{eq:Himag}
\end{equation}
Next, note that since the coefficients of the polynomial $p(\lambda)$
are real, it follows from the symmetry of the cycles $a_k$ in this
special homology basis on $X$ that for any path $c$ on the cut plane,
\begin{equation}
\int_{c^*}\Omega(P)=-\left(\int_c\Omega(P)\right)^*\,,
\end{equation}
where $\Omega(P)$ is the meromorphic differential defined by
(\ref{eq:Omegadef}) and (\ref{eq:meromorphicnormalization}).  Using
this relation together with the symmetry relations
(\ref{eq:symmetries}) and the normalization conditions
(\ref{eq:meromorphicnormalization}) defining the differential
$\Omega(P)$, we find
\begin{equation}
\oint_{b_j}\Omega(P)=2\int_{\frac{1}{2}b_j}\Omega(P)=
-2\int_{\frac{1}{2}b_j^*}\Omega(P)=2\left(\int_{\frac{1}{2}b_j}\Omega(P)
\right)^*=\left(\oint_{b_j}\Omega(P)\right)^*\,,
\end{equation}
where the first and last integrals in the chain of equalities are loop
integrals on the Riemann surface $X$, and the intermediate integrals
are all taken on concrete paths in the cut plane according to
Figure~\ref{fig:cutintegrals}.  This calculation shows that the
components of the vector ${\bf U}$ are all real.

The next gauge symmetry \index{gauge!symmetry} we break is the
invariance with respect to choice of base point $P_0$.  For several
reasons, it is convenient in the hyperelliptic context to choose $P_0$
to be a branch point; here we take $P_0=\lambda_0$.  One advantage of
this choice is that the Riemann constant vector takes a particularly
simple form.  Using the fact that with our choice of homology basis,
the hyperelliptic (sheet exchanging) involution \index{hyperelliptic
involution} of $X$ takes each $a$ cycle into its opposite ({\em i.e.}
into the same loop with opposite orientation), one finds that
\begin{equation}
{\bf K}\equiv \tilde{\bf K} \pmod \Lambda\,,
\end{equation}
where the components of $\tilde{\bf K}$ are given
by
\begin{equation}
\tilde{K}_k=\pi i + \frac{H_{kk}}{2}\,,\hspace{0.2 in}k=1,\dots,G\,.
\label{eq:simpleRiemannconstant}
\end{equation}
By the observation (\ref{eq:Himag}) about the Riemann matrix ${\bf
H}$, we see that the imaginary parts of the components of $\tilde{\bf
K}$ are all integer multiples of $\pi$.

For concreteness we will now choose vectors ${\bf V}_m\in {\mathbb
C}^G$ for $m=1,2$ so that ${\bf A}({\cal D}_m;\lambda_0)+{\bf K} =
{\bf V}_m \pmod\Lambda$.  Before doing this, however, we will first
select a specific path of integration used to define the Abel-Jacobi
mapping itself.  Given a point $P\in X$, this is done by specifying a
path from $P_0=\lambda_0$ to $P$ modulo homotopy.  
\begin{definition}
Let $P\in X$.  By $C_P$ we mean any element of the homotopy equivalence
class of paths from $\lambda_0$ to $P$ on $X$ such that the following
three things are true:
\begin{enumerate}
\item
Each point on $C_P$ lies on the
same sheet as $P$ (being a branch point the base point is considered
to lie on both sheets).
\item
$C_P$ completely avoids the
whole portion of the contour from $\lambda_G^*$
through to $\lambda_G$. 
\item
$C_P$ begins on the ``$+$''
side of the base point $\lambda_0$ on the contour.  
\end{enumerate}
Note that to define the path $C_P$ it is essential to view $X$ as two
copies of the plane cut along the jump contour of the model Riemann-Hilbert
Problem \ref{rhp:model}.
\label{def:path}
\end{definition}
Using the path $C_P$ in the Abel-Jacobi mapping, and recalling
that the same path is used in the integration of the differential
$\Omega$, we see that
\begin{equation}
\begin{array}{rcll}
\displaystyle
\int_{C_P}\Omega(P')&=&h(\lambda(P))-h_+(\lambda_0)\,,&
P\,\,\,\mbox{on the first sheet of}\,\,\,X\,,\\\\
\displaystyle
\int_{C_P}\Omega(P')&=&-h(\lambda(P))+h_+(\lambda_0)\,,&
P\,\,\,\mbox{on the second sheet of}\,\,\,X\,.
\end{array}
\end{equation}
Note that it is sufficient here for $C_P$ to be defined modulo
homotopy because the meromorphic differential $\Omega(P)$ has no
residues. To give representatives ${\bf V}_m$ for ${\bf A}({\cal
D}_m;\lambda_0)+{\bf K}$, we first choose to represent the Riemann
constant vector ${\bf K}$ in ${\mathbb C}^G$ {\em exactly} by the
vector $\tilde{\bf K}$ defined by (\ref{eq:simpleRiemannconstant}).
To represent ${\bf A}({\cal D}_1;\lambda_0)$ and ${\bf A}({\cal
D}_2;\lambda_0)$ in ${\mathbb C}^G$, it suffices by Abel's Theorem to
represent respectively ${\bf A}({\cal D}^0;\lambda_0)-{\bf
A}(\infty_2;\lambda_0)$ and ${\bf A}({\cal D}^0;\lambda_0)-{\bf
A}(\infty_1;\lambda_0)$ ({\em cf.}  (\ref{eq:AbelsTheorem})).  To
begin with, we define ${\bf A}(\infty_m;\lambda_0)$ for $m=1$ and
$m=2$ by setting
\begin{equation}
A_k(\infty_m,\lambda_0)=\int_{C_{\infty_m}}\nu_k(P)\,,
\end{equation}
with the path $C_{\infty_m}$ on $X$ chosen according to
Definition~\ref{def:path}.

It then follows that for these representatives ${\bf
A}(\infty_2;\lambda_0)=-{\bf A}(\infty_1;\lambda_0)$.  Finally, to
represent ${\bf A}({\cal D}^0;\lambda_0)$, we associate with each
branch point in ${\cal D}^0$ a point on the first sheet of $X$
immediately on the ``$+$'' side of the contour, and compute ${\bf
A}({\cal D}^0;\lambda_0)$ as a sum of integrals with paths determined
according to Definition~\ref{def:path}.  Each such integral may be
realized as lying on the ``$+$'' side of the contour, and by the
scheme described in Figure~\ref{fig:cutintegrals}, can be identified
with a specific half-period in $\Lambda/2$, where the lattice
$\Lambda$ is defined by (\ref{eq:Lambdadef}).  To be quite precise,
let ${\bf A}^{\rm cut}(\lambda)$ denote the Abel mapping with base
point and contour $C_P$ chosen according to Definition~\ref{def:path}
for the point $P$ on the first sheet of $X$ for which
$\lambda(P)=\lambda$.  Then
\begin{equation}
\begin{array}{rcl}
V_{1,k}&=&\displaystyle
\left(A^{\rm cut}_k(\lambda_{1+}^*)+A^{\rm cut}_k(\lambda_{2+})+A^{\rm cut}_k
(\lambda_{3+}^*)+\dots+A^{\rm cut}_k(\lambda_{G+})\right)
+A^{\rm cut}_k(\infty) +\pi i + \frac{H_{kk}}{2}\,,\\\\
V_{2,k}&=&\displaystyle
\left(A^{\rm cut}_k(\lambda_{1+}^*)+A^{\rm cut}_k(\lambda_{2+})+
A^{\rm cut}_k(\lambda_{3+}^*)+\dots+A^{\rm cut}_k(\lambda_{G+})\right)
-A^{\rm cut}_k(\infty) +\pi i + \frac{H_{kk}}{2}\,,
\end{array}
\end{equation}
where $k$ varies between $1$ and $G$.

In terms of these gauge choices\index{gauge!choices}, we find
\begin{equation}
y_m(P)=\left\{\begin{array}{ll} \displaystyle \frac{\Theta({\bf
A}^{\rm cut}(\lambda(P))-{\bf V}_m+i{\bf U}/\hbar)} {\Theta({\bf
A}^{\rm cut}(\lambda(P))-{\bf V}_m)}e^{ih(\lambda(P))/\hbar}
e^{-ih_+(x,t,\lambda_0)/\hbar}\,, &P\,\,\,\mbox{on first sheet}\,,
\\\\ \displaystyle \frac{\Theta(-{\bf A}^{\rm
cut}(\lambda(P))-{\bf V}_m+i{\bf U}/\hbar)} {\Theta(-{\bf A}^{\rm
cut}(\lambda(P))-{\bf V}_m)}e^{-ih(\lambda(P))/\hbar}
e^{ih_+(\lambda_0)/\hbar}\,,& P\,\,\,\mbox{on second sheet}\,.
\end{array}\right.
\label{eq:sheetwisey}
\end{equation}
The normalizing constants $N_m$ defined by
(\ref{eq:normalizingconstants}) are easily obtained from
(\ref{eq:sheetwisey}), since $p(\lambda)=h(\lambda)+\bo (1/\lambda)$
as $\lambda$ tends to infinity.  Thus,
\begin{equation}
\begin{array}{rcl}
N_1&=&\displaystyle
\frac{\Theta({\bf A}^{\rm cut}(\infty)-{\bf V}_1+i{\bf U}/\hbar)}
{\Theta({\bf A}^{\rm cut}(\infty)-{\bf V}_1)}e^{-ih_+(\lambda_0)/\hbar}\,,
\\\\
N_2&=&\displaystyle
\frac{\Theta(-{\bf A}^{\rm cut}(\infty)-{\bf V}_2+i{\bf U}/\hbar)}
{\Theta(-{\bf A}^{\rm cut}(\infty)-{\bf V}_2)}e^{ih_+(\lambda_0)/\hbar}\,.
\end{array}
\end{equation}
Combining these with the sheetwise formula (\ref{eq:sheetwisey}) for
$y_m(P)$ gives a similar sheetwise formula for the functions $z_m(P)$.
From these, one obtains sheetwise formulae for the functions $v_m(P)$
defined originally in (\ref{eq:vPdef}) by introducing 
\begin{equation}
b^\pm(\lambda)=\frac{R(\lambda)\pm (\lambda-\lambda_0^*)(\lambda-\lambda_1)\dots (\lambda-\lambda_G^*)}{2R(\lambda)}\,,
\end{equation}
and then observing that for $P$ on the first sheet of $X$,
$f_1(P)=b^-(\lambda(P))$ and $f_2(P)=b^+(\lambda(P))$, while for $P$
on the second sheet of $X$, $f_1(P)=b^+(\lambda(P))$ and
$f_2(P)=b^-(\lambda(P))$.  Since according to (\ref{eq:vPdef}), the
first (respectively second) column of $\bf Q$ is simply the vector
$(v_1,v_2)$ restricted to the first (respectively second) sheet of
$X$, we have thus obtained an explicit representation of the matrix
$\bf Q$.  By the elementary transformations at the beginning of this
section, we see that we have proved:
\begin{theorem}
For even $G>0$, the unique solution of the outer model Riemann-Hilbert
Problem~\ref{rhp:model} is given by the formulae:
\begin{equation}
\begin{array}{rcl}
\tilde{O}_{11}(\lambda)&=&\displaystyle
\frac{b^-(\lambda)}{\beta(\lambda)}
\frac{\Theta({\bf A}^{\rm cut}(\infty)-{\bf V}_1)}
{\Theta({\bf A}^{\rm cut}(\lambda)-{\bf V}_1)}
\frac{\Theta({\bf A}^{\rm cut}(\lambda)-{\bf V}_1+i{\bf U}/\hbar)}
{\Theta({\bf A}^{\rm cut}(\infty)-{\bf V}_1+i{\bf U}/\hbar)}\,,\\\\
\tilde{O}_{12}(\lambda)&=&\displaystyle
\frac{b^+(\lambda)}{\beta(\lambda)}
e^{2ih_+(\lambda_0)/\hbar}
\frac{\Theta({\bf A}^{\rm cut}(\infty)-{\bf V}_1)}
{\Theta(-{\bf A}^{\rm cut}(\lambda)-{\bf V}_1)}
\frac{\Theta(-{\bf A}^{\rm cut}(\lambda)-{\bf V}_1+i{\bf U}/\hbar)}
{\Theta({\bf A}^{\rm cut}(\infty)-{\bf V}_1+i{\bf U}/\hbar)}\,,\\\\
\tilde{O}_{21}(\lambda)&=&\displaystyle
\frac{b^+(\lambda)}{\beta(\lambda)}
e^{-2ih_+(\lambda_0)/\hbar}
\frac{\Theta(-{\bf A}^{\rm cut}(\infty)-{\bf V}_2)}
{\Theta({\bf A}^{\rm cut}(\lambda)-{\bf V}_2)}
\frac{\Theta({\bf A}^{\rm cut}(\lambda)-{\bf V}_2+i{\bf U}/\hbar)}
{\Theta(-{\bf A}^{\rm cut}(\infty)-{\bf V}_2+i{\bf U}/\hbar)}\,,\\\\
\tilde{O}_{22}(\lambda)&=&\displaystyle
\frac{b^-(\lambda)}{\beta(\lambda)}
\frac{\Theta(-{\bf A}^{\rm cut}(\infty)-{\bf V}_2)}
{\Theta(-{\bf A}^{\rm cut}(\lambda)-{\bf V}_2)}
\frac{\Theta(-{\bf A}^{\rm cut}(\lambda)-{\bf V}_2+i{\bf U}/\hbar)}
{\Theta(-{\bf A}^{\rm cut}(\infty)-{\bf V}_2+i{\bf U}/\hbar)}\,.
\end{array}
\end{equation}
\end{theorem}

Note that from the jump relations for $h(\lambda)$, 
\begin{equation}
2h_+(\lambda_0)=-\theta_1-\alpha_0\,.
\end{equation}
The matrix $\tilde{\bf O}(\lambda)$ therefore has the property
that it is uniformly bounded as $\hbar$ tends to zero in any fixed
closed set that does not contain an endpoint $\lambda_k$ or
$\lambda_k^*$.  This kind of behavior is crucial for controlling the
error of these approximations in \S\ref{sec:error}.  Moreover, away
from the endpoints all derivatives with respect to $\lambda$ are
uniformly bounded as $\hbar$ tends to zero.  The $\hbar$ dependence is
totally explicit, and contributes only global phase oscillations.

\subsection{Properties of the semiclassical 
solution of the nonlinear Schr\"odinger equation.}  Consider the
function $\tilde{\psi}$ defined from the solution of the outer model
Riemann-Hilbert Problem~\ref{rhp:model} by
\begin{equation}
\tilde{\psi}:=2i\lim_{\lambda\rightarrow\infty} 
\lambda\tilde{O}_{12}(\lambda)\,.
\end{equation}
Since by direct computation,
\begin{equation}
b^+(\lambda)=\lambda^{-1}\sum_{k=0}^G \frac{i}{2}(-1)^{k+1}\Im(\lambda_k) +
\bo(\lambda^{-2})\,,\hspace{0.3 in}\lambda\rightarrow\infty\,,
\end{equation}
we find
\begin{equation}
\tilde{\psi}=ae^{iU_0/\hbar}\frac{\Theta({\bf
Y}+i{\bf U}/\hbar)}{\Theta({\bf Z}+i{\bf
U}/\hbar)}\,,
\label{eq:psitildeGgtzero}
\end{equation}
where
\begin{equation}
a=\frac{\Theta({\bf Z})}
{\Theta({\bf Y})}
\sum_{k=0}^G(-1)^k\Im(\lambda_k)\,,
\end{equation}
and
\begin{equation}
U_0=-(\theta_1+\alpha_0)\,,
\end{equation}
with
\begin{equation}
{\bf Y}=-{\bf A}^{\rm cut}(\infty)-{\bf V}_1\,,\hspace{0.3 in}
{\bf Z}={\bf A}^{\rm cut}(\infty)-{\bf V}_1\,.
\end{equation}

Subject to finding an appropriate complex phase function
$g^\sigma(\lambda)$ as described in \S\ref{sec:conditions}, we will
prove in \S\ref{sec:error} that the function $\tilde{\psi}$ captures
the leading order behavior of the true solution $\psi$ of the
nonlinear Schr\"odinger equation as $\hbar$ tends to zero.  Here, we
show simply that this asymptotic solution is locally a slowly
modulated $G+1$ phase wavetrain\index{multiphase wavetrain!slowly
modulated}.  To see this, set $x=x_0+\hbar
\hat{x}$ and $t=t_0+\hbar\hat{t}$, and expand $\tilde{\psi}$ using
Taylor series for small $\hbar$, recalling that all quantities depend
parametrically on $x$ and $t$:
\begin{equation}
\tilde{\psi}
=
a^0e^{iU_0^0/\hbar}
e^{i(k_0^0\hat{x}-w_0^0\hat{t})}
\frac{\Theta({\bf Y}^0+
i{\bf U}^0/\hbar+
i({\bf k}^0\hat{x}-{\bf w}^0\hat{t}))}
{
\Theta({\bf Z}^0+
i{\bf U}^0/\hbar+
i({\bf k}^0\hat{x}-{\bf w}^0\hat{t}))}
\cdot(1+\bo(\hbar))\,,
\end{equation}
where
\begin{equation}
k_n=\partial_x U_n\,,\hspace{0.3 in}
w_n=-\partial_t U_n\,,\hspace{0.3 in}
n=0,\dots,G\,,
\end{equation}
and the superscript $0$ indicates evaluation for $x=x_0$ and $t=t_0$.
As a generalization of the exponential function, the theta function is
$2\pi$ periodic in each imaginary direction in $\mathbb{C}^G$.
Therefore, for fixed $x_0$ and $t_0$, we see that the leading order
approximation is a multiphase wavetrain with wavenumbers\index{multiphase wavetrain!wavenumbers of}
$k_0^0,\dots,k_G^0$ and frequencies\index{multiphase wavetrain!frequencies of}
$w_0^0,\dots,w_G^0$ with respect to the variables
$\hat{x}$ and $\hat{t}$.  It is a simple consequence of the definition
of the wavenumbers and frequencies that the modulations of the
waveform due to variations in $x_0$ and $t_0$ are constrained by {\em
conservation of waves}\index{conservation of waves}:
\begin{equation}
\partial_t k_n +\partial_x w_n=0\,.
\end{equation}

\subsection{Genus zero.}
For $G=0$ the whole construction given in this section degenerates
somewhat and theta functions are not required.  For completeness we
give all details here for this special case.  The function
$k(\lambda)$ can be evaluated explicitly by residues:
\begin{equation}
k(\lambda)=-\frac{\alpha_0}{2R(\lambda)}\,.
\end{equation}
Therefore
\begin{equation}
h(\lambda)=-\frac{1}{2}\alpha_0=h_0\,,
\end{equation}
and $p(\lambda)$ is simply a constant (with respect to $\lambda$)
function $h_0$.  It follows that the functions $v_1(P)$ and
$v_2(P)$ that we seek on the compact Riemann surface $X$ of genus
zero are in fact meromorphic functions on the whole of $X$.  These
functions each have a single simple pole at the point $\lambda_0$, and
then $v_1(\infty_2)=0$ while $v_2(\infty_1)=0$.  To give
explicit formulae for $v_1(P)$ and $v_2(P)$, we use the
``lifting'' $R_X(P)$ of the function $R(\lambda)$ to $X$.
This meromorphic function satisfies $R_X(P)\sim -\lambda(P)$ as
$P\rightarrow\infty_1$ and $R_X(P)\sim \lambda(P)$ as
$P\rightarrow\infty_2$.  We then have
\begin{equation}
v_1(P)=-\frac{1}{2}e^{ih_0/\hbar}
\left[\frac{\lambda(P)-\lambda_0^*}{R_X(P)}-1\right]\,,\hspace{0.2 in}
v_2(P)=\frac{1}{2}e^{-ih_0/\hbar}
\left[\frac{\lambda(P)-\lambda_0^*}{R_X(P)}+1\right]\,.
\end{equation}
Restricting respectively to the first and second sheets of $X$ gives
an explicit formula for the matrix ${\bf Q}(\lambda)$ in the cut plane:
\begin{equation}
{\bf Q}(\lambda)=\frac{1}{2R(\lambda)}\exp(i\sigma_3 h_0/\hbar)
\left[\begin{array}{cc}
-\lambda+\lambda_0^*+R(\lambda) & \lambda-\lambda_0^*+R(\lambda)\\\\
\lambda-\lambda_0^*+R(\lambda) & -\lambda+\lambda_0^*+R(\lambda)
\end{array}\right]\,.
\end{equation}
Finally, in terms of the function $\beta(\lambda)$ we see that we have
proved the following.
\begin{theorem}
For $G=0$, the unique solution of the outer model Riemann-Hilbert
Problem~\ref{rhp:model} is given explicitly by
\begin{equation}
\tilde{\bf O}(\lambda)
=
\frac{1}{2R(\lambda)\beta(\lambda)}
\left[\begin{array}{cc}
-\lambda+\lambda_0^*+R(\lambda) & 
(\lambda-\lambda_0^*+R(\lambda))\exp(-i\alpha_0/\hbar)\\\\
(\lambda-\lambda_0^*+R(\lambda))\exp(i\alpha_0/\hbar) &
-\lambda+\lambda_0^*+R(\lambda)\end{array}\right]\,.
\end{equation}
\end{theorem}

It is then a direct matter to compute the corresponding semiclassical
asymptotic description of the solution of the nonlinear Schr\"odinger
equation:
\begin{equation}
\tilde{\psi}:=2i\lim_{\lambda\rightarrow\infty}\lambda
\tilde{O}_{12}(\lambda)=
\Im(\lambda_0)e^{-i\alpha_0/\hbar}\,,
\label{eq:psitildeGeqzero}
\end{equation}
where we recall that generally $\lambda_0$ and $\alpha_0$ will depend
on $x$ and $t$.

\subsection{The outer approximation for ${\bf N}^\sigma(\lambda)$.}
\label{sec:outerparametrix}
\index{outer approximation}
As a final step in this section, we use the solution of the outer
model problem to construct an approximation of ${\bf N}^\sigma$.
This approximation is obtained from $\tilde{\bf O}$ by
redefining the matrix within the lenses on either side of the bands,
and is given explicitly by:
\begin{equation}
\hat{\bf N}^\sigma_{\rm out}(\lambda) := \tilde{\bf
O}(\lambda) {\bf D}^\sigma(\lambda)^{-1}\,,
\label{eq:outerparametrixdef}
\end{equation}
where ${\bf D}^\sigma(\lambda)$ is the explicit piecewise analytic
``lens transformation'' \index{lens transformation} relating
$\tilde{\bf N}^\sigma(\lambda)$ and ${\bf O}^\sigma(\lambda)$:
\begin{equation}
{\bf O}^\sigma(\lambda)=\tilde{\bf N}^\sigma(\lambda)
{\bf D}^\sigma(\lambda)\,.
\end{equation}
Recall that the matrix ${\bf D}^\sigma(\lambda)$ is equal to the
identity outside of all lenses.  In between the contours $C_{k+}^+$ and
$I_k^+$, 
\begin{equation}
{\bf D}^\sigma(\lambda):=\sigma_1^{\frac{1-J}{2}}
\left[\begin{array}{cc}1 & -i\exp(-iJ\alpha_k/\hbar)\exp(-ir_k(\lambda)/\hbar)\\0 & 1\end{array}\right]\sigma_1^{\frac{1-J}{2}}\,,
\label{eq:ldef1}
\end{equation}
while in between the contours $I_k^+$ and $C_{k-}^+$,
\begin{equation}
{\bf D}^\sigma(\lambda):=\sigma_1^{\frac{1-J}{2}}
\left[\begin{array}{cc}1 & 
i\exp(-iJ\alpha_k/\hbar)\exp(ir_k(\lambda)/\hbar)\\ 0 & 1\end{array}\right]
\sigma_1^{\frac{1-J}{2}}\,,
\label{eq:ldef2}
\end{equation}
and for all $\lambda$ in the lower half-plane, ${\bf
D}^\sigma(\lambda)=\sigma_2{\bf D}^\sigma(\lambda^*)^*\sigma_2$.
Here, the functions $r_k(\lambda)$ are defined by
Definition~\ref{def:qr}.  It is easy to check that from the properties
of $\tilde{\bf O}(\lambda)$, the approximation of ${\bf N}^\sigma(\lambda)$
defined above is analytic on the real axis.  Also, we have the
following useful fact, whose proof is immediate.
\begin{lemma}
The outer approximation $\hat{\bf N}_{\rm out}^\sigma(\lambda)$ is
analytic in the complex $\lambda$-plane except for $\lambda\in C\cup
C^*$ and the boundaries of the lenses $C_{k\pm}^+$ and $C_{k\pm}^-$.
In each closed subinterval of the interior of any one band or gap of
$C_\sigma$ or $[C^*]_\sigma$, $\hat{\bf N}_{\rm out}^\sigma(\lambda)$
takes on continuous boundary values.  In each band these boundary
values satisfy exactly the jump relation
relation
\begin{equation} \hat{\bf N}^\sigma_{{\rm out},+}(\lambda)= \hat{\bf
N}^\sigma_{{\rm out},-}(\lambda){\bf v}^\sigma_{\tilde{\bf
N}}(\lambda)\,,
\end{equation} 
whereas in the gaps, 
\begin{equation} \hat{\bf N}^\sigma_{{\rm out},+}(\lambda)= \hat{\bf
N}^\sigma_{{\rm out},-}(\lambda)\exp(iJ\theta^\sigma(\lambda)\sigma_3/\hbar)
\,,
\end{equation} 
where we recall that in each gap the function $\theta^\sigma(\lambda)$
is a real constant.  On the lens boundaries the jump relation is
\begin{equation}
\begin{array}{rcll}
\hat{\bf N}^\sigma_{{\rm out},+}(\lambda)&=& \hat{\bf
N}^\sigma_{{\rm out},-}(\lambda){\bf D}^\sigma_-(\lambda)\,,&\hspace{0.2 in}
\lambda\in C^\pm_{k+}\,, \\
\hat{\bf N}^\sigma_{{\rm out},+}(\lambda)&=& \hat{\bf
N}^\sigma_{{\rm out},-}(\lambda){\bf
D}^\sigma_+(\lambda)^{-1}\,,&\hspace{0.2 in}
\lambda\in C^\pm_{k-}\,.
\end{array}
\end{equation}
Finally, for $\lambda$ in any $\hbar$-independent closed set that does
not contain $\lambda_0,\dots,\lambda_G$ or
$\lambda_0^*,\dots,\lambda_G^*$, $\hat{\bf N}_{\rm
out}^\sigma(\lambda)$ is uniformly bounded as $\hbar$ tends to zero.
\label{lemma:outerjumps}
\end{lemma}

\section[Inner Approximations]{Inner approximations.}
\label{sec:inner}
As pointed out in \S\ref{sec:conditions}, the {\em ad hoc}
approximations made in obtaining the outer model problem from the
original Riemann-Hilbert problem for ${\bf N}^\sigma(\lambda)$ given a
complex phase function $g^\sigma(\lambda)$ clearly break down in the
neighborhood of each endpoint $\lambda_0,\dots,\lambda_G$, its complex
conjugate, and also near $\lambda=0$.  Therefore, we now turn our
attention to these troublesome neighborhoods, and develop inner model
problems to approximate ${\bf N}^\sigma(\lambda)$ locally in each
case.  It will suffice to construct approximations of ${\bf
N}^\sigma(\lambda)$ near $\lambda_0, \lambda_1,\dots,\lambda_G$ and
near $\lambda=0$, because we may then use complex-conjugation symmetry to
obtain approximations near the conjugate points.

In this section, we work under one further assumption about
$g^\sigma(\lambda)$ that will be justified generically (with respect
to the parameters $x$ and $t$) in Chapter~\ref{sec:ansatz}.  Thus we have:
\begin{equation}
\parbox{5 in}{{\bf Working assumption:  }
the density $\rho^\sigma(\lambda)$ vanishes exactly like a square
root, and not to higher order, at each endpoint $\lambda_k$ for
$k=0,\dots,G$.}
\label{eq:workingassumption}
\index{square-root vanishing!assumption of}
\end{equation}
The generic nature of this assumption is clarified somewhat in
Lemma~\ref{lemma:rhowithY}.  Higher-order vanishing of
$\rho^\sigma(\lambda)$ at an endpoint corresponds to the intersection
of two curves in the real $(x,t)$-plane: a curve where the inequality
in a gap fails and a curve where the inequality in an adjacent band
fails.  So in making this assumption we are omitting from
consideration a set of isolated points in the $(x,t)$-plane.  The
nature of the local approximations near the endpoints depends
crucially on the degree of vanishing of $\rho^\sigma(\lambda)$ and we
want to consider here only the most likely case.  For details on
the analogous construction necessary for less generic cases, see section 5 of
\cite{DKMVZ98B}.

\subsection{Local analysis for $\lambda$ near the endpoint 
$\lambda_{2k}$ for $k=0,\dots,G/2$.}
\label{sec:2k}
\begin{figure}[h]
\begin{center}
\mbox{\psfig{file=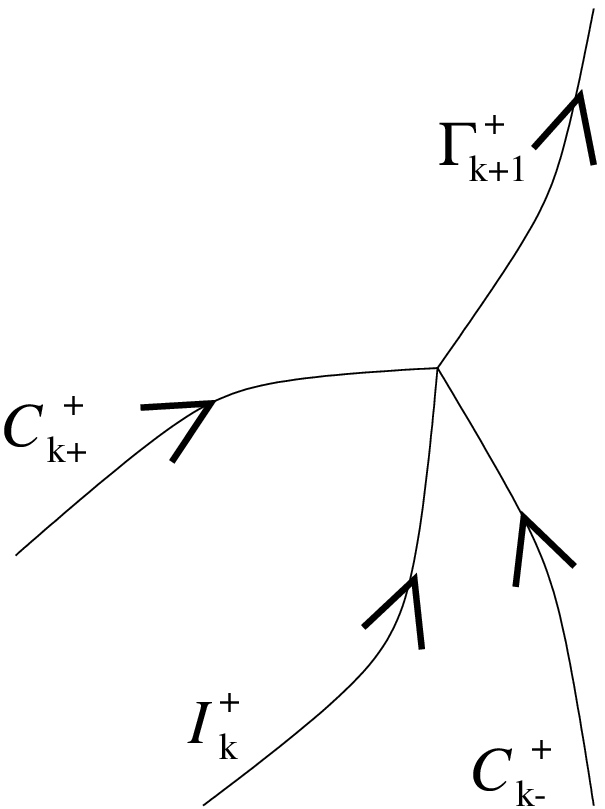,height=2 in}}
\end{center}
\caption{\em The jump matrix near $\lambda_{2k}$ differs from the identity
on a self-intersecting contour with $\lambda_{2k}$ at the intersection point.}
\label{fig:evenendpoint}
\end{figure}

Near an endpoint $\lambda_{2k}$
for $k=0,\dots,G/2$, the approximation of replacing ${\bf
N}^\sigma(\lambda)$ by $\tilde{\bf N}^\sigma(\lambda)$, the
continuum limit, is expected to be valid; the trouble is with the
approximation of the matrix ${\bf O}^\sigma(\lambda)$ by the
matrix $\tilde{\bf O}(\lambda)$.  Locally, the contour
of the Riemann-Hilbert problem for ${\bf O}^\sigma(\lambda)$
looks like that shown in Figure~\ref{fig:evenendpoint}.
Recall the jump relations for ${\bf O}^\sigma(\lambda)$.  For
$\lambda\in\Gamma_{k+1}^+$,
\begin{equation}
{\bf O}^\sigma_+(\lambda)={\bf
O}^\sigma_-(\lambda)\sigma_1^{\frac{1-J}{2}}
\left[\begin{array}{cc}\exp(iJ\theta_{k+1}/\hbar) & 0 \\
-i\exp(\tilde{\phi}^\sigma(\lambda)/\hbar) &
\exp(-iJ\theta_{k+1}/\hbar)
\end{array}\right]
\sigma_1^{\frac{1-J}{2}}\,,
\end{equation}
for $\lambda\in I_k^+$,
\begin{equation}
{\bf O}^\sigma_+(\lambda)={\bf
O}^\sigma_-(\lambda)\sigma_1^{\frac{1-J}{2}}
\left[\begin{array}{cc} 0 & -i\exp(-iJ\alpha_k/\hbar)\\
-i\exp(iJ\alpha_k/\hbar) & 0
\end{array}\right]
\sigma_1^{\frac{1-J}{2}}\,,
\end{equation}
and for $\lambda\in C^+_{k\pm}$,
\begin{equation}
{\bf O}^\sigma_+(\lambda)={\bf
O}^\sigma_-(\lambda)\sigma_1^{\frac{1-J}{2}}
\left[\begin{array}{cc} 1 & i\exp(-iJ\alpha_k/\hbar)\exp(\mp
ir_k(\lambda)/\hbar)\\ 0 & 1\end{array}\right]
\sigma_1^{\frac{1-J}{2}}\,.
\end{equation}
The constants $\alpha_k$ and $\theta_{k+1}$ are related to the
functions $\tilde{\phi}^\sigma(\lambda)$ and $r_k(\lambda)$ (recall
that the latter is the analytic continuation of
$\theta^\sigma(\lambda)$ off of $I_k^+$ according to
Definition~\ref{def:qr}) by
\begin{equation}
r_k(\lambda_{2k})=J\theta_{k+1}\,,\hspace{0.3 in}
\tilde{\phi}^\sigma(\lambda_{2k})=iJ\alpha_k\,.
\end{equation}

Recall also that by the conditions imposed in
\S\ref{sec:conditions} on the complex phase function 
$g^\sigma(\lambda)$ via its density function $\rho^\sigma(\lambda)$
({\em cf.} Definition~\ref{def:admissiblerho}), we have that
$\Re(\tilde{\phi}^\sigma(\lambda))<0$ for
$\lambda\in\Gamma_{k+1}^+\setminus \{\lambda_{2k}\}$, and similarly
that $\Re(-ir_k(\lambda))<0$ for $\lambda\in
C^+_{k+}\setminus\{\lambda_{2k}\}$ and $\Re(ir_k(\lambda))<0$ for
$\lambda\in C^+_{k-}\setminus\{\lambda_{2k}\}$.  Also, the working
assumption (\ref{eq:workingassumption}) that $\rho^\sigma(\lambda)$
vanishes like a square root at $\lambda=\lambda_{2k}$ implies that the
function $r_k(\lambda)$ differs from $J\theta_{k+1}$ by a quantity
that vanishes like $(\lambda-\lambda_{2k})^{3/2}$.  In fact, the
analytic continuation formulae (\ref{eq:plusside}) and
(\ref{eq:minusside}) imply that
for $\lambda\in C^+_{k+}$ we have
\begin{equation}
\tilde{\phi}^\sigma(\lambda)-iJ\alpha_k=
i(r_k(\lambda)-J\theta_{k+1})\,,
\label{eq:Cplus}
\end{equation}
where $r_k(\lambda)$ is the continuation of
$\theta^\sigma(\lambda)$ from $I_k^+$ to the left, and where
$\tilde{\phi}^\sigma(\lambda)$ is the continuation of the
function with the same name from $\Gamma_{k+1}^+$ to the left.
Similarly, for $\lambda\in C^+_{k-}$ we have
\begin{equation}
\tilde{\phi}^\sigma(\lambda)-iJ\alpha_k=
-i(r_k(\lambda)-J\theta_{k+1})\,,
\label{eq:Cminus}
\end{equation}
where here $r_k(\lambda)$ is the continuation of
$\theta^\sigma(\lambda)$ from $I_k^+$ to the right, and
$\tilde{\phi}^\sigma(\lambda)$ is the continuation of the
function with the same name from $\Gamma_{k+1}^+$ to the right.

\begin{figure}[h]
\begin{center}
\mbox{\psfig{file=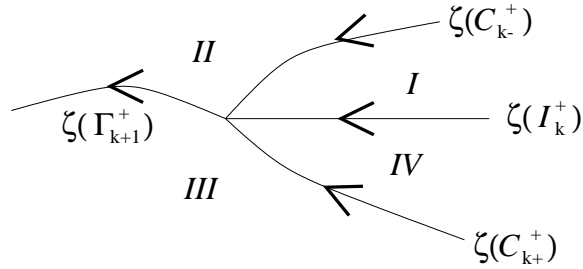,width=3 in}}
\end{center}
\caption{\em The image of the local jump contours in the $\zeta$-plane.
The point $\lambda=\lambda_{2k}$ is mapped to $\zeta=0$, and the contour
$I_k^+$ is mapped to the positive real $\zeta$-axis.}
\label{fig:evenendpointzeta}
\end{figure}

These facts suggest a local change of variables.  Let
$\zeta=\zeta(\lambda)$ be defined by 
\begin{equation}
\zeta(\lambda):=\left(\frac{r_k(\lambda)-J\theta_{k+1}}{\hbar}\right)^{2/3}\,,
\label{eq:zetadef}
\end{equation}
and note that $\zeta\in{\mathbb R}_+$ when $\lambda\in I_k^+$.  Since
by assumption $r_k(\lambda)$ approaches its value at $\lambda_{2k}$
like $(\lambda-\lambda_{2k})^{3/2}$, this change of variables is an
{\em invertible analytic map} of some sufficiently small (but with
size independent of $\hbar$) neighborhood of $\lambda_k$ containing no
other endpoints into the $\zeta$-plane.  In terms of this change of
variables, the discussion preceeding (\ref{eq:plusside}) and
(\ref{eq:minusside}) implies that for $\lambda\in
\Gamma_{k+1}^+$,
\begin{equation}
\frac{\tilde{\phi}^\sigma(\lambda)
-iJ\alpha_k}{\hbar}=-(-\zeta)^{3/2}\,.
\end{equation}
The transformation $\lambda\mapsto\zeta(\lambda)$ maps the local
contour diagram shown in Figure~\ref{fig:evenendpoint} into the
$\zeta$-plane as shown in Figure~\ref{fig:evenendpointzeta}.
We now center a disk $D_{2k}$ in the $\lambda$-plane at
$\lambda=\lambda_{2k}$, and we choose the radius of the disk to be
sufficiently small so that $D_{2k}$ contains no other endpoints and so
that the map $\zeta(\lambda)$ is a biholomorphic map of $D_{2k}$ to
the $\zeta$-plane.  We consider this radius to be independent of
$\hbar$.  We also exploit the fact that, as remarked in
\S\ref{sec:conditions}, the contours $\Gamma_{k+1}^+$ and $C^+_{k\pm}$
are not specifically determined, to {\em choose} them within the disk
$D_{2k}$ (taken here to be sufficiently small independent of $\hbar$)
so that $\zeta(\Gamma_{k+1}^+\cap D_{2k})$ lies on the negative real
$\zeta$-axis, and $\zeta(C^+_{k\pm}\cap D_{2k})$ lies on the straight
ray on which $\arg(\zeta)=\mp\pi/3$.  This choice straightens out the
contours shown in Figure~\ref{fig:evenendpointzeta}.  The image
$\zeta(D_{2k})$ is a domain containing $\zeta=0$ and {\em expanding}
as $\hbar$ tends to zero.

For expressing the exact jump conditions of the matrix ${\bf
O}^\sigma(\lambda)$ in terms of the new variable $\zeta$, it
is convenient to introduce a matrix ${\bf S}_{2k}(\lambda)$ defined by:
\begin{equation}
{\bf S}_{2k}(\zeta):=\left\{\begin{array}{ll} {\bf
O}^\sigma(\lambda(\zeta))\sigma_1^{\frac{1-J}{2}}
\exp(iJ\sigma_3(\theta_{k+1}-\alpha_k)/(2\hbar))\,,&
\zeta\in I\cup II\,,\\\\ {\bf
O}^\sigma(\lambda(\zeta))\sigma_1^{\frac{1-J}{2}}
\exp(-iJ\sigma_3(\theta_{k+1}+\alpha_k)/(2\hbar))\,,&
\zeta\in III\cup IV\,.
\end{array}\right.
\label{eq:Sdef}
\end{equation}
Then, the exact jump relations for ${\bf O}^\sigma(\lambda)$ become
quite simple.
For $\zeta\in\zeta(\Gamma_{k+1}^+)$,
\begin{equation}
{\bf S}_{2k+}(\zeta)={\bf S}_{2k-}(\zeta)
\left[\begin{array}{cc}1 & 0 \\
-i\exp(-(-\zeta)^{3/2}) & 1\end{array}\right]\,,
\end{equation}
for $\zeta\in\zeta(I_k^+)$,
\begin{equation}
{\bf S}_{2k+}(\zeta)={\bf S}_{2k-}(\zeta)
\cdot(-i\sigma_1)\,,
\end{equation}
and for $\zeta\in\zeta(C^+_{k\pm})$,
\begin{equation}
{\bf S}_{2k+}(\zeta)={\bf S}_{2k-}(\zeta)
\left[\begin{array}{cc}
1 & i\exp(\mp i\zeta^{3/2})\\0 & 1\end{array}\right]\,.
\end{equation}
We want to view this as a Riemann-Hilbert problem to be solved exactly
in $\zeta(D_{2k})$, but to pose this problem correctly,
we need to include auxiliary conditions to ensure that the local
solution matches well onto that of the outer model problem.

By way of comparison to the solution $\tilde{\bf O}(\lambda)$
of the outer model problem obtained in \S\ref{sec:outersolve}, we may
introduce an analogous local representation of $\tilde{\bf
O}(\lambda)$ in terms of the variable $\zeta$.  Define the
matrix $\tilde{\bf S}_{2k}(\zeta)$ by
\begin{equation}
\tilde{\bf S}_{2k}(\zeta):=\left\{\begin{array}{ll} \tilde{\bf
O}(\lambda(\zeta))\sigma_1^{\frac{1-J}{2}}
\exp(iJ\sigma_3(\theta_{k+1}-\alpha_k)/(2\hbar))\,,&
\zeta\in I\cup II\,,\\\\ \tilde{\bf
O}(\lambda(\zeta))\sigma_1^{\frac{1-J}{2}}
\exp(-iJ\sigma_3(\theta_{k+1}+\alpha_k)/(2\hbar))\,,&
\zeta\in III\cup IV\,.
\end{array}\right.
\end{equation}
Clearly, this matrix is analytic in $\zeta(D_{2k})$ except for positive
real $\zeta$, where it has continuous boundary values for $\zeta\neq 0$
that satisfy the 
jump relation $\tilde{\bf
S}_{2k+}(\zeta)=\tilde{\bf S}_{2k-}(\zeta)\cdot(-i\sigma_1)$.
\begin{lemma}
The matrix $\tilde{\bf
S}_{2k}(\zeta)$ determined from the solution of the outer model problem has
a unique representation
\begin{equation}
\tilde{\bf S}_{2k}(\zeta) = \tilde{\bf
S}_{2k}^{\rm hol}(\zeta)\tilde{\bf S}^{\rm loc, even}(\zeta)\,,
\label{eq:representationofouter}
\end{equation}
where
\begin{equation}
\tilde{\bf S}^{\rm loc, even}(\zeta):=(-\zeta)^{\sigma_3/4}
\left[\begin{array}{cc}1/\sqrt{2} & -1/\sqrt{2}\\1/\sqrt{2} &
1/\sqrt{2}
\end{array}\right]\,,
\label{eq:tildesloc}
\end{equation}
and where $\tilde{\bf S}_{2k}^{\rm hol}(\zeta)$ is holomorphic in the
interior of $\zeta(D_{2k})$.
\end{lemma}

\begin{proof}
Observe that by direct calculation, the matrix $\tilde{\bf S}^{\rm
loc, even}(\zeta)$ is analytic for all $\zeta$ except on the positive
real axis, where it satisfies the jump relation $\tilde{\bf S}^{\rm
loc, even}_+(\zeta)=
\tilde{\bf S}^{\rm loc, even}_-(\zeta)\cdot(-i\sigma_1)$.  
Since both $\tilde{\bf S}_{2k}(\zeta)$ and $\tilde{\bf S}^{\rm loc,
even}(\zeta)$ have determinant one and have smooth boundary values
except at $\zeta=0$, it follows that the quotient $\tilde{\bf
S}_{2k}(\zeta)\tilde{\bf S}^{\rm loc, even}(\zeta)^{-1}$ is analytic in
$\zeta(D_{2k})\setminus \{0\}$.  But by construction, the matrix
$\tilde{\bf O} (\lambda)$ obtained in \S\ref{sec:outersolve} is
$\bo ((\lambda-\lambda_{2k})^{-1/4})$, and consequently $\tilde{\bf
S}(\zeta)$ is $\bo (|\zeta|^{-1/4})$ at the origin since
$\zeta(\lambda)$ is an analytic mapping.  Therefore, the quotient is
bounded at $\zeta=0$ and hence analytic throughout the interior of
$\zeta(D_{2k})$.
\end{proof}

The main idea of this result is that the matrix $\tilde{\bf
S}_{2k}(\zeta)$ has a representation in terms of an analytic piece
that contains all of the complicated global information and is defined
only in $\zeta(D_{2k})$ and a local piece that is actually defined for
almost all $\zeta\in {\mathbb C}$ and is of a canonical form,
satisfying very simple jump relations.  In particular, the local piece
$\tilde{\bf S}^{\rm loc, even}(\zeta)$ does not depend on $\hbar$ even
though $\tilde{\bf S}_{2k}(\zeta)$ does.  We now seek a similar
decomposition of the matrix ${\bf S}_{2k}(\zeta)$.

\begin{figure}[h]
\begin{center}
\mbox{\psfig{file=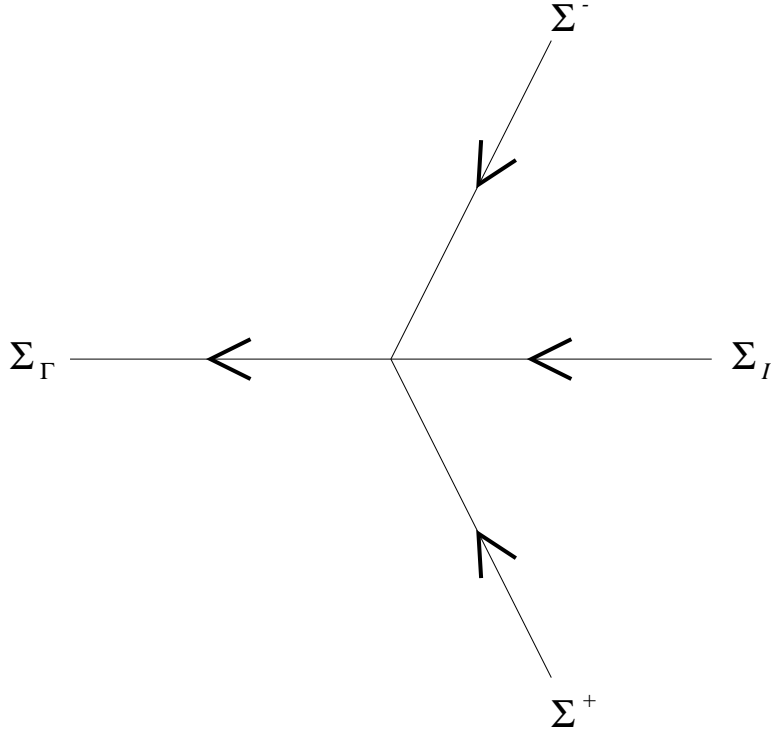,width=4 in}}
\end{center}
\caption{\em The oriented contour $\Sigma^{\rm loc}$.  All rays extend to
$\zeta=\infty$.}
\label{fig:sloc}
\end{figure}

Let $\Sigma_I$ denote the positive real axis in the $\zeta$-plane,
oriented from infinity into the origin; let $\Sigma_\Gamma$ denote
the negative real axis in the $\zeta$-plane, oriented from the origin
to infinity; finally let $\Sigma^\pm$ denote the rays with angles $
\arg(\zeta)=\mp \pi/3$, both oriented from infinity to the origin.
Denote the union of these contours by $\Sigma^{\rm loc}$.
See Figure~\ref{fig:sloc}.
Consider the following Riemann-Hilbert problem.
\begin{rhp}[Local model for even endpoints]
\index{Riemann-Hilbert problem!local model for even endpoints}
Find a matrix ${\bf S}^{\rm loc, even}(\zeta)$ satisfying
\begin{enumerate}
\item
{\bf Analyticity:} ${\bf S}^{\rm loc, even}(\zeta)$ is analytic for
$\zeta\in{\mathbb C}\setminus\Sigma^{\rm loc}$.
\item
{\bf Boundary behavior:} ${\bf S}^{\rm loc, even}(\zeta)$ assumes
continuous boundary values from within each sector of ${\mathbb
C}\setminus\Sigma^{\rm loc}$, with continuity holding also at the
point of self-intersection.
\item
{\bf Jump conditions:}
The boundary values taken on $\Sigma^{\rm loc}$ satisfy
\begin{equation}
\begin{array}{rcll}
{\bf S}^{\rm loc, even}_+(\zeta)&=&{\bf S}^{\rm loc, even}_-(\zeta) 
\left[\begin{array}{cc} 1 & 0\\-i\exp(-(-\zeta)^{3/2}) & 1\end{array} \right]
\,,&\zeta\in \Sigma_\Gamma\,,
\\\\ 
{\bf S}^{\rm loc, even}_+(\zeta)&=& {\bf S}^{\rm loc, even}_-(\zeta)
\left[\begin{array}{cc}1 & i\exp(\mp i\zeta^{3/2})\\ 0 &1\end{array}\right]
\,,&\zeta\in \Sigma^\pm\,,
\\\\
{\bf S}^{\rm loc, even}_+(\zeta)&=&{\bf S}^{\rm loc, even}_-(\zeta)(-i\sigma_1)\,,
&\zeta\in \Sigma_I\,.
\end{array}
\end{equation}
\item
{\bf Normalization:} ${\bf S}^{\rm loc, even}(\zeta)$ is similar to
$\tilde{\bf S}^{\rm loc, even}(\zeta)$ at $\zeta=\infty$, where
$\tilde{\bf S}^{\rm loc, even}(\zeta)$ is defined by
(\ref{eq:tildesloc}).  Precisely,
\begin{equation}
\lim_{\zeta\rightarrow\infty}
{\bf S}^{\rm loc, even}(\zeta)\tilde{\bf S}^{\rm loc, even}(\zeta)^{-1}
 = 
{\mathbb I}\,,
\end{equation}
with the limit being uniform with respect to direction.
\end{enumerate}
\label{rhp:Sloceven}
\end{rhp}

\begin{lemma}
The Riemann-Hilbert Problem~\ref{rhp:Sloceven} has a unique solution,
with the additional property that there exists a constant $M>0$ such
the estimate
\begin{equation}
\|
{\bf S}^{\rm loc, even}(\zeta)\tilde{\bf S}^{\rm loc, even}(\zeta)^{-1} - 
{\mathbb I}\| \le M|\zeta|^{-1}\,,
\label{eq:evenendpointestimate}
\end{equation}
holds for all sufficiently large $|\zeta|$.  The solution ${\bf S}^{\rm loc,
even}(\zeta)$ is universal in the sense that it does not depend on
$\hbar$.
\label{lemma:SlocExist}
\end{lemma}

\begin{proof}
We first introduce an auxiliary Riemann-Hilbert problem.  Let
$\Sigma_{\bf L}$ be the oriented contour illustrated in
Figure~\ref{fig:sigmaL}.
\begin{figure}[h]
\begin{center}
\mbox{\psfig{file=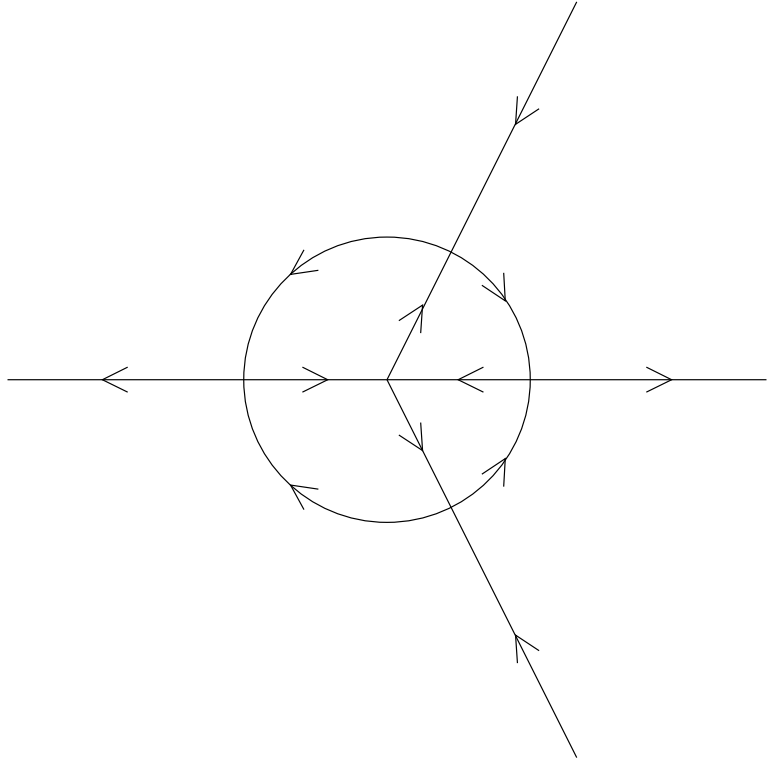,width=4 in}}
\end{center}
\caption{\em The contour $\Sigma_{\bf L}$ is the union of a circle 
of radius $R=1/2$, the real axis, and the rays
$\arg(\zeta)=\pm(\pi/3)$, oriented as shown.}
\label{fig:sigmaL}
\end{figure}
For $\zeta\in\Sigma_{\bf L}\setminus
\{0,1/2,-1/2,\exp(i\pi/3)/2,\exp(-i\pi/3)/2\}$, we define a jump matrix ${\bf v}_{\bf
L}(\zeta)$ as follows.  For $0<|\zeta|<1/2$, set
\begin{equation}
{\bf v}_{\bf L}(\zeta):=\left\{
\begin{array}{ll}
-i\sigma_1\,, & \arg(\zeta)=0\,,\\\\
\left[\begin{array}{cc} 1 & -i\exp(\pm i\zeta^{3/2})\\
0 & 1\end{array}\right]\,, &\arg(\zeta)=\pm\pi/3\,,\\\\
\left[\begin{array}{cc}
1 & 0 \\ i\exp(- (-\zeta)^{3/2}) & 1\end{array}\right]\,,&\arg(\zeta)=\pi \,,
\end{array}\right.
\label{eq:vLdef1}
\end{equation}
for $|\zeta|>1/2$, set
\begin{equation}
{\bf v}_{\bf L}(\zeta):=\left\{\begin{array}{ll}
{\mathbb I}\,,&\arg(\zeta)=0\,,\\\\
\tilde{\bf S}^{\rm loc, even}(\zeta)
\left[\begin{array}{cc} 1 & i\exp(\pm i\zeta^{3/2})\\
0 & 1\end{array}\right]\tilde{\bf S}^{\rm loc, even}(\zeta)^{-1}\,,&
\arg(\zeta)=\pm \pi/3\,,\\\\
\tilde{\bf S}^{\rm loc, even}(\zeta)
\left[\begin{array}{cc}
1 & 0 \\ -i\exp(- (-\zeta)^{3/2}) & 1\end{array}\right]
\tilde{\bf S}^{\rm loc, even}(\zeta)^{-1}\,, &
\arg(\zeta)=\pi\,,
\end{array}\right.
\label{eq:vLdef2}
\end{equation}
and, for $|\zeta|=1/2$, set
\begin{equation}
{\bf v}_{\bf L}(\zeta):= \left\{\begin{array}{ll}
\tilde{\bf S}^{\rm loc, even}(\zeta)^{-1}\,, &
0<\arg(\zeta)<\pi/3 \mbox{ and }-\pi<\arg(\zeta)<-\pi/3\,,\\
\tilde{\bf S}^{\rm loc, even}(\zeta)\,, &
\pi/3<\arg(\zeta)<\pi \mbox{ and } -\pi/3<\arg(\zeta)<0\,.
\end{array}\right.
\label{eq:vLdef3}
\end{equation}
Consider the following problem.
\begin{rhp}[Auxiliary local problem]
\index{Riemann-Hilbert problem!auxiliary local problem}
Find a matrix ${\bf L}(\zeta)$ satisfying:
\begin{enumerate}
\item
{\bf Analyticity:} ${\bf L}(\zeta)$ is analytic for $\zeta\in{\mathbb
C}\setminus\Sigma_{\bf L}$ and takes continuous boundary values on
$\Sigma_L$ including self-intersection points.
\item
{\bf Boundary behavior:} ${\bf L}(\zeta)$ takes continuous boundary
values from each connected component of ${\mathbb
C}\setminus\Sigma_{\bf L}$, with continuity holding also at corner
points corresponding to self-intersections of $\Sigma_{\bf L}$.
\item
{\bf Jump conditions:} The boundary values taken on
$\Sigma_{\bf L}\setminus\{\mbox{self-intersection points}\}$ satisfy
\begin{equation}
{\bf L}_+(\zeta)={\bf L}_-(\zeta){\bf v}_{\bf L}(\zeta)\,,
\end{equation}
with the jump matrix ${\bf v}_{\bf L}(\zeta)$ defined by (\ref{eq:vLdef1}),
(\ref{eq:vLdef2}), and (\ref{eq:vLdef3}).
\item
{\bf Normalization:} ${\bf L}(\zeta)$ is normalized at infinity:
\begin{equation}
{\bf L}(\zeta)\rightarrow {\mathbb I}\mbox{ as }\zeta\rightarrow\infty\,,
\end{equation}
uniformly with respect to direction.
\end{enumerate}
\label{rhp:localAiry}
\end{rhp}

Observe that the jump matrix ${\bf v}_{\bf L}(\zeta)$ has the
following properties:
\begin{enumerate}
\item 
${\bf v}_{\bf L}(\zeta)$ has determinant one for all $\zeta\in\Sigma_{\bf L}$.
\item
${\bf v}_{\bf L}(\zeta)$ is smooth on each open arc, and in particular is
Lipschitz.
\item
At each point $\zeta_0$ of self-intersection of $\Sigma_{\bf L}$, let the 
intersecting arcs be enumerated in counterclockwise order (beginning with
any arc) as $\Sigma_{\bf L}^{(1)},\dots,\Sigma_{\bf L}^{(n)}$ where
$n$ is even.  The limits ${\bf v}_{\bf L}^{(k)}:=\lim_{\zeta\rightarrow\zeta_0,
\zeta\in\Sigma_{\bf L}^{(k)}} {\bf v}_{\bf L}(\zeta)$ exist and satisfy
\begin{equation}
{\bf v}_{\bf L}^{(1)}{\bf v}_{\bf L}^{(2)-1}{\bf v}_{\bf L}^{(3)}\dots
{\bf v}_{\bf L}^{(n-1)}{\bf v}_{\bf L}^{(n)-1}={\mathbb I}\,.
\end{equation}
\item
${\bf v}_{\bf L}(\zeta)-{\mathbb I}=\bo (|\zeta|^{-1})$ as $|\zeta|\rightarrow
\infty$.  In fact, the decay is exponentially fast in $|\zeta|$.
\end{enumerate}
The first two conditions are obvious.  Checking the third condition is a direct
computation that we omit, and the fourth condition follows from the
fact that the $|\zeta|^{1/4}$ growth of the conjugating factors
$\tilde{\bf S}^{\rm loc, even}(\zeta)$ and $\tilde{\bf S}^{\rm
loc, even}(\zeta)^{-1}$ is controlled easily by the exponential decay of
$\exp(\pm i\zeta^{3/2})$ for $\arg(\zeta)=\pm\pi/3$, and of
$\exp(-(-\zeta)^{3/2})$ for $\arg(\zeta)=\pi$.

It then follows from Theorem~\ref{theorem:LocalAlternative} proved in
the appendix that there will exist a unique solution ${\bf L}(\zeta)$ 
of the Riemann-Hilbert problem~\ref{rhp:localAiry} that additionally
satisfies:
\begin{enumerate}
\item
${\bf L}(\zeta)$ is uniformly bounded and satisfies $\|{\bf
L}(\zeta)-{\mathbb I}\|=\bo (|\zeta|^{-\mu})$ as
$|\zeta|\rightarrow\infty$ for all $\mu<1$, and,
\item
the boundary values ${\bf L}_\pm(\zeta)$ taken on each component of
${\mathbb C}\setminus
\Sigma_{\bf L}$ are H\"older continuous for all exponents
strictly less than 1,	
\end{enumerate}
if and only if the corresponding homogeneous Riemann-Hilbert problem
has only the trivial solution, {\em i.e.} the Fredholm alternative
applies.  If there exists a solution ${\bf L}(\zeta)$, then define
${\bf S}^{\rm loc, even}(\zeta)$ by
\begin{equation}
{\bf S}^{\rm loc, even}(\zeta):={\bf L}(\zeta)\cdot\left\{\begin{array}{ll}
{\mathbb I}\,, & |\zeta|<1/2\,,\\
\tilde{\bf S}^{\rm loc, even}(\zeta)\,, & |\zeta| >1/2\,.
\end{array}\right.
\label{eq:SlocevenL}
\end{equation}
The function so defined has a holomorphic extension through the circle
$|\zeta|=1/2$, since it takes boundary values there from both sides
that are continuous and equal.  It is easy to check that it solves the
Riemann-Hilbert Problem~\ref{rhp:Sloceven}.  Uniqueness follows from
the analogous property of ${\bf L}(\zeta)$.

So, we must now show that such a matrix ${\bf L}(\zeta)$ exists by
proving that all solutions of the homogeneous problem are trivial.  
Let us define this homogeneous problem.
\begin{rhp}[Homogeneous auxiliary local problem]
\index{Riemann-Hilbert problem!homogeneous auxiliary local problem}
Let $\mu\in (0,1)$ be given.  Find a matrix ${\bf L}_0(\zeta)$ satisfying
\begin{enumerate}
\item
{\bf Analyticity:} ${\bf L}_0(\zeta)$ is analytic for
$\zeta\in{\mathbb C}\setminus\Sigma_{\bf L}$.
\item
{\bf Boundary behavior:} ${\bf L}_0(\zeta)$ takes boundary values
from each connected component of its domain of analyticity that are
H\"older continuous with exponent $\mu$, including at self-intersection
(corner) points.
\item
{\bf Jump conditions:} The boundary values ${\bf L}_{0\pm}(\zeta)$ that
${\bf L}_0(\zeta)$ assumes on any smooth oriented component
of $\Sigma_{\bf L}\setminus\{\mbox{self-intersection points}\}$ satisfy
\begin{equation}
{\bf L}_{0+}(\zeta)={\bf L}_{0-}(\zeta){\bf v}_{\bf L}(\zeta)\,.
\end{equation}
\item
{\bf Homogeneous normalization:} The matrix function ${\bf L}_0(\zeta)$
vanishes for large $\zeta$, satisfying the precise estimate
\begin{equation}
\|{\bf L}_0(\zeta)\|\le M|\zeta|^{-\mu}\,,
\end{equation}
holding for some $M>0$ and all sufficiently large $|\zeta|$.
\end{enumerate}
\label{rhp:localvanishing}
\end{rhp}

Thus, a solution of the homogeneous Riemann-Hilbert
Problem~\ref{rhp:localvanishing}, is similar to ${\bf L}(\zeta)$, but
vanishes for large $\zeta$.  The identity matrix in the normalization
condition for ${\bf L}(\zeta)$ is replaced with the zero matrix.  Note
that, according to the discussion following the statement of
Theorem~\ref{theorem:LocalAlternative} in the appendix, it suffices to
find a $\mu_0<1$ such that all nontrivial solutions of the homogeneous
Riemann-Hilbert Problem~\ref{rhp:localvanishing} with exponents $\mu>\mu_0$
can be ruled out.  Unfortunately, the jump matrix ${\bf v}_{\bf
L}(\zeta)$ lacks the symmetry needed to apply the general theory
described in the appendix, so we must construct a specific argument.
We will suppose that the H\"older exponent satisfies $\mu>\mu_0=3/4$.
Let ${\bf L}_0(\zeta)$ be a corresponding solution of the homogeneous
Riemann-Hilbert Problem~\ref{rhp:localvanishing}.  First, set
\begin{equation}
{\bf S}^{\rm loc, even}_0(\zeta):={\bf L}_0(\zeta)\cdot
\left\{\begin{array}{ll}{\mathbb I}\,, &
\hspace{0.2 in} |\zeta|<1/2\,,\\
\tilde{\bf S}^{\rm loc, even}(\zeta)\,, & \hspace{0.2 in}
|\zeta|>1/2\,.
\end{array}\right.
\end{equation}
This matrix is analytic in each sector for $|\zeta|=1/2$, and on the
real axis and the rays $\arg(\zeta)=\pm \pi/3$ satisies the same jump
conditions as ${\bf S}^{\rm loc, even}(\zeta)$.  As $\zeta\rightarrow\infty$,
we have for some $M>0$ the estimate
\begin{equation}
\|{\bf S}^{\rm loc, even}_0(\zeta)\|\le \|{\bf L}_0(\zeta)\|\cdot\|
\tilde{\bf S}^{\rm loc, even}(\zeta)\|\le
M|\zeta|^{1/4 - \mu}\,.
\end{equation}
Next, set
\begin{equation}
{\bf A}(\zeta)={\bf S}^{\rm loc, even}_0(\zeta)\cdot\left\{
\begin{array}{ll}
{\mathbb I}\,, & 
\hspace{0.2 in} -\pi < \arg(\zeta) < -\pi/3\,,\\\\
\sigma_1\,, & 
\hspace{0.2 in} \pi/3 < \arg(\zeta) < \pi\,,\\\\
\left[\begin{array}{cc}1 & -i\exp(i\zeta^{3/2})\\0 & 1\end{array}\right]
\sigma_1\,, & \hspace{0.2 in} 0 < \arg(\zeta)< \pi/3\,,\\\\
\left[\begin{array}{cc}1 & i\exp(-i\zeta^{3/2})\\0 & 1\end{array}\right]\,,
&\hspace{0.2 in} -\pi/3<\arg(\zeta)<0\,.
\end{array}
\right.
\end{equation}
Since the matrices multiplying ${\bf S}^{\rm loc, even}_0(\zeta)$ above 
are uniformly bounded, ${\bf A}(\zeta)$ retains the decay
properties of ${\bf S}^{\rm loc, even}_0(\zeta)$.  Also, ${\bf A}(\zeta)$ is
analytic for $\zeta\in {\mathbb C}\setminus{\mathbb R}$.  On the real axis,
oriented from right to left, there is the jump condition
\begin{equation}
{\bf A}_+(\zeta)={\bf A}_-(\zeta)\left\{\begin{array}{ll}
\left[\begin{array}{cc} -i\exp(-(-\zeta)^{3/2}) & 1 \\ 1 & 0\end{array}
\right]\,,&\hspace{0.2 in} \zeta\in {\mathbb R}_-\,,\\\\
\left[\begin{array}{cc} -i & \exp(-i\zeta^{3/2}) \\ \exp(i\zeta^{3/2}) & 0
\end{array}\right]\,,&\hspace{0.2 in}\zeta\in{\mathbb R}_+\,.
\end{array}
\right.
\label{eq:Ajumps}
\end{equation}

Now, the matrix function ${\bf Q}(\zeta):={\bf A}(\zeta){\bf
A}(\zeta^*)^\dagger$ is also analytic for $\zeta\in {\mathbb
C}\setminus{\mathbb R}$, and since $\|{\bf A}(\zeta)\|=\bo
(|\zeta|^{1/4-\mu})$ for large $|\zeta|$, we can apply Cauchy's theorem
\index{Cauchy's theorem} for all $\mu>\mu_0=3/4$ to deduce that
\begin{equation}
i\int_{-\infty}^\infty {\bf Q}_+(\zeta)\,d\zeta=0\,.
\label{eq:CauchyQ}
\end{equation}
Since for $\zeta$ real, ${\bf Q}_+(\zeta)={\bf A}_+(\zeta){\bf
A}_-(\zeta)^\dagger$, (\ref{eq:CauchyQ}) becomes, using the relations 
(\ref{eq:Ajumps})
\begin{equation}
\int_{-\infty}^0{\bf A}_-(\zeta)\left[\begin{array}{cc}
\exp(-(-\zeta)^{3/2}) & i \\ i & 0\end{array}\right]{\bf A}_-(\zeta)^\dagger
\,d\zeta +
\int_0^\infty{\bf A}_-(\zeta)\left[\begin{array}{cc}
1 & i\exp(-i\zeta^{3/2}) \\ i\exp(i\zeta^{3/2}) & 0\end{array}\right]
{\bf A}_-(\zeta)^\dagger\,d\zeta = 0\,.
\end{equation}
Adding this equation to its conjugate-transpose, and looking at the $(1,1)$
entry of the resulting matrix equation, one finds
\begin{equation}
\int_{-\infty}^0 \|{\bf A}^{(1)}_-(\zeta)\|_2^2\exp(-(-\zeta)^{3/2})\,d\zeta
+
\int_0^\infty \|{\bf A}^{(1)}_-(\zeta)\|_2^2\,d\zeta = 0\,,
\end{equation}
where ${\bf A}^{(k)}(\zeta)$ is the $k$th column of ${\bf A}(\zeta)$,
and consequently ${\bf A}^{(1)}(\zeta)\equiv 0$ for $\Im(\zeta)\ge 0$.
From (\ref{eq:Ajumps}), it then follows immediately that ${\bf
A}^{(2)}(\zeta)\equiv 0$ for $\Im(\zeta)<0$.  The jump relations
(\ref{eq:Ajumps}) then relate the boundary values of the remaining,
possibly nonzero, entries of ${\bf A}(\zeta)$ by
\begin{equation}
{\bf A}^{(1)}_+(\zeta)=\left\{
\begin{array}{ll}
{\bf A}^{(2)}_-\,, &\hspace{0.2 in}
\zeta\in{\mathbb R}_-\,,\\
\exp(i\zeta^{3/2}) {\bf A}^{(2)}_-(\zeta)\,, &\hspace{0.2 in}\zeta\in
{\mathbb R}_+\,.\end{array}\right.
\end{equation}
So, defining scalar functions $a_k(\zeta)$ for $\zeta\in{\mathbb C}
\setminus{\mathbb R}$ by
\begin{equation}
a_k(\zeta):=\left\{\begin{array}{ll}{\bf A}^{(1)}_k(\zeta)\,,
&\hspace{0.2 in}\Im(\zeta)<0\,,\\
{\bf A}^{(2)}_k(\zeta)\,,&\hspace{0.2 in}\Im(\zeta)>0\,,
\end{array}\right.
\end{equation}
we see that both functions are analytic for $\zeta\in{\mathbb C}\setminus
{\mathbb R_+}$, both are $\bo(|\zeta|^{1/4-\mu})$ for large $|\zeta|$, and
both take continuous boundary values on ${\mathbb R}_+$, where they satisfy
\begin{equation}
a_{k+}(\zeta)=\exp(i\zeta^{3/2})a_{k-}(\zeta)\,,
\end{equation}
with the ray considered oriented from infinity to the origin.

We now show that necessarily $a_k(\zeta)\equiv 0$.  Given
$a(\zeta):=a_k(\zeta)$ satisfying the above properties, define
a scalar function $b(\zeta)$ that is analytic in the {\em extended}
plane $-\pi/3 <\arg(\zeta) < 2\pi + \pi/3$ by setting
\begin{equation}
b(\zeta):=\left\{ \begin{array}{ll}
a(\zeta)\,, & \hspace{0.2 in} 0 \le \arg(\zeta) \le  2\pi\,,\\
a([\zeta])\exp(i[\zeta]^{3/2})\,, &\hspace{0.2 in} 
2\pi \le \arg(\zeta) \le 2\pi+\pi/3\,,\hspace{0.2 in}
[\zeta]:=|\zeta|\exp(i(\arg(\zeta)-2\pi))\,,\\
a([\zeta])\exp(-i[\zeta]^{3/2})\,, & \hspace{0.2 in}
-\pi/3 \le \arg(\zeta) \le 0\,,\hspace{0.2 in}
[\zeta]:=|\zeta|\exp(i(\arg(\zeta)+2\pi))\,.
\end{array}\right.
\end{equation}
We are using the notation $[\zeta]$ for the class representative of
$\zeta$ with $0<\arg([\zeta])<2\pi$.  From the jump relation for
$a(\zeta)$ it follows that $b(\zeta)$ is analytic for $\arg(\zeta)=0$
and $\arg(\zeta)=2\pi$.  In the extended plane where
$b(\zeta)\not\equiv a(\zeta)$, we have $|b(\zeta)|<|a([\zeta])|$;
moreover, from the mere algebraic decay of $a(\zeta)$ for large
$|\zeta|$, we find that $b(\zeta)$ decays {\em exponentially} for
large $|\zeta|$ in these regions, and in particular on the boundaries
$\arg(\zeta)=-\pi/3$ and $\arg(\zeta)=2\pi+\pi/3$.  Finally, define an
analytic function of $w$ for $\Re(w)\ge 0$ by
\begin{equation}
c(w):=b(-w^{8/3})\,.
\end{equation}
From the exponential decay of $b(\zeta)$ for $\arg(\zeta)=-\pi/3$ and
$\arg(\zeta)=2\pi + \pi/3$, it follows that $|c(iy)|\le M\exp(-y^4)\le
M'\exp(-|y|)$ for all $y\in {\mathbb R}$.  Also $c(w)$ is uniformly bounded
for all $\Re(w)\ge 0$.  Now, we recall
\begin{quote}
{\em Carlson's theorem} \cite{RS78}:
\index{Carlson's theorem}
Suppose that $f(z)$ is a complex-valued function defined and continuous 
for $\Re(z)\ge 0$ and analytic for $\Re(z)> 0$.  Suppose that
$|f(z)|\le M\exp(A|z|)$ for $\Re(z)\ge 0$ and $|f(iy)|
\le M\exp(-B|y|)$ for all $y\in{\mathbb R}$, where $B>0$.  Then $f(z)$ is
identically zero.
\end{quote}
Applying this result of complex analysis for $A=0$ and $B=1$, we
deduce that $c(w)\equiv 0$.  This in turn implies that
$a_k(\zeta)\equiv 0$, and in conjunction with our eariler results that
${\bf A}(\zeta)\equiv 0$.  Consequently we find that ${\bf
L}_0(\zeta)\equiv 0$.  Therefore all solutions of the homogeneous
Riemann-Hilbert Problem~\ref{rhp:localvanishing} with H\"older
exponents $\mu>\mu_0=3/4$ are trivial, and the required function ${\bf
L}(\zeta)$ exists by the Fredholm alternative ({\em cf.}
Theorem~\ref{theorem:LocalAlternative}).  Because ${\bf L}(\zeta)$ has
H\"older continuous boundary values, the matrix ${\bf S}^{\rm loc,
even}(\zeta)$ defined by (\ref{eq:SlocevenL}) is a solution of the
Riemann-Hilbert Problem \ref{rhp:Sloceven} taking uniformly continuous
boundary values on $\Sigma^{\rm loc}$.

Finally, we notice that since the jump matrix ${\bf v}_{\bf L}(\zeta)$
is analytic on each ray of $\Sigma_{\bf L}$ and decays exponentially
to the identity as $\zeta\rightarrow\infty$, all of the order
$\zeta^{-1}$ moments vanish, and it follows from
Theorem~\ref{theorem:decay} that $\|{\bf L}(\zeta)-{\mathbb I}\|$ is
uniformly order $|\zeta|^{-1}$ for large $\zeta$.  Using the formula
(\ref{eq:SlocevenL}), we see that this in turn implies the decay
estimate (\ref{eq:evenendpointestimate}), which completes the proof of
Lemma~\ref{lemma:SlocExist}.
\end{proof}

\begin{remark}
While it is sufficient for our purposes to present an argument for the
existence of the matrix function ${\bf S}^{\rm loc, even}(\zeta)$ based on
abstract Fredholm theory as done here, the solution to the
Riemann-Hilbert Problem~\ref{rhp:Sloceven} can even be given {\em
explicitly} in terms of Airy functions\index{Airy functions}.  See
\cite{DKMVZ98A,DKMVZ98B,D99} for these formulae.  In those papers,
it was essential to have an explicit accurate description of the local
behavior near the endpoint, whereas we require only qualitative
properties sufficient to establish ultimately that the explicit
approximation afforded by the solution of the outer model
Riemann-Hilbert Problem \ref{rhp:model} is part of a uniformly valid
approximation to ${\bf N}^\sigma(\lambda)$.  Indeed, it is only the
expansion of ${\bf N}^\sigma(\lambda)$ for sufficiently large
$\lambda$ that we need to compute asymptotics for the focusing
nonlinear Schr\"odinger equation.

On the other hand, using the explicit solution, it is possible to
refine the decay estimate (\ref{eq:evenendpointestimate}) even
further, to $\bo(|\zeta|^{-3/2})$.  This decay estimate can also be
obtained by working with the differential equation that the matrix
${\bf S}^{\rm loc, even}(\zeta)$ satisfies by virtue of an explicit
transformation that reduces the jump matrices to constants.  Then it
follows that the ratio of ${\bf S}^{\rm loc, even}(\zeta)$ with its
derivative with respect to $\zeta$ is a ratio of entire functions,
which implies a linear differential equation for ${\bf S}^{\rm
loc, even}(\zeta)$ from which one may obtain an asymptotic expansion for
large $\zeta$.  We leave this calculation to the interested reader.
\end{remark}

As was the case with the matrix $\tilde{\bf S}^{\rm loc,
even}(\zeta)$, the matrix ${\bf S}^{\rm loc, even}(\zeta)$ solving the
Riemann-Hilbert Problem~\ref{rhp:Sloceven} is independent of all
parameters $\hbar$, $x$, and $t$, of our asymptotic analysis.  We now
propose a factorized representation of an approximation to ${\bf
S}_{2k}(\zeta)$ by setting for $\zeta\in
\zeta(D_{2k})$
\begin{equation}
\hat{\bf S}_{2k}(\zeta):=\tilde{\bf S}_{2k}^{\rm hol}(\zeta)
{\bf S}^{\rm loc, even}(\zeta)\,.
\end{equation}
This matrix depends on $x$, $t$, and $\hbar$ through the holomorphic
prefactor.  It satisfies {\em exactly} the same jump relations within
$\zeta(D_{2k})$ as does ${\bf S}_{2k}(\zeta)$.

Finally, we use the matrix $\hat{\bf S}_{2k}(\zeta)$ to construct a
local approximation \index{local approximation!near even endpoints} of
${\bf N}^\sigma(\lambda)$ valid within $D_{2k}$.  First, we apply to
the matrix $\hat{\bf S}_{2k} (\zeta)$ the change of variables
(\ref{eq:zetadef}) and (\ref{eq:Sdef}) connecting ${\bf
S}_{2k}(\zeta)$ to ${\bf O}^\sigma(\lambda)$.  This yields a matrix
that exactly satisfies the jump relations for ${\bf
O}^\sigma(\lambda)$, and that by construction matches well onto the
matrix $\tilde{\bf O}(\lambda)$ at the boundary of $D_{2k}$.  To
recover the approximation for ${\bf N}^\sigma(\lambda)$ one multiplies
by the explicit triangular factors ${\bf D}^\sigma(\lambda)$ relating,
by definition, the matrices $\tilde{\bf N}^\sigma(\lambda)$ and ${\bf
O}^\sigma(\lambda)$ in the lens halves, when $\zeta(\lambda)$ is in
regions I and IV of the $\zeta$-plane.  Thus, for $\lambda\in D_{2k}$
the local approximation is defined as follows.  For $\zeta(\lambda)$
in region I of the $\zeta$-plane,
\begin{equation}
\begin{array}{l}
\hat{\bf N}_{2k}^\sigma(\lambda):=\tilde{\bf S}_{2k}^{\rm hol}(\zeta(\lambda))
{\bf S}^{\rm loc, even}(\zeta(\lambda))
\exp(-iJ\sigma_3(\theta_{k+1}-\alpha_k)/(2\hbar))\times\\\\
\hspace{0.4 in}\left[\begin{array}{cc}1 & 
-i\exp(-iJ\alpha_k/\hbar)\exp(ir_k(\lambda)/\hbar)\\
0 & 1\end{array}\right]\sigma_1^{\frac{1-J}{2}}\,, 
\end{array}
\label{eq:Nhat2kI}
\end{equation}
for $\zeta(\lambda)$ in region II of the $\zeta$-plane,
\begin{equation}
\hat{\bf N}_{2k}^\sigma(\lambda):=\tilde{\bf S}_{2k}^{\rm hol}(\zeta(\lambda))
{\bf S}^{\rm loc, even}(\zeta(\lambda))
\exp(-iJ\sigma_3(\theta_{k+1}-\alpha_k)/(2\hbar))
\sigma_1^{\frac{1-J}{2}}\,, 
\label{eq:Nhat2kII}
\end{equation}
for $\zeta(\lambda)$ in region III of the $\zeta$-plane,
\begin{equation}
\hat{\bf N}_{2k}^\sigma(\lambda):=\tilde{\bf S}_{2k}^{\rm hol}(\zeta(\lambda))
{\bf S}^{\rm loc, even}(\zeta(\lambda))
\exp(iJ\sigma_3(\theta_{k+1}+\alpha_k)/(2\hbar))
\sigma_1^{\frac{1-J}{2}}\,, 
\label{eq:Nhat2kIII}
\end{equation}
and for $\zeta(\lambda)$ in region IV of the $\zeta$-plane,
\begin{equation}
\begin{array}{l}
\hat{\bf N}_{2k}^\sigma(\lambda):=\tilde{\bf S}_{2k}^{\rm hol}(\zeta(\lambda))
{\bf S}^{\rm loc, even}(\zeta(\lambda))
\exp(iJ\sigma_3(\theta_{k+1}+\alpha_k)/(2\hbar))\times\\\\
\hspace{0.4 in}\left[\begin{array}{cc}
1 & i\exp(-iJ\alpha_k/\hbar)\exp(-ir_k(\lambda)/\hbar)\\
0 & 1\end{array}\right]
\sigma_1^{\frac{1-J}{2}}\,.
\end{array}
\label{eq:Nhat2kIV}
\end{equation}
Here the function $r_k(\lambda)$ is defined in
Definition~\ref{def:qr}.

We finish our local analysis near the endpoint $\lambda_{2k}$ by
recording several crucial properties of this local approximation.
\begin{lemma}
The local approximation $\hat{\bf N}_{2k}^\sigma(\lambda)$ is analytic
for $\lambda\in D_{2k}\setminus (D_{2k}\cap C)$, and takes continuous
boundary values on $C_\sigma$ that satisfy exactly $\hat{\bf
N}_{2k+}^\sigma(\lambda)=
\hat{\bf N}_{2k-}^\sigma(\lambda){\bf v}_{\tilde{\bf N}}^\sigma(\lambda)$.
\label{lemma:jumpsinD2k}
\end{lemma}

\begin{proof}
This is an elementary consequence of the fact that for all
$\zeta\in\zeta(D_{2k})$, $\hat{\bf S}_{2k}(\zeta)$ satisfies the exact same
jump relations as ${\bf S}_{2k}(\zeta)$.
\end{proof}

\begin{lemma}
There exists some $M>0$ such that for all $\lambda\in D_{2k}$ and all
sufficiently small $\hbar$, 
\begin{equation}
\|\hat{\bf N}^\sigma_{2k}(\lambda)\|
\le M\hbar^{-1/3}\,.
\end{equation}
The same estimate holds for the inverse, since the local approximation has
determinant one.
\label{lemma:holobound}
\end{lemma}

\begin{proof}
Recall the exact representation (\ref{eq:representationofouter}) of
the function $\tilde{\bf S}_{2k}(\zeta)$ related to the outer solution
$\tilde{\bf O}(\lambda)$ obtained in \S\ref{sec:outersolve} by
a change of variables.  It follows from the construction in
\S\ref{sec:outersolve} that $\tilde{\bf S}_{2k}(\zeta(\lambda))$ blows up
like $(\lambda-\lambda_{2k})^{-1/4}$ with a leading coefficient that
is uniformly bounded as $\hbar$ tends to zero.  Now for
$\lambda-\lambda_{2k}$ small, the Taylor expansion for the analytic
map $\zeta(\lambda)$ gives
$\zeta(\lambda)=M'\hbar^{-2/3}(\lambda-\lambda_{2k}) + \bo
((\lambda-\lambda_k)^2)$, where $M'$ is a constant that is bounded as
$\hbar$ goes to zero.  Approximating $\zeta(\lambda)$ on the
right-hand side of (\ref{eq:representationofouter}) by such a formula,
one sees that the holomorphic prefactor $\tilde{\bf S}_{2k}^{\rm
hol}(\zeta(\lambda))$ must be of the form
\begin{equation}
\tilde{\bf S}_{2k}^{\rm hol}(\zeta(\lambda))={\bf T}(\lambda)\hbar^{-\sigma_3/6}\,,
\end{equation}
with ${\bf T}(\lambda)$ being a matrix analytic in $D_{2k}$ that is
uniformly bounded as $\hbar$ tends to zero.  Now, since ${\bf
S}^{\rm loc, even}(\zeta)$ is bounded only by $|\zeta|^{1/4}$ for
large $\zeta$, we get a uniform estimate for all $\lambda\in D_{2k}$
of the form $\|{\bf S}^{\rm loc,
even}(\zeta(\lambda))\|=\bo(\hbar^{-1/6})$ as well.  These bounds,
along with the definition of $\hat{\bf N}^\sigma_{2k}(\lambda)$ yield
the desired estimate.
\end{proof}

\begin{lemma}
There exists some $M>0$ such that for all $\lambda\in \partial
D_{2k}$, and for all sufficiently small $\hbar$, 
\begin{equation}
\|\hat{\bf N}_{2k}^\sigma(\lambda)
\hat{\bf N}_{\rm out}^\sigma(\lambda)^{-1}-{\mathbb I}\|\le M\hbar^{1/3}\,,
\end{equation}
where $\hat{\bf N}_{\rm out}^\sigma(\lambda)$ is defined by
(\ref{eq:outerparametrixdef}) in \S\ref{sec:outerparametrix}.
\label{lemma:matcheven}
\end{lemma}

\begin{proof}
By definition, we have for all $\lambda\in D_{2k}$,
\begin{equation}
\hat{\bf N}_{2k}^\sigma(\lambda)\hat{\bf N}_{\rm out}^\sigma(\lambda)^{-1}=
\tilde{\bf S}_{2k}^{\rm hol}(\zeta(\lambda)){\bf S}^{\rm loc, even}(\zeta(\lambda))
\tilde{\bf S}^{\rm loc, even}(\zeta(\lambda))^{-1}
\tilde{\bf S}_{2k}^{\rm hol}(\zeta(\lambda))^{-1}\,.
\end{equation}
Now, as $\hbar$ tends to zero, $\zeta(\lambda)\rightarrow\infty$ for
all $\lambda\in \partial D_{2k}$; in particular $|\zeta|\sim
\hbar^{-2/3}$ for all $\lambda\in\partial D_{2k}$.  Therefore,
directly from the large $\zeta$ asymptotic properties of the matrix
${\bf S}^{\rm loc, even}(\zeta)$ in the estimate
(\ref{eq:evenendpointestimate}), we have for $\lambda\in \partial
D_{2k}$,
\begin{equation}
\hat{\bf N}_{2k}^\sigma(\lambda)\hat{\bf N}_{\rm out}^\sigma(\lambda)^{-1}=
{\mathbb I} + \tilde{\bf S}_{2k}^{\rm hol}(\zeta(\lambda))
\left[\bo (|\zeta(\lambda)|^{-1}) \right]
\tilde{\bf S}_{2k}^{\rm hol}(\zeta(\lambda))^{-1}\,.
\label{eq:intermediatestep}
\end{equation}
From the proof of Lemma~\ref{lemma:holobound}, the conjugating factors
are each uniformly bounded for $\lambda\in D_{2k}$ by $\bo (\hbar^{-1/6})$.  
Using this fact in (\ref{eq:intermediatestep}) yields the desired bound.
\end{proof}

\subsection{Local analysis for $\lambda$ near the endpoint 
$\lambda_{2k-1}$ for $k=1,\dots,G/2$.}  
\label{sec:2km1}
\begin{figure}[h]
\begin{center}
\mbox{\psfig{file=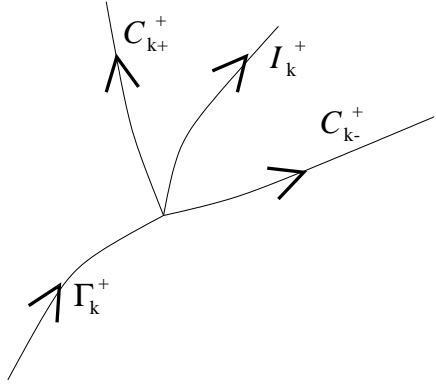,height=2 in}}
\end{center}
\caption{\em The jump matrix for ${\bf O}^\sigma(\lambda)$
near $\lambda_{2k-1}$ differs from the identity on a self-intersecting
contour with $\lambda_{2k-1}$ at the intersection point.}
\label{fig:oddendpoint}
\end{figure}

The analysis near $\lambda_{2k-1}$ proceeds in a similar manner,
beginning with the exact jump relations for the matrix ${\bf
O}^\sigma(\lambda)$.  The local contour structure is illustrated in
Figure~\ref{fig:oddendpoint}.
The exact jump relations for ${\bf O}^\sigma(\lambda)$ in a
neighborhood of $\lambda_{2k-1}$ are
\begin{equation}
{\bf O}^\sigma_+(\lambda)={\bf
O}^\sigma_-(\lambda)\sigma_1^{\frac{1-J}{2}}
\left[\begin{array}{cc} \exp(iJ\theta_k/\hbar) & 0 \\
-i\exp(\tilde{\phi}^{J,\omega}(\lambda)/\hbar) &
\exp(-iJ\theta_k/\hbar)
\end{array}\right]\sigma_1^{\frac{1-J}{2}}
\end{equation}
for $\lambda\in\Gamma_k^+$,
\begin{equation}
{\bf O}^\sigma_+(\lambda)={\bf O}^\sigma_-(\lambda)
\sigma_1^{\frac{1-J}{2}}
\left[\begin{array}{cc}0 & -i\exp(-iJ\alpha_k/\hbar) \\
-i\exp(iJ\alpha_k/\hbar) & 0\end{array}\right]
\sigma_1^{\frac{1-J}{2}}
\end{equation}
for $\lambda\in I_k^+$, and
\begin{equation}
{\bf O}^\sigma_+(\lambda)={\bf
O}^\sigma_-(\lambda)\sigma_1^{\frac{1-J}{2}}
\left[\begin{array}{cc} 1 & i\exp(-iJ\alpha_k/\hbar) \exp(\mp
ir_k(\lambda)/\hbar)\\ 0 &
1\end{array}\right]\sigma_1^{\frac{1-J}{2}}
\end{equation}
for $\lambda\in C^+_{k\pm}$.  Here, we have the identifications
\begin{equation}
r_k(\lambda_{2k-1})=J\theta_k\,,\hspace{0.3
in}\tilde{\phi}^\sigma (\lambda_{2k-1})=iJ\alpha_k\,.
\end{equation}
Once again, analytic continuation arguments using the complex phase
function $g^\sigma(\lambda)$ yield useful relations between
the function $\tilde{\phi}^\sigma(\lambda)$ continued from
$\Gamma_k^+$ and the function $r_k(\lambda)$ continued from $I_k^+$.
One finds that for $\lambda\in C^+_{k+}$,
\begin{equation}
\tilde{\phi}^\sigma(\lambda)-iJ\alpha_k=
i(r_k(\lambda)-J\theta_k)\,,
\end{equation}
where $\tilde{\phi}^\sigma(\lambda)$ and $r_k(\lambda)$
are continued respectively from $\Gamma_k^+$ and $I_k^+$ to the left, and
for $\lambda\in C^+_{k-}$,
\begin{equation}
\tilde{\phi}^\sigma(\lambda)-iJ\alpha_k=
-i(r_k(\lambda)-J\theta_k)\,,
\end{equation}
where here $\tilde{\phi}^\sigma(\lambda)$ and
$r_k(\lambda)$ are continued respectively from $\Gamma_k^+$ and
$I_k^+$ to the right.

\begin{figure}[h]
\begin{center}
\mbox{\psfig{file=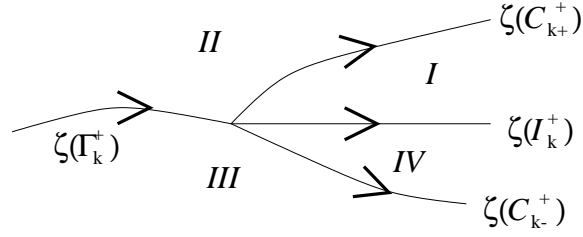,width=3 in}}
\end{center}
\caption{\em The image of the local contours in the $\zeta$-plane.  The
intersection point is $\zeta=0$, and the image of $I_k^+$ is the
positive real $\zeta$-axis.}
\label{fig:oddendpointzeta}
\end{figure}

The appropriate analytic change of variables
$\lambda\mapsto\zeta(\lambda)$ suggested by these continuation facts
and the degree of vanishing of $r_k(\lambda)-J\theta_k$ at
$\lambda=\lambda_{2k-1}$ now is specified by 
\begin{equation}
\zeta(\lambda):=\left(-\frac{r_k(\lambda)-J\theta_k}{\hbar}\right)^{2/3}\,,
\end{equation}
and we note that $\zeta\in{\mathbb R}_+$ when
$\lambda\in I_k^+$.  It follows from the analytic continuation
properties described above that
\begin{equation}
\frac{\tilde{\phi}^\sigma(\lambda)-iJ\alpha_k}{\hbar}=
-(-\zeta)^{3/2}\,.
\end{equation}
The transformation $\zeta(\lambda)$ takes the local contours
illustrated in Figure~\ref{fig:oddendpoint} into the $\zeta$-plane as
shown in Figure~\ref{fig:oddendpointzeta}.
We fix a disk $D_{2k-1}$ centered at $\lambda_{2k-1}$ of sufficiently
small radius independent of $\hbar$.  As before, we choose the
contours $\Gamma_k^+$, $C^+_{k+}$, and $C^+_{k-}$ within $D_{2k-1}$ so
that their images in $\zeta(D_{2k-1})$ lie respectively on the
straight rays $\arg(\zeta)=\pi$, $\arg(\zeta)=\pi/3$, and
$\arg(\zeta)=-\pi/3$.  These choices straighten out the contours in
Figure~\ref{fig:oddendpointzeta} within the expanding neighborhood
$\zeta(D_{2k-1})$.

We make the change of dependent variable
\begin{equation}
{\bf S}_{2k-1}(\zeta):=\left\{\begin{array}{ll} {\bf
O}^\sigma(\lambda(\zeta))\sigma_1^{\frac{1-J}{2}}\exp(iJ\sigma_3
(\theta_k-\alpha_k)/(2\hbar))\,, & \zeta\in I\cup II\,,\\\\
{\bf
O}^\sigma(\lambda(\zeta))\sigma_1^{\frac{1-J}{2}}\exp(-iJ\sigma_3
(\theta_k+\alpha_k)/(2\hbar))\,, & \zeta\in III\cup IV\,.
\end{array}\right.
\end{equation}
Consequently, the matrix ${\bf S}_{2k-1}(\zeta)$ satisfies locally 
simple jump relations. 
For $\zeta\in \zeta(\Gamma_k^+)$,
\begin{equation}
{\bf S}_{2k-1,+}(\zeta)={\bf S}_{2k-1,-}(\zeta)
\left[\begin{array}{cc}
1 & 0 \\-i\exp(-(-\zeta)^{3/2}) & 1\end{array}\right]\,,
\end{equation}
for $\zeta\in\zeta(I_k^+)$,
\begin{equation}
{\bf S}_{2k-1,+}(\zeta)={\bf S}_{2k-1,-}(\zeta)
\cdot(-i\sigma_1)\,,
\end{equation}
and for $\zeta\in\zeta(C^+_{k\pm})$,
\begin{equation}
{\bf S}_{2k-1,+}(\zeta)={\bf S}_{2k-1,-}(\zeta)
\left[\begin{array}{cc}
1 & i\exp(\pm i\zeta^{3/2}) \\ 0 & 1\end{array}\right]\,.
\end{equation}
Along with this, we consider the matrix $\tilde{\bf S}_{2k-1}(\zeta)$ defined
for $\zeta\in\zeta(D_{2k-1})$ in terms of the solution $\tilde{\bf
O}(\lambda)$, obtained in \S\ref{sec:outersolve}, of the outer
model problem by
\begin{equation}
\tilde{\bf S}_{2k-1}(\zeta):=\left\{\begin{array}{ll} \tilde{\bf
O}(\lambda(\zeta))\sigma_1^{\frac{1-J}{2}}\exp(iJ\sigma_3
(\theta_k-\alpha_k)/(2\hbar))\,, & \zeta\in I\cup II\,,\\\\
\tilde{\bf
O}(\lambda(\zeta))\sigma_1^{\frac{1-J}{2}}\exp(-iJ\sigma_3
(\theta_k+\alpha_k)/(2\hbar))\,, & \zeta\in III\cup IV\,.
\end{array}\right.
\end{equation}
As before, this matrix is analytic in $\zeta(D_{2k-1})$ except for
$\zeta\in{\mathbb R}_+$, where it takes continuous boundary
values for all $\zeta\neq 0$ that satisfy $\tilde{\bf S}_{2k-1,+}(\zeta)=
\tilde{\bf S}_{2k-1,-}(\zeta)\cdot(-i\sigma_1)$.
Although the jump relation for $\tilde{\bf S}_{2k-1}(\zeta)$ is
formally the same as for $\tilde{\bf S}_{2k}(\zeta)$ in
\S\ref{sec:2k}, one should keep in mind here that according to
Figure~\ref{fig:oddendpointzeta}, the orientation of ${\mathbb R}_+$
has been reversed, and is oriented here from the origin to
$\zeta=\infty$.  As before, one can prove the following decomposition
result.
\begin{lemma}
The matrix $\tilde{\bf
S}_{2k-1}(\zeta)$ determined from the solution of the outer model problem has
a unique representation
\begin{equation}
\tilde{\bf S}_{2k-1}(\zeta) = \tilde{\bf
S}_{2k-1}^{\rm hol}(\zeta)\tilde{\bf S}^{\rm loc, odd}(\zeta)\,,
\label{eq:representationofouterodd}
\end{equation}
where
\begin{equation}
\tilde{\bf S}^{\rm loc, odd}(\zeta):=
(-\zeta)^{-\sigma_3/4}\left[\begin{array}{cc}
1/\sqrt{2} & -1/\sqrt{2} \\
1/\sqrt{2} & 1/\sqrt{2}
\end{array}\right]\,,
\label{eq:tildeslocodd}
\end{equation}
and where $\tilde{\bf S}_{2k-1}^{\rm hol}(\zeta)$ is holomorphic in the
interior of $\zeta(D_{2k-1})$.
\end{lemma}
This exact local representation of the outer model problem solution is
thus written in terms of a complicated analytic part and a simple
explicit local part that is a function of
$\zeta$ alone (and is in particular independent of $\hbar$).

We obtain a similar factorization of ${\bf S}_{2k-1}(\zeta)$ as follows.
Recall the oriented contour $\Sigma^{\rm loc}$ defined in
\S\ref{sec:2k} and illustrated in Figure~\ref{fig:sloc}.  Consider
the following Riemann-Hilbert problem.
\begin{rhp}[Local model for odd endpoints]
\index{Riemann-Hilbert problem!local model for odd endpoints}
Find a matrix ${\bf S}^{\rm loc, odd}(\zeta)$ satisfying
\begin{enumerate}
\item
{\bf Analyticity:} ${\bf S}^{\rm loc, odd}(\zeta)$ is analytic for
$\zeta\in{\mathbb C}\setminus\Sigma^{\rm loc}$.
\item
{\bf Boundary behavior:} ${\bf S}^{\rm loc, odd}(\zeta)$ assumes
continuous boundary values from within each sector of ${\mathbb
C}\setminus\Sigma^{\rm loc}$, with continuity holding also at the
point of self-intersection.
\item
{\bf Jump conditions:}  The boundary values taken on $\Sigma^{\rm loc}$
satisfy
\begin{equation}
\begin{array}{rcll}
{\bf S}^{\rm loc, odd}_+(\zeta)&=&{\bf S}^{\rm loc, odd}_-(\zeta) 
\left[\begin{array}{cc} 1 & 0\\i\exp(-(-\zeta)^{3/2}) & 1\end{array} \right]
\,,&\zeta\in \Sigma_\Gamma\,,
\\\\ 
{\bf S}^{\rm loc, odd}_+(\zeta)&=& {\bf S}^{\rm loc, odd}_-(\zeta)
\left[\begin{array}{cc}1 & -i\exp(\mp i\zeta^{3/2})\\ 0 &1\end{array}\right]
\,,&\zeta\in \Sigma^\pm\,,
\\\\
{\bf S}^{\rm loc, odd}_+(\zeta)&=&{\bf S}^{\rm loc, odd}_-(\zeta)(i\sigma_1)\,,
&\zeta\in \Sigma_I\,.
\end{array}
\end{equation}
\item
{\bf Normalization:} ${\bf S}^{\rm loc, odd}(\zeta)$ is similar to
$\tilde{\bf S}^{\rm loc, odd}(\zeta)$ at $\zeta=\infty$, where
$\tilde{\bf S}^{\rm loc, odd}(\zeta)$ is defined by (\ref{eq:tildeslocodd}).
Precisely,
\begin{equation}
\lim_{\zeta\rightarrow\infty}
{\bf S}^{\rm loc, odd}(\zeta)\tilde{\bf S}^{\rm loc, odd}(\zeta)^{-1} = 
{\mathbb I}\,,
\end{equation}
with the limit being uniform with respect to direction.
\end{enumerate}
\label{rhp:Slocodd}
\end{rhp}

\begin{lemma}
The Riemann-Hilbert Problem~\ref{rhp:Slocodd} has a unique solution,
with the additional property that there exists some $M>0$ such that
the estimate
\begin{equation}
\|
{\bf S}^{\rm loc, odd}(\zeta)\tilde{\bf S}^{\rm loc, odd}(\zeta)^{-1} - 
{\mathbb I}\| \le M|\zeta|^{-1}\,,
\end{equation}
holds for all sufficiently large $|\zeta|$.  The solution ${\bf
S}^{\rm loc, odd}(\zeta)$ is universal in the sense that it does not
depend on $\hbar$.
\label{lemma:SlocExistodd}
\end{lemma}

\begin{proof}
Rather than repeating similar arguments to those used in the proof of
Lemma~\ref{lemma:SlocExist}, we simply use the matrix ${\bf S}^{\rm
loc, even}(\zeta)$ whose existence is guaranteed by that same lemma to
construct a solution ${\bf S}^{\rm loc, odd}(\zeta)$ of the
Riemann-Hilbert Problem~\ref{rhp:Slocodd}.  For $\zeta\in{\mathbb
C}\setminus
\Sigma^{\rm loc}$, set
\begin{equation}
{\bf S}^{\rm loc, odd}(\zeta):=(i\sigma_1)\cdot{\bf S}^{\rm loc, even}(\zeta)
\cdot (i\sigma_3)\,.
\end{equation}
It is a direct matter to check that the jump relations and
normalization condition for ${\bf S}^{\rm loc, even}(\zeta)$, along
with the smoothness and decay of the boundary values given in
Lemma~\ref{lemma:SlocExist} imply that the matrix so-defined is a
solution of the Riemann-Hilbert Problem~\ref{rhp:Slocodd} with the
desired properties.  Uniqueness follows from the uniform boundedness
for finite $\zeta$, continuity of the boundary values, and Liouville's
theorem.
\end{proof}

We now propose an approximation to ${\bf S}_{2k-1}(\zeta)$ defined for
$\zeta\in\zeta(D_{2k-1})$ by
\begin{equation}
\hat{\bf S}_{2k-1}(\zeta):=\tilde{\bf S}_{2k-1}^{\rm hol}(\zeta){\bf
S}^{\rm loc, odd}(\zeta)\,.
\end{equation}
When we take into account the fact that the contour $\Sigma^{\rm loc}$
is the union of $\zeta(I_k^+)$, $\zeta(\Gamma_k^+)$, $\zeta(C_{k+}^+)$ and
$\zeta(C_{k-}^+)$ {\em with the orientation reversed}, we see that $\hat{\bf
S}_{2k-1}(\zeta)$ satisfies exactly the same jump relations as ${\bf
S}_{2k-1}(\zeta)$.

As before, we may use this matrix to define a local approximation 
\index{local approximation!near odd endpoints} of ${\bf N}^\sigma(\lambda)$ valid for
$\lambda\in D_{2k-1}$.  We define this approximation as follows.  For
$\zeta(\lambda)$ in region I, set
\begin{equation}
\begin{array}{l}
\hat{\bf N}_{2k-1}^\sigma(\lambda):=\tilde{\bf S}_{2k-1}^{\rm hol}(\zeta(\lambda))
{\bf S}^{\rm loc, odd}(\zeta(\lambda))\exp(-iJ\sigma_3
(\theta_k-\alpha_k)/(2\hbar))\times\\\\
\hspace{0.4 in}\left[\begin{array}{cc}
1 & i\exp(-iJ\alpha_k/\hbar)\exp(-ir_k(\lambda)/\hbar)\\
0 & 1\end{array}\right]\sigma_1^{\frac{1-J}{2}}\,,
\end{array}
\label{eq:Nhat2km1I}
\end{equation}
for $\zeta(\lambda)$ in region II, set
\begin{equation}
\hat{\bf N}_{2k-1}^\sigma(\lambda):=\tilde{\bf S}_{2k-1}^{\rm hol}(\zeta(\lambda))
{\bf S}^{\rm loc,
odd}(\zeta(\lambda))\exp(-iJ\sigma_3(\theta_k-\alpha_k)/(2\hbar))
\sigma_1^{\frac{1-J}{2}}\,,
\label{eq:Nhat2km1II}
\end{equation}
for $\zeta(\lambda)$ in region III, set
\begin{equation}
\hat{\bf N}_{2k-1}^\sigma(\lambda):=\tilde{\bf S}_{2k-1}^{\rm hol}(\zeta(\lambda))
{\bf S}^{\rm loc, odd}(\zeta(\lambda))
\exp(iJ\sigma_3(\theta_k+\alpha_k)/(2\hbar))
\sigma_1^{\frac{1-J}{2}}\,,
\label{eq:Nhat2km1III}
\end{equation}
and for $\zeta(\lambda)$ in region IV, set
\begin{equation}
\begin{array}{l}
\hat{\bf N}_{2k-1}^\sigma(\lambda):=\tilde{\bf S}_{2k-1}^{\rm hol}(\zeta(\lambda))
{\bf S}^{\rm loc, odd}(\zeta(\lambda))
\exp(iJ\sigma_3(\theta_k+\alpha_k)/(2\hbar))\times\\\\
\hspace{0.4 in}
\left[\begin{array}{cc}
1 & -i\exp(-iJ\alpha_k/\hbar)\exp(ir_k(\lambda)/\hbar)\\
0 & 1\end{array}\right]\sigma_1^{\frac{1-J}{2}}\,.
\end{array}
\label{eq:Nhat2km1IV}
\end{equation}

As in \S\ref{sec:2k}, we can characterize the local approximation of
${\bf N}^\sigma(\lambda)$ near the endpoint $\lambda_{2k-1}$ by the
following results, all of which are proved in exactly the same manner
as their analogues for the corresponding approximations valid near
$\lambda_{2k}$.
\begin{lemma}
The local approximation $\hat{\bf N}_{2k-1}^\sigma(\lambda)$ is analytic
for $\lambda\in D_{2k-1}\setminus (D_{2k-1}\cap C)$, and takes continuous
boundary values on $C_\sigma$ that satisfy exactly $\hat{\bf
N}_{2k-1,+}^\sigma(\lambda)=
\hat{\bf N}_{2k-1,-}^\sigma(\lambda){\bf v}_{\tilde{\bf N}}^\sigma(\lambda)$.
\label{lemma:jumpsinD2km1}
\end{lemma}

\begin{lemma}
There exists some $M>0$ such that for all $\lambda\in D_{2k-1}$ and
all sufficiently small $\hbar$,
\begin{equation}
\|\hat{\bf N}^\sigma_{2k-1}(\lambda)\|\le M\hbar^{-1/3}\,.
\end{equation}
The same estimate holds for the inverse matrix, since the local approximation
has determinant one.
\label{lemma:holoboundodd}
\end{lemma}

\begin{lemma}
There exists some $M>0$ such that for all $\lambda\in \partial
D_{2k-1}$, and for all sufficiently small $\hbar$
\begin{equation}
\|\hat{\bf N}_{2k-1}^\sigma(\lambda)
\hat{\bf N}_{\rm out}^\sigma(\lambda)^{-1}-{\mathbb I}\|\le M\hbar^{1/3}\,.
\end{equation}
\label{lemma:matchodd}
\end{lemma}

\subsection{Local analysis for $\lambda$ near the origin.}
\label{sec:origin}
Near the origin, the {\em ad hoc} replacement of ${\bf
N}^\sigma(\lambda)$ with the ``continuum limit'' approximation
$\hat{\bf N}^\sigma(\lambda)$ breaks down for two reasons.  First, at
the level of the Riemann-Hilbert problem for the matrix ${\bf
O}^\sigma(\lambda)$, the function $\theta^\sigma(\lambda)$ is not
analytic at $\lambda=0$ and therefore the origin must lie at the
junction of two lenses, one corresponding to the band $I_0^+$
connecting the origin to $\lambda_0$, and the second being $I_0^-$,
the reflection of $I_0^+$ in the real axis.  Furthermore, the terminal
portion of the loop contour $C_\sigma$, namely the gap
$\Gamma_{G/2+1}^+$, and its complex conjugate $\Gamma_{G/2+1}^-$ meet
at the origin.  Although $\Re(\tilde{\phi}^\sigma(\lambda))$ is
negative by assumption on the interior of this gap, it always vanishes
at the origin, which means that significant errors may be introduced
by simply replacing the jump matrix on $\Gamma_{G/2+1}^+$ by the
identity on a neighborhood of the origin.  The breakdown of the
approximations leading to the outer model Riemann-Hilbert
Problem~\ref{rhp:model} by these mechanisms is thus similar to the
corresponding breakdown near the endpoints
$\lambda_0,\dots,\lambda_G$.

On the other hand, a second mechanism for failure of our formal
approximations at the origin is unlike what happens at the nonzero
endpoints.  There is additional difficulty at the origin entering at
the level of the ``discrete'' (referring to a discrete WKB eigenvalue
measure $d\mu_{\hbar_N}^{\rm WKB}$ in the logarithmic integral)
Riemann-Hilbert Problem~\ref{rhp:N} for ${\bf N}^\sigma(\lambda)$.
Namely, the replacement of the function $\phi^\sigma(\lambda)$ by
$\tilde{\phi}^\sigma(\lambda)$ is not valid in any neighborhood of the
origin.  Here, an additional contribution coming from the function
$W(w)$ defined by (\ref{eq:Wdef}) must be included in any uniformly
valid approximation.

The situation near the origin is more complicated than near the
endpoints $\lambda_0,\dots,\lambda_G$ because
analytic continuation properties of the functions appearing in the
jump matrix do not favor the sort of convenient change of variables
that yields a model Riemann-Hilbert problem that does not involve
$\hbar$ and yet captures the asymptotic behavior of the solution in a
local neighborhood of fixed size independent of $\hbar$.  Thus, we are
led to work in a shrinking neighborhood of the origin, and to
introduce a less elegant local change of variables.  

\begin{figure}[h]
\begin{center}
\mbox{\psfig{file=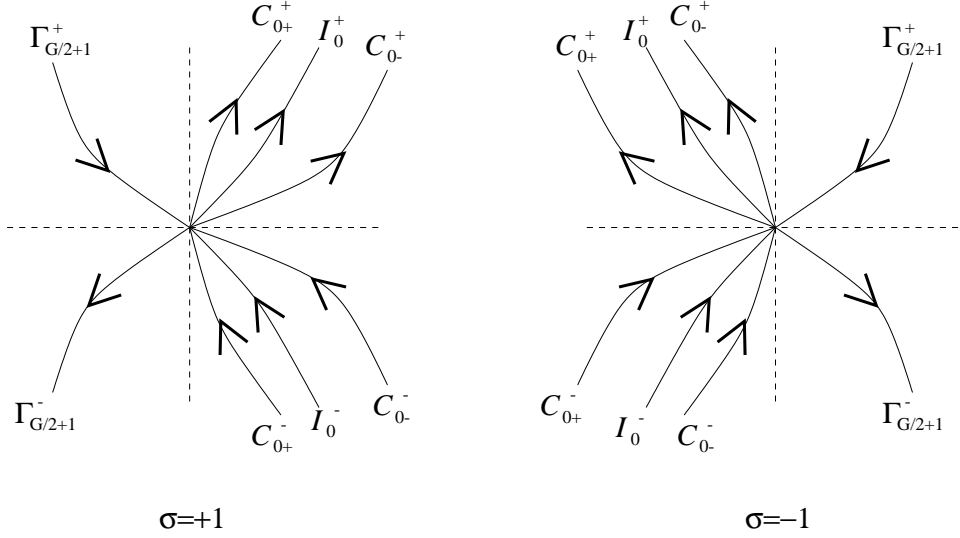,width=5 in}}
\end{center}
\caption{\em The support of the jump matrix for ${\bf E}^{\sigma, \rm
loc}(\lambda)$ near $\lambda=0$.  The picture depends on
the index $\sigma$.  Left: $\sigma=+1$.  Right: $\sigma=-1$.  The real
and imaginary axes of the $\lambda$-plane are shown with dashed
lines.}
\label{fig:zeroN}
\end{figure}

The procedure we use is to consider the local error \index{local
error} between the matrix ${\bf N}^\sigma(\lambda)$ satisfying the
phase-conjugated Riemann-Hilbert Problem~\ref{rhp:N} and its outer
approximation $\hat{\bf N}^\sigma_{\rm out}(\lambda)$ defined in
\S\ref{sec:outersolve} by (\ref{eq:outerparametrixdef}).  Thus, near the 
origin, set
\begin{equation}
{\bf E}^{\sigma, \rm loc}(\lambda):={\bf N}^\sigma(\lambda)
\hat{\bf N}^\sigma_{\rm out}(\lambda)^{-1}\,.
\end{equation}
Near the origin, this matrix is analytic except on the contours shown in
Figure~\ref{fig:zeroN}.
On these contours, we have jump relations of the form 
\begin{equation}
{\bf E}^{\sigma, \rm loc}_+(\lambda)=
{\bf E}^{\sigma, \rm loc}_-(\lambda)
\left(\tilde{\bf O}_-(\lambda){\bf D}^\sigma_-(\lambda)^{-1}
{\bf v}^\sigma_{\bf N}(\lambda)
{\bf D}^\sigma_+(\lambda)
\tilde{\bf O}_+(\lambda)^{-1}\right)\,.
\label{eq:jumpsforEloc}
\end{equation}
Recall that $\tilde{\bf O}(\lambda)$ is the matrix obtained as the
solution of the outer model Riemann-Hilbert Problem \ref{rhp:model} in
\S\ref{sec:outersolve}, and ${\bf D}^\sigma(\lambda)$ is defined to be
the identity outside all ``lenses'' while inside the lenses is given
by (\ref{eq:ldef1}) and (\ref{eq:ldef2}).  Therefore, the boundary
values of ${\bf D}^\sigma(\lambda)$ are equal for $\lambda\in
\Gamma_{G/2+1}^\pm$ and the boundary values of $\tilde{\bf
O}(\lambda)$ only differ for $\lambda\in I_0^\pm$.  Also,
recall that ${\bf v}^\sigma_{\bf N}(\lambda)$ differs from the
identity matrix only for $\lambda\in I_0^\pm$ and
$\lambda\in\Gamma_{G/2+1}^\pm$.  Finally, the jump matrix in the lower
half-plane is determined from that in the upper half-plane by the
symmmetry that ${\bf E}^{\sigma, \rm loc}(\lambda)$ satisfies by
construction: ${\bf E}^{\sigma, \rm loc}(\lambda^*) = \sigma_2 {\bf
E}^{\sigma, \rm loc}(\lambda)^*\sigma_2$.

Let $U_\hbar$ be a sufficiently small disk neighborhood of the origin,
whose radius we will specify later.  For $\lambda\in U_\hbar$, we
introduce a change of variables of the form
\begin{equation}
{\bf F}^\sigma(\lambda):={\bf C}^\sigma(\lambda)^{-1}{\bf E}^{\sigma, \rm loc}
(\lambda){\bf C}^\sigma(\lambda)\,.
\end{equation}
The conjugating factors are specified as follows.  
Let $U^+_\hbar$
(respectively $U^-_\hbar$) denote the part of $U_\hbar$ lying to the
left (respectively right) of $I_0^+\cup I_0^-$.  
Then, we set
\begin{equation}
\begin{array}{rcl}
{\bf C}^\sigma(\lambda)&:=&\displaystyle
\tilde{\bf O}(\lambda)
\sigma_1^\frac{1-J}{2}\cdot\left\{
\begin{array}{ll}
{\mathbb I}\,, &\hspace{0.2 in}\lambda\in U^\sigma_\hbar\,,\\\\
\left[\begin{array}{cc} 0 & -i\sigma\exp(-iJ\alpha_0/\hbar)\\
-i\sigma\exp(iJ\alpha_0/\hbar) & 0\end{array}\right]\,, &\hspace{0.2 in}
\lambda\in U_\hbar^{-\sigma}\,.
\end{array}\right\}\\\\
&&\hspace{0.2 in}\times\,\,\exp(-i(J \alpha_0+\sigma r_0(0))\sigma_3/(2\hbar))
\,.
\end{array}
\end{equation}
Recall that $r_0(0)$ and $\alpha_0$ are purely real constants, and
that the function $r(\lambda)$ is defined in Definition~\ref{def:qr}.
This, along with the properties of the solution $\tilde{\bf
O}(\lambda)$ of the outer model problem developed in
\S\ref{sec:outersolve}, implies that {\em the matrices ${\bf
C}^\sigma(\lambda)$ and their inverses are analytic and uniformly
bounded in any sufficiently small neighborhood of the origin.}  

The exact jump relations for ${\bf F}^+(\lambda)$ (by which we mean
${\bf F}^\sigma(\lambda)$ for $\sigma=+1$) on the contours in the
upper half-plane are as follows.  For $\lambda\in
\Gamma_{G/2+1}^+$,
\begin{equation}
{\bf F}^+_+(\lambda) = {\bf F}^+_-(\lambda)\left[\begin{array}{cc} 1 &
0
\\-i\exp(\delta^+/\hbar)
\exp((\tilde{\phi}^+(\lambda)-\tilde{\phi}^+(0))/\hbar)(1-d^+(\lambda))
& 1\end{array}\right]\,,
\label{eq:fplusgamma}
\end{equation}
for $\lambda\in C_{0+}^+$, 
\begin{equation}
{\bf F}^+_+(\lambda) = {\bf F}^+_-(\lambda)\left[\begin{array}{cc}
1 & i\exp(-i(r_0(\lambda)-r_0(0))/\hbar) \\
0 & 1\end{array}\right]\,,
\end{equation}
for $\lambda\in C_{0-}^+$, 
\begin{equation}
{\bf F}^+_+(\lambda) = {\bf F}^+_-(\lambda)\left[\begin{array}{cc}
1 & 0 \\
i\exp(i(r_0(\lambda)-r_0(0))/\hbar) & 1\end{array}
\right]\,,
\end{equation}
and for $\lambda\in I_0^+$,
\begin{equation}
{\bf F}^+_+(\lambda) = {\bf F}^+_-(\lambda)\left[\begin{array}{cc}
1-d^+(\lambda) &
i\exp(-i(r_0(\lambda)-r_0(0))/\hbar)d^+(\lambda)\\
i\exp(i(r_0(\lambda)-r_0(0))/\hbar)d^+(\lambda) &
1+d^+(\lambda)
\end{array}\right]\,.
\end{equation}
These are expressed in terms of the quantities
\begin{equation}
d^\sigma(\lambda):=
1-\exp((\phi^\sigma(\lambda)-\tilde{\phi}^\sigma(\lambda))/\hbar)\,,
\end{equation}
and 
\begin{equation}
\delta^\sigma:=\tilde{\phi}^\sigma(0)-iJ\alpha_0-i\sigma r_0(0)\,,
\end{equation}
where by $\tilde{\phi}^\sigma(0)$ we mean the limit as $\lambda\rightarrow 0$
in the gap $\Gamma^+_{G/2+1}$, and where we recall that $r_0(\lambda)$ is
defined as the analytic continuation of $\theta^\sigma(\lambda)$ from the
band $I_0^+$.

Similarly, the jump relations for ${\bf F}^-(\lambda)$ ({\em i.e.} for
${\bf F}^\sigma(\lambda)$ in the case when $\sigma=-1$) in the upper
half-plane are as follows.  For $\lambda\in \Gamma_{G/2+1}^+$,
\begin{equation}
{\bf F}^-_+(\lambda)={\bf F}^-_-(\lambda)\left[\begin{array}{cc} 1 & 0
\\-i\exp(\delta^-/\hbar)\exp((\tilde{\phi}^-(\lambda)-\tilde{\phi}^-(0))
/\hbar)(1-d^-(\lambda))
& 1\end{array}\right]\,,
\end{equation}
for $\lambda\in C_{0+}^+$,
\begin{equation}
{\bf F}^-_+(\lambda)={\bf F}^-_+(\lambda)\left[\begin{array}{cc}
1 & 0 \\ i\exp(-i(r_0(\lambda)-r_0(0))/\hbar) & 1
\end{array}\right]\,,
\end{equation}
for $\lambda\in C_{0-}^+$, 
\begin{equation}
{\bf F}^-_+(\lambda)={\bf F}^-_+(\lambda)\left[\begin{array}{cc}
1 & i\exp(i(r_0(\lambda)-r_0(0))/\hbar) \\
0 & 1\end{array}\right]\,,
\end{equation}
and for $\lambda\in I_0^+$,
\begin{equation}
{\bf F}^-_+(\lambda)={\bf F}^-_+(\lambda)\left[\begin{array}{cc}
1+d^-(\lambda)&
i\exp(i(r_0(\lambda)-r_0(0))/\hbar)d^-(\lambda)\\
i\exp(-i(r_0(\lambda)-r_0(0))/\hbar)d^-(\lambda) &
1-d^-(\lambda)\end{array}\right]\,.
\end{equation}
In both cases, the jump relations on the
corresponding contours in the lower half-plane are obtained by the
symmetry ${\bf F}^\sigma(\lambda^*) = \sigma_2{\bf
F}^\sigma(\lambda)^*\sigma_2$.

We now observe a consequence of the fact that we are considering only
values of $\hbar$ in the ``quantum'' sequence
$\hbar=\hbar_N$\index{quantum sequence of values of $\hbar$}, for
$N=1,2,3,\dots$ ({\em cf.} the definition (\ref{eq:quantumsequence})
of $\hbar_N$).  Consider first the case $\sigma=+1$.  We know that in
this case the function
$\theta^\sigma(\lambda)-i\tilde{\phi}^\sigma(\lambda)$ is analytic on
the bounded interior of the loop $C$, except on the support of the
asymptotic eigenvalue measure $\rho^0(\eta)\,d\eta$, namely the
imaginary interval $[0,iA]$.  If we orient this interval from the
origin to $iA$, then we can calculate the explicit jump relation:
\begin{equation}
(\theta^\sigma(\lambda)-i\tilde{\phi}^\sigma(\lambda))_+ - 
(\theta^\sigma(\lambda)-i\tilde{\phi}^\sigma(\lambda))_- = 
2\pi\int_\lambda^{iA}\rho^0(\eta)\,d\eta\,.
\end{equation}
Applying this relation to the limiting values of
$\theta^\sigma(\lambda)$ and $\tilde{\phi}^\sigma(\lambda)$ taken as
$\lambda\rightarrow 0$ along the boundary in either $I_0^+$ or
$\Gamma^+_{G/2+1}$, we take advantage of the fact that throughout
$I_0^+$, we have the identity $\tilde{\phi}^\sigma(\lambda)\equiv
iJ\alpha_0$, and throughout $\Gamma^+_{G/2+1}$, we have the identity
$\theta^\sigma(\lambda)\equiv 0$.  Along with similar reasoning
for the case $\sigma=-1$, we finally obtain the formula
\begin{equation}
\delta^\sigma= 2\pi i\sigma\int_0^{iA}\rho^0(\eta)\,d\eta =
-2\pi i\sigma N\hbar_N\,,
\end{equation}
where the second equality follows from the definition
(\ref{eq:quantumsequence}) of the quantum sequence of values of
$\hbar$.  Consequently, whenever $\hbar=\hbar_N$ for any $N=1,2,3,\dots$, 
we conclude that $\exp(\delta^\sigma/\hbar)\equiv 1$.

As in the local analysis near the nonzero endpoints, we again use the
freedom of placement of the contours $\Gamma_{G/2+1}^+$, and
$C^+_{0\pm}$ to ensure that in some {\em fixed} disk neighborhood $U$
of the origin, these contours are radial straight lines in the
$\lambda$-plane with slopes independent of $\hbar$.  Let
$I_0^{+\prime}$ and $\Gamma_{G/2+1}^{+\prime}$ respectively denote the
tangent lines to $I_0^+$ and $\Gamma_{G/2+1}^+$ at the origin. Note
that the tangent line $\Gamma_{G/2+1}^{+\prime}$ is confined to some
sector for the inequality $\Re(\tilde{\phi}^\sigma(\lambda))<0$ to be
satisfied, but is otherwise arbitrary, while the tangent line
$I_0^{+\prime}$ is not free, being fixed by the measure reality
condition.  For concreteness, we choose the contours so that for
$\sigma=+1$, $C^+_{0+}\cap U$ bisects the sector between
$I_0^{+\prime}$ at the origin and the positive imaginary axis while
$C^+_{0-}\cap U$ bisects the sector between the positive real axis and
$I_0^{+\prime}$ at the origin.  For $\sigma=-1$, we arrange that
$C^+_{0+}\cap U$ bisects the sector between the negative real axis and
$I_0^{+\prime}$, while $C^+_{0-}\cap U$ bisects the sector between
$I_0^{+\prime}$ and the positive imaginary axis.  Let $\kappa$ denote
$\arg(I_0^{+\prime})$, and $\xi$ denote
$\arg(\Gamma_{G/2+1}^{+\prime})$.  For $\sigma=+1$, we have
$0<\kappa<\pi/2$ and $\pi/2<\xi<\pi$, while for $\sigma=-1$, we have
$0<\xi<\pi/2$ and $\pi/2 <\kappa<\pi$.

Our strategy is to approximate the exact jump relations for the
matrices ${\bf F}^\sigma(\lambda)$ in terms of a crude rescaled local
variable $\zeta=-i\rho^0(0)\lambda/\hbar$, combining careful
asymptotic analysis of $d^\sigma(\lambda)$ with elementary Taylor
approximations of $r_0(\lambda)-r_0(0)$ and
$\tilde{\phi}^\sigma(\lambda)-
\tilde{\phi}^\sigma(0)$.  First, note that the definitions of
$\phi^\sigma(\lambda)$ and $\tilde{\phi}^\sigma(\lambda)$ imply
\begin{equation}
1-d^\sigma(\lambda)
=
\left[\prod_{n=0}^{N-1}\frac{\lambda-\lambda_{\hbar_N,n}^{{\rm WKB}*}}
{\lambda-\lambda_{\hbar_N,n}^{\rm WKB}}\right]
\exp\left(-\frac{1}{\hbar_N}\left[
\int_{0}^{iA}L_\eta^0(\lambda)\rho^0(\eta)\,d
\eta
+\int_{-iA}^0L_\eta^0(\lambda)\rho^0(\eta^*)^*\,d\eta\right]
\right)\,,
\end{equation}
so that in particular we see that $d^\sigma(\lambda)$ is independent
of $\sigma$.  Recall now Theorem~\ref{theorem:inner}, 
which gives
\begin{equation}
1-d^\sigma(\lambda)=\frac{1}{W(-i\zeta)}(1+\bo(\hbar_N^{1/3}))\,,
\end{equation}
uniformly for bounded $\lambda$ outside any sector including the
imaginary axis, or equivalently for $\zeta=\bo(\hbar_N^{-1})$.  Now,
the function $W(w)$ defined by (\ref{eq:Wdef}) has a cut on the
positive real $w$ axis, which corresponds to the positive imaginary
$\zeta$ axis.  Thus, we find that
\begin{equation}
1-d^\sigma(\lambda)=\frac{\Gamma(1/2+i\zeta)}{\Gamma(1/2-i\zeta)}
(-i\zeta)^{-2i\zeta}\exp(2i\zeta)\left(1+\bo(\hbar_N^{1/3})\right)
\left\{\begin{array}{lrcccl}
\exp(\pi\zeta)\,,&0&<&\arg(\zeta)&<&\displaystyle\frac{\pi}{2}\,,\\\\
\exp(-\pi\zeta)\,,&\displaystyle\frac{\pi}{2}&<&\arg(\zeta)&<&\pi\,.
\end{array}\right.
\label{eq:oneminusdasymp}
\end{equation}
To compactly express these asymptotics we define the analytic
functions $h^\sigma(\zeta)$ for $\Im(\zeta)\ge 0$ and $\arg(\zeta)\neq
\pi/2$ by setting
\begin{equation}
h^\sigma(\zeta):=1-\frac{\Gamma(1/2 + i\zeta)}{\Gamma(1/2- i\zeta)}
(-i\zeta)^{-2i\zeta}\exp((2i+\sigma\pi)\zeta)\,.
\end{equation}
These functions are uniformly bounded if $\zeta$ is bounded away from
a sector containing the positive imaginary axis.  From Stirling's
formula\index{Stirling's formula}, we deduce their asymptotic behavior
for large $\zeta$ in the upper half-plane:
\begin{equation}
\begin{array}{rcl}
h^+(\zeta)&=&\displaystyle
\left\{\begin{array}{lrcccl}
\displaystyle \frac{1}{12i\zeta} + \bo(|\zeta|^{-2})\,,&0&<&\arg(\zeta)&<&\displaystyle\frac{\pi}{2}\,,\\\\
1+\bo(\exp(2\pi\Re(\zeta)))\,,&\displaystyle\frac{\pi}{2}&<&\arg(\zeta)&<&\pi\,,
\end{array}\right.\\\\
h^-(\zeta)&=&\displaystyle
\left\{\begin{array}{lrcccl}1+\bo(\exp(-2\pi\Re(\zeta)))\,,&
0&<&\arg(\zeta)&<&\displaystyle\frac{\pi}{2}\,,\\\\
\displaystyle\frac{1}{12i\zeta}+\bo(|\zeta|^{-2})\,,
&\displaystyle\frac{\pi}{2}&<&\arg(\zeta)&<&\pi\,.
\end{array}\right.
\end{array}
\label{eq:decayofh}
\end{equation}

Next, define the constants $u$ and $v$ by:
\begin{equation}
u:=\frac{1}{-i\rho^0(0)}\cdot
\lim_{\lambda\rightarrow 0,\lambda\in\Gamma_{G/2+1}^+}
\frac{d\tilde{\phi}^\sigma}{d\lambda}(\lambda)\,,\hspace{0.3 in}
v:=\frac{1}{-i\rho^0(0)}\cdot
\lim_{\lambda\rightarrow 0,\lambda\in I_0^+}
\frac{d r_0}{d\lambda}(\lambda)=\pi \lim_{\lambda\rightarrow 0,\lambda\in I_0^+}
\rho^\sigma(\lambda)\,.
\end{equation}
These constants are of course independent of $\hbar$.  For $\lambda\in
I_0^{+\prime}$, $v\zeta=-i\rho^0(0) v\lambda/\hbar_N$ is real and negative.
Likewise, for $\lambda\in\Gamma_{G/2+1}^{+\prime}$,
$u\zeta=-i\rho^0(0)u\lambda/\hbar_N$ is real and negative.  For
$\lambda\in\Gamma_{G/2+1}^+$, we therefore have
\begin{equation}
\exp((\tilde{\phi}^\sigma(\lambda)-\tilde{\phi}^\sigma(0))/\hbar_N)=
\exp(u\zeta)\exp(\bo(\lambda^2/\hbar_N))\,.
\label{eq:Taylor1}
\end{equation}
Similarly, for $\lambda\in I_0^+\cup C_{0+}^+\cup C_{0-}^+$,
\begin{equation}
\exp(\pm i(r_0(\lambda)-r_0(0))/\hbar_N)=
\exp(\pm iv\zeta)\exp(\bo(\lambda^2/\hbar_N))\,.
\label{eq:Taylor2}
\end{equation}

We are now going to use these results to propose a model
Riemann-Hilbert problem for an approximation to ${\bf
F}^\sigma(\lambda)$.  The range of validity of the asymptotic formulae
(\ref{eq:Taylor1}) and (\ref{eq:Taylor2}) places restrictions on the
size of the neighborhood $U_\hbar$.  In fact, the radius of $U_\hbar$
must {\em shrink} as $\hbar\downarrow 0$.  If $R$ is the radius
$U_\hbar$, we will need to have $R^2/\hbar\ll 1$ in order for the
error factors in the Taylor approximations (\ref{eq:Taylor1}) and
(\ref{eq:Taylor2}) above to be negligible for $\lambda\in U_\hbar$.
On the other hand, to characterize the local behavior in a universal
way (so that $\hbar$ enters into the local approximation in the form
of a simple scaling) we will need the image in the $\zeta$-plane of
the boundary $\partial U_\hbar$ to be {\em expanding} as
$\hbar\downarrow 0$.  This requires $R/\hbar\gg 1$.  The radius $R$ is
therefore asymptotically bounded above and below: $\hbar\ll R\ll
\hbar^{1/2}$.  
Thus, let $\delta$ be a number between $1/2$ and $1$, and fix the size
of $U_\hbar$ by setting $R=\hbar^\delta$.  We reserve the choice of a
particular value of $\delta$ for later optimization of our estimates.
Note that for all sufficiently small $\hbar$, $U_\hbar$ is contained
in the fixed neighborhood $U$.

\begin{figure}[h]
\begin{center}
\mbox{\psfig{file=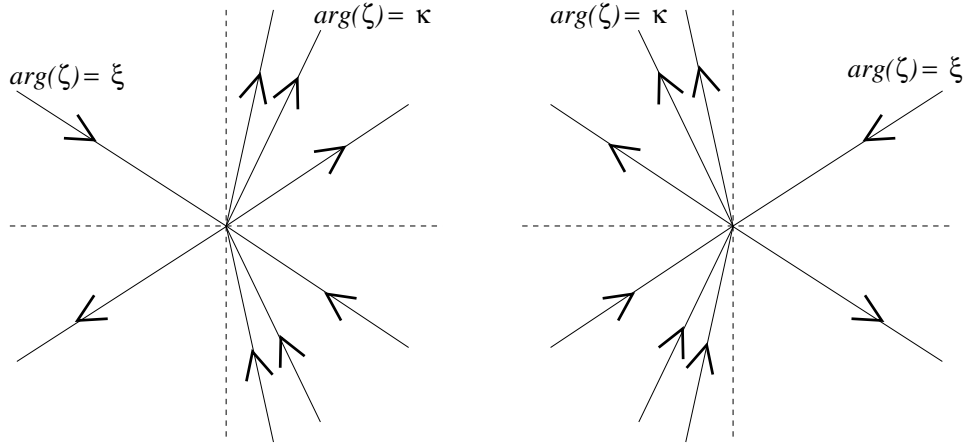,width=5 in}}
\end{center}
\caption{\em The contours $\Sigma^\sigma_{\hat{\bf F}}$ for the model Riemann-Hilbert
problems near the origin.  Left: $\sigma=+1$.  Right: $\sigma=-1$.  In
the upper half-plane, the unmarked contours bisect the angles between
$\arg(\zeta)=\kappa$ and the boundaries of the quadrant.  Although the
orientation is as shown, both contours are symmetric with respect to
complex-conjugation as sets of points.  All rays extend to $\zeta=\infty$.}
\label{fig:originmodel}
\end{figure}

By keeping the leading
terms of the jump matrices for ${\bf F}^\sigma(\lambda)$ in an
expansion for $\zeta$ held fixed as $\hbar$ tends to zero, we are led
to propose a local model Riemann-Hilbert problem.  First,
we introduce a contour.  Let $\Sigma^\sigma_{\hat{\bf F}}$ be the
oriented contour shown in Figure~\ref{fig:originmodel} for both signs of
$\sigma$.
Next, we define on the contour a jump matrix ${\bf v}_{\hat{\bf
F}}^\sigma(\zeta)$.  For $\sigma=+1$, and $\zeta\in \Sigma_{\hat{\bf F}}^+$
with $\Im(\zeta)>0$, set
\begin{equation}
{\bf v}^+_{\hat{\bf F}}(\zeta):=
\left\{\begin{array}{ll}
\left[\begin{array}{cc}
1 & 0 \\
-i(1-h^+(\zeta))\exp((u-2\pi)\zeta)
 & 1
\end{array}\right]\,, &\hspace{0.2 in}
\arg(\zeta)=\xi\,,\\\\
\left[\begin{array}{cc}
1 & i\exp(-iv\zeta) \\
0 & 1\end{array}\right]\,,
&\hspace{0.2 in}\arg(\zeta)=\kappa/2 + \pi/4\,,\\\\
\left[\begin{array}{cc}
1 & 0\\
i\exp(iv\zeta) 
 & 1\end{array}\right]\,,&\hspace{0.2 in}\arg(\zeta)=\kappa/2\,,\\\\
\left[\begin{array}{cc}
1-h^+(\zeta) & ih^+(\zeta)\exp(-iv\zeta)\\
ih^+(\zeta)\exp(iv\zeta) & 1+h^+(\zeta)
\end{array}\right]\,,&\hspace{0.2 in}
\arg(\zeta)=\kappa\,.
\end{array}\right.
\label{eq:fhatplusjump}
\end{equation}
For $\zeta\in \Sigma^+_{\hat{\bf F}}$ with $\Im(\zeta)<0$, we set
${\bf v}^+_{\hat{\bf F}}(\zeta):=\sigma_2{\bf v}^+_{\hat{\bf
F}}(\zeta^*)^*\sigma_2$.  For the opposite parity, $\sigma=-1$,
define for $\zeta\in\Sigma_{\hat{\bf F}}^-$ and $\Im(\zeta)>0$,
\begin{equation}
{\bf v}^-_{\hat{\bf F}}(\zeta):=
\left\{\begin{array}{ll}
\left[\begin{array}{cc}
1 & 0 \\
-i(1-h^-(\zeta))\exp((u+2\pi)\zeta)
 & 1
\end{array}\right]\,,&\hspace{0.2 in}
\arg(\zeta)=\xi\,,\\\\
\left[\begin{array}{cc}
1 & 0 \\
i\exp(-iv\zeta) & 1
\end{array}\right]\,,&\hspace{0.2 in}
\arg(\zeta)=\kappa/2+\pi/2\,,\\\\
\left[\begin{array}{cc}
1 & i\exp(iv\zeta) \\
0 & 1 \end{array}\right]\,,&\hspace{0.2 in}
\arg(\zeta)=\kappa/2+\pi/4\,,\\\\
\left[\begin{array}{cc}
1+h^-(\zeta) & ih^-(\zeta)\exp(iv\zeta)\\
ih^-(\zeta)\exp(-iv\zeta) & 1-h^-(\zeta)\end{array}\right]\,,&\hspace{0.2 in}
\arg(\zeta)=\kappa\,.
\end{array}\right.
\end{equation}
Again, for $\zeta\in\Sigma^-_{\hat{\bf F}}$ with $\Im(\zeta)<0$, we
set ${\bf v}^-_{\hat{\bf F}}(\zeta):=\sigma_2{\bf v}^-_{\hat{\bf
F}}(\zeta^*)^*\sigma_2$.

\begin{rhp}[Local model for the origin]
\index{Riemann-Hilbert problem!local model for the origin}
Find a matrix function $\hat{\bf F}^\sigma(\zeta)$ satisfying
\begin{enumerate}
\item
{\bf Analyticity:} $\hat{\bf F}^\sigma(\zeta)$ is analytic for $\zeta\in
{\mathbb C}\setminus \Sigma_{\hat{\bf F}}^\sigma$.
\item
{\bf Boundary behavior:} $\hat{\bf F}^\sigma(\zeta)$ assumes continuous
boundary values on $\Sigma_{\hat{\bf F}}^\sigma$ from each sector of
the complement, with continuity also at the self-intersection point.
\item
{\bf Jump condition:} On the oriented contour $\Sigma_{\hat{\bf
F}}^\sigma$ minus the origin, the boundary values satisfy
\begin{equation}
\hat{\bf F}^\sigma_+(\zeta)=
\hat{\bf F}^\sigma_-(\zeta){\bf v}_{\hat{\bf F}}^\sigma(\zeta)\,.
\end{equation}
\item
{\bf Normalization:} $\hat{\bf F}^\sigma(\zeta)$ is normalized at infinity:
\begin{equation}
\hat{\bf F}^\sigma(\zeta)\rightarrow{\mathbb I}\mbox{ as }
\zeta\rightarrow\infty\,.
\end{equation}
\end{enumerate}
\label{rhp:Fhat}
\end{rhp}

\begin{lemma}
Let the parameters $u$, $v$, $\kappa$, $\xi$, and $\sigma$ be fixed.
Then, the Riemann-Hilbert Problem~\ref{rhp:Fhat} has a unique
solution with the additional property that there exists a constant $M>0$ 
such that the estimate
\begin{equation}
\|\hat{\bf F}^\sigma(\zeta)-{\mathbb I}\|\le M|\zeta|^{-1}\,,
\label{eq:origindecay}
\end{equation}
holds for all sufficiently large $|\zeta|$.  The solution is
universal in the sense that it is independent of $\hbar$.
\label{lemma:originexist}
\end{lemma}

\begin{proof} 
We begin by introducing an auxiliary Riemann-Hilbert problem.  Let
$\Sigma^\sigma_{{\bf L}}$ be the contour illustrated in
Figure~\ref{fig:originmodelL} for both choices of the parity $\sigma$.
\begin{figure}[h]
\begin{center}
\mbox{\psfig{file=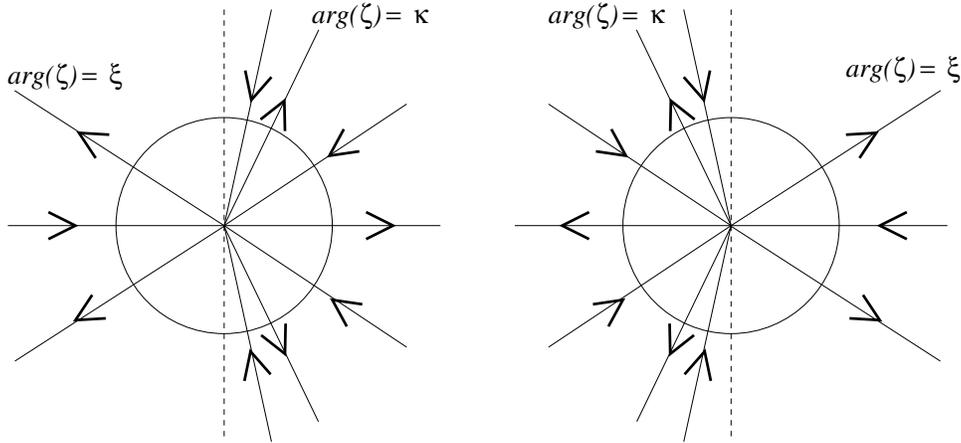,width=5 in}}
\end{center}
\caption{\em The contour $\Sigma^\sigma_{{\bf L}}$ for $\sigma=+1$ (left)
and $\sigma=-1$ (right).  In each case, the contour contains a circle
of radius $1/2$ as well as the real axis.  Outside the circle, the
rays extend to $\zeta=\infty$ and the orientation is as shown.  Inside
and on the circle, the orientation is determined so that the contour
forms the positively oriented boundary of a multiply connected open
region and the negatively oriented boundary of the closure of its
complement in ${\mathbb C}$.  }
\label{fig:originmodelL}
\end{figure}
We define a jump matrix for $\zeta\in\Sigma^\sigma_{{\bf
L}}\setminus\{\mbox{self-intersection points}\}$ as follows.
\begin{equation}
{\bf v}_{\bf L}^\sigma(\zeta):=\left\{\begin{array}{ll}
{\mathbb I}\,,&\zeta\in{\mathbb R}\mbox{ or } |\zeta|=1/2\,,\\\\
{\bf v}_{\hat{\bf F}}^\sigma(\zeta)\,,&
\arg(\zeta)=\kappa \mbox{ and } |\zeta|>1/2\,,\\\\
{\bf v}_{\hat{\bf F}}^\sigma(\zeta)^{-1}\,,&
\arg(\zeta)\neq\kappa \mbox{ and } |\zeta|>1/2\mbox{ and }\Im(\zeta)>0\,,\\\\
{\bf v}_{\hat{\bf F}}^\sigma(\zeta)^{-1}\,,&
\arg(\zeta)=\kappa\mbox{ and }|\zeta|<1/2\,,\\\\
{\bf v}_{\hat{\bf F}}^\sigma(\zeta)\,,&
\arg(\zeta)\neq \kappa\mbox{ and } |\zeta|<1/2\mbox{ and }\Im(\zeta)>0\,,\\\\
\left[\sigma_2{\bf v}_{\bf L}^\sigma(\zeta^*)^*\sigma_2\right]^{-1}\,,&
\Im(\zeta)<0\,.
\end{array}
\right.
\end{equation}

\begin{rhp}[Reoriented local model for the origin]
\index{Riemann-Hilbert problem!reoriented local model for the origin}
Find a matrix ${\bf L}^\sigma(\zeta)$ satisfying
\begin{enumerate}
\item
{\bf Analyticity:} ${\bf L}^\sigma(\zeta)$ is analytic for $\zeta\in
{\mathbb C}\setminus \Sigma_{\bf L}^\sigma$.
\item
{\bf Boundary behavior:} ${\bf L}^\sigma(\zeta)$ assumes continuous
boundary values from each connected component of ${\mathbb
C}\setminus\Sigma_{\bf L}^\sigma$ that are continuous, including
corner points corresponding to self-intersections.
\item
{\bf Jump condition:} On the oriented contour $\Sigma_{\bf
L}^\sigma$ minus the origin, the boundary values satisfy
\begin{equation}
{\bf L}^\sigma_+(\zeta)=
{\bf L}^\sigma_-(\zeta){\bf v}_{\bf L}^\sigma(\zeta)\,.
\end{equation}
\item
{\bf Normalization:} ${\bf L}^\sigma(\zeta)$ is normalized at infinity:
\begin{equation}
{\bf L}^\sigma(\zeta)\rightarrow{\mathbb I}\mbox{ as }
\zeta\rightarrow\infty\,.
\end{equation}
\end{enumerate}
\label{rhp:Lsigma}
\end{rhp}

The Riemann-Hilbert Problem~\ref{rhp:Lsigma} differs from the Riemann-Hilbert
Problem~\ref{rhp:Fhat} of interest only in the orientation of the contour
and the introduction of some contour components supporting identity jump
matrices.  Therefore, the solutions of these two problems are in one-to-one
correspondence:  $\hat{\bf F}^\sigma(\zeta)\equiv{\bf L}^\sigma(\zeta)$.
The re-orientation of the contour and introduction of the real axis and the
circle are simply to rewrite the problem in precisely the form to which
the general results from the appendix can be applied. 

We now proceed to apply the H\"older theory developed in the appendix.
First observe that on smooth component of $\Sigma^\sigma_{{\bf L}}$,
and for each $\nu<1$, the jump matrix ${\bf v}^\sigma_{{\bf
L}}(\zeta)$ is H\"older continuous with exponent $\nu$.  Indeed, in
the interior of each component, the jump matrix is analytic, and it is
easy to see that the only obstruction to arbitrary smoothness is in
the limiting behavior at the origin.  Here, the term that determines
the smoothness is the factor $(-i\zeta)^{-2i\zeta}$ in
$h^\sigma(\zeta)$.  But at $\zeta=0$, this term is in all H\"older
classes with exponents $\nu$ {\em strictly} less than one.  Next, note
that on each ray of $\Sigma^\sigma_{{\bf L}}$, the jump matrix decays
to the identity matrix as $\zeta\rightarrow\infty$ at least as fast as
$\bo(|\zeta|^{-1})$.  Indeed, from the asymptotic formulae
(\ref{eq:decayofh}), we see that for $\arg(\zeta)=\kappa$, the decay
to the identity is $\bo(|\zeta|^{-1})$.  On the two rays on either
side of $\arg(\zeta)=\kappa$ in the same quadrant, the decay of the
jump matrix to the identity is exponential for large $|\zeta|$ since
$v\zeta$ is real and negative for $\arg(\zeta)=\kappa$.  For
$\arg(\zeta)=\xi$, one sees from (\ref{eq:decayofh}) that the jump
matrix decays exponentially to the identity like $\bo(\exp(u\zeta))$.
The symmetry that determines the jump matrix in the lower half-plane
in terms of that in the upper half-plane ensures that similar decay
properties hold in the lower half-plane, and of course on the real
axis the jump matrix is exactly the identity.

Next, we observe that the jump matrices are consistent at the origin,
in the following sense.  If we number the rays in counter-clockwise order
starting with the positive real axis as $\Sigma^{(1)},\dots,\Sigma^{(10)}$,
and define ${\bf v}^{(k)}:=\lim_{\zeta\rightarrow 0, \zeta\in\Sigma^{(k)}}
{\bf v}^\sigma_{\hat{\bf F}}(\zeta)$, then the cyclic relation
\begin{equation}
{\bf v}^{(1)}{\bf v}^{(2)-1}{\bf v}^{(3)}{\bf v}^{(4)-1}\dots
{\bf v}^{(9)}{\bf v}^{(10)-1}={\mathbb I}
\end{equation}
holds for all values of the parameters $u$, $v$, $\kappa$, $\xi$, and
$\sigma$.  It is easy to see that the same property holds at each
intersection point $\zeta_0$ of the circle with a ray ({\em i.e.}
$|\zeta_0|=1/2$), since by definition,
\begin{equation}
\lim_{\epsilon\downarrow 0}{\bf v}^\sigma_{{\bf L}}((1+\epsilon)\zeta_0)\cdot
{\mathbb I}^{-1}
{\bf v}^\sigma_{{\bf L}}((1-\epsilon)\zeta_0){\mathbb I}^{-1}={\mathbb I}\,.
\end{equation}

Finally, we observe that for all $\zeta\in\Sigma^\sigma_{{\bf L}}$ with
$\Im(\zeta)\neq 0$, the relation ${\bf v}^\sigma_{{\bf L}}(\zeta^*)=
\left[\sigma_2{\bf v}^\sigma_{{\bf L}}(\zeta)^*\sigma_2\right]^{-1}$
implies that
\begin{equation}
{\bf v}^\sigma_{{\bf L}}(\zeta^*)=
{\bf v}^\sigma_{{\bf L}}(\zeta)^\dagger\,.
\end{equation}
This is not a general fact, but a consequence of the special structure
of the jump matrices for this Riemann-Hilbert problem.  Also, since
the jump matrix is the identity on the real axis, ${\bf v}^\sigma_{{\bf
L}}(\zeta)+ {\bf v}^\sigma_{{\bf L}}(\zeta)^\dagger$ is
strictly positive definite for real $\zeta$.

These facts allow us to apply Theorem~\ref{theorem:LocalSolvability}
proved in the appendix to deduce the existence of a matrix ${\bf
L}^\sigma(\zeta)$ that is analytic in ${\mathbb C}\setminus
\Sigma^\sigma_{{\bf L}}$, that for each $\mu<\nu$ takes on boundary
values on $\Sigma^\sigma_{{\bf L}}$ that are uniformly bounded and
H\"older continuous with exponent $\mu$, that for each $\mu<\nu$
satisfies ${\bf L}^\sigma(\zeta)-{\mathbb I}=\bo(|\zeta|^{-\mu})$ as
$\zeta\rightarrow\infty$, and whose boundary values satisfy the jump
relation ${\bf L}^\sigma_+(\zeta)={\bf L}^\sigma_-(\zeta){\bf v}^\sigma_{{\bf
L}}(\zeta)$.  Note that the meaning of the subscripts ``$+$''
and ``$-$'' in regard to the boundary values of ${\bf
L}^\sigma(\zeta)$ refer to the contour $\Sigma^\sigma_{{\bf L}}$
oriented as shown in Figure~\ref{fig:originmodelL}.  Since the
boundary values are uniformly continuous and since ${\bf v}^\sigma_{{\bf
L}}(\zeta)={\mathbb I}$ for all real $\zeta\neq 0$ and for all
$\zeta$ on the circle of radius $1/2$ (except at the self-intersection
points where the jump matrix is not defined), the matrix function
${\bf L}^\sigma(\zeta)$ is in fact analytic at these points.

The function defined by $\hat{\bf F}^\sigma(\zeta):={\bf
L}^\sigma(\zeta)$ is easily seen to be the unique solution of the
Riemann-Hilbert Problem~\ref{rhp:Fhat}, since it is analytic in
${\mathbb C}\setminus
\Sigma^\sigma_{\hat{\bf F}}$, and since it has H\"older continuous and
uniformly bounded boundary values that 
satisfy $\hat{\bf F}_+^\sigma(\zeta)=
\hat{\bf F}_-^\sigma(\zeta){\bf v}^\sigma_{\hat{\bf F}}(\zeta)$. Note that here
the subscripts ``$+$'' and ``$-$'' of the boundary values refer to the
orientation of $\Sigma^\sigma_{\hat{\bf F}}$ as shown in
Figure~\ref{fig:originmodel}.  

It remains only to obtain the decay estimate (\ref{eq:origindecay}).
For this, we return to the auxiliary problem for ${\bf
L}^\sigma(\zeta)$, and we compute the moments ({\em cf.} formula
(\ref{eq:momentdefine}) in the appendix) of the jump matrix ${\bf
v}^\sigma_{\bf L}(\zeta)$.  First, observe that as $\zeta$ tends to
infinity along any ray of $\Sigma^\sigma_{\bf L}$ except those
satisfying $\arg(\zeta)=\pm\kappa$,
\begin{equation}
{\bf v}^\sigma_{\bf L}(\zeta)={\mathbb I} + \mbox{ exponentially small}\,.
\end{equation}
Now when $\sigma=+1$ we can use (\ref{eq:decayofh}) to obtain 
\begin{equation}
{\bf v}^+_{\bf L}(\zeta)=\left\{\begin{array}{ll}
\displaystyle
{\mathbb I} - \frac{1}{12i\zeta}\sigma_3 + \bo(|\zeta|^{-2})\,,
&\hspace{0.2 in}\zeta\rightarrow\infty \mbox{ with }\arg(\zeta)=\kappa\,,\\\\
\displaystyle{\mathbb I} + \frac{1}{12i\zeta}\sigma_3 + \bo(|\zeta|^{-2})\,,
&\hspace{0.2 in}\zeta\rightarrow\infty \mbox{ with }\arg(\zeta)=-\kappa\,.
\end{array}\right.
\end{equation}
On the other hand, when $\sigma=-1$, (\ref{eq:decayofh}) implies that
\begin{equation}
{\bf v}^-_{\bf L}(\zeta)=\left\{\begin{array}{ll}
\displaystyle{\mathbb I}+\frac{1}{12i\zeta}\sigma_3 + \bo(|\zeta|^{-2})\,,
&\hspace{0.2 in}\zeta\rightarrow\infty \mbox{ with }\arg(\zeta)=\kappa\,,\\\\
\displaystyle{\mathbb I}-\frac{1}{12i\zeta}\sigma_3 + \bo(|\zeta|^{-2})\,,
&\hspace{0.2 in}\zeta\rightarrow\infty \mbox{ with }\arg(\zeta)=-\kappa\,.
\end{array}\right.
\end{equation}
Therefore in both cases, the higher-order compatibility condition
(\ref{eq:cancellation}) holds, and since the jump matrix on each ray
is uniformly analytic in a sufficiently thin parallel strip
surrounding the ray, Theorem~\ref{theorem:decay} gives a uniform decay
rate for $\|{\bf L}^\sigma(\zeta)-{\mathbb I}\|$ of $\bo(|\zeta|^{-1})$.
This implies the decay estimate (\ref{eq:origindecay}) and completes
the proof.
\end{proof}

\begin{remark}
Note that the construction of $\hat{\bf F}^\sigma(\zeta)$ presumes
that the angle $\kappa$ is not equal to $\pi/2$.  Ultimately this is
because the asymptotic approximation of the discrepancy between the
discrete sum and the integral given in
Chapter~\ref{sec:scatteringdata} is not uniformly valid in any sector
containing $\arg(\zeta)=\pi/2$ ({\em cf.} Theorem
\ref{theorem:inner}).
\end{remark}

\begin{figure}[h]
\begin{center}
\mbox{\psfig{file=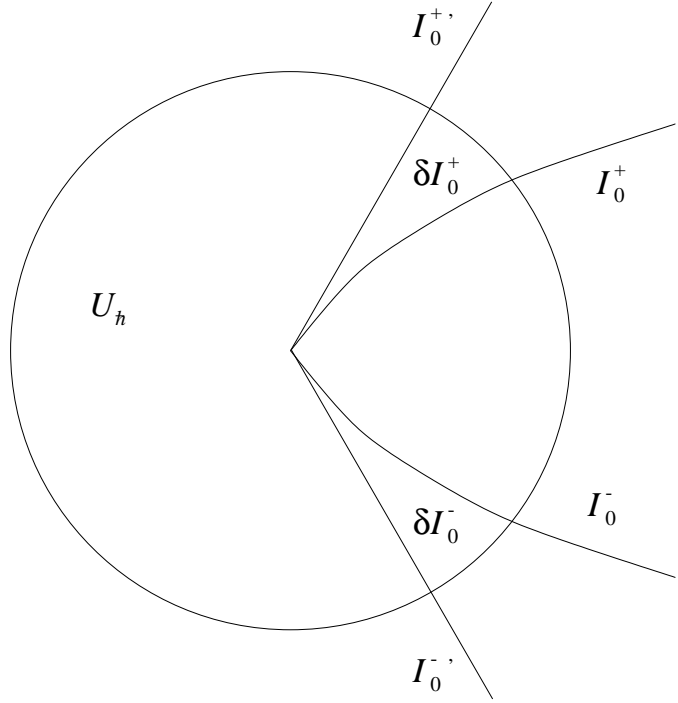,width=3.5 in}}
\end{center}
\caption{\em The regions $\delta I_0^+$ and $\delta I_0^-$ as defined for
sufficiently small $\hbar$.}
\label{fig:deltaIzero}
\end{figure}

Now, we use the matrix $\hat{\bf F}^\sigma(\zeta)$ to build an
approximation $\hat{\bf N}_{\rm origin}^\sigma(\lambda)$ of the matrix ${\bf
N}^\sigma(\lambda)$ for $\lambda\in U_\hbar$.  Since $\hat{\bf
F}^\sigma(\zeta)$ has jumps on the rays $\arg(\zeta)=\pm\kappa$, which
do not quite coincide with the contours $I_0^+$ and $I_0^-$ in the
$\lambda$-plane, we want to make one final deformation.  For
sufficiently small $\hbar$, the tangent $I_0^{+\prime}$ meets the
contour $I_0^+$ within $U_\hbar$ only at the origin.  Denote the
wedge-shaped subset of $U_\hbar$ between these two curves by $\delta
I_0^+$.  See Figure~\ref{fig:deltaIzero}.
The jump matrix ${\bf v}^\sigma_{\hat{\bf
F}}(-i\rho^0(0)\lambda/\hbar_N)$ for $\arg(\lambda)=\kappa$ has an
analytic continuation to $\lambda\in
\delta I_0^+$ for $\hbar$ sufficiently small, which we denote
by ${\bf u}^\sigma(\zeta)$.
If $I_0^+$ lies to the
right of the ray $\arg(\lambda)=\kappa$ in $U_\hbar$ for $\hbar$ small
enough, then we set
\begin{equation}
\hat{\bf G}^\sigma(\lambda):=\hat{\bf F}^\sigma(-i\rho^0(0)\lambda/\hbar_N)
\cdot
\left\{\begin{array}{ll}
{\bf u}^\sigma(-i\rho^0(0)\lambda/\hbar_N)\,,&\hspace{0.2
in}\lambda\in\delta I_0^+\,,\\
\sigma_2{\bf u}^\sigma(-i\rho^0(0)\lambda^*/\hbar_N)^*\sigma_2\,,
&\hspace{0.2 in}
\lambda\in\delta I_0^-\,,\\
{\mathbb I}\,, &\hspace{0.2 in}\lambda\in U_\hbar\setminus (\delta
I_0^+
\cup\delta I_0^-)\,,
\end{array}\right.
\end{equation}
and if $I_0^+$ lies to the left of the ray $\arg(\lambda)=\kappa$ in
$U_\hbar$ for $\hbar$ small enough, then we set
\begin{equation}
\hat{\bf G}^\sigma(\lambda):=\hat{\bf F}^\sigma(-i\rho^0(0)\lambda/\hbar_N)
\cdot
\left\{\begin{array}{ll}
{\bf u}^\sigma (-i\rho^0(0)\lambda/\hbar_N)^{-1}\,,&\hspace{0.2
in}\lambda\in\delta I_0^+\,,\\
\left[\sigma_2{\bf u}^\sigma
(-i\rho^0(0)\lambda^*/\hbar_N)^*\sigma_2\right]^{-1}\,,&\hspace{0.2 in}
\lambda\in\delta I_0^-\,,\\
{\mathbb I}\,, &\hspace{0.2 in}\lambda\in U_\hbar\setminus (\delta I_0^+
\cup\delta I_0^-)\,.
\end{array}\right.
\end{equation}
Note that the constant $-i\rho^0(0)$ is always manifestly real
according to (\ref{eq:WKBdensity}).

We now record several consequences of this definition.
\begin{lemma}
There exists some $M>0$ such that for all $\lambda\in U_\hbar$ and all
sufficiently small $\hbar$,
\begin{equation}
\|\hat{\bf G}^\sigma(\lambda)
\|\le M\,.
\end{equation}
\label{lemma:GhatBound}
\end{lemma}

\begin{proof}
It follows from Lemma~\ref{lemma:originexist} that the factor
$\hat{\bf F}^\sigma(\lambda/\hbar)$ is uniformly bounded for {\em all}
$\lambda$.  Next, we study the behavior of ${\bf
u}^\sigma(\lambda/\hbar)$ in $\delta I_0^+$.  Along with its inverse,
this matrix is bounded for $\arg(\lambda)=\kappa$.  To deform into
$\delta I_0^+$, we need only consider the behavior of the exponential
factors $\exp(\pm iv\zeta)$, since $h^\sigma(\zeta)$ is uniformly
bounded.  Now, since $I_0^+$ is smooth and tangent to the ray
$\arg(\lambda)=\kappa$, it will be the case for sufficiently small
$\hbar$ that
$\Im(-i\rho^0(0)v\lambda/\hbar_N)=\bo(\hbar_N^{2\delta-1})$ uniformly
for $\lambda\in
\delta I_0^+$ because the radius of $U_\hbar$ is $\hbar_N^\delta$, and
$\Im(-i\rho^0(0)v\lambda/\hbar_N)=0$ for $\arg(\lambda)=\kappa$.  Therefore,
uniformly for $\lambda\in \delta I_0^+$ we have $\exp(\pm
\rho^0(0)v\lambda/\hbar_N)=(1+\bo(\hbar_N^{2\delta-1}))\exp(\pm i\Re(-i\rho^0(0)v\lambda/\hbar_N))$.
Similar arguments hold for $\lambda\in \delta I_0^-$, and the proof is
complete since we always have $\delta>1/2$.
\end{proof}

\begin{lemma}
The matrix $\hat{\bf G}^\sigma(\lambda)$ has the same domain of
analyticity for $\lambda\in U_\hbar$ as ${\bf F}^\sigma(\lambda)$,
bounded by the contours $I_0^+$, $\Gamma_{G/2+1}^+$, $C^+_{0\pm}$, and
the corresponding contours in the lower half-plane as illustrated in
Figure~\ref{fig:zeroN}.  Define jump matrices for $\lambda\in
U_\hbar$ relative to the oriented contour of Figure~\ref{fig:zeroN} by
${\bf F}^\sigma_+(\lambda)={\bf F}^\sigma_-(\lambda){\bf v}^\sigma_{{\bf
F}}(\lambda)$ and $\hat{\bf G}^\sigma_+(\lambda)=\hat{\bf G}^\sigma_-
(\lambda){\bf v}^\sigma_{\hat{\bf G}}(\lambda)$.
Then, there exists some $M>0$ such that for all $\lambda\in U_\hbar$ and
all sufficiently small $\hbar$,
\begin{equation}
\|{\bf v}^\sigma_{{\bf F}}(\lambda){\bf v}^\sigma_{\hat{\bf G}}(\lambda)^{-1}
-{\mathbb I}\|\le M\hbar_N^{\min(1/3,2\delta-1)}\,.
\label{eq:FGhatEstimate}
\end{equation}
Recall that the neighborhood $U_\hbar$ has radius $\hbar_N^\delta$ for
$1/2<\delta<1$.
\label{lemma:FGhatEstimate}
\end{lemma}

\begin{proof}
From the definition of $\hat{\bf G}^\sigma(\lambda)$, we have
$\hat{\bf G}^\sigma_+(\lambda)=
\hat{\bf G}^\sigma_-(\lambda)$ for $\arg(\lambda)=\kappa$.
The continuity of the boundary values then implies that $\hat{\bf
G}^\sigma(\lambda)$ is analytic for $\arg(\lambda)=\kappa$ in
$U_\hbar$.  This latter statement assumes that $I_0^+$ does not agree
identically with its tangent line, in which case the regions $\delta
I_0^\pm$ would be empty and we would have $\hat{\bf
G}^\sigma(\lambda)\equiv \hat{\bf
F}^\sigma(-i\rho^0(0)\lambda/\hbar_N)$.  The remaining contours of
nonanalyticity for ${\bf F}^\sigma(\lambda)$ have all been taken to be
straight line segments within the fixed disk neighborhood $U$, and
therefore agree with the corresponding contours for $\hat{\bf
F}^\sigma(-i\rho^0(0)\lambda/\hbar_N)$.  The statement carries over to
$\hat{\bf G}^\sigma(\lambda)$ because for sufficiently small $\hbar$
the regions $\delta I_0^\pm$ where $\hat{\bf G}^\sigma(\lambda)$
differs from $\hat{\bf F}^\sigma(-i\rho^0(0)\lambda/\hbar_N)$ will
only meet these contours at the origin.  This proves that the domains
of analyticity for ${\bf F}^\sigma(\lambda)$ and $\hat{\bf
G}^\sigma(\lambda)$ agree within $U_\hbar$.

The estimate (\ref{eq:FGhatEstimate}) follows from our asymptotic
analysis of the jump matrix ${\bf v}^\sigma_{{\bf F}}(\lambda)$.  On
the straight line segments $C_{0\pm}^+$ and their images in the lower
half-plane, we have $\hat{\bf G}^\sigma(\lambda)\equiv\hat{\bf
F}^\sigma (-i\rho^0(0)\lambda/\hbar_N)$, and therefore the simple
Taylor approximations (\ref{eq:Taylor1}) and (\ref{eq:Taylor2}) give
\begin{equation}
\|{\bf v}^\sigma_{{\bf F}}(\lambda){\bf v}^\sigma_{\hat{\bf G}}(\lambda)^{-1}
-{\mathbb I}\|=
\|{\bf v}^\sigma_{{\bf F}}(\lambda){\bf v}^\sigma_{\hat{\bf F}}
(-i\rho^0(0)\lambda/\hbar_N)^{-1} -{\mathbb
I}\|=\exp(-|\Im(-i\rho^0(0)v\lambda)|/\hbar_N)\bo(\lambda^2/\hbar_N)\,,
\end{equation}
which is $\bo(\hbar^{2\delta-1})$ for all $\lambda\in U_\hbar$.  On
the two straight line segments $\Gamma_{G/2+1}^\pm$, we again find
that $\hat{\bf G}^\sigma(\lambda)\equiv\hat{\bf
F}^\sigma(-i\rho^0(0)\lambda/\hbar_N)$, but we have an additional
contribution from the asymptotic approximation of $d^\sigma(\lambda)$
to take into account.  The errors that dominate are determined by the
choice of $\delta$ since for $\sigma=+1$ and
$\lambda\in\Gamma_{G/2+1}^+\cap U_\hbar$, we have (using
(\ref{eq:fplusgamma}), (\ref{eq:oneminusdasymp}), (\ref{eq:Taylor1}),
and (\ref{eq:fhatplusjump}))
\begin{equation}
\|{\bf v}^\sigma_{{\bf F}}(\lambda){\bf v}^\sigma_{\hat{\bf G}}(\lambda)^{-1}
-{\mathbb I}\|=(1-h^+(-i\rho^0(0)\lambda/\hbar_N))\exp(-i\rho^0(0)(u-2\pi)\lambda/\hbar_N)(1+\bo(\hbar_N^{1/3})+\bo(\hbar_N^{2\delta-1}))\,,
\end{equation}
with the first error term coming from Theorem~\ref{theorem:inner} and
the second error term coming from the Taylor approximations
(\ref{eq:Taylor1}) and (\ref{eq:Taylor2}).  Similarly, for $\sigma=-1$
and $\lambda\in\Gamma_{G/2+1}^+\cap U_\hbar$, we have
\begin{equation}
\|{\bf v}^\sigma_{{\bf F}}(\lambda){\bf v}^\sigma_{\hat{\bf G}}(\lambda)^{-1}
-{\mathbb I}\|=(1-h^-(-i\rho^0(0)\lambda/\hbar_N))\exp(-i\rho^0(0)(u+2\pi)
\lambda/\hbar_N)(1+\bo(\hbar_N^{1/3})+\bo(\hbar_N^{2\delta-1}))\,.
\end{equation}
That both of these estimates are uniformly small in $U_\hbar$ now
follows from the boundedness and asymptotic properties of
$h^\pm(\zeta)$.  By symmetry of the definition of the jump matrices,
both of these estimates also hold on $\Gamma_{G/2+1}^-\cap U_\hbar$.
Finally, we note that to verify the estimate (\ref{eq:FGhatEstimate})
for $\lambda\in I_0^\pm\cap U_\hbar$ requires the same sort of
analysis as above, with the additional observation that the jump
matrix for $\hat{\bf G}^\sigma(\lambda)$, denoted by ${\bf
u}^\sigma(-i\rho^0(0)\lambda/\hbar_N)$ above, is uniformly bounded in
$U_\hbar$ according to the arguments in the proof of
Lemma~\ref{lemma:GhatBound}.
\end{proof}

Now, we give the local approximation \index{local approximation!near
the origin} for ${\bf N}^\sigma$ valid near the origin.  We define,
for $\lambda\in U_\hbar$,
\begin{equation}
\hat{\bf N}^\sigma_{\rm origin}(\lambda):={\bf C}^\sigma(\lambda)\hat{\bf G}^\sigma
(\lambda){\bf C}^\sigma(\lambda)^{-1}\hat{\bf N}^\sigma_{\rm
out}(\lambda)\,.
\label{eq:Nhatorigin}
\end{equation}
The essential properties of this approximation are the following.
\begin{lemma}
$\hat{\bf N}^\sigma_{\rm origin}(\lambda)$ is uniformly bounded independent
of $\hbar$ for $\lambda\in U_\hbar$.  Since $\det(\hat{\bf N}^\sigma_{\rm origin}(\lambda))=1$, this property is also held by the inverse matrix.
\label{lemma:boundednessnearzero}
\end{lemma}

\begin{proof}
The factors ${\bf C}^\sigma(\lambda)$, ${\bf C}^\sigma(\lambda)^{-1}$,
and $\hat{\bf N}^\sigma_{\rm out}(\lambda)$ are clearly all uniformly
bounded in $U_\hbar$, essentially since the solution $\tilde{\bf
O}(\lambda)$ of the outer model Riemann-Hilbert Problem
\ref{rhp:model} obtained in
\S\ref{sec:outersolve} has bounded boundary values at $\lambda=0$
on $I_0$.  To analyze $\hat{\bf G}^\sigma(\lambda)$, we observe that
the factor $\hat{\bf F}^\sigma(-i\rho^0(0)\lambda/\hbar_N)$ is
uniformly bounded for all $\zeta=-i\rho^0(0)\lambda/\hbar_N$, and in
particular for $\lambda\in U_\hbar$.  The remaining factor in the
definition of $\hat{\bf G}^\sigma(\lambda)$ is controlled as described
above, essentially because of the small size of the neighborhood
$U_\hbar$.  The fact that $\det(\hat{\bf N}^\sigma_{\rm
origin}(\lambda))=1$ follows from the analogous properties of the
individual factors, and that the determinant is unchanged by
conjugation by ${\bf C}^\sigma(\lambda)$.
\end{proof}

\begin{lemma}
Let the jump matrix for $\hat{\bf N}_{\rm origin}^\sigma(\lambda)$ for
$\lambda$ near the origin on the oriented contour shown in
Figure~\ref{fig:zeroN} be denoted ${\bf v}_{\rm
origin}^\sigma(\lambda)$.  Then, for some $M>0$,
\begin{equation}
\|{\bf v}^\sigma_{\bf N}(\lambda){\bf v}^\sigma_{\rm origin}(\lambda)^{-1}-
{\mathbb I}\|\le M\hbar_N^{\min(1/3,2\delta-1)}\,,
\label{eq:vNvoriginEstimate}
\end{equation}
for all $\lambda$ on the contour in $U_\hbar$, and for all
sufficiently small $\hbar$.
\label{lemma:jumpsnearzero}
\end{lemma}

\begin{proof}
Using the fact that ${\bf C}^\sigma(\lambda)$ is analytic in $U_\hbar$,
we find
\begin{equation}
{\bf v}^\sigma_{\bf N}(\lambda){\bf v}^\sigma_{\rm origin}(\lambda)^{-1}=
{\bf C}^\sigma(\lambda)\hat{\bf G}^\sigma_-(\lambda)
{\bf v}^\sigma_{{\bf F}}(\lambda)
{\bf v}^\sigma_{\hat{\bf G}}(\lambda)^{-1}
\hat{\bf G}^\sigma_-(\lambda)^{-1}
{\bf C}^\sigma(\lambda)^{-1}\,.
\end{equation}
The estimate (\ref{eq:vNvoriginEstimate}) then follows from 
Lemma~\ref{lemma:FGhatEstimate}, Lemma~\ref{lemma:GhatBound},
and the uniform boundedness in $U_\hbar$ of ${\bf C}^\sigma(\lambda)$
and its inverse.
\end{proof}

\begin{lemma}
There exists a constant $M$ such that for all
$\lambda\in\partial U_\hbar$, and for all sufficiently small $\hbar$, 
\begin{equation}
\|\hat{\bf N}^\sigma_{\rm origin}(\lambda)\hat{\bf N}^\sigma_{\rm out}(
\lambda)^{-1}-{\mathbb I}\|\le M\hbar_N^{1-\delta}\,.
\label{eq:NinoutEstimate}
\end{equation}
\label{lemma:NinoutEstimate}
\end{lemma}

\begin{proof}
From the definition of the local approximation, we have for
$|\lambda|=\hbar_N^\delta$,
\begin{equation}
\begin{array}{rcl}
\|\hat{\bf N}^\sigma_{\rm origin}(\lambda)\hat{\bf N}^\sigma_{\rm out}(
\lambda)^{-1}-{\mathbb I}\|&=&\|{\bf C}^{\sigma}(\lambda)\cdot[
\hat{\bf G}^\sigma(\lambda)-{\mathbb I}]\cdot{\bf C}^\sigma(\lambda)^{-1}\|\\
&\le &\|{\bf C}^\sigma(\lambda)\|\cdot \|{\bf C}^\sigma(\lambda)^{-1}\|\cdot
\|\hat{\bf G}^\sigma(\lambda)-{\mathbb I}\|\\
&\le & M'\cdot\|\hat{\bf G}^\sigma(\lambda)-{\mathbb I}\|\,,
\end{array}
\label{eq:NinoutIntermediate}
\end{equation}
for some $M'>0$ where we have used the uniform boundedness of ${\bf
C}^\sigma(\lambda)$ and its inverse.  Now, for all $\lambda\in\partial
U_\hbar$ that are not on the boundary of $\delta I_0^\pm$, we have
$\hat{\bf G}^\sigma(\lambda)\equiv \hat{\bf
F}^\sigma(-i\rho^0(0)\lambda/\hbar_N)$, and then from the decay
property (\ref{eq:origindecay}) of $\hat{\bf F}^\sigma(\zeta)$
established in Lemma~\ref{lemma:originexist} the estimate
(\ref{eq:NinoutEstimate}) follows from (\ref{eq:NinoutIntermediate}).
For the other values of $\lambda\in\partial U_\hbar$ where $
\hat{\bf G}^\sigma(\lambda)\not\equiv\hat{\bf F}^\sigma(-i\rho^0(0)
\lambda/\hbar_N)$,
we have to take into account an additional factor.  But for
$|\lambda|=\hbar_N^\delta$ on the boundary of $\delta I_0^\pm$, it
follows from the asymptotic behavior (\ref{eq:decayofh})
of $h^\sigma(\zeta)$ and the
geometry of the region $\delta I_0^\pm$ that this factor is uniformly
${\mathbb I}+\bo(\hbar_N^{1-\delta})$, and again
(\ref{eq:NinoutIntermediate}) yields the required estimate
(\ref{eq:NinoutEstimate}).
\end{proof}

\section[Estimating the Error]{Estimating the error.}
\label{sec:error}
We have now completed our analysis of the Riemann-Hilbert problem for
the matrix ${\bf N}^\sigma(\lambda)$ corresponding to arbitrary
soliton ensembles connected with initial data $\psi_0(x)=A(x)$ via
formal WKB theory.  We are now in a position to {\em prove} that the
various approximations we have constructed in different parts of the
complex $\lambda$-plane are indeed valid.  We begin by patching
together the various approximations to obtain a global approximation
that we will establish is indeed uniformly valid.  Such a global
approximation is called a {\em parametrix}\index{parametrix}.

\subsection{Defining the parametrix.}
\begin{figure}[h]
\begin{center}
\mbox{\psfig{file=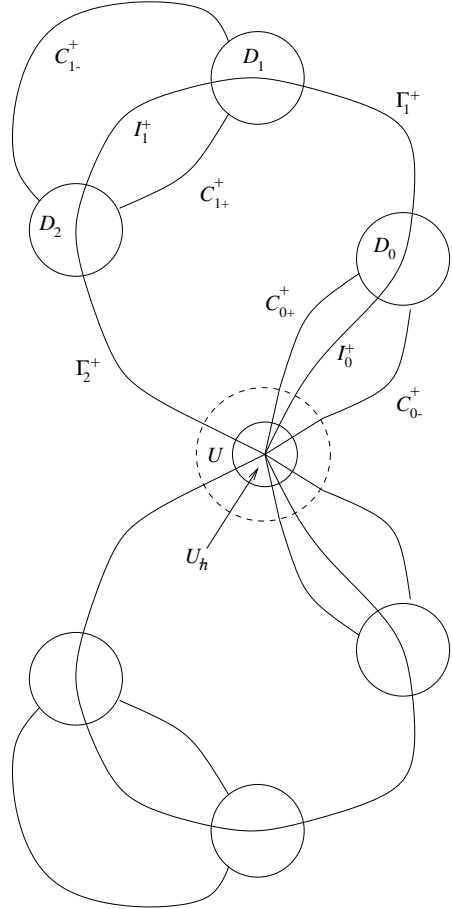,width=2.3 in}}
\end{center}
\caption{\em The domain of analyticity of the 
parametrix $\hat{\bf N}^\sigma(\lambda)$ is ${\mathbb
C}\setminus\Sigma_0$.  The dashed circle is the boundary of the disk $U$
and is not considered to be part of $\Sigma_0$.  The contour $\Sigma_0$ is
oriented as follows.  The loop components $C$ and $C^*$ have
orientation $\sigma$, the lens boundaries $C_{k\pm}^+$ (respectively
$C_{k\pm}^-$) are all oriented 
parallel to $C_\sigma$ (respectively $[C^*]_\sigma$), and all disk boundaries
are oriented in the clockwise direction.}
\label{fig:error}
\end{figure}

We define the global parametrix, a model for ${\bf N}^\sigma(\lambda)$ in
each part of the complex plane, as follows:
\begin{equation}
\hat{\bf N}^\sigma(\lambda):=\left\{\begin{array}{ll}
\hat{\bf N}^\sigma_{2k}(\lambda)\,,&\hspace{0.2 in}
\lambda\in D_{2k}\,,\hspace{0.2 in}
k=0,\dots,G/2\mbox{ ({\em cf.} (\ref{eq:Nhat2kI})--(\ref{eq:Nhat2kIV}))}\,,\\\\
\sigma_2\hat{\bf N}^\sigma_{2k}(\lambda^*)^*\sigma_2\,, &
\hspace{0.2 in}\lambda\in D_{2k}^*\,,\hspace{0.2 in} k=0,\dots,G/2\,,\\\\
\hat{\bf N}^\sigma_{2k-1}(\lambda)\,, &\hspace{0.2 in}
\lambda\in D_{2k-1}\,,\hspace{0.2 in}
k=1,\dots,G/2+1\mbox{ ({\em cf.} (\ref{eq:Nhat2km1I})--(\ref{eq:Nhat2km1IV}))}\,,\\\\
\sigma_2\hat{\bf N}^\sigma_{2k-1}(\lambda^*)^*\sigma_2\,,&\hspace{0.2 in}
\lambda\in D_{2k-1}^*\,, \hspace{0.2 in}
k=1,\dots,G/2+1\,,\\\\
\hat{\bf N}^\sigma_{\rm origin}(\lambda)\,,&\hspace{0.2 in}
\lambda\in U_\hbar
\mbox{ ({\em cf.} (\ref{eq:Nhatorigin}))}\,,\\\\
\hat{\bf N}^\sigma_{\rm out}(\lambda)\,,&\hspace{0.2 in}
\mbox{otherwise ({\em cf.} (\ref{eq:outerparametrixdef}))}\,.
\end{array}\right.
\label{eq:parametrix}
\end{equation}
The parametrix $\hat{\bf N}^\sigma(\lambda)$ so-defined is analytic
except on a complicated union of contours that we refer to simply as
$\Sigma_0$.  The contour $\Sigma_0$ is illustrated qualitatively in
Figure~\ref{fig:error} for $\sigma=+1$ and $G=2$.  See the caption
for details about the orientation.
By contrast, we recall that the matrix ${\bf N}^\sigma(\lambda)$ is
analytic for all $\lambda\in {\mathbb C}\setminus C\cup C^*$.  

Along with the contour $\Sigma_0$, we consider a ``completed'' contour
\index{completed contour} $\Sigma$ defined as follows.  
As a set of points, $\Sigma$ is the
union of $\Sigma_0$ with the restrictions of the lens boundaries
$C_{k\pm}^+$ and $C_{k\pm}^-$ to the interior of the disks $D_k$ and
$D_k^*$.  The contour $\Sigma$ now can easily be seen to divide the
plane into two disjoint regions $\Omega^+$ and $\Omega^-$, and
consequently the orientation of each smooth arc of $\Sigma$ may be
chosen so that $\Sigma$ forms the positively oriented boundary of
$\Omega^+$ and therefore also the negatively oriented boundary of
$\Omega^-$.  The orientation of corresponding arcs of $\Sigma$ and
$\Sigma_0$ will not necessarily agree.  The parametrix $\hat{\bf
N}^\sigma(\lambda)$ may also be considered to be analytic for
$\lambda\in{\mathbb C}\setminus \Sigma$, with a jump matrix on
$\Sigma\setminus\Sigma_0$ equal to the identity matrix.

\subsection{Asymptotic validity of the parametrix.}
The {\em error} \index{error} in replacing ${\bf N}^\sigma(\lambda)$
by its parametrix is the matrix defined for $\lambda\in{\mathbb
C}\setminus\Sigma$ by
\begin{equation}
{\bf E}(\lambda):={\bf N}^\sigma(\lambda)\hat{\bf
N}^\sigma(\lambda)^{-1}\,.
\label{eq:errordef}
\end{equation}
This matrix depends on $\sigma$, as does the contour $\Sigma$, but we
suppress this dependence in this section.  Note that for each
$\hbar_N$, ${\bf E}(\lambda)$ assumes boundary values on $\Sigma$ that
are uniformly continuous on the boundary of each connected component
of ${\mathbb C}\setminus\Sigma$, since this is a property of both
${\bf N}^\sigma(\lambda)$ and of all approximations making up the
sectionally holomorphic definition (\ref{eq:parametrix}) of the
parametrix.  Therefore, we may define a jump matrix relative to the
oriented contour $\Sigma$ by setting
\begin{equation}
{\bf v}(\lambda):={\bf E}_-(\lambda)^{-1}{\bf
E}_+(\lambda)\,,
\hspace{0.2 in}\lambda\in\Sigma\,.
\label{eq:errorjump}
\end{equation}
Under the conditions we assumed in the analysis of \S\ref{sec:outersolve} and
\S\ref{sec:inner}, we can easily establish the following.
\begin{lemma}
The matrix ${\bf v}(\lambda)$ defined by (\ref{eq:errorjump}) is uniformly
bounded on $\Sigma$ and there exists an $M>0$ such that for
all sufficiently small $\hbar_N$ we have the estimate
\begin{equation}
\sup_{\lambda\in\Sigma}\|{\bf v}(\lambda)-{\mathbb I}\|\le M
\hbar_N^{1/3}\,.
\end{equation}
\label{lemma:uniformlysmall}
\end{lemma}

\begin{proof} 
We begin with the smaller oriented contour $\Sigma_0$ and consider its
smooth arc components one type at a time.  For all
$\lambda\in\Sigma_0$, let ${\bf v}_0(\lambda)$ denote the jump matrix
for ${\bf E}(\lambda)$ relative to the orientation of $\Sigma_0$
rather than that of $\Sigma$.  On each arc of $\Sigma_0$, we have
either ${\bf v}(\lambda)={\bf v}_0(\lambda)$ or ${\bf v}(\lambda)={\bf
v}_0(\lambda)^{-1}$.

First, consider the clockwise-oriented boundaries of the circular
disks $U_\hbar$, $D_0,\dots,D_G$, and $D_0^*,\dots,D_G^*$.  On these
boundaries, the jump matrix is simply
\begin{equation}
{\bf v}_0(\lambda)=
\hat{\bf N}^\sigma_-(\lambda){\bf v}^\sigma_{{\bf N}}(\lambda)
{\bf v}^\sigma_{\hat{\bf N}}(\lambda)^{-1}\hat{\bf N}^\sigma_-(\lambda)^{-1}=
\hat{\bf N}^\sigma_-(\lambda)\hat{\bf N}^\sigma_+(\lambda)^{-1}\,,
\end{equation}
since ${\bf N}^\sigma(\lambda)$ is analytic at these boundaries and thus
the corresponding jump matrix is the identity.  The plus boundary value refers
here to the exterior of the disk and therefore by Lemma~\ref{lemma:matcheven}
and Lemma~\ref{lemma:matchodd}, we obtain 
\begin{equation}
\|{\bf v}_0(\lambda)-{\mathbb I}\|\le M\hbar_N^{1/3}\,, \hspace{0.2 in}
\lambda\in \partial D_0\cup\dots\cup \partial D_G\,,
\label{eq:est1}
\end{equation}
with similar estimates holding for the clockwise-oriented boundaries of
$D_0^*,\dots,D_G^*$.  Likewise, by Lemma~\ref{lemma:NinoutEstimate},
\begin{equation}
\|{\bf v}_0(\lambda)-{\mathbb I}\|\le M\hbar_N^{1-\delta}\,, \hspace{0.2 in}
\lambda\in\partial U_\hbar\,.
\label{eq:est2}
\end{equation}

Next, we consider the parts of $\Sigma_0$ in the interiors of the
various circular disks.  First consider one of the fixed disks $D_j$.
Here we obtain
\begin{equation}
\begin{array}{rcl}
\|{\bf v}_0(\lambda)-{\mathbb I}\|&=&\|\hat{\bf N}_-^\sigma(\lambda)
{\bf v}^\sigma_{{\bf N}}(\lambda){\bf v}^\sigma_{\hat{\bf N}}(\lambda)^{-1}
\hat{\bf N}_-^\sigma(\lambda)^{-1}-{\mathbb I}\|\\\\
&\le & \|\hat{\bf N}_-^\sigma(\lambda)\|\cdot
\|\hat{\bf N}_-^\sigma(\lambda)^{-1}\|\cdot \|
{\bf v}^\sigma_{{\bf N}}(\lambda){\bf v}^\sigma_{\hat{\bf N}}(\lambda)^{-1}
-{\mathbb I}\|\\\\
&=&
\|\hat{\bf N}_-^\sigma(\lambda)\|\cdot
\|\hat{\bf N}_-^\sigma(\lambda)^{-1}\|\cdot \|
{\bf v}^\sigma_{{\bf N}}(\lambda){\bf v}^\sigma_{\tilde{\bf N}}(\lambda)^{-1}
-{\mathbb I}\|\\\\
&\le & M\hbar_N^{-2/3} 
\|
{\bf v}^\sigma_{{\bf N}}(\lambda){\bf v}^\sigma_{\tilde{\bf N}}(\lambda)^{-1}
-{\mathbb I}\|\,,
\end{array}
\end{equation}
where we have used Lemma~\ref{lemma:jumpsinD2k} and
Lemma~\ref{lemma:holobound} (or their counterparts for odd numbered
endpoints Lemma~\ref{lemma:jumpsinD2km1} and
Lemma~\ref{lemma:holoboundodd}).  Now, since each disk is bounded away
from the origin as $\hbar\downarrow 0$, it follows from
Theorem~\ref{theorem:outer} that
\begin{equation}
\|{\bf v}^\sigma_{{\bf N}}(\lambda){\bf v}^\sigma_{\tilde{\bf N}}(\lambda)^{-1}
-{\mathbb I}\|\le M\hbar_N\,, \hspace{0.2 in}
\lambda\in \Sigma_0\cap D_j\,.
\end{equation}
Consequently, we have 
\begin{equation}
\|{\bf v}_0(\lambda)
-{\mathbb I}\|\le M\hbar_N^{1/3}\,, \hspace{0.2 in}
\lambda\in \Sigma_0\cap D_j\,,
\label{eq:est3}
\end{equation}
with same estimate holding in $D_j^*$.  
Next, we examine the interior of the shrinking disk $U_\hbar$.
For $\Sigma_0\cap U_\hbar$, we find, applying first
Lemma~\ref{lemma:boundednessnearzero} and then
Lemma~\ref{lemma:jumpsnearzero} that
\begin{equation}
\begin{array}{rcl}
\|{\bf v}_0(\lambda)-{\mathbb I}\|&=&\|\hat{\bf N}_-^\sigma(\lambda)
{\bf v}^\sigma_{{\bf N}}(\lambda){\bf v}^\sigma_{\hat{\bf N}}(\lambda)^{-1}
\hat{\bf N}_-^\sigma(\lambda)^{-1}-{\mathbb I}\|\\\\
&\le &
\|\hat{\bf N}_-^\sigma(\lambda)\|\cdot
\|\hat{\bf N}_-^\sigma(\lambda)^{-1}\|\cdot \|
{\bf v}^\sigma_{{\bf N}}(\lambda){\bf v}^\sigma_{\hat{\bf
N}}(\lambda)^{-1} -{\mathbb I}\|\\\\ &\le & M^2 \| {\bf
v}^\sigma_{{\bf N}}(\lambda){\bf v}^\sigma_{\hat{\bf N}}(\lambda)^{-1}
-{\mathbb I}\|\\\\ &\le &M'\hbar_N^{\min(1/3,2\delta-1)}\,,\hspace{0.2
in}\lambda\in\Sigma_0\cap U_\hbar\,.
\end{array}
\label{eq:est4}
\end{equation}

It remains to examine the components of $\Sigma_0$ that lie strictly
outside the closures of all disks.  We call this exterior component
$\Sigma_0^{\rm out}$.  In all of $\Sigma_0^{\rm out}$ we have a uniform
bound on $\hat{\bf N}^\sigma(\lambda)=\hat{\bf N}^\sigma_{\rm
out}(\lambda)$, and in particular on the boundary values, that is
independent of $\hbar$.  Therefore, we have
\begin{equation}
\|{\bf v}_0(\lambda)-{\mathbb I}\|\le M\|{\bf v}^\sigma_{{\bf N}}(\lambda)
{\bf v}^\sigma_{\hat{\bf N}}(\lambda)^{-1}-{\mathbb I}\|\,,\hspace{0.2 in}
\lambda\in\Sigma_0^{\rm out}\,.
\end{equation}
For $\lambda$ in a band of $(C\cup C^*)\cap \Sigma_0^{\rm out}$, 
we have from
Lemma~\ref{lemma:outerjumps} that
\begin{equation}
\|{\bf v}_0(\lambda)-{\mathbb I}\|\le M\|{\bf v}^\sigma_{{\bf N}}(\lambda)
{\bf v}^\sigma_{\tilde{\bf N}}(\lambda)^{-1}-{\mathbb I}\|\,,\hspace{0.2 in}
\lambda\in I_k^\pm\cap \Sigma_0^{\rm out}\,.
\end{equation}
This quantity is $\bo(\hbar)$ for fixed $\lambda$, but becomes larger
near the outer boundary of the shrinking disk $U_\hbar$.  From
Theorem~\ref{theorem:outer}, we see that the error achieved at the
cost of the shrinking radius of $U_\hbar$ is the estimate
\begin{equation}
\|{\bf v}_0(\lambda)-{\mathbb I}\|\le M'\hbar_N^{1-\delta}\,,\hspace{0.2 in}
\lambda\in I_k^\pm\cap \Sigma_0^{\rm out}\,.
\label{eq:est5}
\end{equation}
For $\lambda$ in a gap of $(C\cup C^*)\cap\Sigma_0^{\rm out}$, 
we have from Lemma~\ref{lemma:outerjumps} that
\begin{equation}
\|{\bf v}_0(\lambda)-{\mathbb I}\|\le M\|{\bf v}^\sigma_{{\bf N}}(\lambda)
\exp(-iJ\theta^\sigma(\lambda)\sigma_3/\hbar_N)-{\mathbb I}\|\,,\hspace{0.2 in}
\lambda\in \Gamma_k^\pm\cap\Sigma_0^{\rm out}\,.
\end{equation}
The function $\theta^\sigma(\lambda)$ evaluates to a real constant in each
gap.  Using the definition of ${\bf v}_{\bf N}^\sigma(\lambda)$, we
find that
\begin{equation}
{\bf v}^\sigma_{\bf N}(\lambda)\exp(-iJ\theta^\sigma(\lambda)\sigma_3/\hbar_N)-
{\mathbb I}=e(\lambda)\cdot
\sigma_1^\frac{1-J}{2}\left[\begin{array}{cc}0 & 0 \\ 
1 & 0\end{array}\right]\sigma_1^\frac{1-J}{2}\,,
\end{equation}
where the scalar $e(\lambda)$ is defined by
\begin{equation}
e(\lambda):=i\exp(-i\theta^\sigma(\lambda)/\hbar_N)
\exp(\tilde{\phi}^\sigma(\lambda)/\hbar_N)
\exp((\phi^\sigma(\lambda)-\tilde{\phi}^\sigma(\lambda))/\hbar_N)\,.
\end{equation}
Now $\exp(-i\theta^\sigma(\lambda)/\hbar_N)$ is uniformly bounded, and
by Theorem~\ref{theorem:outer}, we find that for
$\lambda\in\Sigma_0^{\rm out}$ we have
$\exp((\phi^\sigma(\lambda)-\tilde{\phi}^\sigma(\lambda))/\hbar_N)=1+
\bo(\hbar_N^{1-\delta})$.
The term $\exp(\tilde{\phi}^\sigma(\lambda)/\hbar_N)$ is, however,
exponentially small with its maximum size $\bo(\exp(-M'\hbar_N^{\delta-1}))$
being attained on the outside boundary of $U_\hbar$.  Consequently, we
have the estimate
\begin{equation}
\|{\bf v}_0(\lambda)-{\mathbb I}\|\le M
\exp(-M'\hbar_N^{\delta-1})\,,\hspace{0.2 in}
\lambda\in \Gamma_k^\pm\cap\Sigma_0^{\rm out}\,,
\label{eq:est6}
\end{equation}
for some $M'>0$.  Similarly, on the boundaries of the ``lenses'' in
$\Sigma_0^{\rm out}$, the matrix ${\bf N}^\sigma(\lambda)$ is analytic,
so we have
\begin{equation}
\begin{array}{rcl}
\|{\bf v}_0(\lambda)-{\mathbb I}\|&\le & M\|
{\bf v}^\sigma_{\hat{\bf N}}(\lambda)^{-1}-{\mathbb I}\|\\\\
&=&
M\|
{\bf D}^\sigma_+(\lambda){\bf D}^\sigma_-(\lambda)^{-1}-
{\mathbb I}\|\,,\hspace{0.2 in}
\lambda\in \{\mbox{lens boundaries}\}\cap \Sigma_0^{\rm out}\,.
\end{array}
\end{equation}
Here, ${\bf D}^\sigma(\lambda)$ is the explicit ``lens transformation''
matrix which differs from the identity matrix only inside the lenses,
where it is defined by (\ref{eq:ldef1}) and (\ref{eq:ldef2}).  These
factors are exponentially small perturbations of the identity matrix
away from the endpoints and the origin.  It is again the proximity of
the origin at the outside boundary of $U_\hbar$ that dominates this
error; ultimately we obtain an estimate of the form
\begin{equation}
\|{\bf v}_0(\lambda)-{\mathbb I}\|\le M\exp(-M'\hbar_N^{\delta-1})
\,,\hspace{0.2 in}
\lambda\in \{\mbox{lens boundaries}\}\cap \Sigma_0^{\rm out}\,,
\label{eq:est7}
\end{equation}
for some $M'>0$.  

We now combine the estimates (\ref{eq:est1}), (\ref{eq:est2}),
(\ref{eq:est3}), (\ref{eq:est4}), (\ref{eq:est5}), (\ref{eq:est6}),
and (\ref{eq:est7}).  The overall bound is optimized by taking
$\delta=2/3$, which determines the radius of the neighborhood
$U_\hbar$ as $R=\hbar^{2/3}$.  With this choice, all bounds except for
the exponentially small contributions (\ref{eq:est6}) and
(\ref{eq:est7}) are $\bo(\hbar_N^{1/3})$.  Finally, we note that the
required estimate for the jump matrix ${\bf v}(\lambda)$ for $\lambda$
in the completed contour $\Sigma$ follows from the corresponding
result for $\Sigma_0$ along with the facts that $\det({\bf
v}_0(\lambda))=1$ for all $\lambda\in\Sigma_0$ and ${\bf
v}(\lambda)\equiv {\mathbb I}$ for all
$\lambda\in\Sigma\setminus\Sigma_0$.  This completes the proof.
\end{proof}

Using the matrix ${\bf v}(\lambda)$ and the contour $\Sigma$, we can pose
a final Riemann-Hilbert problem.
\begin{rhp}[Global Error Problem]
\index{Riemann-Hilbert problem!global error problem}
Find a matrix function ${\bf R}(\lambda)$ satisfying
\begin{enumerate}
\item
{\bf Analyticity:}  ${\bf R}(\lambda)$ is analytic for
$\lambda\in {\mathbb C}\setminus \Sigma$.
\item
{\bf Boundary behavior:}  ${\bf R}(\lambda)$ assumes continuous
boundary values on $\Sigma$ from each component of the complement, with
continuity also at self-intersection points.
\item
{\bf Jump condition:} On the oriented contour
$\Sigma\setminus\{\mbox{self-intersection points}\}$, the boundary
values satisfy
\begin{equation}
{\bf R}_+(\lambda)={\bf R}_-(\lambda){\bf v}(\lambda)\,.
\end{equation}
\item
{\bf Normalization:}  ${\bf R}(\lambda)$ is normalized at infinity:
\begin{equation}
{\bf R}(\lambda)\rightarrow {\mathbb I} \mbox{ as }\lambda
\rightarrow\infty\,.
\end{equation}
\end{enumerate}
\label{rhp:error}
\end{rhp}

On the one hand, we already ``have'' the solution to this
Riemann-Hilbert problem.
\begin{lemma}
The global error Riemann-Hilbert Problem \ref{rhp:error} has a unique
solution, namely
\begin{equation}
{\bf R}(\lambda)\equiv {\bf E}(\lambda)\,,
\end{equation}
where ${\bf E}(\lambda)$ is defined by (\ref{eq:errordef}).
\end{lemma}

\begin{proof}
The analyticity, boundary behavior, and normalization properties
follow directly from the definition (\ref{eq:errordef}).  The jump
condition is equivalent to the definition (\ref{eq:errorjump}) of the
matrix ${\bf v}(\lambda)$.  Uniqueness of the solution follows from
the continuity of the boundary values and Liouville's theorem.
\end{proof}

But on the other hand, we can consider constructing the solution ${\bf
R}(\lambda)$ of the global error Riemann-Hilbert Problem
\ref{rhp:error} directly, using {\em only} the available information about
the the jump matrix ${\bf v}(\lambda)$ contained in Lemma
\ref{lemma:uniformlysmall}.  Whatever we learn about ${\bf R}(\lambda)$
in this pursuit is then trivially a fact about the error matrix ${\bf
E}(\lambda)$.

\begin{figure}[h]
\begin{center}
\mbox{\psfig{file=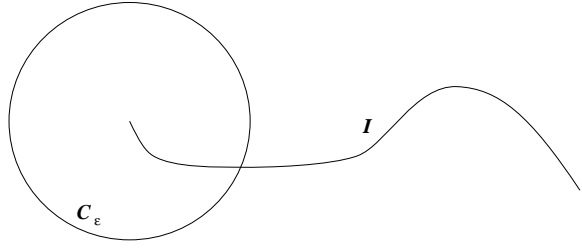,width=3 in}}
\end{center}
\caption{\em The contours of Lemma~\ref{lemma:apple}.}
\label{fig:apple}
\end{figure}

In order to construct ${\bf R}(\lambda)$ and obtain the desired
estimates, we need to establish a uniformity result concerning the
$\hbar$-dependence Cauchy integral operators \index{Cauchy integral
operators} on the contour $\Sigma$ entering through the shrinking
circle $\partial U_\hbar$ of radius $\hbar^\delta$.  The main result
we need is Lemma~\ref{lemma:CplusCminusBound}, but in order to prove
this we will first need the following technical lemma.
\begin{lemma}
Let $I(s)$, $0\le s\le s_{\rm max}<\infty$ be a $C^2$ curve in the
complex plane parametrized by arc length, and suppose that $I(0)=0$
and $I(s_{\rm max})\neq 0$.  Let $C_\epsilon$ denote the
clockwise-oriented circle centered at the origin with radius
$\epsilon$.  See Figure~\ref{fig:apple}.  Then, for ${\bf f}\in
L^2(C_\epsilon,|dz|)$ and $w\in I$, the Cauchy integral
\begin{equation}
(\op{C}^{C_\epsilon}{\bf f})(w):=\frac{1}{2\pi i}
\int_{C_\epsilon} (z-w)^{-1}{\bf f}(z)\,dz
\end{equation}
defines a bounded operator from $L^2(C_\epsilon,|dz|)$ to $L^2(I,ds)$
with a norm that is uniformly bounded above by a constant for all
sufficiently small $\epsilon$.  Similarly, for ${\bf f}\in L^2(I,ds)$
and $w\in C_\epsilon$, the Cauchy integral
\begin{equation}
(\op{C}^{I}{\bf f})(w):=\frac{1}{2\pi i}
\int_0^{s_{\rm max}} (I(s)-w)^{-1}{\bf f}(I(s))\frac{dI}{ds}(s)\,ds
\end{equation}
is a bounded operator from $L^2(I,ds)$ to $L^2(C_\epsilon,|dz|)$ with
a norm that is uniformly bounded above by a constant for all
$\epsilon$ sufficiently small.
\label{lemma:apple}
\end{lemma}

\begin{proof}  For $\epsilon$ sufficiently small there is a unique 
$s_\epsilon>0$ such that $|I(s_\epsilon)|=\epsilon$.  This value
satisfies $s_\epsilon/\epsilon\rightarrow 1$ as $\epsilon\downarrow
0$.  Let $\phi:=\arg(I'(0))$.  We divide each contour into two pieces
as follows.  Let $I^{\rm in}_\epsilon$ denote the contour parametrized
by $I(s)$ for $s\in [0,s_\epsilon]$ and $I^{\rm out}_\epsilon$ denote
the contour parametrized by $I(s)$ for $s\in [s_\epsilon,s_{\rm
max}]$.  Then, let $C_\epsilon^+$ denote the part of $C_\epsilon$ with
$\phi-\pi\le \arg(z)\le \arg(s_\epsilon)$ and $C_\epsilon^-$ the part
of $C_\epsilon$ with $\arg(s_\epsilon)\le \arg(z)\le \phi+\pi$.  Note
that $\arg(s_\epsilon)$ tends to $\phi$ as $\epsilon$ tends to zero,
so for small $\epsilon$ we are nearly dividing the circle in half
along the tangent line to $I$ at the origin.  These subdivisions of
the contours are illustrated in Figure~\ref{fig:apple_ii}.
\begin{figure}[h]
\begin{center}
\mbox{\psfig{file=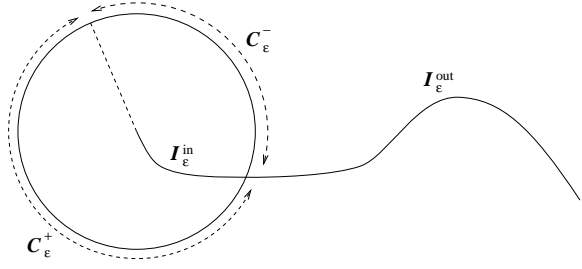,width=3 in}}
\end{center}
\caption{\em The subdivided contours.}
\label{fig:apple_ii}
\end{figure}

Let $\chi[\Sigma](z)$ denote the characteristic function of a contour
segment $\Sigma$; then for ${\bf f}\in L^2(C_\epsilon)$ and $w\in I$,
the Cauchy operator $\op{C}^{C_\epsilon}$ is decomposed as:
\begin{equation}
\begin{array}{rcl}
(\op{C}^{C_\epsilon}{\bf f})(w)&=&\chi[I_\epsilon^{\rm out}](w)\cdot
(\op{C}^{C_\epsilon}(\chi[C_\epsilon^+](\cdot){\bf f}))(w) +
\chi[I_\epsilon^{\rm out}](w)\cdot
(\op{C}^{C_\epsilon}(\chi[C_\epsilon^-](\cdot){\bf f}))(w)\\\\
&&\hspace{0.2 in} +\,\,
\chi[I_\epsilon^{\rm in}](w)\cdot
(\op{C}^{C_\epsilon}(\chi[C_\epsilon^+](\cdot){\bf f}))(w) +
\chi[I_\epsilon^{\rm in}](w)\cdot
(\op{C}^{C_\epsilon}(\chi[C_\epsilon^-](\cdot){\bf f}))(w) \,.
\end{array}
\label{eq:circleintegral}
\end{equation}
Likewise, for ${\bf f}\in L^2(I,ds)$ and $w\in C_\epsilon$, we have
\begin{equation}
\begin{array}{rcl}
(\op{C}^{I}{\bf f})(w)&=&\chi[C_\epsilon^+](w)\cdot
(\op{C}^{I}(\chi[I_\epsilon^{\rm out}](\cdot){\bf f}))(w) +
\chi[C_\epsilon^+](w)\cdot
(\op{C}^{I}(\chi[I_\epsilon^{\rm in}](\cdot){\bf f}))(w)\\\\
&&\hspace{0.2 in} +\,\,
\chi[C_\epsilon^-](w)\cdot
(\op{C}^{I}(\chi[I_\epsilon^{\rm out}](\cdot){\bf f}))(w) +
\chi[C_\epsilon^-](w)\cdot
(\op{C}^{I}(\chi[I_\epsilon^{\rm in}](\cdot){\bf f}))(w) \,.
\end{array}
\label{eq:stemintegral}
\end{equation}
It is therefore sufficient to prove the uniform boundedness of each
operator on the right-hand side of each expression.  Below, we will
give details only for the terms involving $C_\epsilon^+$, since the
reader will see that the bounds involving $C_\epsilon^-$ are obtained 
in exactly the same way.

We now introduce arc length parametrizations of $C_\epsilon^+$,
$I_\epsilon^{\rm in}$ and $I_\epsilon^{\rm out}$, and at the same time
use the invariance of the Cauchy integral under translations and
rotations to bring the intersection point $I(s_\epsilon)$ to the
origin and to make $C_\epsilon^+$ tangent to the positive real axis
there.  Therefore, using tildes to denote the translation and rotation
(see Figure~\ref{fig:apple_iii}), 
\begin{figure}[h]
\begin{center}
\mbox{\psfig{file=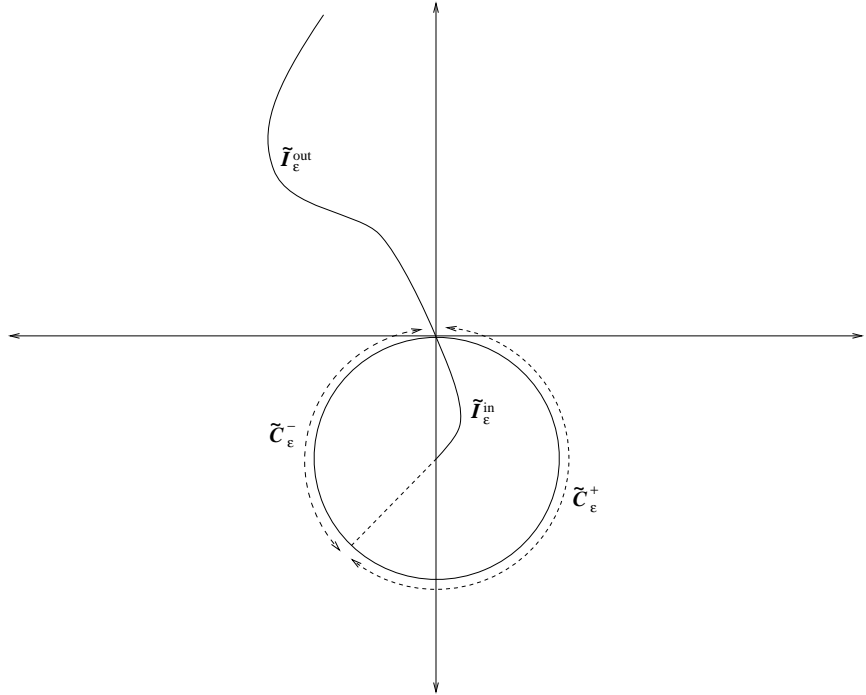,width=4.5 in}}
\end{center}
\caption{\em After rotation and translation.}
\label{fig:apple_iii}
\end{figure}
for $\tilde{C}_\epsilon^+$ we
have the parametrization:
\begin{equation}
\tilde{C}_\epsilon^+(y):=
i\epsilon (\exp(-iy/\epsilon)-1)\,,\hspace{0.3 in}
0\le y\le \epsilon\xi_\epsilon^+\,,
\end{equation}
where $\xi_\epsilon^+:=\pi+\arg(I(s_\epsilon))-\phi$ is an angle converging
to $\pi$ as $\epsilon\downarrow 0$. 
For $\tilde{I}_\epsilon^{\rm in}$ we have
\begin{equation}
\tilde{I}_\epsilon^{\rm in}(x):=
(I(s_\epsilon-x)-I(s_\epsilon))i\exp(-i\arg(I(s_\epsilon)))\,,
\hspace{0.3 in}
0\le x\le s_\epsilon\,,
\end{equation}
and for $\tilde{I}_\epsilon^{\rm out}$ we have
\begin{equation}
\tilde{I}_\epsilon^{\rm out}(x):=
(I(s_\epsilon+x)-I(s_\epsilon))i\exp(-i\arg(I(s_\epsilon)))\,,
\hspace{0.3 in}
0\le x\le S_\epsilon\,,
\end{equation}
where $S_\epsilon:=s_{\rm max}-s_\epsilon$ converges to $s_{\rm
max}>0$ as $\epsilon\downarrow 0$.  Note that at the origin,
$\tilde{I}_\epsilon^{\rm in}$ is tangent to a ray making an angle of
$\zeta_\epsilon^{\rm in}:=-\pi/2
+\arg(I'(s_\epsilon))-\arg(I(s_\epsilon))$ with the positive real
axis, and $\tilde{I}_\epsilon^{\rm out}$ is tangent to a ray making an
angle of $\zeta_\epsilon^{\rm
out}:=\pi/2+\arg(I'(s_\epsilon))-\arg(I(s_\epsilon))$ with the
positive real axis.  These angles converge to $-\pi/2$ and $\pi/2$
respectively as $\epsilon\downarrow 0$.  The reader should note that
the scales have not been changed by these transformations.  The circle
$\tilde{C}^+_\epsilon\cup\tilde{C}^-_\epsilon$ still has radius
$\epsilon$, and the subsequent estimates will be uniform for
$\epsilon$ sufficiently small.

For ${\bf f}\in L^2(\tilde{C}_\epsilon^+,dy)$ and $0\le x\le s_\epsilon$,
we define a kernel $K^{\rm in}(x,y)$ by writing
\begin{equation}
\begin{array}{rcl}
\displaystyle
\frac{1}{2\pi i}
\int_0^{\epsilon\xi_\epsilon^+}
(\tilde{C}_\epsilon^+(y)-\tilde{I}_\epsilon^{\rm in}(x))^{-1}
{\bf f}(\tilde{C}_\epsilon^+(y))\tilde{C}_\epsilon^{+\prime}(y)\,dy &=& 
\displaystyle
\frac{1}{2\pi i}
\int_0^{\epsilon\xi_\epsilon^+}
(y-x\exp(i\zeta_\epsilon^{\rm in}))^{-1}
(\op{U}_1{\bf f})(y)\,dy  \\\\
&&\displaystyle\hspace{0.2 in}+\,\,
\frac{1}{2\pi i}\int_0^{\epsilon\xi_\epsilon^+}
K^{\rm in}(x,y)(\op{U}_1{\bf f})(y)
\,dy\,,  
\end{array}
\label{eq:circlesplit_iii}
\end{equation}
where the map defined by $(\op{U}_1{\bf f})(y):={\bf
f}(\tilde{C}_\epsilon^+(y))
\tilde{C}_\epsilon^{+\prime}(y)$ 
is a unitary isomorphism between $L^2(\tilde{C}_\epsilon^+,dy)$ and
$L^2([0,\epsilon\xi_\epsilon^+],dy)$.  The reader will observe that
the left-hand side of (\ref{eq:circlesplit_iii}) is the third term on
the right-hand side of (\ref{eq:circleintegral}).  Therefore we have
written this term as a sum of an explicit Cauchy integral in the new
coordinate system, plus an error term involving the kernel $K^{\rm
in}(x,y)$.  Then, for ${\bf f}\in L^2(\tilde{I}_\epsilon^{\rm in},dx)$
and $0\le y\le
\epsilon\xi_\epsilon^+$, we have an expression in terms of the same
kernel of the ``reciprocal'' Cauchy integral\index{Cauchy integral!reciprocal}:
\begin{equation}
\begin{array}{rcl}
\displaystyle\frac{1}{2\pi i}\int_0^{s_\epsilon} 
(\tilde{I}_\epsilon^{\rm in}(x)-\tilde{C}_\epsilon^+(y))^{-1} {\bf
f}(\tilde{I}_\epsilon^{\rm in}(x))\tilde{I}_\epsilon^{{\rm
in}\prime}(x)\,dx &=&\displaystyle
\frac{1}{2\pi i}\int_0^{s_\epsilon}(x\exp(i\zeta_\epsilon^{\rm in})-y)^{-1}
(\op{U}_2{\bf f})(x)\,dx\\\\
&&\displaystyle\hspace{0.2 in}-\,\,\frac{1}{2\pi i}
\int_0^{s_\epsilon}K^{\rm in}(x,y)
(\op{U}_2{\bf f})(x)\,dx\,,
\end{array}
\label{eq:stemsplit_ii}
\end{equation}
where the map $(\op{U}_2{\bf f})(x):={\bf f}(\tilde{I}_\epsilon^{\rm
in}(x))\tilde{I}_\epsilon^{{\rm in}\prime}(x)$ is a unitary
isomorphism between $L^2(\tilde{I}_\epsilon^{\rm in},dx)$ and
$L^2([0,s_\epsilon],dx)$.  The left-hand side of
(\ref{eq:stemsplit_ii}) is just the second term on the right-hand side
of (\ref{eq:stemintegral}) written in the new coordinates, which has
similarly been split into a Cauchy integral and a remainder term
involving $K^{\rm in}(x,y)$.  

Similarly, we define a kernel $K^{\rm out}(x,y)$ by writing for ${\bf
f}\in L^2(\tilde{C}_\epsilon^+,dy)$ and $0\le x\le S_\epsilon$
\begin{equation}
\begin{array}{rcl}
\displaystyle
\frac{1}{2\pi i}
\int_0^{\epsilon\xi_\epsilon^+}
(\tilde{C}_\epsilon^+(y)-\tilde{I}_\epsilon^{\rm out}(x))^{-1}
{\bf f}(\tilde{C}_\epsilon^+(y))\tilde{C}_\epsilon^{+\prime}(y)\,dy &=& 
\displaystyle
\frac{1}{2\pi i}
\int_0^{\epsilon\xi_\epsilon^+}
(y-x\exp(i\zeta_\epsilon^{\rm out}))^{-1}
(\op{U}_1{\bf f})(y)\,dy  \\\\
&&\displaystyle\hspace{0.2 in}+\,\,
\frac{1}{2\pi i}\int_0^{\epsilon\xi_\epsilon^+}
K^{\rm out}(x,y)(\op{U}_1{\bf f})(y)
\,dy\,,  
\end{array}
\label{eq:circlesplit_i}
\end{equation}
from which we then obtain for the reciprocal Cauchy integral, for 
${\bf f}\in L^2(\tilde{I}_\epsilon^{\rm out},dx)$ and $0\le y\le\epsilon
\xi_\epsilon^+$,
\begin{equation}
\begin{array}{rcl}
\displaystyle\frac{1}{2\pi i}\int_0^{S_\epsilon} 
(\tilde{I}_\epsilon^{\rm out}(x)-\tilde{C}_\epsilon^+(y))^{-1}
{\bf f}(\tilde{I}_\epsilon^{\rm out}(x))\tilde{I}_\epsilon^{{\rm
out}\prime}(x)\,dx &=&\displaystyle
\frac{1}{2\pi i}\int_0^{S_\epsilon}
(x\exp(i\zeta_\epsilon^{\rm out})-y)^{-1}
(\op{U}_3{\bf f})(x)\,dx\\\\
&&\displaystyle\hspace{0.2 in}-\,\,\frac{1}{2\pi i}\int_0^{S_\epsilon}
K^{\rm out}(x,y)
(\op{U}_3{\bf f})(x)\,dx\,,
\end{array}
\label{eq:stemsplit_i}
\end{equation}
where $\op{ U}_3$ denotes the unitary isomorphism from
$L^2(\tilde{I}_\epsilon^{\rm out},dx)$ to $L^2([0,S_\epsilon],dx)$
defined by $(\op{U}_3{\bf f})(x):= {\bf f}(\tilde{I}_\epsilon^{\rm
out}(x))\tilde{I}_\epsilon^{{\rm out}\prime}(x)$.  The decomposition
(\ref{eq:circlesplit_i}) is a representation of the first term on the
right-hand side of (\ref{eq:circleintegral}).  Likewise,
(\ref{eq:stemsplit_i}) is a representation of the first term on the
right-hand side of (\ref{eq:stemintegral}).

Having represented each term involving $C^+_\epsilon$ on the
right-hand sides of (\ref{eq:circleintegral}) and
(\ref{eq:stemintegral}) as a sum of a Cauchy integral operator in the
new coordinate system and a remainder type integral operator, we now
need to prove that these are in fact all bounded operators.  The
Cauchy operators are handled by an argument of Beals,
Deift, and Tomei
\cite{BDT88} that uses the theory of Mellin transforms 
\index{Mellin transform} in $L^2$ spaces on
straight rays.  Their methods show that regardless of the value of
$\epsilon$,
\begin{equation}
\begin{array}{rcll}
\displaystyle
\int_0^{s_\epsilon}\left\|\frac{1}{2\pi i}\int_0^{\epsilon\xi_\epsilon^+}
\frac{dy}{y-x\exp(i\zeta_\epsilon^{\rm in})}(\op{U}_1{\bf f})(y)\right\|^2\,dx
&\le &\displaystyle
\int_0^{\epsilon\xi_\epsilon^+}
\|{\bf f}(\tilde{C}_\epsilon^+(y))\|^2\,dy\,,&\displaystyle
\forall {\bf f}\in L^2(\tilde{C}_\epsilon^+,dy)
\\\\
\displaystyle
\int_0^{S_\epsilon}
\left\|\frac{1}{2\pi i}\int_0^{\epsilon\xi_\epsilon^+}
\frac{dy}{y-x\exp(i\zeta_\epsilon^{\rm out})}(\op{U}_1{\bf f})(y)
\right\|^2\,dx
&\le &\displaystyle
\int_0^{\epsilon\xi_\epsilon^+}
\|{\bf f}(\tilde{C}_\epsilon^+(y))\|^2\,dy\,,&\displaystyle
\forall {\bf f}\in L^2(\tilde{C}_\epsilon^+,dy)
\\\\
\displaystyle
\int_0^{\epsilon\xi_\epsilon^+}\left\|\frac{1}{2\pi i}
\int_0^{s_\epsilon} \frac{dx}{x\exp(i\zeta_\epsilon^{\rm in})-y}
(\op{U}_2{\bf f})(x)\right\|^2\,dy&\le &\displaystyle
\int_0^{s_\epsilon}\|{\bf f}(\tilde{I}_\epsilon^{\rm in}(x))\|^2\,dx\,,
&\displaystyle
\forall {\bf f}\in L^2(\tilde{I}_\epsilon^{\rm in},dx)
\\\\
\displaystyle
\int_0^{\epsilon\xi_\epsilon^+}\left\|\frac{1}{2\pi i}
\int_0^{S_\epsilon} \frac{dx}{x\exp(i\zeta_\epsilon^{\rm out})-y}
(\op{U}_3{\bf f})(x)\right\|^2\,dy&\le &\displaystyle
\int_0^{S_\epsilon}
\|{\bf f}(\tilde{I}_\epsilon^{\rm out}(x))\|^2\,dx\,,
&\displaystyle
\forall {\bf f}\in L^2(\tilde{I}_\epsilon^{\rm out},dx)\,,
\end{array}
\end{equation}
as long as $\zeta_\epsilon^{\rm in}$ and $\zeta_\epsilon^{\rm out}$
are both nonzero.  Since these angles converge to $-\pi/2$ and $\pi/2$
respectively as $\epsilon\downarrow 0$, this will be the case for all
sufficiently small $\epsilon$.  Thus, the Cauchy integral operators
\index{Cauchy integral operators} appearing as the first terms on the
right-hand sides of (\ref{eq:circlesplit_iii}),
(\ref{eq:stemsplit_ii}), (\ref{eq:circlesplit_i}), and
(\ref{eq:stemsplit_i}) are bounded, with bounds that are independent
of $\epsilon$.

We now turn to the estimation of the ``remainder'' operators
\index{remainder operators} with kernels $K^{\rm in}(x,y)$ and $K^{\rm
out}(x,y)$.  In this connection, we first note that these kernels,
which are explicitly written as
\begin{equation}
K^{\rm in}(x,y):=
\frac{1}{\tilde{C}_\epsilon^+(y)-
\tilde{I}_\epsilon^{\rm in}(x)}-
\frac{1}{y-x\exp(i\zeta_\epsilon^{\rm in})}\,,\hspace{0.2 in}
K^{\rm out}(x,y):=
\frac{1}{\tilde{C}_\epsilon^+(y)-
\tilde{I}_\epsilon^{\rm out}(x)}-
\frac{1}{y-x\exp(i\zeta_\epsilon^{\rm out})}\,,
\end{equation}
are bounded functions on their respective domains of definition, and 
\begin{equation}
\limsup_{x,y\rightarrow 0}|K^{\rm in}(x,y)|<\infty\,,\hspace{0.2 in}
\limsup_{x,y\rightarrow 0}|K^{\rm out}(x,y)|<\infty\,.
\end{equation}
Therefore, for all ${\bf f}\in L^2(\tilde{C}_\epsilon^+,dy)$, 
\begin{equation}
\int_0^{s_\epsilon}\left\|\int_0^{\epsilon\xi_\epsilon^+}K^{\rm in}(x,y)
(\op{U}_1{\bf f})(y)\,dy\right\|^2\,dx\le
\left(\int_0^{s_\epsilon}dx\int_0^{\epsilon\xi_\epsilon^+}dy\,
 |K^{\rm in}(x,y)|^2
\right)
\int_0^{\epsilon\xi^+}\|{\bf f}(\tilde{C}_\epsilon^+(y))\|^2\,dy\,,
\end{equation}
and
\begin{equation}
\int_0^{S_\epsilon}\left\|\int_0^{\epsilon\xi_\epsilon^+}K^{\rm out}(x,y)
(\op{U}_1{\bf f})(y)\,dy\right\|^2\,dx\le
\left(\int_0^{S_\epsilon}dx\int_0^{\epsilon\xi_\epsilon^+}dy\,
 |K^{\rm out}(x,y)|^2
\right)
\int_0^{\epsilon\xi^+}\|{\bf f}(\tilde{C}_\epsilon^+(y))\|^2\,dy\,,
\end{equation}
while for ${\bf f}\in L^2(\tilde{I}_\epsilon^{\rm in},dx)$,
\begin{equation}
\int_0^{\epsilon\xi_\epsilon^+}\left\|\int_0^{s_\epsilon}K^{\rm in}(x,y)
(\op{U}_2{\bf f})(x)\,dx\right\|^2\,dy\le
\left(\int_0^{s_\epsilon}dx\int_0^{\epsilon\xi_\epsilon^+}dy\,
|K^{\rm in}(x,y)|^2\right)
\int_0^{s_\epsilon}\|{\bf f}(\tilde{I}_\epsilon^{\rm in}(x))\|^2\,dx\,,
\end{equation}
and for ${\bf f}\in L^2(\tilde{I}_\epsilon^{\rm out},dx)$,
\begin{equation}
\int_0^{\epsilon\xi_\epsilon^+}\left\|\int_0^{S_\epsilon}K^{\rm out}(x,y)
(\op{U}_3{\bf f})(x)\,dx\right\|^2\,dy\le
\left(\int_0^{S_\epsilon}dx\int_0^{\epsilon\xi_\epsilon^+}dy\,
|K^{\rm out}(x,y)|^2\right)
\int_0^{S_\epsilon}\|{\bf f}(\tilde{I}_\epsilon^{\rm out}(x))\|^2\,dx\,.
\end{equation}
These norm estimates are finite for all $\epsilon>0$, and we must control
their dependence on $\epsilon$ as $\epsilon\downarrow 0$.  

We now claim that
\begin{equation}
\begin{array}{rcccl}
\displaystyle
\lim_{\epsilon\downarrow 0}\int_0^{s_\epsilon}dx\int_0^{\epsilon\xi_\epsilon^+}
dy\,|K^{\rm in}(x,y)|^2 &=&\displaystyle 
\int_0^\pi dz\int_0^1 dw\,
\left|\frac{1}{i(\exp(-iz)-1)+iw}-\frac{1}{z+iw}\right|^2&<&\infty\\\\
\displaystyle
\lim_{\epsilon\downarrow 0}\int_0^{S_\epsilon}dx\int_0^{\epsilon\xi_\epsilon^+}
dy\,|K^{\rm out}(x,y)|^2 &=&\displaystyle
\int_0^\pi dz\int_0^\infty dw\,
\left|\frac{1}{i(\exp(-iz)-1)-iw}-\frac{1}{z-iw}\right|^2&<&\infty\,.
\end{array}
\label{eq:glimits}
\end{equation}
These limits are finite because the integrands are bounded near
$z=w=0$ (in particular they are both less than $1$ for all $z<1$),
decay like $w^{-4}$ for large $w$, and are uniformly bounded
elsewhere.  To prove the claim, first rescale in both integrals by
setting $z=y/\epsilon$ and $w=x/\epsilon$.  This modifies the
integrand through the Jacobian by multiplication by $\epsilon^2$.  For
the $K^{\rm in}(x,y)$ integral, the region of integration tends to the
fixed rectangle $[0,1]\times[0,\pi]$, while for the $K^{\rm out}(x,y)$
integral, the region of integration tends to the semi-infinite strip
$[0,\infty]\times [0,\pi]$.  Using the fact that the curve $I(s)$ is
twice differentiable, one sees that the integrands $\epsilon^2 |K^{\rm
in}(\epsilon w,\epsilon z)|^2$ and $\epsilon^2 |K^{\rm out}(\epsilon
w,\epsilon z)|^2$ converge pointwise as $\epsilon\downarrow 0$ to the
integrands on the right-hand side of (\ref{eq:glimits}).  The
convergence is in fact uniform for the $K^{\rm in}(x,y)$ integral, and
therefore this part of the claim follows immediately.  For the $K^{\rm
out}(x,y)$ integral the claim follows from a dominated convergence
argument.

Since the limits (\ref{eq:glimits}) exist, the operators having
kernels $K^{\rm in}(x,y)$, $-K^{\rm in}(y,x)$, $K^{\rm out}(x,y)$ and
finally $-K^{\rm out}(y,x)$ are bounded in the appropriate $L^2$
spaces, uniformly as $\epsilon\downarrow 0$.  Combining these
estimates with the Beals-Deift-Tomei estimates of the Cauchy kernels
and the results of a parallel analysis involving the circular arc
$C_\epsilon^-$ completes the proof of the lemma.
\end{proof}

For each fixed value of $\hbar$, we can define operators
$\op{C}^\Sigma_\pm$ on $L^2(\Sigma)$ taking a function ${\bf f}(z)$ to
the boundary values taken on each oriented segment of $\Sigma$ from
the $+$ and $-$ sides respectively of the Cauchy contour integral
\begin{equation}
(\op{C}^\Sigma{\bf f})(w):=\frac{1}{2\pi i}\int_\Sigma (z-w)^{-1}{\bf f}(z)\,dz\,.
\end{equation}
With a suitable interpretation of convergence to the boundary
values, for each fixed $\hbar$, these operators are bounded on
$L^2(\Sigma)$.  
\begin{lemma} There exists an $M>0$ such that for all sufficiently 
small $\hbar$, 
\begin{equation}
\|\op{C}^\Sigma_+\|_{L^2(\Sigma)} < M\,, 
\hspace{0.2 in}\mbox{and}\hspace{0.2 in}
\|\op{C}^\Sigma_-\|_{L^2(\Sigma)} < M\,.
\end{equation}
\label{lemma:CplusCminusBound}
\end{lemma}

\begin{proof}
We first note that, modulo self-intersection points, the contour
$\Sigma$ can be written as a union of an $\hbar$-independent part
$\Sigma\setminus \partial U_\hbar$, and several arcs making up the
shrinking circle $\partial U_\hbar$.  Let ${\bf f}\in L^2(\Sigma)$,
and decompose it into a sum ${\bf g}+{\bf h}$ where the support of
${\bf g}$ is contained in $\Sigma\setminus\partial U_\hbar$ and that
of ${\bf h}$ is contained in $\partial U_\hbar$.  Then, for almost
every $z\in\Sigma\setminus\partial U_\hbar$,
\begin{equation}
(\op{C}^\Sigma_\pm{\bf f})(z)=
(\op{C}^{\Sigma\setminus\partial U_\hbar}_\pm{\bf g})(z) +
(\op{C}^{\partial U_\hbar}{\bf h})(z)\,,
\end{equation}
and for almost every $z\in\partial U_\hbar$, 
\begin{equation}
(\op{C}^\Sigma_\pm{\bf f})(z)=
(\op{C}^{\partial U_\hbar}_\pm{\bf h})(z)+
(\op{C}^{\Sigma\setminus\partial U_\hbar}{\bf g})(z)\,.
\end{equation}
Integrating to compute the norm, we first estimate
\begin{equation}
\int_\Sigma \|(\op{C}^\Sigma_\pm{\bf f})(z)\|^2\,|dz|\le
I_{gg}+I_{gh}+I_{hg}+I_{hh}\,,
\end{equation}
where
\begin{equation}
\begin{array}{rclrcl}
I_{gg}&:=&\displaystyle \int_{\Sigma\setminus\partial U_\hbar}
\|(\op{C}^{\Sigma\setminus\partial U_\hbar}_\pm{\bf g})(z)\|^2\,|dz| \,,
&I_{gh}&:=&\displaystyle
\int_{\Sigma\setminus\partial U_\hbar}
\|(\op{C}^{\partial U_\hbar}{\bf h})(z)\|^2\,|dz| \,,\\\\
I_{hg}&:=&\displaystyle
\int_{\partial U_\hbar} 
\|(\op{C}^{\Sigma\setminus\partial U_\hbar}{\bf g})(z)\|^2\,|dz| \,,
&I_{hh}&:=&\displaystyle
\int_{\partial U_\hbar} 
\|(\op{C}^{\partial U_\hbar}_\pm{\bf h})(z)\|^2\,|dz|\,.
\end{array}
\end{equation}

First, we estimate the ``diagonal'' terms $I_{gg}$ and $I_{hh}$.
Because the contour $\Sigma\setminus\partial U_\hbar$ is independent
of $\hbar$, there is some $\hbar$-independent constant $C_{gg}>0$ such
that
\begin{equation}
I_{gg}\le C_{gg}\int_{\Sigma\setminus\partial U_\hbar}\|{\bf g}(z)\|^2\,|dz|
\le C_{gg}\int_{\Sigma}\|{\bf f}(z)\|^2\,|dz|\,.
\end{equation}
Now for the integral $I_{hh}$ the contour depends on $\hbar$, but in a simple
way that can be scaled out.  Thus, rescaling, 
\begin{equation}
I_{hh}=\hbar^{\delta}\int_{\partial U_1}\left\|\frac{1}{2\pi i}
\int_{\partial U_1}(t-w_\pm)^{-1}{\bf h}(\hbar^{\delta}t)\,dt\right\|^2\,|dw|\,.
\end{equation}
With the contour rescaled to a radius independent of $\hbar$, we see that
there exists an $\hbar$-independent constant $C_{hh}>0$ such that
\begin{equation}
I_{hh}\le \hbar^{\delta}C_{hh}\int_{\partial U_1}\|{\bf h}(\hbar^{\delta}t)\|^2\,|dt| = C_{hh}\int_{\partial U_\hbar}\|{\bf h}(z)\|^2\,|dz|\le
C_{hh}\int_{\Sigma}\|{\bf f}(z)\|^2\,|dz|\,.
\end{equation}

Next, we turn to the estimation of the ``cross terms'' $I_{gh}$ and
$I_{hg}$.  For this purpose, we again decompose the
$\hbar$-independent contour $\Sigma\setminus\partial U_\hbar$ into two
$\hbar$-independent parts by cutting it at the boundary of the fixed
but small disk $U$ (this disk is illustrated in
Figure~\ref{fig:error}).  Thus, let $\Gamma^{\rm in}$ (respectively
$\Gamma^{\rm out}$) denote the part of $\Sigma\setminus\partial
U_\hbar$ inside (respectively outside) $U$.  With this decomposition,
we have by Cauchy-Schwarz\index{Cauchy-Schwarz},
\begin{equation}
\begin{array}{rcl}
I_{gh}&=&\displaystyle
\int_{\Gamma^{\rm in}}\|(\op{C}^{\partial U_\hbar}{\bf h})(z)\|^2\, |dz|
+\int_{\Gamma^{\rm out}}\|(\op{C}^{\partial U_\hbar}{\bf h})(z)\|^2 \, |dz|\\\\
&\le &\displaystyle
\int_{\Gamma^{\rm in}}\|(\op{C}^{\partial U_\hbar}{\bf h})(z)\|^2\, |dz|
+
\frac{\hbar^{\delta}|\Gamma^{\rm out}|}{2\pi}\sup_{s\in\Gamma^{\rm out},z\in \partial U_\hbar} |s-z|^{-2}\int_{\partial U_\hbar}\|{\bf h}(s)\|^2\,|ds|\,,
\end{array}
\end{equation}
where $|\Gamma^{\rm out}|$ is the $\hbar$-independent total arc length
of $\Gamma^{\rm out}$.  The supremum in the last line is bounded as
$\hbar$ tends to zero because distance between $\Gamma^{\rm out}$ and
$\partial U_\hbar$ increases as $\hbar$ decreases.  Therefore, there
is an $\hbar$-independent constant $C_{\rm cross}^{\rm out}>0$ such
that
\begin{equation}
I_{gh}\le 
\int_{\Gamma^{\rm in}}\|(\op{C}^{\partial U_\hbar}{\bf h})(z)\|^2\, |dz|
+\hbar^{\delta}C_{\rm cross}^{\rm out}\int_{\Sigma}\|{\bf f}(z)\|^2\,|dz|\,,
\end{equation}
for all sufficiently small $\hbar$.  We momentarily delay the
estimation of the term involving $\Gamma^{\rm in}$.  Applying the same
decomposition to $I_{hg}$ and using Cauchy-Schwarz, we find with the
same constant $C_{\rm cross}^{\rm out}$
\begin{equation}
\begin{array}{rcl}
I_{hg}&\le &\displaystyle
\int_{\partial U_\hbar}\|(\op{C}^{\Gamma^{\rm in}}{\bf g})(z)\|^2\,|dz|
 +
\int_{\partial U_\hbar}\|(\op{C}^{\Gamma^{\rm out}}{\bf g})(z)\|^2\,|dz|
\\\\
&\le &\displaystyle
\int_{\partial U_\hbar}\|(\op{C}^{\Gamma^{\rm in}}{\bf g})(z)\|^2\,|dz|
 +
\hbar^{\delta}C_{\rm cross}^{\rm out}\int_{\Sigma}\|{\bf f}(z)\|^2\,|dz|\,.
\end{array}
\end{equation} 

Finally, to estimate the remaining terms in $I_{gh}$ and $I_{hg}$, we
appeal to Lemma~\ref{lemma:apple}.  Note that $\Gamma^{\rm in}$ is a
union of eight smooth curve segments (in fact all but two of the
segments are exactly straight ray segments) meeting at the origin.
Therefore the lemma applies to the interaction of each curve segment
with the circle $\partial U_\hbar$ of radius $\epsilon=\hbar^{\delta}$.
Summing the $\hbar$-independent estimates guaranteed by Lemma~\ref{lemma:apple}
finally gives constants $C^{\rm in}_{gh}>0$ and $C^{\rm in}_{hg}>0$ such that
\begin{equation}
\int_{\Gamma^{\rm in}}\|(\op{C}^{\partial U_\hbar}{\bf h})(z)\|^2\,|dz|
\le C^{\rm in}_{gh}\int_{\partial U_\hbar}\|{\bf h}(z)\|^2\,|dz|\le
C^{\rm in}_{gh}\int_{\Sigma}\|{\bf f}(z)\|^2\,|dz|\,,
\end{equation}
and
\begin{equation}
\int_{\partial U_\hbar}\|(\op{C}^{\Gamma^{\rm in}}{\bf g})(z)\|^2\,|dz|
\le C_{hg}^{\rm in}\int_{\Gamma^{\rm in}}\|{\bf g}(z)\|^2\,|dz|\le
C_{hg}^{\rm in}\int_{\Sigma}\|{\bf f}(z)\|^2\,|dz|\,.
\end{equation}
Assembling the estimates of the diagonal terms and the cross terms finally
completes the proof.
\end{proof}

With these results in hand, we can estimate the error ${\bf
E}(\lambda)$ under the condition that we have found a complex phase
function $g^\sigma(\lambda)$.  We do this by constructing the solution
${\bf R}(\lambda)$ of the global error Riemann-Hilbert Problem
\ref{rhp:error} directly from its jump matrix ${\bf v}(\lambda)$, for
which we have a uniform estimate from
Lemma~\ref{lemma:uniformlysmall}.
\begin{theorem}[Conditional Error Bound]
Given the existence of a complex phase function $g^\sigma(\lambda)$,
there exists some $M>0$ such that for all sufficiently small
$\hbar_N$, the error satisfies the estimate
\begin{equation}
\|{\bf E}(\lambda)-{\mathbb I}\|\le M\hbar_N^{1/3}\sup_{s\in\Sigma}
|s-\lambda|^{-1}\,.  
\label{eq:uniformerrorestimate}
\end{equation} 
\label{theorem:errorestimate}
\end{theorem} 

\begin{proof} 
For $\lambda\in\Sigma$, set ${\bf w}^+(\lambda):={\bf
v}(\lambda)-{\mathbb I}$ and ${\bf w}^-(\lambda):={\bf 0}$.  Then from
Lemma~\ref{lemma:uniformlysmall} and
Lemma~\ref{lemma:CplusCminusBound}, we observe that the conditions of
Lemma~\ref{lemma:L2RHP} in the appendix are met.  This guarantees the
existence of a matrix ${\bf R}(\lambda)$ satisfying the global error
Riemann-Hilbert Problem \ref{rhp:error} in the $L^2(\Sigma)$ sense.
According to Lemma~\ref{lemma:L2RHP}, this solution ${\bf R}(\lambda)$
satisfies the error estimate (\ref{eq:L2estimate}), which by virtue of
the bound on $\|{\bf v}(\lambda)-{\mathbb I}\|$ afforded by
Lemma~\ref{lemma:uniformlysmall} and the fact that the total length of
the contour $\Sigma$ remains uniformly bounded as $\hbar_N$ tends to
zero implies (\ref{eq:uniformerrorestimate}) via the equivalence ${\bf E}(\lambda)\equiv {\bf R}(\lambda)$.
\end{proof}
\begin{remark}
In principle, the representation ${\bf E}(\lambda)\equiv{\bf
R}(\lambda)$ and the asymptotic control on ${\bf w}^\pm(\lambda)$
implies not only the estimate given here, but also an explicit series
representation of ${\bf R}(\lambda)$ obtained via the solution of the
associated singular integral equation ({\em cf.}
Lemma~\ref{lemma:L2Invertibility}) by Neumann series\index{Neumann
series}.  Making the calculation of correction terms in the expansion
${\bf R}(\lambda)={\mathbb I}+\dots$ effective requires explicit
knowledge of the matrices ${\bf w}^\pm(\lambda)$.  Because the matrix
${\bf v}(\lambda)$ is constructed in terms of boundary values of the
parametrix $\hat{\bf N}^\sigma(\lambda)$ on $\Sigma$, the practical
computation of higher-order corrections \index{higher-order
corrections} for ${\bf R}(\lambda)$ amounts to representing the
various approximations of ${\bf N}^\sigma(\lambda)$ in terms of known
functions.  Such a representation is of course available for the
``outer'' approximation $\hat{\bf N}^\sigma_{\rm out}(\lambda)$.
Also, as remarked in
\S\ref{sec:2k} and
\S\ref{sec:2km1} it is possible 
to obtain explicit formulae for the approximations $\hat{\bf
N}^\sigma_{2k}(\lambda)$ and $\hat{\bf N}^\sigma_{2k-1}(\lambda)$ in
terms of Airy functions \index{Airy functions}(although we did not
pursue this particular path).  Therefore, the obstruction to
calculating explicit higher-order corrections to ${\bf R}(\lambda)$ is
really an {\em explicit} special function \index{special functions}
representation of the matrix $\hat{\bf F}^\sigma(\lambda)$ used to
build the local approximation $\hat{\bf N}^\sigma_{\rm
origin}(\lambda)$ near the origin.
\end{remark}

Recall the definition (\ref{eq:fieldatinfinity}) of the exact solution
$\psi$ of the nonlinear Schr\"odinger equation corresponding to the
semiclassical soliton ensemble that is the nonlinear superposition of
the individual solitons indexed by the WKB eigenvalues
$\lambda_{\hbar_N,n}^{\rm WKB}$, as well as the definitions
(\ref{eq:psitildeGgtzero}) and (\ref{eq:psitildeGeqzero}) of the
explicit approximation $\tilde{\psi}$ obtained in
\S\ref{sec:outersolve}.  Using the relation between ${\bf
N}^\sigma(\lambda)$ and ${\bf M}(\lambda)$ given in
(\ref{eq:changeofvariables}), we obtain
\begin{theorem}[Conditional strong asymptotics for semiclassical soliton ensembles]
\index{semiclassical soliton ensemble!strong asymptotics for}
Given the existence of a complex phase function $g^\sigma(\lambda)$,
there exists a positive constant $M$ such that for all sufficiently
small $\hbar_N$,
\begin{equation}
|\psi-\tilde{\psi}|\le M\hbar_N^{1/3}\,.
\end{equation}
The asymptotics are strong because $x$ and $t$ are held fixed.
\label{theorem:psiestimate}
\end{theorem}

\begin{proof}
Start with the exact representation of ${\bf M}(\lambda)$:
\begin{equation}
{\bf M}(\lambda)={\bf E}(\lambda)\hat{\bf N}^\sigma(\lambda)\exp(g^\sigma(\lambda)\sigma_3/\hbar_N)\,,
\end{equation}
which implies for the $(1,2)$ entry
\begin{equation}
M_{12}(\lambda)=\hat{N}^\sigma_{12}(\lambda)\exp(-g^\sigma(\lambda)/\hbar_N) +
\left[({\bf E}(\lambda)-{\mathbb I})\cdot\hat{\bf N}^\sigma(\lambda)
\right]_{12}\exp(-g^\sigma(\lambda)/\hbar_N)\,.
\end{equation}
Multiplying by $2i\lambda$ and passing to the limit
$\lambda\rightarrow\infty$ for fixed $\hbar_N$ using the fact that for
large $\lambda$, $\hat{\bf N}^\sigma(\lambda)\equiv\hat{\bf
N}^\sigma_{\rm out}(\lambda)$ and the fact that
$g^\sigma(\lambda)=\bo(|\lambda|^{-1})$ yields
\begin{equation}
\psi = \tilde{\psi}+\lim_{\lambda\rightarrow\infty}
\left(2i\lambda\exp(-g^\sigma(\lambda)/\hbar_N)\left[({\bf E}(\lambda)
-{\mathbb I})\cdot\hat{\bf N}^\sigma(\lambda)\right]_{12}\right)\,.
\end{equation}
The theorem is thus established upon using the estimate stated in
Theorem~\ref{theorem:errorestimate}.
\end{proof}

\begin{remark}
Theorem~\ref{theorem:psiestimate} presumes the existence for fixed $x$
and $t$ of an appropriate complex phase function $g^\sigma(\lambda)$
characterized by an admissible density function \index{density
function!admissible} $\rho^\sigma(\eta)$ as in
Definition~\ref{def:admissiblerho} for which
\begin{enumerate}
\item
The contour $C$ is smooth except at the origin, where it has well-defined
tangents on both sides, neither vertical nor horizontal.
\item
The density $\rho^\sigma(\eta)$ vanishes like a square root at the endpoints
$\lambda_0$,\dots,$\lambda_G$.
\end{enumerate}
Under these conditions, Theorem~\ref{theorem:psiestimate} shows that
the function $\tilde{\psi}$ provides explicit, strong
asymptotics for $\psi$ at the particular $x$ and $t$ values under
consideration.  Uniformity with respect to $x$ and $t$
in certain compact sets can be obtained from continuity properties of
$g^\sigma(\lambda)$, which will soon be established.
\end{remark}

The utility of these results in the semiclassical analysis of
semiclassical soliton ensembles rests upon finding the complex phase
function \index{complex phase function} $g^\sigma(\lambda)$, a task we
now pursue.

\chapter{Direct Construction of the Complex Phase}
\label{sec:ansatz}
In this chapter, we turn to the question of reducing the
Riemann-Hilbert Problem~\ref{rhp:M} for the matrix ${\bf M}(\lambda)$
ultimately to the simple form of the outer model Riemann-Hilbert
Problem~\ref{rhp:model} for the matrix $\tilde{\bf O}(\lambda)$ which
was solved exactly in
\S\ref{sec:outersolve}.  Achieving the reduction required finding an
appropriate complex phase function $g^\sigma(\lambda)$ on an
appropriate contour $C\cup C^*$.  We will describe below a
construction of good ``candidate'' complex phase
functions\index{complex phase function!candidate for}.  It is then
often possible to prove directly that an appropriate candidate
actually satisfies all of the criteria ({\em cf.}
Definition~\ref{def:admissiblerho}) required for reducing the
Riemann-Hilbert problem to an analytically tractable form as in
Chapter~\ref{sec:asymptoticanalysis}.

\section[Postponing the Inequalities]{Postponing inequalities.  General considerations.}
\label{sec:postponing}
We will try to construct a complex phase function
$g^\sigma(\lambda)$ on the basis of two ideas:
\begin{enumerate}
\item 
Suppose the number of bands that will ultimately lie on the yet-to-be
determined contour $C$ is given as $G/2+1$ for some even integer $G\ge
0$.  Suppose further that the initial part of $C$ coming out of
$\lambda=0$ is part of a band, and that the final part of $C$ going
into $\lambda=0$ is part of a gap.  Here, ``initial'' and ``final''
are defined by the index $\sigma$.
\item
Ignore, for the moment, {\em all inequalities} that go into the
specification of $g^\sigma(\lambda)$.  They will be checked
later.
\end{enumerate}
So, for each fixed $x$ and $t$ we can try to use the remaining
conditions on $g^\sigma(\lambda)$ to construct a ``candidate'' phase
function for each nonnegative even integer $G$.  It will soon be clear
that this construction of the candidates is completely systematic.
Later, we will need to determine which of the candidates, if any,
satisfy the necessary inequalities that we are about to throw out.
Motivated by the exact solution of the outer model problem presented
in \S\ref{sec:outersolve} which involved hyperelliptic Riemann
surfaces of genus $G$, we will refer to a candidate complex phase
function corresponding to some even integer $G$ as a {\em genus $G$
ansatz}\index{genus $G$ ansatz}.

\subsection{Collapsing the loop contour $C$.}
\begin{figure}[h]
\begin{center}
\mbox{\psfig{file=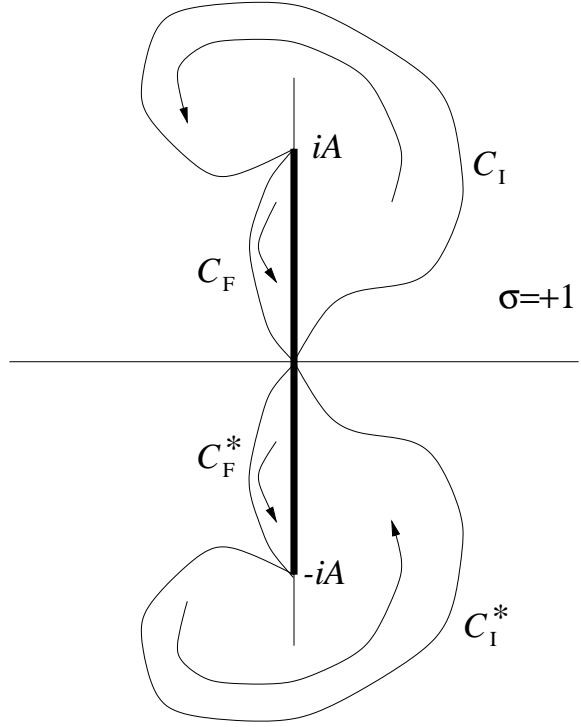,width=3 in}}
\end{center}
\caption{\em The contour $C$ is temporarily assumed to pass through $\lambda=iA$.}
\label{fig:Cpinched}
\end{figure}

The first step in constructing a genus $G$ ansatz for the complex
phase is to temporarily assume that the contour $C$ passes through the
point $\lambda=iA$, which is the top of the support of the asymptotic
eigenvalue measure $\rho^0(\eta)\,d\eta$ on the imaginary axis.  This
effectively divides $C$ into two halves: an ``initial'' half $C_I$ and
a ``final'' half $C_F$.  See Figure~\ref{fig:Cpinched}.
We will suppose for the time being that the support of the function
$\rho^\sigma(\eta)$ on $C$ is contained in $C_I$.  This means that the
entire contour $C_F\cup C_F^*$ is being assumed to be part of a gap of
$C\cup C^*$.  The gap condition $\rho^\sigma(\eta)\equiv 0$ in
$C_F\cup C_F^*$ is satisfied by assumption, and the remaining gap
condition $\Re(\tilde{\phi}^\sigma)<0$ is an inequality that we are
temporarily putting aside.  Therefore, until we restore the
inequalities, we will concern ourselves strictly with the contour
$C_I\cup C_I^*$.  {\em Note that $C_I$ will always be considered to be
oriented from $0$ to $iA$, and that $C_I^*$ will be oriented from
$-iA$ to $0$.}

The function that is our main concern is defined for $\lambda\in C_I$ as
\begin{equation}
\begin{array}{rcl}
\displaystyle\tilde{\phi}^\sigma(\lambda)&=&\displaystyle
\int_{0}^{iA}
L_\eta^0(\lambda)\rho^0(\eta)\,d\eta  +
\int_{-iA}^0 L_\eta^0(\lambda)\rho^0(\eta^*)^*\,d\eta\\\\
&&\displaystyle\,\,+\,\,
J\left(2i\lambda x + 2i\lambda^2 t -(2K+1)i \pi  \int_\lambda^{iA}
\rho^0(\eta)\,d\eta
-g^\sigma_+(\lambda)-
g^\sigma_-(\lambda)\right)\,.
\end{array}
\label{eq:tildephiagain}
\end{equation}
If we define for $\lambda\in C_I$
\begin{equation}
L^{C,\sigma}_{\eta\pm}(\lambda)=
\lim_{\mu\rightarrow\lambda_\pm}L^{C,\sigma}_\eta(\mu)\,,
\end{equation}
indicating the nontangential boundary values from the left ($+$) and
right ($-$) sides of $C_\sigma$, and set
\begin{equation}
\overline{L^{C,\sigma}_{\eta}}(\lambda)=\frac{1}{2}
(L^{C,\sigma}_{\eta+}(\lambda)+
L^{C,\sigma}_{\eta-}(\lambda))\,,
\end{equation}
to denote the average, then for $\lambda\in C_I$ we can use the
presumed complex-conjugation symmetry of $\rho^\sigma(\eta)$ to write
\begin{equation}
\begin{array}{rcl}
\displaystyle\tilde{\phi}^\sigma(\lambda)&=&\displaystyle
\int_{0}^{iA} L_\eta^0(\lambda)\rho^0(\eta)\,d\eta  
-\int_{C_I}
\overline{L^{C,\sigma}_{\eta}}(\lambda)\rho^\sigma(\eta)\,d\eta\\\\
&&\displaystyle\,\,+\,\,
\int_{-iA}^0 L^0_\eta(\lambda)\rho^0(\eta^*)^*\,d\eta -
\int_{C_I^*}\overline{L^{C,\sigma}_\eta}(\lambda)\rho^\sigma(\eta^*)^*\,d\eta\\\\
&&\displaystyle\,\,+\,\, J\left(
2i\lambda x+ 2i\lambda^2 t -(2K+1)i\pi\int_\lambda^{iA}\rho^0(\eta)\,d\eta
\right)\,.
\end{array}
\end{equation}

Now, we need to suppose that {\em the asymptotic eigenvalue density
$\rho^0(\eta)$ is analytic in the region bounded by the imaginary
interval $[0,iA]$ and the curve $C_I$}.  From the formula
(\ref{eq:WKBdensity}), this amounts to the condition that the function
$A(x)$ characterizing the initial data for (\ref{eq:IVP}) is
real-analytic (in addition to being bell-shaped and even, and having
nonzero curvature at the peak and sufficient decay in the tails).
Singularities of $\rho^0(\eta)$ off of the imaginary interval $(0,iA)$
represent obstructions to the free positioning of the contour $C$.
However, we choose to neglect the possibility that these singularities
could constrain our analysis at this point; of course there are
nontrivial cases (like the case of the Satsuma-Yajima soliton
ensemble, for which $\rho^0(\eta)\equiv i$) where $\rho^0(\eta)$ is
{\em entire}.  With the assumption that $\rho^0(\eta)$ is analytic, we
may rewrite the integrals over the imaginary axis for $\lambda\in C_I$
as follows:
\begin{equation}
\int_{0}^{iA}L^0_\eta(\lambda)\rho^0(\eta)\,d\eta = 
\left\{
\begin{array}{ll}
\displaystyle\int_{C_I}
L^{C,\sigma}_{\eta-}(\lambda)\rho^0(\eta)\,d\eta\,, &\hspace{0.2 in}
\sigma=+1\,,\\\\
\displaystyle
\int_{C_I} L^{C,\sigma}_{\eta +}(\lambda)\rho^0(\eta)\,d\eta\,,
&\hspace{0.2 in} \sigma=-1\,,
\end{array}\right.
\end{equation}
and similarly, by symmetry,
\begin{equation}
\int_{-iA}^{0}L^0_\eta(\lambda)\rho^0(\eta^*)^*\,d\eta = 
\left\{
\begin{array}{ll}
\displaystyle\int_{C_I^*}
L^{C,\sigma}_{\eta-}(\lambda)\rho^0(\eta^*)^*\,d\eta\,, &\hspace{0.2 in}
\sigma=+1\,,\\\\
\displaystyle
\int_{C_I^*} L^{C,\sigma}_{\eta +}(\lambda)\rho^0(\eta^*)^*\,d\eta\,,
&\hspace{0.2 in} \sigma=-1\,,
\end{array}\right.
\end{equation}
Next, note that
$L^{C,\sigma}_{\eta+}(\lambda)=L^{C,\sigma}_{\eta-}(\lambda)$ for all
$\eta\in C_I\cup C_I^*$ ``below'' $\lambda\in C_I$ (that is,
all $\eta\in C_I^*$ and all $\eta$ in the oriented portion of
$C_I$ from $0$ to $\lambda$), and at the same time
$L^{C,\sigma}_{\eta_+}(\lambda)= 2\pi i +
L^{C,\sigma}_{\eta-}(\lambda)$ for $\eta\in C_I$ ``above'' $\lambda$
(that is, in the oriented portion of $C_I$ from $\lambda$ to $iA$).
This means that for $\lambda\in C_I$, 
\begin{equation}
\begin{array}{l}
\displaystyle
\int_{C_I} L^{C,\sigma}_{\eta\pm}(\lambda)\rho^0(\eta)\,d\eta +
\int_{C_I^*} L^{C,\sigma}_{\eta\pm}(\lambda)\rho^0(\eta^*)^*\,d\eta =
\\\\
\displaystyle
\hspace{0.4 in}
\int_{C_I} \overline{L^{C,\sigma}_\eta}(\lambda)\rho^0(\eta)\,d\eta +
\int_{C_I^*}\overline{L^{C,\sigma}_\eta}(\lambda)\rho^0(\eta^*)^*\,
d\eta 
\pm\pi i\int_\lambda^{iA}\rho^0(\eta)\,d\eta\,,
\end{array}
\end{equation}
with the final integral being taken along $C_I$.  Assembling these
results gives the expression
\begin{equation}
\begin{array}{l}
\displaystyle
\tilde{\phi}^\sigma(\lambda)=
\int_{C_I}\overline{L^{C,\sigma}_\eta}(\lambda)\overline{\rho}^\sigma(\eta)
\,d\eta +
\int_{C_I^*}\overline{L^{C,\sigma}_\eta}(\lambda)\overline{\rho}^\sigma(\eta^*)^*\,d\eta\\\\
\displaystyle
\hspace{0.4 in}+\,\,J(2i\lambda x + 2i\lambda^2 t) -(J(2K+1)+\sigma)i\pi
\int_\lambda^{iA}\rho^0(\eta)\,d\eta\,,
\end{array}
\end{equation}
valid for $\lambda\in C_I$, where we have introduced the {\em
complementary density} \index{density function!complementary} for
$\eta\in C_I$:
\begin{equation}
\overline{\rho}^\sigma(\eta):=\rho^0(\eta)-\rho^\sigma(\eta)\,.
\label{eq:rhobardef}
\end{equation}
As a matter of future convenience, we now determine the arbitrary integer
$K$ that indexes the interpolants of the proportionality constants by
choosing
\begin{equation}
K:=-\frac{1}{2}(J\sigma+1)\,.
\label{eq:mselect}
\end{equation}
Since $J=\pm 1$ and $\sigma=\pm 1$, one is taking either $K=0$ or
$K=-1$.  Consequently the function $\tilde{\phi}$ becomes simply
\begin{equation}
\tilde{\phi}^\sigma(\lambda)=\int_{C_I}\overline{L_\eta^{C,\sigma}}(\lambda)
\overline{\rho}^\sigma(\eta)\,d\eta +
\int_{C_I^*}\overline{L_\eta^{C,\sigma}}(\lambda)
\overline{\rho}^\sigma(\eta^*)^*\,d\eta + 
J(2i\lambda x + 2i\lambda^2 t)\,.
\label{eq:phitilderewrite}
\end{equation}

\begin{remark}
Note that the relation
(\ref{eq:mselect}) implies that for all four combinations of values
for $J$ and $\sigma$, we have
\begin{equation}
i^J(-1)^K\sigma = -i\,,
\end{equation}
which we assumed in stating the outer model Riemann-Hilbert
Problem~\ref{rhp:model}.
\end{remark}

Until we restore our interest in the inequalities, our task is to find
a system of $G/2+1$ bands and $G/2+1$ gaps on $C_I$ such that
\begin{enumerate}
\item
The initial part of $C_I$ is contained in a band, and the final part
of $C_I$ is contained in a gap,
\item
In each band of $C_I$, $\tilde{\phi}^\sigma(\lambda)$ is
pure imaginary and independent of $\lambda$, while
$\rho^\sigma(\eta)\,d\eta$ is a real differential, and
\item
In each gap of $C_I$, $\rho^\sigma(\eta)\equiv 0$, or equivalently,
$\overline{\rho}^\sigma(\eta)\equiv \rho^0(\eta)$\,.
\end{enumerate}

\subsection{The scalar boundary value problem for genus $G$.
Moment conditions.}
\label{sec:bvp}
Let us begin by imposing just two of the conditions:
\begin{equation}
\begin{array}{rcll}
\displaystyle\frac{\partial\tilde{\phi}^\sigma}
{\partial\lambda}(\lambda)
&\equiv& 0\,,&\lambda \mbox{ in a band of }C_I\,,\\\\
\rho^\sigma(\lambda)&\equiv & 0\,, & \lambda \mbox{ in a gap of }C_I\,.
\end{array}
\label{eq:rawconditions}
\end{equation}
In the first condition, by the derivative with respect to $\lambda$ we
mean the derivative along the contour $C_I$.  
We want to think of these as equations for the unknown function
$\overline{\rho}^\sigma(\eta)$ for $\eta\in C_I$.  What auxiliary
properties do we demand of any solution
$\overline{\rho}^\sigma(\eta)$?  We recall that the analysis in
Chapter~\ref{sec:asymptoticanalysis} required of $\rho^\sigma(\eta)$ that
\begin{enumerate}
\item
$\rho^\sigma(\eta)$ should admit analytic continuation to the left
and right of any band or gap (this is of course trivial in the gaps
where $\rho^\sigma(\eta)\equiv 0$), 
\item
$\rho^\sigma(\eta)$ should take a finite value in the limit
$\eta\rightarrow 0$ for $\eta\in C_I$, and
\item
$\rho^\sigma(\eta)$ should vanish exactly like a square root at each
band endpoint $\lambda_k$ and $\lambda_k^*$.
\end{enumerate}
{\em We now recall the fact that our assumption that the function
$A(x)$ decays sufficiently rapidly for large $|x|$ guarantees via the
definition (\ref{eq:WKBdensity}) that the function $\rho^0(\eta)$ is
bounded as $\eta\rightarrow 0$ for $\eta\in C_I$.}  Then, conditions
1--3 above on $\rho^\sigma(\eta)$ imply in particular that
\begin{equation}
\overline{\rho}^\sigma(\eta) \mbox{ is uniformly H\"older 
continuous on $C_I$ with exponent $1/2$.}
\label{eq:HolderHalf}
\end{equation}
We further
impose the condition that the limit as $\eta$ tends to zero along
$C_I$ of $\overline{\rho}^\sigma(\eta)$ is real:
\begin{equation}
\overline{\rho}^\sigma(0)\in{\mathbb R}\,.
\label{eq:Continuity}
\end{equation}
In this case, $\overline{\rho}^\sigma(\eta)$ extends by the definition
$\overline{\rho}^\sigma(\eta^*)^*$ for $\eta\in C_I^*$ to a function
that satisfies the H\"older condition with exponent $1/2$ on the whole
contour $C_I\cup C_I^*$.

Suppose that a function $\overline{\rho}^\sigma(\eta)$ is given for
$\eta\in C_I$ satisfying (\ref{eq:HolderHalf}), (\ref{eq:Continuity}),
and for which the corresponding function $\tilde{\phi}^\sigma(\lambda)$
defined by (\ref{eq:phitilderewrite}) satisfies (\ref{eq:rawconditions}).
Let $F(\lambda)$ be the Cauchy integral:
\begin{equation}
F(\lambda):=\int_{C_I}\frac{\overline{\rho}^\sigma(\eta)\,d\eta}{
\lambda-\eta} +
\int_{C_I^*}\frac{\overline{\rho}^\sigma(\eta^*)^*
\,d\eta}{\lambda-\eta}\,.
\label{eq:Cauchy}
\end{equation}
Then, from the Plemelj-Sokhotski formula \index{Plemelj-Sokhotski
formula} the function $\overline{\rho}^\sigma(\eta)$ is recovered as:
\begin{equation}
\overline{\rho}^\sigma(\eta)=
-\frac{1}{2\pi i}(F_+(\eta)-F_-(\eta))\,,
\label{eq:hatrhofromF}
\end{equation}
and $F(\lambda)$ has the following properties:
\begin{enumerate}
\item
$F(\lambda)$ is analytic for 
$\lambda\in{\mathbb C}\setminus(C_I\cup C_I^*)$.
\item
$F(\lambda)$ satisfies the decay condition
\begin{equation}
F(\lambda)=\bo(1/\lambda)\,,\hspace{0.3 in}\lambda\rightarrow\infty\,.
\label{eq:milddecay}
\end{equation}
\item
For all $\lambda$ in the domain of analyticity, $F(\lambda)$
satisfies the symmetry property
\begin{equation}
F(\lambda^*)=-F(\lambda)^*\,.
\label{eq:Fsymmetry}
\end{equation}
\item
The boundary values taken by $F(\lambda)$ on both sides of
$C_I\cup C_I^*$ are H\"older continuous with exponent $1/2$ and for
$\lambda\in C_I$ satisfy
\begin{equation}
\begin{array}{rclcll}
F_+(\lambda)&+&
F_-(\lambda)&=&-4iJ(x+2\lambda
t)\,,& \lambda \mbox{ in a band}\,,\\\\
F_+(\lambda)&-&F_-(\lambda)&=&-2\pi
i\rho^0(\lambda)\,, &\lambda \mbox{ in a gap}\,.
\end{array}
\label{eq:BVP}
\end{equation}
\end{enumerate}
These properties follow
from well-known properties of
Cauchy integrals for H\"older continuous functions \cite{M53}
and from the Plemelj-Sokhotski formula.

Conversely, we define a boundary-value problem as follows.  Seek a
function $F(\lambda)$ that satisfies conditions $1$, $2$, and
$4$ above, and additionally takes H\"older $1/2$ boundary
values for $\lambda\in C_I^*$ that satisfy the conjugate boundary
conditions:
\begin{equation}
\begin{array}{rclcrcll}
F_+(\lambda)&+&F_-(\lambda)&=&-(F_+(\lambda^*)^*&+&
F_-(\lambda^*)^*)\,,&\lambda^* \mbox{ in a band}\,,\\\\
F_+(\lambda)&-&F_-(\lambda)&=&-(F_+(\lambda^*)^*&-&
F_-(\lambda^*)^*)\,,&\lambda^* \mbox{ in a gap}\,.
\end{array}
\end{equation}
We call this the {\em scalar boundary-value problem for genus
$G$}\index{scalar boundary-value problem for genus $G$}.

\begin{remark}
Although the complex phase function $g^\sigma(\lambda)$ depends on the
index $\sigma$, this parameter enters into the properties of
$F(\lambda)$ only in that for $\sigma=+1$ (respectively $\sigma=-1$)
the contour $C_I$ lies to the right (respectively left) of the imaginary
interval $[0,iA]$.  Thus to simplify notation we will not reproduce the 
superscript $\sigma$ on the function $F(\lambda)$.
\end{remark}

\begin{lemma}
There exists at most one solution of the scalar boundary-value problem for
genus $G$.  If it exists, the solution satisfies the
symmetry property (\ref{eq:Fsymmetry}), and the function
$\overline{\rho}^\sigma(\eta)$ defined for $\lambda\in C_I$ by
(\ref{eq:hatrhofromF}) gives rise to a function
$\tilde{\phi}^\sigma(\lambda)$ via (\ref{eq:phitilderewrite}) that
satisfies (\ref{eq:rawconditions}) and $\overline{\rho}^\sigma(0)\in
{\mathbb R}$.
\label{lemma:uniqueF}
\end{lemma}

\begin{proof}  The uniqueness and the symmetry condition (\ref{eq:Fsymmetry})
are proved in exactly the same way.  Consider the related boundary
value problem of seeking a function $Z(\lambda)$ that is analytic in
$C_I\cup C_I^*$ and satisfies (\ref{eq:milddecay}), and takes H\"older
continuous boundary values on both sides of $C_I\cup C_I^*$ with exponent
$1/2$ that satisfy $Z_+(\lambda)+Z_-(\lambda)\equiv 0$ for $\lambda$ or
$\lambda^*$ in a band of $C_I$ and $Z_+(\lambda)-Z_-(\lambda)\equiv 0$ for
$\lambda$ or $\lambda^*$ in a gap of $C_I$.  
Recall the function $R(\lambda)$, first used in
\S\ref{sec:outersolve}, that
satisfies
\begin{equation}
R(\lambda)^2 = \prod_{k=0}^G (\lambda-\lambda_k)(\lambda-\lambda_k^*)\,,
\end{equation}
is analytic except at the bands of $C_I\cup C_I^*$ (where the branch
cuts are placed), and satisfies $R(\lambda)\sim -\lambda^{G+1}$ as
$\lambda\rightarrow\infty$.  With this choice, the boundary value $R_+(0)$ relative to the oriented contour $C_I\cup C_I^*$ is the positive
value 
\begin{equation}
R_+(0)=\prod_{k=0}^G|\lambda_k|^2\,.
\label{eq:Rpluszero}
\end{equation}
The function defined from any solution of this boundary value problem
by the formula $W(\lambda):=Z(\lambda)/R(\lambda)$ is again analytic
in ${\mathbb C}\setminus (C_I\cup C_I^*)$ with at worst inverse square-root
singularities at the endpoints of the bands and gaps on $C_I\cup C_I^*$.
The boundary values on $C_I\cup C_I^*$, continuous except possibly at
the isolated band and gap endpoints, satisfy $W_+(\lambda)=W_-(\lambda)$.
The function $W(\lambda)$ decays like $\bo(\lambda^{-(G+2)})$ as $\lambda
\rightarrow\infty$.  The relatively mild nature of the singularities,
together with the agreement of the boundary values where they are continuous
implies that $W(\lambda)$ is actually entire, and then by Liouville's theorem
the decay condition implies that $W(\lambda)\equiv 0$.  It then follows
that $Z(\lambda)\equiv 0$, so that all solutions are trivial.  To use this
result to prove uniqueness for $F(\lambda)$, one considers $Z(\lambda)$
to be the difference of two solutions and finds that this difference satisfies
the above problem and is therefore zero.  Similarly, to prove the symmetry
property (\ref{eq:Fsymmetry}), one considers the function $Z(\lambda)=
F(\lambda)+F(\lambda^*)^*$ and again finds that $Z\equiv 0$.

Now, the function defined for $\eta\in C_I\cup C_I^*$ by 
$u(\eta):=-(F_+(\eta)-F_-(\eta))/(2\pi i)$ is by construction
H\"older continuous on $C_I\cup C_I^*$ with exponent $1/2$ and
satisfies $u(\eta^*)^*=u(\eta)$.  It follows that $u(0)\in{\mathbb R}$.  The
proof is complete upon observing that by continuity of boundary values
and decay at infinity, $F(\lambda)$ necessarily
agrees with the Cauchy integral of the function $u(\eta)$.  That is,
the relation
\begin{equation}
F(\lambda)=\int_{C_I\cup C_I^*}\frac{u(\eta)\,d\eta}{\lambda-\eta}
\end{equation}
is a consequence of Liouville's theorem.  It follows that the function
$\overline{\rho}^\sigma(\eta):=u(\eta)$ for $\eta\in C_I$ leads to a
solution of (\ref{eq:rawconditions}) as required.
\end{proof}

So, it is sufficient to seek a solution to the scalar boundary-value
problem for genus $G$, and we know that the solution will be unique
{\em if it exists}.  This will only be the case if the endpoints of
the bands and gaps satisfy certain explicit conditions.
\begin{lemma}
Fix $J=\pm 1$, $\sigma=\pm 1$, and an even integer $G\ge 0$.  Let a
contour $C_I$ be given in the upper half-plane that connects $0$ to
$iA$ and whose interior points all lie to the right (respectively
left) of the vertically oriented segment $[0,iA]$ for $\sigma=+1$
(respectively $\sigma=-1$), and let points $\lambda_0,\dots,\lambda_G$
be given in order on $C_I$.  Let the bands and gaps on $C_I\cup C_I^*$
separated by these points be denoted according to the scheme
illustrated in Figure~\ref{fig:jumpsforO}, and let $\Gamma_I$ denote
$(\cup_k \Gamma_k^\pm)\cap (C_I\cup C_I^*)$.  Then, there exists a
unique solution to the scalar boundary value problem for genus $G$ if
and only if for $p=0,\dots,G$, the endpoints satisfy the real
equations
\begin{equation}
M_p:=J\int_{\cup_kI_k^\pm}\frac{2ix+4i\eta t}{R_+(\eta)}\eta^p\,d\eta +
2\Re\left(\int_{\Gamma_I\cap C_I}\frac{\pi i\rho^0(\eta)}{R(\eta)}\eta^p\,d\eta\right)=0\,.
\label{eq:momenteqns}
\end{equation}
Moreover the function $\overline{\rho}^\sigma(\eta)$ defined from the solution
$F(\lambda)$ by (\ref{eq:hatrhofromF}) is
analytic in the interior of each band $I_k^\pm$ and each component of
$\Gamma_I$.
\label{prop:moments}
\end{lemma}

\begin{proof}
The proof is by direct construction.  In the scalar boundary-value problem 
for genus $G$, use $R(\lambda)$ to make the
change of variables 
\begin{equation}
H(\lambda):=\frac{F(\lambda)}
{R(\lambda)}\,.
\label{eq:HF}
\end{equation}
Note that $R(\lambda^*)=R(\lambda)^*$, and since $R(\lambda)$ is analytic
in the gaps while in the bands satisfies $R_+(\lambda)+R_-(\lambda)=0$,
the conditions satisfied by the boundary values of $H(\lambda)$
on the oriented contour $C_I$ are then
\begin{equation}
H_+(\lambda)-H_-(\lambda)=\left\{
\begin{array}{ll}
\displaystyle
\frac{-4iJ(x+2\lambda t)}
{R_+(\lambda)}
\,,& \lambda \mbox{ in a band}\,,\\\\
\displaystyle
-\frac{2\pi i\rho^0(\lambda)}
{R(\lambda)}\,,&\lambda \mbox{ in a gap}\,,
\end{array}\right.
\label{eq:Hjumpsup}
\end{equation}
while for $\lambda\in C_I^*$, the conjugate boundary conditions hold:
\begin{equation}
H_+(\lambda)-H_-(\lambda)=-(H_+(\lambda^*)^*-
H_-(\lambda^*)^*)\,.
\label{eq:Hjumpsdown}
\end{equation}

Since the quotient $H(\lambda)$ defined by (\ref{eq:HF})
necessarily decays at infinity if $F(\lambda)$, exists, and has
at worst inverse square-root singularities at the isolated endpoints
$\lambda_k$ and $\lambda_k^*$, by the same kind of reasoning as in the
proof of Lemma~\ref{lemma:uniqueF} it follows that $H(\lambda)$
must agree with the Cauchy integral of the difference of its boundary
values (\ref{eq:Hjumpsup}) and (\ref{eq:Hjumpsdown}).  That is, 
$H(\lambda)$ must be given by
\begin{equation}
H(\lambda)=
\frac{1}{\pi i} \int_{\cup_k
I_k^\pm}\frac{2iJ(x + 2\eta t)}{(\lambda-\eta)R_+(\eta)}\, d\eta +
\frac{1}{\pi i}\int_{\Gamma_I\cap C_I}
\frac{\pi i\rho^0(\eta)}{(\lambda-\eta)R(\eta)}\,
d\eta
+
\frac{1}{\pi i}\int_{\Gamma_I\cap C_I^*}
\frac{\pi i\rho^0(\eta^*)^*}{(\lambda-\eta)R(\eta)}\,
d\eta\,.
\label{eq:Hsolution}
\end{equation}

Now, the function $F(\lambda)$ defined from such a solution
$H(\lambda)$ by (\ref{eq:HF}) will only satisfy the decay condition
$F(\lambda)=\bo(1/\lambda)$ as $\lambda\rightarrow\infty$ if
$H(\lambda)=\bo(
\lambda^{-(G+2)})$ in the same limit.
Expanding the explicit formula (\ref{eq:Hsolution}) for large
$\lambda$ gives a Laurent series whose leading term is $\bo
(1/\lambda)$.  Therefore, for existence it is necessary that the first
$G+1$ coefficients in this Laurent series vanish identically.  These
coefficients are computed as {\em moments} of the densities.
Expanding $(\lambda-\eta)^{-1}$ inside the integrals in geometric
series, one sees that the Laurent series of $H(\lambda)$,
convergent for $|\lambda|$ sufficiently large, is
\begin{equation}
H(\lambda)=\frac{1}{\pi i}\sum_{p=0}^\infty \frac{M_p}{\lambda^{p+1}}\,,
\label{eq:Laurent}
\end{equation}
where the quantities $M_p$, easily seen to be real-valued by using the
complex-conjugation symmetry of the contours, are defined in
(\ref{eq:momenteqns}).  Thus, the conditions for existence are exactly
the {\em moment conditions} \index{moment conditions} recorded in
(\ref{eq:momenteqns}).

When the moment conditions are satisfied, the function $F(\lambda):=
H(\lambda)R(\lambda)$ is the solution (unique, by
Lemma~\ref{lemma:uniqueF}) of the scalar boundary-value problem for
genus $G$.  The claimed analyticity of $\overline{\rho}^\sigma(\eta)$
defined by (\ref{eq:hatrhofromF}) then follows from the explicit
formula (\ref{eq:Hsolution}) for $H(\lambda)$ and the analyticity of
the boundary values of $R(\lambda)$.
\end{proof}

\begin{remark}
Given the endpoints $\lambda_0,\dots,\lambda_G$, the moments $M_p$
have the same value for any contour $C_I$ that connects $0$ to $iA$ in
the domain of analyticity of $\rho^0(\eta)$ and interpolates these
points in order.  This follows because the integrands have analytic
continuations to either side of the contour $C_I\cup C_I^*$.
Therefore, the moments $M_p$ are functions of the {\em ordered
sequence} of endpoints $\lambda_0,\dots,\lambda_G$ alone and thus {\em
the moment conditions are only constraints on the endpoints} and not
on the interpolating contour.  This statement will be strengthened
in \S\ref{sec:symmetry}, when we show that the ordering is irrelevant;
the moments will in fact be seen to be symmetric functions of the endpoints.
\end{remark}

\begin{remark}
The moment conditions amount to $G+1$ real equations for $G+1$ complex
unknowns.  Intuitively, one expects the solution space is expected to
be $G+1$ real-dimensional.  On the other hand, proving the existence
of solutions is a nontrivial analytical issue.  Indeed, in the
simplest of cases solutions can fail to exist for given $x$ and $t$
except for certain values of the parameters $J$ and $\sigma$.
\end{remark}

\begin{remark}
Even when the endpoints do not satisfy the moment conditions
(\ref{eq:momenteqns}), we will still consider a function
$F(\lambda)$ to be defined by the relation
$F(\lambda):=H(\lambda)R(\lambda)$ with
$H(\lambda)$ given by the explicit formula
(\ref{eq:Hsolution}).  Exactly the same arguments as in the proof of
Lemma~\ref{lemma:uniqueF} then show that the function so-defined for
arbitrary endpoints is the unique solution of the corresponding
boundary-value problem in which the decay condition
(\ref{eq:milddecay}) is replaced with the growth condition
\begin{equation}
F(\lambda)=\bo(\lambda^G)\,,\hspace{0.3 in}\lambda\rightarrow\infty\,,
\end{equation}
with all other conditions on $F(\lambda)$ exactly as before.
The moment conditions (\ref{eq:momenteqns}) are then interpreted as the
conditions that must be satisfied in order for $F(\lambda)$ to
have the form of a Cauchy integral (\ref{eq:Cauchy}) corresponding
to the complementary density $\overline{\rho}^\sigma(\eta)$.
\end{remark}

The solution $\overline{\rho}^\sigma(\lambda)$ given by
(\ref{eq:hatrhofromF}) can be calculated explicitly.  One can check
that, as specified, the formula gives
$\overline{\rho}^\sigma(\lambda)\equiv \rho^0(\lambda)$ for all
$\lambda$ in $\Gamma_I\cap C_I$.  For $\lambda$ in any band $I_k^+$ of
$C_I$, the same formula (\ref{eq:hatrhofromF}) gives
\begin{equation}
\overline{\rho}^\sigma(\lambda)=\frac{4Jt}{\pi}R_+(\lambda)\delta_{G,0}-
\frac{R_+(\lambda)}{\pi
i}\int_{\Gamma_I\cap C_I}\frac{\rho^0(\eta)\,d\eta}
{(\lambda-\eta)R(\eta)}
-
\frac{R_+(\lambda)}{\pi
i}\int_{\Gamma_I\cap C_I^*}\frac{\rho^0(\eta^*)^*\,d\eta}
{(\lambda-\eta)R(\eta)}\,.
\label{eq:rhobarbandformula}
\end{equation}
Using the symmetry relation (\ref{eq:Rpluszero}), it follows from this
formula that the condition (\ref{eq:Continuity}) holds.

\begin{remark}
This latter calculation takes advantage of the useful fact that:
\begin{equation}
\int_{\cup_kI_k^\pm}\frac{f(\eta)\,d\eta}{R_+(\eta)} = 
-\frac{1}{2}\oint \frac{f(\eta)\,d\eta}{R(\eta)}\,,
\end{equation}
for all meromorphic $f(\eta)$ where the integral on the right hand
side is over a closed counterclockwise-oriented contour surrounding
all the bands.  This allows such integrals to be evaluated in many
cases exactly by residues.  Such a procedure can be applied to the
first term in $M_p$, for example.  
\end{remark}

From these explicit formulae for the complementary density
$\overline{\rho}^\sigma(\lambda)$, we can of course write down the
corresponding formulae for the density $\rho^\sigma(\lambda)$.
In the gaps of $C_I$ one verifies that $\rho^\sigma(\lambda)\equiv 0$,
while in the bands of $C_I$,
\begin{equation}
\rho^\sigma(\lambda)=
\rho^0(\lambda)-\frac{4Jt}{\pi}R_+(\lambda)\delta_{G,0}+
\frac{R_+(\lambda)}{\pi
i}\int_{\Gamma_I\cap C_I}\frac{\rho^0(\eta)\,d\eta}
{(\lambda-\eta)R(\eta)}
+
\frac{R_+(\lambda)}{\pi
i}\int_{\Gamma_I\cap C_I^*}\frac{\rho^0(\eta^*)^*\,d\eta}
{(\lambda-\eta)R(\eta)}\,.
\label{eq:rhoformula}
\end{equation}
For $\lambda$ on the conjugate contour $C_I^*$, the function
$\rho^\sigma(\lambda)$ is defined by conjugation:
$\rho^\sigma(\lambda):=
\rho^\sigma(\lambda^*)^*$. 

This formula has the following useful property.
\begin{lemma}
The formula (\ref{eq:rhoformula}) can be written in the form
\begin{equation}
\rho^\sigma(\lambda)=R_+(\lambda)Y(\lambda)\,,
\label{eq:rhowithY}
\end{equation}
where $Y(\lambda)$ is analytic in a simply-connected region of the
upper half-plane containing all bands $I_k^+$ in $C_I$, including the
endpoints $\lambda_0,\dots,\lambda_G$, provided $\lambda_G\neq iA$.
\label{lemma:rhowithY}
\end{lemma}

\begin{proof}
It suffices to show that the sum of the first and third terms in
(\ref{eq:rhoformula}) have this property.  Consider the integral
\begin{equation}
I(\lambda):=\frac{1}{\pi i}\int_{\Gamma_I\cap C_I}\frac{\rho^0(\eta)\,d\eta}
{(\lambda-\eta)R(\eta)}\,.
\end{equation}
This integral defines an analytic function of $\lambda\in {\mathbb C}\setminus
(\Gamma_I\cap C_I)$.  Therefore for $\lambda$ in a band $I_k^+$, we have
in particular
\begin{equation}
I(\lambda)=\lim_{\mu\rightarrow\lambda} I(\mu)\,,
\label{eq:mutolambda}
\end{equation}
where for concreteness we suppose $\mu$ to lie to the left of the band
$I_k^+$.  For such $\mu$, we can augment the contour of integration by
writing
\begin{equation}
I(\mu)=\frac{1}{\pi i}\int_{\Gamma_I\cap C_I}\frac{\rho^0(\eta)\,d\eta}
{(\mu-\eta)R(\eta)} 
+ \frac{1}{2\pi i}\int_{\cup_k I_k^+}\frac{\rho^0(\eta)\,d\eta}{(\mu-\eta)R_+(\eta)}
+ \frac{1}{2\pi i}\int_{\cup_k I_k^+}\frac{\rho^0(\eta)\,d\eta}{(\mu-\eta)R_-(\eta)}\,,
\end{equation}
since $\mu$ is not contained in any band $I_k^+$ and since
$R_+(\eta)+R_-(\eta)=0$.  By standard analyticity deformations, this
expression can be written as
\begin{equation}
I(\mu)=\frac{1}{2\pi i}\int_{C_{I+}\cup C_{I-}}\frac{\rho^0(\eta)\,d\eta}
{(\mu-\eta)R(\eta)}\,,
\end{equation}
where $C_{I\pm}$ are contours lying just to the left and right of $C_I$ (see
Figure~\ref{fig:CIpm}).
\begin{figure}[h]
\begin{center}
\mbox{\psfig{file=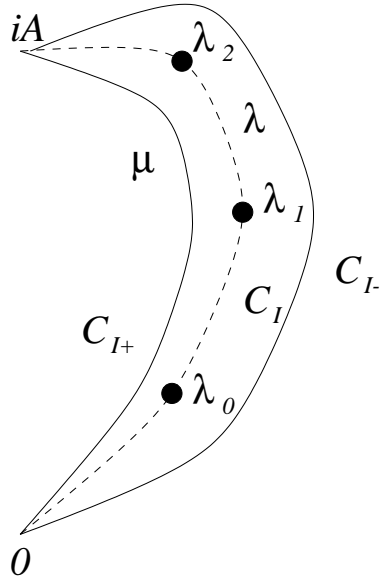,width=2 in}}
\end{center}
\caption{\em The contours $C_{I\pm}$.  The point $\mu$ is outside the enclosed
region and $\lambda\in I_k^+$ is inside.  All three contours $C_I$, $C_{I+}$, 
and $C_{I-}$ are oriented from $0$ to $iA$.}
\label{fig:CIpm}
\end{figure}
Now, passing to the limit (\ref{eq:mutolambda}), there is a residue
contribution as $\mu$ crosses $C_{I+}$, and one obtains the formula
\begin{equation}
I(\lambda)=-\frac{\rho^0(\lambda)}{R_+(\lambda)} +
\frac{1}{2\pi i}\int_{C_{I+}\cup C_{I-}}\frac{\rho^0(\eta)\,d\eta}
{(\lambda-\eta)R(\eta)}\,.
\end{equation}
Using this expression for $I(\lambda)$ in the formula (\ref{eq:rhoformula})
for $\rho^\sigma(\lambda)$, we finally obtain the desired representation
with
\begin{equation}
Y(\lambda):= -\frac{4Jt}{\pi}\delta_{G,0} + 
\frac{1}{2\pi i}\int_{C_{I+}\cup C_{I-}}\frac{\rho^0(\eta)\,d\eta}
{(\lambda-\eta)R(\eta)} +
\frac{1}{\pi i}\int_{\Gamma_I\cap C_I^*}\frac{\rho^0(\eta^*)^*\,d\eta}
{(\lambda-\eta)R(\eta)}\,.
\label{eq:Ydef}
\end{equation}
This function is clearly analytic for all $\lambda$ in between 
$C_{I+}$ and $C_{I-}$, which completes the proof.  Note that the final
term in this formula for $Y(\lambda)$ can be rewritten:
\begin{equation}
\frac{1}{\pi i}\int_{\Gamma_I\cap C_I^*}\frac{\rho^0(\eta^*)^*\,d\eta}
{(\lambda-\eta)R(\eta)}=\frac{1}{2\pi i}\int_{C_{I+}^*\cup C_{I-}^*}
\frac{\rho^0(\eta^*)^*\,d\eta}{(\lambda-\eta)R(\eta)}\,,
\end{equation}
where the conjugate contours are presumed to be oriented from $-iA$
toward the origin.  With this subsitution, the function $Y(\lambda)$
is also defined and analytic for $\lambda$ in the lower half-plane between
$C_{I+}$ and $C_{I-}$, where it satisfies $Y(\lambda^*)=Y(\lambda)^*$.
\end{proof}

\begin{remark}
Since it will turn out that when $x=t=0$ the $G=0$ ansatz satisfies
$\lambda_0=iA$, this representation of the density
$\rho^\sigma(\lambda)$ does not hold at this point in the
$(x,t)$-plane.
\end{remark}

From the formula (\ref{eq:rhowithY}), it is clear that
$\rho^\sigma(\lambda)$ vanishes at least like a square root at the
band endpoints.  Higher-order vanishing of $\rho^\sigma(\lambda)$ at
an endpoint $\lambda_k$ corresponds to a zero of the analytic function
$Y(\lambda)$ at $\lambda=\lambda_k$.  Since $\rho^\sigma(\lambda)$ is
identically zero in the gaps, the candidate density \index{density
function!candidate} $\rho^\sigma(\lambda)$ is continuous at the
nonzero endpoints $\lambda_0,\dots,\lambda_G$.  These formulae
therefore desingularize the expression (\ref{eq:rhoformula}) for the
candidate density.

Finally, recall that $\overline{\rho}^\sigma(\lambda)$ satisfies
(\ref{eq:Continuity}) and therefore extends by the definition
$\overline{\rho}^\sigma(\lambda^*)=\overline{\rho}^\sigma(\lambda)^*$
to a continuous function on $C_I\cup C_I^*$.  Similarly, being a
density function for a real measure on the imaginary axis, the
function $\rho^0(\lambda)$ satisfies $\rho^0(0)\in i{\mathbb R}$ and
therefore does not generally extend in this way to all of $C_I\cup
C_I^*$.  Consequently, the symmetric extension of $\rho^\sigma(\lambda)$ to
the contour $C_I\cup C_I^*$ is generally discontinuous at $\lambda=0$.
The origin is its {\em only} point of discontinuity.

\begin{remark}
An important observation that goes back to the papers of Lax and Levermore
\cite{LL83} is that the partial derivatives of the function $F(\lambda)$
with respect to the parameters $x$ and $t$ satisfy very simple
boundary value problems.  As an application we will refer to in
Chapter~\ref{sec:genuszero}, let us obtain simple formulae for the functions
$\partial F/\partial x$ and $\partial F/\partial t$
valid for the assumption of $G=0$.  

First consider $\partial F/\partial x$.  By differentiating the
jump relations (\ref{eq:BVP}), one observes that $\partial
F/\partial x$ is analytic in the whole $\lambda$-plane except
for $\lambda\in I_0$, where we have
\begin{equation}
\frac{\partial F_+}{\partial x}(\lambda)+\frac{\partial F_-}
{\partial x}
(\lambda) \equiv -4iJ\,.
\end{equation}
Also, $\partial F/\partial x$ has at worst inverse square-root
singularities at $\lambda_0$ and $\lambda_0^*$, and decays like
$1/\lambda$ for large $|\lambda|$.  This simple problem is solved by
defining a new unknown $X(\lambda)$ according to
\begin{equation}
\frac{\partial F}{\partial x}(\lambda) = \frac{X(\lambda)}
{R(\lambda)}\,,
\end{equation}
where $R(\lambda)$ is the square-root function first defined in
\S\ref{sec:outersolve}.  Then,
$X(\lambda)$ is analytic except for $\lambda\in I_0$ where it
satisfies the jump relation
\begin{equation}
X_+(\lambda)-X_-(\lambda)=-4iJR_+(\lambda)\,.
\end{equation}
This problem is solved by a Cauchy integral that can be evaluated explicitly
by residues:
\begin{equation}
X(\lambda)=
\frac{1}{2\pi i}\int_{I_0}\frac{4iJR_+(\eta)\,d\eta}{\lambda-\eta}
=-2iJR(\lambda) +iJ (\lambda_0+\lambda_0^* - 2\lambda)\,.
\end{equation}
Here, the term proportional to $R(\lambda)$ comes from a residue at
$\eta=\lambda$, and the remaining terms come from a residue at $\eta=\infty$
tailored for the special case of genus $G=0$.  Therefore, we find the formula
\begin{equation}
\frac{\partial F}{\partial x}(\lambda)=
-2iJ + \frac{2iJ}{R(\lambda)}\left(
\frac{\lambda_0+\lambda_0^*}{2}-\lambda\right)\,.
\end{equation}
It is easy to verify that this simple formula satisfies the jump
relations exactly.  Now, using the explicit formula for $R(\lambda)$
valid for genus $G=0$, one sees that in fact, 
\begin{equation}
\frac{\partial F}{\partial x}(\lambda)=-2iJ\left(1+\frac{\partial
R}{\partial\lambda}(\lambda)\right)\,.  
\end{equation}

Next, consider $\partial F/\partial t$ for $G=0$, which is
analytic except for $\lambda\in I_0$, where
\begin{equation}
\frac{\partial F_+}{\partial t}(\lambda)+\frac{\partial F_-}
{\partial t}(\lambda)\equiv -8iJ\lambda\,.
\end{equation}
Again, $\partial F/\partial t$ can have at worst inverse square-root
singularities at $\lambda_0$ and $\lambda_0^*$ and must decay like $1/\lambda$.
One solves this problem in a similar way, introducing a new unknown
$T(\lambda)$ by 
\begin{equation}
\frac{\partial F}{\partial t}(\lambda)=\frac{T(\lambda)}
{R(\lambda)}\,,
\end{equation}
and then expressing $T(\lambda)$ in terms of a Cauchy integral
over $I_0$ that one evaluates by residues.  The final result one
obtains for $G=0$ is the formula
\begin{equation}
\frac{\partial F}{\partial t}(\lambda)=-iJ\left(4\lambda +
\frac{\partial}{\partial\lambda}
\left[(2\lambda+\lambda_0+\lambda_0^*)R(\lambda)\right]\right)\,.
\end{equation}

These formulae show that the partial derivatives of $F$ with
respect to $x$ and $t$ actually decay faster at infinity than
originally supposed; they are both $\bo(1/\lambda^2)$. By following
similar reasoning, explicit formulae may be obtained easily for
derivatives of $F$ with respect to $x$ and $t$ for larger
values of $G$.
\end{remark}
 
\subsection{Ensuring $\Re(\tilde{\phi}^\sigma)=0$ in the bands.
Vanishing conditions.}

We now turn to the question of determing what additional constraints
are required to ensure that the constant value of
$\tilde{\phi}^\sigma(\lambda)$ in each band $I_k^\pm$ is in fact
purely imaginary.  Since $\overline{\rho}^\sigma(\eta)$ satisfies the
condition (\ref{eq:Continuity}), it follows from
(\ref{eq:phitilderewrite}) that $\tilde{\phi}^\sigma(\lambda)$ has a
finite limit as $\lambda$ tends to zero in $C_I$.  It then follows
directly from (\ref{eq:phitilderewrite}) that this limiting value is
purely imaginary.  Consequently throughout the band $I_0=I_0^+\cup
I_0^-$, $\tilde{\phi}^\sigma(\lambda)$ is automatically a purely
imaginary constant.  If one is constructing a genus zero ansatz, then
there are no more bands to consider.  However, generally one must
enforce the condition $\Re(\tilde{\phi}^\sigma)= 0$ in the other
bands.  This amounts to further conditions on the endpoints
$\lambda_0,\dots,\lambda_G$, conditions we refer to as {\em vanishing
conditions}\index{vanishing conditions}.

\begin{lemma}
Assume all the conditions of Lemma~\ref{prop:moments} and suppose also
that the endpoints $\lambda_0,\dots,\lambda_G$ satisfy the additional
constraints
\begin{equation}
V_k:=\Re\left(\int_{\lambda_{2k}}^{\lambda_{2k+1}}\left[2iJx+4iJ\lambda t +
\frac{1}{2}(F_+(\lambda)+F_-(\lambda))
\right]\,d\lambda\right) = 0\,,
\hspace{0.2 in}k=0,\dots,G/2-1\,.
\label{eq:vanishing}
\end{equation}
Then, the complementary density function
$\overline{\rho}^\sigma(\eta)$ characterized by
Lemma~\ref{prop:moments} has the property that the
associated function $\tilde{\phi}^\sigma(\lambda)$ defined by
(\ref{eq:phitilderewrite}) agrees with a purely imaginary constant in
each band of $C_I$.
\label{prop:vanishing}
\end{lemma}

\begin{proof}
Starting from a terminal endpoint $\lambda_{2k}$ of a band $I_k^+$ in
which the condition is satisfied, we can ensure that the condition is
satisfied as well in the next band along $C_I$ by integrating
$d\tilde{\phi}^\sigma$ along the intermediate gap $\Gamma_{k+1}^+$,
and insisting that the real part vanish.  It is easy to see that since
$\tilde{\phi}^\sigma(\lambda)$ is defined in terms of the average of
the boundary values of a logarithmic integral, its derivative with
respect to $\lambda$ is the average of boundary values of a Cauchy
integral, which explains the integrand in the expression
(\ref{eq:vanishing}).  Although the integral is taken over the gap
$\Gamma_{k+1}^+$ of the contour $C_I$ between the endpoints
$\lambda_{2k}$ and $\lambda_{2k+1}$, the conditions
(\ref{eq:vanishing}) depend only on the ordered sequence of endpoints
$\lambda_0,\dots,\lambda_G$ and not on the particular contour gaps
$\Gamma^+_{k+1}$.  This is because the integrand has an analytic
continuation from each gap of $C_I$ to either side; using the jump
condition satisfied by the boundary values of $F(\lambda)$ in
the gaps, one finds that
\begin{equation}
\frac{1}{2}(F_+(\lambda)+F_-(\lambda))=
F_\pm(\lambda)\pm\pi i\rho^0(\lambda)\,.
\end{equation}
Therefore, the integrand continues to the left as $2iJx+4iJ\lambda
t+F(\lambda)+\pi i\rho^0(\lambda)$ and to the right as
$2iJx+4iJ\lambda t +F(\lambda)-\pi i\rho^0(\lambda)$.  
\end{proof}

\begin{remark}
Taken together, the moment conditions and the vanishing conditions place 
$3G/2 + 1$ real constraints on the $G+1$ complex endpoints $\lambda_0,\dots,
\lambda_G$.  Intuitively, one expects the set of admissible endpoints  
to be $G/2+1$ real-dimensional.  Existence of solutions remains to be
shown, but certainly at this point the solution is not expected to be
unique without imposing further conditions.
\end{remark}

\subsection{Determination of the contour bands.  Measure reality
conditions.}  Given $G$, and the choices of $J=\pm 1$ and $\sigma=\pm
1$, there still remains some freedom in specifying the endpoints, and
still the contour bands and gaps that connect the endpoints remain
completely unspecified.  Now, we turn to the question of identifying
further constraints sufficient to ensure that the differential measure
$\rho^\sigma(\eta)\,d\eta$ is real-valued in the bands, where it does
not vanish identically.  As remarked in the discussion of the
conditions on the complex phase function in \S\ref{sec:conditions},
the reality of this differential is as much a condition on the contour
bands (making up its support) through the differential $d\eta$ as
on the function $\rho^\sigma(\eta)$ itself.  We therefore expect
that we may have to choose the bands carefully in order to achieve the
required reality.  In fact, the reality condition further constrains
the endpoints as well.

The function $\rho^\sigma(\lambda)$ is defined in the bands of
$C_I$.  But in each separate oriented band $I_k^+$, it has
an analytic continuation to the left and right sides.  To extend to
the left, write
\begin{equation}
\begin{array}{rcl}
\rho^\sigma(\lambda)&=&\displaystyle \rho^0(\lambda)+\frac{1}{2\pi i}
(F_+(\lambda)-F_-(\lambda))\\\\
&=&\displaystyle\rho^0(\lambda)+\frac{1}{\pi i}F_+(\lambda)-
\frac{1}{2\pi i}(F_+(\lambda)+F_-(\lambda))\\\\
&=&\displaystyle\rho^0(\lambda)+\frac{1}{\pi i}F_+(\lambda)+
\frac{2Jx}{\pi}+
\frac{4J\lambda t}{\pi}\,,
\end{array}
\label{eq:leftextend}
\end{equation}
while to extend to the right, 
\begin{equation}
\begin{array}{rcl}
\rho^\sigma(\lambda)&=&\displaystyle \rho^0(\lambda)+\frac{1}{2\pi i}
(F_+(\lambda)-F_-(\lambda))\\\\
&=&\displaystyle\rho^0(\lambda)-\frac{1}{\pi i}
F_-(\lambda)+\frac{1}{2\pi i}(F_+(\lambda)+
F_-(\lambda))\\\\
&=&\displaystyle\rho^0(\lambda)-
\frac{1}{\pi i}F_-(\lambda)-\frac{2Jx}{\pi}-
\frac{4J\lambda t}{\pi}\,.
\end{array}
\label{eq:rightextend}
\end{equation}
These calculations use the fact that $F_+(\lambda)+F_-(\lambda)$ is
given in (\ref{eq:BVP}) as an explicit analytic function in the bands.
Since $\rho^\sigma(\lambda)$ is the same analytic function no matter
what precise contour is taken to be the band $I_k^+$, the reality
condition
\begin{equation}
\Im(\rho^\sigma(\eta)\,d\eta)=
\Im\left(\rho^\sigma(u+iv)\,(du+i\,dv)\right)=0\,,
\label{eq:ODE}
\end{equation}
{\em may be viewed as a differential equation for the band contour
$I_k^+$ in the real $(u,v)$ plane}\index{band!differential equation
for}.  Given the endpoints, the differential equation (\ref{eq:ODE})
is explicit, with $\rho^\sigma(\eta)$ being given by
(\ref{eq:rhoformula}).

\begin{figure}[h]
\begin{center}
\mbox{\psfig{file=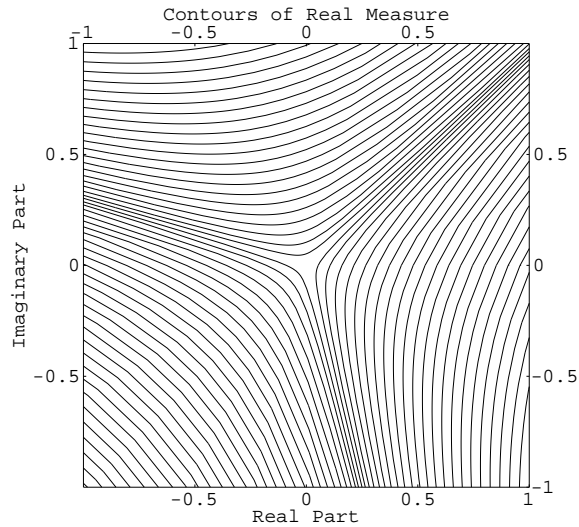,width=3 in}}
\end{center}
\caption{\em The local orbit structure of the differential equation
(\ref{eq:ODE}) for the bands near a nonzero band endpoint.}
\label{fig:trinity}
\end{figure}

The obstruction to using this differential equation to define the
bands $I_k^+$ (and thus the bands $I_k^-$ by complex-conjugation) is
simply that the two endpoints of the band $I_k^+$ might not lie on the
same integral curve of (\ref{eq:ODE}).  However, we can try to exploit
the remaining degrees of freedom in the choice of the endpoints
$\lambda_k$ to solve a kind of {\em connection problem}
\index{connection problem} for the vector field (\ref{eq:ODE}) in the
$(u,v)$ plane.  Thus, we want to further constrain the set of
endpoints $\lambda_0,\dots,\lambda_G$ precisely so that for each band
there exists an integral curve of (\ref{eq:ODE}) that joins the two
endpoints of the band.  Each nonzero endpoint of a band is a fixed
point of the vector field (\ref{eq:ODE}).  Assuming the generic case
of exact square root vanishing of the candidate density
$\rho^\sigma(\lambda)$ given by (\ref{eq:rhoformula}) at each
endpoint, there are locally three orbits of (\ref{eq:ODE}) that meet
the fixed point at $120^\circ$ angles.  See Figure~\ref{fig:trinity}.
If the endpoints are chosen correctly, each band in the upper
half-plane except for $I_0^+$ is then a heteroclinic orbit
\index{heteroclinic orbits} of the system (\ref{eq:ODE}) in the $(u,v)$ 
phase plane.  The band $I_0^+$ is exceptional because one of the
endpoints is $\lambda=0$, which is not a fixed point of
(\ref{eq:ODE}).  Thus, this band can be found in principle by the
process of {\em shooting}\index{shooting}.  That is, one solves
(\ref{eq:ODE}) as a well-defined initial value problem starting from
$u=v=0$ with the correct limiting value of $\rho^\sigma(\eta)$ taken
as $\eta\rightarrow 0$ within $I_0^+$ and insists that the orbit
terminate (in infinite ``time'') at $\lambda_0$.

\begin{remark}
Note that this problem is also known in geometric function theory\index{geometric function theory} as the characterization of {\em trajectories of 
quadratic differentials}\index{quadratic differentials!trajectories of}.
\end{remark}

\begin{remark}
Note that if a {\em single orbit} of the vector field (\ref{eq:ODE})
connects two endpoints, then not only is the candidate measure
$\rho^\sigma(\eta)\,d\eta$ real on the orbit, but {\em it is also
strictly of one sign on the whole interior of the orbit}.  This
follows from the fact that internal zeros of the candidate measure
necessarily correspond to fixed points of the vector field
(\ref{eq:ODE}).  Therefore, it is possible to replace the analytical
constraint that the inequality $\rho^\sigma(\eta)\,d\eta<0$ hold
strictly throughout the interior of each band with a kind of
topological constraint on the orbits of (\ref{eq:ODE}), together with
the verification of the inequality at an isolated interior point.  In
this sense, for steepest-descents analysis of Riemann-Hilbert problems
where the correct contour must be determined as part of the solution
of the problem, the correct generalization of the notion of strictness
of inequality is the notion of {\em connectivity}\index{connectivity}.
We will revisit this theme again shortly.
\end{remark}

How does one find the endpoint configurations for which the integral
curves of (\ref{eq:ODE}) form the bands in practice?  By analyticity
of $\rho^\sigma(\lambda)$, for each band $I_k^+\subset C_I$,
we can define an analytic function by the formulae
\begin{equation}
B_0(\lambda) = \int_0^\lambda
\rho^\sigma(\eta)\,d\eta\,,\hspace{0.2 in}
B_k(\lambda)=\int_{\lambda_{2k-1}}^\lambda
\rho^\sigma(\eta)\,d\eta\,,
\hspace{0.2 in}k=1,\dots,G/2\,,
\label{eq:Bs}
\end{equation}
where $\rho^\sigma(\eta)$ is given by (\ref{eq:rhoformula}).  The main
point is that {\em the contours along which the measure
$\rho^\sigma(\eta)\,d\eta$ is real are exactly the zero level sets of
the real analytic functions $\Im(B_k(\lambda))$}.  By construction,
the ``lower'' endpoint of each band lies on the corresponding zero
level.  A necessary condition for there to exist a single branch of
the zero level curve that connects this endpoint of each band to its
partner is that the {\em measure reality conditions}\index{measure
reality condition}
\begin{equation}
R_k := \Im(B_k(\lambda_{2k}))=0\,,\hspace{0.3 in} k=0,\dots,G/2
\label{eq:realitycond}
\end{equation}
hold.  Once again, it is clear by analyticity that the integrals are
path independent, and therefore the $R_k$ are manifestly real-valued
functions of the endpoints $\lambda_0,\dots,\lambda_G$ alone.  

We therefore have the following.
\begin{lemma}
Suppose the conditions of Lemma~\ref{prop:moments} and
Lemma~\ref{prop:vanishing} are satisfied.  Then a necessary
condition for the candidate measure $\rho^\sigma(\eta)\,d\eta$
to be real in its support is that the endpoints
$\lambda_0,\dots,\lambda_G$ satisfy the reality conditions
$R_k(\lambda_0,\dots,\lambda_G)=0$, for $k=0,\dots,G/2$.  
\label{prop:reality}
\end{lemma}

\begin{remark}
With the addition of the reality conditions to the moment conditions
and the vanishing conditions, we at last have $2G+2$ real equations in
$2G+2$ real unknowns, the real and imaginary parts of
$\lambda_0,\dots,\lambda_G$.  
Intuitively, one expects the set of solutions to be discrete.
Given values of $J$ and $\sigma$, the equations for the endpoints
involve $x$ and $t$ as real parameters.  If for some $x$ and $t$ 
a solution exists and the
Jacobian matrix of derivatives of the conditions with respect to
the endpoints and their complex conjugates is nonsingular, then
by the implicit function theorem the endpoints will locally be
continuous functions of $x$ and $t$.  This property is then inherited
by the function $\rho^\sigma(\eta)$.
\end{remark}

Note that it is by no means clear that the reality condition $R_k=0$
guarantees the existence of a real level connecting the endpoints of
the band $I_k^+$.  To establish the existence we would need to have
some discrete topological information in addition, like the
connectedness of the real level set of the analytic function
$B_k(\lambda)$.  Furthermore, if $R_k=0$, then there may be more than
one contour connecting the two endpoints of the interval, since
locally there are three possible real paths emerging from each nonzero
endpoint.  These are the central difficulties in the characterization
of trajectories of quadratic differentials \index{quadratic
differentials!trajectories of} in geometric function
theory\index{geometric function theory}.  We do not pursue these
questions further in the general context, since it will be clear from
examples to follow how the procedure works in practice.  We will in
fact be able to find the precise contours $I_k^+$ connecting the band
endpoints, as long as $J=\pm 1$ and $\sigma=\pm 1$ are chosen
correctly.  Note that if the measure $\rho^\sigma(\eta)\,d\eta$ is
real in $I_k^+$, then the conjugate measure
$\rho^\sigma(\eta^*)^*\,d\eta$ is automatically real and of the
opposite sign in $I_k^-=I_k^{+*}$ with the orientation of $C_I^*$.

\subsection{Restoring the loop contour $C$.}
Let us suppose that this construction has been successful, so that we
have found a set of admissible endpoints $\lambda_0,\dots,\lambda_G$
that satisfy the $2G+2$ real conditions we have imposed, and we have
shown that for each band $I_k^+$, the two endpoints of the band
(including $\lambda=0$ for the band $I_0^+$) are contained in the {\em
same connected component} of the level set $\Im(B_k(\lambda))=0$.  We
have then constructed a genus $G$ ansatz.  At this point, the bands of
the contour $C_I\cup C_I^*$ are completely specified as contours in
the complex plane.  But the gaps have not been constrained at all by
this construction.  In particular, as long as 
$\lambda_G\neq iA$, the final portion of $C_I$ is part of a gap
that may be chosen freely.  We will now show that under this generic
condition the temporary assumption we made at the beginning of this
chapter --- that the contour $C$ passes through $\lambda=iA$ --- can
be removed.

\begin{figure}[h]
\begin{center}
\mbox{\psfig{file=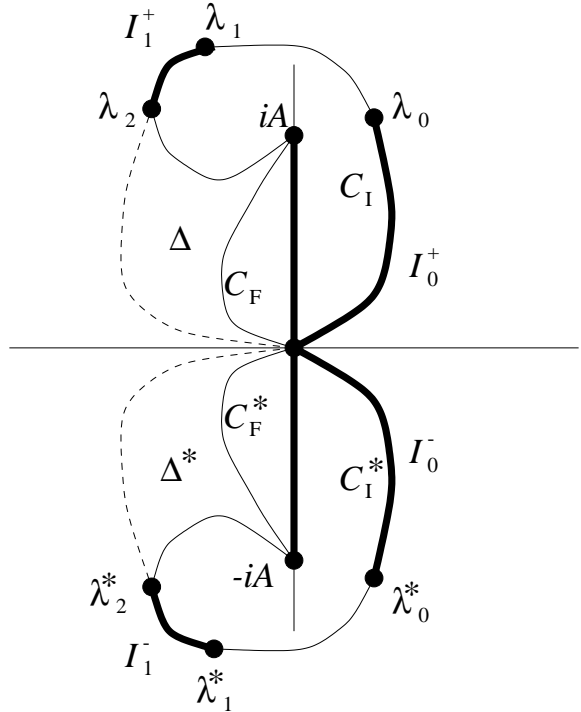,width=3 in}}
\end{center}
\caption{\em A sketch of how the matrix $\tilde{\bf
N}^\sigma(\lambda)$ is redefined as $\tilde{\bf
N}^\sigma(\lambda){\bf v}_{\tilde{\bf N}}^\sigma(\lambda)$ in
the region $\Delta$.  This sketch is for a genus $G=2$ ansatz with
orientation $\sigma=+1$.  In the region $\Delta^*$, the matrix is
redefined to preserve the original reflection symmetry of $\tilde{\bf
N}^\sigma(\lambda)$.}
\label{fig:Delta}
\end{figure}

Without changing the value of $\tilde{\phi}^\sigma(\lambda)$
on the contour $C_I$, we may rewrite it in its original form:
\begin{equation}
\begin{array}{rcl}
\displaystyle
\tilde{\phi}^\sigma(\lambda)&=&\displaystyle
\int_{0}^{iA}L^0_\eta(\lambda)\rho^0(\eta)\,d\eta  +
\int_{-iA}^0 L^0_\eta(\lambda)\rho^0(\eta^*)^*\,d\eta\\\\
&&\displaystyle\,\,-\,\,\int_{\cup_k I_k^+}
\overline{L^{C,\sigma}_\eta}(\lambda)\rho^\sigma(\eta)\,d\eta -
\int_{\cup_k I_k^-}\overline{L^{C,\sigma}_\eta}(\lambda)\rho^\sigma(\eta^*)^*\,d\eta\\\\
&&\displaystyle\,\,+\,\, J(2i\lambda x + 2i\lambda^2t) + i\pi \sigma
\int_\lambda^{iA}\rho^0(\eta)\,d\eta
\,.
\end{array}
\label{eq:phitildeagain}
\end{equation}
But now, it is clear that the jump matrix ${\bf
v}_{\tilde{\bf N}}^\sigma(\lambda)$ is analytic in $\lambda$ from the final
interval endpoint $\lambda_G\in C_I$, along $C_I$ into $\lambda=iA$,
and then out again from $\lambda=iA$ down along $C_F$ to $\lambda=0$.
This means that in a distorted triangular region $\Delta$ (see
Figure~\ref{fig:Delta})
with corners at $\lambda=\lambda_G$, $\lambda=iA$, and $\lambda=0$, it
is possible to redefine the matrix $\tilde{\bf
N}^\sigma(\lambda)$ by the analytic transformation
\begin{equation}
\tilde{\bf N}^\sigma(\lambda)\rightarrow \tilde{\bf
N}^\sigma(\lambda) {\bf
v}_{\tilde{\bf N}}^\sigma(\lambda)\,,\hspace{0.3 in} \lambda\in \Delta\,.
\end{equation}
In the region $\Delta^*$, the transformation that preserves the
complex-conjugation symmetry of $\tilde{\bf N}^\sigma(\lambda)$ is
used; for $\lambda\in\Delta^*$ we set $\tilde{\bf N}^\sigma(\lambda):=
\sigma_2\tilde{\bf N}^\sigma(\lambda^*)^*\sigma_2$.  As a consequence, 
the jump matrix is restored to the identity on the final gap portions
of $C_I$ and $C_I^*$, and also along all of $C_F\cup C_F^*$.  On the
third boundary curve of $\Delta$, there is now a jump, which is given
by exactly the same formula for ${\bf v}_{\tilde{\bf
N}}^\sigma(\lambda)$.  This third boundary curve, that is shown dashed
in Figure~\ref{fig:Delta}, is now the final gap $\Gamma_{G/2+1}^+$ on
the contour $C$ that genuinely encircles the imaginary interval
$[0,iA]$.  The corresponding conjugated contour in the lower
half-plane is $\Gamma_{G/2+1}^-$.

Thus, all reference to a contour $C$ that is required to pass through
the point $\lambda=iA$ disappears.  It is important to correctly
interpret this fact in the context of the various integrals that have
been introduced to characterize the conditions on the endpoints
$\lambda_0,\dots,\lambda_G$ and to provide a formula for the candidate
density $\rho^\sigma(\lambda)$.  In each case that we have
integrated over the set $\Gamma_I\cap C_I$, we simply have a sum of integrals
\begin{equation}
\lambda_0 \rightarrow \lambda_1\,,
\lambda_2 \rightarrow \lambda_3\,,\dots\,,
\lambda_G \rightarrow  iA\,,
\end{equation}
and when we have integrated over $\Gamma_I\cap C_I^*$, 
we have a sum of integrals
\begin{equation}
-iA\rightarrow \lambda_G^*\,,\lambda_{G-2}^*\rightarrow \lambda_{G-3}^*
\,,\dots \,,
\lambda_1^*\rightarrow \lambda_0\,.
\end{equation}
These integrals may be taken over any paths in ${\mathbb
C}\setminus [-iA,iA]$ that when combined with the precisely specified band
contours $I_k^\pm$ make up a non-self-intersecting contour $C_I$.

\section[Imposing the Inequalities]{Imposing the inequalities.  Local and global continuation theory.}
\label{sec:continuation}
In principle, the above algorithm can be carried out for any even $G$,
and it results in a discrete number of real candidate measures
$\rho^\sigma(\eta)\,d\eta$ supported on a system of
well-defined bands in the complex plane with endpoints $(0,\lambda_0),
(\lambda_2,\lambda_3), \dots,(\lambda_{G-1},\lambda_G)$ and their
complex conjugates.  Within each band,
$\tilde{\phi}^\sigma(\lambda)$ is an imaginary constant.  If
one of these candidate measures is to generate a complex phase
function $g^\sigma(\lambda)$ that asymptotically simplifies the
Riemann-Hilbert problem for $\tilde{\bf N}^\sigma(\lambda)$, then two
additional conditions need to be satisfied by the candidate density:
\index{density function!candidate}
\begin{enumerate}
\item
The candidate measure $\rho^\sigma(\eta)\,d\eta$ must be strictly
negative in the interior of each band $I_0^+,\dots,I_{G/2}^+$ of
the loop contour $C$.  
\item
It must be possible to choose the gaps
$\Gamma_1^\pm,\dots,\Gamma_{G/2+1}^\pm$ so that in the interior of
each the real part of the function $\tilde{\phi}^\sigma(\lambda)$
constructed from the candidate measure $\rho^\sigma(\eta)\,d\eta$ is
strictly negative.  
\end{enumerate}
If these additional conditions can be satisfied for some genus $G$
ansatz, and for some values of $J$ and $\sigma$, then
$\rho^\sigma(\eta)\,d\eta$ is promoted from a candidate measure to a
{\em bona fide} measure that generates a true complex phase function
$g^\sigma(\lambda)$, and the rigorous asymptotic analysis described in
Chapter~\ref{sec:asymptoticanalysis} is valid. 
This in turn yields a rigorous pointwise asymptotic description, in
terms of genus $G$ Riemann theta functions, of the sequence of
solutions $\psi(x,t)$ of the nonlinear Schr\"odinger equation that for
each value $\hbar_N$ of $\hbar$ is the soliton ensemble connected with
the WKB approximations $\lambda_{\hbar_N,n}^{\rm WKB}$ of the true
discrete eigenvalues for the initial condition $\psi(x,0)=A(x)$.  One
expects that there will be only one value of $G$ (possibly depending
on the parameters $x$ and $t$) for which a candidate satisfies the
inequalities.

In this section, we will show that under certain conditions the
existence of a genus $G$ ansatz that satisfies all inequalities for
some $x$ and $t$ implies that a successful genus $G$ ansatz in fact
exists for nearby values of $x$ and $t$.  Let ${\mathbb H}$ denote the
open upper half-plane minus the imaginary interval $[0,iA]$.  First,
we will show that, essentially, the existence of band contours
connecting pairs of endpoints is an open condition in the
$(x,t)$-plane.
\begin{lemma}
Fix $\sigma$ and $J$.  Let $x_0$ and $t_0$ be given in the
$(x,t)$-plane such that:
\begin{enumerate}
\item
For each $(x,t)$ in some disk $E$ centered at $(x_0,t_0)$, there is a
solution $\Lambda(x,t):=\{\lambda_0(x,t),\dots,\lambda_G(x,t)\}$ of
the moment conditions \index{moment conditions}(\ref{eq:momenteqns}),
the vanishing conditions \index{vanishing
conditions}(\ref{eq:vanishing}), and the measure reality conditions
\index{measure reality condition}(\ref{eq:realitycond}), for which
\begin{itemize}
\item
each $\lambda_k(x,t)$ lies in ${\mathbb H}$ for all $(x,t)\in E$,
\item
each $\lambda_k(x,t)$ is continuously differentiable in $E$,
\item
the $\lambda_k(x,t)$ are distinct for all $(x,t)\in E$.
\end{itemize}
Let the candidate density \index{density
function!candidate} constructed from the endpoints $\Lambda(x,t)$ for
$(x,t)\in E$ be denoted by $\rho^\sigma(\eta;x,t)$.
\item
The function $\rho^\sigma(\eta;x_0,t_0)$ admits for all
$k=1,\dots,G/2$ a smooth orbit $I_k^+(x_0,t_0)$ of the differential
equation (\ref{eq:ODE}) connecting the pair of consecutive endpoints
$\lambda_{2k-1}(x_0,t_0)$ and $\lambda_{2k}(x_0,t_0)$ and lying
entirely in ${\mathbb H}$, as well as a smooth orbit $I_0^+(x_0,t_0)$
of (\ref{eq:ODE}) connecting the origin to $\lambda_0(x_0,t_0)$ and
lying in ${\mathbb H}\cup\{0\}$.  Furthermore, the function
$\rho^\sigma(\eta;x_0,t_0)$ is nonzero in the interior of each band
$I_k^+(x_0,t_0)$ with $\rho^\sigma(\eta;x_0,t_0)\,d\eta$ being a
negative (real) differential and with
\begin{equation}
\inf_{\eta\in I_k^+(x_0,t_0)}\left|\frac{\rho^\sigma(\eta;x_0,t_0)}{R_+(\eta;x_0,t_0)}\right|>0\,,
\end{equation}
where $R(\eta;x_0,t_0)$ denotes the square root function defined
relative to the endpoints $\Lambda(x_0,t_0)$ and the given bands
$I_k^\pm(x_0,t_0)$.  That is, $\rho^\sigma(\eta;x_0,t_0)$ vanishes exactly
like a square root at the endpoints, and not to higher order.
\end{enumerate}
Then, there exists a disk $D\subset E$ centered at $(x_0,t_0)$ such
that for all $(x,t)\in D$ and corresponding to the candidate density
function $\rho^\sigma(\eta;x,t)$, there is for each $k=1,\dots,G/2$ a
smooth orbit $I_k^+(x,t)$ of (\ref{eq:ODE}) connecting
$\lambda_{2k-1}(x,t)$ to $\lambda_{2k}(x,t)$ and lying in ${\mathbb
H}$, as well as a smooth orbit $I_0^+(x,t)$ of (\ref{eq:ODE})
connecting the origin to $\lambda_0(x,t)$ and lying in ${\mathbb
H}\cup\{0\}$.  Moreover, the differential $\rho^\sigma(\eta;x,t)\,d\eta$
is negative in each band $I_k^+(x,t)$ for all $(x,t)\in D$.
\label{lemma:bandcontinuitygeneral}
\end{lemma}

\begin{proof}
First observe that from the formula (\ref{eq:rhowithY}) for
$\rho^\sigma(\eta;x,t)$ in terms of the function $Y$ ({\em cf.}
Lemma~\ref{lemma:rhowithY}), the continuity of the endpoint functions
$\lambda_k(x,t)$ in $E$ and the continuous dependence of the analytic
function $Y$ on the endpoints, we immediately find that for $(x,t)$ in
some sufficiently small disk $D_{\rm sr}\subset E$, the statement that
$\rho^\sigma(\eta;x_0,t_0)$ vanishes exactly like a square root at all
endpoints $\lambda_0(x_0,t_0),\dots,
\lambda_G(x_0,t_0)$ carries over to $\rho^\sigma(\eta;x,t)$ as well.

Consider the deformation of a band $I^+_k(x_0,t_0)$ for
$k=1,\dots,G/2$.  We seek a map $\tau_k(\eta):
I_k^+(x_0,t_0)\rightarrow I_k^+(x,t)$ that satisfies the implicit
relation
\begin{equation}
\int_{\lambda_{2k}(x,t)}^{\tau_k(\eta)}\rho^\sigma(\zeta;x,t)\,d\zeta
= 
\alpha_k(x_0,t_0,x,t)
\int_{\lambda_{2k}(x_0,t_0)}^\eta\rho^\sigma(\zeta;x_0,t_0)\,d\zeta \,,
\label{eq:implicitrelation}
\end{equation}
where $\alpha_k(x_0,t_0,x,t)$ is a real constant chosen so that 
$\tau_k(\lambda_{2k-1}(x_0,t_0))=\lambda_{2k-1}(x,t)$, that is,
\begin{equation}
\alpha_k(x_0,t_0,x,t):=\frac{\displaystyle
\int_{\lambda_{2k}(x,t)}^{\lambda_{2k-1}(x,t)}\rho^\sigma(\zeta;x,t)
\,d\zeta
}{\displaystyle
\int_{\lambda_{2k}(x_0,t_0)}^{\lambda_{2k-1}(x_0,t_0)}\rho^\sigma(\zeta;x_0,t_0)
\,d\zeta}\,.
\end{equation}
Also, we restrict attention to maps for which
$\tau_k(\lambda_{2k}(x_0,t_0))=
\lambda_{2k}(x,t)$.  

First, we show that for $(x-x_0)^2 + (t-t_0)^2$ sufficiently small, 
\begin{equation}
|\alpha_k(x_0,t_0,x,t)-1|\le C_{\alpha,k}\sqrt{(x-x_0)^2 + (t-t_0)^2}\,,
\label{eq:Calpha}
\end{equation}
for some $C_{\alpha,k}>0$.  Since the denominator of $\alpha_k$ is real and
strictly nonzero by assumption, this will follow if we can argue that
the numerator of $\alpha_k$ is differentiable at $(x,t)=(x_0,t_0)$.
Differentiating the numerator with respect to $x$ or $t$, we may take
the derivative operator inside the integral, since the integrand
vanishes at the endpoints for $x=x_0$ and $t=t_0$.  The derivatives of
the integrand with respect to $x$ and $t$ have contributions from
explicit $x$ and $t$ dependence and from $x$ and $t$ dependence
through the endpoints $\Lambda(x,t)$.  From the explicit formula
(\ref{eq:rhoformula}), it is easy to see that the partial derivatives
with respect to $x$ and $t$ are both integrable in $I_k^+(x_0,t_0)$
for $x=x_0$ and $t=t_0$.  Then, the chain rule terms are integrable in
$I_k^+(x_0,t_0)$ because the endpoints are continuously differentiable
by assumption and because the derivatives of $\rho^\sigma(\zeta;x,t)$ with
respect to the endpoints are integrable in $I_k^+(x_0,t_0)$, although
they blow up like inverse square roots at the two endpoints of the
contour of integration.

Next, introduce the change of variables
\begin{equation}
\mu_k(\eta):=\frac{\tau_k(\eta)-B_k(x_0,t_0,x,t)}{A_k(x_0,t_0,x,t)}\,,
\end{equation}
where
\begin{equation}
\begin{array}{rcl}
A_k(x_0,t_0,x,t)&:=&\displaystyle
\frac{\lambda_{2k}(x,t)-\lambda_{2k-1}(x,t)}
{\lambda_{2k}(x_0,t_0)-\lambda_{2k-1}(x_0,t_0)}\,,\\\\
B_k(x_0,t_0,x,t)&:=&\displaystyle
\frac{\lambda_{2k}(x_0,t_0)\lambda_{2k-1}(x,t)-
\lambda_{2k-1}(x_0,t_0)\lambda_{2k}(x,t)}
{\lambda_{2k}(x_0,t_0)-\lambda_{2k-1}(x_0,t_0)}\,.
\end{array}
\end{equation}
Note that from the differentiability and distinctness properties of
the endpoints in the neighborhood $E$, we have the estimates
\begin{equation}
\begin{array}{rcl}
|A_k(x_0,t_0,x,t)-1|&\le &C_{A,k}\sqrt{(x-x_0)^2+(t-t_0)^2}\,,\\\\
|B_k(x_0,t_0,x,t)|&\le & C_{B,k}\sqrt{(x-x_0)^2+(t-t_0)^2}
\end{array}
\label{eq:AB}
\end{equation}
for some positive constants $C_{A,k}$ and $C_{B,k}$ and all sufficiently small
$(x-x_0)^2+(t-t_0)^2$.  The implicit relation (\ref{eq:implicitrelation})
therefore becomes
\begin{equation}
\int_{\lambda_{2k}(x_0,t_0)}^{\mu_k(\eta)}\rho^\sigma(A_k\zeta+B_k;x,t)\,d\zeta =
\frac{\alpha_k}{A_k}\int_{\lambda_{2k}(x_0,t_0)}^\eta
\rho^\sigma(\zeta;x_0,t_0)\,d\zeta
\,.
\label{eq:implicitrelationbecomes}
\end{equation}

Now, consider the function $h_k(\eta)$ defined by the integral
\begin{equation}
h_k(\eta):=\int_{\lambda_{2k}(x_0,t_0)}^\eta \rho^\sigma(\zeta;x_0,t_0)\,d\zeta
\,,
\end{equation}
for $\eta$ in a lens-shaped neighborhood of $I_k^+(x_0,t_0)$.  If the
neighborhood is sufficiently thin, then the map $h_k(\eta)$ is
one-to-one, since by assumption $\rho^\sigma(\eta;x_0,t_0)$ is strictly
nonzero in the interior of $I_k^+(x_0,t_0)$.  By the reality condition
satisfied by $I_k^+(x_0,t_0)$, the image of the lens-shaped neighborhood
of $I_k^+(x_0,t_0)$ is a lens-shaped neighborhood of the open real interval
\begin{equation}
h_k(I_k^+(x_0,t_0))=\left(0,-\int_{I_k^+(x_0,t_0)}\rho^\sigma(\eta;x_0,t_0)\,
d\eta\right):= (0,h^{\rm max}_k)\,.
\end{equation}
The inverse function $h_k^{-1}(\cdot)$ is defined and analytic in
the open real interval $(0,h^{\rm max}_k)$.  Near
the endpoints, we have
$h_k^{-1}(\phi)\sim C_1\phi^{2/3}$ near $\phi=0$ and
$h_k^{-1}(\phi)\sim C_2+C_3(\phi-h_k^{\rm max})^{2/3}$ near
$\phi=h_k^{\rm max}$ for some constants $C_1$, $C_2$, and $C_3$.
Letting $M:=h_k(\mu_k(\eta))$ and $H:=h_k(\eta)$, the relation
(\ref{eq:implicitrelationbecomes}) can be rewritten as
\begin{equation}
M=H +\int_0^M U_k(\phi)\,d\phi
:=T_k(M)\,.
\label{eq:TofM}
\end{equation}
where
\begin{equation}
U_k(\phi):=
\left[
\rho^\sigma(h_k^{-1}(\phi);x_0,t_0)-
\frac{A_k}{\alpha_k}\rho^\sigma(A_kh_k^{-1}(\phi)+B_k;x,t)\right]
\frac{dh_k^{-1}(\phi)}{d\phi}\,.
\label{eq:squarebrackets}
\end{equation}
We want to consider solving this equation for $M=M(H)$ by fixed-point
iteration\index{fixed-point iteration}, {\em i.e.} by choosing some
$M_0$ and constructing the sequence $\{M_n\}$ by the recursion
$M_n:=T_k(M_{n-1})$.  If this sequence converges, then we have a
solution.

For $\epsilon_k>0$ consider the rectangular region $R_k$ with corner points
$-2\epsilon_k\pm 2i\epsilon_k$ and $h_k^{\rm max}+2\epsilon_k \pm
2i\epsilon_k$.  The interval $h_k(I_k^+(x_0,t_0))$ is contained in
$R_k$.  We claim that for $\epsilon_k>0$ sufficiently small the
function $U_k(\phi)$ in the integrand of (\ref{eq:TofM}) has an
analytic extension as a function of $\phi$ to some open set containing
$R_k$ for which
\begin{equation}
\lim_{x\rightarrow x_0, t\rightarrow t_0}\left[\sup_{\phi\in R_k}|U_k(\phi)|
\right]= 0\,.
\label{eq:smallintegrand}
\end{equation}

To show the analyticity of $U_k(\phi)$ it suffices to examine the
endpoints $\phi=0$ and $\phi=h_k^{\rm max}$.  On the one hand,
$dh_k^{-1}/d\phi$ in (\ref{eq:squarebrackets}) blows up exactly like a
negative one-third power at each endpoint.  But on the other hand, the
inverse map $h_k^{-1}(\cdot)$ vanishes like a two-thirds power at each
endpoint, and since we are working in $D_{\rm sr}$, the function
$\rho^\sigma(\eta;x,t)$ vanishes like a square root at each endpoint;
thus, the function of $\phi$ in square brackets in
(\ref{eq:squarebrackets}) vanishes exactly like a one-third power at
each endpoint.  Analyticity at the endpoints thus follows for the
product $U_k(\phi)$.  Next, to establish (\ref{eq:smallintegrand}), we
note that by analyticity in $R_k$, there exists a uniform bound and
the only question is its behavior as $(x,t)\rightarrow (x_0,t_0)$.
Clearly, $U_k(\phi)$ converges pointwise to zero in this limit for all
$\phi\in R_k$ except possibly at the endpoints $\phi=0$ and
$\phi=h_k^{\rm max}$.  But by analyticity at the endpoints and
compactness of $R_k$, the convergence to zero is in fact uniform for
$\phi\in R_k$, and the result follows.

Note that for all $H\in h_k(I_k^+(x_0,t_0))$ 
the disk $|\phi-H|<\epsilon_k$ is contained
in $R_k$ .  We claim that for
sufficiently small $(x-x_0)^2+(t-t_0)^2$, the transformation $T_k(\phi)$
maps this disk into itself.  Indeed
\begin{equation}
\begin{array}{rcl}
|T_k(\phi)-H|&=&\displaystyle
\left|\int_0^{H+(\phi-H)} U_k(\phi')\,d\phi'\right|\\\\
&\le&\displaystyle
(H + \epsilon_k) \sup_{\phi\in R_k}|U_k(\phi)|\,,
\end{array}
\label{eq:containment}
\end{equation}
which can be made arbitrarily small and in particular less than $\epsilon_k$
for $x$ and $t$ close enough to $x_0$ and $t_0$ respectively in view
of (\ref{eq:smallintegrand}).  

Let $\tilde{D}_k\subset E$ denote the disk in the $(x,t)$-plane
centered at $(x_0,t_0)$ in which the last line is bounded above by
$\epsilon_k$.  Then, for $\phi_1$ and $\phi_2$ both in the disk
$|\phi-H|<\epsilon_k$, we also have for $(x,t)\in \tilde{D}_k$,
\begin{equation}
\begin{array}{rcl}
|T_k(\phi_2)-T_k(\phi_1)|&=&\displaystyle
\left|\int_{\phi_1}^{\phi_2} U_k(\phi')\,d\phi'\right|\\\\
&\le&\displaystyle
|\phi_2-\phi_1|\sup_{\phi\in R_k}|U_k(\phi)|\\\\
&<&\displaystyle \frac{\epsilon_k}{H+\epsilon_k} |\phi_2-\phi_1|\\\\
&<&|\phi_2-\phi_1|\,.
\end{array}
\label{eq:contraction}
\end{equation}

From (\ref{eq:containment}) and (\ref{eq:contraction}), the
contraction mapping theorem guarantees that the iteration
$M_n:=T_k(M_{n-1})$ will converge when $(x,t)\in \tilde{D}_k$ whenever
$M_0$ is taken in the disk $|M_0-H|<\epsilon_k$.  Moreover, the limit
$M=\lim_{n\rightarrow\infty}M_n$ is the unique solution of the
equation (\ref{eq:TofM}) in this disk.  We therefore have a function
$M=M(H)$ defined for all $H$ in the closure of the open interval
$h_k(I_k^+(x_0,t_0))$.  This function is continuously
differentiable in the closed interval of its definition since for all
$M(H)$ defined above and for $(x,t)\in \tilde{D}_k$, we have
\begin{equation}
|U_k(M)|\le \frac{\epsilon_k}{H+\epsilon_k}<1\,,
\end{equation}
and consequently
\begin{equation}
\frac{\partial}{\partial M}[M-T_k(M)] = 1-U_k(M)\neq 0\,,
\end{equation}
holding even at the endpoints.  At these endpoints, we know that the
unique solution in the disk is given simply by $M(0)=0$ and
$M(h_k^{\rm max})=h_k^{\rm max}$.  The curve $M(H)$ is
therefore homotopic to the closed real interval $[0,h_k^{\rm
max}]$.

The function $h_k^{-1}(\cdot)$ is defined on the closed real interval
$[0,h_k^{\rm max}]$ and has a unique analytic continuation to the
curve $M(H)$.  The inverse function so defined on $M(H)$ is
continuous, and we then obtain
\begin{equation}
\mu_k(\eta)= h_k^{-1}(M(h_k(\eta))\,.
\end{equation}
For each $x$ and $t$ in $\tilde{D}_k$, we therefore obtain a curve with
the same endpoints, $\lambda_{2k-1}(x_0,t_0)$ and
$\lambda_{2k}(x_0,t_0)$.  By our estimates, the curves contract
uniformly to $I_k^+(x_0,t_0)$ as $(x,t)\rightarrow (x_0,t_0)$.
Finally, set
\begin{equation}
\tau_k(\eta):=A_k(x_0,t_0,x,t)\cdot h_k^{-1}(M(h_k(\eta))+
B_k(x_0,t_0,x,t)\,.
\end{equation}
This is a continuous function of $\eta\in I_k^+(x_0,t_0)$.  Each point
in the image satisfies (\ref{eq:implicitrelation}) and consequently
the image is a smooth curve $I_k^+(x,t)$ connecting
$\lambda_{2k-1}(x,t)$ to $\lambda_{2k}(x,t)$.  Moreover, by
continuity, $I_k^+(x,t)$ will lie in the set ${\mathbb H}$ for
$(x-x_0)^2+(t-t_0)^2$ sufficiently small, and to achieve this, we
restrict $x$ and $t$ to some slightly smaller disk
$D_k\subset\tilde{D}_k$.  Finally, to see that for all $(x,t)\in D_k$
the differential $\rho^\sigma(\eta;x,t)\,d\eta$ is nonvanishing in
$I_k^+(x,t)$ and of the same sign as
$\rho^\sigma(\eta;x_0,t_0)\,d\eta$ in $I_k^+(x_0,t_0)$, simply
differentiate (\ref{eq:implicitrelation}) to obtain
\begin{equation}
\rho^\sigma(\tau_k;x,t)\,d\tau_k = \rho^\sigma(\tau_k(\eta);x,t)\tau'_k(\eta)\,d\eta = \alpha_k(x_0,t_0,x,t)\rho^\sigma(\eta;x_0,t_0)\,d\eta\,,
\end{equation}
from which the required result follows from the estimate
(\ref{eq:Calpha}).

To verify the continuity of the exceptional band
$I_0^+$, one repeats the above arguments,
substituting
zero everywhere for $\lambda_{2k-1}$.  Thus, one uses
\begin{equation}
\alpha_0(x_0,t_0,x,t):=\frac{\displaystyle \int_{\lambda_0(x,t)}^0
\rho^\sigma(\zeta;x,t)\,d\zeta}{\displaystyle
\int_{\lambda_0(x_0,t_0)}^0\rho^\sigma(\zeta;x_0,t_0)\,d\zeta}\,,
\end{equation}
and obtains an estimate analogous to (\ref{eq:Calpha}).  Also, one
takes $B_0(x_0,t_0,x,t):=0$ and 
\begin{equation}
A_0(x_0,t_0,x,t):=\left[\frac{\lambda_0(x_0,t_0)}{\lambda_0(x,t)}-1\right]^{-1}\,,
\end{equation}
and obtains estimates analogous to (\ref{eq:AB}).  By similar
arguments based on contraction mapping, one verifies the continuity of
$I_0^+(x,t)$ for $x$ and $t$ in some sufficiently small disk
neighborhood $D_0$ of $(x_0,t_0)$.  Finally, we restrict
$(x,t)$ to lie in $D$ where
\begin{equation}
D=D_{\rm sr}\cap \left[\bigcap_{k=0}^{G/2} D_k \right]\,,
\end{equation}
which is nonempty for finite $G$.  This completes the proof.
\end{proof}

The existence of contour segments in which the gap inequalities may 
be satisfied is also an open condition in the $(x,t)$-plane.
\begin{lemma}
Assume all the conditions of Lemma~\ref{lemma:bandcontinuitygeneral},
and let $\tilde{\phi}^\sigma(\lambda;x,t)$ denote the function
corresponding to the candidate density $\rho^\sigma(\eta;x,t)$ for
$(x,t)\in E$ via (\ref{eq:tildephiagain}) with $K$ chosen according to
(\ref{eq:mselect}).  Furthermore, suppose that the bands
$I_k^\pm(x_0,t_0)$ are complemented by a system of gap contours
$\Gamma_k^\pm(x_0,t_0)$ making up a loop contour $C(x_0,t_0)\subset
{\mathbb H}\cup \{0\}$ such that
$\Re(\tilde{\phi}^\sigma(\lambda;x_0,t_0))<0$ strictly in the interior
of all gaps $\Gamma_k^+(x_0,t_0)$.  Then, there exists a disk
$D'\subset D\subset E$ centered at $(x_0,t_0)$ in the $(x,t)$-plane
such that for all $(x,t)\in D'$, smooth gap contours may be chosen in
${\mathbb H}\cup\{0\}$ for which the relevant inequality persists.
That is, there exist smooth paths $\Gamma_k^+(x,t)$ in ${\mathbb H}$
connecting $\lambda_{2k-2}(x,t)$ to $\lambda_{2k-1}(x,t)$ for
$k=1,\dots,G/2$ and a path $\Gamma_{G/2+1}^+(x,t)$ in ${\mathbb
H}\cup\{0\}$ connecting $\lambda_{G}(x,t)$ to the origin such that for
all $\lambda$ in the interior of a path $\Gamma_k^+(x,t)$, the strict
inequality $\Re(\tilde{\phi}^\sigma(\lambda;x,t))<0$ holds.
\label{lemma:gapcontinuitygeneral}
\end{lemma}

\begin{proof}
Using the general relations (\ref{eq:plusside}) and
(\ref{eq:minusside}), we see that the function
$\tilde{\phi}^\sigma(\lambda;x,t)$ may be expressed in terms of an
integral of the corresponding candidate density
$\rho^\sigma(\eta;x,t)$.  In this connection, the desingularized
representation (\ref{eq:rhowithY}) of the density is useful.  Let
$R(\lambda;x,t)$ and $Y(\lambda;x,t)$ denote the square root
function $R$ (defined in \S\ref{sec:outersolve}) and the analytic
function $Y$ ({\em cf.} (\ref{eq:Ydef})) defined from the endpoints
$\Lambda(x,t)$ for $(x,t)\in E$.  Consider an ``internal'' gap
$\Gamma_k^+(x,t)$ intended to connect the endpoints
$\lambda_{2k-2}(x,t)$ and $\lambda_{2k-1}(x,t)$ for $k=1,\dots,G/2$.
Since $D'\subset D$, the results of
Lemma~\ref{lemma:bandcontinuitygeneral} hold and throughout $D'$ the
band contours $I_k^+(x,t)$ exist as smooth curves.  Therefore, in
$\Gamma_k^+(x,t)$, the function $\tilde{\phi}^\sigma(\lambda;x,t)$ may be
written as
\begin{equation}
\begin{array}{rcl}
\tilde{\phi}^\sigma(\lambda;x,t)&=&\displaystyle
\tilde{\phi}^\sigma(\lambda;x,t)\Big|_{\lambda\in I_{k-1}^+(x,t)} +
i\pi\int_{\lambda_{2k-2}(x,t)}^\lambda R(\eta;x,t)Y(\eta;x,t)\,d\eta
\\\\
&=&\displaystyle
\tilde{\phi}^\sigma(\lambda;x,t)\Big|_{\lambda\in I_{k}^+(x,t)} +
i\pi\int_{\lambda_{2k-1}(x,t)}^\lambda R(\eta;x,t)Y(\eta;x,t)\,
d\eta\,.
\end{array}
\end{equation}
where we recall that by construction
$\tilde{\phi}^\sigma(\lambda;x,t)$ is an imaginary constant when
restricted to each band $I_k^+(x,t)$.

Let $\lambda=w(s)$ for $0\le s \le 1$ be a parametrization of
the given gap path $\Gamma_k^+(x_0,t_0)$.  Therefore
$w(0)=\lambda_{2k-2}(x_0,t_0)$ and
$w(1)=\lambda_{2k-1}(x_0,t_0)$.  We will show that for all $(x,t)$ in
the sufficiently small disk $D'$, the path parametrized by
\begin{equation}
\lambda=w(s;x,t):=A_k(x_0,t_0,x,t)w(s)+B_k(x_0,t_0,x,t)\,,
\end{equation}
where
\begin{equation}
\begin{array}{rcl}
A_k(x_0,t_0,x,t)&:= &\displaystyle
\frac{\lambda_{2k-1}(x,t)-\lambda_{2k-2}(x,t)}
{\lambda_{2k-1}(x_0,t_0)-\lambda_{2k-2}(x_0,t_0)}\,,\\\\
B_k(x_0,t_0,x,t)&:= &\displaystyle
\frac{\lambda_{2k-1}(x_0,t_0)\lambda_{2k-2}(x,t)-\lambda_{2k-2}(x_0,t_0)
\lambda_{2k-1}(x,t)}{\lambda_{2k-1}(x_0,t_0)-\lambda_{2k-2}(x_0,t_0)}\,,
\end{array}
\end{equation}
admits the relevant strict inequality for all $s\in (0,1)$.  By
continuity of the endpoints in $x$ and $t$, this is a near-identity
linear transformation.  We are given that
$\Re(\tilde{\phi}^\sigma(w(s);x_0,t_0))<0$ and must show that
$\Re(\tilde{\phi}^\sigma(A_kw(s)+B_k);x,t)<0$ for $D'$ sufficiently
small.

First, we consider a neighborhood of the endpoint $s=0$.  Since by
assumption we are working in the neighborhood $D$ of Lemma~\ref{lemma:bandcontinuitygeneral}, and therefore in the bigger neighborhood $D_{\rm sr}$ ({\em cf.} the proof of Lemma~\ref{lemma:bandcontinuitygeneral}), the integrand
$R(\lambda;x,t)Y(\lambda;x,t)$ vanishes at
$\lambda_{2k-2}(x,t)$ exactly like
$(\lambda-\lambda_{2k-2}(x,t))^{1/2}$ for all $(x,t)\in D'$.  This
implies that in a sufficiently small (independent of $x$ and $t$)
neighborhood $V_0$ in the complex plane that contains
$\lambda=\lambda_{2k-2}(x,t)$ for all $(x,t)$ close enough to
$(x_0,t_0)$, the region where $\Re(\tilde{\phi}^\sigma(\lambda;x,t))<0$
holds is a generalized sector whose boundary curves have tangents at
$\lambda_{2k-1}(x,t)$ that meet at an angle of $2\pi/3$.  The band
contour $I_{k-1}^+(x,t)$ has a tangent at its upper endpoint
$\lambda_{2k-2}(x,t)$ that bisects this angle.  Without loss of
generality, we now suppose that the given gap contour
$\Gamma_k^+(x_0,t_0)$ has a tangent at $\lambda_{2k-2}(x_0,t_0)$ whose
angle lies {\em strictly} between the tangents to the boundary curves.
Indeed, if this is not true of the given gap contour, it may be
achieved sacrificing neither smoothness nor the inequality
$\Re(\tilde{\phi}^\sigma(w(s);x_0,t_0))<0$ by a small deformation near
$s=0$.  Now, the boundary curves in $V_0$ satisfy
\begin{equation}
\Im\left(\int_{\lambda_{2k-2}(x,t)}^\lambda R(\eta;x,t)Y(\eta;x,t)\,
d\eta\right)=0\,,
\end{equation}
and from the fixed point theory used in the proof of
Lemma~\ref{lemma:bandcontinuitygeneral} it follows that these boundary
curves deform continuously in $(x,t)$ near $(x_0,t_0)$.  Since the
same is true of the path $\lambda=w(s;x,t)$ by construction, it is clear
that the disk $D'$ can be taken small enough that the inequality is
satisfied on $\Gamma_k^+(x,t)\cap V_0$ for all $(x,t)\in D'$.  

To handle the other endpoint, $s=1$, choose an analogous fixed
neighborhood $V_1$ containing $\lambda_{2k-1}(x,t)$ for all $(x,t)$
sufficiently close to $(x_0,t_0)$.  Then, a similar argument can be
used to show that, possibly by replacing $D'$ with a smaller disk, the
inequality is satisfied on $\Gamma_k^+(x,t)\cap V_1$ for all $(x,t)\in
D'$.

Let $s_0(x,t)$ be defined so that the interval $(0,s_0(x,t))$
parametrizes the curve $\Gamma_k^+(x,t)\cap V_0$ by the function
$w(s;x,t)$.  Similarly, let $s_1(x,t)$ be defined so that $(s_1(x,t),1)$
parametrizes $\Gamma_k^+(x,t)\cap V_1$.  Let 
\begin{equation}
s_0:=\inf_{(x,t)\in D'}s_0(x,t) > 0\,, \hspace{0.3 in}
s_1:=\sup_{(x,t)\in D'}s_1(x,t) < 1\,.
\end{equation}
It remains to verify the inequality (again, possibly by replacing $D'$
with a smaller disk) for $s\in [s_0,s_1]$.  Now, because we are
avoiding the endpoints, there exists some $\epsilon <0$ depending only
on $s_0$, $s_1$, $x_0$, and $t_0$ such that in this closed interval,
we have $\Re(\tilde{\phi}^\sigma(w(s);x_0,t_0))\le \epsilon$.  Consequently,
it is sufficient to show that $|\Re(\tilde{\phi}^\sigma(w(s;x,t);x,t))-
\Re(\tilde{\phi}^\sigma(w(s);x_0,t_0))|<\epsilon$ for $(x,t)$ close enough
to $(x_0,t_0)$.  We have
\begin{equation}
\begin{array}{l}
|\Re(\tilde{\phi}^\sigma(w(s;x,t);x,t))-
\Re(\tilde{\phi}^\sigma(w(s);x_0,t_0))|\le\\\\
\hspace{0.3 in}
\displaystyle
\pi\Bigg|
\int_{\lambda_{2k-2}(x_0,t_0)}^{w(s)}
\Big[A_kR(A_k\eta+B_k;x,t)Y(A_k\eta+B_k;x,t)
-R(\eta;x_0,t_0)Y(\eta;x_0,t_0)
\Big]\,d\eta\Bigg|\le
\\\\
\hspace{0.6 in}\displaystyle
\pi \sup_{s\in[s_0,s_1]}|w(s)-\lambda_{2k-2}(x_0,t_0)|\\\\
\hspace{0.8 in}\cdot\,\,
\sup_{s\in[s_0,s_1]}
|A_kR(w(s;x,t);x,t)Y(w(s;x,t);x,t)-R(w(s);x_0,t_0)Y(w(s);x_0,t_0)|\,.
\end{array}
\end{equation}
The first factor is uniformly bounded, and by simple continuity
arguments using the fact that the map $w(s)\rightarrow w(s;x,t)$ is a
near-identity transformation, the second factor can be made
arbitrarily small for $(x,t)$ near $(x_0,t_0)$ and in particular the
product can be made less than $\epsilon$.  This completes the proof
of existence of the ``internal'' gap $\Gamma_k^+(x,t)$.

Having established the persistence of the gaps connecting pairs of
endpoints in $\Lambda$, we must now show that the ``final'' gap
$\Gamma_{G/2+1}^+(x,t)$, which must connect $\lambda_G(x,t)$ to the
origin, also persists for $(x,t)$ near $(x_0,t_0)$.  In this case, the
near-identity transformation of the path $\Gamma_{G/2+1}^+(x_0,t_0)$
parametrized by $w(s)$ is given simply by
\begin{equation}
w(s;x,t):=\frac{\lambda_G(x,t)}{\lambda_G(x_0,t_0)}w(s)\,.
\end{equation}
The local analysis near $s=0$ corresponding to the endpoint
$\lambda=\lambda_{G}(x,t)$ goes through exactly as before.  

For the local analysis near $s=1$ corresponding to $\lambda=0$ for all
$(x,t)$, we first consider the definition (\ref{eq:tildephiagain}) of
the function $\tilde{\phi}^\sigma(\lambda;x,t)$.  For $\lambda$ in the
interior of the gap $\Gamma_{G/2+1}^+(x,t)$, we can use the
analyticity of the given eigenvalue density $\rho^0(\eta)$ to rewrite
the formula (\ref{eq:tildephiagain}) for
$\tilde{\phi}^\sigma(\lambda;x,t)$ in the form
\begin{equation}
\begin{array}{rcl}
\tilde{\phi}^\sigma(\lambda;x,t)&=&\displaystyle
\int_{C_I}
L_\eta^{C,\sigma}(\lambda)\overline{\rho}^\sigma(\eta;x,t)\,d\eta 
+\int_{C_I^*}
L_\eta^{C,\sigma}(\lambda)\overline{\rho}^\sigma(\eta^*;x,t)^*\,d\eta \\\\
&&\displaystyle\,\,+\,\, 
2iJ(\lambda x+\lambda^2 t) + i\pi \sigma \int_\lambda^{iA}\rho^0(\eta)\,
d\eta\,,
\end{array}
\label{eq:phitildedeform}
\end{equation}
where $\overline{\rho}^\sigma(\eta;x,t)$ is the complementary density
function corresponding to $\rho^\sigma(\eta;x,t)$ via
(\ref{eq:rhobardef}).  Now it follows from the boundary value problem
(\ref{eq:BVP}) that the function $\overline{\rho}^\sigma(\eta;x,t)$
extended by complex conjugation $\overline{\rho}^\sigma(\eta^*;x,t)^*$
to $C_I\cup C_I^*$ is analytic at $\eta=0$.  Therefore, the first two
integrals on the right-hand side of (\ref{eq:phitildedeform}) can be
combined and the path of integration may be deformed slightly either
to the right (for $\sigma=+1$) or left (for $\sigma=-1$) in a small
neighborhood of the origin.  Thus, we deduce that
$\tilde{\phi}^\sigma(\lambda;x,t)$ extends analytically to a
neighborhood of the final endpoint $\lambda=0$ of the gap
$\Gamma_{G/2+1}^+(x,t)$.  

Now, for $\lambda$ real, it follows from reality of the logarithm that
\begin{equation}
\Re(\tilde{\phi}^\sigma(\lambda;x,t))=\pi\sigma\lambda\,,\hspace{0.3 in}
\lambda\in{\mathbb R}\,.
\end{equation}
Since for $\sigma=+1$ (respectively $\sigma=-1$) the portion of
$\Gamma_{G/2+1}^+(x_0,t_0)$ near the origin necessarily lies in the
second (respectively first) quadrant, this together with the
analyticity of $\tilde{\phi}^\sigma(\lambda;x,t)$ at $\lambda=0$ shows
that in some neighborhood $V_1$ of the origin, the given gap contour
$\Gamma_{G/2+1}^+(x_0,t_0)$ lies in some generalized sector bounded by
the real axis and some boundary curve that makes a nonzero angle with
the real axis at $x=x_0$ and $t=t_0$ (recall that
$\Re(\tilde{\phi}^\sigma(0;x,t))=0$).  Without loss of generality, we
may assume that $\Gamma_{G/2+1}^+(x_0,t_0)$ has a tangent line at the
origin making a nonzero angle with both the real axis and the tangent
line of the boundary curve.  Then, since the boundary curve again
satisfies
\begin{equation}
\Im\left(\int_{\lambda_G(x,t)}^\lambda R(\eta;x,t)Y(\eta;x,t)\,d\eta\right)=0\,,
\end{equation}
the fixed point theory predicts smooth deformation of this curve with
respect to $x$ and $t$ near $x=x_0$ and $t=t_0$, which in conjunction
with the continuity of the near-identity map $w(s)\rightarrow w(s;x,t)$
gives the necessary inequality in $V_1$.  This concludes the analysis near $s=1$ corresponding to the origin in the $\lambda$-plane.  

With the endpoints taken care of in this way, the argument that the
inequality holds on parts of $\Gamma^+_{G/2+1}(x,t)$ that are bounded
away from the two endpoints $\lambda_G$ and $0$ is analogous to the
corresponding argument we used in proving the persistence of the
``internal'' gaps.  Therefore, for all $(x,t)$ in the sufficiently
small disk $D'$, the ``final'' gap contour $\Gamma_{G/2+1}^+(x,t)$
exists as well.  This completes the proof.
\end{proof}

Passing from the local to the global, these continuation arguments can
be developed into a partial characterization of the boundary of the
region of existence of a successful genus $G$ ansatz in the
$(x,t)$-plane.  Given $(x_0,t_0)$ and a continuous branch of the
collection of endpoint functions $\Lambda(x,t)$ such that the
conditions of Lemma~\ref{lemma:bandcontinuitygeneral} and
Lemma~\ref{lemma:gapcontinuitygeneral} are met, let $U$ be the
intersection of the largest open set in the $(x,t)$-plane where the
selected branch of $\Lambda(x,t)$ is differentiable and the largest
open set containing $(x_0,t_0)$ where the genus $G$ ansatz
corresponding to these endpoints satisfies all of the inequalities.
Let $(x_{\rm crit},t_{\rm crit})$ be a boundary point of $U$.  It is
necessary that at this boundary point at least one of the conditions
of either Lemma~\ref{lemma:bandcontinuitygeneral} or
Lemma~\ref{lemma:gapcontinuitygeneral} fails.  Otherwise, the open set
$D'$ guaranteed to exist by these results would contain $(x_{\rm
crit},x_{\rm crit})$ and be contained in $U$ --- a contradiction.

To catalog the possible modes of failure of the ansatz at the boundary
\index{genus $G$ ansatz!failure of}
of $U$ is a task complicated by the geometry of the cut upper
half-plane ${\mathbb H}$.  It is possible for a point on the boundary
of $U$ to correspond to an ansatz for which one of the band contours
meets $\partial{\mathbb H}$ at a point, or for which a gap contour is
``forced'' to meet $\partial{\mathbb H}$ because the boundary of the
region where $\Re(\tilde{\phi}^\sigma)<0$ does so.  However, there are
also modes of failure that do not involve the contour $C$ meeting
$\partial{\mathbb H}$.  These modes can be characterized by equations
for curves in the $(x,t)$-plane.

The onset of failure of the inequality for the bands can correspond to a
point $\lambda$ on one of the bands $I_k^+$ (including endpoints) for
which the function $\rho^\sigma(\lambda;x_{\rm crit},t_{\rm
crit})/R(\lambda;x_{\rm crit}, t_{\rm crit})$, analytic on the closure
of each band, has a zero.  Here, $\rho^\sigma$ is given by the
formula (\ref{eq:rhoformula}) valid in the bands.  Therefore, if the
ansatz fails by this mechanism at the point $(x_{\rm crit},t_{\rm crit})$,
then for some $k=0,\dots,G/2$, the following conditions hold for some 
$\lambda\in{\mathbb H}$:
\begin{equation}
\Im\left(\int_{\lambda_{2k}(x_{\rm crit},t_{\rm crit})}^\lambda
\rho^\sigma(\eta;x_{\rm crit},t_{\rm crit})\,d\eta\right) = 0\,,\hspace{0.2 in}
\frac{\rho^\sigma(\lambda;x_{\rm crit},t_{\rm crit})}
{R(\lambda;x_{\rm crit},t_{\rm crit})}=0\,,
\label{eq:bandfail}
\end{equation}
and $\lambda$ is on the band $I_k^+$.  We note here that, neglecting
the topological condition that $\lambda\in I_k^+$ (which amounts to
the selection of a particular branch of the first relation above), and
upon elimination of $\lambda$, these relations imply one real relation
satisfied by $x_{\rm crit}$ and $t_{\rm crit}$, a curve in the
$(x,t)$-plane.  If in addition, $\lambda$ is actually on the band
$I_k^+$, then these conditions imply that the band $I_k^+$ has the
interpretation of a chain of (at least) two connected heteroclinic
orbits of the vector field (\ref{eq:ODE}).

The onset of failure of inequality for the gaps can correspond to the
pinching off of a narrow ``isthmus'' in the region
$\Re(\tilde{\phi}^\sigma)<0$ in the $\lambda$-plane through which a gap
curve is forced to pass.  
Exactly at onset,
when the inequality first fails, the boundary curve where
$\tilde{\phi}^\sigma(\lambda)$ is purely imaginary becomes singular.
The existence of a singular point on the imaginary level can be
expressed by the equations
\begin{equation}
\frac{\partial\tilde{\phi}^\sigma}
{\partial\lambda}(\lambda;x_{\rm crit},t_{\rm crit})=0\,,\hspace{0.2
in} \Re(\tilde{\phi}^\sigma(\lambda;x_{\rm crit},t_{\rm crit}))=0\,.
\label{eq:gapfail}
\end{equation}
Here, $\tilde{\phi}^\sigma$ refers to the expression valid in the
gaps.  Again, observe that if $\lambda\in {\mathbb H}$ may be
eliminated between these two equations, what remains is a single real
equation in the two unknowns $x_{\rm crit}$ and $t_{\rm crit}$.  These
relations thus describe a union of curves in the real $(x,t)$ plane.

\begin{remark}
It is a consequence of the duality of the function
$\tilde{\phi}^\sigma(\lambda)$ evaluated in the gaps with the function
$\theta^\sigma(\lambda)$ evaluated in the bands ({\em cf.} equations
(\ref{eq:plusside}) and (\ref{eq:minusside})) that the conditions
(\ref{eq:bandfail}) and (\ref{eq:gapfail}) are essentially {\em
equivalent}.  They result in the same curves in the $(x,t)$-plane.
\end{remark}

\begin{remark}
The point $(x_{\rm crit},t_{\rm crit})$ being a solution of either
(\ref{eq:bandfail}) or (\ref{eq:gapfail}) is neither necessary nor
sufficient for $(x_{\rm crit},t_{\rm crit})$ to lie on the boundary of
$U$.  Even if $(x_{\rm crit},t_{\rm crit})$ satisfies
(\ref{eq:bandfail}), the value of $\lambda$ establishing the
consistency might not lie on the band $I_k^+$, instead being contained
in another of the three curves emanating from the band endpoint (see
Figure~\ref{fig:trinity}) or even in a curve branch that is not
connected to the endpoint at all.  Similarly if $(x_{\rm crit},t_{\rm
crit})$ satisfies (\ref{eq:gapfail}), the bottleneck that is created
might not actually constrain any gap contours to pass through the
point $\lambda$; the pinching might occur in an irrelevant part of the
region where $\Re(\tilde{\phi}^\sigma(\lambda))<0$.  On the other
hand, even if it is known that $(x_{\rm crit},t_{\rm crit})$ is on the
boundary of $U$, the failure of the ansatz may correspond to contact
of the contour with $\partial{\mathbb H}$, a mode of failure that is
not captured by the conditions (\ref{eq:bandfail}) or
(\ref{eq:gapfail}).  What may be said with precision is:
{\em if the point $(x_{\rm crit},t_{\rm crit})$ is known to
be a point of failure of the genus $G$ ansatz and the contour may be
taken to avoid $\partial{\mathbb H}$, then $(x_{\rm crit},t_{\rm
crit})$ is contained in the union of solution curves of
(\ref{eq:bandfail}) and (\ref{eq:gapfail}) in the real $(x,t)$-plane.}
\end{remark}

One expects that for points in the $(x,t)$-plane on the other side of
the boundary of $U$, the inequalities can be satisfied by choosing an
ansatz corresponding to a different genus $G$.  For example, in
Chapter~\ref{sec:genus2} we will prove that when the condition
(\ref{eq:gapfail}) holds for a genus zero ansatz, the curve defined by
(\ref{eq:gapfail}) in the $(x,t)$-plane is a boundary between values
of $x$ and $t$ where the genus zero ansatz is valid and values of $x$
and $t$ where a genus two ansatz is valid.  It then follows from the
analysis in Chapter~\ref{sec:asymptoticanalysis} that the asymptotic
behavior of the solution $\psi(x,t)$ of the nonlinear Schr\"odinger
equation will be qualitatively different for $(x,t)$ on opposite sides
of the boundary of $U$, being described by Riemann theta functions of
different genera.  The boundary curve may thus be given the physical
interpretation of a {\em phase transition}.  Such sharp transitions
are indeed clearly visible in computer reconstructions of the
Satsuma-Yajima ensemble
\cite{MK98}, for example.  They have also been seen in recent simulations
of (\ref{eq:IVP}) \cite{BK99,CM00} for more general initial data.

\section[Modulation Equations]{Modulation equations.}
\label{sec:modulation} Here, we will show that if the endpoints
$\lambda_0,\dots,\lambda_G$ satisfy the $2G+2$ real equations
contained in (\ref{eq:momenteqns}), (\ref{eq:vanishing}), and
(\ref{eq:realitycond}) --- equations in which $x$ and $t$ appear
analytically as explicit parameters --- then it turns out that the
endpoints considered as functions of the independent variables $x$ and
$t$ also satisfy a quasilinear system of partial differential
equations.  This system has no explicit dependence on $x$ and $t$ in
its coefficients, and also the system takes the same form regardless
of the function $A(x)$ that approximates the initial data for
(\ref{eq:IVP}).  The equations making up this quasilinear system are
the Whitham \index{Whitham equations} or modulation equations 
\index{modulation equations} associated with genus $G$
wavetrain solutions of the focusing nonlinear Schr\"odinger equation.
They are elliptic, which makes the initial-value problem for them
ill-posed\index{ill-posed}.

We will begin by returning to the function $F(\lambda)$ guaranteed to
exist by Lemma~\ref{prop:moments} because the moment conditions
(\ref{eq:momenteqns}) are among those satisfied by the endpoints.  The
first observation that we make about the function $F(\lambda)$ is the
following.

\begin{lemma}
Whenever the endpoints satisfy the measure reality conditions
(\ref{eq:realitycond}), the function $F(\lambda)$ satisfies
$F(\lambda)=\bo (\lambda^{-2})$ as $\lambda\rightarrow \infty$.
\label{lemma:higherorder}
\end{lemma}

\begin{proof}
From the Cauchy integral representation (\ref{eq:Cauchy}) of 
$F(\lambda)$, the result will follow if it is true that
\begin{equation}
\int_{C_I}\overline{\rho}^\sigma(\eta)\,d\eta + \int_{C_I^*}
\overline{\rho}^\sigma(\eta^*)^*\,d\eta =0\,.
\end{equation}
Now, using the conjugation symmetry of the contours, the definition
(\ref{eq:rhobardef}), and analyticity of $\rho^0(\eta)$, this is
equivalent to the condition
\begin{equation}
\Im\left(\int_0^{iA}\rho^0(\eta)\,d\eta\right) - \Im\left(
\int_{C_I}\rho^\sigma(\eta)\,d\eta\right) = 0\,.
\end{equation}
The first term then vanishes because the given
asymptotic eigenvalue measure is real on the imaginary axis, and the
second term is equivalent to a sum of integrals of
$\rho^\sigma(\eta)\,d\eta$ over the bands $I_k^+$ of $C_I$.  The
reality of each of these integrals is exactly the content of the
equations (\ref{eq:realitycond}), which proves the lemma.
\end{proof}

By assumption, the endpoints satisfy the moment conditions $M_p=0$ for
$p=0,\dots,G$.  If the endpoints also satisfy the measure reality
conditions, then slightly more is true.

\begin{lemma}
Whenever the endpoints satisfy the moment conditions
(\ref{eq:momenteqns}) for $p=0,\dots,G$ and the measure reality
conditions (\ref{eq:realitycond}), then 
\begin{equation}
M_{G+1}=0
\end{equation}
as well.
\label{lemma:moremoments}
\end{lemma}

\begin{proof}
This follows immediately from the series expansion (\ref{eq:Laurent})
for the function $H(\lambda)$, along with the fact that
$F(\lambda)=R(\lambda)H(\lambda)$ where $R(\lambda)\sim
-\lambda^G$ for large $\lambda$, and the large $\lambda$ asymptotic
behavior of $F(\lambda)$ guaranteed by
Lemma~\ref{lemma:higherorder}.
\end{proof}

\begin{remark}
Lemma~\ref{lemma:moremoments} means that with the use of the measure
reality conditions (\ref{eq:realitycond}) we can deduce one additional
moment condition.  We now make the correspondence between the reality
conditions and the moment conditions more precise.  Summing up the
integrals $R_\ell$, we find
\begin{equation}
\sum_{\ell = 0}^{G/2}R_{\ell} = \Im\left(\int_{\cup_{k}I_{k}^{+}}
\rho^\sigma(\eta)\, d \eta\right)\,.
\label{eq:addemup}
\end{equation}
Now for $\eta\in I_k^+$, $\rho^\sigma(\eta)$ can be expressed in terms
of $\rho^0(\eta)$ and the difference of boundary values of $F(\eta)$.
Therefore, (\ref{eq:addemup}) becomes
\begin{equation}
\sum_{\ell = 0}^{G/2}R_{\ell} = 
\Im\left(\int_{\cup_kI_k^+}\rho^0(\eta)\,d\eta\right) -\frac{1}{2\pi}\Re
\left(\int_{\cup_kI_k^+}\Big(F_+(\eta)-F_-(\eta)\Big)\,d\eta
\right)\,.
\label{eq:addemupagain}
\end{equation}
Now for $\eta \in \Gamma_I\cap C_I$, we
have that  
$F_{+}(\eta) - F_{-}(\eta)  = 
-2\pi i \rho^{0}(\eta)$.
Thus we may rewrite (\ref{eq:addemupagain}) as 
\begin{equation}
\sum_{\ell = 0}^{G/2}R_{\ell} = 
\Im\left(\int_{C_I}\rho^0(\eta)\,d\eta\right)-\frac{1}{2\pi}
\Re\left(\int_{C_I}\Big(F_+(\eta)-F_-(\eta)\Big)\,d\eta
\right)\,,
\end{equation}
or, using analyticity to deform the path in the first integral to the
imaginary interval $[0,iA]$ and exploiting reality of the asymptotic eigenvalue
measure $\rho^0(\eta)\,d\eta$ on that path,
\begin{equation}
\sum_{\ell = 0}^{G/2}R_{\ell} = 
-\frac{1}{2\pi}
\Re\left(\int_{C_I}\Big(F_+(\eta)-F_-(\eta)\Big)\,d\eta
\right)\,.
\end{equation}
Finally, since $F(\lambda)$ satisfies the symmetry (\ref{eq:Fsymmetry}),
we can write this as
\begin{equation}
\sum_{\ell = 0}^{G/2}R_{\ell} = -\frac{1}{4\pi}\int_{C_I\cup C_I^*}
\Big(F_+(\eta)-F_-(\eta)\Big)\,d\eta\,.
\end{equation}
Now since $F(\lambda)$ is analytic in ${\mathbb C} \setminus (C_{I}
\cup C_{I}^{*})$, we may express this integral as a contour integral on
any counter-clockwise oriented loop $L$ completely 
encircling the contour $C_{I}\cup C_{I}^{*}$:
\begin{equation}
\sum_{\ell = 0}^{G/2}R_{\ell} = \frac{1}{4\pi}\oint_L F(\eta)\,d\eta\,. 
\end{equation}
Using the residue theorem to evaluate the integral, assuming only that
$F(\lambda)$ decays like $1/\lambda$ at infinity, we find at
last
\begin{equation}
\sum_{\ell = 0}^{G/2}R_{\ell} =
\frac{i}{2}\lim_{\lambda\rightarrow\infty}\lambda F(\lambda) = -\frac{M_{G+1}}{2\pi}\,.
\end{equation}
The last equality follows from the formula
$F(\lambda)=H(\lambda)R(\lambda)$, the asymptotic behavior $R(\lambda)\sim -\lambda^{G+1}$, and the fact that the moments $M_0$ through $M_G$
are presumed to vanish.  This establishes the fact that any one of the
reality conditions (\ref{eq:realitycond}) may be replaced with the additional
moment condition $M_{G+1}=0$.
\end{remark}

Next, we consider computing derivatives of the moments.  We begin with the
following.
\begin{lemma}
The moments $M_{j}$, for $j = 1, 2, \ldots$, satisfy the
following differential equations:
\begin{eqnarray}
\label{eq:momentderiv}
\frac{\partial M_j}{\partial \lambda_{k}}& =& \frac{1}{2} M_{j-1} +
\lambda_{k} \frac{\partial M_{j-1}}{\partial \lambda_{k}} \,, \\
\label{eq:momentderivconjg}
\frac{\partial M_j}{\partial \lambda^*_{k}} &=&
\frac{1}{2} M_{j-1} + \lambda^*_{k} \frac{\partial M_{j-1}}{\partial
\lambda^*_{k}}\,.
\end{eqnarray}
Furthermore, the function $F(\lambda)$ 
satisfies the following equations, valid for $\lambda
\in {\mathbb C} \setminus \left( C_{I}
\cup C_{I}^{*} \right)$, with
appropriate boundary values taken on $C_{I}\cup C_{I}^{*}$.
\begin{eqnarray}
\label{eq:Fderiv}
\frac{\partial F}{\partial \lambda_{j}} &=&
\frac{1}{\pi i}
\cdot\frac{R(\lambda)}{\lambda - \lambda_{j}}
\cdot\frac{\partial M_0}{\partial \lambda_{j}}\,, 
\\
\label{eq:Fderivconjg}
\frac{\partial F}{\partial \lambda^*_{j}}& = &
\frac{1}{\pi i}
\cdot\frac{R(\lambda)}{\lambda - \lambda^*_{j}}
\cdot\frac{\partial M_0}{\partial \lambda^*_{j}}\,.
\end{eqnarray}
\end{lemma}

\begin{proof}
To prove (\ref{eq:momentderiv}), note that 
\begin{equation}
\begin{array}{rcl}
M_{j} - \lambda_{k} M_{j-1} &=&\displaystyle
 J\int_{\cup_{\ell}I_{\ell}^\pm}
\frac{2 i x + 4 i \eta t}{R_{+}(\eta)} \eta^{j-1}( \eta - \lambda_{k})\, d
\eta \\\\
&&\displaystyle\,\,+\,\,  
\int_{\Gamma_{I}\cap C_I} \frac{ i \pi \rho^{0}(\eta)}
{R(\eta)} \eta^{j-1} ( \eta - \lambda_{k} )\,d \eta +
\int_{\Gamma_{I}\cap C_I^*} \frac{ i \pi \rho^{0}(\eta^*)^*}
{R(\eta)} \eta^{j-1} ( \eta - \lambda_{k} )\,d \eta
\,.
\end{array}
\label{eq:clevercombination}
\end{equation}
Differentiating (\ref{eq:clevercombination}) with respect to
$\lambda_{k}$, we find
\begin{equation}
\frac{\partial M_{j}}{\partial \lambda_{k}} - M_{j-1} - \lambda_{k} 
\frac{\partial M_{j-1}}{\partial \lambda_{k}}  = -\frac{1}{2} M_{j-1}\,,
\end{equation}
and we have proved (\ref{eq:momentderiv}).  To prove
(\ref{eq:momentderivconjg}), replace $\lambda_{k}$ with
$\lambda^*_{k}$ in (\ref{eq:clevercombination}), and differentiate
with respect to $\lambda^*_{k}$.

To prove (\ref{eq:Fderiv}), we use the formula $F(\lambda)=H(\lambda)R(\lambda)$ and the Laurent series representation (\ref{eq:Laurent})
for $H(\lambda)$ and differentiate with respect
to $\lambda_{k}$:
\begin{equation}
\begin{array}{rcl}
\displaystyle\frac{\partial F}{\partial \lambda_{k}} &=
&\displaystyle
\frac{R(\lambda)}{\pi i \lambda} \sum_{j=0}^{\infty} \left(
-\frac{1}{\lambda - \lambda_{k}} \cdot\frac{M_{j}}{2} + \frac{\partial
M_j}{\partial \lambda_{k}} \right) \lambda^{-j} \\\\ &=&\displaystyle
-\frac{R(\lambda)}{2\pi i \lambda} \cdot\frac{1}{\lambda -
\lambda_{k}} \left[ -2 \lambda \frac{\partial M_0}{\partial \lambda_{k}}
+ \sum_{j=0}^{\infty} \left( - 2 
\frac{\partial M_{j+1}}{\partial \lambda_{k}}  +  2 \lambda_{k} 
\frac{\partial M_j}{\partial \lambda_{k}}  + M_{j} \right) \lambda^{-j}
\right]\,, 
\end{array}
\end{equation}
where the second equality results from factoring out $(\lambda-\lambda_k)^{-1}$
and rearranging the sum.  
The relation (\ref{eq:Fderiv}) then follows by using (\ref{eq:momentderiv}).
To prove (\ref{eq:Fderivconjg}), one
differentiates with respect to $\lambda^*_{k}$,
and uses (\ref{eq:momentderivconjg}).
\end{proof}

Now, we will show that we can use the equations $M_p=0$ taken for
$p=0,\dots,G+1$,
together with the gap conditions (\ref{eq:vanishing}), and the measure
reality conditions $R_k=0$ taken for $k=1, \ldots, G/2$,
to derive the elliptic modulation equations.  First, observe
that if we evaluate (\ref{eq:momentderiv}) and (\ref{eq:momentderivconjg}) 
for a set of endpoints $
\lambda_{0},\dots,\lambda_G$ chosen to satisfy these $2G+2$ real conditions,
then we have
for all $j=1,\dots,G+1$ and $k=0,\dots,G$, 
\begin{equation}
\frac{\partial M_j}{\partial \lambda_{k}} = \lambda_{k}^{j}
\frac{\partial M_0}{\partial \lambda_{k}}\,,\hspace{0.3 in} 
\frac{\partial M_j}{\partial \lambda^*_{k}}  = 
\lambda_{k}^{*j} \frac{\partial M_0}{\partial \lambda^*_{k}}\,. 
\label{eq:momentderivsevaluate}
\end{equation}

Second, the formulae (\ref{eq:Fderiv})-(\ref{eq:Fderivconjg}) yield
rather simple representations for the derivatives of the functions
$V_{j}$ defined in (\ref{eq:vanishing}), and $R_{\ell}$ defined in
(\ref{eq:realitycond}), with respect to $\lambda_{k}$ and
$\lambda^*_{k}$.  Let us first consider the function $V_{j}$, for any
$j$, $0 \le j \le G/2 -1$.  From the formula (\ref{eq:vanishing}), and
the symmetry (\ref{eq:Fsymmetry}) of the function $F(\lambda)$,
we have the following representation of the function
$V_{j}$:
\begin{equation}
V_{j} = \frac{1}{2} \int_{\Gamma_{j+1}^+\cup\Gamma_{j+1}^-} 
\left[  2 i Jx + 4 i J\eta
t + \frac{1}{2}\Big(F_+(\eta)+F_-(\eta)\Big)\right]
\,d \eta\,.
\label{eq:Vjsymmetric}
\end{equation}
Recall that by definition $\Gamma_{j+1}^+$ is oriented from
$\lambda_{2j}$ to $\lambda_{2j+1}$, and $\Gamma_{j+1}^-$ is oriented
from $\lambda_{2j+1}^*$ to $\lambda_{2j}^*$.  Since the boundary
values of $F(\eta)$ are H\"older continuous with exponenent
$1/2$, and hence uniformly continuous, it follows from the boundary
conditions satisfied by $F(\eta)$ on $C_I\cup C_I^*$ that
the integrand in (\ref{eq:Vjsymmetric}) vanishes at the endpoints of the two
gaps of integration.  Therefore, differentiating (\ref{eq:Vjsymmetric}) with
respect to $\lambda_{k}$, one finds simply
\begin{equation}
\frac{\partial V_j}{\partial \lambda_{k}} = \frac{1}{4} 
\int_{\Gamma_{j+1}^+\cup\Gamma_{j+1}^-}
 \frac{\partial}{\partial
\lambda_{k}} \left( F_{+}(\eta) + F_{-}(\eta) 
\right)\, d \eta\,.
\label{eq:Vjderiv}
\end{equation}
Now inserting (\ref{eq:Fderiv}) into (\ref{eq:Vjderiv}), we find
\begin{equation}
\frac{\partial V_j}{\partial \lambda_{k}} = \frac{1}{2 \pi i}
\left(
\frac{\partial M_0}{\partial \lambda_{k}} \right) 
\int_{\Gamma_{j+1}^+\cup\Gamma_{j+1}^-}
\frac{R(\eta)}{\eta - \lambda_{k}}\, d \eta\,.
\label{eq:Vjderivsimp}
\end{equation}
Repeating the above calculations, but differentiating with respect to
$\lambda^*_{k}$, one may easily verify 
\begin{equation}
\frac{\partial V_j}{\partial \lambda^*_{k}}  = \frac{1}{2 \pi i}
\left(
\frac{\partial M_0}{\partial \lambda^*_{k}}  \right) 
\int_{\Gamma_{j+1}^+\cup\Gamma_{j+1}^-}
\frac{R(\eta)}{\eta - \lambda^*_{k}}\, d \eta\,.
\label{eq:Vjderivsimpconjg}
\end{equation}

To obtain the derivatives of the functions $R_{j}$, $1 \le j \le G/2$
with respect to $\lambda_0,\dots,\lambda_G$ and
$\lambda_0^*,\dots,\lambda_G^*$, we start with the following formula
for $R_{j}$:
\begin{equation}
R_j = \frac{1}{2i}\int_{I_j^+}\left[\rho^0(\eta)+\frac{1}{2\pi i}
\Big(F_+(\eta)-F_-(\eta)\Big)\right]\,d\eta +
\frac{1}{2i}\int_{I_j^-}\left[\rho^0(\eta^*)^* +\frac{1}{2\pi i}
\Big(F_+(\eta)-F_-(\eta)\Big)\right]\,d\eta\,,
\label{eq:Rjsymmetric}
\end{equation}
obtained by representing $\rho^\sigma(\eta)$ in the bands $I_j^+$ in terms
of the asymptotic eigenvalue density $\rho^0(\eta)$ and $F(\eta)$
and using the symmetry property (\ref{eq:Fsymmetry}).  Recall that 
by definition for $j>0$, 
the band $I_j^+$ is oriented from $\lambda_{2j-1}$ to 
$\lambda_{2j}$ while the conjugate band $I_j^-$ is oriented from
$\lambda_{2j}^*$ to $\lambda_{2j-1}^*$.
Also, by the same arguments as above in our discussion of the quantities
$V_j$, the integrand vanishes at the endpoints.
Therefore, differentiating (\ref{eq:Rjsymmetric}) with
respect to $\lambda_{k}$, we find
\begin{equation}
\frac{ \partial R_j}{ \partial \lambda_{k}}  = -\frac{1}{4\pi}
\int_{I_{j}^+\cup I_j^-} \frac{ \partial }{ \partial \lambda_{k}} 
\left( F_{+}(\eta) - F_{-}(\eta) \right) \, d \eta\,.
\label{eq:Rjderiv}
\end{equation}
Now inserting (\ref{eq:Fderiv}) into (\ref{eq:Rjderiv}), we find
\begin{equation}
\label{eq:Rderivsimp}
\frac{ \partial R_j}{ \partial \lambda_{k}} = \frac{i}{2 \pi^{2}}
\left(
\frac{\partial M_0}{\partial \lambda_{k}} \right)
\int_{I_{j}^+\cup I_j^-}
\frac{R_{+}(\eta)}{ \eta - \lambda_{k}}\,  d \eta\,.
\end{equation}
Repeating the above calculations, but differentiating with respect to
$\lambda^*_{k}$, one may easily derive
\begin{equation}
\label{eq:Rderivsimpconjg}
\frac{ \partial R_j}{ \partial \lambda^*_{k}} = 
\frac{i}{2 \pi^{2}} 
\left(
\frac{\partial M_0}{\partial \lambda^*_{k}}  \right)
\int_{I_{j}^+\cup I_j^-}
\frac{R_{+}(\eta)}{ \eta - \lambda^*_{k}}\,  d \eta\,.
\end{equation}
We have proved the following.

\begin{lemma} 
The partial derivatives of the quantities $M_j$, $V_j$, and $R_j$ with
respect to the endpoints $\lambda_0,\dots,\lambda_G$ and their
complex conjugates satisfy
a set of canonical formulae whenever 
the endpoints solve the equations:
\begin{equation}
\begin{array}{ll}
M_j=0\,,&
j=0,\dots,G+1\,,\\\\
V_j=0\,,&j=0,\dots,G/2-1\,,\\\\
R_j=0\,,&j=1,\dots,G/2\,.
\end{array}
\label{eq:conditionsrewrite}
\end{equation}
These formulae are:
\begin{equation}
\frac{\partial M_j}{\partial \lambda_{k}} = 
\lambda_{k}^{j}
\frac{\partial M_0}{\partial \lambda_{k}}\,,\hspace{0.3 in}
\frac{\partial M_j}{\partial \lambda^*_{k}}  = 
\lambda_{k}^{*j} \frac{\partial M_0}{\partial \lambda^*_{k}}\,,
\label{eq:summary1}
\end{equation}
for $j = 1, \ldots, G+1$ and $k = 0, \ldots, G$,
\begin{equation}
\frac{\partial V_j}{\partial \lambda_{k}} =
\frac{1}{2 \pi i}
\frac{\partial M_0}{\partial \lambda_{k}}  
\int_{\Gamma_{j+1}^+\cup\Gamma_{j+1}^-}
\frac{R(\eta)}{\eta - \lambda_{k}}\, d \eta\,,\hspace{0.3 in}
\frac{\partial V_j}{\partial \lambda^*_{k}}  = 
\frac{1}{2 \pi i}
\frac{\partial M_0}{\partial \lambda^*_{k}} 
\int_{\Gamma_{j+1}^+\cup\Gamma_{j+1}^-}
\frac{R(\eta)}{\eta - \lambda^*_{k}}\, d \eta\,,
\label{eq:summary2}
\end{equation}
for $j = 0, \ldots, G/2 - 1$ and $k = 0, \ldots, G$, and
\begin{equation}
\frac{ \partial R_j}{ \partial \lambda_{k}} =
 \frac{i}{2 \pi^{2}}
\frac{\partial M_0}{\partial \lambda_{k}} 
\int_{I_{j}^+\cup I_j^-}
\frac{R_{+}(\eta)}{ \eta - \lambda_{k}}\,  d \eta\,, \hspace{0.3 in}
\frac{ \partial R_k}{ \partial \lambda^*_{k}} = 
 \frac{i}{2 \pi^{2}}
\frac{\partial M_0}{\partial \lambda^*_{k}} 
\int_{I_{j}^+\cup I_j^-}
\frac{R_{+}(\eta)}{ \eta - \lambda^*_{k}} \, d \eta\,,
\label{eq:summary3}
\end{equation}
for $j = 1, \ldots, G/2$ and $k = 0, \ldots, G$. 
\end{lemma}

Third, we will compute the partial derivatives of $M_{p}$, $V_{k}$,
and $R_{\ell}$ with respect to $x$ and $t$.  For fixed endpoints, a
simple residue calcuation shows that $M_{p}$ satisfies
\begin{equation}
\frac{\partial M_p}{ \partial x} =  0 \,,\hspace{0.3 in}
 p = 0, \ldots,
G-1\,,
\end{equation}
while 
\begin{equation}
\frac{\partial M_G}{ \partial x}  =
  - 2 J \pi\,,\hspace{0.3 in}
\frac{\partial M_{G+1}}{ \partial x}  =
  - J \pi  
\sum_{k=0}^{G} \left( \lambda_{k} + \lambda^*_{k}
\right)\, .
\label{eq:momentxderivs}
\end{equation}
Similarly, one finds
\begin{equation}
\frac{\partial M_p}{ \partial t} =  0\,, \hspace{0.3 in} p = 0, \ldots,
G-2\,, 
\end{equation}
while
\begin{equation}
\begin{array}{c}
\displaystyle\frac{\partial M_{G-1}}{ \partial t}=  -4 \pi J\,, \hspace{0.3 in}
\frac{\partial M_G}{ \partial t} = 
-2 \pi J \sum_{k=0}^{G}
\left( \lambda_{k} + \lambda^*_{k} \right)\,, \\\\
\displaystyle\frac{\partial M_{G+1}}{ \partial t} =
  - \pi J \left[ 
\sum_{0 \le j < k \le G} \left( \lambda_{j} +
\lambda^*_{j} \right)
\left( \lambda_{k} + \lambda^*_{k} \right) 
- \frac{1}{2}\sum_{j=0}^{G} \left( \lambda_{j} -
\lambda^*_{j} \right)^{2} \right]\,.
\end{array}
\label{eq:momenttderivs}
\end{equation}

To compute the partial derivatives of the functions $V_{j}$ and
$R_{j}$ with respect to $x$ and $t$, we first observe that from the
representation $F(\lambda)=H(\lambda)R(\lambda)$ 
and the explicit formula (\ref{eq:Hsolution}) for $H(\lambda)$,
we find
\begin{equation}
\label{eq:Fxfixed}
\frac{\partial F}{\partial x}(\lambda)\Bigg|_{\mbox{endpoints fixed}}
=\frac{2J}{\pi}R(\lambda)\int_{\cup_kI_k^\pm}\frac{d\eta}{(\lambda-\eta)R_+(\eta)} = -2iJ\,,
\end{equation}
and
\begin{equation}
\label{eq:Ftfixed}
\frac{\partial F}{\partial t}(\lambda)\Bigg|_{\mbox{endpoints fixed}}
=\frac{4J}{\pi}R(\lambda)\int_{\cup_kI_k^\pm}\frac{\eta\,d\eta}{(
\lambda-\eta)R_+(\eta)} = -4iJ\lambda-4iR(\lambda)\cdot\delta_{G,0}\,,
\end{equation}
where the integral is evaluated explicitly by residues, and for the
$t$ derivative there is a residue at infinity only for $G=0$ which
explains the Kronecker delta.

\begin{remark}
These partial derivatives of $F(\lambda)$ are computed holding
the endpoints $\lambda_0,\dots,\lambda_G$ and their complex conjugates
fixed.  These formulae therefore do not contradict the discussion in
\S\ref{sec:bvp} which concerned the total variations of
$F(\lambda)$ with respect to $x$ and $t$ when the endpoints are
constrained by the moment conditions (\ref{eq:momenteqns}).
\end{remark}

Combining these with (\ref{eq:Vjsymmetric}),
we find that $V_{j}$ satisfies simply
\begin{equation}
\frac{ \partial V_j}{\partial x}  =0\,,\hspace{0.3 in} 
\frac{ \partial V_j}{\partial t}  =0\,,  
\end{equation}
for $j=0,\dots,G/2-1$.  Observe that for $G=0$, there are no gap
conditions, and in this case the Kronecker delta term in (\ref{eq:Ftfixed})
plays no role.
Similarly, from (\ref{eq:Rjsymmetric}) and the fact
that $\rho^0(\eta)$ is independent of $x$ and $t$, we find simply
\begin{equation}
\frac{ \partial R_j}{\partial x} = 
 0\,, \hspace{0.3 in}
\frac{ \partial R_j}{\partial t} = 
0\,,
\end{equation}
for $j=1,\dots,G/2$.  Again, note that for $G=0$, the Kronecker delta
term in (\ref{eq:Ftfixed}) plays no role because we are considering the
only measure reality condition $R_0=0$ present to be absorbed into the
additional moment condition $M_1=0$.

Finally we indicate how the $2G + 2$ real conditions
(\ref{eq:conditionsrewrite}) imply the elliptic Whitham modulation
equations.  Define the column vector $\vec{ \lambda} := \left(
\lambda_{0}, \lambda^*_{0}, \lambda_{1},
\lambda^*_{1}, \ldots, \lambda_{G},
\lambda^*_{G} \right)^{T}$, and the vector-valued function $ {\cal
G}(\vec{ \lambda} )$ via
\begin{equation}
{\cal G}(\vec{ \lambda} )^{T} =({\cal G}_1,\dots,{\cal G}_{2G+2}):= 
\left( M_{0}, \ldots, M_{G+1}, V_{0}, \ldots, V_{G/2-1}, R_{1}, \ldots,
R_{G/2} \right)\,.
\end{equation}
Then the equations (\ref{eq:conditionsrewrite})
are written compactly as
\begin{equation}
{\cal G}(\vec{ \lambda}) = \vec{  0}\,.
\end{equation}
Differentiating with respect to $x$ and 
$t$, we find
\begin{equation}
{\bf  M }(\vec{\lambda}) \frac{\partial\vec{\lambda}}{\partial x} =
- \frac{ \partial {\cal G}}{\partial x}\,,\hspace{0.3 in}
{\bf  M}(\vec{\lambda})  \frac{\partial\vec{\lambda}}{\partial t}
 =
- \frac{ \partial {\cal G}}{\partial t}\,,
\label{eq:tosolve}
\end{equation}
where 
\begin{equation}
\frac{\partial {\cal G}_j}{\partial x}=0\,,\hspace{0.3 in} \mbox{for}\,\,j=1,
\dots,G\,,\,\,\mbox{and}\,\,j=G+3,\dots,2G+2\,,
\end{equation}
while
\begin{equation}
\frac{ \partial {\cal G}_{G+1}}{\partial x} = - 2 J \pi\,,\hspace{0.3 in}
\frac{ \partial {\cal G}_{G+2}}{\partial x} = - J \pi
\sum_{k=0}^{G} \left( \lambda_{k} + \lambda^*_{k}
\right)\,, 
\end{equation}
and
\begin{equation}
\frac{ \partial {\cal G}_j}{\partial t} = 0\,,\hspace{0.3 in}
\mbox{for}\,\, j = 1,
\dots, G-1\,,\,\, \mbox{and}\,\, j = G+3, \dots, 2G + 2\,,
\end{equation}
while
\begin{equation}
\begin{array}{c}
\displaystyle
\frac{ \partial {\cal G}_{G}}{\partial t} = - 4 J \pi\,, \hspace{0.3 in}
\frac{ \partial {\cal G}_{G+1}}{\partial t}= - 2 J \pi
\sum_{k=0}^{G}
\left( \lambda_{k} + \lambda^*_{k} \right)\,, \\\\
\displaystyle
\frac{ \partial {\cal G}_{G+2}}{\partial t} =\displaystyle - J \pi \left[
\sum_{0 \le j < k \le G} \left( \lambda_{j} +
\lambda^*_{j}\right)
\left( \lambda_{k}+\lambda^*_{k} \right) - \frac{1}{2}
\sum_{j=0}^{G}
\left( \lambda_{j} - \lambda^*_{j} \right)^{2} \right]\,,
\end{array}
\end{equation}
and the (Jacobian) matrix ${\bf M}(\vec{\lambda})$ is defined by
\begin{equation}
{\bf M}(\vec{\lambda}) := \frac{\partial{\cal G}}{\partial\vec{\lambda}}=
\left[
\begin{array}{cccc}
\displaystyle\frac{\partial M_0}{\partial \lambda_{0}} & 
\displaystyle\frac{\partial M_0}{\partial \lambda^*_{0}} & 
\cdots & 
\displaystyle\frac{\partial M_0}{\partial \lambda^*_G} \\
\vdots & \vdots &\vdots & \vdots \\
\displaystyle\frac{\partial M_{G+1}}{\partial \lambda_{0}} & 
\displaystyle\frac{\partial M_{G+1}}{\partial \lambda^*_{0}}  & 
\cdots & 
\displaystyle\frac{\partial M_{G+1}}{\partial \lambda^*_G} \\\\
\displaystyle\frac{\partial V_0}{\partial \lambda_{0}} & 
\displaystyle\frac{\partial V_0}{\partial \lambda^*_{0}} &
\cdots &
\displaystyle\frac{\partial V_0}{\partial \lambda^*_G}  \\
\vdots & \vdots &\vdots & \vdots \\
\displaystyle\frac{\partial V_{G/2-1}}{\partial \lambda_{0}}  & 
\displaystyle\frac{\partial V_{G/2-1}}{\partial \lambda^*_{0}}  &
\cdots &
\displaystyle\frac{\partial V_{G/2-1}}{\partial \lambda^*_G} \\\\
\displaystyle\frac{\partial R_1}{\partial \lambda_{0}} & 
\displaystyle\frac{\partial R_1}{\partial \lambda^*_{0}}  &
\cdots &
\displaystyle\frac{\partial R_1}{\partial \lambda^*_G}  \\
\vdots & \vdots &\vdots & \vdots \\
\displaystyle\frac{\partial R_{G/2}}{\partial \lambda_{0}} & 
\displaystyle\frac{\partial R_{G/2}}{\partial \lambda^*_{0}} &
\cdots &
\displaystyle\frac{\partial R_{G/2}}{\partial \lambda^*_G}
\end{array} \right]\,.
\end{equation}
Now using the relations (\ref{eq:summary1}), (\ref{eq:summary2}), and
(\ref{eq:summary3}), we find that miraculously, the Jacobian ${\bf
M}(\vec{\lambda})$ factors:
\begin{equation}
{\bf M}(\vec{\lambda}) = 
\mbox{diag} \left( 1, \ldots, 1, \frac{1}{2 \pi i}, \ldots,
\frac{1}{2 \pi i}, \frac{i}{2 \pi^{2}}, \ldots, \frac{i}{2 \pi^{2}}
\right) \cdot \tilde{\bf M}(\vec{\lambda})\cdot\frac{\partial M_0}{\partial\vec{\lambda}}\,,
\end{equation}
where
\begin{equation}
\frac{\partial M_0}{\partial\vec{\lambda}}:=
 \mbox{ diag } \left( 
\frac{\partial M_0}{\partial \lambda_{0}}, 
\frac{\partial M_0}{\partial \lambda^*_{0}},
\ldots, 
\frac{\partial M_0}{\partial \lambda_G}, 
\frac{\partial M_0}{\partial \lambda^*_G} \right)\,,
\end{equation}
and where
\begin{equation}
\tilde{\bf M}(\vec{\lambda}) :=
\left[ \begin{array}{cccc}
1 & 1 & \cdots & 1 \\
\lambda_{0} & \lambda^*_{0} & \cdots &\lambda^*_G\\
\vdots & \vdots & \vdots & \vdots \\
\lambda_{0}^{G+1} & \lambda_{0}^{*G+1} & \cdots &
\lambda_G^{*G+1} \\\\
\displaystyle \int_{ \Gamma_1^+\cup\Gamma_1^-} \frac{R(\eta)\, d \eta
}{\eta - \lambda_{0}} & 
\displaystyle \int_{ \Gamma_1^+\cup\Gamma_1^-} \frac{R(\eta)\, d \eta
}{\eta - \lambda^*_{0}} & \cdots &
\displaystyle \int_{\Gamma_1^+\cup\Gamma_1^-} \frac{R(\eta)\, d \eta
}{\eta - \lambda^*_G} \\
\vdots & \vdots & \vdots & \vdots \\
\displaystyle \int_{ \Gamma_{G/2}^+\cup\Gamma_{G/2}^-} 
\frac{R(\eta)\, d
\eta 
}{\eta - \lambda_{0}} & 
\displaystyle \int_{ \Gamma_{G/2}^+\cup\Gamma_{G/2}^-} 
\frac{R(\eta)\, d
\eta 
}{\eta - \lambda^*_{0}} & \cdots &
\displaystyle
 \int_{ \Gamma_{G/2}^+\cup\Gamma_{G/2}^-} 
\frac{R(\eta)\, d
\eta
}{\eta - \lambda^*_G} \\\\
\displaystyle \int_{I_{1}^+\cup I_1^-} 
\frac{R_{+}(\eta)\, d \eta }{\eta - \lambda_{0}} &
\displaystyle \int_{I_{1}^+\cup I_1^-} 
\frac{R_{+}(\eta)\, d \eta }{\eta - \lambda^*_{0}} &
\cdots &
\displaystyle \int_{I_{1}^+\cup I_1^-} 
\frac{R_{+}(\eta)\, d \eta }{\eta - \lambda^*_G} \\
\vdots & \vdots & \vdots & \vdots \\
\displaystyle \int_{I_{G/2}^+\cup I_{G/2}^-} 
\frac{R_{+}(\eta)\, d \eta }{\eta - \lambda_{0}} &
\displaystyle \int_{I_{G/2}^+\cup I_{G/2}^-} 
\frac{R_{+}(\eta)\, d \eta }{\eta - \lambda^*_{0}}
&
\cdots &
\displaystyle \int_{I_{G/2}^+\cup I_{G/2}^-} 
\frac{R_{+}(\eta)\, d \eta }{\eta - \lambda^*_G} 
\end{array} \right]\,.
\end{equation}
The determinant of $\tilde{\bf M}(\vec{\lambda})$ can be calculated
explicitly.  First, one uses the linearity of the determinant in each
row to write
\begin{equation}
\det\tilde{\bf M}(\vec{\lambda}) = 
\int_{\Gamma_1^+\cup\Gamma_1^-}\dots\int_{\Gamma_{G/2}^+\cup\Gamma_{G/2}^-}
\prod_{j=1}^{G/2} R(\eta_j)\,d\eta_j
\int_{I_1^+\cup I_1^-}\dots\int_{I_{G/2}^+\cup I_{G/2}^-}
\prod_{k=G/2+1}^{G} R_+(\eta_k)\,d\eta_k \,\det {\bf S}(\vec{\lambda},\vec{\eta})\,,
\end{equation}
where
\begin{equation}
{\bf S}(\vec{\lambda},\vec{\eta})
=\left[\begin{array}{cccc}
1 & 1 & \cdots & 1 \\
\lambda_{0} & \lambda^*_{0} & \cdots &\lambda^*_G\\
\vdots & \vdots & \vdots & \vdots \\
\lambda_{0}^{G+1} & \lambda_{0}^{*G+1} & \cdots &
\lambda_G^{*G+1} \\\\
\displaystyle \frac{1}{\eta_1 - \lambda_{0}} & 
\displaystyle \frac{1}{\eta_1 - \lambda^*_{0}} & \cdots &
\displaystyle \frac{1}{\eta_1 - \lambda^*_G} \\
\vdots & \vdots & \vdots & \vdots \\
\displaystyle \frac{1}{\eta_{G} - \lambda_{0}} & 
\displaystyle \frac{1}{\eta_{G} - \lambda^*_{0}} & \cdots &
\displaystyle \frac{1}{\eta_{G} - \lambda^*_G}
\end{array}
\right]\,.
\end{equation}
This matrix is a combination of a Vandermonde matrix
\index{Vandermonde matrix} and a Cauchy matrix\index{Cauchy matrix}.  
The determinant $\det{\bf S}(\vec{\lambda},\vec{\eta})$ can be
computed by observing that it is a rational function in each variable
with obvious singularities and with the same number of explicit zeros.
For example, as a function of $\lambda_0$, the determinant has $G$
simple poles at $\eta_1,\dots,\eta_G$, and behaves like
$\lambda_0^{G+1}$ near infinity.  Therefore it has exactly $2G+1$
zeros, and it is easy to see that these occur exactly for
$\lambda_0=\lambda_0^*,
\lambda_1,\dots,
\lambda_G^*$ since each of these choices makes two columns identical.
By Liouville's theorem\index{Liouville's theorem}, this fixes the
determinant up to a constant factor, which may be obtained by similar
considerations viewing the determinant as a function of the other
variables.  In any case, we find
\begin{equation}
\det{\bf S}(\vec{\lambda},\vec{\eta}) = 
\frac{\displaystyle\prod_{j=0}^{G} \prod_{k=0}^{G}
(\lambda^*_{k} - \lambda_{j} ) \prod_{0 \le j < k \le
G} (\lambda_{k}-\lambda_{j})(\lambda^*_{k} -
\lambda^*_{j}) \prod_{1 \le j < k \le G} ( \eta_{k} -
\eta_{j})}
{\displaystyle (-1)^{G} \prod_{j=1}^{G} \prod_{k=0}^{G}
(\eta_{j} - \lambda_{k})(\eta_{j}-\lambda^*_{k})}\,.
\end{equation}

It is straightforward to solve (\ref{eq:tosolve}) for
$\partial\vec{\lambda}/\partial x$ and $\partial\vec{\lambda}/\partial
t$ by Cramer's rule\index{Cramer's rule}.  In doing so, one first
inverts the diagonal prefactor and notes that from the positions of
the only nonzero entries on the right-hand side, (\ref{eq:tosolve}) is
really just
\begin{equation}
\tilde{\bf M}(\vec{\lambda})\cdot\frac{\partial M_0}{\partial
\vec{\lambda}}\cdot
\frac{\partial\vec{\lambda}}{\partial x}=
-\frac{\partial{\cal G}}{\partial x}\,,\hspace{0.3 in}
\tilde{\bf M}(\vec{\lambda})\cdot\frac{\partial M_0}{\partial
\vec{\lambda}}\cdot
\frac{\partial\vec{\lambda}}{\partial t}=
-\frac{\partial{\cal G}}{\partial t}\,.
\end{equation}
By a direct calculation, one can see that none of the partial
derivatives of $M_0$ with respect to an endpoint vanishes identically.
Therefore, inverting the diagonal matrix $\partial
M_0/\partial\vec{\lambda}$ explicitly,
one finds that for $k=1,\dots,2G+2$,
\begin{equation}
\left(\frac{\partial\vec{\lambda}}{\partial t}\right)_k + 
c_k(\vec{\lambda})\left(\frac{\partial\vec{\lambda}}{\partial
x}\right)_k = 0\,,
\label{eq:generalWhitham}
\end{equation}
where 
\begin{equation}
c_k(\vec{\lambda}):=-\frac{\det \tilde{\bf M}^{(k,t)}}{\det\tilde{\bf
M}^{(k,x)}}\,,
\label{eq:Whithamspeed}
\end{equation}
and $\tilde{\bf M}^{(k,x)}$ is the matrix obtained from $\tilde{\bf
M}$ by replacing the $k$th column with $\partial{\cal G}/\partial x$
while $\tilde{\bf M}^{(k,t)}$ is the matrix obtained from $\tilde{\bf
M}$ by replacing the $k$th column with $\partial{\cal G}/\partial t$.
Note that (\ref{eq:generalWhitham}) is a first-order system of
quasilinear partial differential equations in $x$ and $t$ that is
explicitly written in Riemann-invariant form \index{Whitham
equations!Riemann-invariant form of} regardless of the size of the
system (value of $G$).  Also, it is clear from the definition
(\ref{eq:Whithamspeed}) that the characteristic velocities
\index{Whitham equations!characteristic velocities for} 
$c_k(\vec{\lambda})$ have no explicit dependence on $x$ and $t$.

Without belaboring the point, let us observe in passing that in
computing these determinants, we have established that the $2G + 2$
conditions contained in (\ref{eq:momenteqns}), (\ref{eq:vanishing}),
and (\ref{eq:realitycond}) actually imply that the endpoints
$\lambda_0,\dots,\lambda_G$ and their complex conjugates solve the
partial differential equations (\ref{eq:generalWhitham}).  The
characteristic velocities (\ref{eq:Whithamspeed}) are explicitly
expressed in terms of ratios of determinants of matrices whose entries
are all hyperelliptic integrals.  Also, as seen in
\S\ref{sec:outersolve}, the dependent variables $\lambda_0,\dots,
\lambda_G$ and their complex conjugates have the interpretation of moduli of
a hyperelliptic Riemann surface used in the reconstruction of the
asymptotic semiclassical solution in the vicinity of fixed $x$ and
$t$.  The system (\ref{eq:generalWhitham}) is therefore just the set
of Whitham or modulation equations for genus $G$ wavetrain solutions
of the focusing nonlinear Schr\"odinger equation, expressed in
Riemann-invariant form.  See, for example, \cite{FL86} for a formal
derivation of these equations from the starting point of the
assumption of an approximate solution of the focusing nonlinear
Schr\"odinger equation in the form of a slowly modulated
wavetrain\index{multiphase wavetrain!slowly modulated}.

Our formula (\ref{eq:Whithamspeed}) for the characteristic velocities
is not written in exactly the same form as in Forest and Lee's paper
\cite{FL86}.  Making the identification requires identifying the
ratios of determinants in (\ref{eq:Whithamspeed}) with those obtained in
\cite{FL86} by the normalization of the pair of canonical meromorphic
differentials by adding appropriate holomorphic differentials to
achieve zero $a$ cycles.  We do not concern ourselves further with the
aforementioned equivalence, leaving this to the interested reader.  

We want to emphasize that certain steps in obtaining the equations
(\ref{eq:generalWhitham}) from the solution of (\ref{eq:momenteqns}),
(\ref{eq:vanishing}), and (\ref{eq:realitycond}), such as the
nontrivial issue of proving that the matrix $\tilde{\bf M}$ possesses
an inverse, require very delicate analysis.  In any case, from the
point of view of computing rigorous semiclassical asymptotics, the
non-differential relations (\ref{eq:momenteqns}),
(\ref{eq:vanishing}), and (\ref{eq:realitycond}) form a complete
characterization of the endpoints, always containing information about
the approximate initial data $A(x)$ encoded in the asymptotic
eigenvalue density $\rho^0(\eta)$.  From this point of view, the fact
that the system (\ref{eq:generalWhitham}) does not contain any
reference to the initial data via $\rho^0(\eta)$ and yet is satisfied
by solutions of (\ref{eq:momenteqns}), (\ref{eq:vanishing}), and
(\ref{eq:realitycond}), which do depend on $\rho^0(\eta)$, is a happy
coincidence.

To make the derivation of the Whitham equations more concrete, let us
now carry out the above program for the case of genus $G=0$.  We have
the following matrix equation satisfied by $\{\lambda_{0},
\lambda^*_{0}
\}$:
\begin{equation}
\label{M02.062}
\left[\begin{array}{cc}\displaystyle 
\frac{\partial M_{0}}{\partial\lambda_0} &
\displaystyle \frac{\partial M_{0}}{\partial \lambda^*_{0}} \\\\ 
\displaystyle \frac{\partial M_{1}}{\partial \lambda_{0}} &
\displaystyle \frac{\partial M_{1}}{\partial 
\lambda^*_{0}}\end{array}\right]
\left[
\begin{array}{c}
\displaystyle \frac{\partial \lambda_{0}}{\partial x} \\\\
\displaystyle \frac{\partial \lambda^*_{0}}{\partial x}
\end{array}
\right] =  -
\left[
\begin{array}{c}
\displaystyle\frac{\partial M_{0}}{\partial x} \\\\
\displaystyle\frac{\partial M_{1}}{\partial x}
\end{array}
\right]\,.
\end{equation}
Now from (\ref{eq:momentxderivs}) we find that
\begin{equation}
\frac{\partial M_{0}}{\partial x} = - 2 \pi J\,, \hspace{0.3 in}
\frac{\partial M_{1}}{\partial x} = - 2 \pi J a_{0}\,,
\end{equation} 
where $a_0:=(\lambda_0+\lambda_0^*)/2$, and so equation
(\ref{M02.062}) becomes
\begin{equation}
\left[\begin{array}{cc}
\displaystyle\frac{\partial M_{0}}{\partial \lambda_{0}} &
\displaystyle\frac{\partial M_{0}}{\partial \lambda^*_{0}} \\\\ 
\displaystyle\lambda_{0} \frac{\partial M_{0}}{\partial \lambda_{0}} &
\displaystyle \lambda^*_{0} \frac{\partial M_{0}}{\partial \lambda^*_{0}}
\end{array}\right]
\left[
\begin{array}{c}
\displaystyle \frac{\partial \lambda_{0}}{\partial x} \\\\
\displaystyle \frac{\partial \lambda^*_{0}}{\partial x}
\end{array}
\right] =  2 \pi J
\left[
\begin{array}{c}
1 \\\\ a_{0}
\end{array}
\right]\,.
\end{equation}
Here we have also used (\ref{eq:momentderivsevaluate}).
We may simplify this as follows:
\begin{equation}
\left[\begin{array}{cc}1&   1\\\\
\lambda_{0} &
\lambda^*_{0} \end{array}\right]
\left[
\begin{array}{c}
\displaystyle\frac{\partial M_{0}}{\partial \lambda_{0} }\cdot
\frac{\partial \lambda_{0}}{\partial x} \\\\
\displaystyle\frac{\partial M_{0}}{\partial \lambda^*_{0}}\cdot
\frac{\partial \lambda^*_{0}}{\partial x}
\end{array}
\right] =  2 \pi J
\left[
\begin{array}{c}
1 \\\\
a_{0} 
\end{array}
\right]\,,
\end{equation}
whose solution is given by
\begin{equation}
\label{M02.067}
\left[
\begin{array}{c}
\displaystyle\frac{\partial M_{0}}{\partial \lambda_{0} }\cdot
\frac{\partial \lambda_{0}}{\partial x} \\\\
\displaystyle\frac{\partial M_{0}}{\partial \lambda^*_{0}}\cdot
\frac{\partial \lambda^*_{0}}{\partial x}
\end{array}
\right]
= \pi J
\left[ \begin{array}{c}
1\\\\
1
\end{array}
\right]\,.
\end{equation}
Similarly, 
for the $t$ derivatives, we find
\begin{equation}
\left[\begin{array}{cc}\displaystyle
\frac{\partial  M_{0}}{\partial \lambda_{0}} &
\displaystyle\frac{\partial M_{0}}{\partial \lambda^*_{0}} \\\\
\displaystyle\lambda_{0}\frac{\partial M_{0}}{\partial \lambda_{0}} &
\displaystyle \lambda^*_{0}\frac{\partial M_{0}}{\partial \lambda^*_{0}}
\end{array}\right]
\left[
\begin{array}{c}
\displaystyle\frac{\partial\lambda_{0}}{\partial t} \\\\
\displaystyle\frac{\partial \lambda^*_{0}}{\partial t}
\end{array}
\right] =  2 \pi J
\left[
\begin{array}{c}
2 a_{0}  \\\\
2 a_{0}^{2} - b_{0}^{2} 
\end{array}
\right]\,,
\end{equation}
where $b_0:=(\lambda_0-\lambda_0^*)/(2i)$.  This uses
(\ref{eq:momenttderivs}), the fact that $\partial M_{0}/\partial t = -
4 \pi a_{0}$, and $\partial M_{1}/\partial t = - 2 \pi \left( 2
a_{0}^{2} - b_{0}^{2}
\right)$.
From this we find that 
\begin{equation}
\label{M02.069}
\left[
\begin{array}{c}
\displaystyle\frac{\partial M_{0}}{\partial\lambda_{0} }\cdot
\frac{\partial \lambda_{0}}{\partial t} \\\\
\displaystyle \frac{\partial M_{0}}{\partial \lambda^*_{0}}\cdot
\frac{\partial \lambda^*_{0}}{\partial t}
\end{array}
\right]
= \pi J
\left[ \begin{array}{c}
 2 a_{0} + i b_{0} \\\\
2 a_{0} - i b_{0}
\end{array}
\right]\,.
\end{equation}
Combining (\ref{M02.067}) and (\ref{M02.069})
gives at last the following.
\begin{theorem}
Let $G=0$, and let $\lambda_0(x,t)$ be any solution of the moment
equations $M_0=0$ and $M_1=0$ that is differentiable with respect to $x$ 
and $t$ in some open set in the $(x,t)$-plane.  Then the function
$\lambda_0(x,t)$ satisfies the system of partial differential
equations
\begin{equation}
\frac{\partial \lambda_0}{\partial t} + (-2a_0-ib_0)
\frac{\partial \lambda_0}{\partial x}=0\,,\hspace{0.3 in}
\frac{\partial \lambda_0^*}{\partial t} + (-2a_0+ib_0)
\frac{\partial \lambda_0^*}{\partial x}=0\,,
\label{eq:genuszeroWhitham}
\end{equation}
where $\lambda_0(x,t)=a_0(x,t)+ib_0(x,t)$.  This system is exactly the
complex form of the elliptic modulation equations for genus $G=0$.
\label{theorem:G0Whitham}
\end{theorem}

\section[Symmetries of the Equations]{Symmetries of the endpoint equations.}
\label{sec:symmetry}
The relations that determine the endpoints as functions of $x$ and $t$
for an ansatz of a given even genus $G$ involve contour integrals over
paths that are not known {\em a priori}.  In \S\ref{sec:postponing},
it was shown by elementary contour deformation arguments that given an
ordered sequence of complex endpoints $\lambda_0,\dots,\lambda_G$, the
moment conditions (\ref{eq:momenteqns}), vanishing conditions
(\ref{eq:vanishing}) and measure reality conditions
(\ref{eq:realitycond}) have the same value for all contours $C_I$ in
the cut upper half-plane ${\mathbb H}$ connecting the origin to $iA$
via this sequence of points that can be smoothly deformed into each
other while holding the intermediate points
$\lambda_0,\dots,\lambda_G$ fixed.

\begin{figure}[h]
\begin{center}
\mbox{\psfig{file=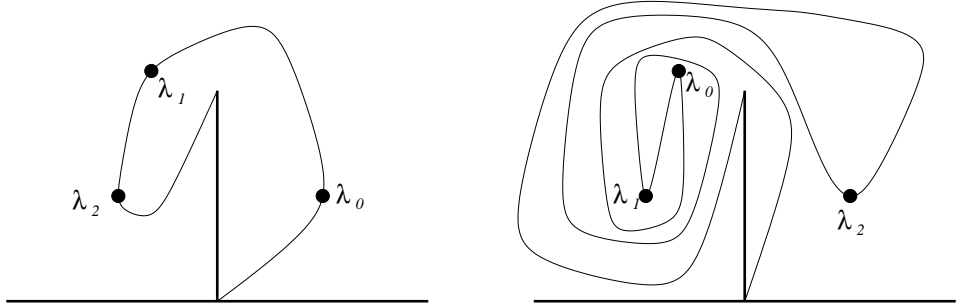,width=5 in}}
\end{center}
\caption{\em Two ways to thread a contour $C_I$ through the same three
points in ${\mathbb H}$.}
\label{fig:Reorder}
\end{figure}

But this fact alone does not provide sufficient invariance.  One would
really like to know that the determination of the endpoints is
completely insensitive to the choice of integration contour and even
the ordering of the endpoints along the contour ({\em i.e.} which
intervals between the endpoints constitute bands and which constitute
gaps).  For example, if the configuration on the left in
Figure~\ref{fig:Reorder} satisfies the endpoint relations for genus $G=2$,
then it should follow that the configuration on the right does as well.
Note that these invariance issues are nontrivial compared with inverse
problems like the zero-dispersion limit \index{zero-dispersion limit}
of the Korteweg-de Vries equation \index{Korteweg-de Vries equation} 
\cite{LL83} and the continuum limit
of the Toda lattice \index{Toda lattice}
\cite{DM98}.  In these selfadjoint problems, the endpoints are totally
ordered because they are necessarily real and similarly there is no
ambiguity whatsoever about paths of integration.

In this section, we explore the symmetries of the equations
(\ref{eq:momenteqns}), (\ref{eq:vanishing}), and
(\ref{eq:realitycond}) in more detail.  According to the
calculations presented in \S\ref{sec:modulation}, we are free to
replace the condition $R_0=0$ with $M_{G+1}=0$, and we do this here.
We begin with the following lemma.

\begin{lemma}
Each moment $M_p$ defined by (\ref{eq:momenteqns}) depends only on the
endpoints $\lambda_0,\dots,\lambda_G$.  Considered as a function of
the independent complex variables
$\lambda_0,\dots,\lambda_G\in{\mathbb H}$ and
$\lambda_0^*,\dots,\lambda_G^*\in{\mathbb H}^*$, it is analytic in
$({\mathbb H}\times{\mathbb H}^*)^{G+1}$ and is symmetric under any
permutation among the endpoints $(\lambda_0,\dots,\lambda_G)$ or,
independently, among $(\lambda_0^*,\dots,\lambda_G^*)$.
\end{lemma}

\begin{proof}
The complexification of the moment $M_p$ is the formula
\begin{equation}
M_p=J\int_{\cup_kI_k^\pm}\frac{2ix+4i\eta t}{R_+(\eta)}\eta^p\,d\eta +
\int_{\Gamma_I}\frac{\pi i\rho^0(\eta)}{R(\eta)}\eta^p\,d\eta +
\int_{\Gamma_I^*}\frac{\pi i\rho^0(\eta^*)^*}{R(\eta)}\eta^p\,d\eta\,,
\end{equation}
that is, when $\lambda_k^*$ is taken to be the complex conjugate of
$\lambda_k$, this formula agrees with (\ref{eq:momenteqns}).  Using
now-familiar contour deformations and the paths $C_{I+}$ and $C_{I-}$
introduced in the proof of Lemma~\ref{lemma:rhowithY} to represent the
function $Y(\lambda)$, one rewrites the moment as
\begin{equation}
M_p=-\frac{J}{2}\oint_L\frac{2ix+4i\eta t}{R(\eta)}\eta^p\,d\eta +
\frac{1}{2}\int_{C_{I+}\cup C_{I-}}\frac{\pi i\rho^0(\eta)}{R(\eta)}
\eta^p\,d\eta +\frac{1}{2}\int_{C_{I+}^*\cup C_{I-}^*}
\frac{\pi i\rho^0(\eta^*)^*}{R(\eta)}\eta^p\,d\eta\,,
\label{eq:momentrewrite}
\end{equation}
where $L$ is an arbitrarily large positively oriented loop contour and
where the conjugate paths $C_{I\pm}^*$ are taken to be oriented from
$-iA$ toward the origin.  Note that the paths $C_{I\pm}$ may be taken
to be the same for all choices of the path $C_I$; the path
$C_{I\sigma}$ may be taken as the imaginary interval $[0,iA]$ and the
path $C_{I(-\sigma)}$ may be deformed toward infinity with the only
obstruction being any points of nonanalyticity of $\rho^0(\eta)$.

With the moment $M_p$ rewritten in this way, the only dependence on
the endpoints enters through the function $R(\eta)$.  Since all cuts
of this function are contained inside the closed contour $L$ and also
between the contours $C_{I+}$ and $C_{I-}$ or between $C_{I+}^*$ and
$C_{I-}^*$, and since the branch of the square root is determined by
asymptotic behavior at infinity, $M_p$ is easily seen to be completely
independent of $C_I$ and an analytic function of the $2G+2$
independent complex variables
$\lambda_0,\dots,\lambda_G,\lambda_0^*,\dots,\lambda_G^*$.  The permutation
symmetry of swapping any pair of endpoints $\lambda_j\leftrightarrow\lambda_k$
or any pair $\lambda_j^*\leftrightarrow\lambda_k^*$ follows from similar
considerations.
\end{proof}

Next, we consider the vanishing conditions (\ref{eq:vanishing}) and
the reality conditions (\ref{eq:realitycond}).  These two apparently
different kinds of conditions are essentially equivalent.  This is
because it follows from differentiating the relation (\ref{eq:plusside}) 
with respect to $\lambda$ that in the gaps of $C_I$,
\begin{equation}
\frac{d\tilde{\phi}^\sigma}{d\lambda}(\lambda)=i\pi\rho^\sigma(\lambda)\,,
\end{equation}
where on the right-hand side the function $\rho^\sigma$ is
analytically continued from the ``$+$'' side of any band.  Using the
formula (\ref{eq:rhowithY}), we see that while the reality functions $R_k$
can be written for $k=1,\dots,G/2$ as
\begin{equation}
R_k=\frac{1}{2i}\left[\int_{I_k^+}R_+(\eta)Y(\eta)\,d\eta +
\int_{I_k^-}R_+(\eta)Y(\eta)\,d\eta\right]\,,
\end{equation}
the vanishing functions $V_k$ can be similarly written for $k=0,\dots,G/2-1$ as
\begin{equation}
V_k=-\frac{\pi}{2i}\left[\int_{\Gamma_{k+1}^+}R(\eta)Y(\eta)\,d\eta +
\int_{\Gamma_{k+1}^-}R(\eta)Y(\eta)\,d\eta\right]\,.
\end{equation}
By passing to the Riemann surface of the square root function
$R(\lambda)$, the functions $\pi R_k$ for $k=1,\dots,G/2$ and $V_j$
for $j=0,\dots,G/2-1$ can be reinterpreted as periods (integrals over
complete homology cycles) {\em of the same
differential}\index{differential!periods of}.  Fix a set of $G+1$
complex endpoints in the cut upper half-plane ${\mathbb H}$, and
consider two different paths $C_I$ and $C_I'$ interpolating these
points, possibly in different order.  Let ${\bf
v}:=(R_1,\dots,R_{G/2},V_0,\dots,V_{G/2-1})^T$ be the vector of
functions corresponding to the path $C_I$, and likewise let ${\bf v}'$
correspond to the path $C_I'$.  Then it is possible to show using
homology arguments that ${\bf v}'={\bf G}_{{\bf v}\rightarrow{\bf v}'}
{\bf v}$, where ${\bf G}_{{\bf v}\rightarrow{\bf v}'}$ is an
invertible matrix with integer entries.  Thus, while each separate
function $R_k$ and $V_j$ undergoes nontrivial monodromy
\index{monodromy} when the path $C_I$ is changed by adding a cycle, or
the branch points are re-ordered, the zero locus of the full set of
equations is invariant.  This leads us to state the following.

\begin{lemma}
The common zero locus of the vanishing conditions (\ref{eq:vanishing})
and the reality conditions (\ref{eq:realitycond}) is independent of
the ordering of the endpoints $\lambda_0,\dots,\lambda_G$ and of the
contour $C_I$.
\end{lemma}

\begin{remark}
These statements about the vanishing conditions and reality conditions
are only valid in the real subspace of ${\mathbb C}^{2G+2}$ when the
variables $\lambda_k$ and $\lambda_k^*$ are linked by complex conjugation.
\end{remark}

\begin{remark}
Unlike the moments $M_p$, the functions $V_j$ and $R_k$ are
multivalued functions.  They are branched when either of the two
endpoints of the corresponding integral coalesces with any another
$\lambda_k$ different from the opposite endpoint.  
\end{remark}

Together these two results imply the main symmetry result.

\begin{theorem}
Consider the equations $M_p=0$ for $p=0,\dots,G+1$, $V_j=0$ for
$j=0,\dots, G/2-1$, and $R_k=0$ for $k=1,\dots,G/2$.  Then the set of
real solutions (that is, where $\lambda_k^*$ is the complex conjugate
of $\lambda_k$) of this system is invariant under permutations of the
endpoints and arbitrary redefinitions of the interpolating contour $C_I$.
\end{theorem}

\chapter{The Genus Zero Ansatz}
\label{sec:genuszero}
\section[Location of the Endpoints]{Location of the endpoints for general data.}
\index{genus $G$ ansatz!for $G=0$} 
For $G=0$, there is only one complex endpoint to determine,
$\lambda_0$.  This endpoint is constrained by one moment condition and
one measure reality condition.  Both conditions are real and taken
together are expected to determine the endpoint up to a discrete
multiplicity of solutions.  The equations that constrain the endpoint
for $G=0$ are:
\begin{equation}
M_0=J\int_{I_0}\frac{2ix+4i\eta t}{R_+(\eta)}\,d\eta+\int_{\Gamma_I\cap C_I}
\frac{\pi i\rho^0(\eta)}{R(\eta)}\,d\eta
+\int_{\Gamma_I\cap C_I^*}\frac{\pi i\rho^0(\eta^*)^*}{R(\eta)}\,d\eta=0\,,
\end{equation}
and
\begin{equation}
R_0=\Im\left(\int_0^{\lambda_0}\rho^\sigma(\eta)\,d\eta
\right)=0\,.
\end{equation}
In the measure reality condition $R_0=0$, we use the formula
(\ref{eq:rhoformula}) for the candidate measure $\rho^\sigma(\eta)$
valid in the band $I_0^+$:
\begin{equation}
\rho^\sigma(\lambda)=\rho^0(\lambda)-\frac{4Jt}{\pi}R_+(\lambda)+
\frac{R_+(\lambda)}{\pi
i}\int_{\Gamma_I\cap C_I}\frac{\rho^0(\eta)\,d\eta}
{(\lambda-\eta)R(\eta)}+
\frac{R_+(\lambda)}{\pi
i}\int_{\Gamma_I\cap C_I^*}\frac{\rho^0(\eta^*)^*\,d\eta}
{(\lambda-\eta)R(\eta)}\,.
\end{equation}
In these formulae, $I_0=I_0^+\cup I_0^-$ is the unknown band
connecting $\lambda_0^*$ in the lower half-plane to
$\lambda_0$ via the origin.  Also, $\Gamma_I\cap C_I$ denotes 
a path from $\lambda_0$ to $iA$ and $\Gamma_I\cap C_I^*$ denotes
a path from $-iA$ to $\lambda_0^*$, both in the complex plane cut at $I_0$.

It is useful to simplify the two conditions $M_0=0$ and $R_0=0$
somewhat.  We begin with the moment condition $M_0=0$, evaluating the
first term by residues by rewriting the integral as a closed loop
around the band $I_0$ as described in
Chapter~\ref{sec:ansatz}.  Thus,
\begin{equation}
M_0=-2J\pi (x+2a_0 t) + 
\int_{\Gamma_I\cap C_I}\frac{\pi i\rho^0(\eta)}{R(\eta)}
\,d\eta +
\int_{\Gamma_I\cap C_I^*}\frac{\pi i\rho^0(\eta^*)^*}{R(\eta)}
\,d\eta\,,
\label{eq:M0general}
\end{equation}
where $a_0=\Re(\lambda_0)$.  Continuing with the reality condition
$R_0=0$, we use similar reasoning as in \S\ref{sec:modulation}
to obtain the representation
\begin{equation}
R_0 = 
\frac{1}{8\pi}\oint_L F(\eta)\,d\eta\,,
\end{equation}
where $L$ is an arbitrarily large counter-clockwise circular loop.
Using the relation $F(\eta)=H(\eta)R(\eta)$ and the
Laurent series expansion (\ref{eq:Laurent}) for $H(\eta)$ along
with the expansion of $R(\eta)$ for genus $G=0$,
\begin{equation}
R(\eta)=-\eta+a_0-\frac{b_0^2}{2\eta}+\bo(\eta^{-2})\,,\hspace{0.2 in}
\eta\rightarrow\infty\,,
\end{equation}
one finds simply
\begin{equation}
R_0=\frac{1}{4\pi}(a_0M_0-M_1)\,.
\end{equation}
Computing the first term in the moment $M_1$ by residues as done for $M_0$
above, we obtain
\begin{equation}
2iR_0 = -2iJtb_0^2 +\int_{\Gamma_I\cap C_I}\frac{(\eta-a_0)\rho^0(\eta)}
{R(\eta)}\,d\eta +
\int_{\Gamma_I\cap C_I^*}\frac{(\eta-a_0)\rho^0(\eta^*)^*}
{R(\eta)}\,d\eta\,.
\end{equation}
Finally, since for $G=0$,
\begin{equation}
\frac{\eta-a_0}{R(\eta)}=\frac{\partial R}{\partial\eta}(\eta)\,,
\end{equation}
the measure reality condition becomes
\begin{equation}
2iR_0 = -2iJtb_0^2 + \int_{\Gamma_I\cap C_I}\rho^0(\eta)\frac{\partial R}
{\partial\eta}(\eta)\,d\eta +
\int_{\Gamma_I\cap C_I^*}\rho^0(\eta^*)^*\frac{\partial R}
{\partial\eta}(\eta)\,d\eta\,.
\label{eq:R0general}
\end{equation} 
Further analysis of these conditions on the endpoint $\lambda_0(x,t)$ 
requires either detailed knowledge of the function $\rho^0(\eta)$ or
a simplifying assumption like $t=0$ or $x=0$. 

\section[Small Time Theory]{Success of the ansatz for general data and small time.
Rigorous small-time asymptotics for semiclassical soliton ensembles.}

\subsection{The genus zero ansatz for $t=0$.  Success of the ansatz
and recovery of the initial data.}
\label{sec:tzero}
When $t=0$, it follows from the fact that the function $\rho^0(\eta)$
is purely imaginary for $\eta$ on the imaginary axis between the
origin and $iA$ that the measure reality condition $R_0=0$ is
satisfied by assuming that the endpoint $\lambda_0$ is purely
imaginary and lies below $\lambda=iA$.  We write $\lambda_0 = ib_0$
for $0<b_0<A$.  Using this information, the moment condition
$M_0=0$ becomes for $t=0$
\begin{equation}
-\int_{b_0}^{A}\frac{i\rho^0(i\nu)}{\sqrt{\nu^2-b_0^2}}\,d\nu
= Jx\,.
\label{eq:M0t0}
\end{equation}
Here, the square-root symbol refers to the principal branch.  Since
the measure $i\rho^0(i\nu)\,d\nu$ is strictly negative ({\em cf.} the
WKB formula (\ref{eq:WKBformula})), this formula is inconsistent unless
we choose the Jost function normalization index $J$ to satisfy
\begin{equation}
J:={\rm sign}(x)\,.
\label{eq:Jostdetermine}
\end{equation}
Inserting the WKB formula (\ref{eq:WKBformula}) for even, single-maximum initial
data $A(x)$ (in which case the symmetry $x_-(\eta)=-x_+(\eta)$ holds)
into (\ref{eq:M0t0}) subject to (\ref{eq:Jostdetermine}) gives
\begin{equation}
\frac{2}{\pi}\int_{b_0}^A d\nu\int_0^{x_+(i\nu)}dx\,\frac{\nu}
{\sqrt{\nu^2 - b_0^2}\sqrt{A(x)^2 - \nu^2}} = |x|\,.
\end{equation}
Exchanging the order of integration, using the fact that $x_+(i\nu)$
is an inverse function to $A(x)$, ({\em i.e.} $A(x_+(i\nu))=\nu$ for
$0\le \nu\le A$), one finds
\begin{equation}
\frac{2}{\pi}\int_0^{x_+(ib_0)}dx\int_{b_0}^{A(x)}d\nu\,
\frac{\nu}{\sqrt{\nu^2 - b_0^2}\sqrt{A(x)^2 - \nu^2}} = |x|\,.
\end{equation}
Let $S(\nu)$ denote the square root function satisfying $S(\nu)^2 =
(\nu^2-b_0^2)(A(x)^2 -\nu^2)$, defined in the $\nu$-plane cut in
the real intervals $[-A(x),-b_0]$ and $[b_0,A(x)]$, and normalized
so that for $\mu\in (b_0,A(x))$,
\begin{equation}
\lim_{\epsilon\downarrow 0} S(\mu+i\epsilon) > 0\,.
\end{equation}
Then, the inner integral can be written as:
\begin{equation}
\begin{array}{rcl}
\displaystyle
\int_{b_0}^{A(x)}\frac{\nu\,d\nu}{\sqrt{\nu^2 - b_0^2}\sqrt{A(x)^2 - 
\nu^2}} &=&\displaystyle
 \frac{1}{2}\lim_{\epsilon\downarrow 0}\left(\int_{-A(x)}^{-b_0}
+\int_{b_0}^{A(x)}\right)\frac{\nu\,d\nu}{S(\nu+i\epsilon)}\\\\
&=&\displaystyle -\frac{1}{4}\oint_L \frac{\nu\,d\nu}{S(\nu)}\,,
\end{array}
\end{equation}
where $L$ is an arbitrarily large counter-clockwise oriented closed loop.
This integral can be evaluated exactly by residues.  Since $S(\nu)=-i\nu^2 +
\bo(\nu)$ for large $\nu$, we obtain simply
\begin{equation}
\int_{b_0}^{A(x)}\frac{\nu\,d\nu}{\sqrt{\nu^2 - b_0^2}\sqrt{A(x)^2 - 
\nu^2}} =\frac{\pi}{2}\,.
\end{equation}
Therefore, the moment condition at $t=0$ becomes 
\begin{equation}
x_+(ib_0) = |x|\,,\mbox{ which implies  } b_0 = b_0(x):= A(x)\,.
\end{equation}

Combining this information about the endpoint with the final remark at
the end of \S\ref{sec:bvp}, we can obtain a useful formula for the
candidate density $\rho^\sigma(\eta)$ for $t=0$ by expressing it as
the integral of its derivative with respect to $x$.  Using the fact
that $\rho^0(\eta)$ is independent of $x$ and the formula for the
total derivative $\partial F/\partial x$ 
({\em i.e.} the derivative including dependence of the endpoints on
$x$ and $t$) , one finds that for any
$x_0$
\begin{equation}
\rho^\sigma(\eta;x)=\rho^\sigma(\eta;x_0)-\frac{2J}{\pi}\int_{x_0}^x
\frac{\partial R_+}{\partial \eta}(\eta;x')\,dx'\,.
\end{equation}
In particular, if $\eta$ is on the imaginary axis between the origin and $
ib_0(x)$, we may choose $x_0=Jx_+(\eta)$.  For this choice, we obtain
\begin{equation}
\rho^\sigma(\eta;x)=-\frac{2J}{\pi}\int_{Jx_+(\eta)}^x
\frac{\partial}{\partial\eta}\sqrt{\eta^2 + b_0(x')^2}\,dx'\,,
\end{equation}
using the explicit formula for $R_+$ written in terms of the standard
branch of the square-root function, valid for $t=0$ and
$\eta\in(0,ib_0(x))$.  Thus, we see that the integrand is always
positive imaginary (the square root function is real and
decreasing in the positive imaginary direction).  From the relation
(\ref{eq:Jostdetermine}), it then follows that $\rho^\sigma(\eta;x)$ is
positive imaginary for all $\eta\in(0,ib_0(x))$.  We have proven the
following.
\begin{lemma}
For $t=0$ and a genus zero ansatz, the oriented contour band $I_0^+$
may be taken to coincide with the vertical segment $[0,ib_0(x)] =
[0,iA(x)]$.  That is, on this segment the differential
$\rho^\sigma(\eta;x)\,d\eta$ is real and negative.  Moreover, for all
$x$ such that the function $x_+(\eta)$ is differentiable at
$ib_0=ib_0(x)=iA(x)$, the function $\rho^\sigma(\eta;x)$ vanishes
exactly like a square root at $\eta=ib_0(x)$ and not to higher order.
\end{lemma}
Note that our monotonicity assumption on the initial data $A(x)$
implies that $x_+(\eta)$ fails to be differentiable only when
$\eta=iA$, and therefore the only value of $x$ where
$\rho^\sigma(\eta;x)$ fails to vanish exactly like a square root at
$\eta=ib_0(x)$ is $x=0$.  In fact when $x=0$, we have
$\rho^\sigma(\eta)\equiv \rho^0(\eta)$, and the latter does not
generally vanish at all in the limit as $\eta$ approaches $iA$ from
below.

This result requires some clarification in the context of the
fundamental assumption that the contour $C$ should be a loop
encircling the imaginary interval $(0,iA)$.  We choose to imagine the
band $I_0^+$ lying infinitessimally either to the right of $(0,iA)$
(for $\sigma=+1$) or to the left of $(0,iA)$ (for $\sigma=-1$).  Using
the relations (\ref{eq:plusside}) and (\ref{eq:minusside}), it is easy
to find a formula for the boundary value of the function
$\tilde{\phi}^\sigma(\lambda;x)$ on the same side of the imaginary
interval $[0,iA]$ above the endpoint.  Let $\lambda\in (ib_0(x),iA)$.
Then,
\begin{equation}
\lim_{\epsilon\downarrow 0}\tilde{\phi}^\sigma(\lambda+\sigma\epsilon;x)
=\tilde{\phi}^\sigma\Big|_{\lambda\in I_0^+}(x)+2iJ\int_{ib_0(x)}^\lambda
d\eta\int_{Jx_+(\eta)}^x dx'\,\frac{\partial R}{\partial\eta}(\eta;x')\,.
\end{equation}
Since by construction $\tilde{\phi}^\sigma$ evaluates to an imaginary constant 
in the band $I_0^+$ for each $x$, using the relation (\ref{eq:Jostdetermine})
along with the fact that $\partial R/\partial\eta$ is negative real 
over the range of integration, we obtain the following result.
\begin{lemma}
For $t=0$ and a genus zero ansatz, and for all $x\neq 0$ and all
$\lambda\in (ib_0(x),iA]$,
\begin{equation}
\lim_{\epsilon\downarrow 0}\Re(\tilde{\phi}^\sigma(\lambda+\sigma\epsilon;x))
<0\,.
\end{equation}
\end{lemma}

The boundary values of the analytic function
$\tilde{\phi}^\sigma(\lambda;x)$ on the interval $[0,iA]$ are related
by ({\em cf.} the definition (\ref{eq:phitildedef}))
\begin{equation}
\lim_{\epsilon\downarrow 0}\tilde{\phi}^\sigma(\lambda+\sigma\epsilon;x) -
\lim_{\epsilon\downarrow 0}\tilde{\phi}^\sigma(\lambda-\sigma\epsilon;x)
= -2\pi i\sigma\int_\lambda^{iA}\rho^0(\eta)\,d\eta\,.
\label{eq:spikejump}
\end{equation}
Because this quantity is purely imaginary, we can immediately extend
the previous result to the boundary value on the other side of the
interval $[0,iA]$ {\em above} the endpoint $\lambda=ib_0(x)$.  This
proves the following.
\begin{lemma}
For $t=0$ and a genus zero ansatz, and for all $x\neq 0$ and all
$\lambda\in (ib_0(x),iA]$,
\begin{equation}
\lim_{\epsilon\downarrow 0}\Re(\tilde{\phi}^\sigma(\lambda-\sigma\epsilon;x))
<0\,.
\end{equation}
\end{lemma}

To complete the analysis on this side of the asymptotic spectral
interval $[0,iA]$, we must obtain a similar inequality valid {\em
below} the endpoint $ib_0(x)$.  Using (\ref{eq:spikejump}) and the
relations (\ref{eq:plusside}) and (\ref{eq:minusside}), we find that
for $\lambda\in (0,ib_0(x))$,
\begin{equation}
\begin{array}{rcl}
\lim_{\epsilon\downarrow 0}\tilde{\phi}^\sigma(\lambda-\sigma\epsilon;x)&=&
\displaystyle
-i\pi\sigma\left[\int_\lambda^{ib_0(x)}\rho^\sigma(\eta;x)\,d\eta -
2\int_\lambda^{iA}\rho^0(\eta)\,d\eta\right]\\\\&=&\displaystyle
i\pi \sigma\left[\int_\lambda^{iA}\overline{\rho}^\sigma(\eta;x)\,d\eta
+\int_\lambda^{iA}\rho^0(\eta)\,d\eta\right]\,,
\end{array}
\label{eq:bdryval}
\end{equation}
where we have used the definition (\ref{eq:rhobardef}) of the
complementary density $\overline{\rho}^\sigma(\eta;x)$.  This boundary
value is purely imaginary.  Now, we will show that for $t=0$ the genus
zero ansatz yields a measure $\overline{\rho}^\sigma(\eta;x)\,d\eta$
that like $\rho^0(\eta)\,d\eta$, is negative real for all $\eta\in
(0,iA)$, taken with upward orientation.  Since for $\eta$ above the
endpoint $ib_0(x)$ on $C$ (against the imaginary axis) the function
$\rho^\sigma(\eta;x)$ vanishes identically, it is only necessary to
check this for $\eta\in (0,ib_0(x))$.  For this purpose, we note that
for $t=0$ and genus zero, with the endpoint $\lambda_0=ib_0(x)$, the
formula (\ref{eq:rhobarbandformula}) that holds for $\eta\in
(0,ib_0(x))$ can be written as
\begin{equation}
\overline{\rho}^\sigma(\eta;x)=\frac{iR_+(\eta;x)}{\pi}
\int_{ib_0(x)}^{iA}\left(\frac{1}{|\xi|-|\eta|}+\frac{1}{|\xi|+|\eta|}
\right)\frac{i\rho^0(\xi)}{R(\xi;x)}\,d\xi\,.
\end{equation}
Since $R_+(\eta;x)$ is positive real and $R(\xi;x)$ is negative imaginary
over the range of integration, and since the measure $\rho^0(\xi)\,d\xi$
is negative real, this formula shows that $\overline{\rho}^\sigma(\eta;x)$
is strictly positive imaginary for $\eta\in (0,ib_0(x))$.  

\begin{remark}  
This result shows that at $t=0$ the measure
$-\rho^\sigma(\eta;x)\,d\eta$ on the contour satisfies an {\em upper
constraint} \index{upper constraint} as well as a positivity
condition, since $0<-\rho^\sigma(\eta;x)\,d\eta <
-\rho^0(\eta)\,d\eta$.  In this respect, the analysis at $t=0$ on the
imaginary axis is very similar to the analysis of Lax and Levermore
\cite{LL83}\index{Lax-Levermore analysis}.
\end{remark}

For $\lambda\in (0,ib_0(x))$, it therefore follows that
$\Re(\tilde{\phi}^\sigma(
\lambda+\sigma 0;x))\equiv 0$ and at the same time that
$\Im(\tilde{\phi}^\sigma(\lambda+\sigma 0;x))$ is increasing
(respectively decreasing) in the positive imaginary direction for
$\sigma=+1$ (respectively for $\sigma=-1$).  An application of the
Cauchy-Riemann equations for the analytic function
$\tilde{\phi}^\sigma(\lambda;x)$ then yields the following.
\begin{lemma}
For $\sigma=+1$ (respectively $\sigma=-1$), there exists a lens-shaped
region to the left (respectively right) of the imaginary interval
$(0,ib_0(x))$ in which $\Re(\tilde{\phi}^\sigma (\lambda;x))<0$.
\end{lemma}

\begin{figure}[h]
\begin{center}
\mbox{\psfig{file=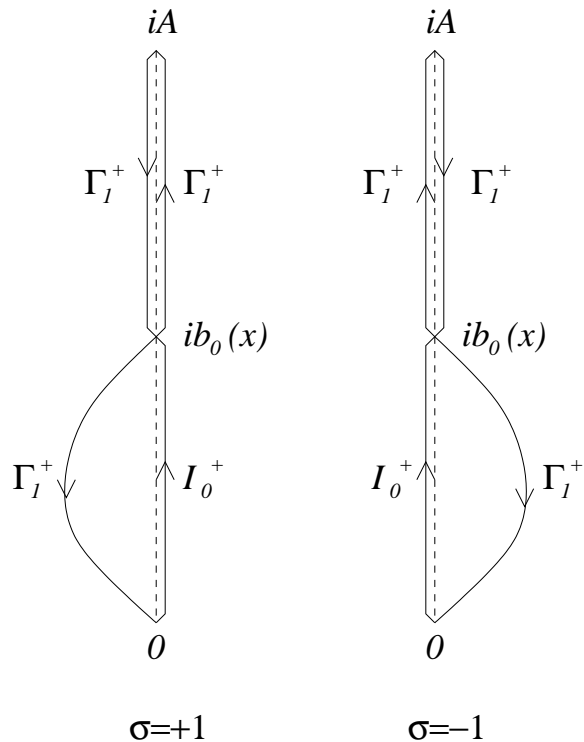,width=3 in}}
\end{center}
\caption{\em The degenerate contours for $t=0$.  For each $x$ there
are two possibilities corresponding to $\sigma=+1$ and $\sigma=-1$.}
\label{fig:tzero}
\end{figure}

\begin{figure}[h]
\begin{center}
\mbox{\psfig{file=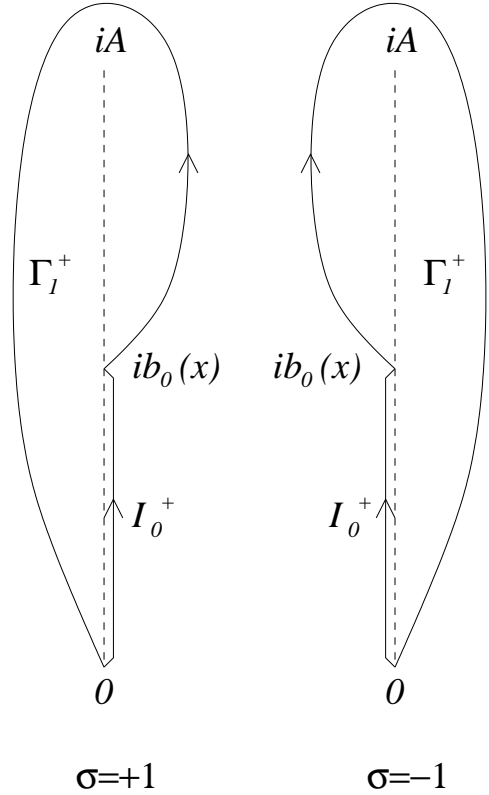,width=2.5 in}}
\end{center}
\caption{\em The contours on which Theorem~\ref{theorem:t0} holds.}
\label{fig:tzeroopen}
\end{figure}

At this point we have proved that at $t=0$, and for each nonzero $x$,
there is a genus zero ansatz for each choice of $\sigma=\pm 1$ that
satisfies the inequalities
$\rho^\sigma(\eta;x)\,d\eta<0$ for $\eta\in I_0^+$ and
$\Re(\tilde{\phi}^\sigma(\lambda;x))\le 0$ for $\lambda\in \Gamma_1^+$,
where $I_0^+$ and $\Gamma_1^+$ are the band and gap components of a
degenerate contour $C$ that barely encircles the imaginary interval $[0,iA]$.
See Figure~\ref{fig:tzero}.
The inequality $\Re(\tilde{\phi}^\sigma(\lambda;x))\le 0$ is in fact
strict at all points interior to $\Gamma_1^+$ {\em except} for 
$\lambda=ib_0(x)-\sigma 0$.  However, using the fact that $\rho^\sigma(\eta;x)=
\bo(|\eta-ib_0(x)|^{1/2})$ for $\eta$ near $ib_0(x)$, the formula
(\ref{eq:bdryval}) gives
\begin{equation}
\lim_{\epsilon\downarrow 0}\tilde{\phi}^\sigma(\lambda-\sigma\epsilon;x)=
2\pi i\sigma\int_{ib_0(x)}^{iA}\rho^0(\eta)\,d\eta -
2\pi i\sigma\rho^0(ib_0(x))\cdot (\lambda-ib_0(x)) + \bo (|\lambda-ib_0(x)|^{3/2})\,,
\end{equation}
in the limit $\lambda\rightarrow ib_0(x)$.  Therefore, since the
function $\rho^0(\eta)$ defined by (\ref{eq:WKBformula}) never
vanishes for $\eta\in(0,iA)$, the linear terms dominate and for all
$\lambda$ in a sufficiently small half-disk centered at
$\lambda=ib_0(x)$ and lying in the left (respectively right)
half-plane for $\sigma=+1$ (respectively $\sigma=-1$), the inequality
$\Re(\tilde{\phi}^\sigma(\lambda;x))<0$ holds.  This means that the
gap contour $\Gamma_1^+$ can be pulled slightly away from the interval
$[0,iA]$ except at the endpoints $\lambda=ib_0(x)+\sigma 0$ and
$\lambda=-\sigma 0$ while achieving {\em strict} inequality on the
interior.  This proves the following theorem.
\begin{theorem}
For $t=0$, the endpoint function may be taken as $\lambda_0 = iA(x)$,
and then for all $x\neq 0$, and both signs of $\sigma$, the genus zero
ansatz corresponding to a contour $C$ for which $I_0^+$ is the
imaginary interval $[\sigma 0,iA(x)+\sigma 0]$, and $\Gamma_1^+$ has
endpoints $\lambda=iA(x)+\sigma 0$ and $-\sigma 0$ and lies in the
slit half-plane ${\mathbb H}$, satisfies:
\begin{itemize}
\item
The differential $\rho^\sigma(\eta;x)\,d\eta$ is strictly negative
in the interior of $I_0^+$ and vanishes exactly like a square root
at the endpoint $\eta=iA(x)+\sigma 0$.
\item
The inequality $\Re(\tilde{\phi}^\sigma(\lambda;x))<0$ holds strictly
in the interior of the contour $\Gamma_1^+$.
\end{itemize}
The contour $C$ is illustrated in Figure~\ref{fig:tzeroopen}.
\label{theorem:t0}
\end{theorem}

Finally, we observe that the genus zero ansatz for $t=0$ formally
reconstructs the initial data for (\ref{eq:IVP}) in the semiclassical
limit $\hbar\downarrow 0$.\index{initial data!asymptotic recovery of}
\begin{theorem}
The function $\tilde{\psi}$ corresponding to the genus zero ansatz for
$t=0$ and defined by (\ref{eq:psitildeGeqzero}) is given by
$\tilde{\psi}\equiv A(x)$.
\end{theorem}

\begin{proof}
Since we have already shown that $\Im(\lambda_0)=A(x)$, it remains to
show that $\alpha_0$, the constant value of the function
$-iJ\tilde{\phi}^\sigma(\lambda)$ in the band $I_0^+$, is identically zero
as a function of $x$.  

To do this, we first establish a general formula, holding for an arbitrary
genus zero ansatz ({\em i.e.} not only for $t=0$), for the derivative of
$-\alpha_0$ with respect to $x$.  In \S\ref{sec:outersolve} we observed that
this quantity had the interpretation of a local wavenumber $k$.
Let us calculate the wavenumber $k$ in terms of the endpoints
$\lambda_0$ and $\lambda_0^*$.  First, from the definition of the constant
$\alpha_0$, we have
\begin{equation}
k=iJ\frac{\partial\tilde{\phi}^\sigma}{\partial x}\Bigg|_{\lambda=0}\,.
\end{equation}
Next, using (\ref{eq:phitildedef}), we find
\begin{equation}
k=iJ\frac{\partial}{\partial x}\int_{C_I}
\overline{L_\eta^{C,\sigma}}(0)
\overline{\rho}^\sigma(\eta)\,d\eta + iJ\frac{\partial}{\partial x}
\int_{C_I^*}\overline{L_\eta^{C,\sigma}}(0)\overline{\rho}^\sigma(\eta^*)^*
\,d\eta
\,,
\label{eq:k1}
\end{equation}
where we recall that the overbar on the logarithm indicates the
average of the two boundary values taken on the contour $C$.  We now
show that the derivative with respect to $x$ can be moved inside the
integral and put onto the complementary density
$\overline{\rho}^\sigma(\eta)$.  Recalling the definition
(\ref{eq:rhobardef}) and using the facts that the logarithmic integral
of the function $\rho^0(\eta)$ is independent of $x$ and that
$\rho^\sigma(\eta)\equiv 0$ for $\eta\in\Gamma_I\cap C_I$,
(\ref{eq:k1}) implies
\begin{equation}
\begin{array}{rcl}
k&=&\displaystyle
-iJ\frac{\partial}{\partial x}\int_{I_0^+}\overline{L_\eta^{C,\sigma}}(0)
\rho^\sigma(\eta)\,d\eta -iJ\frac{\partial}{\partial x}
\int_{I_0^-}\overline{L_\eta^{C,\sigma}}(0)\rho^\sigma(\eta^*)^*\,d\eta\\\\
&=&\displaystyle
-iJ\int_{I_0^+}\overline{L_\eta^{C,\sigma}}(0)
\frac{\partial}{\partial x}\rho^\sigma(\eta)\,d\eta
-iJ\int_{I_0^-}\overline{L_\eta^{C,\sigma}}(0)
\frac{\partial}{\partial x}\rho^\sigma(\eta^*)^*\,d\eta\\\\
&=&\displaystyle
iJ\int_{I_0^+}\overline{L_\eta^{C,\sigma}}(0)
\frac{\partial}{\partial x}\overline{\rho}^\sigma(\eta)\,d\eta
+iJ\int_{I_0^-}\overline{L_\eta^{C,\sigma}}(0)
\frac{\partial}{\partial x}\overline{\rho}^\sigma(\eta^*)^*\,d\eta\,.
\end{array}
\end{equation}
Here, the $x$-derivative can be brought inside the integral since
$\rho^\sigma(\eta)$ vanishes at the moving endpoints, and the last
step follows because the density function $\rho^0(\eta)$ is
independent of $x$.  In terms of the function $F(\lambda)$, we
then have
\begin{equation}
k=-\frac{J}{2\pi}\int_{I_0}\overline{L_\eta^{C,\sigma}}(0)\left(
\frac{\partial F_+}{\partial x}(\eta)-
\frac{\partial F_-}{\partial x}(\eta)\right)\,d\eta\,.
\end{equation}
Using the expression for $\partial F/\partial x$ obtained in
the final remark in \S\ref{sec:bvp}, we find that for $\eta\in I_0$,
\begin{equation}
\frac{\partial F_+}{\partial x}(\eta)
-\frac{\partial F_-}{\partial x}(\eta) = 
-4iJ\frac{\partial R_+}{\partial\eta}(\eta)\,,
\end{equation}
and therefore
\begin{equation}
\begin{array}{rcl}
k&=&\displaystyle
\frac{2i}{\pi}\int_{I_0}\overline{L^{C,\sigma}_\eta}(0)\frac{\partial R_+}{\partial\eta}(\eta)\,d\eta\\\\
&=&\displaystyle
\frac{i}{\pi}\int_{I_0}L^{C,\sigma}_{\eta +}(0)
\frac{\partial R_+}{\partial\eta}(\eta)\,d\eta +
\frac{i}{\pi}\int_{I_0}L^{C,\sigma}_{\eta -}(0)
\frac{\partial R_+}{\partial\eta}(\eta)\,d\eta\,,
\end{array} 
\end{equation}
where we have used the definition of
$\overline{L^{C,\sigma}_\eta}(\lambda)$ as an average of boundary
values.  Now the function $L_{\eta +}^{C,\sigma}(0)$ is the boundary
value of $L_\eta^{C,\sigma}(\lambda)$ as $\lambda$ approaches the
origin in the oriented contour $[C\cup C^*]_\sigma$ from the ``plus''
side.  This means that for $\eta\in [C\cup C^*]_\sigma$, $L_{\eta
+}^{C,\sigma}(0)$ has an analytic extension as a function of $\eta$ to
the ``minus'' side of $[C\cup C^*]_\sigma$.  A similar argument shows
that $L_{\eta -}^{C,\sigma}(0)$ is analytic in $\eta$ on the ``plus''
side of $[C\cup C^*]_\sigma$.  And of course $R_+(\lambda)$ extends
analytically to the ``plus'' side of $[C\cup C^*]_\sigma$ while
$R_-(\lambda)$ extends to the ``minus'' side.  These observations
allow us to move the path of integration away from the integrable
singularity at the origin in each integral.  Namely, if we let $C_a$
be a path from $\lambda_0^*$ to $\lambda_0$ lying to the right of
$I_0$ and $C_b$ be a path from $\lambda_0$ to $\lambda_0^*$ lying to
the left of $I_0$, then it is easy to see that
\begin{equation}
k=
-\frac{i}{\pi}\int_{C_a}L^{C,\sigma}_{\eta +}(0)
\frac{\partial R}{\partial\eta}(\eta)\,d\eta -
\frac{i}{\pi}\int_{C_b}L^{C,\sigma}_{\eta -}(0)
\frac{\partial R}{\partial\eta}(\eta)\,d\eta\,.
\end{equation}
Now, integrate by parts in each integral using the fact that $R$ vanishes
at the endpoints to find
\begin{equation}
k=
\frac{i}{\pi}\int_{C_a}\frac{R(\eta)\,d\eta}{\eta}
+
\frac{i}{\pi}\int_{C_b}\frac{R(\eta)\,d\eta}{\eta}
\,.
\end{equation}
The paths of integration may now be combined into a single
counterclockwise loop surrounding $I_0$ and the singularity at the
origin.  Calculating this loop integral by a residue at infinity
(again using detailed information about the form of $R(\lambda)$ valid
for genus $G=0$), one finds at last that
\begin{equation}
k = -(\lambda_0+\lambda_0^*)=-2\Re(\lambda_0) = -2a_0\,.
\label{eq:wavenumberformula}
\end{equation}

Using the fact that for the genus zero ansatz at $t=0$ the endpoint is
purely imaginary, we see from this general formula that $\alpha_0$ is
independent of $x$.  We now show that with $t=0$,
\begin{equation}
\lim_{x\rightarrow 0}\alpha_0 = 0\,.
\end{equation}
This follows from two observations.  First, for $\eta$ fixed on the
imaginary axis below $iA$, the function $\overline{\rho}^\sigma(\eta)$
converges pointwise to zero as $x\rightarrow 0$.  This can be seen by
noting that for $|x|$ small enough $\eta$ lies in the band $I_0^+$; a
direct estimate of the boundary values of the functions
$H(\eta)$ and $R_+(\eta)$ that vanishes as $x\rightarrow 0$ is
then easily obtained from the exact formula (\ref{eq:Hsolution}).
Next, since as noted above there is an effective upper constraint for
$t=0$ on the measure $\overline{\rho}^\sigma(\eta)\,d\eta$ on the
imaginary axis, it follows from a dominated convergence argument that
the function $\tilde{\phi}^\sigma(\lambda)$ converges pointwise to
zero for $t=0$ as $x\rightarrow 0$.  These results imply that for all
$x$ at $t=0$, $\alpha_0=0$, and the proof is complete.
\end{proof}

\begin{remark}
Although the inequalities are all strict, the fact that the band
$I_0^+$ lies against the imaginary axis when $t=0$ precludes the
application of the asymptotic inverse theory in
Chapter~\ref{sec:asymptoticanalysis} to establish the recovery of the
initial data.  In other words, the fact that our conditions on the
complex phase function selects at $t=0$ a contour that coincides with
polar singularities of the matrix ${\bf m}(\lambda)$ solving the
meromorphic Riemann-Hilbert Problem~\ref{rhp:m} means that the strong
$\bo(\hbar_N^{1/3})$ error estimate we obtained in
Theorem~\ref{theorem:psiestimate} is not uniformly valid in any
neighborhood that should happen to include $t=0$ (at least not without
a modification of the methods we have presented here).  On the other
hand, we know from the Lax-Levermore type analysis carried out by
Ercolani, Jin, Levermore, and MacEvoy \cite{EJLM93} that it is
possible to prove $L^2({\mathbb R})$ convergence of $\psi$ to
$\tilde{\psi}\equiv A(x)$ exactly at $t=0$.  Note however, that at
least for the special case of the Satsuma-Yajima case, this strange
situation is no real obstruction to our analysis since there is no
adjustment of the initial data ({\em e.g.}  neglect of a reflection
coefficient) for $\hbar$ values in the ``quantum'' sequence
$\hbar=\hbar_N$ ({\em cf.}  (\ref{eq:quantumsequence})) and
consequently nothing to prove at $t=0$.
\end{remark}

\subsection{Perturbation theory for small time.}
\label{sec:smalltime}
We begin this section by establishing the existence of the endpoint
$\lambda_0(x,t)$ for small time.
\begin{lemma}
Let the intitial data $A(x)>0$ be real-analytic, even, and monotone 
decreasing in $|x|$. Then, for each fixed $x\neq 0$,
the equations $M_0=0$ and $R_0=0$ have a solution for $|t|$
sufficiently small that is differentiable in $t$ and agrees with
the purely imaginary solution obtained in \S\ref{sec:tzero} upon
setting $t=0$.  
\label{lemma:endpointexist}
\end{lemma}
\begin{proof}
We need to show that the Jacobian determinant
$\partial(M_0,R_0)/\partial(\lambda_0,\lambda_0^*)$ does not vanish.
Since $R_0$ is a linear combination of $M_0$ and $M_1$, it is
equivalent to show that
$\partial(M_0,M_1)/\partial(\lambda_0,\lambda_0^*)\neq 0$.  In
\S\ref{sec:modulation}, it was shown that
\begin{equation}
\frac{\partial(M_0,M_1)}{\partial(\lambda_0,\lambda_0^*)}:=
\det\left[\begin{array}{cc}
\displaystyle \frac{\partial M_0}{\partial \lambda_0} &
\displaystyle \frac{\partial M_0}{\partial \lambda_0^*} \\\\
\displaystyle \frac{\partial M_1}{\partial \lambda_0} &
\displaystyle \frac{\partial M_1}{\partial \lambda_0^*} 
\end{array}\right]=
\det\left[\begin{array}{cc}
\displaystyle \frac{\partial M_0}{\partial \lambda_0} &
\displaystyle \frac{\partial M_0}{\partial \lambda_0^*} \\\\
\displaystyle \lambda_0\frac{\partial M_0}{\partial \lambda_0} &
\displaystyle \lambda_0^*\frac{\partial M_0}{\partial \lambda_0^*} 
\end{array}\right] = 
-(\lambda_0-\lambda_0^*)\frac{\partial M_0}{\partial \lambda_0}
\frac{\partial M_0}{\partial \lambda_0^*}\,.
\end{equation}

To calculate the partial derivatives, we first establish a simple
formula for $M_0$ that involves the initial data $A(x)$.  For this
purpose, we define a quantity $I(\lambda_0,\lambda_0^*)$ from
(\ref{eq:M0general}) by writing $M_0=-2J\pi(x + (\lambda_0 +
\lambda_0^*)t) + I(\lambda_0,\lambda_0^*)$, 
and to calculate $I(\lambda_0,\lambda_0^*)$ we momentarily suppose
that $\lambda_0=i\alpha$ and $\lambda_0^*=i\beta$ with $\alpha$ and
$\beta$ being independent real numbers with $0<\alpha<A$ and
$-A<\beta<0$.  Then, substituting the formula (\ref{eq:WKBformula})
for $\rho^0(\eta)$ into the formula (\ref{eq:M0general}), one
exchanges the order of integration to find
\begin{equation}
\begin{array}{rcl}
I(i\alpha,i\beta) &=&\displaystyle 
\int_0^{x_+(i\alpha)}dx\int_\alpha^{A(x)}d\nu\,
\frac{2\nu}{\sqrt{(\nu-\alpha)(\nu-\beta)}\sqrt{A(x)^2 - \nu^2}} \\\\
&&\displaystyle\,\,-\,\,
\int_0^{x_+(-i\beta)}dx\int_{-A(x)}^\beta d\nu\,
\frac{2\nu}{\sqrt{(\nu-\alpha)(\nu-\beta)}\sqrt{A(x)^2 - \nu^2}}\,.
\end{array}
\end{equation}
Now, define a square-root function $T(\nu)$ satisfying $T(\nu)^2 = 
(\nu-\alpha)(\nu-\beta)(A(x)^2 - \nu^2)$, and defined in the $\nu$-plane
slit along the real intervals $[-A(x),\beta]$ and $[\alpha,A(x)]$ with 
the normalization
\begin{equation}
\lim_{\epsilon\downarrow 0} T(\mu+i\epsilon)>0\,,\hspace{0.3 in}
\mu\in(\alpha,A(x))\,.
\end{equation}
Letting $L_\alpha$ (respectively $L_\beta$) be a small
counter-clockwise oriented loop encircling $[\alpha,A]$ (respectively
encircling $[-A,\beta]$) as shown in Figure~\ref{fig:Lalphabeta}, 
\begin{figure}[h]
\begin{center}
\mbox{\psfig{file=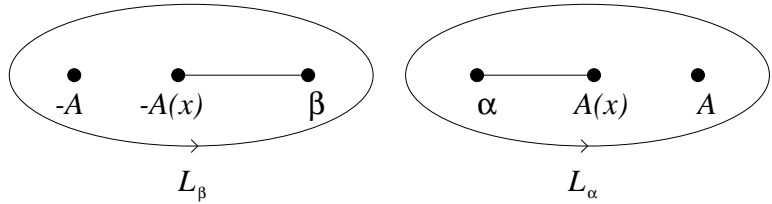,width=4 in}}
\end{center}
\caption{\em The contours $L_\alpha$ and $L_\beta$ surrounding the cuts
of $T(\nu)$.  }
\label{fig:Lalphabeta}
\end{figure}
we may therefore write
\begin{equation}
I(i\alpha,i\beta) = -\int_0^{x_+(i\alpha)}dx\oint_{L_\alpha}
\frac{\nu\,d\nu}{T(\nu)} - \int_0^{x_+(-i\beta)}dx\oint_{L_\beta}
\frac{\nu\,d\nu}{T(\nu)}\,.
\label{eq:Iformula}
\end{equation}
This formula will be analytic in $\lambda_0=-i\alpha$ and
$\lambda_0^*=-i\beta$ for $\lambda_0$ in a complex neighborhood
of the imaginary interval $(0,iA)$ if the initial data $A(x)$ is
analytic.

With the help of the formula (\ref{eq:Iformula}), we may now easily compute
derivatives of $I(\lambda_0,\lambda_0^*)$ with respect to $\lambda_0$
and $\lambda_0^*$ at $\lambda_0=iA(x)$ and $\lambda_0^*=-iA(x)$.
First, note that
\begin{equation}
\begin{array}{rcl}
\displaystyle
\frac{\partial }{\partial\alpha}I(i\alpha,i\beta) &=&\displaystyle 
-\frac{d}{d\alpha}x_+(i\alpha)\cdot\oint_{L_\alpha}\frac{\nu\,d\nu}{T(\nu)}\Bigg|_{x=x_+(i\alpha)} \\\\
&&\displaystyle\,\,-\,\,
\int_0^{x_+(i\alpha)}dx\oint_{L_\alpha}\frac{\nu\,d\nu}{2(\nu-\alpha)T(\nu)}
-\int_0^{x_+(-i\beta)}dx\oint_{L_\beta}\frac{\nu\,d\nu}{2(\nu-\alpha)T(\nu)}\,,
\end{array}
\end{equation}
and
\begin{equation}
\begin{array}{rcl}
\displaystyle
\frac{\partial }{\partial\beta}I(i\alpha,i\beta) &=&\displaystyle 
\frac{d}{d\beta}x_+(-i\beta)\cdot\oint_{L_\beta}\frac{\nu\,d\nu}{T(\nu)}\Bigg|_{x=x_+(-i\beta)} \\\\
&&\displaystyle\,\,-\,\,
\int_0^{x_+(i\alpha)}dx\oint_{L_\alpha}\frac{\nu\,d\nu}{2(\nu-\beta)T(\nu)}
-\int_0^{x_+(-i\beta)}dx\oint_{L_\beta}\frac{\nu\,d\nu}{2(\nu-\beta)T(\nu)}\,.
\end{array}
\end{equation}
Setting $\alpha=A(x)$ and $\beta=-A(x)$, the final two terms in each
of the above formulae can be combined, and the integrand of the $x$
integral then can be calculated by a residue at $\nu=\infty$ which
vanishes identically as a function of $x$.  Thus, only the first term
survives in each case, and these too can be computed by residues.
Using $\partial/\partial\alpha = i\partial/\partial\lambda_0$
and $\partial/\partial\beta = i\partial/\partial \lambda_0^*$, and expressing
the derivatives of the inverse function $x_+(\cdot)$ in terms of
derivatives of $A(\cdot)$, one obtains
\begin{equation}
\frac{\partial I}{\partial\lambda_0}(iA(x),-iA(x)) = \frac{i\pi }{A'(x)}\,,
\hspace{0.3 in}
\frac{\partial I}{\partial\lambda_0^*}(iA(x),-iA(x))=-\frac{i\pi}{A'(x)}\,.
\end{equation}

Finally, we use these formulae to evaluate the Jacobian for $t=0$.  We find
\begin{equation}
\frac{\partial(M_0,M_1)}{\partial(\lambda_0,\lambda_0^*)}
\Bigg|_{t=0}(iA(x),-iA(x)) = -2i\pi^2\frac{A(x)}{A'(x)^2}\,.
\end{equation}
By our monotonicity assumptions on the initial data, this Jacobian is
finite and strictly nonzero for all $x\neq 0$.  Thus, the lemma is
proved by appealing to the implicit function theorem.
\end{proof}

\begin{theorem}
Let the initial data $A(x)$ be real-analytic, even, and monotone
decreasing in $|x|$.  Then for each fixed $x\neq 0$, there is a
$\tau_x>0$ such that for all positive $t<\tau_x$ (respectively negative
$t>-\tau_x$) the genus zero ansatz holds with $\sigma={\rm sign}(x)$
(respectively for $\sigma=-{\rm sign}(x)$) for a loop contour $C$
lying in the cut upper half-plane ${\mathbb H}$.  The value $t=0$ is
excluded only in the sense that as explained in \S\ref{sec:tzero} the
loop $C$ cannot be taken to lie completely in ${\mathbb H}$ for either value
of $\sigma$.
\label{theorem:smalltimeansatz}
\end{theorem}

\begin{proof}
From Lemma~\ref{lemma:endpointexist}, the endpoint function
$\lambda_0(x,t)$ is differentiable in a neighborhood of $t=0$ for each
nonzero $x$.  The proofs of the local continuation results given in
\S\ref{sec:continuation} can easily be applied here to show that, if the
endpoint leaves the imaginary axis by moving into the right half-plane
for a $\sigma=+1$ ansatz or by moving into the left half-plane for a
$\sigma=-1$ ansatz, then a gap contour $\Gamma_1^+(x,t)$ will exist in
${\mathbb H}$ connecting $\lambda_0(x,t)$ to $\lambda=-\sigma 0$, on
the interior points of which the inequality
$\Re(\tilde{\phi}^\sigma(\lambda;x,t))<0$ holds strictly.  Similarly,
these proofs show that for small $|t|$ a band contour $I_0^+(x,t)$
will exist on which the differential $\rho^\sigma(\eta;x,t)\,d\eta$ is
real and strictly negative.  However in this case the difficulty is
that for $t=0$ the band $I_0^+(x,0)$ lies on the boundary of ${\mathbb
H}$ and we must therefore prove that the band $I_0^+(x,t)$ lies
entirely on one side or the other of the imaginary axis for small time.

Note that as $\lambda$ moves along the contour $I_0^+(x,t)$ (whose
existence for small time is guaranteed by the arguments in
\S\ref{sec:continuation}) from $\lambda=0$ to $\lambda=\lambda_0(x,t)$,
the function $B_0(\lambda)$ defined in (\ref{eq:Bs}) is real and
strictly decreasing.  Therefore, by the Cauchy-Riemann
equations\index{Cauchy-Riemann equations}, $\Im(B_0(\lambda))$ is
negative (respectively positive) for all $\lambda$ in a small
lens-shaped region just to the left (respectively right) of
$I_0^+(x,t)$.  We use the expression for the total derivative of the
function $F(\lambda)$ with respect to $t$ obtained in the final remark
in \S\ref{sec:bvp} to calculate
\begin{equation}
\begin{array}{rcl}
\displaystyle\frac{\partial}{\partial t}\Im(B_0(\lambda)) &=& 
\displaystyle
\Im\left(\frac{1}{2\pi i}\int_0^\lambda \left[\frac{\partial F}{\partial t}(\eta)-\frac{\partial F}{\partial t}(\eta)\right]\,d\eta\right)\\\\
&=&\displaystyle
-\frac{2J}{\pi}\Im\left(\int_0^\lambda \frac{\partial}{\partial\eta}\left[\eta
R_+(\eta)\right]\,d\eta\right)\\\\
&=&\displaystyle
-\frac{2J}{\pi}\Im(\lambda R_+(\lambda))\,.
\end{array}
\end{equation}
For the purely imaginary endpoint configuration at $t=0$, this
quantity is strictly nonzero with sign $-J$ for all $\lambda$ on the
positive imaginary axis below the endpoint $\lambda_0(x,0)=iA(x)$.

Using the relation (\ref{eq:Jostdetermine}), we therefore see that for
$x>0$ the interior points of the band $I_0^+(x,t)$ all move into the
right half-plane for $t$ small and positive and into the left
half-plane for $t$ small and negative.  Similarly, for $x<0$ the band
moves to the left for $t>0$ and to the right for $t<0$.  This shows
that the ansatz corresponding to $\sigma={\rm sign}(xt)$ always
deforms for small time so that all inequalities remain valid, which
proves the theorem.
\end{proof}

Combining Lemma~\ref{lemma:endpointexist} with
Theorem~\ref{theorem:G0Whitham} leads to a representation of the
solution of the analytic Cauchy problem for the elliptic genus zero
Whitham equations (\ref{eq:genuszeroWhitham}).
\begin{theorem}[Solution of the analytic Cauchy problem for the Whitham 
equations]
\index{Whitham equations!solution of the initial-value problem for}
The algebraic equations $M_0=0$ and $R_0=0$, with $\rho^0(\eta)$ given
in terms of the even, single-maximum, real analytic function $A(x)$ by
(\ref{eq:WKBformula}), implicitly defines for small $t$ and all $x\neq
0$ the solution $(\lambda_0(x,t),\lambda_0^*(x,t))$ of the Cauchy
(initial-value) problem for the elliptic Whitham system
(\ref{eq:genuszeroWhitham}) corresponding to the initial data
$\lambda_0(x,0)=iA(x)$ and $\lambda_0^*(x,0)=-iA(x)$.
\end{theorem}

So the genus $G=0$ ansatz is sufficient to enable the error
analysis of \S\ref{sec:error} to be valid, as long as $t$ is different
from zero, but is sufficiently small for any given $x$.  Combining
Theorem~\ref{theorem:smalltimeansatz} with Theorem~\ref{theorem:psiestimate}
yields one of our most important results, the rigorous description of the
small-time semiclassical limit of WKB soliton ensembles for the focusing
nonlinear Schr\"odinger equation.
\begin{theorem}[Rigorous small-time asymptotics for semiclassical soliton 
ensembles]
\index{semiclassical soliton ensembles!rigorous small-time asymptotics for}
Let $\psi(x,t)$ be, for each $\hbar=\hbar_N$ the solution of the
focusing nonlinear Schr\"odinger equation that is the WKB soliton
ensemble corresponding to the approximate initial data
$\psi(x,0)=A(x)$.  Then, for each $x\neq 0$ there is an open time
interval $T_x$ containing $t=0$ and independent of $\hbar$ such that
the formula (\ref{eq:psitildeGeqzero}) for $\tilde{\psi}(x,t)$ built
from the genus zero ansatz satisfies $|\psi-\tilde{\psi}|\le
K_{x,t}\hbar_N^{1/3}$ for $\hbar_N$ sufficiently small as long as
$t\in T_x\setminus\{0\}$.  The constant $K_{x,t}$ may vary in $x$ and $t$.
\label{theorem:generalsmalltime}
\end{theorem}

For the special case of the Satsuma-Yajima ensemble, where the
semiclassical soliton ensemble coincides with the solution of the
initial-value problem (\ref{eq:IVP}) because there is no modification
of the initial data by the WKB approximation of the spectrum for
$\hbar=\hbar_N$, we obtain a uniform description of the semiclassical limit
for the focusing nonlinear Schr\"odinger equation.
\begin{corollary}[The semiclassical limit for the Satsuma-Yajima initial data]
\index{Satsuma-Yajima potential!rigorous semiclassical asymptotics for}
Let $\psi(x,t)$ be the solution of the focusing nonlinear
Schr\"odinger equation with initial data $\psi(x,0)=A\,\,{\rm
sech}(x)$.  Then, for each $x\neq 0$ there is an open time interval
$T_x$ containing $t=0$ and independent of $\hbar$ such that the
formula (\ref{eq:psitildeGeqzero}) for $\tilde{\psi}(x,t)$ built from
the genus zero ansatz satisfies $|\psi-\tilde{\psi}|\le
K_{x,t}\hbar_N^{1/3}$ whenever $t\in T_x$.  The error is uniformly
small in compact subsets of the $(x,t)$-plane where the approximation
is valid.
\label{corollary:SYsmalltime}
\end{corollary}

\section[Larger Time Theory]{Larger time analysis for soliton ensembles.}
\label{sec:finitetimeSY}
Here, we consider the genus zero ansatz for larger times.  First, we
establish a simple formula for the solution of the analytic Cauchy
problem for the genus zero Whitham modulation equations
(\ref{eq:genuszeroWhitham}) that holds when $x=0$, that is, in the
center of the symmetric evolution.  Then, we use the concrete example
of the Satsuma-Yajima soliton ensemble, {\em i.e.} we assume
$\rho^0(\eta)=\rho^0_{\rm SY}(\eta)\equiv i$ to show how to determine
the boundary of the genus zero region of the $(x,t)$-plane in
practice.  That is, with the help of numerical calculations, we will
be able to indicate the success of the ansatz in regard to satisfying
the relevant inequalities in a certain region of the $(x,t)$ plane,
and to make concrete the mechanism of failure of the ansatz, as
described in general terms above, at the boundary of this region.

Given values of $x$ and $t$, we may choose the discrete parameters
$J=\pm 1$ and $\sigma=\pm 1$.  This choice will turn out to be
essential in order to treat the whole $(x,t)$ plane; in fact,
different values of $J$ and $\sigma$ will be needed for different
signs of $x$ and $t$.  Of course, for all real-valued, even initial
data, we have the symmetries $\psi(-x,t;\hbar)=\psi(x,t;\hbar)$ and
$\psi(x,-t;\hbar)=\psi(x,t;\hbar)^*$ that allow the solution for all
$x$ and $t$ to be obtained from the solution for $x$ and $t$ positive.
Therefore for the semiclassical soliton ensembles we consider in this
paper it is strictly speaking not necessary to carry out any more
analysis for other signs of $x$ and $t$.  Nonetheless, it will be
useful to document how the other signs of $x$ and $t$ break symmetry
of for more general future applications.

\subsection{The explicit solution of the analytic Cauchy problem for 
the genus zero Whitham equations along the symmetry axis $x=0$.}
We now study the equations $M_0=0$ and $R_0=0$ for the endpoint
$\lambda_0(x,t)$ under the assumption that $x=0$ with $|t|$
sufficiently small.  Our main result is contained in the following
theorem.
\begin{theorem}[Explicit location of the endpoint for $x=0$]
Assume that $A(x)$ is a real-analytic even bell-shaped function
satisfying $A''(0)< 0$.  Then, if $x=0$ and $|t|$ is sufficiently
small, the equations $M_0=0$ and $R_0=0$ are satisfied by a point
$\lambda_0(t)=ib_0(t)$ with $b_0>A$.  The relation between $b_0$ and
$t$ is simply
\begin{equation}
|t|=\frac{2}{\pi b_0}\int_0^{-iA^{-1}(b_0)}E(1-A(iy)^2/b_0^2)\,dy\,,
\label{eq:xiszero}
\end{equation}
where the upper limit of integration is the number $y>0$ for which
$A(iy)=b_0>A$, and $E(m)$ is the complete elliptic integral of the
second kind\index{complete elliptic integral of the second kind}.
The upper limit of integration makes sense for $b_0>A$ because
$A''(0)<0$ implies that $A(iy)$ is an increasing function of $|y|$ for
$y$ real and small enough.
\label{theorem:xiszero}
\end{theorem}

\begin{proof}
To prove this, we need to examine the equations $M_0=0$ and $R_0=0$
for such endpoint configurations, which requires in particular that
$\rho^0(\eta)$ can be defined for $\eta$ on the imaginary axis above
$\eta=iA$.  We will now show that under the condition $A''(0)<0$, the
function $\rho^0(\eta)$ is analytic at $\eta=iA$, and therefore has a
unique analytic continuation for some distance along the imaginary
axis above $\eta=iA$.  We begin with the WKB formula
(\ref{eq:WKBdensity}) that defines $\rho^0(\eta)$ for $\eta$ on the
imaginary axis between $0$ and $iA$.  In this formula, $x_-(\eta)$ and
$x_+(\eta)$ are respectively the negative and positive real roots of
the equation $A(x)^2+\eta^2=0$.  For even bell-shaped functions
$A(x)$, $x_-(\eta)=-x_+(\eta)$ and both functions are well-defined for
$\eta$ in the imaginary interval in question.  To show the analyticity
at $\eta=iA$, we use the fact that $A(x)$ is real-analytic and for
$\eta$ just below $iA$ define a function $B(x,\eta)$ satisfying
$B(x,\eta)^2 = A(x)^2 + \eta^2$ in a neighborhood $U$ of $x=0$
containing only $x_\pm(\eta)$ as roots of $B(x,\eta)^2=0$.
$B(x,\eta)$ is taken to be cut on the real axis between $x_-(\eta)$
and $x_+(\eta)$ and has positive boundary values on the the upper
half-plane side of the cut.  With this normalization, we also have
\begin{equation}
\lim_{x\rightarrow\pm\infty}B(x,\eta) = \mp \eta\,.
\label{eq:norm}
\end{equation}
Then, the WKB density (\ref{eq:WKBdensity}) is rewritten as
\begin{equation}
\rho^0(\eta)=\frac{\eta}{2\pi}\oint_L\frac{dx}{B(x,\eta)}\,,
\label{eq:WKBcontinue}
\end{equation}
where $L$ is a clockwise-oriented loop surrounding the cut of
$B(x,\eta)$ and lying in $U$.  Because we are assuming that
$A''(0)\neq 0$, we can choose the neighborhood $U$ and the loop
contour $L$ to be independent of $\eta$ below but sufficiently near
$\eta=iA$ such that for all such $\eta$ the contour $L$ only ever
encloses the two roots $x_\pm(\eta)$.  If $A''(0)=0$, then more than
two roots would have to coalesce at $x=0$ when $\eta=iA=iA(0)$, and
the contour $L$ would have to shrink as $\eta$ approaches $iA$ in
order to exclude the unwanted roots.  Now, with $A''(0)<0$, the two
roots $x_\pm(\eta)$ coalesce as $\eta$ moves up the axis through $iA$,
and reemerge as a purely imaginary complex-conjugate pair for $\eta$
just above $iA$.  For $\eta$ just above $iA$, we still have only two
roots within $U$ and enclosed by $L$, and we define $B(x,\eta)$ to be
cut along the imaginary axis between these two roots and to be
normalized by the same relation as before (\ref{eq:norm}).  With this
choice, it is then clear that the formula (\ref{eq:WKBcontinue})
defines the analytic continuation of the original formula
(\ref{eq:WKBdensity}) for $\rho^0(\eta)$ through the point $\eta=iA$.

For $\eta$ on the imaginary axis above $iA$, the function $B(x,\eta)$
takes positive imaginary boundary values on the left of the vertical
cut.  Thus, for such $\eta$ we can write (\ref{eq:WKBcontinue}) in the
form
\begin{equation}
\rho^0(\eta)=\frac{2\eta}{\pi}\int_0^{A^{-1}(-i\eta)}\frac{dx}{i\sqrt{-(A(x)^2+\eta^2)}}\,,
\end{equation}
where the positive square root is taken, and where $A^{-1}(\eta)$ is
the positive imaginary number $x$ that satisfies $A(x)=-i\eta$.  Or,
changing variables to $x=iy$,
\begin{equation}
\rho^0(\eta)=\frac{2\eta}{\pi}\int_0^{-iA^{-1}(-i\eta)}\frac{dy}{\sqrt{-(A(iy)^2+\eta^2)}}\,.
\label{eq:continuation}
\end{equation}
This formula, representing the analytic continuation of $\rho^0(\eta)$, is
positive imaginary for $\eta$ above $iA$ on the imaginary axis.

The moment condition $M_0=0$ is satisfied automatically for $x=0$ by any
$G=0$ configuration with endpoint $\lambda_0=ib_0$ on the imaginary axis with
$b_0>A$.  In this situation, the moment $M_0$ explicitly takes the form
\begin{equation}
M_0:=\int_{\Gamma_I\cap C_I}\frac{\pi i\rho^0(\eta)}{R(\eta)}\,d\eta +
\int_{\Gamma_I\cap C_I^*}\frac{\pi i\rho^0(\eta^*)^*}{R(\eta)}\,d\eta\,.
\end{equation}
Now, with the band $I_0^+$ connecting the origin to $ib_0$ in the
first quadrant of the $\eta$-plane for $\sigma=+1$ and the second
quadrant of the $\eta$-plane for $\sigma=-1$ (we are assuming that
$I_0^+$ does not coincide identically with an interval of the positive
imaginary axis), it is easy to see that the function $R(\eta)$
satisfying $R(\eta)^2 = \eta^2+b_0^2$, cut on the band $I_0$ and
normalized to $-\eta$ for large $\eta$, is purely real for $\eta$ in
the imaginary interval $[-ib_0,ib_0]$, and in fact
\begin{equation}
R(\eta)=\sigma\sqrt{\eta^2+b_0^2}\,,
\label{eq:R}
\end{equation}
for such $\eta$, where the positive square root is intended.  The
contour $\Gamma_I\cap C_I$ may be taken to coincide with the interval
$[ib_0,iA]$ oriented from $ib_0$ down to $iA$, and correspondingly,
$\Gamma_I\cap C_I^*$ coincides with the interval $[-iA,-ib_0]$
oriented from $-iA$ down to $-ib_0$.  So, the moment becomes
\begin{equation}
M_0:=\int_{ib_0}^{iA}\frac{\pi
i\rho^0(\eta)}{\sigma\sqrt{\eta^2+b_0^2}}\,d\eta +
\int_{-iA}^{-ib_0}\frac{\pi
i\rho^0(\eta^*)^*}{\sigma\sqrt{\eta^2+b_0^2}}\,d\eta\,.
\end{equation}
Using $\rho^0(\eta^*)^* = -\rho^0(-\eta)$ and changing variables
$\eta\rightarrow -\eta$ in the second term, one sees that $M_0=0$ holds
identically for all $b_0>A$.

We will now show that the reality condition $R_0=0$ will determine the
endpoint $\lambda_0=ib_0$ at $x=0$ as a function of $t$.  Using the
formula (\ref{eq:R}) for $R(\eta)$, and the fact that $\partial
R/\partial\eta = \eta/R(\eta)$, The relevant quantity to consider is
\begin{equation}
R_0:= -Jtb_0^2 + \frac{1}{2i}\int_{ib_0}^{iA}\frac{\eta \rho^0(\eta)}{\sigma\sqrt{\eta^2 + b_0^2}}\,d\eta + \frac{1}{2i}\int_{-iA}^{-ib_0}
\frac{\eta\rho^0(\eta^*)^*}{\sigma\sqrt{\eta^2 + b_0^2}}\,d\eta\,,
\end{equation}
or with $\rho^0(\eta^*)^* = -\rho^0(-\eta)$ and a change of variables 
$\eta\rightarrow -\eta$, 
\begin{equation}
R_0:=-Jtb_0^2 + \int_{ib_0}^{iA}\frac{\eta\rho^0(\eta)}{i\sigma\sqrt{\eta^2+b_0^2}}\,d\eta\,.
\end{equation}
Using (\ref{eq:continuation}) and $\eta=iz$ with $z$ real and positive,
we get
\begin{equation}
R_0:=-Jtb_0^2 + \int_{A}^{b_0}\frac{2z^2}{\pi\sigma\sqrt{b_0^2-z^2}}\int_0^{-iA^{-1}(z)}\frac{dy}{\sqrt{z^2-A(iy)^2}}\,dz\,.
\end{equation}
The equation $R_0=0$ can evidently only have solutions if $\sigma J
t\ge 0$.  In this case we have
\begin{equation}
R_0:=-|t|b_0^2 + \int_{A}^{b_0}\frac{2z^2}{\pi\sqrt{b_0^2-z^2}}\int_0^{-iA^{-1}(z)}\frac{dy}{\sqrt{z^2-A(iy)^2}}\,dz\,.
\end{equation}
We simplify further by exchanging the order of integration using
\begin{equation}
\int_A^{b_0}\left[\int_{0}^{-iA^{-1}(z)} f(y,z)\,dy\right]\,dz =
\int_{0}^{-iA^{-1}(b_0)}\left[\int_{A(iy)}^{b_0} f(y,z)\,dz\right]\,dy\,,
\end{equation}
and thus find
\begin{equation}
R_0:=-|t|b_0^2 + \frac{2}{\pi}\int_0^{-iA^{-1}(b_0)}\left[
\int_{A(iy)}^{b_0}\frac{z^2\,dz}{\sqrt{(b_0^2-z^2)(z^2-A(iy)^2)}}\right]\,dy\,,
\end{equation}
where the positive square root is meant.  The inner integral is
identified as ({\em cf.} page 596 of
\cite{AS65})
\begin{equation}
\int_{A(iy)}^{b_0}\frac{z^2\,dz}{\sqrt{(b_0^2-z^2)(z^2-A(iy)^2)}}=
b_0E(1-A(iy)^2/b_0^2)\,,
\end{equation}
where $E(m)$ denotes the complete elliptic integral of the second kind
with modulus $m$.  Thus, the condition $R_0=0$ becomes the relation
(\ref{eq:xiszero}) which completes the proof of our claim.
\end{proof}

\begin{remark}
If $A''(0)=0$, then the argument used in the proof to show that
$\rho^0(\eta)$ defined by (\ref{eq:WKBdensity}) is analytic at
$\eta=iA$ does not apply, and $\rho^0(\eta)$ simply may not continue
through $\eta=iA$.  We can understand this qualitatively as follows.
From the perspective of one-dimensional quantum mechanics
\index{quantum mechanics} ({\em i.e.}  theory of Schr\"odinger
operators \index{Schr\"odinger operator} in one dimension), the
formula (\ref{eq:WKBdensity}) gives the density of energy levels of
the potential well $-A(x)^2$ in the vicinity of the negative energy
$E=\eta^2$.  If the potential well is too flat near the bottom, then
for energies just above the bottom the well will look like a constant
potential that supports a continuous spectrum of scattering states at
this energy.  So intuitively, one expects the density of states
$\rho^0(\eta)$ to be infinite at $\eta=iA$ if $A''(0) = 0$.
\end{remark}

The solution formula (\ref{eq:xiszero}) is explicit enough to be very
useful in applications for quite general initial data $A(x)$.  In
particular, using (\ref{eq:xiszero}) it is easy to locate the earliest
time where the endpoint function $b_0(t)$ fails to be analytic (at
$x=0$).  This gives an elementary upper bound on the breaking time,
which is the earliest time for which the genus zero ansatz fails, and
more complicated behavior takes over in the semiclassical solution.
This sort of calculation is carried out using the formula
(\ref{eq:xiszero}) in \cite{CM00}, where it is also shown that the
formula (\ref{eq:xiszero}) provides a very accurate approximation to
the square modulus of the numerical solution of (\ref{eq:IVP}) for
small $\hbar$ with initial data $\psi_0(x)=A(x)$.  The formula used in
\cite{CM00} is simply $|\psi(0,t)|^2\sim b_0(t)^2$, where $b_0(t)$ satisfies
(\ref{eq:xiszero}).

\subsection{Determination of the endpoint for the Satsuma-Yajima 
ensemble and general $x$ and $t$.}  Using the explicit formula for the
eigenvalue density function $\rho^0(\eta)\equiv \rho_{\rm
SY}^0(\eta)\equiv i$ appropriate for the Satsuma-Yajima ensemble, let
us obtain more detailed information about the endpoint
$\lambda_0(x,t)$ for finite times.  First consider the moment
condition $M_0=0$.  To evaluate the integral term in $M_0$ ({\em cf.}
(\ref{eq:M0general})) in terms of standard functions, we observe that
on the paths of integration $\Gamma_I$, we have
\begin{equation}
R(\eta)=\sigma\sqrt{(\eta-a_0)^2 + b_0^2}\,,
\end{equation}
where $a_0=\Re(\lambda_0)$, $b_0=\Im(\lambda_0)$, and where the
function $\sqrt{z}$ refers to the principal branch whose cut is the
negative real $z$-axis.  We consider only positive values of $b_0$ and
therefore find that the integrals on the right-hand side of
(\ref{eq:M0general}) can be written as
\begin{equation}
\begin{array}{ll}
\displaystyle
\int_{\Gamma_I\cap C_I}\frac{\pi i\rho_{\rm SY}^0(\eta)}{R(\eta)}\,d\eta
+
\int_{\Gamma_I\cap C_I^*}\frac{\pi i\rho_{\rm SY}^0(\eta^*)^*}{R(\eta)}\,
d\eta\,\,=\\\\
\displaystyle\hspace{1 in}
-\pi\sigma\int_{a_0+ib_0}^{iA}\frac{d\eta}{\sqrt{(\eta-a_0)^2+b_0^2}}+
\pi\sigma\int_{-iA}^{a_0-ib_0}\frac{d\eta}{\sqrt{(\eta-a_0)^2+b_0^2}}\,.
\end{array}
\end{equation}
Performing the quadrature puts the moment condition $M_0=0$ in the
form
\begin{equation}
M_0=-J\pi(2x+4a_0t)+\pi\sigma\left(\mbox{arcsinh}\,
\left(\frac{a_0+iA}{b_0}\right)+
\mbox{arcsinh}\,\left(\frac{a_0-iA}{b_0}\right)\right)=0\,,
\label{eq:momentG0}
\end{equation}
where $\mbox{arcsinh}\,(z)$ is the principal branch whose cuts are on the
imaginary $z$ axis for $|z|\ge 1$.

Next, consider the reality condition $R_0=0$ with $R_0$ given by
(\ref{eq:R0general}).  Since the function $\rho_{\rm SY}^0(\eta)$ is
constant on $\Gamma_I\cap C_I$, the integrals on the right-hand side
of (\ref{eq:R0general}) can be evaluated explicitly:
\begin{equation}
\begin{array}{rcl}
2iR_0&=&\displaystyle
-2iJtb_0^2+\int_{\Gamma_I\cap C_I}\rho_{\rm SY}^0(\xi)\frac{\partial R}{\partial\xi}
(\xi)\,d\xi+
\int_{\Gamma_I\cap C_I^*}\rho_{\rm SY}^0(\xi^*)^*
\frac{\partial R}{\partial\xi}
(\xi)\,d\xi\\\\
&=&-2iJtb_0^2 + iR(iA)+iR(-iA)\,.
\end{array}
\end{equation}
With the observation that, in terms of the principal branch of the
square root, $\sqrt{z}$,
\begin{equation}
R(\pm iA)=\sigma\sqrt{(a_0\mp iA)^2 + b_0^2}\,,
\end{equation}
the reality condition is therefore expressed in terms of standard
functions as:
\begin{equation}
2R_0=\sigma\sqrt{(a_0+iA)^2+b_0^2}+\sigma\sqrt{(a_0-iA)^2+b_0^2}-2Jtb_0^2=0\,.
\label{eq:realityG0}
\end{equation}

\begin{figure}[h]
\begin{center}
\mbox{\psfig{file=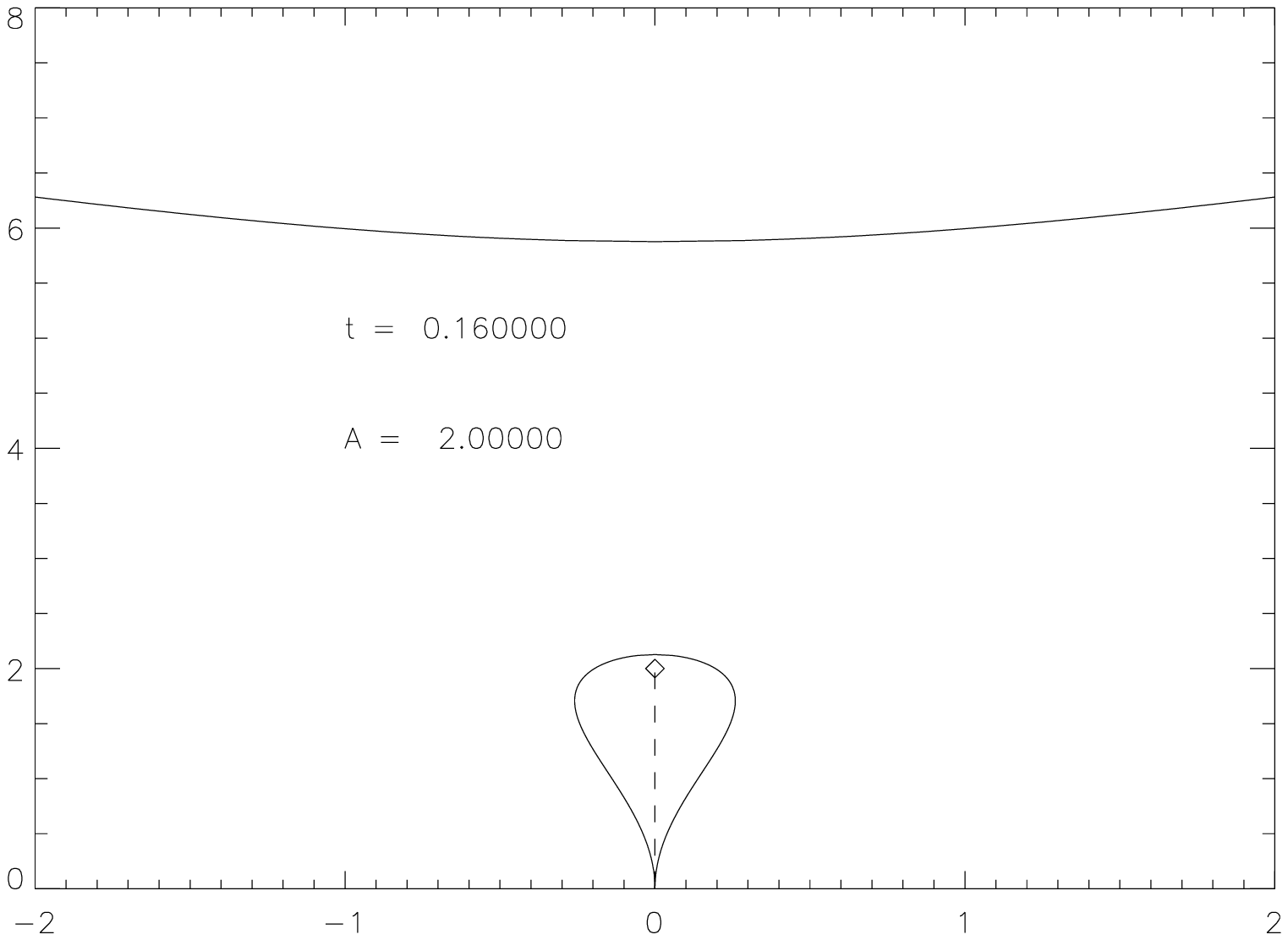,width=1.75 in}
\psfig{file=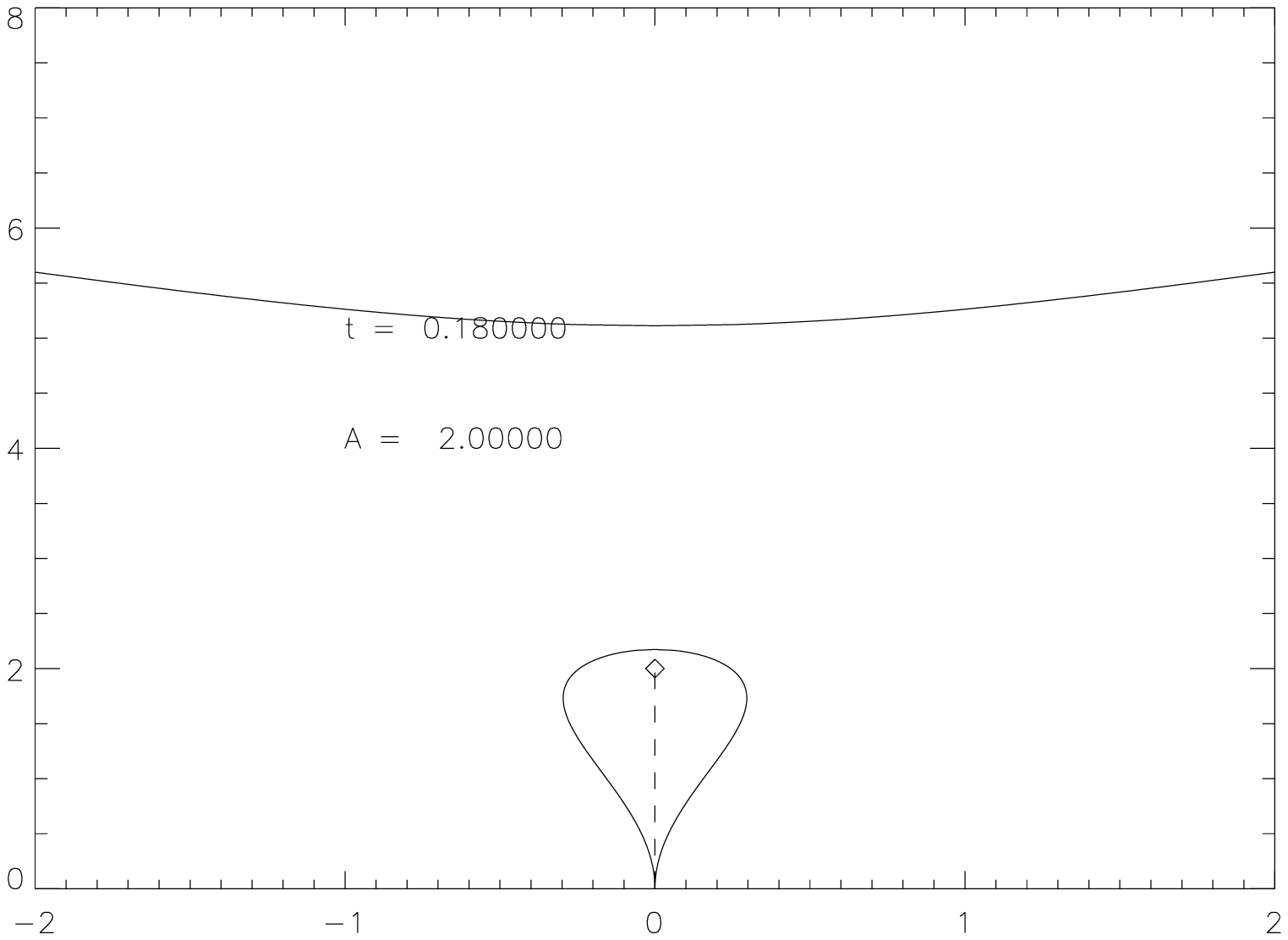,width=1.75 in}
\psfig{file=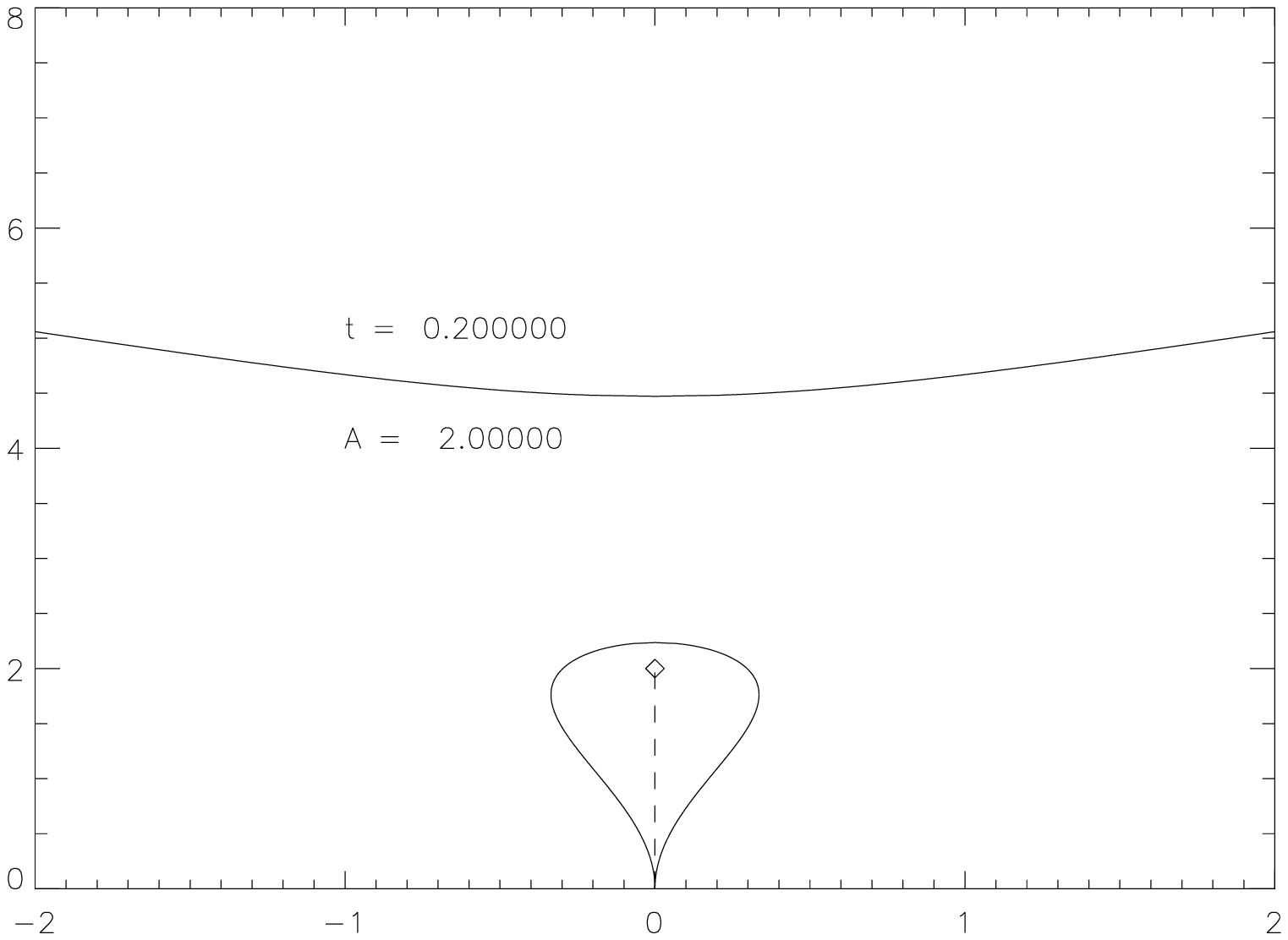,width=1.75 in}}
\mbox{\psfig{file=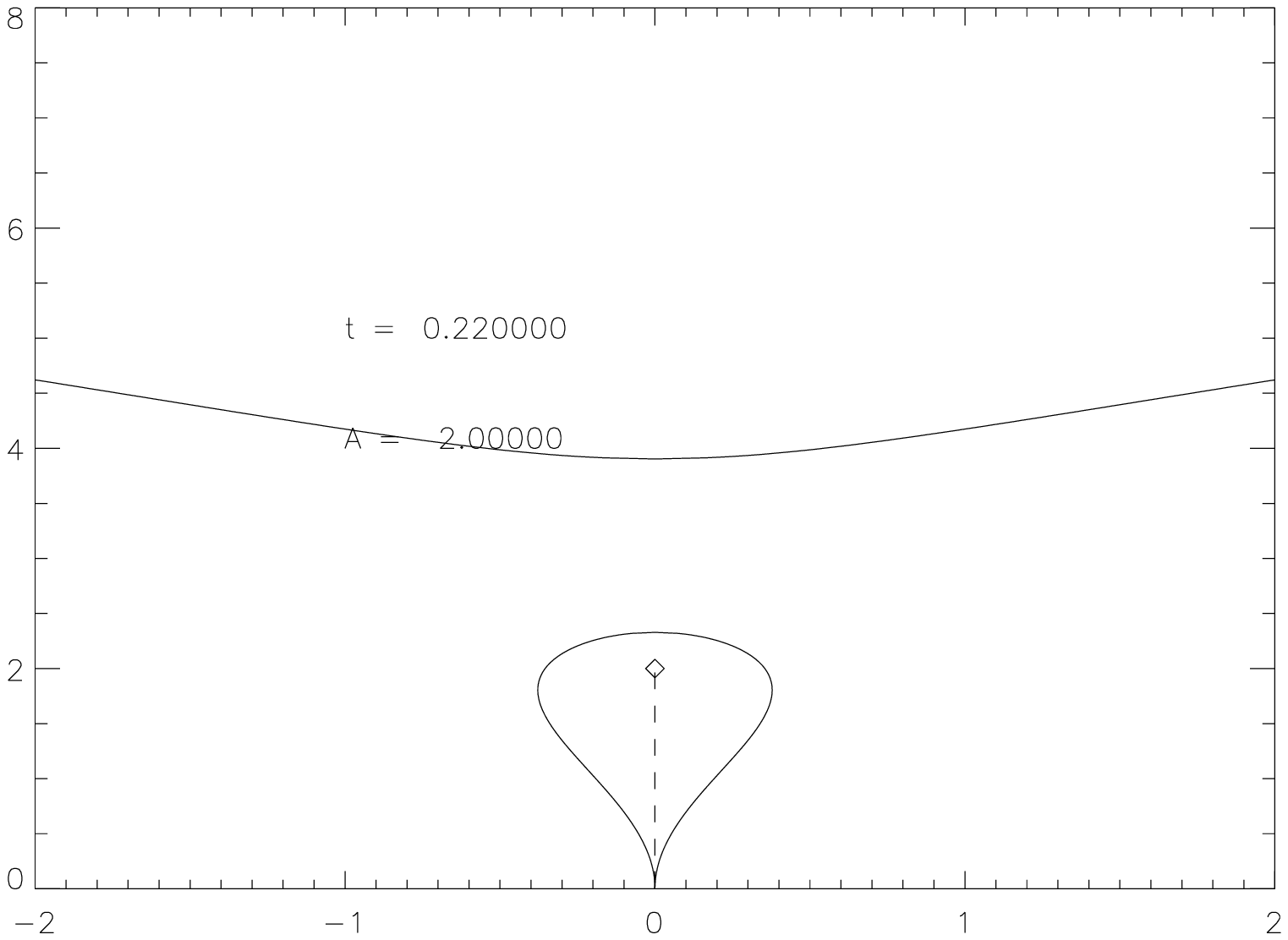,width=1.75 in}
\psfig{file=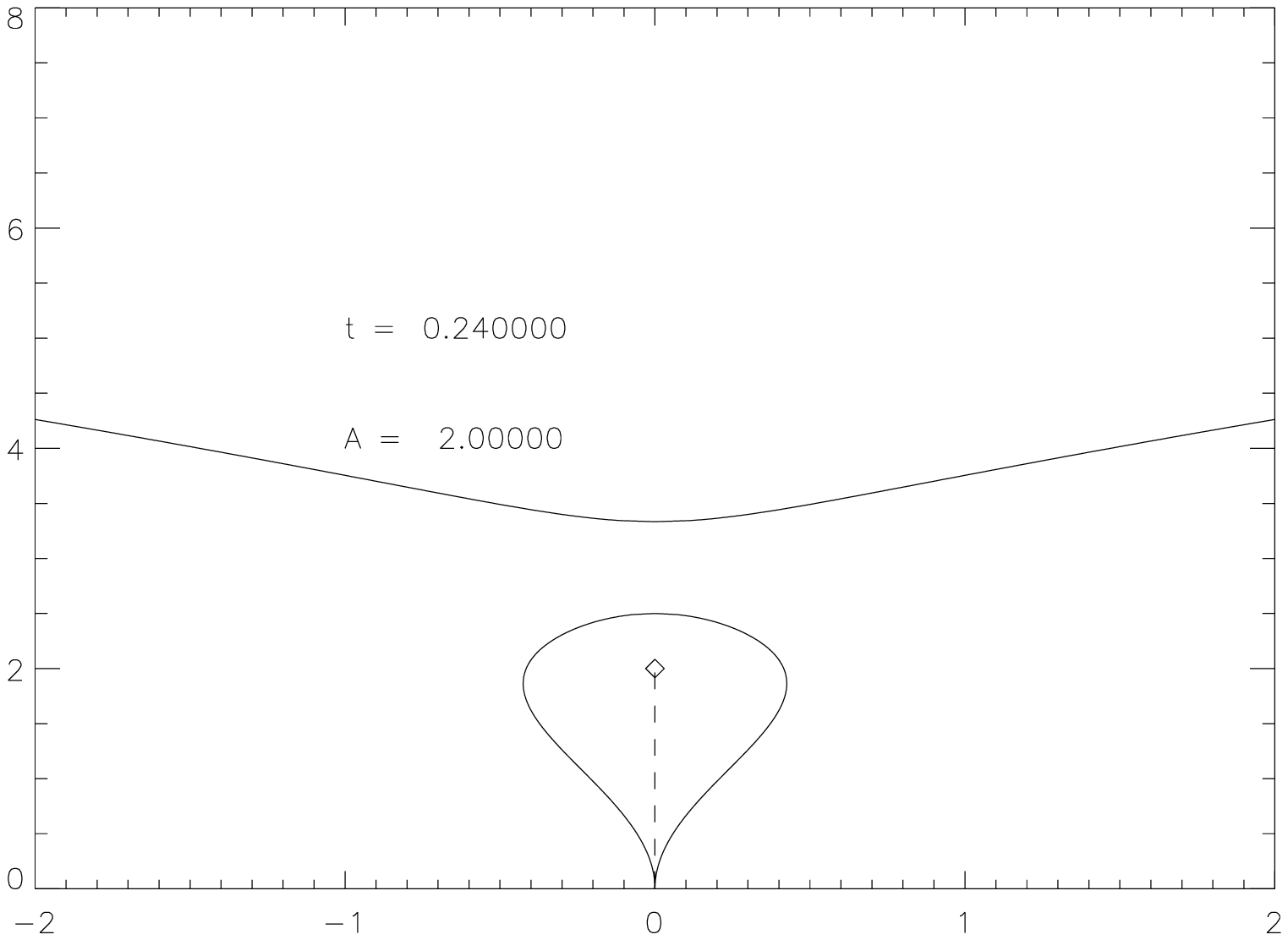,width=1.75 in}
\psfig{file=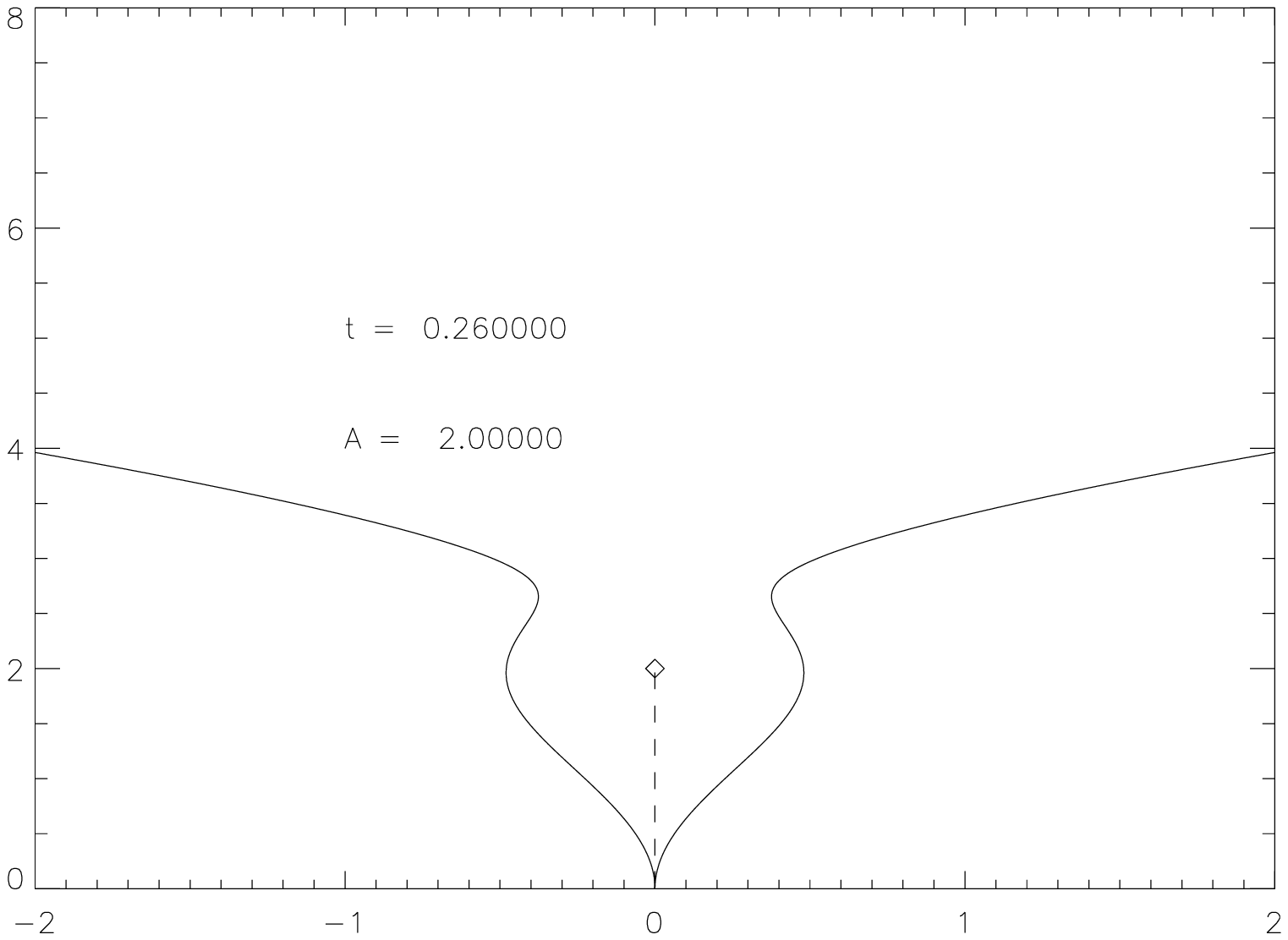,width=1.75 in}}
\end{center}
\caption{\em Plots of the real graph of the reality condition $R_0=0$
for $A=2$.  Top row: $t=0.16$, $t=0.18$, $t=0.20$.  Bottom row:
$t=0.22$, $t=0.24$, $t=0.26$.  The curve develops a double point at
$t=1/(2A)$.  In each plot, the dashed vertical line indicates the
imaginary interval $[0,iA]$.}
\label{fig:realitycurve}
\end{figure}

Let us now investigate the degree to which the two conditions
(\ref{eq:momentG0}) and (\ref{eq:realityG0}) determine the endpoint
$\lambda_0$ as a function of $x$ and $t$.  The first observation is
\begin{lemma}
The reality condition (\ref{eq:realityG0}) is consistent only if
\begin{equation}
\sigma J t \ge 0\,.
\end{equation}
Thus, our options in choosing the parameters
$J=\pm 1$ and $\sigma=\pm 1$ are limited by the sign of $t$.
\end{lemma}
If $J$ and $\sigma$ are chosen so that the reality
condition is consistent, then it may be solved explicitly for $a_0$:
\begin{equation}
a_0=\pm t b_0^2\sqrt{\frac{A^2-b_0^2+t^2b_0^4}{A^2+t^2b_0^4}}\,.
\label{eq:a0}
\end{equation}
For each fixed $t$, the graph of (\ref{eq:a0}) is a curve in the real
$(a_0,b_0)$ plane that contains the endpoint $\lambda_0=a_0+ib_0$.
Several of these curves are plotted in Figure~\ref{fig:realitycurve}.
The curve becomes singular when $|t|=1/(2A)$, developing a double
point at $a_0=0$ and $b_0=A\sqrt{2}$.  

\begin{remark}
The endpoint $\lambda_0$ must of course lie on the graph.  It is
important to note that the bounded component of the graph of
(\ref{eq:a0}), namely the loop encircling the imaginary interval
$[0,iA]$ for $|t|<1/(2A)$, turns out {\em not} to be directly related
to the loop contour $C$.  In particular, the contour $C$ may be
different for different values of $x$, while the graph of
(\ref{eq:a0}) is the same for all $x$.
\end{remark}

We now return to the moment condition $M_0=0$, in which we set $\sigma
J=\mbox{sgn}\,(t)$ for consistency.  Solving (\ref{eq:momentG0}) for
$x$ gives:
\begin{equation}
x = -2a_0t + \frac{\mbox{sgn}\,(t)}{2}\left(
\mbox{arcsinh}\,\left(\frac{a_0+iA}{b_0}\right)+
\mbox{arcsinh}\,\left(\frac{a_0-iA}{b_0}\right)\right)\,.
\end{equation}
For each fixed $|t|<1/(2A)$, this transformation continuously and
invertibly maps the loop enclosing $[0,iA]$ onto the whole real $x$
line, with the point of the loop on the imaginary axis being mapped to
$x=0$.  For $t>0$, the left (respectively right) half of the loop is
mapped to $x<0$ (respectively $x>0$), while for $t<0$, the situation
is reversed.  On this bounded component of the graph, the point
$\lambda=0$ corresponds to $x=\pm\infty$.  The unbounded component of
the graph for $|t|<1/(2A)$ is also placed continuously into one-to-one
correspondence with the real $x$ line.  For $t>0$, the left
(respectively right) half of the graph is mapped to $x>0$
(respectively $x<0$), and for $t<0$ the situation is again reversed.
For $|t|>1/(2A)$, there are two branches of the graph, left and right,
each one unbounded.  Each branch of the graph is placed into
one-to-one correspondence with the real $x$ line.  

We therefore arrive at the result that, given $x$ and $t$ (nonzero),
the equations for the endpoint $\lambda_0=a_0+ib_0$ are only
consistent if we choose $\sigma J=\mbox{sgn}\,(t)$.  This leaves us
free to choose, say, $\sigma=\pm 1$ with $J$ being then determined.
And for each case $\sigma=+1$ and $\sigma=-1$, there are {\em two
distinct solutions of the constraint equations in the upper
half-plane}.  For $|t|<1/(2A)$ there is one solution on the bounded
branch of the graph and one solution on the unbounded branch.  For
$|t|>1/(2A)$, there is one solution on each of the left and right
branches.  Thus, there is always one solution $\lambda_0(x,t)$ in the
right half-plane and one in the left half-plane.  So for each $x$ and
$t$, we still have four possibilities to investigate: $\sigma=\pm 1$
and $\mbox{sgn}\,(a_0)=\pm 1$.

\subsection{Numerical determination of the contour band for the 
Satsuma-Yajima ensemble.}
\index{band!numerical determination of}
\label{sec:findtheband}
At this point, we turn to numerical computations in order to determine
whether there exists a connected component of the graph of
$\Im(B_0(\lambda))=0$ containing $\lambda=0$ and
$\lambda=\lambda_0=a_0+ib_0$, and if so, whether the candidate measure
$\rho^\sigma(\eta)\,d\eta$ supported there is of the correct sign.
Later, we will exploit numerics yet further to verify the possibility
of satisfying the inequality $\Re(\tilde{\phi}^\sigma(\lambda))< 0$ in
the gap.  We search for the band $I_0$ by integrating numerically the
differential equation (\ref{eq:ODE}) for the bands.  We used a
Simpson's rule integrator in conjunction with local changes of
variables to remove integrable inverse square root singularities and
the formula (\ref{eq:rhoformula}) with the desingularization afforded
by the representation (\ref{eq:rhowithY}) to compute the function
$\rho^\sigma(\lambda)$.  Then, we used a fourth-order Runge-Kutta
scheme to integrate numerically the ordinary differential equations
(\ref{eq:ODE}) using arc length as a parameter.  The integration
proceeded from $\lambda=0$ in the direction where
$\rho^\sigma(\eta)\,d\eta$ was negative.

The desired result of this numerical procedure is a path of
integration in the complex plane that emerges in the half-plane
determined by the choice of orientation $\sigma$, avoids the imaginary
interval $[-iA,iA]$ that is the support of the measure $\rho_{\rm
SY}^0(\eta)\,d\eta$, and that ultimately meets the endpoint
$\lambda=\lambda_0$.  However, quite often the numerical integration
revealed other possibilities.  Sometimes, the integral curve starting
at $\lambda=0$ either fails to emerge in correct half-plane as
determined by the choice of orientation $\sigma=\pm 1$, or intersects
the support of the measure $\rho_{\rm SY}^0(\eta)\,d\eta$ (the
imaginary interval $[-iA,iA]$).  And even if this is not the case,
sometimes the integral curve misses the endpoint $\lambda_0$
altogether, being deflected off to infinity in the left or right
half-planes.

\begin{figure}[h]
\begin{center}
\hspace{0.5 in}\mbox{\psfig{file=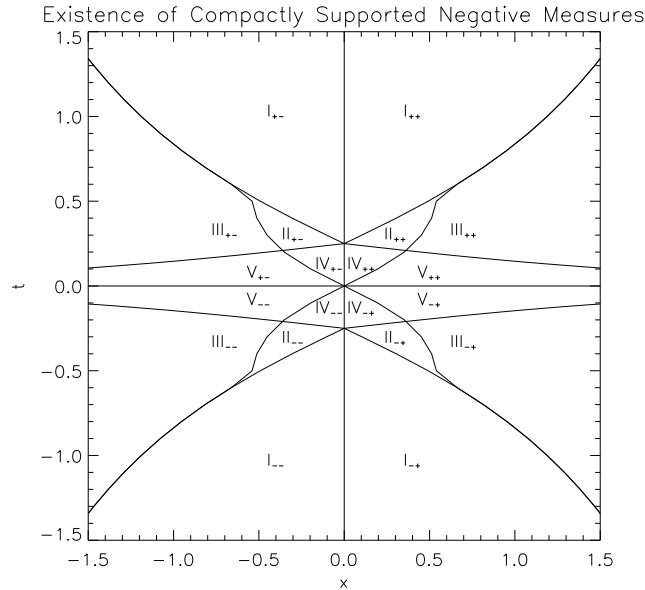,width=4.5 in}}
\end{center}
\caption{\em Regions of the $(x,t)$ plane in which different
assumptions regarding $\sigma=\pm 1$ and $\mbox{sgn}\,(a_0)=\pm 1$
lead to a compact band $I_0^+$ supporting a measure
$\rho^\sigma(\eta)\,d\eta$ of the correct sign.  This picture
is for $A=2$.  See Table~\ref{tab:sectors} and the text for a full
explanation.}
\label{fig:sectors}
\end{figure}

What actually is observed for given choices of $\sigma=\pm 1$ and
$\mbox{sgn}\,(a_0)=\pm 1$ depends on the values of $x$ and $t$.  Our
numerical experiments indicate that the $(x,t)$ plane can be
partitioned into 20 regions in each of which the integral curve
displays qualitatively uniform behavior.  The regions are illustrated
for $A=2$ in Figure~\ref{fig:sectors}, and the meaning of each
region for the integration contour is given in
Table~\ref{tab:sectors}.
\begin{table}[h]
\begin{center}
\begin{tabular}[h]{c||l|l|l|l|}
& $\sigma=+1$, $a_0>0$ & $\sigma=-1$, $a_0>0$ &
$\sigma=+1$, $a_0<0$, & $\sigma=-1$, $a_0<0$ \\
\hline
\hline
I$_{++}$ and I$_{--}$ & connection & left deflection &
right deflection & connection \\
\hline
II$_{++}$ and II$_{--}$ & connection & connection  &
right deflection & connection \\
\hline
III$_{++}$ and III$_{--}$ & connection & intersection &
right deflection & connection  \\
\hline
IV$_{++}$ and IV$_{--}$ & connection & connection &
left deflection & right deflection \\
\hline
V$_{++}$ and V$_{--}$ & connection & intersection &
left deflection & right deflection\\
\hline
\hline
I$_{+-}$ and I$_{-+}$ & connection & left deflection &
right deflection & connection \\
\hline
II$_{+-}$ and II$_{-+}$ & connection & left deflection  &
connection & connection \\
\hline
III$_{+-}$ and III$_{-+}$ & connection & left deflection &
intersection & connection  \\
\hline
IV$_{+-}$ and IV$_{-+}$ & left deflection & right deflection &
connection & connection \\
\hline
V$_{+-}$ and V$_{-+}$ & left deflection & right deflection &
intersection & connection\\
\hline
\hline
\end{tabular}
\end{center}
\caption{\em Regions of the $(x,t)$ plane in which there exist
compactly supported candidate measures of the appropriate sign.  See
Figure~\ref{fig:sectors}.}
\label{tab:sectors}
\end{table}
Note that in Figure~\ref{fig:sectors}, the regions I$_{**}$,
II$_{**}$, and IV$_{**}$ all meet on the $t$-axis at $t=\pm 1/(2A)$.
The meanings of the various scenarios listed in Table~\ref{tab:sectors}
are as follows.
\begin{itemize}
\item
{\em Intersection} means that the contour
of integration meets the imaginary interval $[0,iA]$ either immediately
or after some finite arc length of integration.
\item
{\em Left/Right Deflection} means that
the integration contour
emerges from the origin in the correct quadrant but does not terminate
at the endpoint.  Instead, it misses, and the orbit goes off to infinity 
in the left or right
half-plane.
\item
{\em Connection} means that the path of integration lies completely in the
cut upper half-plane ${\mathbb H}$ and terminates at the
endpoint $\lambda_0$ with finite arc length.
\end{itemize}
Therefore, only the cases labeled as ``connection'' are admissible for 
the asymptotic analysis described in Chapter~\ref{sec:asymptoticanalysis} to
succeed.

For each $x$ and $t$, we see that there is at least one choice of
$\sigma=\pm 1$ and $\mbox{sgn}\,(a_0)=\pm 1$ for which there exists an
a band $I_0$ that connects $\lambda=0$ to $\lambda=\lambda_0$.  In
particular, in each quadrant of the $(x,t)$ plane there is one choice
that always works uniformly throughout all five sub-regions.  Note
also that there are subregions where more than one choice yields a
connecting band $I_0$: in the regions II$_{**}$ there are three
possible choices, each of which yields an admissible band $I_0$.  In
order to distinguish further among these, we need to continue by
checking whether each possible band $I_0$ admits a gap contour
connecting the endpoint to $\lambda=0$ in such a way that the band and
the gap together make a loop encircling the imaginary interval
$[0,iA]$ and such that everywhere on the gap,
$\tilde{\phi}^\sigma(\lambda)$ has a strictly negative real part.

\subsection{Seeking a gap contour on which
$\Re(\tilde{\phi}^\sigma(\lambda))<0$.  
\index{gap!numerical verification of inequality in}
The primary caustic for the Satsuma-Yajima ensemble.}  For given
values of $x$ and $t$, one can choose $\sigma$ and the sign of $a_0$
in one or more ways such that the genus zero ansatz results in a
negative candidate measure $\rho^\sigma(\eta)\,d\eta$ on $I_0^+$, a
contour connecting the origin to $\lambda_0$ that can be determined
numerically as described above.  Given this measure, it is then
possible to compute numerically the real part of the corresponding
function $\tilde{\phi}^\sigma(\lambda)$ as given by the formula
(\ref{eq:phitildeagain}).  This is computationally very efficient,
since the first two terms of (\ref{eq:phitildeagain}) can be
integrated explicitly for the special case $\rho^0(\eta)=\rho^0_{\rm
SY}(\eta)$ ({\em cf.}
\S\ref{sec:ParticularEnsemble}).  Similarly, the real part of the
integral involving $\rho^\sigma(\eta)\,d\eta$ is easy to evaluate
numerically because the candidate measure has already been computed in
the process of finding its support ({\em cf.}
\S\ref{sec:findtheband}), and is real by construction.  Thus, to
calculate this term one simply replaces
$\overline{L_\eta^{C,\sigma}}(\lambda)$ by $\log |\lambda-\eta|$ and
sums over the support weighted by the measure.

\begin{figure}[h]
\begin{center}
\mbox{\psfig{file=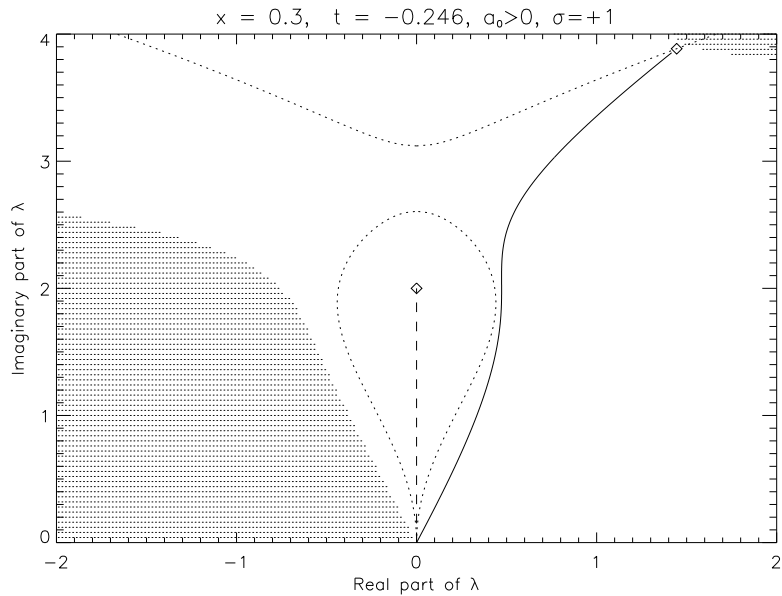,width=4.5 in}}
\end{center}
\caption{\em The impossibility of satisfying the gap inequality for an
incorrectly chosen genus zero ansatz.  Here, $x=0.3$ and $t=-0.246$.
Also, $A=2$.  The dashed curves are the components of the graph of the
reality relation, one of which contains the endpoint
$\lambda_0=a_0+ib_0$, indicated with a diamond.  The imaginary
interval $[0,iA]$ is indicated with a dotted line capped with a
diamond.  The interval $I_0^+$, found by Runge-Kutta integration of
the differential equation (\ref{eq:ODE}) is shown with a solid curve.
The regions of the plane where the real part of the associated function
$\tilde{\phi}^\sigma(\lambda)$ is negative are shaded.  Note
that it does not appear possible to find any path from the endpoint to
zero that lies entirely within the shaded region.}
\label{fig:wrongsidebefore}
\end{figure}

\begin{figure}[h]
\begin{center}
\mbox{\psfig{file=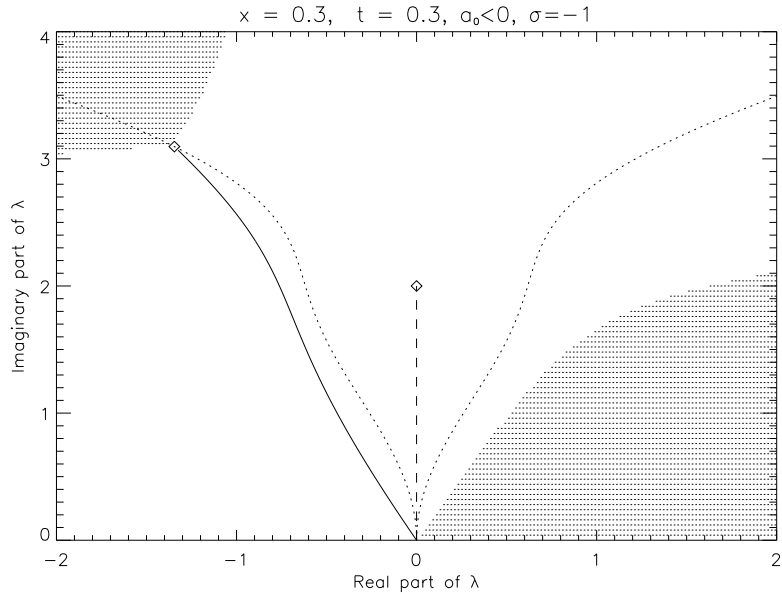,width=4.5 in}}
\end{center}
\caption{\em The impossibility of satisfying the gap inequality.  This
picture is for $x=0.3$ and $t=0.3$, with $A=2$.  As in
Figure~\ref{fig:wrongsidebefore} it appears to be impossible to find a
gap contour on which the relevant inequality is satisfied everywhere.}
\label{fig:wrongsideafter}
\end{figure}

In our numerical investigations, we of course restrict attention at
this point to those cases labeled ``connection'' in
Table~\ref{tab:sectors}.  Our first observation is that it appears
that there can only exist an appropriate gap contour on which
$\Re(\tilde{\phi}^\sigma(\lambda))<0$ everywhere if $x\cdot
t\cdot a_0 > 0$.  This means that the connections given in the final
column of Table~\ref{tab:sectors} for $x\cdot t>0$ and in the first
column of Table~\ref{tab:sectors} for $x\cdot t <0$ do not appear to
admit {\em any} connected path from $\lambda_0$ to zero that closes
the loop and on which the relevant inequality is satisfied everywhere.
In a given quadrant of the $(x,t)$ plane the behavior of this ansatz
is the same in all three regions I$_*$, II$_*$, and III$_*$.
Representative figures showing the region where
$\Re(\tilde{\phi}^\sigma(\lambda))<0$, and thus where a gap
contour might live, are shown in Figures~\ref{fig:wrongsidebefore} and
\ref{fig:wrongsideafter}.

If we accept these numerical results, we see that for each $x$ and
$t$, at most one solution of the equations for the endpoint is
relevant for constructing a genus zero ansatz that satisfies all
necessary inequalities.  We now concern ourselves exclusively with the
unique solution $\lambda_0$ that is in the right half-plane for $x$
and $t$ of the same sign, and in the left half-plane for $x$ and $t$
of opposite signs.

\begin{figure}[h]
\begin{center}
\mbox{\psfig{file=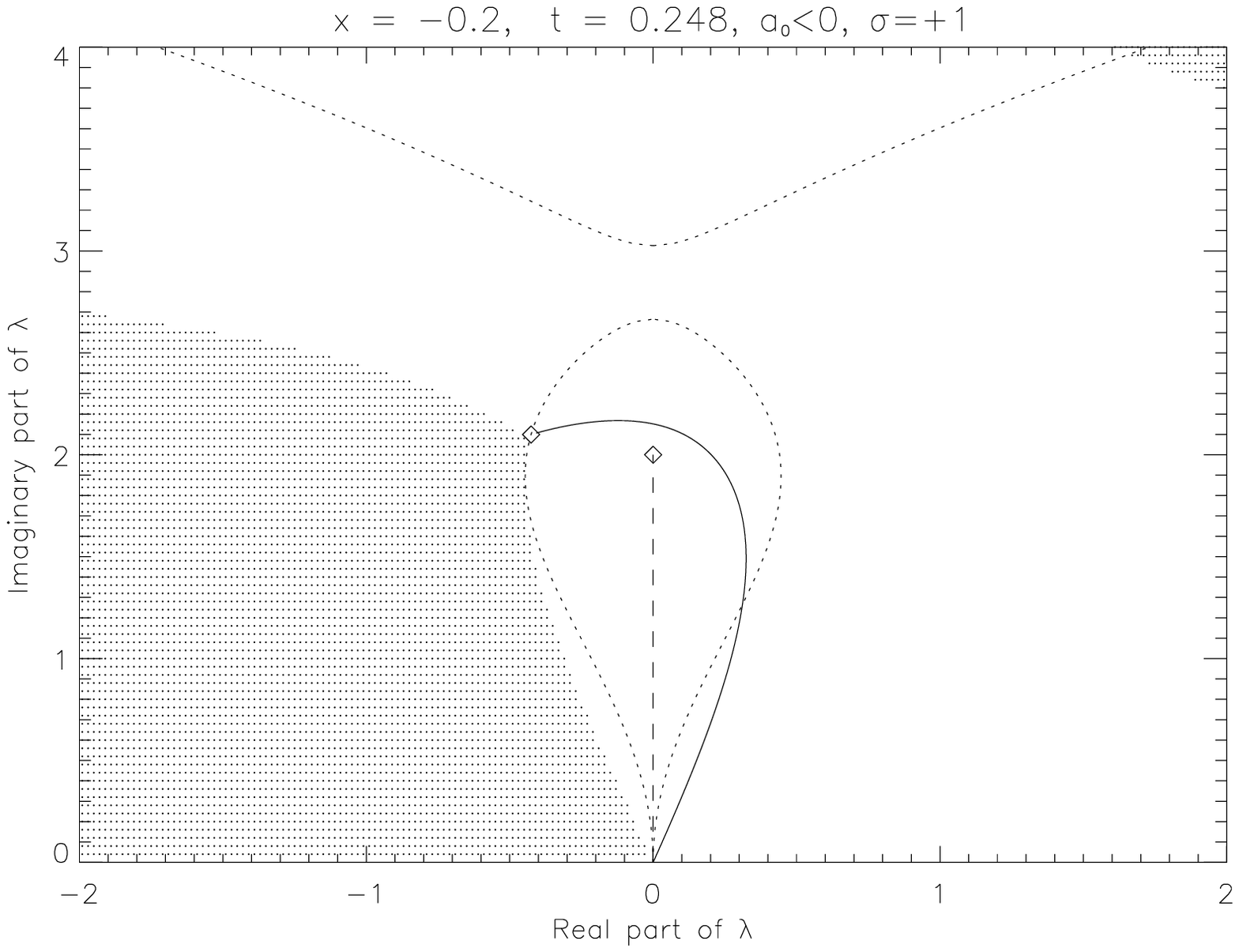,width=4.5 in}}
\end{center}
\caption{\em A case where the inequalities can be satisfied.  Here,
$x=-0.2$ and $t=0.248$, while $A=2$.}
\label{fig:wraparoundbefore}
\end{figure}

\begin{figure}[h]
\begin{center}
\mbox{\psfig{file=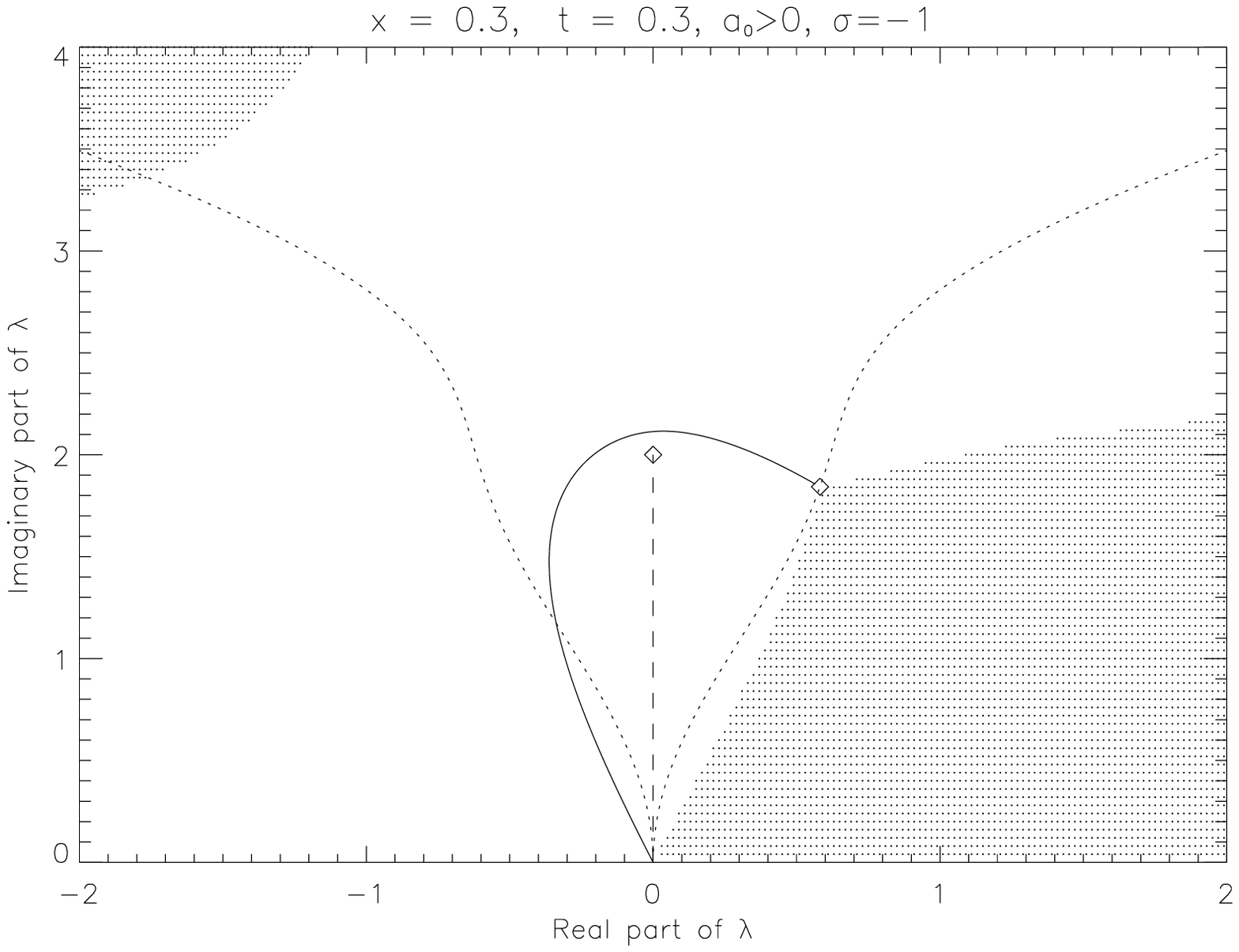,width=4.5 in}}
\end{center}
\caption{\em The inequalities can also be satisfied for this case.  Here,
$x=0.3$ and $t=0.3$, and $A=2$.}
\label{fig:wraparoundafter}
\end{figure}

\begin{figure}[h]
\begin{center}
\mbox{\psfig{file=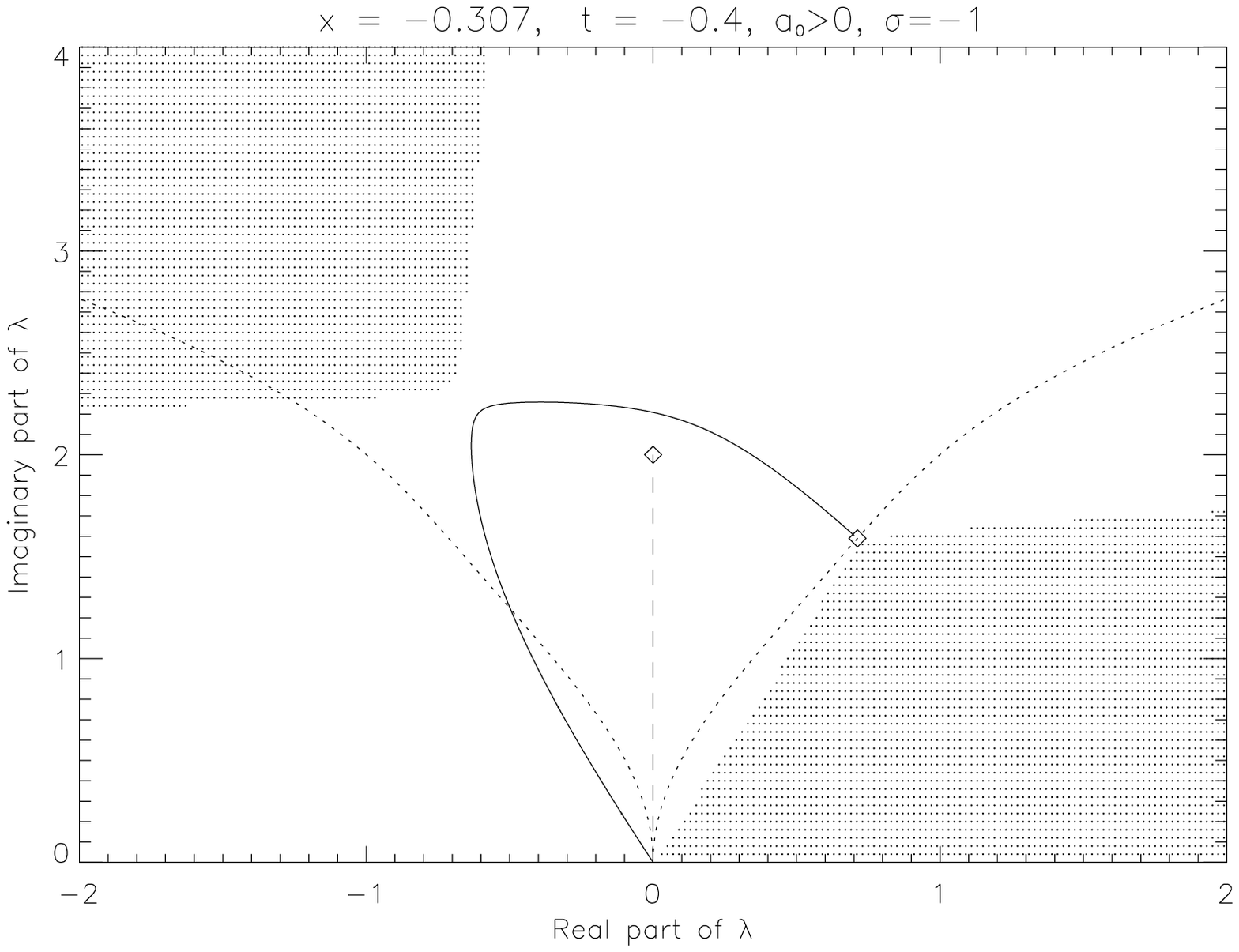,width=4.5 in}}
\end{center}
\caption{\em A case of the successful ``wrap around'' genus zero ansatz very
close to the boundary between region II$_{--}$ and region I$_{--}$.}
\label{fig:wraparoundclose}
\end{figure}

In studying this case, we first consider those cases when the band
$I_0^+$ must ``wrap around'' the imaginary interval $[0,iA]$ because
$\sigma$ and $a_0$ are of opposite signs.  The possible connections
are listed in the second column of Table~\ref{tab:sectors} for $x$ and
$t$ of the same sign and in the third column of
Table~\ref{tab:sectors} for $x$ and $t$ of opposite sign.  These
connections are only possible in the small regions labeled II$_*$ and
IV$_*$ of Figure~\ref{fig:sectors}.  Based on our numerical experiments,
the main observation we want to make
for these cases is that {\em a gap contour may always be found}.  Two
examples of such cases are shown in Figures~\ref{fig:wraparoundbefore}
and \ref{fig:wraparoundafter}.
In each of these figures, we note that there exists a shaded connected
region where $\Re(\tilde{\phi}^\sigma(\lambda))<0$ that contains
many paths connecting the endpoint to zero and completing the loop around
the imaginary interval $[0,iA]$.  For future reference, we also record a
case of this type when the point $(x,t)$ is very close to the boundary of
the region I$_*$.  Such a case is shown in Figure~\ref{fig:wraparoundclose}.
In this figure, the ansatz is close to failure because a complex zero
of the function $\rho^\sigma(\eta)$ is about to move onto the
band contour $I_0^+$.  This is evidenced by the nearly square angle made
by the integral curve of the vector field of (\ref{eq:ODE}) as it makes a 
close approach to this fixed point.  

\begin{figure}[h]
\begin{center}
\mbox{\psfig{file=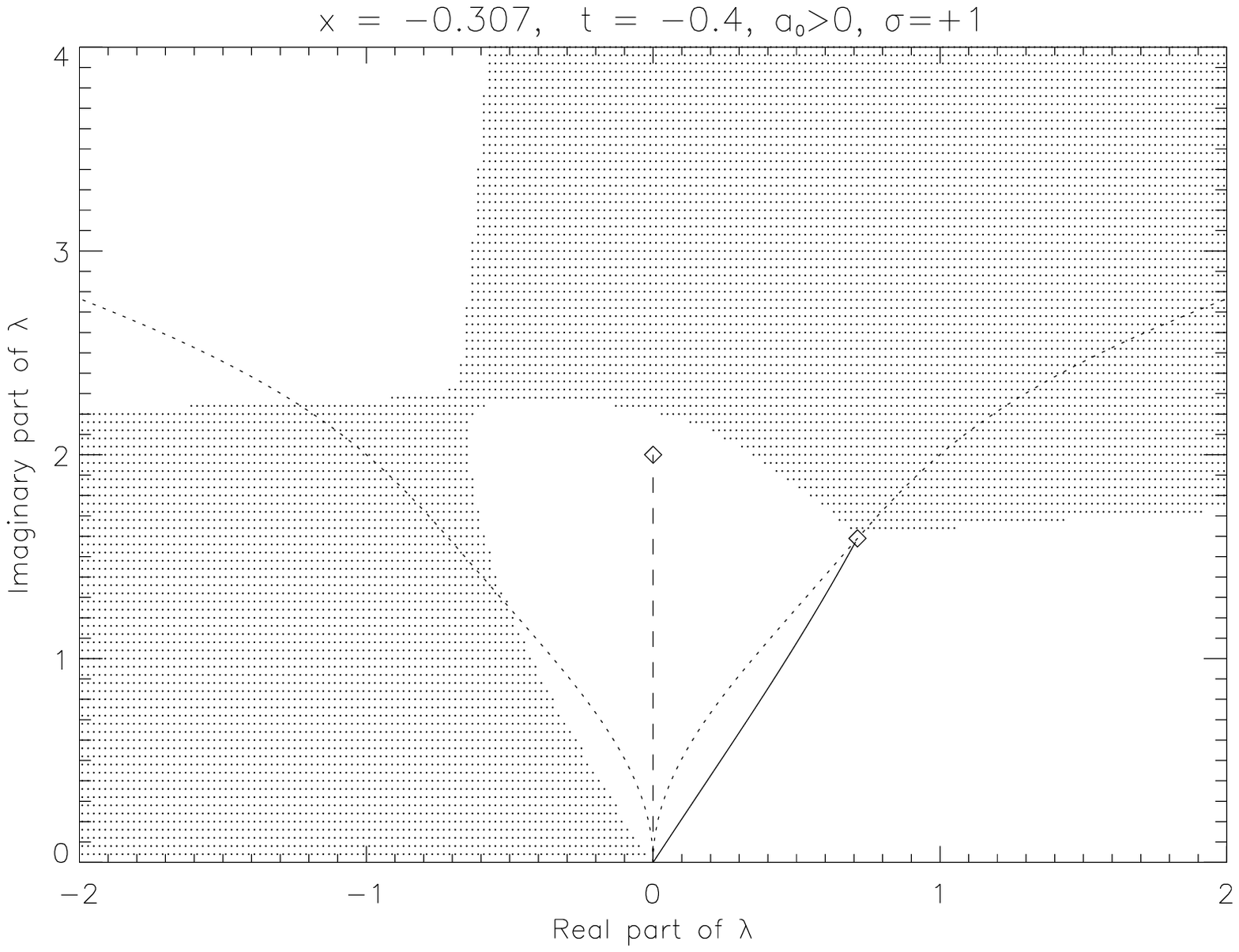,width=4.5 in}}
\end{center}
\caption{\em A case of a barely successful genus zero ansatz.  The
values of $x$ and $t$ are the same as in
Figure~\ref{fig:wraparoundclose}.}
\label{fig:I-II-outside}
\end{figure}

\begin{figure}[h]
\begin{center}
\mbox{\psfig{file=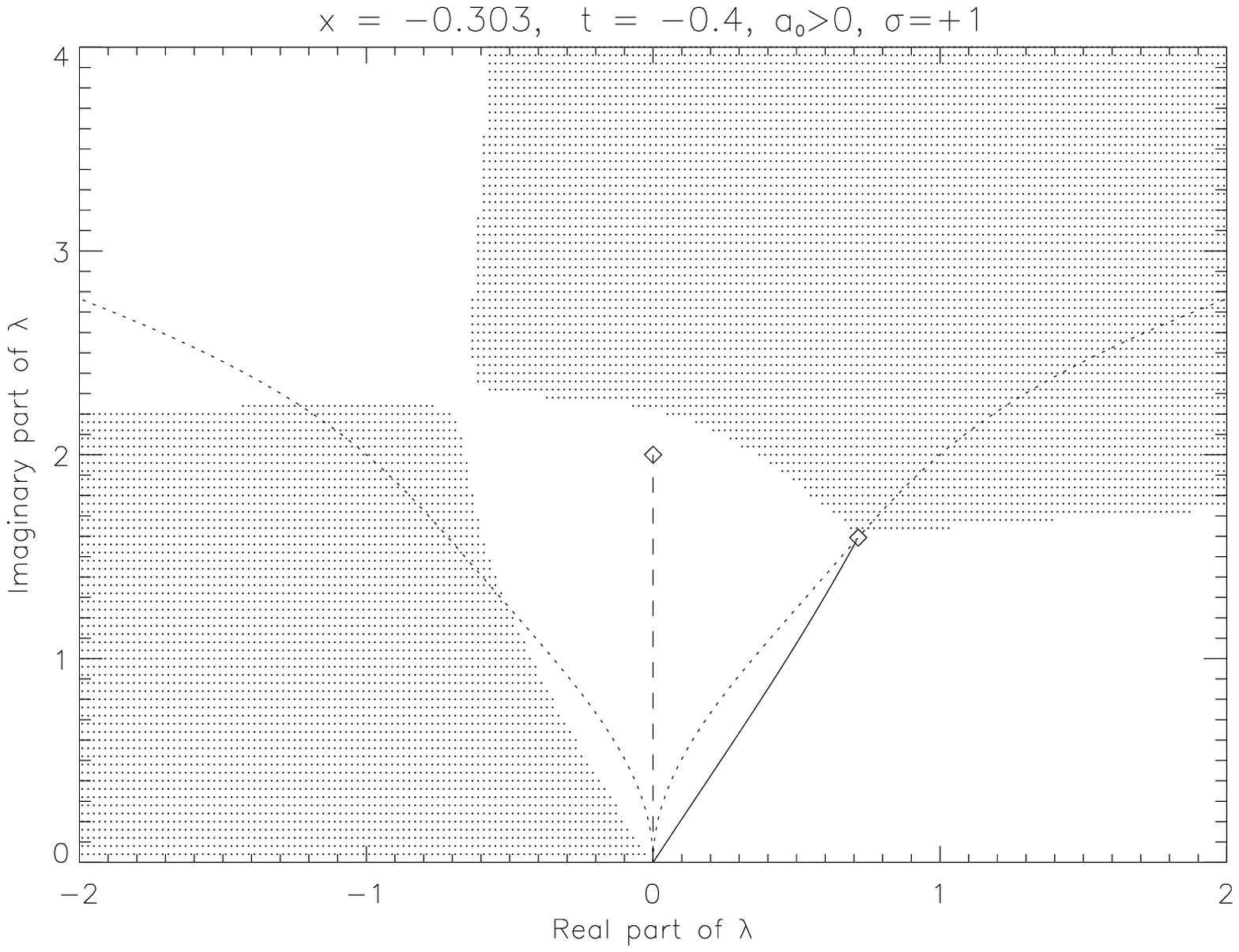,width=4.5 in}}
\end{center}
\caption{\em A case of a barely unsuccessful genus zero ansatz.  The
values of $x$ and $t$ represent a point just on the other side of the
boundary between the regions {\rm I}$_{--}$ and {\rm II}$_{--}$ from
the $x$ and $t$ values used in Figures~\ref{fig:wraparoundclose} and
\ref{fig:I-II-outside}.}
\label{fig:I-II-inside}
\end{figure}

We now study the case represented in the first column of
Table~\ref{tab:sectors} for $x$ and $t$ of the same sign, and in the
last column for $x$ and $t$ of opposite sign.  In this case, the band
$I_0^+$ connecting zero to the endpoint exists for $x$ and $t$ in the
entire quadrant, and we can use plots of the type presented so far to
try and distinguish any regions within a given quadrant where it
appears that the inequality
$\Re(\tilde{\phi}^\sigma(\lambda))<0$ can or cannot be
satisfied on an appropriate gap contour closing the loop around
$[0,iA]$.  The results appear to be that {\em an ansatz of this type
is successful everywhere in the $(x,t)$ plane except in the regions
{\rm I}$_*$ as shown in Figure~\ref{fig:sectors}}.  To illustrate the
mechanism for the breakdown of the ansatz in this case we first look
at two plots corresponding to points just on either side of the
boundary between the regions I$_{--}$ and II$_{--}$.  The first plot,
shown in Figure~\ref{fig:I-II-outside} corresponds to exactly the same
values of $x$ and $t$ as in Figure~\ref{fig:wraparoundclose}.  We see
that a path representing the gap by completing the loop $C$
surrounding the imaginary interval $[0,iA]$ while satisfying
$\Re(\tilde{\phi}^\sigma(\lambda))<0$ everywhere can indeed be found,
albeit barely.  For $x$ and $t$ just on the other side of the
boundary, in region I$_{--}$ the picture that we obtain is given in
Figure~\ref{fig:I-II-inside}.

Figures~\ref{fig:wraparoundclose}, \ref{fig:I-II-outside}, and
\ref{fig:I-II-inside} clearly demonstrate the {\em duality} of the
successful ans\"atze corresponding to two orientations $\sigma=+1$ and
$\sigma=-1$ in the small regions II$_*$ and III$_*$ of the $(x,t)$
plane where they coexist.  The band and gap are dual to each other and
interchangable by reversing orientation while maintaining the same
endpoint.  Also, the band $I_0^+$ for one ansatz coincides with a
connected component of the graph of 
$\Re(\tilde{\phi}^\sigma(\lambda))=0$ for the dual ansatz.
Note that for fixed $x$ and $t$, this interchange requires changing
from $J=+1$ to $J=-1$ and vice-versa, so each choice corresponds to
the asymptotic simplification of a different Riemann-Hilbert problem.
When the mutually dual ans\"atze break down at the boundary with
region I$_*$, they fail {\em at the same point in the complex
$\lambda$-plane}.  When the failure occurs in the gap, the situation
is exactly as described in Chapter~\ref{sec:ansatz}; the ``isthmus'' through
which the gap contour must pass becomes singular at a certain point in
the $\lambda$-plane, and pinches off when $x$ and/or $t$ are tuned
into the region I$_*$.  When the failure occurs in the band, a complex
zero of the candidate density $\rho^\sigma(\lambda)$
approaches the band $I_0^+$ and ultimately meets it at a definite
point in the $\lambda$-plane --- {\em exactly the same point where the
isthmus pinches off in the dual case}.  When $x$ and $t$ are tuned
into the region I$_*$ from the region II$_*$, the zero has crossed the
contour and there is no longer any possibility of finding a band
$I_0^+$ connecting zero to the endpoint supporting
$\rho^\sigma(\eta)\,d\eta$ as a negative real measure.

\begin{figure}[h]
\begin{center}
\mbox{\psfig{file=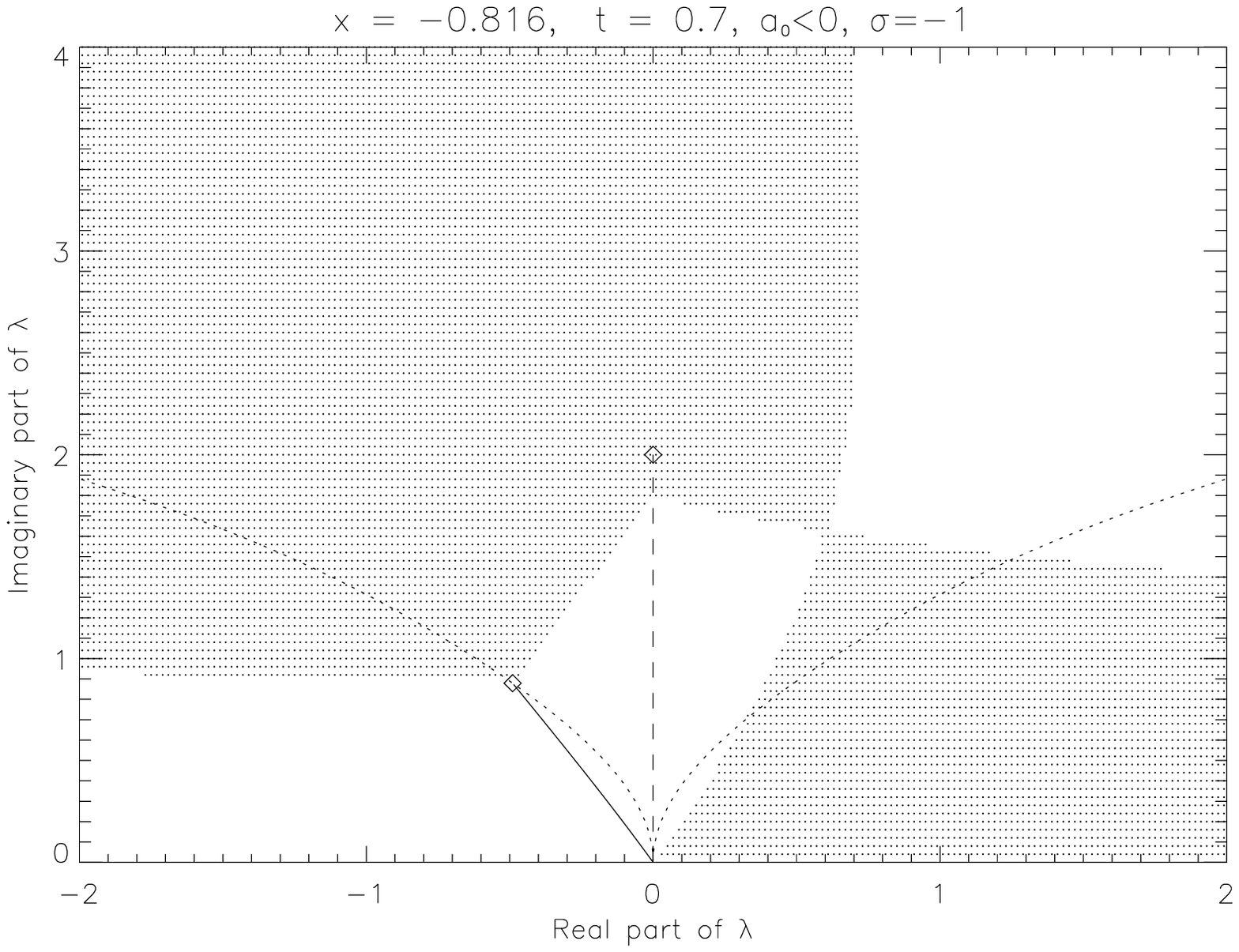,width=4.5 in}}
\end{center}
\caption{\em A barely successful ansatz.  In this plot, like all the
others, $A=2$.  The values of $x$ and $t$ are chosen to be within
region {\rm III}$_{+-}$, but very close to the boundary with region
{\rm I}$_{+-}$.}
\label{fig:I-III-outside}
\end{figure}

\begin{figure}[h]
\begin{center}
\mbox{\psfig{file=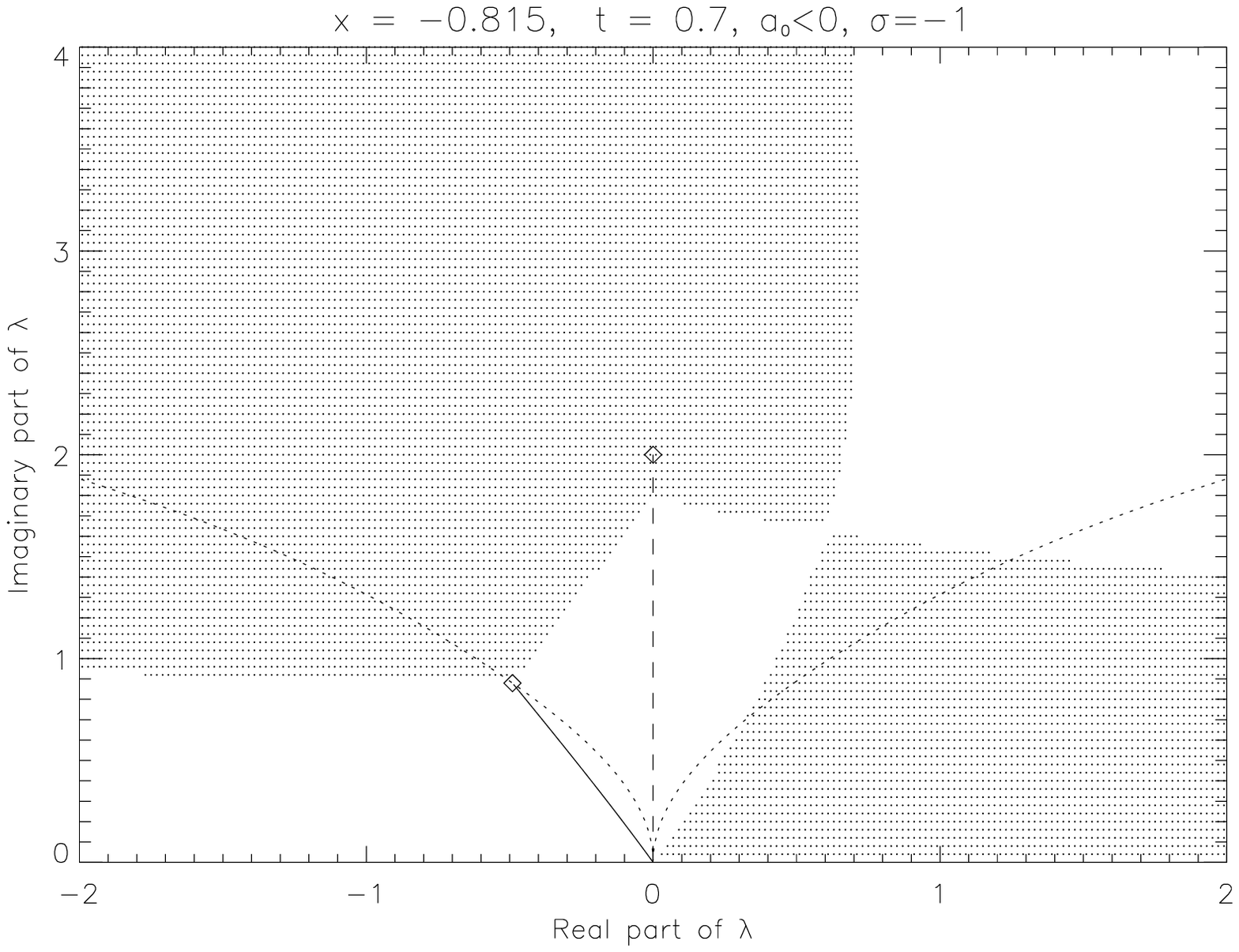,width=4.5 in}}
\end{center}
\caption{\em A barely unsuccessful ansatz.  Again, $A=2$.  The values
of $x$ and $t$ are chosen to be within region {\rm I}$_{+-}$, but
very close to the boundary with region {\rm III}$_{+-}$.}
\label{fig:I-III-inside}
\end{figure}

At the boundary between the regions III$_*$ and I$_*$, the mechanism
of breakdown in the ansatz is similar, although the interpretation in
terms of duality is no longer viable because the ``wrap around''
ansatz is not valid as described above (it fails because the
integration of (\ref{eq:ODE}) from zero intersects the imaginary
interval $[0,iA]$).  Representative diagrams are shown in
Figures~\ref{fig:I-III-outside} and \ref{fig:I-III-inside}.
Figure~\ref{fig:I-III-outside} concerns a point in the $(x,t)$ plane in
region III$_{+-}$ very close to the boundary with region I$_{+-}$.
The isthmus through which the gap must pass is very close to
pinch-off.
In Figure~\ref{fig:I-III-inside}, the value of $x$ has been tuned so
that the point $(x,t)$ is just barely in region I$_{+-}$.  The
pinch-off has occurred and it is no longer possible to find an
admissible gap contour.  

The results we have obtained with our numerical computations are
summarized in Table~\ref{tab:gaps}.
\begin{table}[h]
\begin{center}
\begin{tabular}[h]{c||l|l|l|l|}
& $\sigma=+1$, $a_0>0$ & $\sigma=-1$, $a_0>0$ &
$\sigma=+1$, $a_0<0$, & $\sigma=-1$, $a_0<0$ \\
\hline
\hline
I$_{++}$ and I$_{--}$ & inequality violated & &
 & inequality violated \\
\hline
II$_{++}$ and II$_{--}$ & gap exists & gap exists  &
& inequality violated \\
\hline
III$_{++}$ and III$_{--}$ & gap exists &  &
& inequality violated  \\
\hline
IV$_{++}$ and IV$_{--}$ & gap exists & gap exists &
&  \\
\hline
V$_{++}$ and V$_{--}$ & gap exists &  &
& \\
\hline
\hline
I$_{+-}$ and I$_{-+}$ & inequality violated &  &
 & inequality violated \\
\hline
II$_{+-}$ and II$_{-+}$ & inequality violated &   &
gap exists & gap exists \\
\hline
III$_{+-}$ and III$_{-+}$ & inequality violated &  &
 & gap exists  \\
\hline
IV$_{+-}$ and IV$_{-+}$ &  &  &
gap exists & gap exists \\
\hline
V$_{+-}$ and V$_{-+}$ & & &
& gap exists\\
\hline
\hline
\end{tabular}
\end{center}
\caption{\em Success or failure of the genus zero ansatz in the
regions of the $(x,t)$ plane in which there exist compactly supported
candidate measures of the appropriate sign.  See
Figure~\ref{fig:sectors}.  The blank entries in the table correspond
to ans\"atze for which there is no band $I_0^+$ connecting zero to the
endpoint on which the candidate measure $\rho^\sigma(\eta)\,d\eta$
is negative real.}
\label{tab:gaps}
\end{table}
For all $x$ and $t$ outside the region
I$=$I$_{++}\cup$I$_{+-}\cup$I$_{-+}\cup$I$_{--}$, we have at least one
genus zero ansatz (and sometimes there is also the dual ansatz) that
satisfies all of the conditions required of a complex phase function
$g^\sigma(\lambda)$.  For such $x$ and $t$, it follows from the
rigorous analysis carried out in Chapter~\ref{sec:asymptoticanalysis} that
the corresponding complex phase function therefore correctly captures
the behavior of the solution of the Riemann-Hilbert problem in the
limit $\hbar\downarrow 0$.  In light of the exact solution of the
outer model problem given in
\S\ref{sec:outersolve}, we see that that the Satsuma-Yajima semiclassical
soliton ensemble behaves like a modulated exponential plane-wave
solution of the nonlinear Schr\"odinger equation for all $x$ and $t$
outside of the boundary of region I.  This is in agreement with the
observations made in \cite{MK98}, where the curve in the $(x,t)$ plane
at which the modulated plane-wave behavior was seen to break down was
called the {\em primary caustic} (a secondary caustic was also
observed).  Comparing with the figures in \cite{MK98}, it is clear
that {\em the boundary of the region {\rm I} is exactly the primary
caustic}\index{primary caustic}.

\begin{remark}
These numerical experiments indicate that for the Satsuma-Yajima
soliton ensemble, the failure of the genus zero ansatz can be
essentially captured by the two conditions (\ref{eq:bandfail}) and
(\ref{eq:gapfail}).  Indeed, if we examine the boundary of region I,
we see from Table~\ref{tab:sectors} that whenever for a particular
choice of $\sigma$ and the sign of $a_0$ a band fails to exist upon
crossing this boundary, it is due to a transition from ``connection''
to ``left/right deflection''.  In this sense, the ``intersection''
scenario plays no real role in the breakdown of the ansatz.  The
transition from ``connection'' to ``deflection'' corresponds to the
passage of a zero of the candidate density $\rho^\sigma(\eta)$ through
the band $I_0^+$ and therefore such a transition point $(x_{\rm
crit},t_{\rm crit})$ satisfies the conditions (\ref{eq:bandfail}).
This phenomenon is illustrated in Figure~\ref{fig:wraparoundclose};
since from the relations (\ref{eq:plusside}) and (\ref{eq:minusside})
the curves satisfying $\Re(\tilde{\phi}^\sigma(\lambda))=0$ are also
orbits of the differential equation (\ref{eq:ODE}), the boundary of
the shaded region in the second quadrant is another orbit of
(\ref{eq:ODE}).  The mechanism of failure of the band to exist at
$(x_{\rm crit},t_{\rm crit})$ is therefore the meeting of these
two orbits at a mutual analytic fixed point.  Similarly, at each point
$(x_{\rm crit},t_{\rm crit})$ on the boundary of region I where for a
particular choice of $\sigma$ and the sign of $a_0$ a band $I_0^+$
exists on both sides of the boundary, the failure of a gap to exist is
brought on by the pinching off at some point in the cut upper
half-plane ${\mathbb H}$ of the region where
$\Re(\tilde{\phi}^\sigma(\lambda))<0$.  This phenomenon is clearly
illustrated in, for example, Figures
\ref{fig:I-II-outside} and \ref{fig:I-II-inside}, or Figures
\ref{fig:I-III-outside} and \ref{fig:I-III-inside}.  For such a
transition, the boundary point $(x_{\rm crit},t_{\rm crit})$ satisfies
the gap failure criterion (\ref{eq:gapfail}).
\end{remark}

Although we have verified the existence of a complex phase function
corresponding to the genus zero ansatz all the way to the boundary of
region I by verifying all inequalities numerically ({\em i.e.} using
the data leading to pictures of the sort that have been presented in
this section for a fine grid of values of $x$ and $t$),
we do not have at this point a direct method to verify these
inequalities analytically.  We are, however, compelled to state the
following.
\begin{theorem}
The conclusion of Corollary~\ref{corollary:SYsmalltime}, giving the
rigorous semiclassical limit of the initial-value problem
(\ref{eq:IVP}) for the focusing nonlinear Schr\"odinger equation with
the Satsuma-Yajima initial data $\psi_0(x)=A\,{\rm sech}(x)$, extends
from $t=0$ all the way to the boundary of region {\rm I}.  This
boundary is explicitly characterized by the gap failure criterion
(\ref{eq:gapfail}).
\end{theorem}
An analytical proof of this theorem awaits the development of tools
that generalize the ``integration in $x$'' methods that have proven so
useful for real-line problems like the zero-dispersion limit
\index{zero-dispersion limit} of the Korteweg-de Vries equation
\index{Korteweg-de Vries equation} \cite{LL83} 
and the continuum limit of the Toda lattice \index{Toda lattice}
\cite{DM98} to the complex plane.  Formal WKB theory \cite{M00} may be
of some assistance in this connection, as it identifies certain
significant paths in the complex $x$-plane that may play the role
usually played simply by the real $x$-axis.

\begin{remark}
Note that while the genus zero ansatz breaks down at the boundary of
region I, the endpoint $\lambda_0(x,t)$ is analytic in $x$ and $t$ at
the boundary and indeed continues into region I.  As will be shown
below, the only singular point of $\lambda_0(x,t)$ in common with the
boundary of region I is for $x=0$.
\end{remark}

\begin{remark}
The numerics suggest that the point $x=0$ is not as singular as one
might expect from the analytical small-time analysis.  Indeed, the
ansatz for $x$ small and positive appears to smoothly continue around
$x=0$ to negative values of $x$ on any path with $t\neq 0$.  In fact,
for $t\neq 0$ and $x=0$, the endpoint lies on the imaginary axis {\em
above} $iA$, and the genus zero ansatz appears to be successful for
both $\sigma=+1$ and $\sigma=-1$, although in neither case does the
band $I_0^+$ coincide with the imaginary axis.  For $x=0$ the two 
workable ans\"atze are mutually dual and are mapped into each other
by reflection through the imaginary axis.  Thus, in the regions 
IV$_{++}\cup$IV$_{+-}$ and IV$_{--}\cup$IV$_{-+}$, {\em both} workable
ans\"atze are continuous at $x=0$ for $t\neq 0$ in that as the endpoint
crosses the imaginary axis, the ansatz for which the band $I_0^+$ lies
entirely in one quadrant continuously becomes an ansatz for which the
band $I_0^+$ must ``wrap around'' the point $iA$ in order to meet the
endpoint $\lambda_0(x,t)$.  

The numerical results on the $t$-axis are consistent with the exact
solution of the endpoint equations in terms of the elliptic integral
$E(m)$ given in Theorem~\ref{theorem:xiszero}.
\end{remark}

\section[The Solution of Akhmanov, Sukhorukov, and Khokhlov]{The elliptic modulation equations and the 
particular solution of Akhmanov, Sukhorukov, and Khokhlov for the
Satsuma-Yajima initial data.}  Recall that in
\S\ref{sec:modulation}, it was shown that
whenever the endpoints $\lambda_0(x,t),\dots,\lambda_G(x,t)$ are
obtained for a genus $G$ ansatz as the solution of the three sets of
conditions (\ref{eq:momenteqns}), (\ref{eq:vanishing}), and
(\ref{eq:realitycond}), then they satisfy a first-order quasilinear
coupled system of partial differential equations (the Whitham or
modulation equations) in $x$ and $t$, and that each endpoint and its
complex conjugate is a Riemann invariant of this quasilinear system.
In particular, for genus zero, we found that ({\em cf.}
(\ref{eq:genuszeroWhitham}))
\begin{equation}
\frac{\partial \lambda_0}{\partial t} + (-2a_0-ib_0)
\frac{\partial \lambda_0}{\partial x}=0\,,\hspace{0.3 in}
\frac{\partial \lambda_0^*}{\partial t} + (-2a_0+ib_0)
\frac{\partial \lambda_0^*}{\partial x}=0\,,
\end{equation}
where $a_0:=\Re(\lambda_0)$ and $b_0:=\Im(\lambda_0)$.  These
equations can be given a more direct physical interpretation in view
of the semiclassical solution of the nonlinear Schr\"odinger equation
given by ({\em cf.} (\ref{eq:psitildeGeqzero}))
\begin{equation}
\psi\sim \Im(\lambda_0)e^{-i\alpha_0/\hbar}\,,
\end{equation}
and proved in \S\ref{sec:error} to be uniformly valid in any compact
set in the $(x,t)$-plane where the genus zero ansatz is valid with all
inequalities being strictly satisfied.  For fixed $t$, and for
$x=x_0+\hbar\hat{x}$, this approximate solution is a modulated plane
wave of the form (up to a phase depending on $x_0$)
\begin{equation}
\psi\sim \sqrt{\rho} e^{ik \hat{x}}\,,
\end{equation}
where the ``fluid density'' (not to be confused with the density
function $\rho^\sigma(\eta)$ for the complex phase) is defined by
$\rho:=\Im(\lambda_0)^2$ and the wavenumber is defined by
$k:=-\partial_x \alpha_0$.

With the help of the formula (\ref{eq:wavenumberformula}) expressing
$k$ in terms of the $\lambda_0$, we can rewrite the modulation
equations in terms of the fluid density \index{fluid density} $\rho$
and momentum \index{momentum} $\mu:=\rho k$.  Since $\rho=b_0^2$ and
$\mu = -2a_0 b_0^2$, we find immediately from
(\ref{eq:genuszeroWhitham}) that
\begin{equation}
\begin{array}{rcccl}
\displaystyle\frac{\partial \rho}{\partial t} &+&\displaystyle
\frac{\partial \mu}{\partial x} &=&0\,,\\\\
\displaystyle \frac{\partial \mu}{\partial t}&+&\displaystyle
\frac{\partial}{\partial x}\left(\frac{\mu^2}{\rho}-\frac{\rho^2}{2}
\right) &=& 0\,.
\end{array}
\label{eq:rhomusys}
\end{equation}
This elliptic quasilinear system has been known in this form for some
time in connection with the formal semiclassical theory of nonlinear
Schr\"odinger equations.  In 1966, an exact solution of
(\ref{eq:rhomusys}) was obtained in implicit form by Akhmanov,
Sukhorukov, and Khokhlov \cite{ASK66}, through an application of the
hodograph transform \index{hodograph transform} method.  See also
Whitham \cite{W74} for a discussion of their result.  They defined
$\rho(x,t)$ and $\mu(x,t)$ as a branch of the solution of the
relations
\begin{equation}
\begin{array}{rcl}
\mu &=&\displaystyle -2t\rho^2\tanh\left(\frac{\rho x-\mu t}{\rho}\right)\,,\\\\
\rho &=&\displaystyle (A^2 + t^2\rho^2)\,\mbox{sech}^2\left(
\frac{\rho x-\mu t}{\rho}\right)\,.
\end{array}
\label{eq:implicitguys}
\end{equation}
It is easy to see by setting $t=0$, that the solution obtained in
\cite{ASK66} satisfies the initial condition
\begin{equation}
\rho(x,0)=A^2\,\mbox{sech}^2(x)\,,\hspace{0.3 in}\mu(x,0)\equiv 0\,.
\end{equation}
Clearly, the functions $\rho(x,t)$ and $\mu(x,t)$ obtained from the
endpoints $\lambda_0(x,t)$ and $\lambda_0^*(x,t)$ also satisfy these
initial conditions.  We therefore see that our analysis both reproduces
and makes rigorous a formal result that has been in the literature for
more than thirty years.

It is instructive to study the implicit relations (\ref{eq:implicitguys})
for $x=0$.  In this case, it is easily seen that
there is one solution for which $\mu\equiv 0$ as a function of $t$,
and then from the second equation one finds
\begin{equation}
\rho(x=0,t) = \frac{1\pm\sqrt{1-4A^2t^2}}{2t^2}\,.
\label{eq:SYonaxis}
\end{equation}
Thus, a branch-point singularity develops for $x=0$ when $t=1/(2A)$.
This singularity corresponds exactly to the intersection of the boundary
of region I with the line $x=0$.  The {\em first} point on the boundary
of the region where the genus zero ansatz holds therefore corresponds
to a singularity of the endpoint function $\lambda_0(x,t)$, although
as remarked above the remaining boundary points are no obstruction to
the analytic continuation of the function $\lambda_0(x,t)$.

The same formula (\ref{eq:SYonaxis}) can of course also be obtained
from the general formula (\ref{eq:xiszero}) by using $A(x)=A\,{\rm
sech}(x)$ and setting $\rho(x=0,t)=b_0(t)^2$.

\chapter{The Transition to Genus Two}
\label{sec:genus2}
Recall that in \S\ref{sec:smalltime} it was shown that for each fixed $x
\neq 0$, there exists some choice of the parameters $\sigma$ and $J$ 
such that the $G=0$ ansatz holds for $|t|$ sufficiently small.
Furthermore, it was shown in \S\ref{sec:continuation} that if the pair
$(x_0,t_0)$ is such that the $G=0$ ansatz holds and the endpoint
functions are differentiable, then there is a small neighborhood of
$(x_0,t_0)$ on which the $G=0$ ansatz holds as well, and this allows
us to define a region of the $(x,t)$-plane containing $(x_0,t_0)$
throughout which the $G=0$ ansatz satisfies all inequalities necessary
for the asymptotic analysis of
Chapter~\ref{sec:asymptoticanalysis} to be valid.  

A point $(x_{\rm crit},t_{\rm crit})$ on the boundary of this region
of validity is characterized by one or more of the following six
critical events:
\begin{enumerate}
\item
The endpoint function $\lambda_0(x,t)$ fails to be analytic in $x$ and $t$
at $(x_{\rm crit},t_{\rm crit})$.
\item
The endpoint $\lambda_0(x_{\rm crit},t_{\rm crit})$ lies on the boundary
of the cut upper half-plane ${\mathbb H}$.
\item
The band $I_0^+$ is a smooth orbit of $(\ref{eq:ODE})$ connecting
$\lambda=\sigma 0$ to the endpoint $\lambda_0(x_{\rm crit},t_{\rm
crit})$, but either $I_0^+$ has a point of tangency with the boundary
of ${\mathbb H}$ on the real axis or the imaginary interval $[0,iA)$,
or the point $\lambda=iA$ lies on $I_0^+$.
\item
There is a connected region in the upper half-plane with
$\lambda_0(x_{\rm crit},t_{\rm crit})$ and $\lambda=-\sigma 0$ on the
boundary where $\Re(\tilde{\phi}^\sigma(\lambda))<0$, but $\lambda=iA$ is
also on the boundary and the region is bisected by the segment $[0,iA]$.
\item
The band $I_0^+$ passes through an analytic fixed point
$\hat{\lambda}$ of the vector field (\ref{eq:ODE}) on the way to
$\lambda_0(x_{\rm crit},t_{\rm crit})$ and thus is not smooth at this
point, making an angle of $90^\circ$ (for a simple zero of
$\rho^\sigma(\lambda)$).  Strictly speaking, $I_0^+$ is a union of
three orbits of (\ref{eq:ODE}):  two regular orbits and a fixed point.
In a degenerate situation, the zero $\hat{\lambda}$ may coincide with 
an endpoint of the band.
\item
The closed region where $\Re(\tilde{\phi}^\sigma(\lambda))\le 0$ holds
admits a gap contour $\Gamma_1^+$ in ${\mathbb H}$ connecting
$\lambda_0(x_{\rm crit},t_{\rm crit})$ to $\lambda=-\sigma 0$, but is
pinched off at a point $\hat{\lambda}$ through which $\Gamma_1^+$ must
pass and at which $\Re(\tilde{\phi}^\sigma(\hat{\lambda}))=0$.
\end{enumerate}
The last two of these critical events may be characterized by the
equations (\ref{eq:bandfail}) and (\ref{eq:gapfail}).  The
computer-assisted analysis of the genus zero ansatz for the
Satsuma-Yajima ensemble carried out in \S\ref{sec:finitetimeSY}
indicated that in that particular case the ansatz parameters may be
chosen at $(x_{\rm crit},t_{\rm crit})$ such that the failure is
indeed due either to the conditions (\ref{eq:bandfail}) or the
conditions (\ref{eq:gapfail}).  In fact, it is clear from
Table~\ref{tab:gaps} that for particular choices of the ansatz
parameters it is sufficient to characterize the boundary of the region
where the genus zero ansatz holds by the gap failure condition
(\ref{eq:gapfail}).  Also, as pointed out before, the band failure
condition (\ref{eq:bandfail}) and the gap failure condition
(\ref{eq:gapfail}) are essentially equivalent according to the
relations (\ref{eq:plusside}) and (\ref{eq:minusside}); they generate
the same set of points in the $(x,t)$-plane.

In this chapter we will assume that $(x_{\rm crit}, t_{\rm crit})$ is
such that the sixth condition listed above holds, which implies that
the gap failure conditions (\ref{eq:gapfail}) hold for
$\lambda=\hat{\lambda}\in{\mathbb H}$.  We call this situation a {\em
critical genus zero ansatz}\index{genus $G$ ansatz!critical, for
$G=0$}.  The problem at hand then is to describe what happens as
$(x,t)$ leaves the connected region on which the genus zero ansatz
holds, moving away from $(x_{\rm crit},t_{\rm crit})$.  We want to
establish that for some genus $G\neq 0$ the ansatz (hopefully with the
same parameters $J$ and $\sigma$) will be valid just beyond the
boundary.  Following examples from the integrable systems literature
(see, for example, \cite{LL83}) and the approximation theory
literature (see \cite{DKM98}), we suppose that as $(x,t)$ leaves the
connected component of the $(x,t)$-plane where the genus zero ansatz
holds, the genus jumps from $G=0$ to $G=2$.

\begin{remark}
The reason for supposing that the genus ``skips'' the value $G=1$ is
essentially connected to the complex-conjugation symmetry of the
contour $C\cup C^*$.  Indeed what is expected is that the critical
point $\hat{\lambda}$ will open up into a pair of endpoints of a new
band.  By symmetry, the same thing will happen at the conjugate point
$\hat{\lambda}^*$, so the number of bands (and hence the genus)
increases by two.
\end{remark}

\section[Matching Genus Zero to Genus Two]{Matching the critical $G=0$ ansatz with a degenerate $G=2$ ansatz.}
By a degenerate $G=2$ ansatz\index{genus $G$ ansatz!degenerate, for
$G=2$}, we simply mean one for which two of the three complex
endpoints, say $\lambda_1$ and $\lambda_2$, are equal.  In general,
the three complex endpoints $\lambda_0$, $\lambda_1$, and $\lambda_2$
must satisfy the four real equations $M_p=0$ for $p=0,\dots,3$, along
with the vanishing condition $V_0=0$ and the reality condition
$R_1=0$.  Here we are exchanging the reality condition $R_0=0$ for the
additional moment condition $M_3=0$ as described in
\S\ref{sec:modulation}.  As pointed out in
\S\ref{sec:symmetry}, the moments are analytic and completely
symmetric functions of the endpoints, and are also analytic in $x$ and
$t$.  We begin our analysis of the feasibility of a degenerate genus
$G=2$ ansatz by evaluating the moments on a degenerate set of
endpoints satisfying $\lambda_0=\lambda_0^{\rm crit}$ and
$\lambda_1=\lambda_2=\hat{\lambda}$.  Let $M^{(0)}_p(x,t,\lambda_0^{\rm crit})$
for $p=0,1$ denote the moment functions for the $G=0$ ansatz.  Then we
have the following result.

\begin{lemma}
Evaluating the genus $G=2$ moments on a degenerate set of endpoints
yields the following four relations:
\begin{equation}
M_3(x,t,\lambda_0^{\rm crit},\hat{\lambda},\hat{\lambda})-2\Re(\hat{\lambda})
M_2(x,t,\lambda_0^{\rm crit},\hat{\lambda},\hat{\lambda})+|\hat{\lambda}|^2
M_1(x,t,\lambda_0^{\rm crit},
\hat{\lambda},\hat{\lambda}) = M_1^{(0)}(x,t,\lambda_0^{\rm crit})\,,
\label{eq:momentreduce1}
\end{equation}
\begin{equation}
M_2(x,t,\lambda_0^{\rm crit},\hat{\lambda},\hat{\lambda})-2\Re(\hat{\lambda})
M_1(x,t,\lambda_0^{\rm crit},\hat{\lambda},\hat{\lambda})+|\hat{\lambda}|^2
M_0(x,t,\lambda_0^{\rm crit},
\hat{\lambda},\hat{\lambda}) = M_0^{(0)}(x,t,\lambda_0^{\rm crit})\,,
\label{eq:momentreduce2}
\end{equation}
\begin{equation}
M_1(x,t,\lambda_0^{\rm crit},\hat{\lambda},\hat{\lambda})-\hat{\lambda}^*
M_0(x,t,\lambda_0^{\rm crit},\hat{\lambda},\hat{\lambda}) = 
\pi^2 Y^{(0)}(\hat{\lambda})\,,
\label{eq:momentreduce3}
\end{equation}
\begin{equation}
M_1(x,t,\lambda_0^{\rm crit},\hat{\lambda},\hat{\lambda})-\hat{\lambda}
M_0(x,t,\lambda_0^{\rm crit},\hat{\lambda},\hat{\lambda}) = 
\pi^2 Y^{(0)}(\hat{\lambda}^*)\,,
\label{eq:momentreduce4}
\end{equation}
where $Y^{(0)}(\lambda)$ is the function given by (\ref{eq:Ydef}) for
genus $G=0$ in terms of the single complex endpoint $\lambda_0^{\rm crit}$.
\label{lemma:momentreduce}
\end{lemma}

\begin{proof}
The evaluation of the moments on the degenerate solution is completely
straightforward using the formula (\ref{eq:momentrewrite}) that makes
clear the analytic dependence of the moments on the endpoints.  
Thus, for any $p\ge 0$, we find
\begin{equation}
\begin{array}{rcl}
M_{p+2}-2\Re(\hat{\lambda})M_{p+1}+|\hat{\lambda}|^2M_{p}&=&\displaystyle
-\frac{J}{2}\oint_L\frac{2ix+4i\eta t}{R(\eta)}\eta^p(\eta-\hat{\lambda})
(\eta-\hat{\lambda}^*)\,d\eta \\\\
&&\displaystyle\,\,+\,\,
\frac{1}{2}\int_{C_{I+}\cup C_{I-}}\frac{\pi i\rho^0(\eta)}{R(\eta)}
\eta^p(\eta-\hat{\lambda})(\eta-\hat{\lambda}^*)\,d\eta\\\\
&&\displaystyle\,\, +\,\,
\frac{1}{2}\int_{C_{I+}^*\cup C_{I-}^*}\frac{\pi i\rho^0(\eta^*)^*}{R(\eta)}
\eta^p(\eta-\hat{\lambda})(\eta-\hat{\lambda}^*)\,d\eta\,.
\end{array}
\end{equation}
Since for the degenerate set of endpoints
$R(\eta)=(\eta-\hat{\lambda})(\eta-\hat{\lambda})R^{(0)}(\eta)$, where
$R^{(0)}(\eta)$ is the square root function for genus $G=0$
corresponding to the single complex endpoint $\lambda_0^{\rm crit}$, we can again
use the formula (\ref{eq:momentrewrite}) to identify the right-hand
side as exactly $M_p^{(0)}(x,t,\lambda_0^{\rm crit})$, which in particular proves
(\ref{eq:momentreduce1}) and (\ref{eq:momentreduce2}).  Similarly,
using the formula (\ref{eq:momentrewrite}) for the moments along with
degenerate form of the square root function for
$\lambda_1=\lambda_2=\hat{\lambda}$ as described above, we find
\begin{equation}
\begin{array}{rcl}
M_1(x,t,\lambda_0^{\rm crit},\hat{\lambda},\hat{\lambda})-\hat{\lambda}^*
M_0(x,t,\lambda_0^{\rm crit},\hat{\lambda},\hat{\lambda}) &=&\displaystyle
-\frac{J}{2}\oint_L\frac{2ix+4i\eta
t}{(\eta-\hat{\lambda})R^{(0)}(\hat{\lambda})}\,d\eta \\\\
&&\displaystyle\,\,+\,\,\frac{1}{2}\int_{C_{I+}\cup C_{I-}}
\frac{\pi i\rho^0(\eta)}{(\eta-\hat{\lambda})R^{(0)}(\eta)}\,d\eta\\\\
&&\displaystyle\,\,+\,\,\frac{1}{2}\int_{C_{I+}^*\cup C_{I-}^*}
\frac{\pi i\rho^0(\eta^*)^*}{(\eta-\hat{\lambda})R^{(0)}(\eta)}\,d\eta\,.
\end{array}
\end{equation}
Evaluating the first integral exactly by residues, and identifying the
result with the formula (\ref{eq:Ydef}) proves
(\ref{eq:momentreduce3}).  Finally, (\ref{eq:momentreduce4}) is proved
either by repeating the above arguments, or by simply noting the
symmetry $Y^{(0)}(\lambda^*)=Y^{(0)}(\lambda)^*$ and using the reality
of the moments.
\end{proof}

\begin{corollary}
Suppose that for some $x$ and $t$ the endpoint $\lambda_0^{\rm crit}$ satisfies
the genus $G=0$ equations $M^{(0)}_0=0$ and $M^{(0)}_1=0$, and that
$\hat{\lambda}\neq \lambda_0^{\rm crit}$ 
is chosen so that $\Im(\hat{\lambda})\neq 0$
and
\begin{equation}
\frac{d\tilde{\phi}^{(0)\sigma}}{d\lambda}(\hat{\lambda})=0\,,
\end{equation}
where by $\tilde{\phi}^{(0)\sigma}(\lambda)$ we mean the function
$\tilde{\phi}^\sigma(\lambda)$ constructed for genus $G=0$ using the endpoint
$\lambda_0^{\rm crit}$.  Then, the genus $G=2$ moment equations
$M_p(x,t,\lambda_0^{\rm crit},\hat{\lambda},\hat{\lambda})=0$ are satisfied for
$p=0,1,2,3$.
\label{corollary:degeneratemoments}
\end{corollary}

\begin{proof}
As described in \S\ref{sec:symmetry}, it follows from (\ref{eq:plusside})
and the representation (\ref{eq:rhowithY}) of $\rho^\sigma(\lambda)$
that
\begin{equation}
\frac{d\tilde{\phi}^{(0)\sigma}}{d\lambda}(\lambda)=i\pi R^{(0)}(\lambda)
Y^{(0)}(\lambda)\,.
\label{eq:phiprimeYR}
\end{equation}
Since $\hat{\lambda}\neq\lambda_0^{\rm crit}$, we have
$R^{(0)}(\hat{\lambda})\neq 0$, and consequently the right-hand sides
of (\ref{eq:momentreduce3}) and (\ref{eq:momentreduce4}) vanish.
Since $M^{(0)}_0=M^{(0)}_1=0$, the right-hand sides of
(\ref{eq:momentreduce1}) and (\ref{eq:momentreduce2}) vanish as well.
The determinant of the left-hand side of these relations is
$-(\hat{\lambda}-\hat{\lambda}^*)\neq 0$, and the corollary is proved.
\end{proof}

In a degenerate situation, one can select a contour $C_I$ passing
through the point $\hat{\lambda}$ and one might consider solving the
scalar boundary-value problem for {\em both} genera, $G=0$ and $G=2$,
as described in \S\ref{sec:bvp}, defining two different functions
analytic in ${\mathbb C}\setminus (C_I\cup C_I^*)$.  Under the
conditions of Corollary~\ref{corollary:degeneratemoments} it follows
from Lemma~\ref{prop:moments} that the scalar boundary-value problems
for $G=0$ and $G=2$ both have unique solutions.  Let
$F^{(0)}(\lambda)$ denote the unique solution of the scalar
boundary-value problem for $G=0$ corresponding to the endpoint
$\lambda_0^{\rm crit}$ and the contour $C_I$, and let $F^{(2)}_{\rm
deg}(\lambda)$ denote the unique solution of the scalar boundary-value
problem for $G=2$ corresponding to the endpoints $\lambda_0^{\rm crit}$,
$\lambda_1=\hat{\lambda}$ and $\lambda_2=\hat{\lambda}$ and the same
contour $C_I$.  These two functions are given by apparently different
explicit formulae:
\begin{equation}
\begin{array}{rcl}
F^{(0)}(\lambda)&=&\displaystyle
\frac{R^{(0)}(\lambda)}{\pi i}\Bigg[
\int_{I_0^+\cup I_0^-}\frac{J(2ix+4i\eta t)}{(\lambda-\eta)R^{(0)}_+(\eta)}
\,d\eta \\\\
&&\displaystyle\,\,+\,\,
\int_{\Gamma_I\cap C_I}\frac{i\pi\rho^0(\eta)}{(\lambda-\eta)
R^{(0)}(\eta)}\,d\eta +\int_{\Gamma_I\cap C_I^*}\frac{i\pi\rho^0(\eta^*)^*}
{(\lambda-\eta)R^{(0)}(\eta)}\,d\eta\Bigg]\,,
\end{array}
\end{equation}
\begin{equation}
\begin{array}{rcl}
F^{(2)}_{\rm deg}(\lambda)&=&\displaystyle\frac{R^{(0)}(\lambda)}{\pi i}
(\lambda-\hat{\lambda})(\lambda-\hat{\lambda}^*)\Bigg[
\int_{I_0^+\cup I_0^-}\frac{J(2ix+4i\eta t)}{(\lambda-\eta)(\eta-\hat{\lambda})
(\eta-\hat{\lambda}^*)R^{(0)}_+(\eta)}\,d\eta \\\\
&&\displaystyle\,\,+\,\,\mbox{P.V.}\int_{\Gamma_I\cap C_I}
\frac{i\pi\rho^0(\eta)}
{(\lambda-\eta)(\eta-\hat{\lambda})(\eta-\hat{\lambda}^*)R^{(0)}(\eta)}
\,d\eta \\\\
&&\displaystyle\,\,+\,\,
\mbox{P.V.}\int_{\Gamma_I\cap C_I^*}
\frac{i\pi\rho^0(\eta^*)^*}
{(\lambda-\eta)(\eta-\hat{\lambda})(\eta-\hat{\lambda}^*)R^{(0)}(\eta)}
\,d\eta\Bigg] \,.
\end{array}
\label{eq:Fdeg}
\end{equation}
In both of these formulae, the paths $\Gamma_I\cap C_I$ and
$\Gamma_I\cap C_I^*$ pass respectively through $\eta=\hat{\lambda}$
and $\eta=\hat{\lambda}^*$.  To verify (\ref{eq:Fdeg}), one uses the
degeneration of the square root function
$R(\eta)=(\eta-\hat{\lambda})(\eta-\hat{\lambda}^*)R^{(0)}(\eta)$ and
notes that in the general genus $G=2$ expression for the function
$F(\lambda)$ given by $F(\lambda)=H(\lambda)R(\lambda)$ with
$H(\lambda)$ given by (\ref{eq:Hsolution}) for $G=2$, the paths of
integration $\Gamma_I\cap C_I$ and $\Gamma_I\cap C_I^*$ may be
replaced respectively by $(C_{I+}\cup C_{I-})/2$ and its conjugate
path, where $C_{I\pm}$ are taken to lie closer to $C_I$ than
$\lambda$.  This version of the formula allows one to evaluate
$F(\lambda)$ when $\lambda_1=\lambda_2=\hat{\lambda}$, after which the
contours may be collapsed to $C_I$ and $C_I^*$ resulting in the
principal value interpretation of the singular integrals via the
Plemelj formula.  In this sense, the formula (\ref{eq:Fdeg}) is the
limit of the nondegenerate genus $G=2$ formula as the endpoints
$\lambda_1$ and $\lambda_2$ coalesce at $\hat{\lambda}$. Despite
appearances, the differences in these two formulae are superficial.
We have the following.
\begin{lemma}
Assume the conditions of Corollary~\ref{corollary:degeneratemoments}.
Then,
\begin{equation}
F^{(0)}(\lambda)\equiv F^{(2)}_{\rm deg}(\lambda)\,,
\end{equation}
for all $\lambda\in {\mathbb C}\setminus (C_I\cup C_I^*)$.
\label{lemma:Fsame}
\end{lemma}

\begin{proof}
This essentially follows from the uniqueness result for the scalar
boundary-value problem for $G=0$ described in
Lemma~\ref{lemma:uniqueF}.  Indeed, both functions satisfy the same
decay conditions at infinity, and by the explicit formulae are
analytic in ${\mathbb C}\setminus (C_I\cup C_I^*)$ with boundary
values that are H\"older continuous with exponent $1/2$.  The fact
that both functions satisfy the same boundary conditions almost
everywhere on $C_I\cup C_I^*$ follows from taking the limit
$\lambda_1\rightarrow\hat{\lambda}$ and
$\lambda_2\rightarrow\hat{\lambda}$ in the scalar boundary-value
problem for genus $G=2$.  Therefore both functions satisfy the scalar
boundary-value problem for $G=0$ and are equal by
Lemma~\ref{lemma:uniqueF}.
\end{proof}

\begin{remark}
The arguments in Lemma~\ref{lemma:Fsame}
make no
explicit reference to the genus.  Thus the
argument actually shows that whenever a band
closes up, then the function $F(\lambda)$ reduces to the
solution of the scalar boundary-value problem
for genus $G-2$.
\end{remark}

With the help of this result, we may now study the functions $V_0$ and
$R_1$ on a degenerate endpoint configuration for genus $G=2$.  

\begin{lemma}
Assume the conditions of Lemma~\ref{lemma:Fsame}.  The reality
condition $R_1=0$ is automatically satisfied by any degenerate genus
$G=2$ configuration with $\lambda_1=\lambda_2=\hat{\lambda}$.
\label{lemma:R1limit}
\end{lemma}

\begin{proof}
From Lemma~\ref{lemma:Fsame}, the candidate density function
$\rho^\sigma(\eta)$ obtained for the degenerate genus $G=2$
configuration from $F^{(2)}_{\rm deg}(\lambda)$ agrees with the
nondegenerate genus $G=0$ candidate density function.  Since this
function is H\"older continuous it is bounded.  This implies that
in the degenerate limit, $\rho^\sigma(\eta)$ remains uniformly bounded
on $I_1^+$, the path of
integration from $\lambda_1$ to $\lambda_2$.
Since $\lambda_1$ and $\lambda_2$ both converge to $\hat{\lambda}$
in the degenerate limit, the 
result follows from the
definition of the function $R_1$ by the 
formula (\ref{eq:realitycond}) as an integral 
along the shrinking band $I_1^+$.
\end{proof}

\begin{lemma}
Assume the conditions of Lemma~\ref{lemma:Fsame}.  Then for any degenerate
genus $G=2$ configuration with $\lambda_1=\lambda_2=\hat{\lambda}$,
we have
\begin{equation}
V_0 = \Re(\tilde{\phi}^{(0)\sigma}(\hat{\lambda}))\,,
\end{equation}
where $\tilde{\phi}^{(0)\sigma}(\lambda)$ corresponds to the $G=0$
candidate density function with the single complex endpoint $\lambda_0^{\rm crit}$.
\label{lemma:V0limit}
\end{lemma}

\begin{proof}
By the definition (\ref{eq:vanishing}), $V_0$ is the real part of the
integral of $d\tilde{\phi}^\sigma/d\lambda$ from $\lambda_0$ to
$\lambda_1$, where the analytic function
$\tilde{\phi}^\sigma(\lambda)$ corresponds to the endpoints
$\lambda_0$, $\lambda_1$ and $\lambda_2$.  By construction,
$\Re(\tilde{\phi}^\sigma(\lambda_0))=0$, so equivalently we have
$V_0=\Re(\tilde{\phi}^\sigma(\lambda_1))$.  According to
Lemma~\ref{lemma:Fsame}, the $G=2$ function
$\tilde{\phi}^\sigma(\lambda)$ agrees with
$\tilde{\phi}^{(0)\sigma}(\lambda)$ when
$\lambda_1=\lambda_2=\hat{\lambda}$ since they are both derived from
the same unique solution of the scalar boundary value for $G=0$ with
$\lambda_0=\lambda_0^{\rm crit}$.  Evaluating for $\lambda_1=\hat{\lambda}$
completes the proof.
\end{proof}

\begin{remark} Both of the results contained in Lemma~\ref{lemma:R1limit}
and Lemma~\ref{lemma:V0limit} will be strengthened shortly when we provide
more detailed asymptotics near the degenerate configuration.
\end{remark}

Combining the results of this section, we have proved the following.
\begin{theorem}
Suppose that $x=x_{\rm crit}$, $t=t_{\rm crit}$, and $\lambda_{0}^{(0)}$ and
$\hat{\lambda}$ are such that the two $G=0$ moment conditions
$M^{(0)}_0=0$ and $M^{(0)}_1=0$ hold, and the three real conditions
contained in (\ref{eq:gapfail}) hold true at $\lambda=\hat{\lambda}$.  Then
holding $x_{\rm crit}$ and $t_{\rm crit}$ fixed, the complex endpoint
configuration $\lambda_0=\lambda_{0}^{(0)}$, $\lambda_1=\hat{\lambda}$ and
$\lambda_2=\hat{\lambda}$ represents a degenerate solution of the
equations $M_p=0$ for $p=0,\dots,3$ along with $R_1=0$ and $V_0=0$.
Moreover, if we suppose that the genus $G=0$ ansatz is successful in
the sense that the band $I_0^+$ exists connecting the origin to
$\lambda_0^{\rm crit}$ on which the differential $\rho^\sigma(\eta)\,d\eta$ is
negative real and the corresponding gap contour exists passing through
$\hat{\lambda}$ such that $\Re(\tilde{\phi}^{(0)\sigma})\le 0$ with
inequality being strict except at the endpoints and at
$\hat{\lambda}$, then the degenerate genus $G=2$ ansatz is successful
in the same sense, with exactly the same contour.
\label{theorem:degenerate}
\end{theorem}

\begin{proof}
The fact that the degenerate triple of endpoints satisfies the $G=2$
endpoint equations follows from the chain of results already presented
in this section.  It remains to verify the final claim: that the
inequalities persist under reinterpretation of the $G=0$ configuration
subject to the additional conditions (\ref{eq:gapfail}) at
$\lambda=\hat{\lambda}$ as a degenerate $G=2$ configuration.  But this
follows from Lemma~\ref{lemma:Fsame}, which implies that the functions
$\rho^\sigma$ and $\tilde{\phi}^\sigma$ are exactly the same in both cases.
\end{proof}

\section[Perturbation Theory at the Boundary]{Perturbing the degenerate $G=2$ ansatz.  Opening the band $I_1^+$
by varying $x$ near $x_{\rm crit}$.}

Let $t=t_{\rm crit}$ be fixed.  We now want to consider the
possibility that the $G=2$ ansatz exists ({\em i.e.} the endpoint
equations can be solved) for $x$ near $x_{\rm crit}$ but in the region
of the $(x,t)$-plane beyond the primary caustic where the inequalities
fail for the $G=0$ ansatz due to the pinching-off of the region where
$\Re(\tilde{\phi}^{(0)\sigma})<0$ at the point $\hat{\lambda}$.
Unfortunately, a direct application of the implicit function theorem
fails to establish existence, because it can be shown using the
explicit formula given in \S\ref{sec:modulation} that the
corresponding Jacobian determinant vanishes when evaluated on the
degenerate $G=2$ solution.  In a sense, this is not surprising since
we know from \S\ref{sec:symmetry} that the endpoints are only
determined by the constraint equations up to permutation, and
consequently the double-point $\lambda_1=\lambda_2=\hat{\lambda}$
cannot be unfolded uniquely.  However, the difficulties also run
deeper, with the appearance of logarithms in the perturbation
expansion arising from the multivaluedness (monodromy) of the function
$V_0$ described in \S\ref{sec:symmetry}.

In this section, we begin with the assumption of the existence of the
degenerate $G=2$ ansatz.  Therefore, we assume that for $x=x_{\rm
crit}$ and $t=t_{\rm crit}$ the single complex endpoint $\lambda_0^{\rm crit}$
satisfies the two real equations $M_0^{(0)}=M_1^{(0)}=0$, and for some
non-real $\hat{\lambda}\neq\lambda_0^{\rm crit}$ in ${\mathbb H}$, we have
\begin{equation}
\frac{d\tilde{\phi}^{(0)\sigma}}{d\lambda}(\hat{\lambda})=0\,,
\hspace{0.3 in}
\Re(\tilde{\phi}^{(0)\sigma}(\hat{\lambda}))=0\,.
\label{eq:gapfailagain}
\end{equation}
We further assume the following to rule out higher-order degeneracy:
\begin{enumerate}
\item
The $G=0$ endpoint equations $M_0^{(0)}=0$ and $M_1^{(0)}=0$ can be
solved for the single complex endpoint as a function of $(x,t)$ in a
neighborhood of $(x_{\rm crit},t_{\rm crit})$.
\item
The critical point is simple, so that
\begin{equation}
\frac{d^2\tilde{\phi}^{(0)\sigma}}{d\lambda^2}(\hat{\lambda})\neq 0\,.
\label{eq:nothigher}
\end{equation}
\end{enumerate}
Under these conditions, we can reduce the size of the problem somewhat.

\begin{lemma}
The four $G=2$ moment conditions $M_0=M_1=M_2=M_3=0$ can be solved for
$\lambda_0$, $\lambda_0^*$, $\lambda_2$, and $\lambda_2^*$ as analytic
functions of $x$, $\lambda_1$, and $\lambda_1^*$ in a complex
neighborhood of the degenerate solution.  The linear terms in the
implicitly defined functions are:
\begin{equation}
\begin{array}{rcl}
\lambda_0(x,\lambda_1,\lambda_1^*)&=&\displaystyle
\lambda_0^{\rm crit} +
J\pi\left(\frac{\partial M^{(0)}_0}{\partial \lambda_0}\Bigg|_{\rm crit}
\right)^{-1}
(x-x_{\rm crit}) +\dots\,,\\\\
\lambda_0^*(x,\lambda_1,\lambda_1^*)&=&
\displaystyle
\lambda_0^{{\rm crit},*} +
J\pi 
\left(\frac{\partial M^{(0)}_0}{\partial \lambda_0^*}\Bigg|_{\rm crit}
\right)^{-1}
(x-x_{\rm crit}) +\dots\,,\\\\
\lambda_2(x,\lambda_1,\lambda_1^*)&=&\displaystyle
\hat{\lambda}-2iJ 
\left(
\frac{d^2\tilde{\phi}^{(0)\sigma}}{d\lambda^2}(\hat{\lambda})\right)^{-1}
\frac{\lambda_0^{\rm crit}+\lambda_0^{{\rm crit},*}-2\hat{\lambda}}
{R^{(0)}(\hat{\lambda})}
(x-x_{\rm crit})
\\\\&&\,\,-\,\,(\lambda_1-\hat{\lambda}) +\dots\,,\\\\
\lambda_2^*(x,\lambda_1,\lambda_1^*)&=&\displaystyle
\hat{\lambda}^*
+2iJ
\left(
\frac{d^2\tilde{\phi}^{(0)\sigma}}{d\lambda^2}(\hat{\lambda})^*\right)^{-1}
\frac{\lambda_0^{\rm crit}+\lambda_0^{{\rm crit},*}-2\hat{\lambda}^*}
{R^{(0)}(\hat{\lambda})^*}
(x-x_{\rm crit})\\\\
&&\,\,-\,\,(\lambda_1^*-\hat{\lambda}^*) +\dots\,,
\end{array}
\label{eq:linearterms}
\end{equation}
where the derivatives of the $G=0$ moment $M^{(0)}_0$ are evaluated on the
critical $G=0$ ansatz.  The above coefficients that do not vanish identically 
are finite and strictly nonzero by our assumptions.
\end{lemma}

\begin{remark}
Note that in (\ref{eq:linearterms}) we have written down {\em all} of
the linear terms.  So in particular, the dependence of $\lambda_0$ and 
$\lambda_0^*$ on $\lambda_1-\hat{\lambda}$ and $\lambda_1^*-\hat{\lambda}^*$
is higher order.
\end{remark}

\begin{proof}
We begin the proof by computing the partial derivatives of the $G=2$
moment $M_0$ with respect to the endpoints, as well as the partial
derivatives of the first four moments with respect to $x$, and
evaluating them on the degenerate solution.

First, from the formula (\ref{eq:momentrewrite}), one finds that for the
moment $M_p$ corresponding to a
general genus $G$ ansatz
\begin{equation}
\begin{array}{rcl}
\displaystyle
\frac{\partial M_p}{\partial\lambda_k} &=&\displaystyle 
-\frac{J}{4}\oint_L\frac{2ix+4i\eta t}{(\eta-\lambda_k)R(\eta)}\eta^p\,d\eta +
\frac{1}{4}\int_{C_{I+}\cup C_{I-}}\frac{\pi i\rho^0(\eta)\eta^p\,d\eta}
{(\eta-\lambda_k)
R(\eta)} +
\frac{1}{4}\int_{C_{I+}^*\cup C_{I-}^*}\frac{\pi i\rho^0(\eta^*)^*\eta^p
\,d\eta}
{(\eta-
\lambda_k)R(\eta)}\,,\\\\
\displaystyle
\frac{\partial M_p}{\partial\lambda_k^*} &=&\displaystyle 
-\frac{J}{4}\oint_L\frac{2ix+4i\eta t}{(\eta-\lambda_k^*)R(\eta)}\eta^p
\,d\eta +
\frac{1}{4}\int_{C_{I+}\cup C_{I-}}\frac{\pi i\rho^0(\eta)\eta^p\,d\eta}
{(\eta-\lambda_k^*)
R(\eta)} +
\frac{1}{4}\int_{C_{I+}^*\cup C_{I-}^*}\frac{\pi i\rho^0(\eta^*)^*\eta^p
\,d\eta}
{(\eta-
\lambda_k^*)R(\eta)}\,.
\end{array}
\label{eq:explicitmomentderivs}
\end{equation}
Recall that the contours $C_{I+}$ and $C_{I-}$ are bounded away from
all of the endpoints $\lambda_k$ (see Figure~\ref{fig:CIpm}).  From
this and the relation
$R(\eta)=(\eta-\hat{\lambda})(\eta-\hat{\lambda}^*)R^{(0)}(\eta)$
holding for the degenerate $G=2$ configuration, we find by taking
linear combinations,
\begin{equation}
\begin{array}{rcl}
\displaystyle
\frac{\partial}{\partial\lambda_0}(M_2-2\Re(\hat{\lambda})M_1+
|\hat{\lambda}|^2M_0)\Bigg|_{\rm deg} &=&\displaystyle
\frac{\partial M^{(0)}_0}{\partial\lambda_0}\Bigg|_{\rm crit}\,,\\\\
\displaystyle
\frac{\partial}{\partial\lambda_0^*}(M_2-2\Re(\hat{\lambda})M_1+
|\hat{\lambda}|^2M_0)\Bigg|_{\rm deg} &=&\displaystyle
\frac{\partial M^{(0)}_0}{\partial\lambda_0^*}\Bigg|_{\rm crit}\,,
\end{array}
\end{equation}
where on the left-hand side the derivatives of genus $G=2$ moments are
evaluated on the degenerate configuration and on the right-hand side
the derivatives of $M^{(0)}_0$ are evaluated on the corresponding
critical $G=0$ configuration.  Using the relations developed in
\S\ref{sec:modulation} that allow derivatives 
of higher moments to be expressed in terms of derivatives of $M_0$, 
these relations imply
\begin{equation}
\begin{array}{rcl}
\displaystyle\frac{\partial M_0}{\partial\lambda_0}\Bigg|_{\rm deg}&=&
\displaystyle \frac{1}{(\lambda_0^{\rm crit}-\hat{\lambda})(\lambda_0^{\rm crit}-\hat{\lambda}^*)}\frac{\partial M_0^{(0)}}{\partial \lambda_0}\Bigg|_{\rm crit}\,,\\\\
\displaystyle\frac{\partial M_0}{\partial\lambda_0^*}\Bigg|_{\rm deg}&=&
\displaystyle \frac{1}{(\lambda_0^{{\rm crit},*}-\hat{\lambda})
(\lambda_0^{{\rm crit},*}-\hat{\lambda}^*)}
\frac{\partial M_0^{(0)}}{\partial \lambda_0^*}\Bigg|_{\rm crit}\,.
\end{array}
\end{equation}

Second, from the permutation invariance of the moments ({\em cf.}
\S\ref{sec:symmetry}) a chain rule calculation shows that
\begin{equation}
\frac{\partial M_p}{\partial\lambda_1}\Bigg|_{\rm deg} =
\frac{\partial M_p}{\partial\lambda_2}\Bigg|_{\rm deg} =
\frac{1}{2}\frac{\partial}{\partial\hat{\lambda}}\left(M_p\Bigg|_{\rm deg}
\right)\,,
\end{equation}
where on the right-hand side $M_p$ is first evaluated on the
degenerate endpoint configuration, and then differentiated with
respect to $\hat{\lambda}$.  Therefore, using
Lemma~\ref{lemma:momentreduce}, we find
\begin{equation}
\left(\frac{\partial}{\partial\lambda_{1,2}}(M_1-\hat{\lambda}^*M_0)
\right)\Bigg|_{\rm deg} = \frac{\pi^2}{2}\frac{dY^{(0)}(\hat{\lambda})}
{d\hat{\lambda}} = 
\frac{-i\pi}{2R^{(0)}(\hat{\lambda})}\frac{d^2\tilde{\phi}^{(0)\sigma}}
{d\lambda^2}(\hat{\lambda})\,,
\end{equation}
where we have simplified the result with the help of
(\ref{eq:phiprimeYR}) and (\ref{eq:gapfailagain}).  At the same time,
the left-hand side can be expressed in terms of derivatives of $M_0$
by the reasoning of \S\ref{sec:modulation}, yielding
\begin{equation}
\left(\frac{\partial}{\partial\lambda_{1,2}}(M_1-\hat{\lambda}^*M_0)
\right)\Bigg|_{\rm deg} = (\hat{\lambda}-\hat{\lambda}^*)\frac{\partial
M_0}{\partial\lambda_{1,2}}\Bigg|_{\rm deg}\,.
\end{equation}
Putting these results together along with a similar calculation involving
derivatives with respect to $\lambda^*_{1,2}$ and the linear combination
$M_1-\hat{\lambda}M_0$, one finds
\begin{equation}
\frac{\partial M_0}{\partial \lambda_{1,2}}\Bigg|_{\rm deg} = 
\frac{-i\pi}{2(\hat{\lambda}-\hat{\lambda}^*)R^{(0)}(\hat{\lambda})}
\frac{d^2\tilde{\phi}^{(0)\sigma}}{d\lambda^2}(\hat{\lambda})=
\left(\frac{\partial M_0}{\partial\lambda_{1,2}^*}\Bigg|_{\rm deg}\right)^*\,.
\label{eq:M0l1l2}
\end{equation}
Note that this identity, together with condition (\ref{eq:nothigher}),
implies that 
\begin{equation}
\frac{\partial M_0}{\partial\lambda_{1,2}}\Bigg|_{\rm deg}\neq 0\,.
\end{equation}

Third, to calculate the partial derivatives of the moments $M_p$ with respect
to $x$, we use (\ref{eq:momentrewrite}) and the expansion of $R(\eta)=(\eta-\hat{\lambda})(\eta-\hat{\lambda}^*)R^{(0)}(\eta)$ for large $\eta$ to find
\begin{equation}
\begin{array}{rcl}
\displaystyle
\frac{\partial M_p}{\partial x}\Bigg|_{\rm deg} &=&\displaystyle
 iJ\oint_L \eta^{p-3} \left(1+\frac{
2\hat{\lambda}+2\hat{\lambda}^*+\lambda_0^{\rm crit}+\lambda_0^{{\rm crit},*}}{2\eta}+\dots\right)\,d\eta\\\\& =&\displaystyle 
\left\{\begin{array}{ll}
0\,, & p=0,1\,,\\
-2\pi J\,, & p=2\,,\\
-\pi J(2\hat{\lambda}+2\hat{\lambda}^*+\lambda_0^{\rm crit}+\lambda_0^{{\rm crit},*})\,, &p=3\,.\end{array}\right.
\end{array}
\end{equation}

Now, according to the calculations presented in
\S\ref{sec:modulation}, the relevant Jacobian matrix for eliminating $\lambda_0$, $\lambda_0^*$, $\lambda_2$, and $\lambda_2^*$ is a product of a
Vandermonde matrix and a diagonal matrix:
\begin{equation}
\frac{\partial(M_0,M_1,M_2,M_3)}{\partial(\lambda_0,\lambda_0^*,\lambda_2,
\lambda_2^*)}=
\left[\begin{array}{cccc}
1 & 1 & 1 & 1\\\\
\lambda_0 & \lambda_0^* & \lambda_2 & \lambda_2^*\\\\
\lambda_0^2 & \lambda_0^{*2} & \lambda_2^2 & \lambda_2^{*2}\\\\
\lambda_0^3 & \lambda_0^{*3} & \lambda_2^3 & \lambda_2^{*3}
\end{array}\right]\cdot{\rm diag}\,\left(
\frac{\partial M_0}{\partial \lambda_0},
\frac{\partial M_0}{\partial\lambda_0^*},
\frac{\partial M_0}{\partial\lambda_2},
\frac{\partial M_0}{\partial\lambda_2^*}\right)\,.
\end{equation}
Evaluating the Jacobian determinant on the degenerate configuration,
we find that the Vandermonde determinant factor is nonzero because
$\lambda_0^{\rm crit}$, $\lambda_0^{{\rm crit},*}$, $\hat{\lambda}$,
and $\hat{\lambda}^*$ are all distinct, and the diagonal determinant
is nonzero according to the above calculations and the assumption
(\ref{eq:nothigher}).  It follows from the implicit function theorem
that we can solve for $\lambda_0$, $\lambda_0^*$, $\lambda_2$, and
$\lambda_2^*$.  The partial derivatives of these implicitly-defined
functions with respect to the remaining independent variables $x$,
$\lambda_1$, and $\lambda_1^*$, are then obtained by Cramer's rule,
using the expressions for the derivatives of the moments obtained above.
This yields the expressions (\ref{eq:linearterms}) and completes 
the proof.
\end{proof}

We have therefore reduced our problem to the study of the two
equations $R_1=0$ and $V_0=0$ involving $x$, $\lambda_1$, and
$\lambda_1^*$.  As we will no longer use analyticity properties in any
essential way, we work from now on in the real subspace where
$\lambda_1^*$ is the complex conjugate of $\lambda_1$, and seek a
solution near $\lambda_1=\hat{\lambda}$ for $x$ near $x_{\rm crit}$.
Note that it is clear that the functions $R_1$ and $V_0$ can be
defined for $\lambda_1$ near $\hat{\lambda}$; the question is only
about their local behavior.  Let $\epsilon\ge 0$ be a small parameter.
Dominant balance considerations ultimately justified by the proof of
Lemma~\ref{lemma:converge} to follow below suggest the following
scalings of the remaining variables:
\begin{equation}
\lambda_1=\hat{\lambda}+\epsilon r e^{i\theta}\,,\hspace{0.3 in}
\lambda_1^*=\hat{\lambda}^*+\epsilon r e^{-i\theta}\,,\hspace{0.3 in}
x=x_{\rm crit}+\epsilon^2{\rm Log}(\epsilon^{-1})\cdot \chi\,.
\label{eq:scalings}
\end{equation}
Our new variables will therefore be $r>0$, $\theta$, and $\chi$, all real.
Define real parameters $P>0$, $\alpha$, and $c$ by
\begin{equation}
Pe^{i\alpha}
:=\frac{i}{2}\frac{d^2\tilde{\phi}^{(0)\sigma}}{d\lambda^2}(\hat{\lambda})
\,,\hspace{0.3 in}
c:=2J\Im(R^{(0)}(\hat{\lambda}))\,,
\end{equation}
and consider the functions
\begin{equation}
\begin{array}{rcl}
\displaystyle
R_1^{\rm model}(r,\theta,\chi)&:=&\displaystyle
Pr^2\left[
\sin(\alpha)\cdot\sin(2\theta)-\cos(\alpha)\cdot\cos(2\theta)\right]\,,\\\\
\displaystyle V_0^{\rm model}(r,\theta,\chi)&:=&\displaystyle
c\chi+Pr^2\left[\cos(\alpha)\cdot\sin(2\theta)+
\sin(\alpha)\cdot\cos(2\theta)\right]\,.
\end{array}
\end{equation}

\begin{lemma}
Let $r$, $\theta$, and $\chi$ be fixed.  Then, as $\epsilon\downarrow 0$,
\begin{equation}
\epsilon^{-2}R_1\rightarrow R_1^{\rm model}(r,\theta,\chi)\,,\hspace{0.3 in}
\left[\epsilon^2{\rm Log}(\epsilon^{-1})\right]^{-1}V_0\rightarrow
V_0^{\rm model}(r,\theta,\chi)\,.
\end{equation}
The partial derivatives with respect to $r$ and $\theta$ also converge
pointwise:
\begin{equation}
\begin{array}{rclrcl}
\displaystyle
\frac{\partial}{\partial r}\left(\epsilon^{-2}R_1\right)&\rightarrow &
\displaystyle
\frac{\partial R^{\rm model}_1}{\partial r}(r,\theta,\chi)\,,&
\displaystyle
\frac{\partial}{\partial r}\left(\left[\epsilon^2{\rm Log}(\epsilon^{-1})\right]^{-1}V_0\right)&\rightarrow &
\displaystyle
\frac{\partial V^{\rm model}_0}{\partial r}(r,\theta,\chi)\,,
\\\\
\displaystyle
\frac{\partial}{\partial \theta}\left(\epsilon^{-2}R_1\right)&\rightarrow &
\displaystyle
\frac{\partial R^{\rm model}_1}{\partial \theta}(r,\theta,\chi)\,,
&
\displaystyle
\frac{\partial}{\partial \theta}\left(\left[\epsilon^2{\rm Log}(\epsilon^{-1})\right]^{-1}V_0\right)&\rightarrow &
\displaystyle
\frac{\partial V^{\rm model}_0}{\partial \theta}(r,\theta,\chi)\,.
\end{array}
\label{eq:derivsconverge}
\end{equation}
For fixed $\chi$, the convergence is uniform in any finite annulus
$0<r_{\rm min}\le r\le r_{\rm max}<\infty$.
\label{lemma:converge}
\end{lemma}

\begin{proof}
Using the relation (\ref{eq:rhowithY}) and rewriting the integral over
$I_1^\pm$ as half of a loop integral around the band, we find
\begin{equation}
\epsilon^{-2}R_1=-\frac{1}{4i\epsilon^2}\oint_{L(\hat{\lambda})}
R(\eta)Y(\eta)\,d\eta -\frac{1}{4i\epsilon^2}\oint_{L(\hat{\lambda}^*)}
R(\eta)Y(\eta)\,d\eta\,,
\label{eq:e2R1}
\end{equation}
where $L(\lambda)$ denotes a sufficiently small positively oriented contour
surrounding $\lambda$ that is held fixed as $\epsilon$ tends to zero.
On both paths of integration, the following approximation for $R(\eta)$
holds uniformly:
\begin{equation}
\begin{array}{rcl}
R(\eta)&=&\displaystyle
\frac{1}{2}R^{(0)}(\eta)\cdot(\eta-\hat{\lambda})(\eta-\hat{\lambda}^*)\\\\
&&\,\,\times\,\,\displaystyle
\Bigg[2-\epsilon^2{\rm Log}(\epsilon^{-1})\Bigg(
\frac{\partial\lambda_0}{\partial x}\Bigg|_{\rm
deg}\cdot\frac{\chi}{\eta-\lambda_0^{\rm crit}} +
\frac{\partial\lambda_0^*}{\partial x}\Bigg|_{\rm deg}
\cdot\frac{\chi}{\eta-\lambda_0^{{\rm crit},*}} \\\\
&&\displaystyle\hspace{0.5 in}+\,\,
\frac{\partial\lambda_2}{\partial x}\Bigg|_{\rm deg}
\cdot\frac{\chi}{\eta-\hat{\lambda}} +
\frac{\partial\lambda_2^*}{\partial x}\Bigg|_{\rm deg}
\cdot\frac{\chi}{\eta-\hat{\lambda}^*}\Bigg)\\\\
&&\displaystyle\hspace{0.3 in}
-\,\,\epsilon^2\left(\frac{r^2e^{2i\theta}}{(\eta-\hat{\lambda})^2}+
\frac{r^2e^{-2i\theta}}{(\eta-\hat{\lambda}^*)^2}\right)
+\bo(\epsilon^3{\rm Log}(\epsilon^{-1}))\Bigg]\,,
\end{array}
\label{eq:rootexpansion}
\end{equation}
where the partial derivatives with respect to $x$ may be obtained
explicitly from (\ref{eq:linearterms}) if desired.  But in fact, they
are not necessary for the present calculation; since $R^{(0)}(\eta)$
and $Y(\eta)$ are analytic and uniformly bounded inside each loop as
$\epsilon$ tends to zero ($L(\hat{\lambda})$ and $L(\hat{\lambda}^*)$
taken small enough to exclude $\lambda_0^{\rm crit}$ and
$\lambda_0^{{\rm crit},*}$ for all sufficiently small $\epsilon$), we
find from the residue theorem that
\begin{equation}
\epsilon^{-2}R_1 = \frac{\pi r^2e^{2i\theta}}{4}
(\hat{\lambda}-\hat{\lambda}^*)
R^{(0)}(\hat{\lambda})Y(\hat{\lambda}) + \frac{\pi r^2e^{-2i\theta}}{4}
(\hat{\lambda}^*-\hat{\lambda})R^{(0)}(\hat{\lambda}^*)Y(\hat{\lambda}^*)
+\bo(\epsilon{\rm Log}(\epsilon^{-1}))\,.
\end{equation}
Now, as $\epsilon\downarrow 0$, by analytic dependence on the
endpoints, $Y(\hat{\lambda})$ and $Y(\hat{\lambda}^*)$ converge to the
corresponding quantities evaluated on the degenerate $G=2$ endpoint
configuration.  That is, from formula (\ref{eq:Ydef}) with
$\lambda=\hat{\lambda}$ and using the degenerate configuration,
\begin{equation}
\begin{array}{rcl}
Y(\hat{\lambda})&\rightarrow &\displaystyle
\frac{-1}{2\pi i}\int_{C_{I+}\cup C_{I-}}
\frac{\rho^0(\eta)\,d\eta}{(\eta-\hat{\lambda})^2(\eta-\hat{\lambda}^*)R^{(0)}(\eta)} +\frac{-1}{2\pi i}\int_{C_{I+}^*\cup C_{I-}^*}\frac{\rho^0(\eta^*)^*\,d\eta}{(\eta-\hat{\lambda})^2(\eta-\hat{\lambda}^*)R^{(0)}(\eta)}\,,\\\\
Y(\hat{\lambda}^*)&\rightarrow &\displaystyle
\frac{-1}{2\pi i}\int_{C_{I+}\cup C_{I-}}
\frac{\rho^0(\eta)\,d\eta}{(\eta-\hat{\lambda})(\eta-\hat{\lambda}^*)^2R^{(0)}(\eta)} +\frac{-1}{2\pi i}\int_{C_{I+}^*\cup C_{I-}^*}\frac{\rho^0(\eta^*)^*\,d\eta}{(\eta-\hat{\lambda})(\eta-\hat{\lambda}^*)^2R^{(0)}(\eta)}\,.
\end{array}
\end{equation}
Comparing with (\ref{eq:explicitmomentderivs}) for $k=1$ or $k=2$ evaluated
on the degenerate configuration, we find
\begin{equation}
Y(\hat{\lambda})\rightarrow  \frac{2}{\pi^2}\frac{\partial M_0}
{\partial\lambda_{1,2}}\Bigg|_{\rm deg}\,,\hspace{0.3 in}
Y(\hat{\lambda}^*)\rightarrow \frac{2}{\pi^2}\frac{\partial M_0}
{\partial\lambda^*_{1,2}}\Bigg|_{\rm deg}\,.
\end{equation}
Finally, using the explicit representation of these derivatives of $M_0$
given in (\ref{eq:M0l1l2}), we find
\begin{equation}
\begin{array}{rcl}
\epsilon^{-2}R_1 &=&\displaystyle
 -\frac{i}{4}\frac{d^2\tilde{\phi}^{(0)\sigma}}{d\lambda^2}(\hat{\lambda})
\cdot r^2e^{2i\theta} + \mbox{complex conjugate} + o(1)\\\\
&=&R^{\rm model}_1(r,\theta,\chi)+o(1)\,,
\end{array}
\end{equation}
where we have used the definition of the parameters $P$ and $\alpha$.
Passing to the limit $\epsilon\downarrow 0$ then completes the first
part of the proof.  

To establish the convergence of the partial derivatives with respect
to $r$ and $\theta$, 
we note that since $\lambda_0$, $\lambda_0^*$, $\lambda_2$ and
$\lambda_2^*$ depend differentially on $\lambda_1$, $\lambda_1^*$, and
$x$, and since the functions $R$ and $Y$ depend analytically on
$\lambda_0$, $\lambda_0^*$, $\lambda_2$, $\lambda_2^*$, and $x$, it
follows from formula (\ref{eq:e2R1}) that $\epsilon^{-2}R_1$ is
differentiable uniformly in $\epsilon$.  This yields the convergence
of derivatives of $\epsilon^{-2}R_1$ expressed in
(\ref{eq:derivsconverge}).

Now we carry out a similar analysis of the function $V_0$, which is a
bit more complicated due to a logarithmic divergence.  First, we show
that $V_0$ is differentiable with respect to $x$ at $x=x_{\rm crit}$,
$\lambda_1=\hat{\lambda}$ and $\lambda_1^*=\hat{\lambda}^*$.  Using
the chain rule and the expressions for the partial derivatives of
$V_0$ with respect to endpoints obtained in \S\ref{sec:modulation}, we
find
\begin{equation}
\begin{array}{rcl}
\displaystyle
\frac{\partial V_0}{\partial x}&=&\displaystyle
\frac{1}{2\pi i}\int_{\Gamma_1^+\cup
\Gamma_1^-}\frac{(\eta-\lambda_1)(\eta-\lambda_1^*)}{R(\eta)} 
\\\\
&&\displaystyle\hspace{0.2 in}\times\,\,
\Bigg(\frac{\partial M_0}{\partial\lambda_0}\frac{\partial\lambda_0}
{\partial x} (\eta-\lambda_0^*)(\eta-\lambda_2)(\eta-\lambda_2^*) +
\frac{\partial M_0}{\partial\lambda_0^*}\frac{\partial\lambda_0^*}{\partial x}
(\eta-\lambda_0)(\eta-\lambda_2)(\eta-\lambda_2^*)\\\\
&&\displaystyle\hspace{0.4 in}+\,\,
\frac{\partial M_0}{\partial\lambda_2}\frac{\partial\lambda_2}
{\partial x}(\eta-\lambda_0)(\eta-\lambda_0^*)(\eta-\lambda_2^*)+
\frac{\partial M_0}{\partial\lambda_2^*}\frac{\partial\lambda_2^*}
{\partial x}(\eta-\lambda_0)(\eta-\lambda_0^*)(\eta-\lambda_2)\Bigg)\,d\eta\,.
\end{array}
\end{equation}
One can solve for the products $(\partial M_0/\partial
\lambda_k)(\partial \lambda_k/\partial x)$ explicitly (it is an exact
Vandermonde system), and evaluate the result on the degenerate
configuration, yielding
\begin{equation}
\frac{\partial V_0}{\partial x}\Bigg|_{\rm deg}=\frac{1}{2\pi i}\left(\int_{\lambda_0^{\rm crit}}^{\hat{\lambda}} +\int_{\hat{\lambda}^*}^{\lambda_0^{{\rm crit},*}}\right)
\frac{J\pi (2\eta-\lambda_0^{\rm crit}-\lambda_0^{{\rm crit},*})}
{R^{(0)}(\eta)}\,d\eta = 2J\Im(R^{(0)}(\hat{\lambda}))\,.
\end{equation}
It follows that as $\epsilon\downarrow 0$,
\begin{equation}
\frac{\partial}{\partial \chi}\left([\epsilon^2{\rm Log}(\epsilon^{-1})]^{-1}
V_0\right)=2J\Im(R^{(0)}(\hat{\lambda}))+o(1)\,,
\end{equation}
with the error being uniformly small for $\chi$ in any compact neighborhood of
$\chi=0$.  Therefore, we have
\begin{equation}
\begin{array}{rcl}
[\epsilon^2{\rm Log}(\epsilon^{-1})]^{-1}V_0 &= &\displaystyle
[\epsilon^2{\rm Log}(\epsilon^{-1})]^{-1}V_0 \Bigg|_{\chi=0}+
\int_0^\chi\left(2J\Im(R^{(0)}(\hat{\lambda}))+o(1)\right)\,d\chi'\\\\
& =&\displaystyle 
[\epsilon^2{\rm Log}(\epsilon^{-1})]^{-1}V_0 \Bigg|_{\chi=0}+2J\Im(R^{(0)}(\hat{\lambda}))\cdot\chi +o(1)\,.
\end{array}
\end{equation}

Thus, it remains to analyze $[\epsilon^2{\rm Log}(\epsilon^{-1})]^{-1}V_0$ with
$\chi=0$.  Recall from \S\ref{sec:symmetry} the formula for $V_0$ in
terms of the functions $R$ and $Y$:
\begin{equation}
V_0=\frac{ i\pi}{2}\int_{\lambda_0(\epsilon)}^{\lambda_1(\epsilon)}
R(\eta)Y(\eta)\,d\eta + \mbox{complex conjugate}\,.
\label{eq:Vformula}
\end{equation}
We note several asymptotic properties of $R$ and $Y$ that follow from the
form of the linear terms in the series expansions of the endpoints in
positive powers of $\epsilon$ (for $\chi=0$).  For fixed $\lambda$,
we have 
\begin{equation}
R(\eta) = (\eta-\hat{\lambda})(\eta-\hat{\lambda}^*)R^{(0)}(\hat{\lambda}) +
\bo(\epsilon^2)\,,
\label{eq:RApproxOuter}
\end{equation}
where the order $\epsilon$ terms cancel because the symmetric contributions
from $\lambda_1(\epsilon)$ and $\lambda_2(\epsilon)$ at this order cancel
exactly ({\em cf.} (\ref{eq:linearterms})).  The error is uniformly small
for $\eta$ in any compact set not containing the points $\lambda_0^{\rm crit}$,
$\hat{\lambda}$ or their conjugates.  Using this result in the formula
(\ref{eq:Ydef}) for $Y$, we see from the fact that the contours of integration
lie a fixed distance from these points that for {\em all} $\eta$ in between
the contours $C_{I+}$ and $C_{I-}$,
\begin{equation}
Y(\eta)=Y_{\rm deg}(\eta) + \bo(\epsilon^2)\,,
\end{equation}
where $Y_{\rm deg}(\eta)$ means the $G=2$ function $Y$ constructed for
$\epsilon=0$, {\em i.e.} on the degenerate configuration.  While the
approximation (\ref{eq:RApproxOuter}) of $R(\eta)$ holds for
intermediate points on the contour of integration in the formula
(\ref{eq:Vformula}) for $V_0$, it fails near both limits.  To give
an approximation uniformly valid near the lower limit of integration, let
$T_\epsilon(\eta)$ be the function defined by the relation
$T_\epsilon(\eta)^2=\eta-\lambda_0(\epsilon)$, cut along the
band $I_0^+$ and the negative imaginary axis, and normalized so that
for $\eta-\lambda_0(\epsilon)$ sufficiently large and positive real,
$T_\epsilon(\eta)$ is positive real.  Then we have
\begin{equation}
R(\eta)=T_\epsilon(\eta)\left[-(\eta-\hat{\lambda})(\eta-\hat{\lambda}^*)
T_0(\eta^*)^*+\bo(\epsilon^2)\right]\,,
\label{eq:RT}
\end{equation}
holding uniformly in a sufficiently small (but fixed as
$\epsilon\downarrow 0$) neighborhood of $\eta=\lambda_0^{\rm crit}$.
To approximate the square root near the upper limit of integration,
let $S_\epsilon(\eta)$ be the function defined by the relation
$S_\epsilon(\eta)^2=(\eta-\lambda_1(\epsilon))(\eta-\lambda_2(\epsilon))$,
cut along the shrinking band $I_1^+$ and normalized so that for large
$\eta$, $S_\epsilon(\eta)\sim \eta$.  Then,
\begin{equation}
R(\eta)=S_\epsilon(\eta)\left[(\eta-\hat{\lambda}^*)R^{(0)}(\eta)+\bo(\epsilon^2)\right]\,,
\label{eq:RS}
\end{equation}
holding uniformly in a sufficiently small fixed neighborhood of
$\eta=\hat{\lambda}$.  We stress that these expansions are only valid
when $\chi=0$.  For $\chi\neq 0$ larger terms come into play that we
have already taken into account by computing the derivative with
respect to $\chi$.

We take the path of integration in (\ref{eq:Vformula}) to pass through
two points $q_0$ and $q_1$ that are fixed as $\epsilon\downarrow 0$ and
lie respectively in the regions of validity of (\ref{eq:RT}) and (\ref{eq:RS}).
Then, we have
\begin{equation}
\begin{array}{rcl}
\displaystyle
V_0\Bigg|_{\chi=0}&=&\displaystyle
-\frac{i\pi}{2}\int_{\lambda_0(\epsilon)}^{q_0}
T_\epsilon(\eta)\left[(\eta-\hat{\lambda})(\eta-\hat{\lambda}^*)
Y_{\rm deg}(\eta)T_0(\eta^*)^*\right]\,d\eta \\\\
&&\displaystyle\,\,+\,\,
\frac{i\pi}{2}\int_{q_0}^{q_1}(\eta-\hat{\lambda})(\eta-\hat{\lambda}^*)R^{(0)}(\eta)Y_{\rm deg}(\eta)\,d\eta \\\\
&&\displaystyle\,\,+\,\, 
\frac{i\pi}{2}\int_{q_1}^{\lambda_1(\epsilon)} S_\epsilon(\eta)\left[(\eta-\hat{\lambda}^*)R^{(0)}(\eta)Y_{\rm deg}(\eta)\right]\,d\eta \\\\
&&\displaystyle\,\,+\,\, \mbox{complex conjugate} + \bo(\epsilon^2)\,.
\end{array}
\end{equation}
To handle the first term make the change of variables $\tau=T_0(\eta)$:
\begin{equation}
\begin{array}{rcl}
W_0&:=&\displaystyle
-\frac{i\pi}{2}\int_{\lambda_0(\epsilon)}^{q_0}T_\epsilon(\eta)
\left[(\eta-\hat{\lambda})(\eta-\hat{\lambda}^*)Y_{\rm deg}(\eta)
T_0(\eta^*)^*\right]\,d\eta \\\\ &=&\displaystyle
-i\pi\int_{T_0(\lambda_0(\epsilon))}^{T_0(q_0)}T_\epsilon(\tau^2+\lambda_0^{\rm
crit})\\\\
&&\displaystyle\,\,\times\,\,
\left[\tau (\tau^2-(\hat{\lambda}-\lambda_0^{\rm crit}))
(\tau^2-(\hat{\lambda}^*-\lambda_0^{\rm crit}))Y_{\rm deg}
(\tau^2+\lambda_0^{\rm crit}) T_0(\tau^{*2}+\lambda_0^{{\rm
crit},*})^*\right]\,d\tau\,.
\end{array}
\end{equation}
The quantity in square brackets has a convergent expansion in odd
powers of $\tau$ with coefficients $c_n$ that are indpendent of
$\epsilon$.  Similarly, $T_\epsilon(\tau^2+\lambda_0^{\rm crit})$
has the convergent expansion
\begin{equation}
T_\epsilon(\tau^2+\lambda_0^{\rm crit})=\tau\sum_{m=0}^\infty s_m
\left(\frac{\lambda_0^{\rm crit}-\lambda_0(\epsilon)}{\tau^2}\right)^m\,,
\end{equation}
where $s_m$ are the Taylor coefficients of $\sqrt{1+x}$.  Since the
convergence is uniform on the path of integration, the order of integration
and summation may be exchanged:
\begin{equation}
\begin{array}{rcl}
W_0&=&\displaystyle
-i\pi\sum_{m=0}^\infty\sum_{n=0}^\infty c_{2n+1}s_m (\lambda_0^{\rm crit}-\lambda_0(\epsilon))^m \int_{T_0(\lambda_0(\epsilon))}^{T_0(q_0)}
\tau^{2(n-m+1)}\,d\tau\\\\
&=&\displaystyle
-i\pi\sum_{m=0}^\infty\sum_{n=0}^\infty \frac{c_{2n+1}s_m}{2n-2m+3}\left[
(\lambda_0^{\rm crit}-\lambda_0(\epsilon))^mT_0(q_0)^{2n-2m+3}
-(-1)^mT_0(\lambda_0(\epsilon))^{2n+3}\right]\\\\
&=&\displaystyle
-\pi i\sum_{n=0}^\infty 
\frac{c_{2n+1}}{2n+3}T_0(q_0)^{2n+3} + \bo(\epsilon^2)\,,
\end{array}
\end{equation}
since for $\chi=0$, $\lambda_0(\epsilon)-\lambda_0^{\rm crit}=\bo(\epsilon^2)$.
Upon dividing by $\epsilon^2{\rm Log}(\epsilon^{-1})$, the main contribution
must necessarily come from the third term, 
\begin{equation}
W_1:=\frac{i\pi}{2}\int_{q_1}^{\lambda_1(\epsilon)} S_\epsilon(\eta)
\left[(\eta-\hat{\lambda}^*)R^{(0)}(\eta)Y_{\rm deg}(\eta)\right]\,d\eta\,.
\end{equation}
Here, the quantity in square brackets has a uniformly convergent
expansion in positive powers of $\eta-\hat{\lambda}$, with
coefficients $d_n$ that are independent of $\epsilon$, while $S_\epsilon(\eta)$
has the uniformly convergent Laurent expansion:
\begin{equation}
S_\epsilon(\eta)=(\eta-\hat{\lambda})\sum_{m=0}^\infty s_m\sum_{p=0}^m
\left(\begin{array}{c}m\\p\end{array}\right)\frac{
B_{m,p}(\epsilon)}{(\eta-\hat{\lambda})^{2m-p}}\,,
\end{equation}
where
\begin{equation}
B_{m,p}(\epsilon):=
[(\hat{\lambda}-\lambda_1(\epsilon))+(\hat{\lambda}-\lambda_2(\epsilon))]^p
(\hat{\lambda}-\lambda_1(\epsilon))^{m-p}
(\hat{\lambda}-\lambda_2(\epsilon))^{m-p}\,.
\end{equation}
Note that for $\chi=0$, the terms proportional to $\epsilon$ in the
square brackets cancel ({\em cf.} (\ref{eq:linearterms})), and
therefore $B_{m,p}(\epsilon)$ is a quantity of order
$\bo(\epsilon^{2m})$.  Exchanging the order of summation and
integration by uniform convergence, we have
\begin{equation}
W_1=\frac{i\pi}{2}\sum_{n=0}^\infty\sum_{m=0}^\infty\sum_{p=0}^m
\left(\begin{array}{c}m\\p\end{array}\right)d_ns_mB_{m,p}(\epsilon)
\int_{q_1}^{\lambda_1(\epsilon)}(\eta-\hat{\lambda})^{1+n+p-2m}\,d\eta\,.
\end{equation}
As long as $1+n+p-2m\neq -1$, 
\begin{equation}
\begin{array}{rcl}
\displaystyle
B_{m,p}(\epsilon)\int_{q_1}^{\lambda_1(\epsilon)}
(\eta-\hat{\lambda})^{1-n+p-2m}\,d\eta &=&\displaystyle 
\frac{B_{m,p}(\epsilon)}{(\lambda_1(\epsilon)-\hat{\lambda})^{2m}}
\cdot\frac{(\lambda_1(\epsilon)-\hat{\lambda})^{2+n+p}}
{2+n+p-2m}\\\\
&&\displaystyle\hspace{0.4 in}-\,\,\frac{B_{m,p}(\epsilon)
(q_1-\hat{\lambda})^{2+n+p-2m}}{2+n+p-2m}\\\\
&=&\bo(\epsilon^{n+p+2})-\bo(\epsilon^{2m})\,.
\end{array}
\end{equation}
These terms give constant contributions only for $m=0$, with all other
terms being order at least $\bo(\epsilon^2)$.
On the other hand, if $1+n+p-2m=-1$, then there are logarithmic
contributions.  Thus, using $s_1=1/2$, we have
\begin{equation}
\begin{array}{rcl}
W_1&=&\displaystyle
-\frac{i\pi}{2}\sum_{n=0}^\infty d_n \frac{(q_1-\hat{\lambda})^{n+2}}{n+2}
\\\\&&\displaystyle
\,\,+\,\,\frac{i\pi}{2}\left[\frac{d_0}{2}B_{1,0}(\epsilon) + 
\sum_{m=2}^\infty\sum_{n=m-2}^{2m-2}\left(\begin{array}{c}m\\2m-2-n\end{array}
\right)d_ns_mB_{m,2m-2-n}(\epsilon)\right]\int_{q_1}^{\lambda_1(\epsilon)}
\frac{d\eta}{\eta-\hat{\lambda}}\\\\
&&\,\, +\,\, \bo(\epsilon^2)\,.
\end{array}
\end{equation}
Regardless of the path of integration, as $\epsilon\downarrow 0$,
\begin{equation}
\int_{q_1}^{\lambda_1(\epsilon)}\frac{d\eta}{\eta-\hat{\lambda}}=
-{\rm Log}(\epsilon^{-1}) + \bo(1)\,,
\end{equation}
and consequently
\begin{equation}
W_1=-\frac{i\pi}{2}\sum_{n=0}^\infty d_n\frac{(q_1-\hat{\lambda})^{n+2}}
{n+2} + \frac{i\pi}{2}\left[\frac{1}{2}(\hat{\lambda}-\hat{\lambda}^*)
R^{(0)}(\hat{\lambda})Y_{\rm deg}(\hat{\lambda})r^2e^{2i\theta} + \bo(\epsilon)\right]\epsilon^2{\rm Log}(\epsilon^{-1}) + \bo(\epsilon^2)\,.
\end{equation}
When we combine this expression with the other components of $V_0$ at $\chi=0$,
we recall that the sum of the constant terms in $V_0$ vanishes because
$\Re(\tilde{\phi}^{(0)\sigma}(\hat{\lambda}))=0$, and thus we find
\begin{equation}\begin{array}{rcl}
[\epsilon^2{\rm Log}(\epsilon^{-1})]^{-1}V_0 &=& 
\displaystyle
c\chi + \left[-\frac{i}{2}Pr^2e^{i\alpha}e^{2i\theta} + 
\mbox{complex conjugate}\right] + o(1)\,,\\\\
&=&V_0^{\rm model}(r,\theta,\chi)+o(1)\,,
\end{array}
\end{equation}
as $\epsilon\downarrow 0$.  This establishes the desired convergence
of $V_0$.  

To verify the convergence of the corresponding partial derivatives, we
use the following exact formula which can be obtained by direct
application of the chain rule and substitution from the Vandermonde-type
system used to eliminate $\lambda_0$, $\lambda_0^*$, $\lambda_2$, and
$\lambda_2^*$:
\begin{equation}
\frac{dV_0}{d\lambda_1}=\frac{1}{2\pi i}\cdot
\frac{\partial M_0}{\partial\lambda_1}\cdot
\frac{\det {\bf V}(\lambda_0,\lambda_0^*,\lambda_1,\lambda_2,\lambda_2^*)}
{\det {\bf V}(\lambda_0,\lambda_0^*,\lambda_2,\lambda_2^*)}
\int_{\Gamma_1^+\cup\Gamma_1^-}\frac{\eta-\lambda_1^*}{R(\eta)}\,d\eta\,,
\end{equation}
where ${\bf V}(a_1,\dots,a_N)$ denotes the $N\times N$ Vandermonde
matrix\index{Vandermonde matrix}.  In this formula, the notation
$dV_0/d\lambda_1$ refers to the derivative after $\lambda_0$,
$\lambda_0^*$, $\lambda_2$, and $\lambda_2^*$ have been eliminated in
favor of $\lambda_1$, $\lambda_1^*$, and $x$.  Evaluating the
determinants explicitly gives
\begin{equation}
\frac{\det {\bf V}(\lambda_0,\lambda_0^*,\lambda_1,\lambda_2,\lambda_2^*)}
{\det {\bf V}(\lambda_0,\lambda_0^*,\lambda_2,\lambda_2^*)}
=(\lambda_0-\lambda_1)(\lambda_0^*-\lambda_1)(\lambda_1-\lambda_2)(\lambda_1-\lambda_2^*)\,.
\end{equation}
Substituting the expansions of the endpoints in terms of $\epsilon$, one
finds that as $\epsilon\downarrow 0$, 
\begin{equation}
[\epsilon{\rm Log}(\epsilon^{-1})]^{-1}\frac{dV_0}{d\lambda_1}=
-iPre^{i\alpha}e^{i\theta} + o(1)\,.
\end{equation}
Combining this relation with its complex conjugate, and using the 
chain rule relations
\begin{equation}
\frac{\partial}{\partial r}=\epsilon
e^{i\theta}\frac{\partial}{\partial \lambda_1} + \epsilon
e^{-i\theta}\frac{\partial}{\partial\lambda_1^*}\,, \hspace{0.3 in}
\frac{\partial}{\partial\theta}=i\epsilon r
e^{i\theta}\frac{\partial}{ \partial\lambda_1} -i\epsilon
re^{-i\theta}\frac{\partial}{\partial\lambda_1^*}\,, 
\label{eq:ComplexToReal}
\end{equation}
one obtains the desired convergence of the partial derivatives of
$[\epsilon^2{\rm Log}(\epsilon^{-1})]^{-1}V_0$ with respect to $r$ and
$\theta$.

We complete the proof with a simple remark about the uniformity of these
limits.  The statement that for fixed $\chi$ the region of uniform
validity is an arbitrary fixed annulus in the $(r,\theta)$ polar plane is
mirrored in the expressions for the functions $R^{\rm model}_1$ and
$V^{\rm model}_0$, which become meaningless if $r$ tends to infinity
or zero.  
\end{proof}

The model equations $R^{\rm model}_1(r,\theta,\chi)=0$ and $V^{\rm
model}_0(r,\theta,\chi)=0$ are easily solved.  The graph of $R^{\rm
model}_1(r,\theta,\chi)=0$ in the $(r,\theta)$ polar plane is
independent of $\chi$ and is simply the union of two perpendicular
lines through the origin:
\begin{equation}
\theta = \theta_n:=
\frac{\pi}{4}-\frac{\alpha}{2} + \frac{n\pi}{2}\,,\hspace{0.2 in}
n\in{\mathbb Z}\,.
\end{equation}
We then have
\begin{equation}
V_0^{\rm model}(r,\theta_n,\chi)=c\chi + (-1)^nPr^2\,.
\end{equation}
When $c\chi>0$, this equation is consistent only for $n\in 2{\mathbb Z}-1$
while for $c\chi<0$ we must have $n\in 2{\mathbb Z}$.  Under this condition,
the radial coordinate is uniquely determined:
\begin{equation}
r(\chi)=\sqrt{\frac{|c\chi|}{P}}\,.
\end{equation}
So, for each $\chi$ there are two opposite solutions $(r,\theta)$ with
$-\pi<\theta\le\pi$.  As $\chi$ moves through zero, these two solutions
coalesce at the origin and reemerge moving in the perpendicular direction.

\begin{theorem}
Let $t=t_{\rm crit}$ be fixed, and suppose that for $x=x_{\rm crit}$
there is a simply degenerate $G=2$ ansatz (in the sense of the two
additional assumptions given at the beginning of this section).  Then
for each $x$ with $|x-x_{\rm crit}|$ sufficiently small, there exists a
nondegenerate solution of the $G=2$ endpoint equations that is unique
up to permutation of the endpoints and is continuous in $x$.
\end{theorem}

\begin{proof}
By Lemma~\ref{lemma:momentreduce}, it is sufficient to prove that the
equations $R_1=0$ and $V_0=0$ can be solved for $\lambda_1$ and
$\lambda_1^*$ when $x$ is near $x_{\rm crit}$.  Fix an arbitrary
$\chi\neq 0$, and for all $\epsilon>0$ define
\begin{equation}
\begin{array}{rcl}
\displaystyle
R_1^{\rm family}(r,\theta;\epsilon)&:=&
\displaystyle
\epsilon^{-2}R_1(x_{\rm crit}+\epsilon^2{\rm Log}(\epsilon^{-1})\cdot\chi,\hat{\lambda}+\epsilon re^{i\theta},\hat{\lambda}^*+\epsilon re^{-i\theta})\,,\\\\
V_1^{\rm family}(r,\theta;\epsilon)&:=&
\displaystyle
[\epsilon^2{\rm Log}(\epsilon^{-1})]^{-1}V_0(
x_{\rm crit}+\epsilon^2{\rm Log}(\epsilon^{-1})\cdot\chi,\hat{\lambda}+\epsilon re^{i\theta},\hat{\lambda}^*+\epsilon re^{-i\theta})\,.
\end{array}
\end{equation}
By Lemma~\ref{lemma:converge}, these functions are differentiable with
respect to $r$ and $\theta$ and the partial derivatives are continuous
down to $\epsilon=0$.  Of course when $\epsilon=0$, we have
\begin{equation}
R_1^{\rm family}(r,\theta;0)=R_1^{\rm model}(r,\theta,\chi)\,,\hspace{0.3 in}
V_0^{\rm family}(r,\theta;0)=V_0^{\rm model}(r,\theta,\chi)\,.
\end{equation}
When $\epsilon=0$, the Jacobian determinant of these relations is
\begin{equation}
\left|\begin{array}{cc}
\partial R^{\rm model}_1/\partial r & \partial R^{\rm model}_1/\partial\theta\\\\\partial V^{\rm model}_0/\partial r & \partial V^{\rm model}_0/\partial\theta
\end{array}\right|=-4P^2r^3\,,
\end{equation}
which is not zero when evaluated on either of the two explicit
solutions of the model equations for $\chi\neq 0$.  It follows from
the implicit function theorem that for each of the two solutions of
the model problem and for sufficiently small positive $\epsilon$ there
is a solution $r(\epsilon)$ and $\theta(\epsilon)$, continuous in
$\epsilon$, of the equations $R^{\rm family}_1(r,\theta;\epsilon)=
V^{\rm family}_0(r,\theta;\epsilon)=0$.  The corresponding solution
of the endpoint equations for genus $G=2$ is given in terms of these
functions for $x=x_{\rm crit}+\epsilon^2{\rm Log}(\epsilon^{-1})\cdot\chi$ by
\begin{equation}
\begin{array}{rcl}
\lambda_1&=&\hat{\lambda}+\epsilon r(\epsilon)e^{i\theta(\epsilon)}\,,\\\\
\lambda_0&=&\lambda_0(x_{\rm crit}+\epsilon^2{\rm Log}(\epsilon^{-1})\cdot\chi,
\hat{\lambda}+\epsilon r(\epsilon)e^{i\theta(\epsilon)},
\hat{\lambda}^*+\epsilon r(\epsilon)e^{-i\theta(\epsilon)})\,,\\\\
\lambda_2&=&
\lambda_2(x_{\rm crit}+\epsilon^2{\rm Log}(\epsilon^{-1})\cdot\chi,
\hat{\lambda}+\epsilon r(\epsilon)e^{i\theta(\epsilon)},
\hat{\lambda}^*+\epsilon r(\epsilon)e^{-i\theta(\epsilon)})\,,
\end{array}
\end{equation}
with similar formulae for the complex conjugates.
\end{proof}

Having established the existence of a nondegenerate genus $G=2$
endpoint configuration for all $x$ in a sufficiently small deleted
neighborhood of $x_{\rm crit}$, we now consider whether the necessary
inequalities can be satisfied by the $G=2$ ansatz.  For $x_{\rm
crit}\neq 0$, the local unfolding will take place for $x$ values
totally of one sign or the other.  We suppose from now on that the
critical $G=0$ ansatz and the degenerate $G=2$ ansatz that agrees with
it both correspond to the choice $J={\rm sign}(x_{\rm crit})$.  The
unfolding of the degenerate ansatz corresponds to the same value of
$J$ for all $x$ under consideration.  Of course, we found that this
choice of $J$ was necessary for the small-time existence theory of the
$G=0$ ansatz ({\em cf.} \S\ref{sec:smalltime}), and even in the global
analysis carried out with the help of the computer, we found this choice
to lead to a workable ansatz right up to the primary caustic.

\begin{figure}[h]
\begin{center}
\mbox{\psfig{file=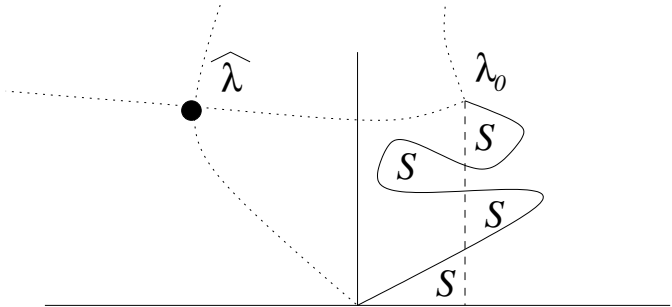,width=3.5 in}}
\end{center}
\caption{\em The shadow $S$ of a complicated band $I_0^+$ can have several
components.  The critical point $\hat{\lambda}$ is always assumed to
lie outside the shadow.}
\label{fig:NotAllowed}
\end{figure}

In order to
proceed, we make one further assumption about the degenerate
ansatz at $x=x_{\rm crit}$ and $t=t_{\rm crit}$:  
\begin{equation}
\mbox{\bf Shadow condition:  } \Im(R^{(0)}(\hat{\lambda}))<0\,.
\label{eq:OutsideShadow}
\index{shadow condition}
\end{equation}
By virtue of the normalization condition
$R^{(0)}(\lambda)\sim-\lambda$ for $\lambda$ near infinity, this
condition holds for all $\hat{\lambda}\in {\mathbb H}$ outside a
bounded region that is the ``shadow'' of the band $I_0^+$, the region
enclosed by $I_0^+$ and the vertical segment descending from the
endpoint $\lambda_0^{\rm crit}$ to the real axis.  See
Figure~\ref{fig:NotAllowed}.
We note that the computer plots presented in
Chapter~\ref{sec:genuszero} indicate that for the Satsuma-Yajima
initial data the critical point $\hat{\lambda}$ indeed always lies
outside the shadow of the band $I_0^+$ and therefore
(\ref{eq:OutsideShadow}) is always satisfied at the primary caustic.

Under these assumptions, we see that the constant $c$ appearing in the
model equation $V_0^{\rm model}(r,\theta,\chi)=0$ always has the
opposite sign of $x_{\rm crit}$.  From the exact solution of the model
problem, and the fact that it is a good approximation for small
$\epsilon$ (equivalently for $x$ sufficiently close to $x_{\rm crit}$)
to the true dynamics of the endpoint
$\lambda_1(\epsilon)=\hat{\lambda}+\epsilon
r(\epsilon)e^{i\theta(\epsilon)}$, we can easily deduce that when $x$
is tuned away from $x_{\rm crit}$ toward $x=0$, the two endpoints move
apart in the direction of steepest descent of the function
$\Re(\tilde{\phi}^{(0)\sigma}(\lambda))$ at the saddle point
$\hat{\lambda}$.  On the other hand, when $x$ is tuned away from
$x_{\rm crit}$ in the direction of increasing $|x|$, the endpoints
separate in the direction of sharpest increase of this function.

\begin{figure}[h]
\begin{center}
\mbox{\psfig{file=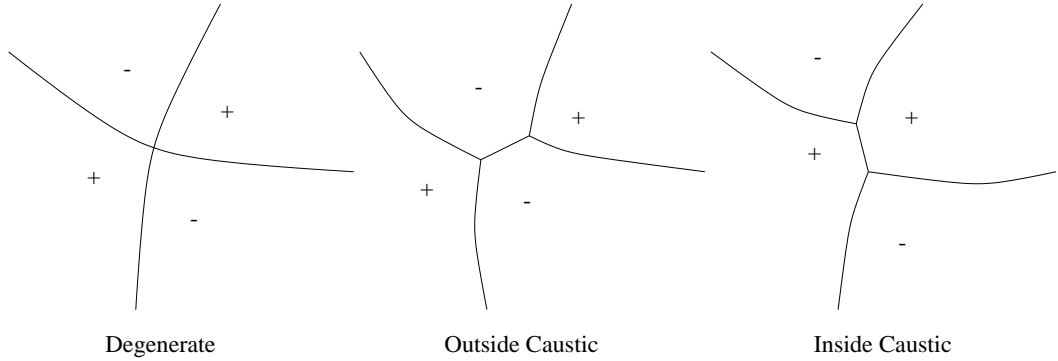,width=5.5 in}}
\end{center}
\caption{\em The two ways the degenerate $G=2$ endpoint configuration 
in the neighborhood of $\lambda=\hat{\lambda}$ unfolds for $x$ near
$x_{\rm crit}$.  The curves are the zero level sets of
$\Re(\tilde{\phi}^\sigma(\lambda))$, whose sign is also indicated.}
\label{fig:BandOpen}
\end{figure}

Once separation has occurred, the assumption (\ref{eq:nothigher})
ensures that for $|x-x_{\rm crit}|$ sufficiently small the function
$R(\lambda)Y(\lambda)$ vanishes {\em exactly} like a square root at
both endpoints emerging from the critical point $\hat{\lambda}$.  For
small $|x-x_{\rm crit}|$, the function $Y(\lambda)$ can be
approximated locally by the constant value $Y_{\rm
deg}(\hat{\lambda})$, and in a rescaled $\epsilon$-neighborhood of
$\hat{\lambda}$ the function $R(\lambda)$ takes on a canonical form.
These facts allow fixed-point arguments similar to those used in the
proofs of the local continuation theorems in \S\ref{sec:continuation}
to be used to prove that as $x$ passes through $x_{\rm crit}$, the 
local orbit structure of the differential equation (\ref{eq:ODE}) switches
between the two cases illustrated in Figure~\ref{fig:BandOpen}.

Furthermore, the continuation arguments show that for $x$ just inside
the primary caustic ({\em i.e.}  for $|x|<|x_{\rm crit}|$), exactly
one of the two new endpoints born from $\hat{\lambda}$ lies on a
trajectory of (\ref{eq:ODE}) connecting to $\lambda_0(x,t_{\rm
crit})$.  We break symmetry by calling this endpoint
$\lambda_1(x,t_{\rm crit})$, which makes the other endpoint
$\lambda_2(x,t_{\rm crit})$.  Clearly, the genus $G=0$ gap contour
$\Gamma_1^+$ that connects $\lambda_0^{\rm crit}$ to $\lambda=-\sigma
0$ and passes through the critical point $\hat{\lambda}$ at $x=x_{\rm
crit}$ can be taken to split into two new gap contours $\Gamma_1^+$
connecting $\lambda_0(x,t_{\rm crit})$ to $\lambda_1(x,t_{\rm crit})$,
and $\Gamma_2^+$ connecting $\lambda_2(x,t_{\rm crit})$ to
$\lambda=-\sigma 0$.  Both of these contours can be chosen for
$|x_{\rm crit}|-|x|$ sufficiently small and positive so that the
inequality $\Re(\tilde{\phi}^\sigma(\lambda))<0$ holds except at the
endpoints.

With the function $R(\lambda)$ taken to be cut between
$\lambda_1(x,t_{\rm crit})$ and $\lambda_2(x,t_{\rm crit})$ along the
zero level $I_1^+$ of $\Re(\tilde{\phi}^\sigma(\lambda))$, it remains
to verify the inequality $\rho^\sigma(\eta)\,d\eta\in {\mathbb R}_-$
in this newly born band.  It follows from the fixed point argument for
the existence of this small orbit of (\ref{eq:ODE}) for small
$|x-x_{\rm crit}|$ that the band $I_1^+$ is smooth; this implies that
there are no internal zeros of $\rho^\sigma(\eta)$ and therefore that
the differential $\rho^\sigma(\eta)\,d\eta$ is necessarily real and of
one sign in $I_1^+$.  From the sign table for
$\Re(\tilde{\phi}^\sigma(\lambda))$ shown for the configuration in the
right-hand plot of Figure~\ref{fig:BandOpen}, and the relation between
the functions $\tilde{\phi}^\sigma(\lambda)$ and
$\rho^\sigma(\lambda)$ we can easily compute the sign of the
differential $\rho^\sigma(\eta)\,d\eta$ for all $\eta\in I_1^+$.
Select the sign of the differential $d\eta$ according to the
orientation of $I_1^+$ starting at $\lambda_1(x,t_{\rm crit})$ and
ending at $\lambda_2(x,t_{\rm crit})$.  Just beyond the endpoint
$\lambda_2(x,t_{\rm crit})$ in the direction tangent to $I_1^+$, we
see that the differential $[d\tilde{\phi}^\sigma/d\eta]\,d\eta$ is
negative real because $d\eta$ is oriented in the direction of steepest
decrease of $\Re(\tilde{\phi}^\sigma(\eta))$ and
$\Im(\tilde{\phi}^\sigma(\eta))=0$ along this same trajectory.  Since
$d\tilde{\phi}^\sigma/d\eta = \pi i R(\eta)Y(\eta)$ with $Y(\eta)$
analytic and $R(\eta)$ vanishing like a simple square root at
$\lambda_2(x,t_{\rm crit})$, this formula continues analytically
around the endpoint in the counter-clockwise direction to the cut
$I_1^+$ as $\pi i R_+(\eta)Y(\eta)$ which we identify as $\pi
i\rho^\sigma(\eta)$.  In the process of continuing around the
square-root branch point, a factor of $i$ is contributed:
\begin{equation}
R_+(\lambda_2(x,t_{\rm crit})-d\eta) \approx iR(\lambda_2(x,t_{\rm
crit})+d\eta)\,.
\end{equation}
Therefore, we have
\begin{equation}
\begin{array}{rcl}
\rho^\sigma(\lambda_2(x,t_{\rm crit})-d\eta)\,d\eta &=&
 R_+(\lambda_2(x,t_{\rm crit})-d\eta)Y(\lambda_2(x,t_{\rm crit})-d\eta)\,d\eta
\\\\
&\approx &
iR(\lambda_2(x,t_{\rm crit})+d\eta)Y(\lambda_2(x,t_{\rm crit})+d\eta)\,d\eta\\\\
&= &
\displaystyle\frac{1}{\pi}\frac{d\tilde{\phi}^\sigma}{d\eta}(\lambda_2(x,t_{\rm crit})+d\eta)\,d\eta\,,
\end{array}
\end{equation}
which is negative real, as desired.  We have therefore proved the
following.

\begin{theorem}
If the genus $G=0$ ansatz undergoes a failure at $(x_{\rm crit},t_{\rm
crit})$ characterized by the (simple) pinching off of the gap contour 
at a point $\hat{\lambda}$ not in the shadow of the band $I_0^+$, then
there exists a genus $G=2$ ansatz for $|x_{\rm crit}|-|x|$ small and
positive that satisfies all inequalities and becomes degenerate at
$x=x_{\rm crit}$ with the closing of the band $I_1^+$ where it matches
onto the critical $G=0$ solution.
\end{theorem}

\begin{remark}
The shadow condition (\ref{eq:OutsideShadow}) may seem somewhat
artificial.  However, it is equivalent to the statement that the
region where $|x|>|x_{\rm crit}|$ corresponds to a genus zero
unfolding.  If it is known {\em a priori} that this region ``outside''
the primary caustic corresponds to a genus zero ansatz that first
becomes critical when $|x|$ is decreased to $|x_{\rm crit}|$, then the
shadow condition (\ref{eq:OutsideShadow}) must {\em automatically} be
satisfied in order for the unfolding that occurs to be consistent.
Under these conditions, the shadow condition need not be checked at
all.  \end{remark}

\begin{remark}
The scalings (\ref{eq:scalings}) of the variables
$\lambda_1-\hat{\lambda}$, $\lambda_1^*-\hat{\lambda}^*$, and
$x-x_{\rm crit}$ are significant in that they determine the size of
the new band that opens up in the complex $\lambda$-plane as $x$ is
tuned into the genus two region.  In particular, to obtain a band of
length $|\lambda_2-\lambda_1|\sim\epsilon$, one must have $|x-x_{\rm
crit}|\sim\epsilon^2{\rm Log}(\epsilon^{-1})$.
\end{remark}

\chapter{Variational Theory of the Complex Phase}
\label{sec:variational}
Apart from relying heavily on the analyticity of the function
$\rho^0(\eta)$ characterizing the asymptotic density of eigenvalues on
the imaginary interval $[0,iA]$ by the WKB formula
(\ref{eq:WKBformula}), in the direct construction of the complex phase
function $g^\sigma(\lambda)$ presented in Chapter~\ref{sec:ansatz} there was
no way to determine {\em a priori} the value of the genus $G$ for
which a successful ansatz could be constructed for given values of $x$
and $t$, nor indeed whether such a finite $G$ exists at all.  To begin
to address these issues, we need to reformulate the conditions for an
admissible density function \index{density function!admissible}
$\rho^\sigma(\eta)$ for generating a complex phase function
$g^\sigma(\lambda)$ given in Definition~\ref{def:admissiblerho} in a
more abstract form.

The Green's function \index{Laplace's equation!Green's function for}
for Laplace's equation in the upper half-plane $\mathbb{C}_+$ with
Dirichlet boundary conditions \index{Dirichlet boundary conditions} on
the real axis is
\begin{equation}
G(\lambda;\eta)=\log\left|\frac{\lambda-\eta^*}{\lambda-\eta}
\right|\,,
\end{equation}
for $\lambda$ and $\eta$ in $\mathbb{C}_+$.  For
$\lambda\in\mathbb{C}_+$ and also in the domain of analyticity of
$\rho^0(\lambda)$ (which of course is the whole upper half-plane for
the special case of the Satsuma-Yajima ensemble, when
$\rho^0(\lambda)\equiv \rho^0_{\rm SY}(\lambda)\equiv i$), define the
``external field''\index{external field}
\begin{equation}
\varphi^\sigma(\lambda)=-\Re\left(
\int_{0}^{iA}L^0_\eta(\lambda)\rho^0(\eta)\,d\eta +
\int_{-iA}^0L^0_\eta(\lambda)\rho^0(\eta^*)^*\,d\eta
+i\pi\sigma\int_\lambda^{iA}\rho^0(\eta)\,d\eta + 2iJ(\lambda x +
\lambda^2 t)\right)\,.
\end{equation}
Note that this field is a sum of a harmonic part and a subharmonic
potential part.  Let $d\mu^0(\eta)$ be the nonnegative measure
$-\rho^0(\eta)\,d\eta$ on the segment $[0,iA]$ oriented from $0$ to
$iA$.  Then we can write:
\begin{equation}
\varphi^\sigma(\lambda)=-\int G(\lambda;\eta)d\mu^0(\eta) - \Re\left(
i\pi\sigma\int_\lambda^{iA}\rho^0(\eta)\,d\eta + 2iJ(\lambda x +
\lambda^2 t)\right)\,,
\end{equation}
which displays the field $\varphi^\sigma(\lambda)$ as a sum of a
Green's potential of a system of fixed {\em negative} charges
distributed on the segment $[0,iA]$ and an ``ambient'' harmonic
contribution.  This is the explicit Riesz decomposition \index{Riesz
decomposition} of the superharmonic function
$-\varphi^\sigma(\lambda)$ in the upper half-plane \cite{ST97}.

Let $d\mu$ be a nonnegative Borel measure \index{Borel measure} with
support contained in the closure of ${\mathbb C}_+$, and consider the
weighted energy functional\index{weighted energy functional}
\begin{equation}
E[d\mu]:=\frac{1}{2}\int d\mu(\lambda)\int
G(\lambda;\eta)\,d\mu(\eta) + \int \varphi^\sigma(\lambda)\,d\mu(\lambda)\,.
\end{equation}
This can be interpreted physically as the potential energy of a given
system of {\em positive} charges with distribution $d\mu$ in the upper
half-plane with the real-axis as a conducting boundary, in the
presence of the external potential field $\varphi^\sigma(\lambda)$.
The first term in $E[d\mu]$ is the self-energy of the charge
distribution $d\mu$, and the second term is the interaction energy
with the field $\varphi^\sigma(\lambda)$.  From the remarks above, one
term in this interaction energy is the Green's energy of interaction
between the positive charges in $d\mu$ and the fixed negative charges
with distribution $-d\mu^0$.

\begin{theorem}
Let $\rho^\sigma(\eta)$ be an admissible density function on the oriented loop
contour $C_\sigma$ lying in ${\mathbb H}$ as in
Definition~\ref{def:admissiblerho}.  Then
\begin{equation}
E[-\rho^\sigma(\eta)\,d\eta] = \inf_{d\mu\in {\cal B}_+(C)}E[d\mu]\,,
\label{eq:inf}
\end{equation}
where the infimum is taken over ${\cal B}_+(C)$, the set of all
nonnegative Borel measures with support in the closure of $C$ having
finite total mass and finite Green's energy, that is, measures for
which
\begin{equation}
\int d\mu(\lambda)<\infty\mbox{    and    }
\int d\mu(\lambda)\int G(\lambda;\eta)
d\mu(\eta)
<\infty\,.
\end{equation}
\label{theorem:min}
\end{theorem}

\begin{proof}
With the orientation $\sigma$ of the contour $C$, the admissible
differential $-\rho^\sigma(\eta)\,d\eta$ is a real nonnegative Borel
measure on $C$ with finite mass.  Let $d\mu\in{\cal B}_+(C)$.  Then
\begin{equation}
E[d\mu]-E[-\rho^\sigma(\eta)\,d\eta] = \frac{1}{2}\int d\Delta(\lambda)
\int d\Delta(\eta) G(\lambda;\eta) +\int d\Delta(\lambda)\left[
\varphi^\sigma(\lambda)+\int_{C_\sigma} G(\lambda;\eta)\rho^\sigma(\eta)\,d\eta \right]\,,
\end{equation}
where $d\Delta(\eta):=d\mu(\eta)+\rho^\sigma(\eta)\,d\eta$ with
$d\eta$ defined on $C_\sigma$ by the orientation $\sigma$.  First,
note that the term that is quadratic in $d\Delta$ is always
nonnegative, being the Green's energy of a signed measure with finite
positive and negative parts, each of which has finite Green's energy.
Indeed, the nonnegativity of the Green's energy for such measures is,
for example, the content of Theorem II.5.6 in \cite{ST97}.  Next,
observe that for $\lambda\in C$, and with the value of the interpolant
index $K$ chosen according to (\ref{eq:mselect}), we have
\begin{equation}
\Re(\tilde{\phi}^\sigma(\lambda)) = -\left[\varphi^\sigma(\lambda) +
\int_{C_\sigma}G(\lambda;\eta)\rho^\sigma(\eta)\,d\eta\right]\,.
\end{equation}
Thus we have
\begin{equation}
E[d\mu]-E[-\rho^\sigma(\eta)\,d\eta] \ge -\int 
\Re(\tilde{\phi}^\sigma(\lambda))\,d\Delta(\lambda)\,.
\end{equation}
Since according to Definition~\ref{def:admissiblerho} we have
$\Re(\tilde{\phi}^\sigma(\lambda))\equiv 0$ for $\lambda$ in the
support of $\rho^\sigma(\eta)\,d\eta$, the integral on the right-hand
side may be taken over the gaps of $C$.  Therefore,
\begin{equation}
E[d\mu]-E[-\rho^\sigma(\eta)\,d\eta]\ge -\int_{\cup_k\Gamma_k^+}
\Re(\tilde{\phi}^\sigma)\,d\mu\ge 0\,,
\end{equation}
because $d\mu$ is a nonnegative measure and according to 
Definition~\ref{def:admissiblerho} we have
$\Re(\tilde{\phi}^\sigma(\lambda))\le 0$ for $\lambda$ in the gaps of
$C$.
\end{proof}

\begin{remark}
Note that the weaker condition that $\Re(\tilde{\phi}^\sigma(\lambda))\le 0$
in the gaps suffices in the proof of the theorem.  Therefore, 
$-\rho^\sigma(\eta)\,d\eta$ is a minimizer even if the inequality is not
strict in the gaps.
\end{remark}

Therefore, the measure $-\rho^\sigma(\eta)\,d\eta$ on the oriented
contour $C_\sigma$ solves the energy minimization problem for positive
charge distributions on the contour $C$.  It is an {\em equilibrium
measure} \index{equilibrium measure} corresponding to the contour $C$,
and the corresponding value of $E$ is the {\em equilibrium energy}
\index{equilibrium energy} $E_{\rm min}[C]$ of $C$.  Although we have
so far only considered contours $C$ that support admissible density
functions $\rho^\sigma(\eta)$, the equilibrium energy $E_{\rm min}[C]$
of an arbitrary loop contour $C$ can be defined by the infimum on the
right-hand side of (\ref{eq:inf}). Note that the equilibrium measure
is by no means unique due to the requirement that the curve $C$ meet
the origin, which lies on the boundary of the domain for the Green's
function.  Thus, the support of a measure $d\mu$ can contain the
origin, and two measures differing only by a Dirac mass \index{Dirac
mass} at the origin always have the same energy because
$\varphi^\sigma(0)=0$.  This is a nontrivial issue because the the
support of $-\rho^\sigma(\eta)\,d\eta$ on $C$ always includes the
origin according to Definition~\ref{def:admissiblerho}.

Next, we consider the variations of the energy as the contour $C$
undergoes small deformations.  
\begin{theorem}
Let $\rho^\sigma(\eta)$ be an admissible density function on an
oriented contour $C_\sigma$ in the sense of Definition~\ref{def:admissiblerho}.
For each function $\kappa(\eta)$ analytic in a
neighborhood of the the support of $-\rho^\sigma(\eta)\,d\eta$ in $C$
and satisfying $\kappa(0)=0$, and for each real $\epsilon$ with
$|\epsilon|$ sufficiently small, define a measure $d\mu_\epsilon^\kappa$ as
follows: the support of $d\mu^\kappa_\epsilon$ is the image of that of
$-\rho^\sigma(\eta)\,d\eta$ under the near-identity map 
\begin{equation}
\nu^\kappa_\epsilon:\eta\mapsto
\eta+\epsilon\kappa(\eta)\,,
\end{equation}
and the measure $\mu^\kappa_\epsilon(M)$ of each measurable subset $M$ of
its support is defined to be the integral of
$-\rho^\sigma(\eta)\,d\eta$ over the inverse image of $M$ under the
map $\nu_\epsilon^\kappa$.  Then, with the function $\kappa(\eta)$ held fixed,
\begin{equation}
\frac{d}{d\epsilon}E[d\mu^\kappa_\epsilon]\Bigg|_{\epsilon=0} = 0\,.
\end{equation}
\label{theorem:Sproperty}
\end{theorem}

\begin{proof}
For each $\kappa(\eta)$, we have $d\mu_0^\kappa(\eta) =
-\rho^\sigma(\eta)\,d\eta$.  By definition of the deformed measure
$d\mu^\kappa_\epsilon$, we have
\begin{equation}
E[d\mu_\epsilon^\kappa]=\frac{1}{2}\int d\mu_0^\kappa(\lambda)\int
d\mu_0^\kappa(\eta)\,G(\nu_\epsilon^\kappa(\lambda);\nu_\epsilon^\kappa(\eta))
 +\int d\mu_0^\kappa(\lambda)\, \varphi^\sigma(\nu_\epsilon^\kappa(\lambda))\,.
\end{equation}

First, we expand the quadratic term for $\epsilon$ small, using the fact
that for any branch of the logarithm, $G(\lambda;\eta)=\Re(\log(\lambda-\eta^*))-\Re(\log(\lambda-\eta))$,
\begin{equation}
\begin{array}{rcl}
\displaystyle
\frac{1}{2}\int d\mu_0^\kappa(\lambda)\int
d\mu_0^\kappa(\eta)\,G(\nu_\epsilon^\kappa(\lambda);\nu_\epsilon^\kappa(\eta))
&=&\displaystyle
\frac{1}{2}\int d\mu_0^\kappa(\lambda)\int d\mu_0^\kappa(\eta)\,
G(\lambda;\eta) \\\\
&&\displaystyle\,\,+\,\, \frac{\epsilon}{2}\Re\left(\int d\mu^\kappa_0(\lambda)
\int d\mu^\kappa_0(\eta)\frac{\kappa(\lambda)-\kappa(\eta)^*}{\lambda-\eta^*}
\right)
\\\\
&&\displaystyle\,\,-\,\,\frac{\epsilon}{2}\Re\left(\int d\mu^\kappa_0(\lambda)
\int d\mu^\kappa_0(\eta)\frac{\kappa(\lambda)-\kappa(\eta)}{\lambda-\eta}
\right)
+\bo(\epsilon^2)\,.
\end{array}
\end{equation}
The second integral proportional to $\epsilon$ above is nonsingular
because $\kappa$ is analytic on the support of $d\mu_0^\kappa$.  Upon
regularization by interpreting one or the other of the iterated
integrals in the sense of the Cauchy principal value, the terms in the
numerator can be separated.  Thus,
\begin{equation}
\begin{array}{rcl}\displaystyle
\int d\mu_0^\kappa(\lambda)\int d\mu_0^\kappa(\nu)
\frac{\kappa(\lambda)-\kappa(\eta)}{\lambda-\eta} &=&
\displaystyle \int d\mu_0^\kappa(\lambda)\,\kappa(\lambda)
\,\mbox{P.V.}\int \frac{d\mu_0^\kappa(\eta)}{\lambda-\eta} \\\\
&&\displaystyle\,\,-\,\, 
\int d\mu_0^\kappa(\eta)\,\kappa(\eta)\,\mbox{P.V.}\int 
\frac{d\mu^\kappa_0(\lambda)}{\lambda-\eta}\\\\
&=&\displaystyle
2\int d\mu_0^\kappa(\lambda)\,\kappa(\lambda)\,\mbox{P.V.}\int
\frac{d\mu_0^\kappa(\eta)}{\lambda-\eta}\\\\
&=&\displaystyle
-2\int d\mu_0^\kappa(\lambda)\,\kappa(\lambda)\,\mbox{P.V.}\int_{C_\sigma}
\frac{\rho^\sigma(\eta)\,d\eta}{\lambda-\eta}\,.
\end{array}
\end{equation}
The first integral proportional to $\epsilon$ can be handled without
regularization.  Here, for the real part we find
\begin{equation}
\begin{array}{rcl}
\displaystyle \Re\left(\int d\mu_0^\kappa(\lambda)\int d\mu_0^\kappa(\eta)
\frac{\kappa(\lambda)-\kappa(\eta)^*}{\lambda-\eta^*}\right) &=&
\displaystyle 
2\Re\left(\int d\mu_0^\kappa(\lambda)\,\kappa(\lambda)\int
\frac{d\mu_0^\kappa(\eta)}{\lambda-\eta^*}\right)\\\\
&=&\displaystyle
2\Re\left(\int d\mu_0^\kappa(\lambda)\,\kappa(\lambda)\int_{[C^*]_\sigma}
\frac{\rho^\sigma(\eta^*)^*\,d\eta}{\lambda-\eta}\right)
\,.
\end{array}
\end{equation}

Next, we expand the linear term in the energy for small $\epsilon$.
We find
\begin{equation}
\frac{d}{d\epsilon}\varphi^\sigma(\nu_\epsilon^\kappa(\lambda))\Bigg|_{\epsilon=0}
= -
\Re\left[\kappa(\lambda)\left(\int_0^{iA}\frac{\rho^0(\eta)\,d\eta}
{\lambda-\eta} + \int_{-iA}^0\frac{\rho^0(\eta^*)^*\,d\eta}{\lambda-\eta}
-i\pi\sigma\rho^0(\lambda) + 2iJ(x+2\lambda t)\right)\right]\,.
\end{equation}
In a now-familiar step ({\em cf.} Chapter~\ref{sec:ansatz}), we introduce a
path of integration $C_I:0\rightarrow iA$ that agrees with $C_\sigma$
in the support of $d\mu^\kappa_0$ and then connects the final point of
support to $iA$.  Then, using analyticity of $\rho^0(\eta)$, this
expression becomes
\begin{equation}
\frac{d}{d\epsilon}\varphi^\sigma(\nu_\epsilon^\kappa(\lambda))\Bigg|_{\epsilon=0} = -
\Re\left[\kappa(\lambda)\left(\mbox{P.V.}\int_{C_I}\frac{\rho^0(\eta)\,d\eta}{\lambda-\eta} + \int_{C_I^*}\frac{\rho^0(\eta^*)^*\,d\eta}{\lambda-\eta} + 2iJ(x+2\lambda t)\right)\right]\,.
\end{equation}

Combining these calculations, and making an identification with the derivative
of $\tilde{\phi}^\sigma(\lambda)$ along the contour $C$, we find that
\begin{equation}
\frac{d}{d\epsilon} E[d\mu^\kappa_\epsilon]\Bigg|_{\epsilon=0} = 
-\int d\mu_0^\kappa(\lambda)\Re\left[\kappa(\lambda)\frac{d}{d\lambda}\tilde{\phi}^\sigma(\lambda)\right]\,,
\end{equation}
where the derivative along the contour is meant.  It is sufficient to
integrate over the support of
$d\mu^\kappa_0=-\rho^\sigma(\eta)\,d\eta$.  By
Definition~\ref{def:admissiblerho}, the function
$\tilde{\phi}^\sigma(\lambda)$ is constant along each component of the support
of $d\mu_0^\kappa$, which proves the theorem.
\end{proof}

\begin{remark}
A contour $C$ for which the variations described in the statement of
Theorem~\ref{theorem:Sproperty} all vanish is said to have the {\em
S-property} \index{S-property} \cite{GR87}.  This terminology appears
in the approximation theory literature where ``S'' stands for
``symmetry''.  Clearly, it might just as well stand for ``stationary''
or, in the context of applications to steepest-descent type asymptotic
analysis of Riemann-Hilbert problems, ``steepest''.
\end{remark}

\begin{remark}
In some applications, it may be enough that the above theorem holds
for a dense subset of analytic functions $\kappa(\eta)$.  For example,
one often restricts attention to {\em Schiffer variations}
\index{Schiffer variations} \cite{S50} in which $\kappa(\eta)$ has the
form of a simple rational function
\begin{equation}
\kappa(\eta)=\frac{\alpha\eta}{\eta-\eta_0}\,,
\end{equation}
for $\alpha\in {\mathbb C}$ and $\eta_0$ not lying on the contour $C$.
The condition that $\kappa(0)=0$ simply fixes the contour to the origin
under deformation.
\end{remark}

The results described in Theorem~\ref{theorem:min} and
Theorem~\ref{theorem:Sproperty} indicate that the conditions that
characterize the complex phase function $g^\sigma(\lambda)$ ({\em cf.}
Definition~\ref{def:admissiblerho}) are equivalent to the existence of
a certain kind of critical point for the energy functional, where
variations with respect to both the measure {\em and the contour of
support} are permitted.  {\em Thus, we have obtained a generalization
of the method of Lax and Levermore \index{Lax-Levermore
analysis!generalization of} \cite{LL83}, who considered the restricted
problem of minimizing the energy of measures supported on a fixed and
given contour}.  In this connection, it is attractive to consider
whether an appropriate variational problem can be well-posed whose
solution is exactly a critical point of the desired type.  This would
effectively complement the ansatz-based construction of
$g^\sigma(\lambda)$ given in Chapter~\ref{sec:ansatz} by allowing techniques
of functional analysis and logarithmic potential theory
\index{logarithmic potential theory} to be applied to determine
properties of the complex phase function, {\em e.g.}  existence,
uniqueness, and genus.

If we suppose that for each given analytic function $\kappa(\lambda)$
as in the statement of Theorem~\ref{theorem:Sproperty} the equilibrium
energy $E_{\rm min}[\nu_\epsilon^\kappa(C)]$ is differentiable with
respect to $\epsilon$, then the existence of a loop contour $C$ for
which
\begin{equation} \frac{d}{d\epsilon}E_{\rm
min}[\nu_\epsilon^\kappa(C)]\Bigg|_{\epsilon=0}=0\,, 
\label{eq:stationarymin}
\end{equation}
implies that $C$ has the S-property.  To see this, let $d\mu$ be an
equilibrium measure for $C$, and consider the corresponding family of
measures $d\mu^\kappa_\epsilon$ supported on the curve
$\nu_\epsilon^\kappa(C)$ as in the statement of
Theorem~\ref{theorem:Sproperty}.  Clearly, $E[d\mu^\kappa_0]=E_{\rm
min}[C]=E_{\rm min}[\nu_0^\kappa(C)]$, and since
$E[d\mu^\kappa_\epsilon]$ is not generally an equilibrium measure for
$\nu_\epsilon^\kappa(C)$ we have $E[d\mu^\kappa_\epsilon]\ge E_{\rm
min}[\nu_\epsilon^\kappa(C)]$.  Now as a function of $\epsilon$,
$E[d\mu^\kappa_\epsilon]$ is clearly differentiable at $\epsilon=0$,
and it follows from its domination of the equilibrium energy that
(\ref{eq:stationarymin}) implies that the derivative of
$E[d\mu^\kappa_\epsilon]$ with respect to $\epsilon$ vanishes at
$\epsilon=0$.

Thus, differentiability of the equilibrium energy \index{equilibrium
energy!differentiability of} implies that the object that may be taken
to be stationary at a curve with the S-property is the equilibrium
energy $E_{\rm min}[C]$ itself as a functional of the loop contour
$C$.  This suggests posing a ``stationary-min'' problem for the energy
functional $E$, with possible special case variants ``min-min'' and
``max-min''.  It is not difficult to argue that the ``min-min''
problem, {\em i.e.} finding a contour $C$ for which the equilibrium
energy $E_{\rm min}[C]$ is minimal, has no solution.  This is because
the external field $\varphi^\sigma(\lambda)$ goes to $-\infty$ as
$\lambda\rightarrow\infty$ in a sector of the upper half-plane
(depending on $x$ and $t$) and consequently the equilibrium energy can
be made arbitrarily negative by considering a sequence of contours
expanding into this sector.  On the other hand, the ``max-min''
problem, {\em i.e.} finding a contour $C$ for which the equilibrium
energy $E_{\rm min}[C]$ is as large as possible, is a version of the
well-studied problem of finding sets of {\em minimal weighted
logarithmic capacity} \index{minimal weighted logarithmic capacity}
satisfying certain geometrical constraints (here, the geometrical
constraint is that the set must be a contour surrounding the imaginary
interval $[0,iA]$ and connecting $0-$ to $0+$).  This sort of problem
is sometimes referred to in the literature as a Chebotarev problem
\index{Chebotarev problem} \cite{GR87}.  In circumstances significantly 
simpler than those of our problem, the minimal capacity problem is
known to have a solution that is unique in the support of the
equilibrium measure.

We pose the ``max-min'' problem \index{max-min problem} in the
following conjecture.
\begin{conjecture}
Suppose for simplicity that $A(x)$ is such that $\rho^0(\eta)$ defined
by (\ref{eq:WKBdensity}) is entire.  Let $\cal C$ be a family of loop
contours $C$ in the cut upper half-plane ${\mathbb H}$ that begin and
end at the origin.  For each contour $C\in {\cal C}$, let $d\mu_C^*$
be a measure minimizing the weighted energy $E$:
\begin{equation}
E[d\mu_C^*] = \inf_{d\mu\ge 0, \,\,{\rm supp}(d\mu)\subset C}E[d\mu]\,.
\end{equation}
Suppose $C^*\in {\cal C}$ can be found such that
\begin{equation}
E[d\mu_{C^*}^*]=\sup_{C\in{\cal C}}E[d\mu_C^*]\,.
\end{equation}
Then, the extremal measure $d\mu_{C^*}^*$ is unique modulo point
masses at the origin, and its support consists of a finite number of
analytic arcs, one of which meets the origin.  Writing
$d\mu_{C^*}^*=-\rho^\sigma(\eta)\,d\eta$ defines a density function
$\rho^\sigma(\eta)$ that is admissible in the sense of Definition
\ref{def:admissiblerho} and thus generates a complex phase function
permitting the asymptotic analysis of the semiclassical soliton
ensemble corresponding to the initial data $A(x)$.
\label{conjecture:maxmin}
\end{conjecture}

Posing the semiclassical limit for the focusing nonlinear
Schr\"odinger equation as a constrained minimum capacity problem thus
closes the circle.  We began our analysis of the inverse problem in
\S\ref{sec:phase} with the observation that what we were essentially
dealing with was a problem of rational interpolation \index{rational
interpolation} of entire functions; indeed the set of minimal weighted
capacity has played a central role in the theory of rational
approximation for several years.  We plan to address these issues more
carefully in the future.

\chapter{Conclusion and Outlook}
\label{sec:conclusion}
The generalized steepest-descents scheme we have described in detail
for analyzing the semiclassical limit of the initial-value problem
(\ref{eq:IVP}) for the focusing nonlinear Schr\"odinger equation
provides what we believe to be the first rigorous result of its kind:
that solutions of a sequence of well-posed problems ({\em i.e.} the
initial-value problem (\ref{eq:IVP}) with WKB-modified initial data
corresponding to true data of the form $\psi_0(x)=A(x)$ for the
sequence $\hbar=\hbar_N$) converge to an object whose macroscopic
properties (weak limits of conserved local densities) are described by
a system of elliptic modulation equations, whose initial-value problem
is significantly less well-behaved.  Our methods allow the convergence
to be effectively analyzed for times that are not necessarily small,
and in particular for times beyond which the solutions become wild and
oscillatory.

From the point of view of the semiclassical limit for the
initial-value problem (\ref{eq:IVP}), the tools we have developed in
the preceding pages will provide an avenue toward the analysis of many
open problems.  For example, questions of the way that limits of
solutions depend on the analyticity properties of the initial data can
be systematically addressed (see \cite{CM00} for some recent
considerations in this direction).

But also from the point of view of other problems that can be attacked
by Riemann-Hilbert methods ({\em e.g.} long-time asymptotics for
integrable partial differential equations, problems in approximation
theory and statistical analysis of random matrix ensembles, and some
related combinatorial problems), the generalization of the steepest
descent method of Deift and Zhou that we have presented here is likely
to be useful as a general technique.  For example, certain problems in
the theory of orthogonal polynomials involving exotic orthogonality
conditions can be treated by our methods.

In this final chapter, we would like to outline several ways that we
would like to consider extending what we have presented.

\section{Generalization for Non-Quantum Values of $\hbar$}
An essential role was played in our work by the assumption that the
semiclassical parameter $\hbar$ should be restricted to a particular
discrete sequence of values as it goes to zero.  Thus, technically
speaking we have only established that the explicit model we have
presented in terms of Riemann theta functions is a strong limit point
for the semiclassical asymptotics.  However unlikely it may seem, we
cannot {\em a priori} rule out the possibility that there could be
other limit points as well, that one might find by considering values
of $\hbar$ intermediate to those in the sequence $\hbar=\hbar_N$.

Therefore, it appears that some advantage would be gained by addressing
the asymptotic behavior of soliton ensembles without the quantization
restriction on $\hbar$.  For the special case of the Satsuma-Yajima
initial data $\psi_0(x)=A(x)$, the {\em exact} spectral data becomes
somewhat more complicated when general values of $\hbar$ are considered,
since there is a nonzero reflection coefficient when $\hbar\neq\hbar_N$
for any $N$.  In fact, the reflection coefficient is not even uniformly
small in any neighborhood of $\lambda=0$ as $\hbar\rightarrow 0$.  

In this special case, however (and also in the recent cases described 
in \cite{TV00}), at least one has an exact formula for the reflection
coefficient, and since it is small except near $\lambda=0$, it could
be taken into account at the level of the local model Riemann-Hilbert
problem \ref{rhp:Fhat} for the matrix ${\bf F}^\sigma(\zeta)$.  In fact,
the reader will observe that without the incorporation of the reflection
coefficient into this Riemann-Hilbert problem, the jump matrices will
not even satisfy the compatibility condition (\ref{eq:compatibility})
for general values of $\hbar$ (we needed to assume that $\hbar=\hbar_N$
to obtain the desired compatibility), and consequently the problem
will be unsolvable.  

As interesting as it will be to see how to handle the intermediate
values of $\hbar$ for the Satsuma-Yajima special case, it is of more
interest to have a scheme that works for general soliton ensembles.
Here, the difficulty is that an accurate WKB approximation is required
for the reflection coefficient in a small enough (shrinking)
neighborhood of the origin.  Formal WKB theory simply predicts the
pointwise (fixed $\lambda$) convergence of the reflection coefficient
to zero, and fails to capture the asymptotic structure of the
coefficient near the origin.  This structure would be needed to
generalize our techniques to general values of $\hbar$ for arbitrary
soliton ensembles.

\section{Effect of Complex Singularities in $\rho^0(\eta)$}
For the Satsuma-Yajima ensemble, the function $\rho^0(\eta)$ defined
by (\ref{eq:WKBdensity}) is entire, and there is no obstruction
whatsoever to the placement of the contour $C$ anywhere in the cut
upper half-plane ${\mathbb H}$.  However, for other real-analytic
bell-shaped initial data, even data with sufficient decay and
curvature at its peak to admit analytic continuation of $\rho^0(\eta)$
to a complex neighborhood of the interval $[0,iA]$, there could be
singularities some distance from $[0,iA]$ in the complex plane that
could ultimately constrain the free motion of the contour $C$
according to the variational conditions.  

Since the presence of complex singularities will be the rule rather
than the exception, it is of some interest to determine the effect of
these on the dynamics of the contour motion.  For example, it may be
the case that the contour $C$ is typically repelled by any
singularities.  On the other hand, if it is possible for the contour
to collide with a singularity of $\rho^0(\eta)$ for finite $x$ and
$t$, what can one expect to happen to the ansatz-based construction of
the complex phase function $g^\sigma(\lambda)$ at such a moment?  Is
it somehow still possible for the phase function to exist, possibly by
passing to a higher genus ansatz?  In other words, is the collision of
the contour with a complex singularity of $\rho^0(\eta)$ a possible
mechanism for phase transitions?  Or might it even be the case that
upon meeting a singularity the support of the equilibrium measure becomes
irregular, with an infinite number of bands and gaps?  

\section{Uniformity of the Error Near $t=0$}
We have noted that for general soliton ensembles, we cannot control
the error of our approximation exactly at $t=0$ because here the
variational conditions select a contour loop $C$ part of which
coincides with the a subset of the imaginary interval $[0,iA]$.  Since
this is the locus of accumulation of the poles in the meromorphic
Riemann-Hilbert Problem~\ref{rhp:m}, the specified contour $C$ does not
do the job of surrounding the poles, and thus our error analysis fails.  

At the same time, however, we know that our approximation remains
valid when $t=0$, at least in the $L^2$ sense, because this
calculation can paradoxically be done directly by Lax-Levermore
methods.  Indeed this was shown explicitly in \cite{EJLM93}.  The
reason we cannot control the error is that we are trying to bound the
error in a uniform approximation of the eigenfunction in the complex
plane, rather than just worrying about the error in the potential,
$|\psi-\tilde{\psi}|$.  The eigenfunction is simply more complicated
when $t=0$ than for nonzero $t$ because there is an endpoint of the
support of the equilibrium measure (near which the local behavior of
the eigenfunction should be described in terms of Airy functions)
superimposed on the locus of accumulation of eigenvalues.  So
approximation of the eigenfunction is just a different problem at
$t=0$ than for nonzero $t$.

Nonetheless, we feel that there would be considerable advantage in
presenting a unified Riemann-Hilbert based approach to semiclassical
asymptotics for (\ref{eq:IVP}) that works for all $t$.  So we view the
development of new methods to model the matrix ${\bf
N}^\sigma(\lambda)$ at $t=0$ as a challenge for the future.

\section{Errors Incurred by Modifying the Initial Data}
If we desire to interpret our completely rigorous results regarding
asymptotics for semiclassical soliton ensembles in the context of
the semiclassical limit for the initial-value problem (\ref{eq:IVP}),
then there is a step missing in our analysis.  Namely, we would need
to provide an estimate for arbitrary $x$ and $t$ of the errors we make
by replacing the $\hbar$-parametrized family of initial-value problems
(\ref{eq:IVP}) by another family of problems in which the initial
data has been modified so that its spectrum is replaced with its
purely discrete WKB approximation.  The latter is what we have been
referring to throughout as a soliton ensemble, and for which we compute
accurate asymptotics.  

This modification of the initial data was also an essential ingredient
in Lax and Levermore's original analysis of the zero-dispersion limit
for the Korteweg-de Vries equation.  It has also been built into each
analogous study of an integrable system since that time.  As pointed
out above, the Lax-Levermore method proves $L^2$ convergence of the
modified initial data to the true initial data.  And in the light of
the local well-posedness of the hyperbolic Whitham equations that they
prove governs the limit, it is a reasonable claim that this convergence
also holds for finite nonzero times.  

But in our study of the focusing nonlinear Schr\"odinger equation, we
find that the limit is governed by Whitham equations that are
elliptic.  Without any local well-posedness for this asymptotic
dynamical model, we must look elsewhere if we wish to control the
errors introduced by modifying the initial data.  

The best bet may be to study more carefully the forward-scattering
problem for the nonselfadjoint Zakharov-Shabat eigenvalue problem in
the $\hbar\downarrow 0$ limit.  If it would be possible to directly
estimate the error of the WKB approximation at the level of the
scattering data, then this error could be built into the
Riemann-Hilbert analysis ({\em cf.} \S\ref{sec:error}) as another
layer of approximation to be expanded in a Neumann series and
consequently controlled.

\section{Analysis of the Max-Min Variational Problem}
At the philosophical heart of our work is
Conjecture~\ref{conjecture:maxmin}, that the ``max-min'' problem for
the weighted Green's energy functional described in
Chapter~\ref{sec:variational} is a natural and appropriate
generalization of the celebrated variational principle of Lax and
Levermore's zero-dispersion analysis, and that it characterizes the
semiclassical limit of the initial-value problem (\ref{eq:IVP}) for
the focusing nonlinear Schr\"odinger equation.  We may even speculate
that a variational principle of this type characterizes the limit
for a class of initial data that is much more general than what we
have considered here.

In the Lax-Levermore method, the variational principle plays a central
role.  It is the equilibrium measure that determines the
zero-dispersion limit for the Korteweg-de Vries equation in general,
and if the initial data is smooth enough then it can be shown
\cite{DKM98} that the support of the equilibrium measure is sufficiently 
regular (and in particular consists of a finite number of bands and
gaps) that it may be constructed by an ansatz-based method of which
what we presented in Chapter~\ref{sec:ansatz} is a
generalization.

The reader will observe that our approach to the semiclassical limit
for the focusing nonlinear Schr\"odinger equation has been quite
different.  We began with analyticity of the initial data, and
constructed the complex phase function $g^\sigma(\lambda)$ directly,
by ansatz.  Then, after the fact, we observed in
Chapter~\ref{sec:variational} that the complex phase function could be
given a variational interpretation.  Indeed, if it turned out that the
max-min problem had a solution for which the support of the
equilibrium measure were severely irregular, we would not know how to
use it in the Riemann-Hilbert analysis to asymptotically reduce the
phase conjugated Riemann-Hilbert Problem~\ref{rhp:N} to a simple form.
In short, the Riemann-Hilbert approach would appear to manifestly
depend on analyticity ({\em e.g.} the deformation of ``opening the
lenses'' is only possible if $\rho^\sigma(\eta)\,d\eta$ is an analytic
measure).

At the same time, we would need to develop existence and regularity
arguments for the max-min problem.  Some results of this type do
indeed exist in the literature for problems that are perhaps not too
different from ours.  Even if it turns out that regularity properties
are somehow built in from the start, it would be useful to have a
sufficiently developed theory of the max-min problem that we could
estimate the genus for given values of $x$ and $t$ in terms of
elementary properties of the initial data.

\section{Initial Data with $S(x)\not\equiv 0$}
Throughout, we have assumed $S(x)\equiv 0$, and considered the
initial-value problem (\ref{eq:IVP}) with purely real initial data.
Dropping the assumption $S(x)\equiv 0$ is expected to require new
alterations in our method.  In this case, the existing numerical
evidence \cite{B96} and formal calculations
\cite{M00} suggest that for analytic potentials the eigenvalues
accumulate as $\hbar$ tends to zero on a union of curves in the
complex plane.  In the examples presented in the literature, the
asymptotic spectrum is distributed on a ``Y-shaped'' curve consisting
of a ``neck'' on the imaginary axis connected to the origin, out of
which are born two ``branches'' on which the eigenvalues have nonzero
real parts.  While there exist analogs of the Bohr-Sommerfeld
quantization rule for these Y-shaped specta \cite{M00}, there
is not an adequate WKB description of the proportionality constant
$\gamma_k$ connected with each discrete eigenvalue.

Even with the forward-scattering analysis in this primitive state of
affairs, it may be possible to apply our methods to semiclassical
soliton ensembles appropriately defined from the formal WKB
approximations for $S(x)\not\equiv 0$.  The basic set-up should be the
same, with a holomorphic Riemann-Hilbert problem being posed relative
to a loop contour that surrounds the whole (possibly Y-shaped) locus
of accumulation of the eigenvalues in the upper half-plane.  The
conditions imposed on the complex phase function $g^\sigma(\lambda)$
in \S\ref{sec:conditions} should be unchanged; similarly the error
analysis should require only technical modifications.  The part of the
method that will need to be rethought almost completely is the
ansatz-based construction of $g^\sigma(\lambda)$ ({\em cf.}
Chapter~\ref{sec:ansatz}), since for an asymptotic eigenvalue measure
$\rho^0(\eta)\,d\eta$ with complicated support properties it is not
clear how to ``push'' the measure by analyticity onto the loop contour
$C$.  We plan to generalize our method to handle these more general
semiclassical soliton ensembles in the near future.

\section{Final Remarks}
Thus, the story of the semiclassical limit for the focusing nonlinear
Schr\"odinger equation is far from complete.  Even for analytic
initial data, much work remains on the front of rigorous WKB theory
for the nonselfadjoint Zakharov-Shabat operator, as well as on the
development of the variational-theoretic aspect of the inverse problem
described in Chapter~\ref{sec:variational} which has the potential to be a
very powerful analytical tool.  And once the problem is rigorously
understood for given analytic initial data, the next task is to
determine the sensitivity of the resulting semiclassical limit within
this analytic class.  How (un)stable are the caustic curves (phase
transitions) separating one kind of local behavior from another with
respect to, say, $L^2$-small analytic perturbations?  Is there sense
in putting a probability measure on some class of initial data and
determining the statistics of the local genus $G(x,t)$ considered as a
random variable?  Sometimes, instability can be mollified by
statistical averaging, and this could be one avenue toward giving
useful physical meaning to the semiclassical limit of the focusing
nonlinear Schr\"odinger equation.

\appendix
\chapter{H\"older Theory of Local Riemann-Hilbert Problems}
\label{sec:A1}
In this appendix we collect together a number of results, some quite
classical, with the aim of rigorously establishing in some generality
the existence, uniqueness, and decay properties of solutions of
``local'' Riemann-Hilbert problems of the type that often arise in
steepest-descent type calculations.  Throughout this appendix, a fixed
norm $\|\cdot\|$ on matrices is assumed.
\section[Local Riemann-Hilbert Problems]{Local Riemann-Hilbert problems.  Statement of results.}
\begin{figure}[h]
\begin{center}
\mbox{\psfig{file=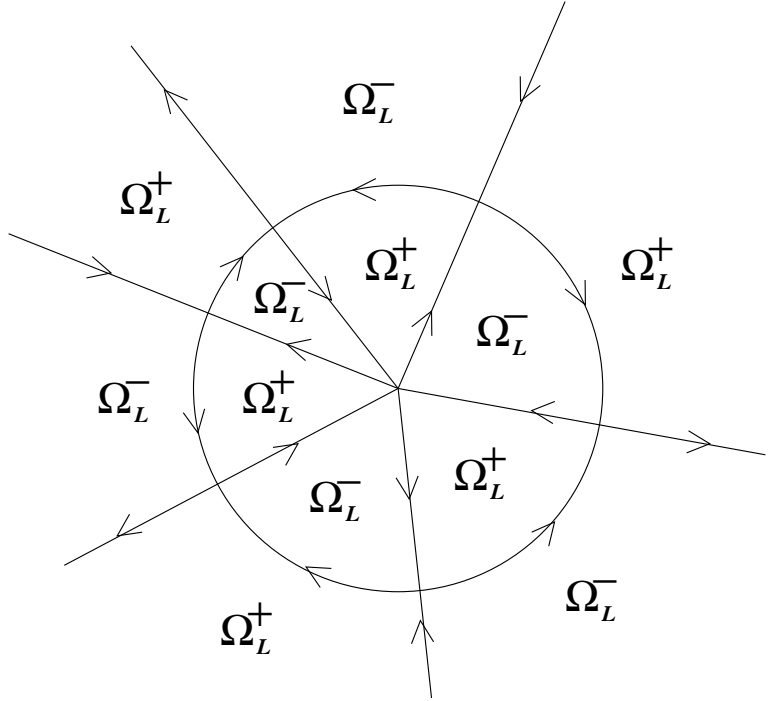,width=4 in}}
\end{center}
\caption{\em A contour for a local Riemann-Hilbert problem.}
\label{fig:FredholmContour}
\end{figure}

Let us define what we mean by a {\em local Riemann-Hilbert problem}.
Let $\Sigma_{\bf L}$ be a union of an even number of straight-line
rays emanating from the origin in the $\zeta$-plane, and a circle of
radius $R\neq 1$ centered at the origin.  Note that $R\neq 1$ can
always be arranged by a simple rescaling of $\zeta$.  The contour
$\Sigma_{\bf L}$ divides the $\zeta$-plane into two disjoint regions,
$\Omega_{\bf L}^+$ and $\Omega_{\bf L}^-$, each of which may comprise
several simply connected regions, such that each ray or circular arc
of $\Sigma_{\bf L}$ forms part of the boundary of both $\Omega_{\bf
L}^+$ and $\Omega_{\bf L}^-$.  Choose some labeling of the regions
consistent with this description; now each is a component of either
$\Omega_{\bf L}^+$ or $\Omega_{\bf L}^-$.  Consider $\Sigma_{\bf L}$
oriented such that it forms the positively oriented boundary of the
region $\Omega_{\bf L}^+$.  See Figure~\ref{fig:FredholmContour}.
The jump matrices we consider are defined as follows.
\begin{definition}[Admissible jump matrices for local problems]
Let $\zeta_1,\dots,\zeta_N$ denote the points of intersection of the
rays of $\Sigma_{\bf L}$ with the circle.  An admissible jump matrix
for a local Riemann-Hilbert problem is a 
matrix-valued function ${\bf v}_{\bf L}(\zeta)$ 
defined for $\zeta\in\Sigma_{\bf
L}\setminus\{0,\zeta_1,\dots,\zeta_N\}$ satisfying for some $0<\nu\le
1$ and some $K>0$ the following conditions:
\begin{enumerate}
\item{\bf Unimodularity:}  For all
$\zeta\in\Sigma_{\bf L}\setminus\{0,\zeta_1,\dots,\zeta_N\}$,
$\det({\bf v}_{\bf L}(z))=1$.
\item{\bf Interior smoothness:}  Whenever $\zeta_1$ and
$\zeta_2$ belong to the same smooth component of $\Sigma_{\bf L}$
(either ray segment or circular arc), the H\"older condition $\|{\bf
v}_{\bf L}(\zeta_2)-{\bf v}_{\bf L}(\zeta_1)\|\le K
|\zeta_2-\zeta_1|^\nu$ holds.
\item{\bf Compatibility at self-intersection points:}  
Let $\zeta_0$ denote any of the points
$0,\zeta_1,\dots,\zeta_N$.  By interior smoothness, it follows that on each
smooth component $\Sigma_{\bf L}^{(k)}$ of $\Sigma_{\bf L}$ meeting
$\zeta_0$, the limit
\begin{equation}
{\bf v}_{\bf L}^{(k)}:=\lim_{\begin{array}{c}
\scriptstyle\zeta\rightarrow \zeta_0\\\scriptstyle\zeta\in\Sigma_{\bf
L}^{(k)}\end{array}}{\bf v}_{\bf L}(\zeta)
\end{equation}
exists.  Let the components $\Sigma_{\bf L}^{(k)}$ be ordered with
increasing $k$ in the counterclockwise direction about $\zeta_0$,
starting with any given contour component.  Then the condition
\begin{equation}
{\bf v}_{\bf L}^{(1)}{\bf v}_{\bf L}^{(2)-1}{\bf v}_{\bf
L}^{(3)}{\bf v}_{\bf L}^{(4)-1}\dots{\bf v}_{\bf L}^{(n-1)}{\bf
v}_{\bf L}^{(n)-1}={\mathbb I}
\label{eq:compatibility}
\index{compatibility at self-intersection points}
\end{equation}
is to be satisfied, where $n$ is the even number of contour components
meeting at $\zeta_0$ (this number is exactly $4$ for
$\zeta_0=\zeta_1,\dots,\zeta_N$ and $N$ for $\zeta_0=0$).
\item{\bf Decay:}  As $\zeta\rightarrow\infty$ on
any ray of $\Sigma_{\bf L}$, $\|{\bf v}_{\bf L}(\zeta)-{\mathbb I}\|={\cal
O}(|\zeta|^{-\nu})$.
\end{enumerate}
\label{def:JL}
\end{definition}

Let ${\bf N}_{\bf L}$ be any constant matrix.  Given the data $(\Sigma_{\bf L},{\bf
v}_{\bf L}(\zeta),{\bf N}_{\bf L})$ we pose the following problem.
\begin{rhp}[Local problem in the H\"older sense]
\index{Riemann-Hilbert problem!local problem in the H\"older sense}
Let ${\bf v}_{\bf L}(\zeta)$ be an admissible jump matrix defined on a
contour $\Sigma_{\bf L}$ as in Figure~\ref{fig:FredholmContour}.  Let
$\nu$ be the H\"older exponent of the jump matrix.  Find a matrix
function ${\bf L}(\zeta)$ with the following properties:
\begin{enumerate}
\item
{\bf Analyticity:}  The matrix function ${\bf L}(\zeta)$ is holomorphic
in ${\mathbb C}\setminus \Sigma_{\bf L}$.
\item
{\bf Boundary Behavior:} For each $\mu<\nu$, ${\bf L}(\zeta)$ assumes
H\"older continuous boundary values from each connected component of
its domain of analyticity, including corner points, with H\"older
exponent $\mu$.  More precisely, for each $\zeta\in\Sigma_{\bf
L}\setminus\{0,\zeta_1,\dots,\zeta_N\}$, the boundary values
\begin{equation}
{\bf L}_\pm(\zeta):=\lim_{\begin{array}{c}
\scriptstyle\lambda\rightarrow\zeta\\
\scriptstyle\lambda\in\Omega_{\bf
L}^\pm\end{array}}{\bf L}(\lambda)
\end{equation}
exist independently of the path of approach.  For all positive
$\mu<\nu$, there exists a constant $K'$ such that for all
$\zeta\in\Sigma_{\bf L}\setminus\{0,\zeta_1,\dots,\zeta_N\}$ and all
$\lambda\in
\Omega_{\bf L}^\pm$, $\|{\bf L}(\lambda)-{\bf L}_\pm(\zeta)\|\le
K'|\lambda-\zeta|^\mu$.  Also, whenever $\zeta_1$ and $\zeta_2$ belong
to the same smooth component of $\Sigma_{\bf L}$, $\|{\bf
L}_\pm(\zeta_2)-{\bf L}_\pm(\zeta_1)\|\le K'|\zeta_2-\zeta_1|^\mu$.
Also, whenever $\Sigma_{\bf L}^{(j)}$ and $\Sigma_{\bf L}^{(k)}$ are
two smooth components of the contour bounding a connected component of
$\Omega_{\bf L}^+$ and meeting at $\zeta=\zeta_0$, we have
\begin{equation}
\lim_{\begin{array}{c}
\scriptstyle\zeta\rightarrow \zeta_0\\\scriptstyle
\zeta\in\Sigma_{\bf L}^{(j)}\end{array}}{\bf
L}_+(\zeta)= \lim_{\begin{array}{c}\scriptstyle
\zeta\rightarrow\zeta_0\\\scriptstyle\zeta\in\Sigma_{\bf L}^{(k)}\end{array}}
{\bf L}_+(\zeta)\,,
\end{equation}
and both limits are finite.  Similarly, if
$\Sigma_{\bf L}^{(j)}$ and $\Sigma_{\bf L}^{(k)}$ meet at $\zeta_0$ and
bound a connected component of
$\Omega_{\bf L}^-$, then
\begin{equation}
\lim_{\begin{array}{c}\scriptstyle
\zeta\rightarrow \zeta_0\\\scriptstyle\zeta\in\Sigma_{\bf
L}^{(j)}\end{array}}{\bf L}_-(\zeta)=
\lim_{\begin{array}{c}\scriptstyle
\zeta\rightarrow \zeta_0\\\scriptstyle\zeta\in\Sigma_{\bf L}^{(k)}\end{array}}
{\bf L}_-(\zeta)\,, 
\end{equation}
with both limits being finite.
\item
{\bf Jump conditions:}  For each
$\zeta\in\Sigma_{\bf L}\setminus\{0,\zeta_1,\dots,\zeta_N\}$, the
boundary values are related by ${\bf L}_+(\zeta)={\bf L}_-(\zeta){\bf
v}_{\bf L}(\zeta)$.
\item
{\bf Normalization:}  As
$\zeta\rightarrow\infty$ in each ray of $\Sigma_{\bf L}$, $\|{\bf
L}_\pm(\zeta)-{\bf N}_{\bf L}\|={\cal O}(|\zeta|^{-\mu})$ for all
positive $\mu<\nu$.  The same estimate with
${\bf L}_\pm(\zeta)$ replaced by ${\bf L}(\zeta)$ holds uniformly in all
other directions.
\end{enumerate}
\label{rhp:local}
\end{rhp}

Note that in formulating any local Riemann-Hilbert problem,
the condition that $\Sigma_{\bf L}$ should consist of an even number
of rays can always be achieved by adjoining a new ray on which the
jump matrix is taken to be the identity matrix.  Also, we emphasize
that in many applications the circle will be absent altogether, which
can also be accomplished by taking the jump to be the identity there.
There is no cost for adding identity jumps to the contour for
Riemann-Hilbert problems posed in the class of H\"older continuous
boundary values.  This follows from an elementary Cauchy integral
argument establishing that whenever the jump matrix is equal to the
identity on a smooth contour segment and the solution takes on
boundary values as above (and in particular in the sense of uniform
continuity) then the solution matrix is in fact analytic {\em on} the
contour segment.

We will establish here the following results.
\begin{theorem}[Local Fredholm Alternative]
\index{Fredholm alternative}
\label{theorem:LocalAlternative}
Suppose that a set of data $(\Sigma_{\bf L},{\bf v}_{\bf
L}(\zeta),{\bf N}_{\bf L})$ where the jump matrix ${\bf v}_{\bf
L}(\zeta)$ satisfies the conditions of Definition~\ref{def:JL}.  Then,
the associated local Riemann-Hilbert Problem~\ref{rhp:local}
has a unique solution if and only if the corresponding
homogeneous problem with data $(\Sigma_{\bf L},{\bf v}_{\bf
L}(\zeta),{\bf 0})$ has only the trivial solution ${\bf
L}(\zeta)\equiv {\bf 0}$.
\end{theorem}

\begin{figure}[h]
\begin{center}
\mbox{\psfig{file=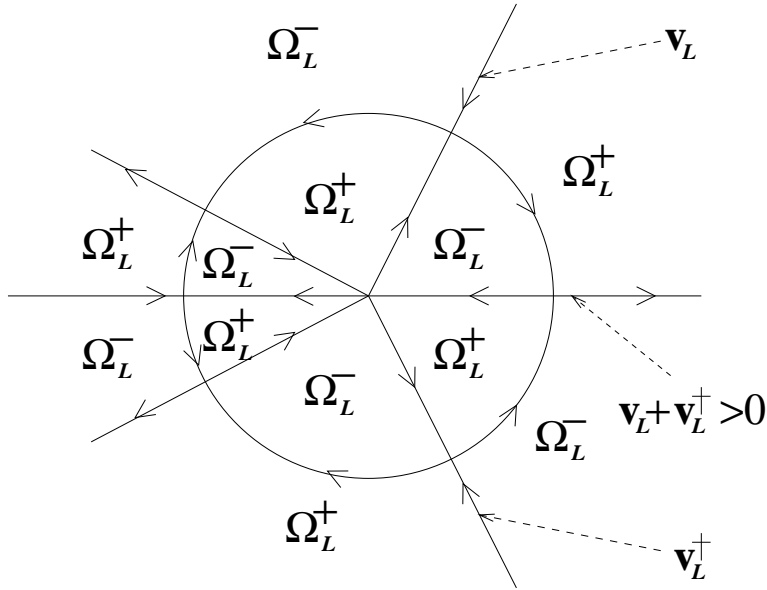,width=4 in}}
\end{center}
\caption{\em A contour $\Sigma_{\bf L}$ with Schwartz reflection symmetry.}
\label{fig:SymmetricContour}
\end{figure}

We note that if a solution ${\bf L}(\zeta)$ exists for some particular
H\"older exponent $\mu_0<\nu$, then ${\bf L}(\zeta)$ also serves as a
solution for all positive H\"older exponents $\mu<\mu_0$ .  Therefore,
to establish solvability with H\"older exponent $\mu$ for {\em all}
$\mu<\nu$, it is sufficient to find a number $\mu_0<\nu$ such that all
homogeneous solutions with H\"older exponent $\mu>\mu_0$ are trivial.
As not every problem is solvable, there is no completely general
method for ruling out nontrivial vanishing solutions, and a variety of
techniques and results from complex analysis are useful.  What appears
to be the most general result holds for contours and jump matrices
that satisfy the following symmetry criterion:
\begin{equation}
\parbox{5 in}{
{\bf Schwartz reflection symmetry:} 
\index{Schwartz reflection symmetry}
The contour $\Sigma_{\bf L}$
contains the real axis, is invariant under complex-conjugation and for
all $\zeta\in\Sigma_{\bf L}$ with $\Im(\zeta)\neq 0$, ${\bf v}_{\bf
L}(\zeta) = {\bf v}_{\bf L}(\zeta^*)^\dagger$ while for all real
$\zeta$, the matrix ${\bf v}_{\bf L}(\zeta)+{\bf v}_{\bf
L}(\zeta)^\dagger$ is strictly positive definite.
}
\label{eq:J5L}
\end{equation}
The condition that the real axis be part of a contour satisfying
(\ref{eq:J5L}) may be always be satisfied without loss of generality
by taking the jump matrix to be the identity there (this obviously
satisfies the positive definiteness condition).  A contour with the
symmetry of (\ref{eq:J5L}) is illustrated in
Figure~\ref{fig:SymmetricContour}.

Then, we have
\begin{theorem}[Local Unique Solvability]
\label{theorem:LocalSolvability}
Suppose that the data set
$(\Sigma_{\bf L},{\bf v}_{\bf
L}(\zeta),{\bf N}_{\bf L})$ satisfies condition (\ref{eq:J5L}) in
addition to the requirements of Theorem~\ref{theorem:LocalAlternative}.  Then
the associated local Riemann-Hilbert Problem~\ref{rhp:local} has
a unique solution.
\end{theorem}

\begin{remark}
The main difficulty lies in proving the existence of solutions of the
required type.  Indeed, uniqueness holds even without the assumption
(\ref{eq:J5L}).  One simply considers the matrix quotient of any two
solutions ${\bf Q}:={\bf L}(\zeta){\bf L}'(\zeta)^{-1}$, which by the
Banach algebra property of the H\"older spaces (see below for precise
definitions of these spaces) again takes on boundary values in the
H\"older (and in particular uniformly continuous) sense.  The boundary
values ${\bf Q}_+(\zeta)$ and ${\bf Q}_-(\zeta)$ are easily seen to be
the same for all points $\zeta\in\Sigma_{\bf L}\setminus\{0\}$, and
${\bf Q}(\zeta)\rightarrow {\mathbb I}$ as $\zeta\rightarrow\infty$.
Given the behavior of the boundary values, an elementary Cauchy
integral argument shows that ${\bf Q}(\zeta)$ is in fact analytic in
the whole plane, and then it follows from Liouville's theorem that
${\bf Q}(\zeta)\equiv {\mathbb I}$.  It is possible to push through
the same argument with less control on the boundary values; in
\cite{D99} this is explained in the $2\times 2$ case when the H\"older
smoothness property of the boundary values is replaced with $L^2$
convergence, so that in particular one admits solutions for which the
boundary values become unbounded.  Therefore if there exists a
solution with H\"older class boundary values, it is unique in much
larger spaces, including at least $L^2$.
\end{remark}

Finally, some refinement of the decay of the solution of the local
Riemann-Hilbert Problem \ref{rhp:local} is possible under certain
additional conditions.
\begin{theorem}[Enhanced Decay]
Suppose that on the interior of each smooth arc of $\Sigma_{\bf L}$
the jump matrix ${\bf v}_{\bf L}(\zeta)$ is analytic, and that from
any ray component $\Sigma_{\bf L}^{(k)}$ of $\Sigma_{\bf L}$ the jump
matrix may be continued to either side within a strip $S_k$ bounded by
two parallel rays such that the moments
\begin{equation}
{\bf v}^{(\infty,k)}:=\lim_{\zeta\rightarrow\infty\atop\zeta\in S_k} 
\zeta({\bf v}_{\bf L}(\zeta)-{\mathbb I})
\label{eq:momentdefine}
\end{equation}
all exist uniformly in $S_k$.  Also assume that the condition
\begin{equation}
\sum_{\mbox{\rm \scriptsize all rays } k}{\bf v}^{(\infty,k)} = {\bf 0}
\label{eq:cancellation}
\end{equation}
is satisfied.  Then if there exists a solution ${\bf L}(\zeta)$ of
Riemann-Hilbert Problem \ref{rhp:local}, there is a constant $M>0$ such
that uniformly for all $\zeta$ sufficiently large,
\begin{equation}
\|{\bf L}(\zeta)-{\mathbb I}\|\le M|\zeta|^{-1}\,.
\end{equation}
\label{theorem:decay}
\end{theorem}

\begin{remark}
The refinement afforded by Theorem \ref{theorem:decay} is significant,
since without the condition (\ref{eq:cancellation}), the typical decay
is only $\bo (|\zeta|^{-1}\log |\zeta|)$.  Note also, that the
condition (\ref{eq:cancellation}) may be viewed as a ``higher-order''
version of the compatibility condition 
\index{compatibility condition at self-intersection points!higher-order version of}(\ref{eq:compatibility}) that
is necessary for solvability.
\end{remark}

\section[Umbilical Riemann-Hilbert Problems]{Umbilical Riemann-Hilbert problems.}
\begin{figure}[h]
\begin{center}
\mbox{\psfig{file=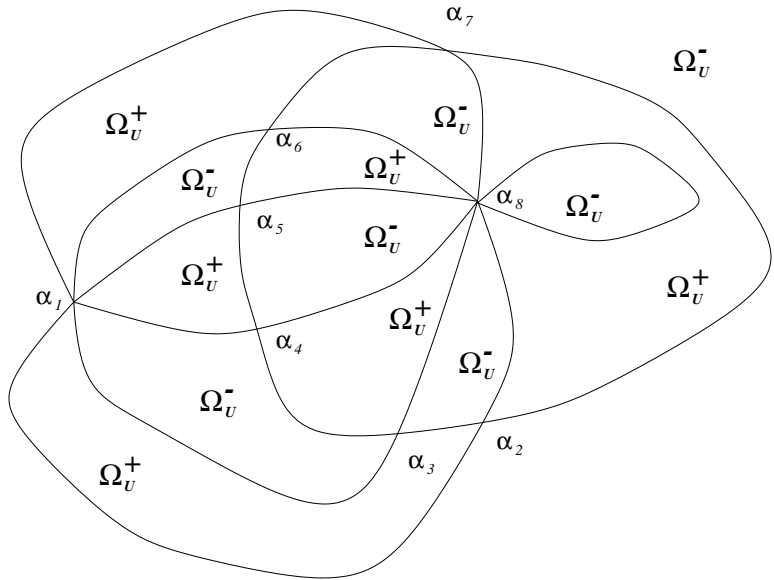,width=4 in}}
\end{center}
\caption{\em An example of an umbilical contour.}
\label{fig:UmbilicalGeneral}
\end{figure}

We now consider posing a certain type of auxiliary Riemann-Hilbert
problem in the $z$-plane on a {\em compact contour}.  Let us now
describe what we mean by an {\em umbilical Riemann-Hilbert problem}.
\index{Riemann-Hilbert problem!umbilical problem} 
Let $\Sigma_{\bf U}$ be any compact contour consisting of a union of a
finite number of smooth closed arcs terminating at self-intersection
points $z=\alpha_k$ such that $\Sigma_{\bf U}$ divides the $z$-plane
into two disjoint regions, $\Omega_{\bf U}^+$ and $\Omega_{\bf U}^-$,
while serving as the positively oriented boundary of $\Omega_{\bf
U}^+$ and at the same time the negatively oriented boundary of
$\Omega_{\bf U}^-$.  This means in particular that an even number of
arcs meet at each intersection point $\alpha_k$.  The geometry of an
umbilical contour is shown in Figure~\ref{fig:UmbilicalGeneral}.
\begin{definition}[Admissible jump matrices for umbilical problems]
An admissible jump matrix for an umbilical Riemann-Hilbert problem
is a matrix-valued function ${\bf v}_{\bf U}(z)$ defined for $z\in \Sigma_{\bf
U}\setminus\{\alpha_k\}$ with the following properties for some $K''>0$
and $0<\nu\le 1$:
\begin{enumerate}
\item
{\bf Unimodularity:}  
For all $z\in\Sigma_{\bf U}\setminus\{\alpha_k\}$, 
$\det({\bf v}_{\bf U}(z))=1$.
\item
{\bf Interior smoothness:} Whenever $z_1$ and $z_2$
belong to the same smooth arc of $\Sigma_{\bf U}$, the H\"older condition
$\|{\bf v}_{\bf U}(z_2)-{\bf v}_{\bf U}(z_1)\|\le K'' |z_2-z_1|^\nu$
holds.
\item
{\bf Compatibility at self-intersection
points:}  Let $z_0$ be any of the points $\alpha_i$ of
self-intersection of $\Sigma_{\bf U}$, let $\Sigma_{\bf
U}^{(1)},\dots,\Sigma_{\bf U}^{(N)}$ be the open arcs meeting at
$z=\alpha_i$ enumerated in counter-clockwise order, and define
\begin{equation}
{\bf v}_{\bf U}^{(k)}:=
\lim_{\begin{array}{c}\scriptstyle
z\rightarrow z_0\\\scriptstyle
z\in\Sigma_{\bf U}^{(k)}\end{array}} {\bf v}_{\bf U}(z)\,.
\end{equation}
Then for each such point $z_0$, the condition 
\begin{equation}
{\bf v}_{\bf
U}^{(1)}{\bf v}_{\bf U}^{(2)-1}{\bf v}_{\bf U}^{(3)}{\bf v}_{\bf
U}^{(4)-1}\dots {\bf v}_{\bf U}^{(N-1)}{\bf v}_{\bf
U}^{(N)-1}={\mathbb I}
\end{equation}
holds.
\end{enumerate}
\label{def:JU}
\end{definition}

Let ${\bf N}_{\bf U}$ be any constant matrix.  The umbilical
Riemann-Hilbert problem associated with the data $(\Sigma_{\bf U},{\bf
v}_{\bf U}(z),{\bf N}_{\bf U})$ is posed as follows.
\begin{rhp}[Umbilical problem]
Find a matrix function
${\bf U}(z)$ with the following properties:
\begin{enumerate}
\item
{\bf Analyticity:}  The matrix function ${\bf U}(z)$ is holomorphic
in ${\mathbb C}\setminus \Sigma_{\bf U}$, and also at $z=\infty$.
\item
{\bf Boundary behavior:} ${\bf U}(z)$ assumes H\"older continuous
boundary values from each connected component of its domain of
analyticity, including self-intersection points, with all exponents
$\mu<\nu$.  More precisely, for each $z\in\Sigma_{\bf
U}\setminus\{\alpha_k\}$, the boundary values 
\begin{equation}
{\bf U}_\pm(z):=\lim_{\begin{array}{c}
\scriptstyle w\rightarrow z\\\scriptstyle w\in\Omega_{\bf U}^\pm\end{array}}
{\bf U}(w)
\end{equation}
exist independently of the path of approach.  For all positive
$\mu<\nu$, there exists a constant $K'$ such that for all
$z\in\Sigma_{\bf U}\setminus\{\alpha_k\}$ and all $w\in \Omega_{\bf
U}^\pm$, $\|{\bf U}(w)-{\bf U}_\pm(z)\|\le K'|w-z|^\mu$.  Also,
whenever $z_1$ and $z_2$ belong to the same arc of $\Sigma_{\bf U}$,
$\|{\bf U}_\pm(z_2)-{\bf U}_\pm(z_1)\|\le K'|z_2-z_1|^\mu$.  Also,
whenever $\Sigma_{\bf U}^{(j)}$ and $\Sigma_{\bf U}^{(k)}$ are two
arcs meeting at a self-intersection point $z=\alpha_i$ and bounding a
connected component of $\Omega_{\bf U}^+$, we have
\begin{equation}
\lim_{\begin{array}{c}\scriptstyle z\rightarrow
\alpha_i\\\scriptstyle
z\in\Sigma_{\bf U}^{(j)}\end{array}}
{\bf U}_+(z)= \lim_{\begin{array}{c}\scriptstyle
z\rightarrow \alpha_i\\\scriptstyle
z\in\Sigma_{\bf U}^{(k)}\end{array}}{\bf
U}_+(z)\,, 
\end{equation}
and both limits are finite.  Similarly, if $\Sigma_{\bf U}^{(j)}$ and
$\Sigma_{\bf U}^{(k)}$ meet at $z=\alpha_i$ and bound a connected component of
$\Omega_{\bf U}^-$, then
\begin{equation}
\lim_{\begin{array}{c}\scriptstyle
z\rightarrow \alpha_i\\\scriptstyle
z\in\Sigma_{\bf U}^{(j)}\end{array}}{\bf U}_-(z)=
\lim_{\begin{array}{c}\scriptstyle
z\rightarrow \alpha_i\\\scriptstyle
z\in\Sigma_{\bf U}^{(k)}\end{array}}{\bf U}_-(z)\,, 
\end{equation}
with both limits being finite.
\item{\bf Jump condition:}  For each $z\in\Sigma_{\bf
U}\setminus\{\alpha_i\}$, the boundary values are related by ${\bf
U}_+(z)={\bf U}_-(z){\bf v}_{\bf U}(z)$.
\item
{\bf Normalization:}  ${\bf U}(\infty)={\bf N}_{\bf U}$.
\end{enumerate}
\label{rhp:umbilical}
\end{rhp}

Each local Riemann-Hilbert problem is equivalent to an umbilical
Riemann-Hilbert problem.  
\begin{lemma}
\label{lemma:LocalToUmbilical}
Consider a local Riemann-Hilbert Problem~\ref{rhp:local} for an
unknown matrix ${\bf L}(\zeta)$ corresponding to the data
$(\Sigma_{\bf L},{\bf v}_{\bf L}(\zeta),{\bf N}_{\bf L})$, with ${\bf
N}_{\bf L}$ invertible and with the jump matrix satisfying
Definition~\ref{def:JL}.  Let $\theta$ be any angle different from
those of all the rays of $\Sigma_{\bf L}$, and introduce the
automorphism of the Riemann sphere \index{Riemann sphere!automorphism of} 
given by
\begin{equation}
z(\zeta)=\frac{\zeta e^{-i\theta}-1}{\zeta e^{-i\theta}+1}\,,
\hspace{0.3 in}
\mbox{with inverse}\hspace{0.3 in}
\zeta(z)=e^{i\theta}\frac{1+z}{1-z}\,.
\end{equation}
This transformation defines an umbilical contour in the $z$-plane by
$\Sigma_{\bf U}:=z(\Sigma_{\bf L})$, preserving orientation of each
smooth component.  On $\Sigma_{\bf U}$, define a jump matrix by ${\bf
v}_{\bf U}(z):={\bf v}_{\bf L}(\zeta(z))$.  Then the jump matrix so
defined satisfies the conditions of Definition~\ref{def:JU},
and the
solution(s) of the umbilical Riemann-Hilbert Problem~\ref{rhp:umbilical}
with data $(\Sigma_{\bf U},{\bf v}_{\bf U}(z),{\bf N}_{\bf U})$
and ${\bf N}_{\bf U}$ invertible are in one-to-one correspondence with
those of the local Riemann-Hilbert Problem~\ref{rhp:local}
with data $(\Sigma_{\bf L},{\bf v}_{\bf L}(\zeta),{\bf N}_{\bf L})$.
\end{lemma}

\begin{proof} Under the transformation $\zeta\rightarrow
z(\zeta)$, the image of the straight line making an angle $\phi$ with
the positive real axis in the $\zeta$-plane is the graph of
\begin{equation}
|z+i\cot(\phi-\theta)|^2 = \csc^2(\phi-\theta)\,,
\end{equation}
a circle for all $\phi\neq \theta$, which always contains the two
points $z=\pm 1$.  This transformation fixes the straight line in the
$\zeta$-plane making an angle $\phi=\theta$ with the positive real
axis, and takes the origin to $z=-1$ and the point at infinity to
$z=+1$.  Furthermore, the real $\zeta$ line is mapped to a complete
circle (for $\theta\neq 0$) and all other ray components of
$\Sigma_{\bf L}$ are mapped to various circular arcs connecting $z=-1$
to $z=+1$.  The circle of radius $R\neq 1$ centered at the origin in
the $\zeta$-plane is mapped to a circle of radius $2R/(R^2-1)$
centered at $z=(R^2+1)/(R^2-1)$.  Since $\theta$ differs from the
angles of all components of $\Sigma_{\bf L}$, and since we are
assuming that $R\neq 1$, the image $\Sigma_{\bf U}:=z(\Sigma_{\bf L})$
of the contour is compact.  For contours $\Sigma_{\bf L}$ with the
additional symmetry (\ref{eq:J5L}), which by definition contain the the
real axis in the $\zeta$-plane, the condition $\theta\neq 0$ is
necessary for compactness of $\Sigma_{\bf U}$.  The image of the
contour $\Sigma_{\bf L}$ under this transformation is illustrated in
Figure~\ref{fig:fractionallinear}.
\begin{figure}[h]
\begin{center}
\mbox{\psfig{file=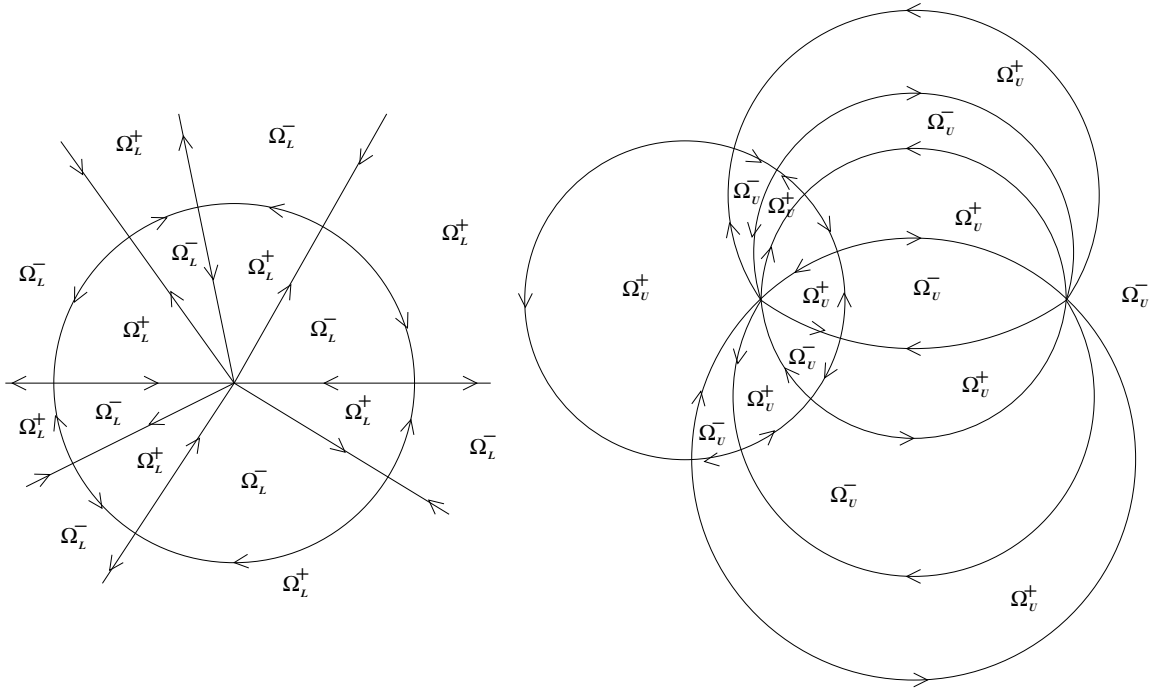,width=6 in}}
\end{center}
\caption{\em Left: The contour $\Sigma_{\bf L}$ in the $\zeta$-plane.
Right: Its image $\Sigma_{\bf U}:=z(\Sigma_{\bf L})$ in the $z$-plane
for some $\theta>0$.  Also shown are the regions $\Omega_{\bf
U}^+:=z(\Omega_{\bf L}^+)$ and $\Omega_{\bf U}^-:=z(\Omega_{\bf
L}^-)$.  }
\label{fig:fractionallinear}
\end{figure}

It is immediate from the definition and the unimodularity of ${\bf
v}_{\bf L}(\zeta)$ that the jump matrix ${\bf v}_{\bf U}(z):={\bf
v}_{\bf L}(\zeta(z))$ is also unimodular.  The fact that ${\bf v}_{\bf
U}(z)$ satisfies the interior smoothness condition follows from the
corresponding local H\"older smoothness of ${\bf v}_{\bf L}(\zeta)$ on
finite parts of $\Sigma_{\bf L}$ which is preserved under composition
with the smooth map $\zeta(z)$, and the local smoothness near $z=1$
follows from the decay condition satisfied by ${\bf v}_{\bf
L}(\zeta)$.  The jump matrix ${\bf v}_{\bf U}(z)$ is compatible
at the self-intersection points
$\alpha_1=-1$ and $\alpha_{k+1}=z(\zeta_k)$ by the corresponding
property of ${\bf v}_{\bf L}(\zeta)$ and the continuity of
the map $z(\zeta)$ at $\zeta=0$ and $\zeta=\zeta_k$.  At the other
self-intersection point $z=+1$, the compatibility condition 
follows from the decay of ${\bf v}_{\bf L}(\zeta)$ at infinity.

The correspondence between solutions of the two Riemann-Hilbert
problems is set up as follows.  Given a matrix ${\bf L}(\zeta)$
solving the local Riemann-Hilbert Problem~\ref{rhp:local}, the matrix
${\bf U}(z):= {\bf N}_{\bf U}{\bf L}(-e^{i\theta})^{-1}{\bf
L}(\zeta(z))$ is a solution of the umbilical Riemann-Hilbert
Problem~\ref{rhp:umbilical}.  Conversely, given a matrix ${\bf U}(z)$
satisfying the umbilical Riemann-Hilbert Problem~\ref{rhp:umbilical},
the matrix ${\bf L}(\zeta):={\bf N}_{\bf L}{\bf U}(1)^{-1}{\bf
U}(z(\zeta))$ satisfies the local Riemann-Hilbert
Problem~\ref{rhp:local}.  These formulae make sense because by taking
determinants of the jump conditions for the two problems and using the
unimodularity of the jump matrices in conjunction with the H\"older
smoothness of the boundary values and Liouville's theorem
\index{Liouville's theorem} that $\det({\bf L}(\zeta))\equiv\det({\bf
N}_{\bf L})\neq 0$ and $\det({\bf U}(z))\equiv\det({\bf N}_{\bf
U})\neq 0$.  A similar argument (taking ratios of matrices and using
the jump relations, H\"older boundary conditions, and Liouville's
theorem) shows that these formulae are injective transformations.
\end{proof}

\vspace{0.2 in}

We now restrict our study to the umbilical Riemann-Hilbert
Problem~\ref{rhp:umbilical} subject to the jump matrix satisfying the
conditions of Definition~\ref{def:JU}.  Before we begin, we need to
review some elementary results.

\section[H\"older Theory for Simple Contours]{Review of H\"older results for simple contours.}
\label{sec:fundamental}
\begin{figure}[h]
\begin{center}
\mbox{\psfig{file=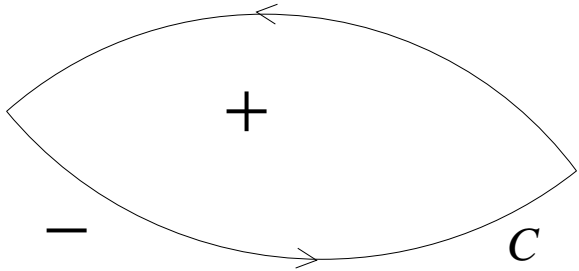,width=3 in}}
\end{center}
\caption{\em The piecewise smooth, closed contour $C$.  It is
given an arbitrary orientation.}
\label{fig:L}
\end{figure}

Here, we recall some classical facts about H\"older spaces of
functions on contours, and the associated Cauchy integral operators.
The basic references are Muskhelishvili \cite{M53} and Pr\"ossdorf
\cite{P78}.  Let $C$ be a piecewise smooth, closed, compact contour,
where the points at which $C$ is not smooth are at worst corner points
with finite angles.  The sort of contour we have in mind for future
applications is shown in Figure~\ref{fig:L}.  
Let $\Lip^\nu(C)$ denote the
space of all H\"older continuous complex matrix-valued functions ${\bf
f}(z)$ on $C$ with exponent $\nu$, equipped with the norm
\begin{equation}
\|{\bf f}\|_{\Lip^\nu(C)} := \sup_{z\in C}\|{\bf f}(z)\| +
\sup_{z_1,z_2\in C}\frac{\|{\bf f}(z_2)-{\bf f}(z_1)\|}{|z_2-z_1|^\nu}\,,
\end{equation}
where $\|\cdot\|$ denotes some matrix norm.  This norm makes
$\Lip^\nu(C)$ a Banach space.  If ${\bf f}\in \Lip^\nu(C)$ for any $\nu>0$,
then ${\bf f}(z)$ is uniformly continuous on $C$.  Also, it is easy to see
that if ${\bf f}\in\Lip^\nu(C)$, then ${\bf f}\in\Lip^\mu(C)$ whenever
$0<\mu<\nu$.  This fact defines an inclusion operator
$\op{I}_{\nu\rightarrow\mu}: \Lip^\nu(C)\rightarrow\Lip^\mu(C)$.  The
remarkable and useful fact about this inclusion map is the following.
\begin{lemma}
Whenever $0<\mu<\nu$, the inclusion operator
$\op{I}_{\nu\rightarrow\mu}:\Lip^\nu(C) \rightarrow\Lip^\mu(C)$ is
compact.
\label{lemma:inclusionsimple}
\end{lemma}
A proof of this statement is given by Pr\"ossdorf (see pages 102--103
of \cite{P78}) in the scalar case when $C$ is a smooth compact
contour.  The proof relies heavily on the Arzel\'a-Ascoli
theorem\index{Arzel\'a-Ascoli theorem}.  There are only cosmetic
differences in extending the result to the matrix case, but
Pr\"ossdorf's proof needs to be extended to admit corner points in the
contour $C$.  The estimates required to carry out this extension
involve bounding arc length distance above and below by the shortest
distance for pairs of points near a corner point of $C$ and can be
found in the appendices of \cite{M53}.

The space $\Lip^\nu(C)$ is also a Banach algebra \index{Banach
algebra} in that it is closed under multiplication and satisfies the
estimate:
\begin{equation}
\|{\bf fg}\|_{\Lip^\nu(C)}\le\|{\bf f}\|_{\Lip^\nu(C)}\|{\bf
g}\|_{\Lip^\nu(C)}\,.
\label{eq:productestimate}
\end{equation}
This follows from simply writing ${\bf f}(z_2){\bf g}(z_2)-{\bf
f}(z_1){\bf g}(z_1)$ as $({\bf f}(z_2)-{\bf f}(z_1)){\bf g}(z_2)+ {\bf
f}(z_1)({\bf g}(z_2)-{\bf g}(z_1))$, the domination of the $L^\infty$
norm by the H\"older norm, and the fact that the
matrices with the norm $\|\cdot\|$ are themselves a Banach algebra
satisfying an estimate of the same form as (\ref{eq:productestimate}).
If ${\bf f}(z)$ is in $\Lip^\nu(C)$ and is invertible for each $z\in
C$ then ${\bf g}(z)={\bf f}(z)^{-1}$ is also in $\Lip^\nu(C)$.  The
estimate (\ref{eq:productestimate}) immediately gives the following
result.
\begin{lemma}
Let ${\bf g}\in \Lip^\nu(C)$, and let $\op{R}_{\bf g}$ and
$\op{L}_{\bf g}$ denote the operators of pointwise right and left
multiplication by ${\bf g}(z)$.  Then $\op{R}_{\bf
g}:\Lip^\nu(C)\rightarrow\Lip^\nu(C)$ and $\op{L}_{\bf
g}:\Lip^\nu(C)\rightarrow\Lip^\nu(C)$ are bounded linear operators.
\label{lemma:multsimple}
\end{lemma}

Consider $C$ to be oriented.
For ${\bf f}\in \Lip^\nu(C)$, the Cauchy contour
integral
\begin{equation}
(\op{C}^C{\bf f})(z):=\frac{1}{2\pi i}\oint_C(s-z)^{-1}{\bf f}(s)\,ds
\end{equation}
defines a holomorphic matrix-valued function in the multiply-connected
domain ${\mathbb C}\setminus C$.  
For $z\in C$, denote the boundary values of $(\op{C}^C{\bf f})(w)$ as $w$
tends to $z$ from the left (respectively right) of $C$ by $(\op{C}^C_+{\bf
f})(z)$ (respectively $(\op{C}^C_-{\bf f})(z)$).  Then, we have
\begin{lemma}[Plemelj-Privalov]
\index{Plemelj-Privalov theorem}
Let ${\bf f}\in\Lip^\nu(C)$.  The boundary values $(\op{C}_+^C{\bf f})(z)$
and $(\op{C}_-^C{\bf f})(z)$ exist independently of the path of approach to
the boundary and are in $\Lip^\nu(C)$ as functions of $z\in C$.
Moreover, the linear operators
$\op{C}_\pm^C:\Lip^\nu(C)\rightarrow\Lip^\nu(C)$ taking ${\bf f}(z)$ to the
boundary values of the Cauchy integral are bounded with respect to the
H\"older norm.
\label{lemma:PPsimple}
\end{lemma}
The proof of this statement is in \S 19 of Muskhelishvili \cite{M53},
with adjustments for the corners as described in his appendices 1 and
2.  It is a scalar result, but the matrix generalization requires only
cosmetic alterations.  The simple nature of the contour $C$ guarantees
that the boundary value operators are complementary projections
\index{complementary projections} so that the Plemelj formula holds:
$(\op{C}^C_+{\bf f})(z)-(\op{C}^C_-{\bf f})(z)={\bf f}(z)$.  Also we
have the operator identity $\op{C}^C_+\circ \op{C}^C_- =
\op{C}^C_-\circ \op{C}^C_+=0$.

The next statement of interest concerns certain commutators.  Let
${\bf g}\in\Lip^\beta(C)$ for $0<\beta < 1$, and consider the
commutators $[\op{C}_+^C,\op{L}_{\bf g}]$, $[\op{C}_-^C,\op{L}_{\bf
g}]$, $[\op{C}_+^C,\op{R}_{\bf g}]$, and $[\op{C}_-^C,\op{R}_{\bf
g}]$.  These operators can be interpreted as nonsingular integral
operators by the formulae:
\begin{equation}
\begin{array}{rcccl}
([\op{C}_+^C,\op{L}_{\bf g}]{\bf f})(z)&=& ([\op{C}_-^C,\op{L}_{\bf g}]{\bf
f})(z)&=&\displaystyle\frac{1}{2\pi i}\oint_C (s-z)^{-1}({\bf
g}(s)-{\bf g}(z)){\bf f}(s)\,ds\,,\\\\
([\op{C}_+^C,\op{R}_{\bf g}]{\bf f})(z)&=& ([\op{C}_-^C,\op{R}_{\bf g}]{\bf
f})(z)&=&\displaystyle\frac{1}{2\pi i}\oint_C (s-z)^{-1}{\bf f}(s)({\bf
g}(s)-{\bf g}(z))\,ds\,.
\end{array}
\end{equation}
These operators are essentially as nice as the function ${\bf g}(z)$
is.  We have the following result.
\begin{lemma}
Let ${\bf g}\in\Lip^\beta(C)$.  The operator $[\op{C}_+^C,\op{L}_{\bf
g}]=[\op{C}_-^C,\op{L}_{\bf g}]$ and the operator $[\op{C}_+^C,\op{R}_{\bf
g}]=[\op{C}_-^C,\op{R}_{\bf g}]$ are bounded operators from $\Lip^\alpha(C)$ to
$\Lip^\beta(C)$ as long as $0<\alpha\le 1$ and $0\le\beta<1$.
\label{lemma:commutatorsimple}
\end{lemma}
This statement is proved for scalar functions (trivially extended to
the matrix case, however) by Pr\"ossdorf in his Lemma 4.1 on page 100
and Corollary 4.2 on page 102 \cite{P78}.  Again, some technical
adjustment as described by Muskhelishvili \cite{M53} will need to be
used in order to admit corner points in the closed compact contour
$C$.  Note that the statement holds even if $\beta>\alpha$.  In this
sense, the commutator can {\em improve} the smoothness of the
functions on which it acts.

\section[Generalization for Umbilical Contours]{Generalization for umbilical contours.}
\label{sec:generalization}

For an umbilical Riemann-Hilbert problem, let the simply-connected
components of $\Omega_{\bf U}^+$ be denoted $\Omega^{+(k)}_{\bf U}$,
and the components of $\Omega^-_{\bf U}$ be denoted
$\Omega^{-(k)}_{\bf U}$.  All but one of these components are bounded
domains.  For any simply connected domain $D$ (possibly unbounded)
with piecewise smooth boundary, let $\partial D$ denote the boundary
oriented with $D$ on the left.  Then the oriented contour for the
umbilical Riemann-Hilbert problem can be written either as
\begin{equation}
\Sigma_{\bf U}=\sum_{k}\partial\Omega^{+(k)}_{\bf U}\,,\hspace{0.2 in}\mbox{or}\hspace{0.2 in}
\Sigma_{\bf U}=-\sum_{k}\partial\Omega^{-(k)}_{\bf U}\,.
\end{equation}
Note that while these formulae hold when we regard the contours as
paths of integration of H\"older class functions, they do not hold in
the set-theoretic sense of disjoint union since the self-intersection
points $z=\alpha_i$ are each counted several times on the right-hand
side and only once on the left; this is reflected in the use of the
sum notation and the use of signs to denote orientation.

\begin{figure}[h]
\begin{center}
\mbox{\psfig{file=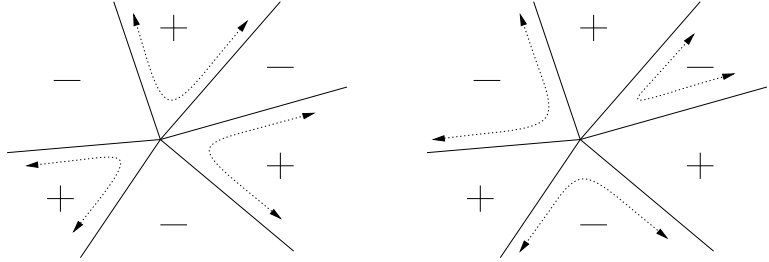,width=4 in}}
\end{center}
\caption{\em Left: the dashed arrows indicate the continuity
properties of a function in $A^\nu_+$ near
a self-intersection point $z=\alpha_i$.  Right: the same for
$A^\nu_-$.}
\label{fig:continuity}
\end{figure}

We begin by introducing some spaces.  Let $A_\pm^\nu$ denote the set
of matrix-valued functions ${\bf f}(z)$ on $\Sigma_{\bf
U}\setminus\{\alpha_i\}$ that for each $k$ may be assigned values at
$z=\alpha_i$ so that the restriction of ${\bf f}(z)$ to
$\partial\Omega^{\pm (k)}_{\bf U}$ is in $\Lip^\nu(\partial\Omega^{\pm
(k)}_{\bf U})$.  Note that at a fixed intersection point $z=\alpha_i$,
the values given to ${\bf f}(z)$ to establish the required continuity
will generally be different for each $k$.  The continuity properties
of functions in these spaces are illustrated in
Figure~\ref{fig:continuity}.
These sets are Banach algebras \index{Banach algebra} with the norm
being taken as the sum of the norms of the individual components:
\begin{equation}
\|{\bf f}\|_{A_\pm^\nu}:=
\sum_k \|{\bf f}\|_{\Lip^\nu(\partial\Omega^{\pm (k)}_{\bf U})}\,,
\end{equation}
where in each term in the sum on the right-hand side, the matrix
function $\bf f$ is considered to be assigned definite values at all
intersection points $z=\alpha_i$ (possibly different in each term) and
subsequently to be restricted to the simple closed contour
$\partial\Omega^{\pm (k)}_{\bf U}$.  Finally, note that every matrix
function $\bf f$ in $A_+^\nu\cap A_-^\nu$ has a unique continuous
extension to the whole closed contour $\Sigma_{\bf U}$, and thus that
$A_+^\nu\cap A_-^\nu$ can be identified with the Banach algebra
$\Lip^\nu(\Sigma_{\bf U})$ of matrix functions ${\bf f}(z)$ defined
for $z\in\Sigma_{\bf U}$ with the norm
\begin{equation}
\|{\bf f}\|_{\Lip^\nu(\Sigma_{\bf U})}:=\sup_{z\in\Sigma_{\bf U}}
\|{\bf f}(z)\| + \sup_{z_1,z_2\in\Sigma_{\bf U}} \frac{\|{\bf
f}(z_2)-{\bf f}(z_1)\|}{|z_2-z_1|^\nu}\,.
\end{equation}
Convergence in $\Lip^\nu(\Sigma_{\bf U})$ is equivalent to
simultaneous convergence in both $A_+^\nu$ and $A_-^\nu$.

Now we use these spaces to state the appropriate generalizations of
the results from \S\ref{sec:fundamental} to the case of the umbilical
contour $\Sigma_{\bf U}$.  First of all,
Lemma~\ref{lemma:inclusionsimple} can be applied to individual
components of the boundary of $\Omega^\pm_{\bf U}$ to prove an
analogous result for these spaces.
\begin{lemma}
For $0<\mu\le \nu$, the inclusion operator
$\op{I}_{\nu\rightarrow\mu}$ can be defined from $A_+^\nu$ to
$A^\mu_+$, from $A_-^\nu$ to $A_-^\mu$, or from $\Lip^\nu(\Sigma_{\bf
U})$ to $\Lip^\mu(\Sigma_{\bf U})$.  It is a compact operator in all
of these instances whenever $\mu <\nu$.
\label{lemma:inclusiongeneral}
\end{lemma}
Similarly, the obvious generalization of Lemma~\ref{lemma:multsimple}
is the following.
\begin{lemma}
Let ${\bf g}\in A^\nu_+$.  Then the multiplication operators
$\op{L}_{\bf g}$ and $\op{R}_{\bf g}$ are bounded on $A^\nu_+$ as well
as from $\Lip^\nu(\Sigma_{\bf U})$ to $A^\nu_+$.  Similarly if ${\bf
g}\in A^\nu_-$, then $\op{L}_{\bf g}$ and $\op{R}_{\bf g}$ are bounded
on $A^\nu_-$ as well as from $\Lip^\nu(\Sigma_{\bf U})$ to $A^\nu_-$.
Finally, if ${\bf g}\in \Lip^\nu(\Sigma_{\bf U})$, then $\op{L}_{\bf
g}$ and $\op{R}_{\bf g}$ are bounded on $\Lip^\nu(\Sigma_{\bf U})$,
$A^\nu_+$, and $A^\nu_-$.
\end{lemma}

For a function ${\bf f}\in A^\nu_+$, we can write the corresponding
Cauchy integral over the whole contour $\Sigma_{\bf U}$ in terms of
Cauchy integrals over simple closed contours as defined in
\S\ref{sec:fundamental}:
\begin{equation}
(\op{C}^{\Sigma_{\bf U}}{\bf f})(z):=\frac{1}{2\pi i}\int_{\Sigma_{\bf
U}} (s-z)^{-1}{\bf f}(s)\,ds = \sum_k
(\op{C}^{\partial\Omega^{+(k)}_{\bf U}}{\bf f})(z)\,,
\label{eq:friendlyplus}
\end{equation}
where on the right-hand side we use the same symbol ${\bf f}$ to
denote the various $\Lip^\nu(\partial\Omega^{+(k)}_{\bf U})$
completions and restrictions to the closed compact contours
$\partial\Omega^{+(k)}_{\bf U}$.  Likewise, for a function ${\bf
f}\in A^\nu_-$, we can write
\begin{equation}
(\op{C}^{\Sigma_{\bf U}}{\bf f})(z)=- \sum_k
(\op{C}^{\partial\Omega^{-(k)}_{\bf U}}{\bf f})(z)\,.
\label{eq:friendlyminus}
\end{equation}
These Cauchy integrals are of course analytic in ${\mathbb
C}\setminus\Sigma_{\bf U}$.  These formulae allow the boundary values
of the Cauchy integral over $\Sigma$ to be expressed in terms the
boundary values of Cauchy integrals of H\"older class functions over
simple closed contours as developed in \S\ref{sec:fundamental}.  This
allows those results to be generalized to the umbilical contour
$\Sigma_{\bf U}$ with self-intersection points.  

An intermediate result that we need is the following.
\begin{lemma}
Let $j\neq k$ and consider the Cauchy operators
$\op{C}^{\partial\Omega^{+(k)}_{\bf U}}|_{\partial\Omega^{+(j)}_{\bf
U}}$ and $\op{C}^{\partial\Omega^{-(k)}_{\bf
U}}|_{\partial\Omega^{-(j)}_{\bf U}}$, that is, the restrictions of
the Cauchy operators over one simple component of
$\partial\Omega^\pm_{\bf U}$ to another.
Then, the first of these is a bounded operator
from $\Lip^\nu(\partial\Omega^{+(k)}_{\bf U})$ to 
$\Lip^\nu(\partial\Omega^{+(j)}_{\bf U})$, and
the second of these is a bounded operator from
$\Lip^\nu(\partial\Omega^{-(k)}_{\bf U})$ to 
$\Lip^\nu(\partial\Omega^{-(j)}_{\bf U})$.
\label{lemma:balayage}
\end{lemma}
For a proof of (a stronger version of) this statement, see
Muskhelishvili \cite{M53}, \S 22.  To apply his result, we need only
to observe that either $\partial\Omega^{+(j)}_{\bf U}\subset
\Omega^{+(k)}_{\bf U}\cup\partial\Omega^{+(k)}_{\bf U}$ or
$\partial\Omega^{+(j)}_{\bf U}\subset {\mathbb
C}\setminus\Omega^{+(k)}_{\bf U}$.  Similarly for
$\partial\Omega^{-(j)}_{\bf U}$.

For $z\in\Sigma_{\bf U}\setminus\{\alpha_i\}$, we denote by
$(\op{C}^{\Sigma_{\bf U}}_\pm{\bf f})(z)$ the boundary value of the
Cauchy integral $(\op{C}^{\Sigma_{\bf U}}{\bf f})(w)$ as $w$ tends to
$z$ from the region $\Omega^\pm_{\bf U}$.  It is a consequence of the
above representations (\ref{eq:friendlyplus}) and
(\ref{eq:friendlyminus}) of the Cauchy integral, along with
Lemma~\ref{lemma:balayage}, that that the following generalization of
Lemma~\ref{lemma:PPsimple} holds.
\begin{lemma}[Generalized Plemelj-Privalov]
\index{Plemelj-Privalov theorem!generalized version of}
The operator $\op{C}^{\Sigma_{\bf U}}_+$ is bounded on $A^\nu_+$, and
is bounded from $A^\nu_-$ to the smaller space $\Lip^\nu(\Sigma_{\bf
U})$.  Similarly, the operator $\op{C}^{\Sigma_{\bf U}}_-$ is bounded on
$A^\nu_-$, and is bounded from
$A^\nu_+$ to $\Lip^\nu(\Sigma_{\bf U})$.  Also, for ${\bf f}\in
A^\nu_+\cup A^\nu_-$ the Plemelj
formula holds:
\begin{equation}
(\op{C}^{\Sigma_{\bf U}}_+{\bf f})(z)-(\op{C}^{\Sigma_{\bf U}}_-{\bf
f})(z) = {\bf f}(z)\,,
\end{equation}
as well as the relation $(\op{C}^{\Sigma_{\bf U}}_+\circ
\op{C}^{\Sigma_{\bf U}}_-{\bf f})(z)=(C^{\Sigma_{\bf U}}_-
\circ C^{\Sigma_{\bf U}}_+{\bf
f})(z)\equiv 0$.
\label{lemma:wholecontour}
\end{lemma}

\begin{proof} Consider first $\op{C}^{\Sigma_{\bf U}}_+$
acting on ${\bf f}\in A^\nu_+$.  Using the representation
(\ref{eq:friendlyplus}), the restriction of the result to a component
$\partial\Omega^{+(j)}_{\bf U}$ of the boundary can be written as:
\begin{equation}
(\op{C}^{\Sigma_{\bf U}}_+{\bf f})|_{\partial\Omega^{+(j)}_{\bf
U}}=\op{C}_+^{\partial\Omega^{+(j)}_{\bf U}}{\bf
f}|_{\partial\Omega^{+(j)}_{\bf U}} + \sum_{k\neq
j}(\op{C}^{\partial\Omega^{+(k)}_{\bf U}}{\bf
f})|_{\partial\Omega^{+(j)}_{\bf U}}\,.
\end{equation}
That this operator is bounded to $\Lip^\nu(\partial\Omega^{+(j)}_{\bf
U})$ is clear from Lemma~\ref{lemma:PPsimple} and
Lemma~\ref{lemma:balayage}.  Summing the norm estimates over the
components $\partial\Omega^{+(j)}_{\bf U}$ then gives the boundedness
of $\op{C}_+^{\Sigma_{\bf U}}:A^\nu_+\rightarrow
A^\nu_+$.  Now consider $\op{C}^{\Sigma_{\bf U}}_+$ acting on ${\bf
f}\in A^\nu_-$.  Using the representation
(\ref{eq:friendlyminus}), the restriction of the result to a component
$\partial\Omega^{-(j)}_{\bf U}$ of the boundary can be written as:
\begin{equation}
(\op{C}^{\Sigma_{\bf U}}_+{\bf f})|_{\partial\Omega^{-(j)}_{\bf
U}}=-\op{C}_-^{\partial\Omega^{-(j)}_{\bf U}}{\bf
f}|_{\partial\Omega^{-(j)}_{\bf U}} - \sum_{k\neq
j}(\op{C}^{\partial\Omega^{-(k)}_{\bf U}}{\bf
f})|_{\partial\Omega^{-(j)}_{\bf U}}\,.
\end{equation}
The boundary value on the right-hand side is ``minus'' because the
approach to the boundary is from the right of
$\partial\Omega^{-(j)}_{\bf U}$ with its orientation (recall that by
convention we are taking the boundary $\partial D$ of a
simply-connected domain $D$ to be oriented with $D$ on the
left).  By similar arguments, it is then clear that
$\op{C}_+^{\Sigma_{\bf U}}$ is bounded on $A^\nu_-$.  However, in this
case, more is true.  If $z=\alpha_i$ is one of the self-intersection points
of $\Sigma_{\bf U}$, then at the point
$\alpha_i\in\partial\Omega^{-(j)}_{\bf U}$ the above formula reads:
\begin{equation}
(\op{C}^{\Sigma_{\bf U}}_+{\bf f})|_{\partial\Omega^{-(j)}_{\bf
U}}(\alpha_i)= - \sum_k(\op{C}_-^{\partial\Omega^{-(k)}_{\bf U}}{\bf
f})|_{\partial\Omega^{-(j)}_{\bf U}}(\alpha_i)\,,
\end{equation}
where the sum is taken over all components $k$, and it is clear that
the value at the mutual corner point $\alpha_i$ is the same in all
components $\partial\Omega^{-(j)}_{\bf U}$.  This, along with the
usual argument (as in \cite{M53}, page 13) that piecewise H\"older
functions that are continuous are globally H\"older implies that
$\op{C}^{\Sigma_{\bf U}}_+$ is actually bounded from $A^\nu_-$ to
$\Lip^\nu(\Sigma_{\bf U})$.  Similar arguments establish the analogous
results for $\op{C}_-^{\Sigma_{\bf U}}$.  
\end{proof}

\vspace{0.2 in}

Similarly, the generalization of Lemma~\ref{lemma:commutatorsimple} is 
the following.
\begin{lemma}
Let $0<\alpha\le 1$ and $0\le\beta <1$.  Suppose that ${\bf g}\in
\Lip^\beta(\Sigma_{\bf U})$.  Then the commutators
$[\op{C}^{\Sigma_{\bf U}}_+,\op{R}_{\bf g}]$ and $[\op{C}^{\Sigma_{\bf
U}}_+,\op{L}_{\bf g}]$ are bounded from $A^\alpha_-$ to
$\Lip^\beta(\Sigma_{\bf U})$ and the commutators $[\op{C}^{\Sigma_{\bf
U}}_-,\op{R}_{\bf g}]$ and $[\op{C}^{\Sigma_{\bf U}}_-,\op{L}_{\bf
g}]$ are bounded from $A^\alpha_+$ to $\Lip^\beta(\Sigma_{\bf U})$.
Now suppose only that ${\bf g}\in A^\beta_+$.  Then the commutators
$[\op{C}_+^{\Sigma_{\bf U}},\op{R}_{\bf g}]$ and
$[\op{C}_+^{\Sigma_{\bf U}},\op{L}_{\bf g}]$ are bounded from
$A^\alpha_+$ to $A^\beta_+$.
Similarly if ${\bf g}\in A^\beta_-$ then
$[\op{C}_-^{\Sigma_{\bf U}},\op{R}_{\bf g}]$ and 
$[\op{C}_-^{\Sigma_{\bf U}},\op{L}_{\bf g}]$ are bounded from
$A^\alpha_-$ to $A^\beta_-$.
\label{lemma:commutatorgeneral}
\end{lemma}

\begin{proof} Again, one proceeds by decomposing the Cauchy
operators $\op{C}^{\Sigma_{\bf U}}_\pm$ according to
(\ref{eq:friendlyplus}) or (\ref{eq:friendlyminus}) depending on the
space, and then applies Lemma~\ref{lemma:commutatorsimple}
componentwise.  When the range of the transformation is
$\Lip^\beta(\Sigma_{\bf U})$, one shows continuity at the intersection
points exactly as in the proof of Lemma~\ref{lemma:wholecontour}.
\end{proof}

\section[Fredholm Alternative]{Fredholm alternative for umbilical Riemann-Hilbert problems.}
In this section, we apply the H\"older theory of Cauchy integrals on
the umbilical contour $\Sigma_{\bf U}$ to establish a Fredholm
alternative theorem for umbilical Riemann-Hilbert
Problems~\ref{rhp:umbilical}.  We begin by factoring the jump matrix
${\bf v}_{\bf U}(z)$ in a straightforward way:
\begin{lemma}
\index{jump matrix!algebraic factorization of}
Let $\Sigma_{\bf U}$ be an umbilical contour, and let there be given
on $\Sigma_{\bf U}\setminus\{\alpha_i\}$ a matrix function ${\bf
v}_{\bf U}(z)$ satisfying the conditions of Definition~\ref{def:JU}.
Then ${\bf v}_{\bf U}(z)$ admits a factorization ${\bf v}_{\bf
U}(z)={\bf b}^-(z)^{-1}{\bf b}^+(z)$ with ${\bf b}^\pm(z)$ invertible,
where ${\bf b}^+\in A^\nu_+$ and ${\bf b}^-\in A^\nu_-$.
\label{lemma:factorization}
\end{lemma}

\begin{proof} We construct such a factorization
algorithmically as follows.  The factorization will be carried out
locally at each intersection point, so first select for each
$\alpha_i$ a number $r_i>0$ sufficiently small that $\Sigma_{\bf
U}\cap B_{r_i}(\alpha_i)$, where $B_{r_i}(\alpha_i)$ is the open disk
of radius $r_i$ centered at $z=\alpha_i$, contains no other
intersection points and that each circle of radius $r<r_i$ centered at
$\alpha_i$ meets each arc terminating at $\alpha_i$ exactly once.  For
$z\in\Sigma_{\bf U}\cap (\cap_k B_{r_k}(\alpha_k)^c)$ where the
superscript $c$ denotes the complement, {\em i.e.} outside all disks,
set ${\bf b}^+(z)\equiv {\bf v}_{\bf U}(z)$ and ${\bf b}^-(z)\equiv
{\mathbb I}$.  Now, letting $\alpha_i$ be an intersection point, we
will specify the factorization for $z\in\Sigma_{\bf U}\cap
B_{r_i}(\alpha_i)$.  Let the open arcs meeting at $\alpha_i$ be
enumerated in counterclockwise order about $\alpha_i$: $\Sigma_{\bf
U}^{(1)},\dots,\Sigma_{\bf U}^{(N)}$ (here $N$ is even but may depend
on $i$).  Let the unique point in $\Sigma_{\bf U}^{(j)}$ common to the
boundary of $B_{r_i}(\alpha_i)$ be denoted $z_j$.  For
$z\in\Sigma_{\bf U}^{(1)}\cap B_{r_i}(\alpha_i)$ begin the
factorization by setting ${\bf b}^-|_{\Sigma_{\bf U}^{(1)}}\equiv
{\mathbb I}$, and therefore ${\bf b}^+|_{\Sigma_{\bf U}^{(1)}}\equiv
{\bf v}_{\bf U}|_{\Sigma_{\bf U}^{(1)}}$.  Now suppose that a
factorization ${\bf b}^+|_{\Sigma_{\bf U}^{(j)}}$ and ${\bf
b}^-|_{\Sigma_{\bf U}^{(j)}}$ has been constructed on $\Sigma_{\bf
U}^{(j)}\cap B_{r_i}(\alpha_i)$.  We now describe how to extend the
factorization to $\Sigma_{\bf U}^{(j+1)}\cap B_{r_i}(\alpha_i)$.
%
Suppose first that the region bounded by $\Sigma_{\bf U}^{(j)}$,
$\Sigma_{\bf U}^{(j+1)}$, and the boundary of the disk is a component
$\Omega^{+(k)}_{\bf U}$ of the ``plus'' region $\Omega^+_{\bf U}$.
Let ${\bf s}(\rho)$ be a $C_1$ map from $[0,r_i]$ into $GL(n,{\mathbb
C})$ with ${\bf s}(0)={\bf b}^+|_{\Sigma_{\bf U}^{(j)}}(\alpha_i)$ and
$ {\bf s}(r_i)={\bf v}_{\bf U}|_{\Sigma_{\bf U}^{(j+1)}}(z_{j+1})$.
Such a map exists because $GL(n,{\mathbb C})$ is arcwise connected
\cite{SattingerWeaver}.  Then, for $z\in\Sigma_{\bf U}^{(j+1)}\cap
B_{r_i}(\alpha_i)$, we set ${\bf b}^+(z)={\bf s}(|z-\alpha_i|)$, and
${\bf b}^-(z)= {\bf s}(|z-\alpha_i|){\bf v}_{\bf U}(z)^{-1}$.
On the other hand, if the region bounded by $\Sigma_{\bf U}^{(j)}$,
$\Sigma_{\bf U}^{(j+1)}$, and the boundary of the disk is a component
of $\Omega^-_{\bf U}$, then we take ${\bf s}(\rho)$ to be a $C_1$ map
from $[0,r_i]$ into $GL(n,{\mathbb C})$ with ${\bf s}(0)={\bf
b}^-|_{\Sigma_{\bf U}^{(j)}} (\alpha_i)$ and ${\bf s}(r_i)={\mathbb
I}$, and then for $z\in\Sigma_{\bf U}^{(j+1)}\cap B_{r_i}(\alpha_i)$
we set ${\bf b}^-(z)={\bf s}(|z-\alpha_i|)$ and then ${\bf
b}^+(z)={\bf s}(|z-\alpha_i|){\bf v}_{\bf U}(z)$.
Using this algorithm, we then construct factorizations on the part of
each open arc $\Sigma_{\bf U}^{(j)}$ within $B_{r_i}(\alpha_i)$
starting from $\Sigma_{\bf U}^{(1)}$ and working counterclockwise
about $z=\alpha_i$.  This construction, when carried out under the
compatibility condition ({\em cf.} Definition~\ref{def:JU}) satisfied
by the limiting jump matrices ${\bf v}_{\bf U}|_{\Sigma_{\bf
U}^{(j)}}$ at each endpoint, guarantees that the restrictions ${\bf
b}^\pm|_{\Sigma_{\bf U}^{(j)}}$ uniformly satisfy the H\"older
continuity condition with exponent $\nu$ on each open arc $\Sigma_{\bf
U}^{(j)}$.  This follows from the interior smoothness condition, the
continuity of the factorization at the disk boundaries, the Banach
algebra property of H\"older continuous functions, and the fact that
$C_1$ functions are H\"older continuous with any exponent less than or
equal to one.  The construction also guarantees that ${\bf
b}^+|_{\partial\Omega^{+(k)}_{\bf U}}$ may be defined at each
intersection point $\alpha_i$ to be continuous at the corner points
for each $k$, and likewise that ${\bf
b}^-|_{\partial\Omega^{-(k)}_{\bf U}}$ may be defined to be continuous
at the corner points for each $k$.  Then it follows (see
\cite{M53}, page 13 and the appendices), that the
functions ${\bf b}^+(z)$ and ${\bf b}^-(z)$ are in $A^\nu_+$ and
$A^\nu_-$ respectively.  Finally, the invertibility (and in fact the
unimodularity) of ${\bf b}^\pm(z)$ follows directly from the above
algorithm and the unimodularity of ${\bf v}_{\bf
U}(z)$.  
\end{proof}

\vspace{0.2 in}

Now, set ${\bf w}^+(z):={\bf b}^+(z)-{\mathbb I}\in A^\nu_+$ and ${\bf
w}^-(z):={\mathbb I}-{\bf b}^-(z)\in A^\nu_-$.  Choose any positive
$\mu$ with $\mu\le\nu$, and define the operator $\op{C}_{\bf w}$ on
the space $\Lip^\mu(\Sigma_{\bf U})$ by
\begin{equation}
\begin{array}{rcl}
\displaystyle (\op{C}_{\bf w}{\bf m})(z) &:= &
\displaystyle
(\op{C}^{\Sigma_{\bf U}}_+ \circ
\op{R}_{{\bf w}^-}{\bf m})(z) + (\op{C}^{\Sigma_{\bf U}}_- \circ
\op{R}_{{\bf w}^+}{\bf m})(z)\\\\
&=&\displaystyle
(\op{C}^{\Sigma_{\bf U}}_+({\bf m}{\bf w}^-))(z) +
(\op{C}^{\Sigma_{\bf U}}_-({\bf m}{\bf w}^+))(z)\,.
\end{array}
\end{equation}
It follows from Lemma~\ref{lemma:wholecontour} that this formula
indeed defines a function in the Banach space $\Lip^\mu(\Sigma_{\bf
U})$.  For example, in the first term, the multiplication operator is
a map from $\Lip^\mu(\Sigma_{\bf U})$ to $A^\mu_-$ (with the function
${\bf w}^-$ reinterpreted under the inclusion map as an element of
$A^\mu_-$ since $\mu\le\nu$), and then the operator
$\op{C}^{\Sigma_{\bf U}}_+$ brings us back from this space to
$\Lip^\mu(\Sigma_{\bf U})$ according to
Lemma~\ref{lemma:wholecontour}.  The second term is understood
similarly.  Moreover, as the composition of bounded operators,
$\op{C}_{\bf w}$ is itself a bounded operator on $\Lip^\mu(\Sigma_{\bf
U})$.

The umbilical Riemann-Hilbert Problem~\ref{rhp:umbilical} can now be
reformulated as a singular integral equation in $\Lip^\mu(\Sigma_{\bf
U})$.
\begin{lemma}
\label{lemma:equivalence}
Consider the umbilical Riemann-Hilbert Problem~\ref{rhp:umbilical}
with data $(\Sigma_{\bf U},{\bf v}_{\bf U}(z),{\bf N}_{\bf U})$.  Let
the normalization matrix ${\bf N}_{\bf U}$ be identified with a
constant function in $\Lip^\mu(\Sigma_{\bf U})$, and suppose the jump
matrix ${\bf v}_{\bf U}(z)$ satisfying the conditions of
Definition~\ref{def:JU} to be factored according to
Lemma~\ref{lemma:factorization}, with $\op{C}_{\bf w}$ being the
corresponding singular integral operator in $\Lip^\mu(\Sigma_{\bf
U})$.  Then, the solutions ${\bf U}(z)$ of the umbilical
Riemann-Hilbert Problem~\ref{rhp:umbilical} are in one-to-one
correspondence with the solutions ${\bf m}\in
\Lip^\mu(\Sigma_{\bf U})$ of the integral equation
\index{Riemann-Hilbert problem!singular integral equation equivalent to}
\begin{equation}
{\bf m}(z)-(\op{C}_{\bf w}{\bf m})(z) = {\bf N}_{\bf U}\,.
\label{eq:inteqn}
\end{equation}
\end{lemma}

\begin{proof} First suppose that we are
given a solution ${\bf m}(z)$ of the integral equation
(\ref{eq:inteqn}) in $\Lip^\mu(\Sigma_{\bf U})$.  For $z\in {\mathbb
C}\setminus\Sigma_{\bf U}$ define
\begin{equation}
\begin{array}{rcl}
\displaystyle
{\bf U}(z;{\bf m}) &:= &\displaystyle
{\bf N}_{\bf U} + (\op{C}^{\Sigma_{\bf U}} \circ
\op{R}_{{\bf w}^+}{\bf m})(z) + (\op{C}^{\Sigma_{\bf U}} \circ
\op{R}_{{\bf w}^-}{\bf m})(z)\\\\
&=&\displaystyle {\bf N}_{\bf U}+
(\op{C}^{\Sigma_{\bf U}} ({\bf m}{\bf w}^+))(z) + (\op{C}^{\Sigma_{\bf U}} ({\bf m}{\bf w}^-))(z)\,.
\end{array}
\label{eq:repn}
\end{equation}
Then, ${\bf U}(z;{\bf m})$ is a solution of the umbilical
Riemann-Hilbert Problem~\ref{rhp:umbilical}.  The analyticity of ${\bf
U}(z;{\bf m})$ in ${\mathbb C}\setminus\Sigma_{\bf U}$ and the
normalization ${\bf U}(\infty;{\bf m})={\bf N}_{\bf U}$ follows
directly from the representation (\ref{eq:repn}) and the properties of
elements of $\Lip^\mu(\Sigma_{\bf U})$.  That the $A^\mu_\pm$ boundary
values of ${\bf U}(z;{\bf m})$ satisfy the jump relations follows from
simply inserting (\ref{eq:repn}) into the jump relations and using the
Plemelj formula in conjunction with (\ref{eq:inteqn}).

To show the injectivity of the map ${\bf m}(z)\rightarrow {\bf
U}(z;{\bf m})$ observe that the Cauchy integral representation
(\ref{eq:repn}) implies that ${\bf U}(z;{\bf m}_2)\equiv {\bf
U}(z;{\bf m}_1)$ if and only if $({\bf m}_2(z)-{\bf m}_1(z))({\bf
w}^+(z)+{\bf w}^-(z))\equiv {\bf 0}$.  At the same time, since ${\bf
m}_1(z)$ and ${\bf m}_2(z)$ both satisfy (\ref{eq:inteqn}), it follows
that 
\begin{equation}
({\bf m}_2(z)-{\bf m}_1(z))-
(\op{C}^{\Sigma_{\bf U}}_+(({\bf m}_2-{\bf m}_1){\bf w}^-))(z)-
(\op{C}^{\Sigma_{\bf U}}_-(({\bf m}_2-{\bf m}_1){\bf w}^+))(z)
\equiv {\bf 0}\,.
\end{equation}
Putting these two together gives
\begin{equation}
\begin{array}{rcl}
{\bf 0}&\equiv &\displaystyle ({\bf m}_2(z)-{\bf m}_1(z))+
(\op{C}^{\Sigma_{\bf U}}_+(({\bf m}_2-{\bf m}_1){\bf w}^+))(z)-
(\op{C}^{\Sigma_{\bf U}}_-(({\bf m}_2-{\bf m}_1){\bf w}^+))(z)\\\\
&=&\displaystyle
({\bf m}_2(z)-{\bf m}_1(z))+
({\bf m}_2(z)-{\bf m}_1(z)){\bf w}^+(z)\\\\
&=&\displaystyle ({\bf m}_2(z)-{\bf m}_1(z)){\bf b}^+(z)\,,
\end{array}
\end{equation}
where we have used the Plemelj formula.  From the invertibility of 
${\bf b}^+(z)$ it follows that ${\bf m}_2(z)\equiv {\bf m}_1(z)$.

On the other hand, suppose we are given a solution ${\bf U}(z)$ of the
umbilical Riemann-Hilbert Problem~\ref{rhp:umbilical}.  For
$z\in\Sigma_{\bf U}\setminus
\{\alpha_i\}$, set
\begin{equation}
{\bf m}(z;{\bf U}):={\bf U}_+(z){\bf b}^+(z)^{-1} = 
{\bf U}_-(z){\bf b}^-(z)^{-1}\,.
\end{equation}
That these two expressions yield the same result follows from the
factorization of the jump matrix established in
Lemma~\ref{lemma:factorization} and the jump conditions satisfied by
${\bf U}(z)$ on $\Sigma_{\bf U}$.  Moreover, it is clear that ${\bf
m}(\cdot;{\bf U})$ is in both $A^\mu_+$ and $A^\mu_-$; therefore it is
an element of $\Lip^\mu(\Sigma_{\bf U})$.  Also, the map ${\bf
U}(z)\rightarrow {\bf m}(z;{\bf U})$ is injective because the matrix
functions ${\bf b}^\pm(z)$ are invertible.  Now observe that
\begin{equation}
\begin{array}{rcl}
\displaystyle {\bf m}(z;{\bf U})-(\op{C}_{\bf w}{\bf m}(\cdot;{\bf
U}))(z) &=& \displaystyle {\bf m}(z;{\bf U}) - (\op{C}^{\Sigma_{\bf
U}}_+ ({\bf U}_-({\bf b}^-)^{-1}{\bf w}^-))(z) -
(\op{C}^{\Sigma_{\bf U}}_-({\bf U}_+({\bf b}^+)^{-1}{\bf w}^+))(z) \\\\
&=&\displaystyle
{\bf m}(z;{\bf U})-(\op{C}^{\Sigma_{\bf U}}_+({\bf U}_-({\bf b}^-)^{-1} - 
{\bf U}_-))(z) - 
(\op{C}_-^{\Sigma_{\bf U}}({\bf U}_+-{\bf U}_+({\bf b}^+)^{-1}))(z)\\\\
&=&\displaystyle
{\bf m}(z;{\bf U})-(\op{C}^{\Sigma_{\bf U}}_+{\bf m}(\cdot;{\bf U}))(z)+
(\op{C}^{\Sigma_{\bf U}}_-{\bf m}(\cdot;{\bf U}))(z)\\\\
&&\hspace{0.3 in}+\,\,(\op{C}^{\Sigma_{\bf U}}_+
{\bf U}_-)(z)-(\op{C}^{\Sigma_{\bf U}}_- {\bf U}_+)(z)\\\\
&=&\displaystyle
(\op{C}^{\Sigma_{\bf U}}_+
{\bf U}_-)(z)-(\op{C}^{\Sigma_{\bf U}}_- {\bf U}_+)(z)\\\\
&=&{\bf N}_{\bf U}\,.
\end{array}
\end{equation}
Here, in the next-to-last step we have used the Plemelj formula, and
in the final step we have used the continuity of the boundary values
and computed a residue at $z=\infty$, which necessarily occurs within
a component of either $\Omega_{\bf U}^+$ or $\Omega_{\bf U}^-$.  

Finally, a similar argument shows that the composition of these two
correspondences is the identity mapping.  Consider (\ref{eq:repn})
evaluated for ${\bf m}(z;{\bf U})$:  
\begin{equation}
\begin{array}{rcl}
{\bf N}_{\bf U} + (\op{C}^{\Sigma_{\bf U}} \circ
\op{R}_{{\bf w}^+}{\bf m}(\cdot;{\bf U}))(z) + (\op{C}^{\Sigma_{\bf U}} \circ
\op{R}_{{\bf w}^-}{\bf m}(\cdot;{\bf U}))(z)&=&{\bf N}_{\bf U}+
(\op{C}^{\Sigma_{\bf U}} {\bf U}_+)(z) + (\op{C}^{\Sigma_{\bf U}}{\bf U}_-)(z)
\\\\
&=&{\bf U}(z)\,,
\end{array}
\end{equation}
with the last equality following from Cauchy's theorem.  
\end{proof}

\vspace{0.2 in}

So solving the umbilical Riemann-Hilbert Problem~\ref{rhp:umbilical}
amounts to inverting the operator $1-\op{C}_{\bf w}$ on the Banach
space $\Lip^\mu(\Sigma_{\bf U})$, or at least defining the inverse on
the subspace of constant functions ${\bf N}_{\bf U}$.  We note for
future use the following corollary of Lemma~\ref{lemma:equivalence}.
\begin{corollary}
\label{corollary:vanishing}
Suppose ${\bf w}^\pm(z)\in A^\nu_\pm$ as above.  Let ${\bf
m}_0\in\,{\rm ker}\,(1-\op{C}_{\bf w})$, with ${\bf m}_0\not\equiv 0$.
Then, ${\bf U}_0(z):=(\op{C}^{\Sigma_{\bf U}}({\bf m}_0{\bf w}^+))(z)+
(\op{C}^{\Sigma_{\bf U}}({\bf m}_0{\bf w}^-))(z)$ is a nontrivial
solution of the homogeneous umbilical Riemann-Hilbert
Problem~\ref{rhp:umbilical} with data $(\Sigma_{\bf U},{\bf v}_{\bf
U}(z),{\bf 0})$.
\end{corollary}

We continue our analysis by studying the integral equation
(\ref{eq:inteqn}).  Let $\tilde{\bf w}^+(z)={\bf b}^+(z)^{-1}-{\mathbb
I}\in A^\nu_+$ and $\tilde{\bf w}^-(z) = {\mathbb I}-{\bf
b}^-(z)^{-1}\in A^\nu_-$.  Along with the operator $\op{C}_{\bf w}$,
we consider also another bounded operator on $\Lip^\mu(\Sigma_{\bf
U})$ defined by
\begin{equation}
(\op{C}_{\tilde{\bf w}}{\bf m})(z) = 
(\op{C}^{\Sigma_{\bf U}}_+ \circ \op{R}_{{\tilde{\bf w}}^-}{\bf m})(z) +
(\op{C}^{\Sigma_{\bf U}}_- \circ \op{R}_{{\tilde{\bf w}}^+}{\bf m})(z)\,.
\end{equation}
On the space $\Lip^\mu(\Sigma_{\bf U})$, we have
\begin{equation}
(1-\op{C}_{\tilde{\bf w}})\circ(1-\op{C}_{\bf w}) =
1+\op{T}_{\tilde{\bf w},{\bf w}}
\label{eq:f1}
\end{equation}
and
\begin{equation}
(1-\op{C}_{\bf w})\circ(1-\op{C}_{\tilde{\bf w}}) = 1+\op{T}_{{\bf w},\tilde{\bf w}}
\label{eq:f2}
\end{equation}
where 
\begin{equation}
\begin{array}{rcl}
(\op{T}_{\tilde{\bf w},{\bf w}}{\bf m})(z) &:= &(\op{C}^{\Sigma_{\bf
U}}_+\circ \op{R}_{\tilde{\bf w}^-}\circ \op{C}^{\Sigma_{\bf
U}}_-\circ \op{R}_{{\bf w}^+} {\bf m})(z) \\\\ &&\hspace{0.2 in}+\,\,
(\op{C}^{\Sigma_{\bf U}}_+\circ \op{R}_{\tilde{\bf w}^-}\circ
\op{C}^{\Sigma_{\bf U}}_-\circ \op{R}_{{\bf w}^-} {\bf m})(z) \\\\
&&\hspace{0.2 in}+\,\, (\op{C}^{\Sigma_{\bf U}}_-\circ
\op{R}_{\tilde{\bf w}^+}\circ \op{C}^{\Sigma_{\bf U}}_+\circ
\op{R}_{{\bf w}^+} {\bf m})(z) \\\\ &&\hspace{0.2 in}+\,\,
(\op{C}^{\Sigma_{\bf U}}_-\circ \op{R}_{\tilde{\bf w}^+}\circ
\op{C}^{\Sigma_{\bf U}}_+\circ \op{R}_{{\bf w}^-} {\bf m})(z)\\\\
&:=&(\op{T}^{(1)}_{\tilde{\bf w},{\bf w}}{\bf m})(z) +
(\op{T}^{(2)}_{\tilde{\bf w},{\bf w}}{\bf m})(z) +
(\op{T}^{(3)}_{\tilde{\bf w},{\bf w}}{\bf m})(z) +
(\op{T}^{(4)}_{\tilde{\bf w},{\bf w}}{\bf m})(z)\,, 
\end{array}
\end{equation}
and $\op{T}_{{\bf w},\tilde{\bf w}}$ is similarly defined with the
roles of ${\bf w}^\pm$ and $\tilde{\bf w}^\pm$ exchanged.  The
formulae (\ref{eq:f1}) and (\ref{eq:f2}) require only the Plemelj
relation and the fact that ${\bf w}^+(z){\tilde{\bf w}^+}(z) =-({\bf
w}^+(z)+{\tilde{\bf w}^+}(z))$ and ${\bf w}^-(z){\tilde{\bf w}^-}(z)
={\bf w}^-(z)+{\tilde{\bf w}^-}(z)$.  The results from
\S\ref{sec:generalization} can now be used to show the following.
\begin{lemma}
With ${\bf w}^+$ and $\tilde{\bf w}^+$ in $A^\nu_+$ and with ${\bf
w}^-$ and $\tilde{\bf w}^-$ in $A^\nu_-$, the operators
$\op{T}_{\tilde{\bf w},{\bf w}}$ and $\op{T}_{{\bf w},\tilde{\bf w}}$
are compact on $\Lip^\mu(\Sigma_{\bf U})$ whenever $\mu<\nu$.
\label{lemma:regularization}
\end{lemma}

\begin{proof} It is sufficient to prove the result for
$\op{T}_{\tilde{\bf w},{\bf w}}$.  Consider the second term.  Because
$\op{C}_+^{\Sigma_{\bf U}} \circ \op{C}_-^{\Sigma_{\bf U}} = 0$ on
$A^\mu_-$, we can write
\begin{equation}
(\op{T}^{(2)}_{\tilde{\bf w},{\bf w}}{\bf m})(z):=(\op{C}_+^{\Sigma_{\bf U}}\circ \op{R}_{\tilde{\bf w}^-}\circ 
\op{C}_-^{\Sigma_{\bf U}}\circ \op{R}_{{\bf w}^-}{\bf m}
)(z) = -(\op{C}_+^{\Sigma_{\bf U}}\circ
[\op{C}_-^{\Sigma_{\bf U}},\op{R}_{\tilde{\bf w}^-}]\circ \op{R}_{{\bf w}^-}
{\bf m})(z)\,.
\label{eq:Twsecond}
\end{equation}
Similarly, because $\op{C}_-^{\Sigma_{\bf U}} \circ
\op{C}_+^{\Sigma_{\bf U}} = 0$ on $A^\mu_+$, the third
term can be written in the form
\label{eq:Twthird}
\begin{equation}
(\op{T}^{(3)}_{\tilde{\bf w},{\bf w}}{\bf m})(z)
:=(\op{C}_-^{\Sigma_{\bf U}}\circ \op{R}_{\tilde{\bf w}^+}\circ
\op{C}_+^{\Sigma_{\bf U}}\circ \op{R}_{{\bf w}^+}{\bf m} )(z) =
-(\op{C}_-^{\Sigma_{\bf U}}\circ [\op{C}_+^{\Sigma_{\bf
U}},\op{R}_{\tilde{\bf w}^+}]\circ \op{R}_{{\bf w}^+}{\bf m})(z)\,.
\end{equation}
To handle the first term, we first use the Plemelj formula to
decompose: $\tilde{\bf w}^-(z)=\tilde{\bf w}^-_+(z)-\tilde{\bf
w}^-_-(z)$ where $\tilde{\bf w}^-_+(z)=(\op{C}_+^{\Sigma_{\bf U}}
\tilde{\bf w}^-)(z)$ has an analytic continuation into $\Omega_{\bf
U}^+$ and $\tilde{\bf w}^-_-(z)=(\op{C}_-^{\Sigma_{\bf U}} \tilde{\bf
w}^-)(z)$ has an analytic continuation into $\Omega_{\bf U}^-$.  In whichever
region contains $z=\infty$ the corresponding function decays like
$1/z$.  Using this decomposition, we find
\begin{equation}
\begin{array}{rcl}
(\op{T}^{(1)}_{\tilde{\bf w},{\bf w}}{\bf m})(z) &:=&
(\op{C}_+^{\Sigma_{\bf U}}\circ \op{R}_{\tilde{\bf w}^-}\circ 
\op{C}_-^{\Sigma_{\bf U}} \circ \op{R}_{{\bf w}^+}
{\bf m})(z) \\\\&=&
 (\op{C}_+^{\Sigma_{\bf U}}\circ \op{R}_{\tilde{\bf w}^-_+}\circ 
\op{C}_-^{\Sigma_{\bf U}} \circ \op{R}_{{\bf w}^+}
{\bf m})(z) 
-
 (\op{C}_+^{\Sigma_{\bf U}}\circ \op{R}_{\tilde{\bf w}^-_-}\circ 
\op{C}_-^{\Sigma_{\bf U}} \circ \op{R}_{{\bf w}^+}
{\bf m})(z) \\\\
&=&(\op{C}_+^{\Sigma_{\bf U}}\circ \op{R}_{\tilde{\bf w}^-_+}\circ 
\op{C}_-^{\Sigma_{\bf U}} \circ \op{R}_{{\bf w}^+}
{\bf m})(z) \,.
\end{array}
\end{equation}
The term that vanishes above does so because it is of the form
$\op{C}_+^{\Sigma_{\bf U}}$ acting on a product of functions each
having an analytic extension to $\Omega_{\bf U}^-$ and decaying
appropriately if $\infty\in\Omega_{\bf U}^-$.  Now, since by
Lemma~\ref{lemma:wholecontour}, $\tilde{\bf w}^-_+\in A^\mu_+$, we can
again use $\op{C}_+^{\Sigma_{\bf U}} \circ \op{C}_-^{\Sigma_{\bf
U}}=0$ on this Banach algebra to finally write the first term of
$\op{T}_{\tilde{\bf w},{\bf w}}$ in the form
\begin{equation}
(\op{T}^{(1)}_{\tilde{\bf w},{\bf w}}{\bf m})(z) =
-(\op{C}_+^{\Sigma_{\bf U}}\circ 
[\op{C}_-^{\Sigma_{\bf U}},\op{R}_{\tilde{\bf w}^-_+}] \circ \op{R}_{{\bf w}^+}
{\bf m})(z) \,.
\label{eq:Twfirst}
\end{equation}
Similarly writing $\tilde{\bf w}^+(z)=\tilde{\bf w}^+_+(z)-\tilde{\bf
w}^+_-(z)$ with $\tilde{\bf w}^+_\pm(z)=(\op{C}^{\Sigma_{\bf U}}_\pm\tilde{\bf
w}^+)(z)$, and applying similar reasoning, one finds that the fourth
term in $\op{T}_{\tilde{\bf w},{\bf w}}$ can be written as
\begin{equation}
(\op{T}^{(4)}_{\tilde{\bf w},{\bf w}}{\bf m})(z) =
+(\op{C}_-^{\Sigma_{\bf U}}\circ 
[\op{C}_+^{\Sigma_{\bf U}},\op{R}_{\tilde{\bf w}^+_-}]\circ
\op{R}_{{\bf w}^-}{\bf m})(z)\,.
\label{eq:Twfourth}
\end{equation}

With each term of $\op{T}_{\tilde{\bf w},{\bf w}}$ written in this
way, it is not hard to see the compactness from more basic results
already summarized.  Consider $\op{T}^{(1)}_{\tilde{\bf
w},{\bf w}}$ written in terms of the commutator as (\ref{eq:Twfirst}).
The multiplication operator $\op{R}_{{\bf w}^+}$ is bounded from
$\Lip^\mu(\Sigma_{\bf U})$ to $A^\mu_+$.  Then, from
Lemma~\ref{lemma:commutatorgeneral}, the commutator is a bounded
operator from $A^\mu_+$ to the better space $\Lip^\nu(\Sigma_{\bf
U})$.  The result of this operation can be reinterpreted as an element
of $\Lip^\mu(\Sigma_{\bf U})$, and from
Lemma~\ref{lemma:inclusiongeneral}, the inclusion map
$I_{\nu\rightarrow\mu}:\Lip^\nu(\Sigma_{\bf U})\rightarrow
\Lip^\mu(\Sigma_{\bf U})$ is a compact operator.  The trivial
inclusion map $\Lip^\mu(\Sigma_{\bf U})\rightarrow A^\mu_+$ is of
course bounded, and finally composition with the operator
$\op{C}_-^{\Sigma_{\bf U}}$, bounded from $A^\mu_+$ to
$\Lip^\mu(\Sigma_{\bf U})$ by Lemma~\ref{lemma:wholecontour} does not
alter the compactness.  The term $\op{T}^{(4)}_{\tilde{\bf w},{\bf
w}}$ expressed by (\ref{eq:Twfourth}) is handled similarly.  The
term $\op{T}^{(2)}_{\tilde{\bf w},{\bf w}}$ given in
(\ref{eq:Twsecond}) is shown to be compact as follows.  The
multiplication operator $\op{R}_{{\bf w}^-}$ is a bounded map from
$\Lip^\mu(\Sigma_{\bf U})$ to $A^\mu_-$.  By
Lemma~\ref{lemma:commutatorgeneral}, the commutator is then a bounded
map from $A^\mu_-$ to the better space $A^\nu_-$.  Again, the
inclusion map $I_{\nu\rightarrow\mu}:A^\nu_-\rightarrow A^\mu_-$ is
compact by Lemma~\ref{lemma:inclusiongeneral}.  Finally, by
Lemma~\ref{lemma:wholecontour} the operator $\op{C}_+^{\Sigma_{\bf
U}}$ is bounded from $A^\mu_-$ to $\Lip^\mu(\Sigma_{\bf U})$, and
compactness is retained.  The term $\op{T}^{(3)}_{\tilde{\bf w},{\bf
w}}$ written in the form (\ref{eq:Twthird}) is handled similarly.  All
four terms of $\op{T}_{\tilde{\bf w},{\bf w}}$ have thus been shown to be
compact on $\Lip^\mu(\Sigma_{\bf U})$.  
\end{proof}

\vspace{0.2 in}

We have the following result.
\begin{lemma}
The bounded operator $1-\op{C}_{\bf w}$, with the matrices ${\bf
w}^\pm(z)\in A^\nu_\pm$ as above, is Fredholm on the Banach space
$\Lip^\mu(\Sigma_{\bf U})$ for all $\mu$ with $0<\mu<\nu$, and has
index zero.
\index{Cauchy integral operators!Fredholm}
\label{lemma:Fredholm}
\end{lemma}

\begin{proof} By Lemma~\ref{lemma:regularization} the operator
$1-\op{C}_{\tilde{\bf w}}$ serves as both a left and a right
pseudoinverse for $1-\op{C}_{\bf w}$, and therefore $\dim {\rm
ker}\,(1-\op{C}_{\bf w})$ and $\dim {\rm coker}\,(1-\op{C}_{\bf w})$
are both finite.  This proves that $1-\op{C}_{\bf w}$ is a Fredholm
operator on $\Lip^\mu(\Sigma_{\bf U})$.  To calculate the index, we
invoke continuity of the index for Fredholm operators with respect to
uniform operator norm in $\Lip^\mu(\Sigma_{\bf U})$.  The same
arguments as above prove that the family of operators $1-\epsilon
\op{C}_{\bf w}$ is Fredholm for all $\epsilon\in{\mathbb C}$, and in
particular for those real $\epsilon$ between $0$ and $1$.  By the
boundedness of $\op{C}_{\bf w}$ on $\Lip^\mu(\Sigma_{\bf U})$, this
family of operators is continuous in operator norm as a function of
$\epsilon\in [0,1]$.  Since for $\epsilon=0$ the ${\rm ind}\,(1-\epsilon
\op{C}_{\bf w})=0$, and since the index is a continuous
integer-valued function over the whole range of $\epsilon\in [0,1]$,
we conclude that ${\rm ind}\,(1-\op{C}_{\bf w})=0$.  
\end{proof}

\begin{corollary}
\label{corollary:kernel}
With ${\bf w}^\pm(z)\in A^\nu_\pm$ as above, $1-\op{C}_{\bf
w}$ has a bounded inverse defined on $\Lip^\mu(\Sigma_{\bf U})$ for
all positive $\mu<\nu$ whenever ${\rm ker}\,(1-\op{C}_{\bf w})=\{0\}$.
\end{corollary}

\begin{proof} Since ${\rm ind}\,(1-\op{C}_{\bf w})=0$, ${\rm
ker}\,(1-\op{C}_{\bf w})=\{0\}$ implies $\dim{\rm
coker}\,(1-\op{C}_{\bf w})=0$, and then $1-\op{C}_{\bf w}$ being
bounded implies via the closed graph theorem that ${\rm
Ran}\,(1-\op{C}_{\bf w})= \Lip^\mu(\Sigma_{\bf U})$.  Therefore the
inverse $(1-\op{C}_{\bf w})^{-1}$ exists and is defined on the whole
space $\Lip^\mu(\Sigma_{\bf U})$.  Since $1-\op{C}_{\bf w}$ is bounded
and hence closed, the inverse is also closed and therefore bounded by
the closed graph theorem.  
\end{proof}

\vspace{0.2 in}

Combining Lemma~\ref{lemma:equivalence} with Corollary~\ref{corollary:kernel}
and Corollary~\ref{corollary:vanishing} gives the main result of this section.
\begin{theorem}[Umbilical Fredholm Alternative]
\label{theorem:UmbilicalAlternative}
Let $\Sigma_{\bf U}$ be an umbilical contour and let ${\bf v}_{\bf
U}(z)$ be a jump matrix for $z\in\Sigma_{\bf U}\setminus\{\alpha_i\}$
satisfying the conditions of Definition~\ref{def:JU}.  Let ${\bf
N}_{\bf U}$ be any constant matrix.  Then the umbilical
Riemann-Hilbert Problem~\ref{rhp:umbilical} with data
$(\Sigma_{\bf U},{\bf v}_{\bf U}(z),{\bf N}_{\bf U})$ has a unique
solution if and only if the corresponding homogeneous problem with
data $(\Sigma_{\bf U},{\bf v}_{\bf U}(z),{\bf 0})$ has only the
trivial solution.
\end{theorem}

\section[Applications to Local Riemann-Hilbert Problems]{Application to local Riemann-Hilbert problems.}
Here, we put together the pieces to establish the proofs of Theorems
\ref{theorem:LocalAlternative}, \ref{theorem:LocalSolvability}, and
\ref{theorem:decay}.

{\it Proof of Theorem~\ref{theorem:LocalAlternative}}.  By
Lemma~\ref{lemma:LocalToUmbilical}, the local Riemann-Hilbert
Problem~\ref{rhp:local} will have a unique solution if and only if the
corresponding umbilical Riemann-Hilbert Problem~\ref{rhp:umbilical}
does.  By Theorem~\ref{theorem:UmbilicalAlternative}, the latter will
be the case if the only solution of the umbilical Riemann-Hilbert
Problem~\ref{rhp:umbilical} that vanishes at $z=\infty$ is the trivial
solution.  If ${\bf U}_0(z)$ is such a nontrivial ``vanishing''
solution, then 
\begin{equation}
{\bf L}_0(\zeta):= (\zeta-e^{i\theta})^{-1}{\bf
U}_0(z(\zeta))
\end{equation}
will be a nontrivial solution of the homogeneous local Riemann-Hilbert
Problem~\ref{rhp:local}, where $z(\zeta)$ is the transformation in
Lemma~\ref{lemma:LocalToUmbilical}.  Conversely, if ${\bf L}_0(\zeta)$
is a nontrivial solution of the homogeneous local Riemann-Hilbert
Problem~\ref{rhp:local}, then with $\zeta(z)$ as in
Lemma~\ref{lemma:LocalToUmbilical}, 
\begin{equation}
{\bf U}_0(z):= (z-1)^{-1}{\bf
L}_0(\zeta(z))
\end{equation}
will be a nontrivial solution of the homogeneous umbilical
Riemann-Hilbert Problem~\ref{rhp:umbilical}.  Therefore, the unique
solvability of the umbilical problem, and therefore of the local
problem by Lemma~\ref{lemma:LocalToUmbilical} is guaranteed by the
condition $\{{\bf L}_0(\zeta)\}=\{\bf 0\}$, which proves the theorem.
$\Box$

\vspace{0.2 in}

{\it Proof of Theorem~\ref{theorem:LocalSolvability}}.  The fact that
under the additional symmetry condition (\ref{eq:J5L}) on the contour
$\Sigma_{\bf L}$ and jump matrix ${\bf v}_{\bf L}(\zeta)$, all
solutions of the homogeneous local Riemann-Hilbert Problem~\ref{rhp:local}
with data $(\Sigma_{\bf L},{\bf v}_{\bf L}(\zeta),{\bf 0})$
are trivial follows from an argument given by Zhou in \cite{Z89}.
This result is combined with Theorem~\ref{theorem:LocalAlternative} to
complete the proof.  Note that Zhou's argument relies on Cauchy's
theorem, and in the $L^2$ context of \cite{Z89} a rational
approximation argument is required to pull the paths of integration
away from the boundary.  But in the H\"older spaces, no such argument
is needed, since the boundary values are automatically uniformly
continuous. $\Box$

\vspace{0.2 in}

{\it Proof of Theorem~\ref{theorem:decay}}.  We begin with the
representation of the solution ${\bf L}(\zeta)$ in terms of the
fractional linear transformation $\zeta(z)$ and the solution ${\bf
m}(z)$ of the integral equation on the corresponding umbilical contour
$\Sigma_{\bf U}$ by means of the formula (\ref{eq:repn}).  Since
$\zeta=\infty$ corresponds to $z=1$, we have
\begin{equation}
{\bf L}(\zeta)-{\mathbb I} = {\bf U}(z(\zeta))-{\bf U}(1)
=\frac{i}{\pi\zeta e^{-i\theta} + \pi}\left(I_1^+(\zeta)+I_1^-(\zeta)+
I_2^+(\zeta)+I_2^-(\zeta)\right)\,,
\label{eq:L-I}
\end{equation}
where
\begin{equation}
I_1^\pm(\zeta):=\int_{\Sigma_{\bf U}}{\bf w}^\pm(s)\frac{ds}{(s-z(\zeta))(s-1)}\,,\hspace{0.2 in}
I_2^\pm(\zeta):=\int_{\Sigma_{\bf U}}({\bf m}(s)-{\mathbb I}){\bf w}^\pm(s)
\frac{ds}{(s-z(\zeta))(s-1)}\,.
\end{equation}
Note that we have both 
\begin{equation}
\|{\bf w}^\pm(z)\|=\bo(|z-1|) \mbox{ and } \|{\bf
m}(z)-{\mathbb I}\|=\bo(|z-1|^\mu) \mbox{ for all } \mu<1\,, 
\label{eq:znear1}
\end{equation}
with the latter estimate following from the H\"older theory of the
integral equation (\ref{eq:inteqn}).

First we will show that the $\bo(|\zeta|^{-1})$ decay estimate holds
subject to the condition (\ref{eq:cancellation}) uniformly for $\zeta$
outside of all of the parallel strips $S_k$ surrounding each ray.  By
(\ref{eq:L-I}), it suffices to show that the four integrals
$I_1^\pm(\zeta)$ and $I_2^\pm(\zeta)$ remain bounded.  Now, changing
variables in the integrals by $\xi=\zeta(s)$, we find
\begin{equation}
I_2^\pm(\zeta)= -\frac{1}{2}\int_{\Sigma_{\bf L}}({\bf m}(z(\xi))-{\mathbb I}){\bf w}^\pm(z(\xi))\frac{\zeta e^{-i\theta}+1}{\xi-\zeta}\,d\xi\,.
\end{equation}
Now, observe that
\begin{equation}
-\frac{1}{2}\frac{\zeta e^{-i\theta}+1}{\xi-\zeta} = \frac{e^{-i\theta}}{2}
-\frac{1}{2}\frac{1}{\xi-\zeta}-\frac{e^{-i\theta}}{2}\xi\frac{1}{\xi-\zeta}\,.
\end{equation}
Since the estimates (\ref{eq:znear1}) imply
\begin{equation}
({\bf m}(z(\xi))-{\mathbb I}){\bf w}^\pm(z(\xi))\in L^1(\Sigma_{\bf
L})\cap L^2(\Sigma_{\bf L})\,,
\end{equation}
and
\begin{equation}
({\bf m}(z(\xi))-{\mathbb I}){\bf w}^\pm(z(\xi))\xi\in L^2(\Sigma_{\bf L})\,,
\end{equation}
and since uniformly for all $\zeta$ outside the union of the strips $S_k$ we
have
\begin{equation}
\frac{1}{\xi-\zeta}\in L^2(\Sigma_{\bf L})\,,
\end{equation}
we can apply the Cauchy-Schwarz inequality to find
\begin{equation}
\|I_2^\pm(\zeta)\|=\bo(1)
\end{equation}
uniformly for $\zeta$ outside all strips $S_k$.  

The terms $I_1^\pm(\zeta)$ will, by contrast, only be bounded if the
condition (\ref{eq:cancellation}) is satisfied.  To see this, make
sure that in solving the integral equation (\ref{eq:inteqn}) the
factorization of the jump matrix ${\bf v}_{\bf U}(z)$ is such that the
limits
\begin{equation}
{\bf w}^{\pm(\infty,k)}:=\lim_{\zeta\rightarrow\infty\atop\zeta\in S_k}
\zeta {\bf w}^\pm(z(\zeta))
\label{eq:newlimits}
\end{equation}
exist uniformly in $S_k$.  Then, changing variables from $s$ to $\xi=\zeta(s)$
in $I_1^\pm(\zeta)$ as before, we find
\begin{equation}
I_1^\pm(\zeta)=-\frac{1}{2}\int_{\Sigma_{\bf L}}{\bf w}^\pm(z(\xi))
\frac{d\xi}{\xi-\zeta} - \frac{e^{-i\theta}}{2}\int_{\Sigma_{\bf L}}
{\bf w}^\pm(z(\xi))\frac{\zeta\,d\xi}{\xi-\zeta}\,.
\end{equation}
Because from (\ref{eq:znear1}), ${\bf w}^\pm(z(\xi))\in
L^2(\Sigma_{\bf L})$, the first term is uniformly bounded for $\zeta$
outside all strips $S_k$ by the Cauchy-Schwarz inequality.  To handle
the second term, we first split off the part $\Sigma_{\bf L}^{\rm in}$
of the contour that is bounded by $|\xi|\le R$ ({\em i.e.} the part
interior to and including the circle component of $\Sigma_{\bf L}$)
and let the component of $\Sigma_{\bf L}\setminus\Sigma_{\bf L}^{\rm
in}$ lying on the $k$th ray be parametrized by $re^{i\phi_k}$ for
$r>R$.  Then,
\begin{equation}
\begin{array}{rcl}
\displaystyle
\int_{\Sigma_{\bf L}}{\bf w}^\pm(z(\xi))\frac{\zeta\,d\xi}{\xi-\zeta} &=&
\displaystyle
\int_{\Sigma_{\bf L}^{\rm in}}{\bf w}^\pm(z(\xi))\frac{\zeta\,d\xi}{\xi-\zeta}
+\sum_{\mbox{\rm \scriptsize all rays }k}{\bf w}^{\pm(\infty,k)}
\int_R^\infty\frac{\zeta\,dr}{r(re^{i\phi_k}-\zeta)} 
\\\\&&\displaystyle \,\,+\,\,
 \sum_{\mbox{\rm \scriptsize all rays }k}
\int_R^\infty({\bf w}^{\pm}(z(re^{i\phi_k}))-
r^{-1}e^{-i\phi_k}{\bf w}^{\pm(\infty,k)})
\frac{\zeta e^{i\phi_k}\,dr}{re^{i\phi_k}-\zeta}\,.
\end{array}
\end{equation}
Now the first term on the right-hand side is clearly bounded for large
$\zeta$ because the contour $\Sigma_{\bf L}^{\rm out}$ is compact.
In the third term the difference in parenthesis is $o(r^{-1})$, and it
follows that similar Cauchy-Schwarz estimates bound this sum independently
of $\zeta$ outside the strips $S_k$.  Now,
\begin{equation}
\int_R^\infty \frac{\zeta\,dr}{r(re^{i\phi_k}-\zeta)}=
\log\left(\frac{R}{R-\zeta e^{-i\phi_k}}\right)=-\log|\zeta| + \bo(1)\,,
\end{equation}
uniformly as $\zeta\rightarrow\infty$ outside all strips $S_k$.  Therefore,
we will have
\begin{equation}
\|I_1^\pm(\zeta)\|=\bo(1)\,,
\end{equation}
uniformly outside all strips if and only if
\begin{equation}
\sum_{\mbox{\rm\scriptsize all rays }k}{\bf w}^{\pm(\infty,k)} = {\bf 0}\,.
\end{equation}
Now, if the factorization of the jump matrix ${\bf v}_{\bf U}(z)={\bf
v}_{\bf L}(\zeta(z))$ in some small neighborhood of $z=1$ is such that
${\bf w}^-(z)\equiv 0$ there, then this condition is equivalent to
(\ref{eq:cancellation}) since ${\bf w}^+(z)\equiv {\bf v}_{\bf
L}(\zeta(z))-{\mathbb I}$ in this special case.  But in fact it can be
seen by expanding the arbitrary factorization ${\bf v}_{\bf
L}(\zeta(z))= ({\mathbb I}-{\bf w}^-(z))^{-1}({\mathbb I}+{\bf
w}^+(z))$ near $z=1$ that no matter how the factorization is done as
long as the limits (\ref{eq:newlimits}) exist,
\begin{equation}
{\bf v}^{(\infty,k)}={\bf w}^{+(\infty,k)}+{\bf w}^{-(\infty,k)}\,.
\end{equation}
Therefore, if the condition (\ref{eq:cancellation}) is satisfied, the
{\em sum} $I_1^+(\zeta)+I_1^-(\zeta)$ is always uniformly bounded
independently of the factorization employed in solving the integral
equation (\ref{eq:inteqn}), even though the individual terms
$I_1^+(\zeta)$ and $I_1^-(\zeta)$ may be logarithmic in $\zeta$.

This proves that the estimate stated in the theorem holds for $\zeta$
bounded away from the contour rays by avoiding the parallel strips.
But now we can use the analyticity of the jump matrix ${\bf v}_{\bf
L}(\zeta)$ along with the uniformity of the limits
(\ref{eq:momentdefine}) defining the moments ${\bf v}^{(\infty,k)}$ in
a simple deformation argument to prove that the estimate in fact holds
right up to the (analytic) boundary values taken on $\Sigma_{\bf L}$,
which completes the proof of the theorem.
$\Box$

\chapter{Near-Identity Riemann-Hilbert Problems in $L^2$}
\label{sec:A2}
In this appendix, we collect together those results from the theory of
Riemann-Hilbert problems and Cauchy integral operators in $L^2$ that
we use in \S\ref{sec:error} to characterize and estimate the error of
our approximations.

Let $\Sigma$ be a compact contour in the complex $\lambda$-plane,
consisting of a union of a finite number of smooth arcs, and that is
oriented so that it forms the positively-oriented boundary of a
multiply-connected open region $\Omega^+$ whose complement is
the disjoint union  $\Sigma\cup\Omega^-$.  Fix
any matrix norm $\|\cdot\|$, and let $L^2(\Sigma)$ denote the space of
matrix-valued functions ${\bf f}(\lambda)$ defined almost everywhere on
$\Sigma$ such that the norm defined by
\begin{equation}
\|{\bf f}\|_{L^2(\Sigma)}^2:=\int_\Sigma \|{\bf f}(\lambda)\|^2\,|d\lambda|
\end{equation}
is finite.  Now we recall some facts \cite{D99} about Cauchy integrals
in this space.  For ${\bf f}\in L^2(\Sigma)$, the Cauchy integral
\begin{equation}
(\op{C}^\Sigma{\bf f})(\lambda):=\frac{1}{2\pi i}\int_\Sigma
(s-\lambda)^{-1}{\bf f}(s)\,ds
\end{equation}
defines a piecewise-analytic function for $\lambda\not\in\Sigma$.  
The left and right boundary values
\begin{equation}
(\op{C}^\Sigma_\pm{\bf f})(\lambda):=\lim_{z\rightarrow\lambda,z\in\Omega^\pm}
(\op{C}^\Sigma{\bf f})(z)
\end{equation}
exist for almost all $\lambda\in\Sigma$ and may be identified with
unique elements of $L^2(\Sigma)$.  The linear operators
$\op{C}^\Sigma_\pm$ thus defined on $L^2(\Sigma)$ are bounded, with
norms depending on the contour $\Sigma$.  

Let ${\bf w}^\pm(\lambda)$
be defined for almost all $\lambda\in\Sigma$ and uniformly bounded.
Define an operator on $L^2(\Sigma)$ by
\begin{equation}
(\op{C}_{\bf w}{\bf m})(\lambda):=(\op{C}^\Sigma_+({\bf m}{\bf w}^-))(\lambda)+
(\op{C}^\Sigma_-({\bf m}{\bf w}^+))(\lambda)\,.
\end{equation}
Then $\op{C}_{\bf w}$ is bounded according to the simple estimate
\begin{equation}
\|\op{C}_{\bf w}\|_{L^2(\Sigma)}\le 
\|\op{C}^\Sigma_+\|_{L^2(\Sigma)}\sup_{s\in\Sigma}\|{\bf w}^-(s)\| +
\|\op{C}^\Sigma_-\|_{L^2(\Sigma)}\sup_{s\in\Sigma}\|{\bf w}^+(s)\|\,.
\end{equation}
Therefore, we have
\begin{lemma}
\label{lemma:L2intsolve}
Assume that
\begin{equation}
\sup_{s\in\Sigma}\|{\bf w}^-(s)\|< \frac{1}{2}\left(
\|\op{C}^\Sigma_+\|_{L^2(\Sigma)}\right)^{-1}\,,
\hspace{0.3 in} 
\sup_{s\in\Sigma}\|{\bf w}^+(s)\|<\frac{1}{2}
\left(\|\op{C}^\Sigma_-\|_{L^2(\Sigma)}\right)^{-1}\,.
\end{equation}
Then, for any ${\bf f}\in L^2(\Sigma)$, the equation
\begin{equation}
{\bf m}(\lambda)-(\op{C}_{\bf w}{\bf m})(\lambda)={\bf f}(\lambda)
\label{eq:L2inteqn}
\end{equation}
has a unique solution ${\bf m}\in L^2(\Sigma)$, and 
\begin{equation}
\|{\bf m}\|_{L^2(\Sigma)}\le 
\frac{\|{\bf f}\|_{L^2(\Sigma)}}
{1-(\|\op{C}^\Sigma_+\|_{L^2(\Sigma)}\sup_{s\in\Sigma}\|{\bf w}^-(s)\| +
\|\op{C}^\Sigma_-\|_{L^2(\Sigma)}\sup_{s\in\Sigma}\|{\bf w}^+(s)\|)}
\,.
\end{equation}
\label{lemma:L2Invertibility}
\end{lemma}

\begin{proof} This follows from standard functional analysis results concerning
the invertibility of bounded perturbations of the identity operator.
The solution of the equation is furnished by Neumann
series\index{Neumann series}.
\end{proof}

\begin{lemma}
Along with the conditions of Lemma~\ref{lemma:L2intsolve}, assume that
${\mathbb I}-{\bf w}^-(\lambda)$ is invertible for $\lambda\in\Sigma$
and set ${\bf v}(\lambda):= ({\mathbb I}-{\bf
w}^-(\lambda))^{-1}({\mathbb I}+{\bf w}^+(\lambda))$.  Let ${\bf
m}(\lambda)$ be the unique solution of (\ref{eq:L2inteqn}) with ${\bf
f}(\lambda)\equiv {\mathbb I}$.  Define, for $\lambda\not\in\Sigma$,
\begin{equation}
{\bf R}(\lambda):={\mathbb I}+(\op{C}^\Sigma({\bf m}({\bf w}^++{\bf w}^-)))(\lambda)\,.
\label{eq:Edef}
\end{equation}
Then ${\bf R}(\lambda)$ has boundary values ${\bf R}_\pm(\lambda)$ in
$L^2(\Sigma)$ that satisfy almost everywhere
\begin{equation}
{\bf R}_+(\lambda)={\bf R}_-(\lambda){\bf v}(\lambda)\,.
\label{eq:Ejumps}
\end{equation}
Also, ${\bf R}(\infty)={\mathbb I}$, and
\begin{equation}
\|{\bf R}(\lambda)-{\mathbb I}\|\le 
\frac{1}{2\pi}\|{\bf m}\|_{L^2(\Sigma)}^2 \cdot |\Sigma|^2\cdot
\sup_{s\in\Sigma}|s-\lambda|^{-1}\cdot
\sup_{s\in\Sigma}\|{\bf w}^+(s)+{\bf w}^-(s)\|\,,
\label{eq:L2estimate}
\end{equation}
where $|\Sigma|$ is the total arc length of $\Sigma$.
\label{lemma:L2RHP}
\end{lemma}

\begin{proof}
It follows directly from the integral equation satisfied by ${\bf
m}(\lambda)$ and the Plemelj formula that everywhere the boundary
values taken on $\Sigma$ by ${\bf R}(\lambda)$ are finite, they
satisfy the jump relation (\ref{eq:Ejumps}).  The estimate
(\ref{eq:L2estimate}) also follows directly from the representation
(\ref{eq:Edef}).
\end{proof}

We say that ${\bf R}(\lambda)$ solves the Riemann-Hilbert problem with 
data $(\Sigma,{\bf v}(\lambda),{\mathbb I})$ in the $L^2(\Sigma)$ sense.
\index{Riemann-Hilbert problem!in the $L^2$ sense}
In \cite{D99}, the following uniqueness result is proved:
\begin{lemma}
Suppose all matrices are $2\times 2$, and that ${\bf w}^\pm(\lambda)$
are smooth functions on each arc of $\Sigma$ for which $\det {\bf
v}(\lambda)=1$.  Then the solution ${\bf R}(\lambda)$ of the
Riemann-Hilbert problem with data $(\Sigma,{\bf v}(\lambda),{\mathbb
I})$, posed in the $L^2(\Sigma)$ sense, is unique if it exists.
\label{lemma:L2Uniqueness}
\end{lemma}

\printindex
\end{document}